%% file: main.tex
\documentclass{article}

\usepackage[citecolor=blue, hidelinks]{hyperref}

\usepackage{geometry}
\usepackage{layout}
\usepackage{enumitem}
\usepackage[utf8]{inputenc} %
\usepackage[T1]{fontenc}    %
\usepackage{hyperref}       %
\usepackage{url}            %
\usepackage{booktabs}       %
\usepackage{amsfonts}       %
\usepackage{nicefrac}       %
\usepackage{microtype}      %
\usepackage{xcolor}         %
\usepackage{amsmath}

\title{A permutation-free kernel two-sample test}
\date{}

\usepackage{authblk}
\author[1]{Shubhanshu Shekhar \thanks{shubhan@umich.edu}}
\author[3]{Ilmun Kim \thanks{ilmun@yonsei.ac.kr}} 
\author[1, 2]{Aaditya Ramdas \thanks{aramdas@stat.cmu.edu}} 
\affil[1]{Department of Statistics and Data Science, Carnegie Mellon University }
\affil[2]{Machine Learning Department, Carnegie Mellon University }
\affil[3]{Department of Statistics and Data Science, Yonsei University }

\usepackage{comment}

\RequirePackage{etex} 
\input{preamble}

\input{newMacros}

\newtheorem{manualtheoreminner}{Theorem}
\newenvironment{manualtheorem}[1]{%
  \renewcommand\themanualtheoreminner{#1}%
  \manualtheoreminner
}{\endmanualtheoreminner}

\usepackage{tikz}

\usepackage{pgfplots}
\pgfplotsset{compat=newest}
\pgfplotsset{scaled y ticks=false}
\usepgfplotslibrary{groupplots}
\usepgfplotslibrary{dateplot}
\tikzstyle{every node}=[font=\small]
\pgfplotsset{
    yticklabel style={/pgf/number format/fixed},  
}

\pgfplotsset{compat=1.11,
 /pgfplots/ybar legend/.style={
 /pgfplots/legend image code/.code={
 \draw[##1,/tikz/.cd,yshift=-0.25em]
 (0cm,0cm) rectangle (3pt,0.8em);},
 },
}

\begin{document}

\maketitle
\begin{abstract}
    The kernel Maximum Mean Discrepancy~(MMD) is a popular multivariate distance metric between distributions that has found utility in two-sample testing. The usual kernel-MMD test statistic is a degenerate U-statistic under the null, and thus it has an intractable limiting distribution. Hence, to design a level-$\alpha$ test, one usually selects the rejection threshold as the $(1-\alpha)$-quantile of the permutation distribution. The resulting nonparametric test has finite-sample validity but suffers from large computational cost, since every permutation takes quadratic time.  We propose the cross-MMD, a new quadratic-time MMD test statistic based on sample-splitting and studentization. We prove that under mild assumptions, the cross-MMD has a limiting standard Gaussian distribution under the null. Importantly, we also show that the resulting test is consistent against any fixed alternative, and when using the Gaussian kernel, it has minimax rate-optimal power against local alternatives.  For large sample sizes, our new cross-MMD  provides a significant speedup over the MMD, for only a slight loss in power.
\end{abstract}

\tableofcontents 

\section{Introduction}
\label{sec:introduction}
    We study the two-sample testing problem: 
    given $\Xsample = (X_1, \ldots, X_n) \overset{\text{i.i.d.}}{\sim} P$ and $\Ysample = (Y_1, \ldots, Y_m) \overset{\text{i.i.d.}}{\sim} Q$, we test the null hypothesis $H_0:P=Q$ against the alternative $H_1: P \neq Q$.   This is a nonparametric hypothesis problem with a composite null hypothesis and a composite alternative hypothesis. It finds applications in diverse areas such as testing microarray data, clinical diagnosis and database attribute matching~\citep{gretton2012kernel}. 
    
    A popular approach to solving this problem is based on the \emph{kernel-MMD} distance between the two empirical distributions~\citep{gretton2006kernel}. Given a positive definite kernel $k$, the kernel-MMD distance between two  distributions $P$ and $Q$ on $\mc{X}$, denoted by $\dmmd(P,Q)$,  is defined as
    \begin{align}
        \dmmd(P, Q) = \|\mu - \nu\|_k, \
        \text{ where } \ \mu(\cdot) = \int_{\mc{X}} k(x, \cdot)dP(x),  \text{ and }  \nu(\cdot) = \int_{\mc{X}} k(x, \cdot)dQ(x).  \label{eq:kernel-mmd-stat} 
    \end{align}
    Above, $\mu$ and $\nu$ are commonly called ``kernel mean maps'', and denote the kernel mean embeddings of the distributions $P$ and $Q$ into the reproducing kernel Hilbert space~(RKHS) associated with the positive-definite kernel $k$, and $\|\cdot\|_{k}$ denotes the corresponding RKHS norm. Under mild conditions on the positive definite kernel $k$~\citep{sriperumbudur2011universality}, $\dmmd$ is a metric on the space of probability distributions. \citet{gretton2006kernel} suggested using an empirical estimate of the squared distance as the test statistic. In particular, given $\Xsample$ and $\Ysample$, define the test statistic
    \begin{align}
        &\mmdhat^2 := \frac{1}{n(n-1)m(m-1)} \sum_{1 \leq i \neq i' \leq n} \sum_{1 \leq j\neq j' \leq m} h(X_i, X_{i'}, Y_j, Y_{j'}), \label{eq:kernel-mmd-0}%
    \end{align}
    where $h(x, x', y, y') \defined k(x, x') - k(x, y') - k(y, x') + k(y, y')$. The above statistic has an alternative form that only takes quadratic time to calculate.
    
    The MMD test rejects the null if $\mmdhat^2$ exceeds a suitable threshold $\tau\equiv\tau(\alpha)$ that ensures the false positive rate is at most $\alpha$.  For  ``characteristic kernels'', this test is  \emph{consistent} against fixed alternatives, meaning the power (the probability of rejecting the null when $P \neq Q$) increases to one as $m,n\to\infty$. 
    
    The difficulty in practically determining $\tau$ will play a key role in this paper.
    It is well known that when $P=Q$, $\mmdhat^2$ is an instance of a ``degenerate two-sample U-statistic'', meaning that:  
     \begin{align}
        \text{Under $H_0$,} \quad \mathbb{E}_P[h(x,X',y,Y')] =\mathbb{E}_Q[h(X,x',Y,y')] = 0.
    \end{align}
    (Above, $x,y,x',y'$ are fixed, and the expectations are over $X,Y,X',Y'\overset{\text{i.i.d.}}{\sim} P$.)
    As a consequence, its (limiting) null distribution is unwieldy; it is an infinite sum of independent $\chi^2$ random variables weighted by the eigenvalues of an operator that depends on the kernel $k$ and the underlying distribution $P$ (see equation~\eqref{eq:null-dist-mmd2} in~\Cref{appendix:background}). Since $P$ is unknown, one cannot explicitly calculate $\tau$. 
    
    In practice, a  permutation-based approach is commonly used, where $\tau$ is set as the $(1-\alpha)$-quantile of the kernel-MMD statistic computed on $B$ permuted versions of the aggregate data $(\Xsample, \Ysample)$. The resulting test has finite-sample validity, but its practical applicability is reduced due to the high computational complexity; if $B=200$ permutations are used, the (permuted) test statistic must be recomputed 201 times, rather than once (usually, $B$ is chosen between 100 and 1000). 
    
    Due to the high computational complexity of the permutation test, some permutation-free alternatives for selecting $\tau$ have been proposed. However, as we discuss in~\Cref{subsec:related-work}, these alternatives are either too conservative in practice (using concentration inequalities), or heuristics with no theoretical guarantees (Pearson curves and Gamma approximation) or are only shown to be consistent in the setting where the kernel $k$ does not vary with $n$~(spectral approximation). We later recap some computationally efficient alternatives to $\mmdhat^2$, but these have significantly lower power.
    
    As far as we are aware, there exists no method in literature based on the kernel-MMD that is  (i) permutation-free (does not require permutations), {(ii)} consistent against any fixed alternative, (iii) achieves minimax rate-optimality against local alternatives, and (iv) is correct for both the fixed kernel setting ($k$ is fixed as $m,n \to \infty$) and the changing kernel setting ($k$ changes as a function of $m,n$, for instance, by selecting the scale parameter of a Gaussian kernel in a sample-size dependent manner). 
    
     Our work delivers a novel and simple test satisfying all four desirable properties. We propose a new variant of the kernel-MMD statistic %
     that~(after studentization) has a standard Gaussian limiting distribution under the null in both the fixed and changing kernel settings, in low- and high-dimensional settings. There is a computation-statistics tradeoff: our permutation-free test loses about a $\sqrt{2}$ factor in power compared to the standard kernel-MMD test, but it is hundreds of times faster.
     \begin{remark}
        \label{remark:two-sample-test-definition}
        Let $\mc{P}(\mc{X})$ denote the set of all probability measures on the observation space $\mc{X}$, where we often use $\mc{X} = \mathbb{R}^d$ for some $d \geq 1$. 
        For simplicity, in the above presentation, the distributions $P,Q$, kernel $k$ and dimension $d$ did not change with sample size, and this is the setting considered in the majority of the literature. Later, we prove several of our results in a significantly more general setting  where $P,Q,d,k$ can vary with $n,m$. Under the null, this provides a much more robust type-I error control in high-dimensional settings, even with sample-size dependent kernels. Under the alternative, this provides a more fine-grained power result. To elaborate on the latter, we assume that for every $n, m$, the pair $(P,Q)=(\Pnm,\Qnm) \in \altclass \subset \mc{P}(\calX)\times \mc{P}(\calX)$ for some sequence $\{\altclass: n,m\geq 2\}$. The class $\altclass$ is such that with increasing $n$ and $m$, it contains pairs $(P', Q')$ that are increasingly closer in some distance measure $\varrho$; thus the alternatives can approach the null and be equal in the limit. That is, $\Delta_{n,m} \defined \inf_{(P',Q')\in \altclass} \varrho(P',Q')$ decreases with $n,m$, and such alternatives are called \emph{local} alternatives (as opposed to \emph{fixed} alternatives). This framework allows us to characterize the \emph{detection boundary} of a test, that is, the smallest perturbation from the null (in terms of $\Delta_{n,m}$) that can be consistently detected by a test. 
    \end{remark}

    \textbf{Paper outline.} We present an overview of our main results in~\Cref{subsec:overview-of-results} and discuss related work in~\Cref{subsec:related-work}. In~\Cref{sec:kernel-mmd-test}, we present the cross-MMD statistic and obtain its limiting null distribution in~\Cref{subsec:asymptotic-gaussian-1}. We demonstrate its consistency against fixed alternatives and  minimax rate-optimality against smooth local alternatives in~\Cref{subsec:consistency}. Section~\ref{sec:experiments} contains numerical experiments that demonstrate our theoretical claims. All our proofs are in the supplement.

    \subsection{Overview of our main results}
    \label{subsec:overview-of-results}
        We propose a variant of the quadratic time kernel-MMD statistic of~\eqref{eq:kernel-mmd-stat} that relies on two key ideas: (i)~\emph{sample splitting} and (ii)~\emph{studentization}. In particular, we split the sample $\Xsample$ of size $n \geq 2$ into $\Xsample_1$ and $\Xsample_2$ of sizes $n_1 \geq 1$ and $n_2 \geq 1$, respectively (and $\Ysample$ of size $m\geq 2$ into $\Ysample_1$ and $\Ysample_2$ of sizes $m_1 \geq 1$ and $m_2 \geq 1$), and define the \emph{two-sample cross kernel-MMD} statistic $\crossmmd$ as follows: 
        \begin{align}
            &\crossmmd := \frac{1}{n_1 m_1 n_2 m_2} \sum_{i =1}^{n_1}\sum_{i'=1}^{n_2} \sum_{j=1}^{m_1} \sum_{j'=1}^{m_2}  h(X_i, X_{i'}, Y_j, Y_{j'}).  \label{eq:cross-U} 
        \end{align}
        Our final test statistic is $\csmmd:= \crossmmd/ \sigmahat$, where $\sigmahat$ is an empirical variance introduced in~\eqref{eq:two-sample-sigma-1}. %
        
        Our first set of results show that quite generally,  $\csmmd$ has an $N(0,1)$ asymptotic null distribution. \Cref{prop:simple-asymptotic-normality} obtains this result in the setting where both the kernel $k$ and null distribution $P$ are fixed. This is then generalized to deal with changing kernels (for instance, Gaussian kernels with bandwidth choices that depend on the sample-size) in~\Cref{theorem:asymptotic-normality-2}. Finally, in~\Cref{theorem:asymptotic-limit} in~\Cref{appendix:background}, we significantly expand the scope of these results by also allowing the null distribution to change with $n$, and also weakening the moment conditions required by~\Cref{theorem:asymptotic-normality-2}.

        \begin{figure}[htb!]
            \def\figwidth{0.35\linewidth}
            \def\figheight{0.35\linewidth} %
            \centering
            \hspace*{-1cm}
        \begin{tabular}{ccc}
            \input{FinalFigs/Null_Dists_d_10_500_n_100_m_150_kernel__Gaussian_RBF_2022_10_13_18_01_37cross}
        &
            \input{FinalFigs/PowerCurve_RBFd_10_eps_0_3}
        &
            \input{FinalFigs/PowerVsTime_RBFd_10_eps_0_2}
        \end{tabular}
        \caption{ The first figure shows the distribution of our proposed statistic $\csmmd$ predicted by~\Cref{theorem:asymptotic-normality-2} under the null for dimensions $d=10$ and $d=100$. The statistic is computed with Gaussian kernel~($k_{s_n}(x,y) = \exp(-s_n\|x-y\|_2^2)$) with scale parameter $s_n$ chosen by the \emph{median} heuristic  for different choices of $n$, $m$ and $d$, and with samples $\Xsample$ and $\Ysample$ drawn from a multivariate Gaussian distribution with identity covariance matrix. The second figure compares the power curves of the two-sample test using the $\csmmd$ statistic with the kernel-MMD permutation test (with $200$ permutations). The final figure plots the power vs computation time for the two tests. The size of the markers are proportional to the sample-size used in the test.}
        \label{fig:general-null-distribution}
       \end{figure}
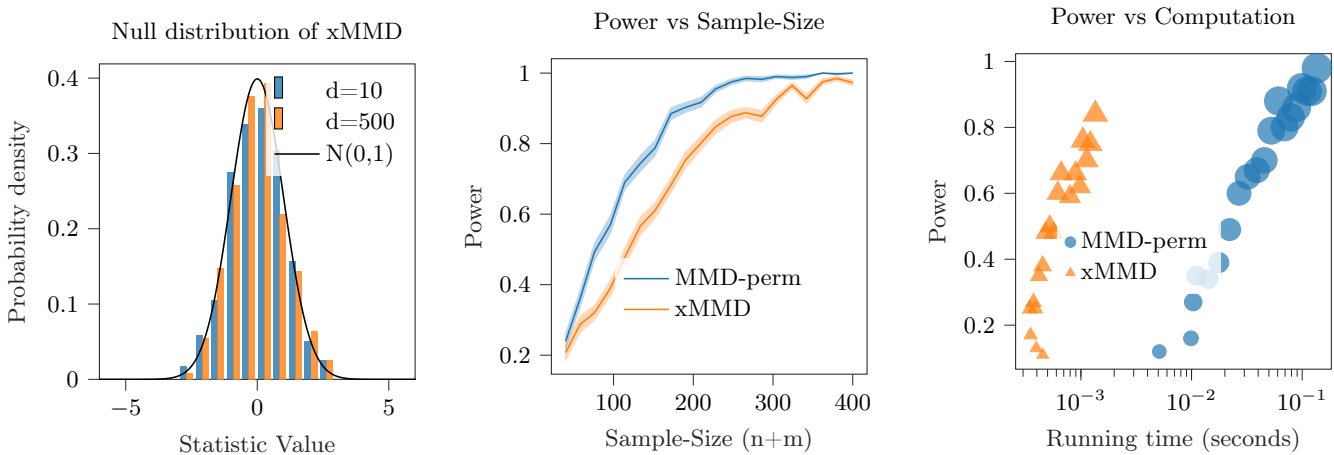  
        
       Our main methodological contribution is the ``$\cmmd$ test'', denoted $\Psi$, which rejects the null if $\csmmd$ exceeds $z_{1-\alpha}$, which is the $(1-\alpha)$-quantile of $N(0,1)$. Formally, 
       \begin{align} \label{eq:test}
           \text{$\cmmd$ test: } \Psi(\Xsample, \Ysample) = \ind_{\csmmd \geq z_{1-\alpha}}.
       \end{align}
       By the previous results, $\Psi$ has type-I error at most $\alpha$, meaning that $\mathbb{E}[\Psi(\Xsample, \Ysample)]\leq \alpha$ under the null.
       
        We next study the power of the $\cmmd$ test $\Psi$ in~\Cref{subsec:consistency}. 
        First, in the fixed alternative case, i.e., when the distributions $P\neq Q$ do not change with $n$, we show in~\Cref{corollary:fixed-alternative}, that the $\cmmd$ test implemented with any characteristic kernel is consistent under a bounded fourth moment condition. Next, we consider the more challenging case of local alternatives, i.e., when the  distributions, $P_n\neq Q_n$, change with $n$. In~\Cref{prop:general-consistency}, we first identify general sufficient conditions for the $\cmmd$ test to be uniformly consistent over a class of alternatives. Then, we specialize this to the case when $P_n$ and $Q_n$ admit  densities $p_n$ and $q_n$ with $\|p_n-q_n\|_{L^2} \geq \Delta_n$ for some $\Delta_n \to 0$.   We show in~\Cref{prop:smooth-alternative}, that the $\cmmd$ test with a Gaussian kernel $k_{s_n}(x,y) = \exp(-s_n\|x-y\|_2^2)$, with scale parameter $s_n$ increasing at an appropriate rate can consistently detect the local alternatives $\{\Delta_n: n\geq 1\}$ decaying at the minimax rate. 
        
        Finally, we note that while our primary focus in the paper is on the special case of kernel-MMD statistic, the ideas involved in defining the $\cmmd$ statistic can be extended to the case of general two-sample U-statistics. We describe this in~\Cref{appendix:proof-general-U-statistic}, and obtain sufficient conditions for asymptotic Gaussian limit of the resulting statistic, possibly of independent interest.

    \subsection{Comparisons to related work}
    \label{subsec:related-work}
        \textbf{Attempts to avoid permutations.} There have been some prior attempts to avoid permutations, but they are either heuristics (no provable type-I error control) or have poor power (higher type-II error).

        The first approach to obtaining a rejection threshold is based on large deviation bounds for the MMD statistic~\citep[\S~3]{gretton2006kernel} or the permuted MMD statistic~\citep[\S~5]{kim2021comparing}. The resulting tests are distribution-free, but they tend to be too conservative (type-I error much less than $\alpha$, resulting in low power). %
        Another approach involves choosing the threshold based on parametric estimates of the limiting null distribution. For example, \citet{gretton2006kernel} suggested fitting to the Pearson family of densities based on the first four moments, while \citet{gretton2009fast} introduced a more computationally efficient method using a two-parameter Gamma approximation. However, both of these methods are heuristic and do not have any consistency guarantees. 
    
        \citet{gretton2009fast} introduced a spectral method for approximating the null distribution using the eigendecomposition of the gram matrix. They showed that the resulting distribution converges to the true null distribution as long as the square roots of the eigenvalues associated with the kernel operator are summable. While this method is asymptotically consistent, the conditions imposed on the kernel are more stringent than that used in our work. Furthermore, this method was shown to be consistent only in the fixed kernel (or low-dimensional) setting. Hence, it is unknown whether the results carry over to the case of kernels varying with sample size or high-dimensional settings. This method is also computationally nontrivial due to the need for a full eigendecomposition. Keeping only the top few eigenvectors is another heuristic, but this introduces an extra hyperparameter and loses theoretical guarantees; as a result this method is rarely used in practice. Our methods are simpler (no extra hyperparameter), faster (closed-form threshold), and more robust (type I error guarantees also hold for changing kernels, and in high-dimensional settings).
        
        \textbf{Changing the statistic: block-MMD and linear-MMD statistics.}
        An idea more closely related to ours is that changing the test statistic itself would help make it cheaper to compute and also yield a tractable limiting distribution. One approach splits the observations into disjoint blocks, compute the kernel-MMD statistic on every block, and the final test statistic averages over all the blocks. If the size of each block is fixed, we get a linear-time kernel-MMD~\citep{gretton2012kernel,gretton2012optimal}. The case of block sizes increasing with $n, m$ was studied by \citet{zaremba2013b,ramdas2015adaptivity}. Depending on the block size ($b$), the computational complexity of block-MMD statistic varies from linear~(constant $b$) to quadratic ($b = \Omega(n)$). Further, if $b=o(n)$, then one gets a Gaussian null distribution as well. 
        
        Our proposed statistic is fundamentally different from the block-MMD statistics,  despite both being incomplete U-statistics~\citep{lee1990ustatistic}. In particular, the block-MMD statistics can be understood as building a block-diagonal approximation of the gram matrix. On the other hand, our proposed  cross-MMD statistic uses the \emph{off-diagonal blocks} of the gram matrix, exactly the blocks that the block-MMD with two blocks  ($b=n/2$) excludes! The reason that this is a sensible thing to do is nontrivial, and our test is motivated quite differently from the block-MMD. In fact, when $b=n/2$, the block-MMD does not have a Gaussian null, but the cross-MMD does. 
        
        For the block-MMD with $b=o(n)$, the Gaussian null distribution is achieved at the cost of suboptimal power, as observed empirically in \citet{zaremba2013b}, and proved by~\citet{reddi2015high} for the case of linear-MMD  and~\citet{ramdas2015adaptivity} for general block-MMD statistics.
        In particular, their power is worse by factors scaling with $n$, which means that they are not minimax rate optimal.
        In contrast, our test uses exactly half the elements of the gram matrix, and its power is about a $\sqrt 2$ worse than the MMD test, independent of $n$ and $d$, and we prove explicitly that it achieves the minimax rate.

            \textbf{Beyond the kernel-MMD.} The literature on two-sample testing is vast, and one can move even further away from the kernel-MMD (than the block-MMD) while retaining some of its intuition. For example, \citet{chwialkowski2015fast} proposed a linear-time test statistic by computing the average squared-distance between the empirical kernel embeddings at $J$ randomly drawn points.  {\citet{jitkrittum2016interpretable} proposed a variant of this statistic in which the $J$ points are selected to maximize a lower bound on the power.}
            In both cases, when the kernel $k$ being used is analytic, in addition to being characteristic and integrable, the authors showed that the limiting distribution under null for this statistic is a combination of $J$ independent $\chi^2$ random variables.  However, similar to the spectral method of~\citet{gretton2009fast}, the high-dimensional behavior of these statistics are unknown. {In fact, some preliminary experiments with $d \approx n$  in~\Cref{appendix:linear-time-tests} suggest that these linear-time statistics have a different null distribution in this regime.} Further, the authors only proved consistency of these tests against fixed alternatives, but their power is not known to be minimax rate optimal. In contrast, our statistic has the same limiting distribution in low- and high-dimensional settings even with changing kernels, and is provably minimax rate optimal for smooth alternatives, and it is a much more direct tweak of the usual MMD test.

        \textbf{{Comparison with \cite{kubler2022witness}.}} As elaborated in \Cref{sec:kernel-mmd-test}, the cross-MMD statistic has the form of two-sample $t$-statistic that compares the sample means of $f(X_1),\ldots,f(X_{n_1})$ and $f(Y_1),\ldots,f(Y_{m_1})$, respectively, for some function $f$. In particular, the function $f$ used in the cross-MMD is proportional to the empirical MMD witness function~\citep{gretton2012kernel} based on $\Xsample_2$ and $\Ysample_2$. A similar form of statistic has been considered by the recent work of \cite{kubler2022witness} where $f$ is set to be a maximizer of an empirical signal-to-noise ratio. In spite of sharing a similar form of statistic, our asymptotic analysis is markedly different from \cite{kubler2022witness}. In their asymptotic arguments, the authors assume that $f$ is fixed, i.e.,~$f$ does not vary with the sample size, under which the studentized statistic is asymptotically Gaussian \emph{conditional} on $f$. In sharp contrast, our asymptotic arguments allow $f$ to freely change with the sample size and the final asymptotic guarantee is \emph{unconditional} on $f$. Moving from the conditional guarantee to the unconditional guarantee is nontrivial, which is one of our main technical contributions. We also point out that in terms of power perspective, it is ideal to increase the size of $\Xsample_2$ and $\Ysample_2$, which violates the assumptions of \cite{kubler2022witness} in their asymptotic analysis. Due to this technical gap, \cite{kubler2022witness} recommend and also implement the permutation test using an unstudentized statistic in their simulation study. On the other hand, our test is completely permutation-free and comes with rigorous theoretical guarantees.

        \textbf{One-sample (goodness-of-fit) testing.} \citet{kim2020dimension} proposed and analyzed a similar studentized cross U-statistic in the simpler one-sample setting. Our work has different motivations: our primary goal in this paper is to design a permutation-free kernel-MMD test that does not significantly sacrifice the power, while \citet{kim2020dimension} pursued the related but different goal of \emph{dimension-agnostic} inference, which means having the same limiting distribution in low-dimensional and high-dimensional settings. Nevertheless, our results can be seen as an extension of their methods to two-sample testing. Our proofs also build on their advances, but we require a more involved analysis since in their case the second distribution is known (making it a point null).

\section{Deriving the cross-MMD test}
\label{sec:kernel-mmd-test}

In this section, we present our test statistic and investigate its limiting distribution. First note that the squared kernel-MMD distance between two probability measures $P$ and $Q$ can be expressed as an inner product, namely $\langle \mu - \nu, \mu - \nu \rangle_k$. The usual kernel-MMD statistic is obtained by plugging the empirical kernel embeddings into this inner product expression and removing the diagonal terms to make it unbiased. Our proposal instead considers pairs of empirical estimates $(\muhat_{1}, \muhat_{2})$ and $(\nuhat_{1}, \nuhat_{2})$ constructed via sample splitting, and use the inner product between $\muhat_{1} - \nuhat_{1}$ and $\muhat_{2} - \nuhat_{2}$ instead. This careful construction allows us to obtain a Gaussian limiting distribution after studentization.  
\begin{figure}
    \centering
    \includegraphics[width=0.3\textwidth]{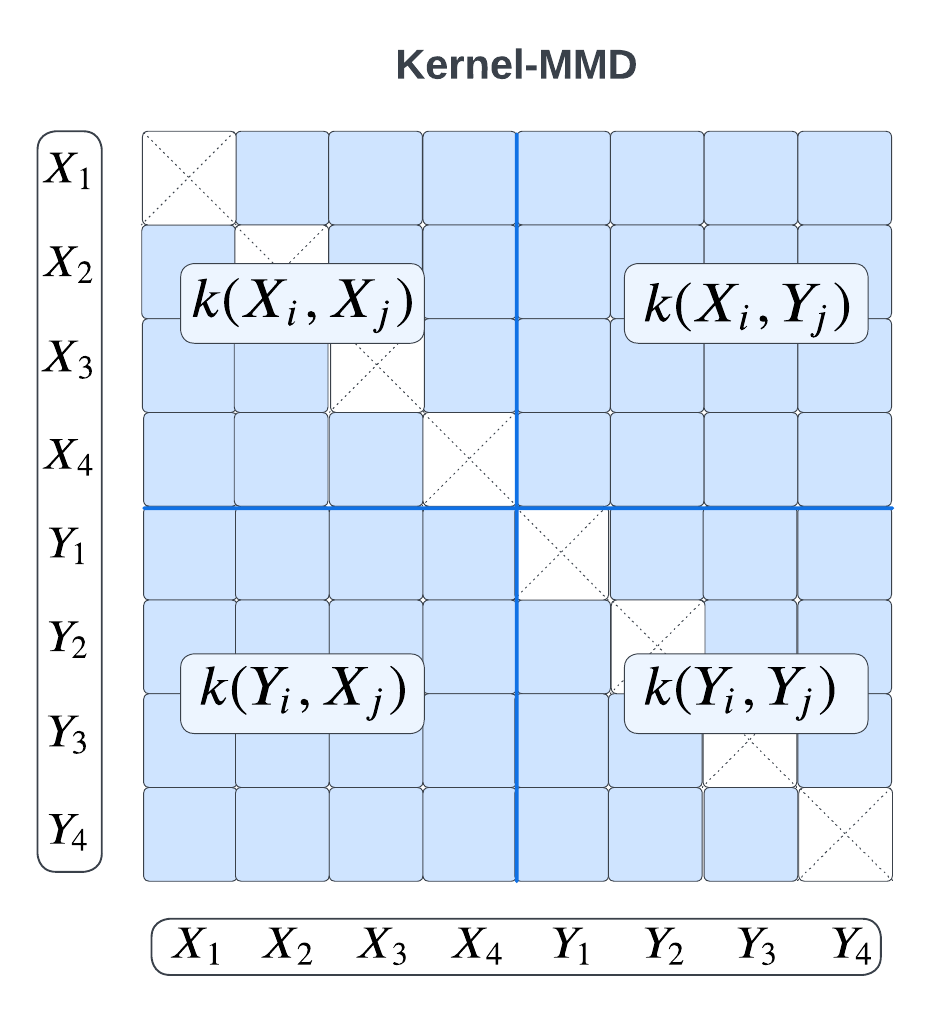}
    \includegraphics[width=0.3\textwidth]{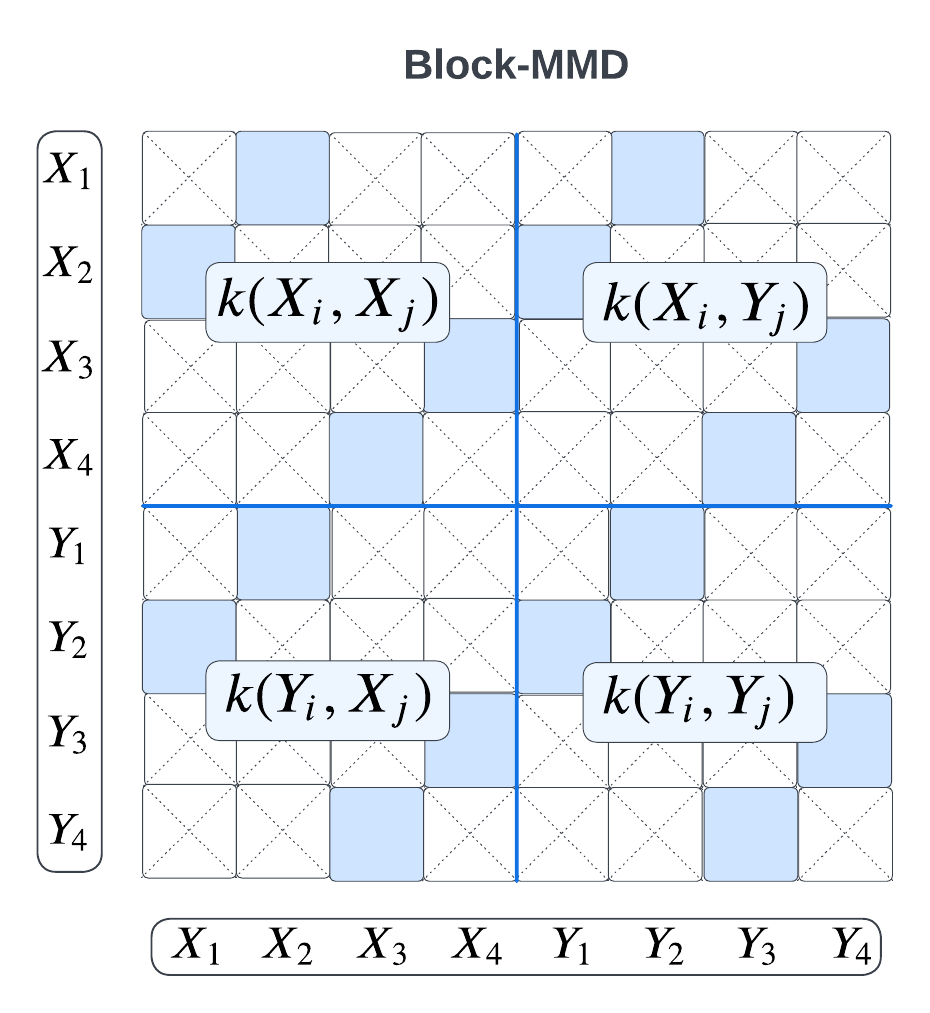}
    \includegraphics[width=0.3\textwidth]{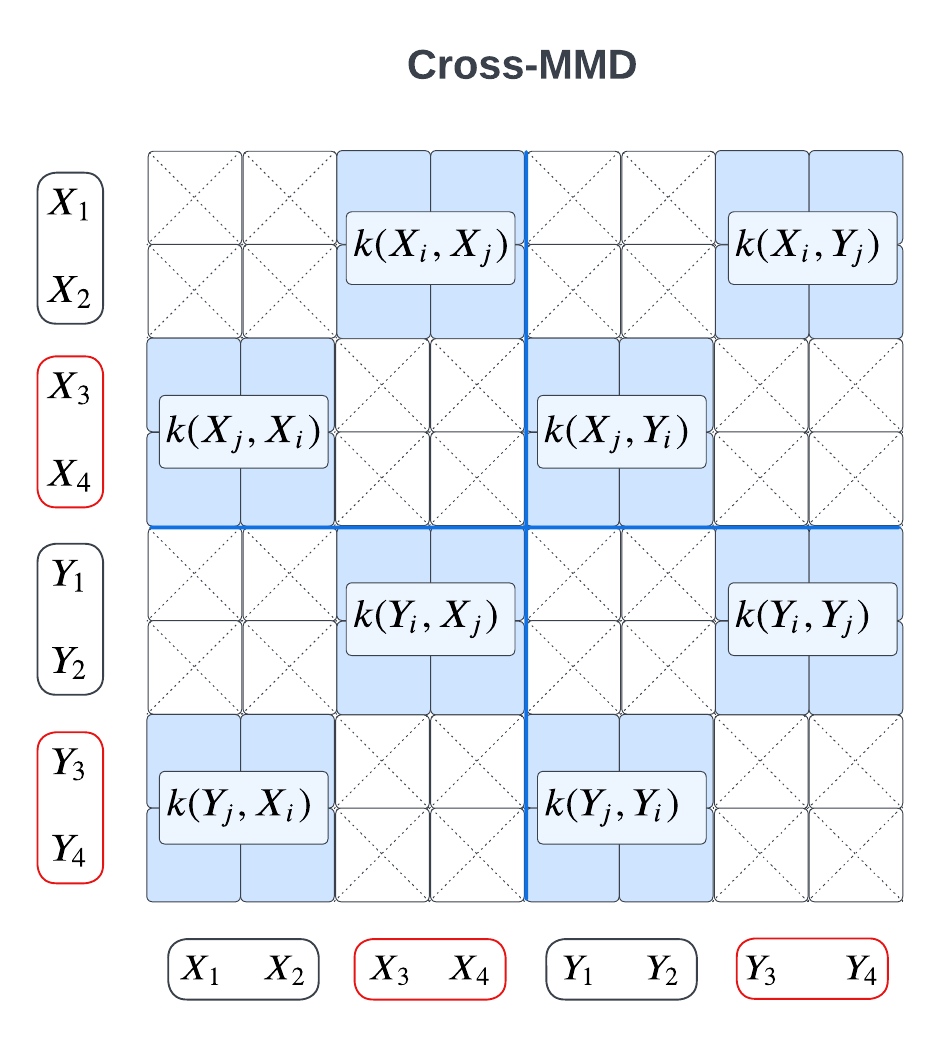}
    \caption{The figures visually  illustrate the main differences in computing the usual quadratic-time kernel-MMD statistic~(left), block-MMD~(center) statistic, and our new cross-MMD statistic.  
    In particular, the quadratic-time kernel-MMD statistic considers all pairwise kernel evaluations, with the exception of the diagonal terms. For block-MMD, we obtain the statistic by partitioning the data into several disjoint blocks; and then taking the average of the kernel-MMD statistic calculated over these disjoint blocks. Finally, our cross-MMD statistic first splits the data into two disjoint parts~(red and black), and then uses the pairwise kernel evaluations with data from different splits. Interestingly, the observation pairs included by our cross-MMD statistic are exactly complementary to those included by the block-MMD statistic.}
    \label{fig:intuition}
\end{figure}
To elaborate, recall from \Cref{subsec:overview-of-results} that we partition $\Xsample$ into $\Xsample_1$ and $\Xsample_2$, and similarly $\Ysample$ into $\Ysample_1$ and $\Ysample_2$. We then compute empirical kernel embeddings based on each partition, yielding $\muhat_{1} := n_1^{-1} \sum_{i=1}^{n_1} k(X_i, \cdot)$, $\muhat_{2} := n_2^{-1} \sum_{i'=1}^{n_2} k(X_{i'}, \cdot)$, $\nuhat_{1} := m_1^{-1} \sum_{j=1}^{m_1} k(Y_j, \cdot)$ and $\nuhat_{2} := m_2^{-1} \sum_{j'=1}^{m_2} k(Y_{j'}, \cdot)$. Using these embeddings coupled with the kernel trick, the cross U-statistic~\eqref{eq:cross-U} can be  written as $\crossmmd =  \langle \muhat_{1} - \nuhat_{1}, \;  \muhat_{2} - \nuhat_{2} \rangle_k$. To further motivate our test statistic, denote $U_{X,i} := \langle k(X_i, \cdot), \; \muhat_{2}  - \nuhat_{2} \rangle_k$ for $i=1,\ldots, n_1$ and $U_{Y,j} := \langle k(Y_j, \cdot), \; \muhat_{2}  - \nuhat_{2} \rangle_k$ for $j=1,\ldots,m_1$. Then the cross U-statistic can be viewed as the difference between two sample means:
	$\crossmmd = \frac{1}{n_1} \sum_{i=1}^{n_1} U_{X,i} - \frac{1}{m_1} \sum_{j=1}^{m_1} U_{Y,j}$. 
Since the summands are independent \emph{conditional on $\Xsample_2$ and $\Ysample_2$}, one may expect that $\crossmmd$ is approximately Gaussian after studentization. Our results in~\Cref{subsec:asymptotic-gaussian-1} formalize this intuition under standard moment conditions, where it takes some care to remove the above conditioning, since we care about the unconditional distribution. 

Let us further denote the sample means of $U_{X,i}$'s and $U_{Y,j}$'s by $\bar{U}_X$ and $\bar{U}_Y$, respectively, and define
\begin{equation}
	\begin{aligned}
		& \sigmahat_X^2 \defined \frac{1}{n_1} \sum_{i = 1}^{n_1} \lp U_{X,i} - \bar{U}_X \rp^2, \ \sigmahat_Y^2 \defined \frac{1}{m_1} \sum_{j=1}^{m_1} \lp U_{Y,j} - \bar{U}_Y \rp^2 \ \text{and} \ \sigmahat^2 \defined \frac{1}{n_1} \sigmahat_X^2 + \frac{1}{m_1} \sigmahat_Y^2. \label{eq:two-sample-sigma-1}
	\end{aligned}
\end{equation}
Now we have completed the description of our studentized cross U-statistic $\csmmd = \crossmmd / \sigmahat$,
and the resulting test $\Psi$ in \eqref{eq:test}. The asymptotic validity of the $\cmmd$ test is guaranteed by Theorem~\ref{theorem:asymptotic-limit} that establishes the asymptotic normality of $\csmmd$ under the null.

\begin{remark}[Computational Complexity] \label{remark: Computational Complexity}
\sloppy The  overall cost of computing the statistic $\csmmd$ is  $\mc{O}\lp (n+m)^2\rp$, and in particular, both  $\crossmmd$ and $\sigmahat$ have quadratic complexity. To see this, note that $\crossmmd$ can be expanded into $\langle \muhat_{1}, \muhat_{2} \rangle_k +\langle \nuhat_{1}, \nuhat_{2} \rangle_k -\langle \muhat_{1}, \nuhat_{2} \rangle_k-\langle \nuhat_{1}, \muhat_{2} \rangle_k$. Each of these terms can be computed in~$\mc{O}\lp (n+m)^2\rp$. 
Similarly, each term in the summations defining $\sigmahat_X^2$ and $\sigmahat_Y^2$ also require $\calO\lp (n+m)^2 \rp$ computation,
implying that the  $\sigmahat$ also has $\calO( (n+m)^2)$ complexity. 
\end{remark}

\begin{remark}
    \label{remark:notation}
    To simplify notation in what follows, we denote $m$ as $m_n$, where $m_n$ is some unknown nondecreasing sequence such that $\lim_{n \to \infty} m_n = \infty$. This still permits $m,n$ to be separate quantities growing to infinity at potentially different rates, but it allows us to index the sequence of problems with the single index $n$ (rather than $m,n$). We will use $k_n$, $d_n$, $\calX_n$, $P_n$ and $Q_n$ to indicate that quantities could (but do not have to) change as $n$ increases, and drop the subscript when they are fixed. 
    Furthermore, unless explicitly stated, we will focus on the balanced splitting scheme, i.e.,~$n_1 = \lfloor n/2 \rfloor$ and $m_1 = \lfloor m/2 \rfloor$ in what follows, because we currently see no apriori reason to split asymmetrically.
\end{remark}

\subsection{Gaussian limiting distribution under the null hypothesis}
\label{subsec:asymptotic-gaussian-1}

As shown in~\Cref{fig:general-null-distribution}, the empirical distribution of $\csmmd$ resembles a standard normal distribution for various choices of $m$, $n$ and dimension $d$ under the null. In this section, we formally prove this statement. 
Recalling the mean embedding $\mu$ from~\eqref{eq:kernel-mmd-stat}, define 
\begin{align}
\label{eq:h1}
    \bark(x,y) \defined \langle k(x, \cdot) - \mu, k(y, \cdot) - \mu \rangle_k.
\end{align}

\begin{theorem}
    \label{prop:simple-asymptotic-normality}
    Suppose that $k$ and $P$ do not change with $n$. If $0< \mathbb{E}_P[\bark(X, X')^4]<\infty$ for $X, X' \overset{\text{i.i.d.}}{\sim} P$, then $\csmmd \convdist N(0,1)$. 
\end{theorem}
We next present a more general result  that implies~\Cref{prop:simple-asymptotic-normality}. 
\begin{theorem}
    \label{theorem:asymptotic-normality-2}
    Suppose $P$ is fixed, but the kernel $k_n$ changes with $n$. If
    \begin{align}
    \label{eq:kernel-condition-weak}
    \lim_{n \to \infty} \frac{\mathbb{E}_P[\bark_n(X_1,X_2)^4]}{ \mathbb{E}_P[\bark_n(X_1, X_2)^2]^2} \lp \frac{1}{n} + \frac{1}{m_n} \rp  = 0,  \quad {\text{and} \quad \lim_{n \to \infty} \frac{\lambda_{1,n}^2}{\sum_{l=1}^{\infty} \lambda_{l,n}^2} \text{ exists}},
    \end{align}
    {where $(\lambda_{l,n})_{l=1}^{\infty}$ denote the eigenvalues of $\bark$ introduced in~\eqref{eq:eigen-decomposition-1}}, then we have $\csmmd \convdist N(0,1)$. 
\end{theorem}

It is easy to check that condition~\eqref{eq:kernel-condition-weak} is trivially satisfied if the kernels $\{k_n: n \geq 1\}$ are uniformly bounded by some constant; prominent examples are the Gaussian or Laplace kernel with a {sample size} dependent bandwidth. Thus, the above condition really exists to handle unbounded kernels and heavy-tailed distributions.
To motivate this requirement, we recall~\citet{bentkus1996berry}~(see~\Cref{fact:lyapunov-clt} in~\Cref{appendix:background}) who proved a studentized CLT for \iid random variables in a triangular array setup: $W_{1,n},W_{2,n},\dots, W_{n,n} \overset{\text{i.i.d.}}{\sim} P_n$. Define $V_n = \sqrt{n} \sum_{i=1}^n  W_{i,n}/ \sqrt{\sum_i (W_{i,n}-\bar{W}_n)^2}$ where $\bar{W}_n = (\sum_iW_{i,n})/n$. They showed that a sufficient condition for the asymptotic normality of $V_n$ is that 
$ \lim_{n \to \infty} \mathbb{E}_{P_n}[W_{1,n}^3]/\sqrt{\mathbb{E}_{P_n}[W_{1,n}^2]^3n} = 0$. 
(This last condition is trivially true if $P_n$ does not change with $n$, meaning that the triangular array setup is irrelevant and $W_{1,n}$ can be replaced by $W_1$.)

Our requirement is slightly stronger: condition~\eqref{eq:kernel-condition-weak} with $\bark_n(X_1,X_2)$ replaced by $W_{1,n}$  implies the previous condition of~\citet{bentkus1996berry}~(details in~\Cref{remark:sufficient-condition-clt} in~\Cref{appendix:background}). We need this stronger condition, because the terms in the definition of $\crossmmd$ are not \iid (indeed, not even independent), and thus we cannot directly apply the result of \citet{bentkus1996berry}. 
 Instead, we take a different route by first conditioning on the second half of data~$(\Xsample_2, \Ysample_2)$, then showing the conditional asymptotic normality of the standardized $\crossmmd$~(i.e., divided by conditional standard deviation instead of empirical), and finally showing that the ratio of conditional and empirical standard deviations converge in probability to $1$ (see~\Cref{appendix:proof-asymptotic-limit}). 

Finally, we note that the result of~\Cref{theorem:asymptotic-normality-2} can be further generalized in several ways: \textbf{(i)} instead of a fixed $P$ and changing $k_n$, we can consider a sequence of pairs $\{(P_n, k_n): n \geq 1\}$ changing with $n$, \textbf{(ii)} we can let $P_n \in \mc{P}_n^{(0)}$, for a class of distributions changing with $n$, and obtain the Gaussian limit uniformly over all elements of $\mc{P}_n^{(0)}$, and finally, \textbf{(iii)} the moment requirements on $\bark_n$ stated in condition in~\eqref{eq:kernel-condition-weak} can also be slightly weakened. We state and prove this significantly more general version of~\Cref{theorem:asymptotic-normality-2} in~\Cref{appendix:proof-asymptotic-limit}. 

\begin{remark}
    {In the statement of the two theorems of this section, the splits $(\Xsample_1, \Ysample_1)$ and $(\Xsample_2, \Ysample_2)$ are assumed to be drawn \iid from the same distribution $P$. However, a closer look at the proof of~\Cref{theorem:asymptotic-normality-2} indicates that the conclusions of the above two theorems hold even when the two splits are independent and drawn \iid from possibly different distributions; that is $(\Xsample_1, \Ysample_1)$ and $(\Xsample_2, \Ysample_2)$ are independent of each other and drawn \iid from distributions $P_1$ and $P_2$ respectively, with $P_1 \neq P_2$. 
    In particular, under this more general condition, the asymptotic normality of $\csmmd$ still holds, and the resulting test $\Psi$ still controls the type-1 error at the desired level. This may be useful for two-sample testing in settings where the entire set of data is not \iid, but two different parts of the data were collected in two different situations. The usual MMD can also handle such scenarios by using a subset of permutations that do not exchange the data across the two situations.}
\end{remark}

\subsection{Consistency against fixed and local alternatives}
\label{subsec:consistency}
    Here, we show that the $\cmmd$ test $\Psi$ introduced in~\eqref{eq:test}  is consistent against a fixed alternative and also has minimax rate-optimal power against smooth local alternatives separated in $L^2$ norm.
    
    We first show that analogous to~\Cref{prop:simple-asymptotic-normality}, $\cmmd$ is consistent against fixed alternatives.
    \begin{theorem}
        \label{corollary:fixed-alternative}
        Suppose $P,Q,k$ do not change with $n$, and $P \neq Q$. If $k$ is a characteristic kernel satisfying $0<\mathbb{E}_P[\bark(X_1, X_2)^4] < \infty$, and $0<\mathbb{E}_Q[\bark(Y_1, Y_2)^4]<\infty$, then the $\cmmd$ test is consistent, meaning it has asymptotic power $1$. 
    \end{theorem}
    The moment conditions required above are mild, and are satisfied trivially, for instance, by bounded kernels such as the Gaussian kernel. The ``characteristic'' condition is also needed for the consistency of the usual MMD test~\citep{gretton2012kernel}, and is also satisfied by the Gaussian kernel.
    
    Recalling Remark~\ref{remark:two-sample-test-definition}, we next consider the more challenging setting where $d_n,k_n$ can change with $n$, and $(P_n,Q_n)$ can vary within a class $\altclass \subset \mc{P}(\mc{X}_n)\times \mc{P}(\mc{X}_n)$ that can also change with $n$. We present a sufficient condition under which the $\cmmd$ test $\Psi$ is consistent \emph{uniformly} over $\altclass$.  Define $\mmdval  \defined \dmmd(\Pnm, \Qnm)$, which is assumed nonzero for each $n$ but could approach zero in the limit. 
   
    \begin{theorem}
        \label{prop:general-consistency} %
        Let $\{\delta_{n}: n \geq 2\}$ denote any positive sequence converging to zero. If
        \begin{align}
            \label{eq:general-consistency-conditions}
            \lim_{n \to \infty} \sup_{(\Pnm, \Qnm)\in \altclass} \frac{ \mathbb{E}_{P_n,Q_n}[\sigmahat^2]}{\delta_{n} \mmdval^4} + \frac{\var_{P_n,Q_n}(\crossmmd)}{\mmdval^4} = 0, \text{ where $\var$ denotes variance, }
        \end{align}
        then
            $\lim_{n \to \infty} \sup_{(\Pnm, \Qnm) \in \altclass} \mathbb{E}_{P_n,Q_n}[1-\Psi(\Xsample, \Ysample)] = 0$, meaning the $\cmmd$ test is consistent. 
\end{theorem} 
Note that while \emph{any} sequence $\{\delta_{n}\}$ converging to zero suffices for the general statement above, the condition~\eqref{eq:general-consistency-conditions} is easiest to satisfy for slowly decaying $\delta_{n}$, such as $\delta_{n} = 1/\log\log n$ for instance. 
    The sufficient conditions for consistency of~$\Psi$ stated in terms of $\sigmahat$ and $\crossmmd$ in~\eqref{eq:general-consistency-conditions} can also be translated into equivalent conditions on the kernel function $k_{n}$, similar to~\eqref{eq:kernel-condition-weak}, and we present the details in~\Cref{appendix:proof-consistency}. Importantly, if $P_n,Q_n,d_n$ are fixed and $k_n$ is bounded, then both $\mathbb{E}[\sigmahat^2]$ and $\var(\crossmmd)$ are $O(1/n)$, and $\gamma_n$ is a constant, so the condition is trivially satisfied, and in fact the above condition is even weaker than the fourth-moment condition of the previous theorem.

\subsection{Minimax rate optimality against smooth local alternatives}
     We now apply the general result of~\Cref{prop:general-consistency} to  the case where the distributions $\Pnm$ and $\Qnm$  admit Lebesgue densities $p_{n}$ and $q_{n}$ that lie in the order $\beta$ Sobolev ball for some $\beta>0$, defined as $\Sobolev(M) \defined \{f:\calX \to \reals \, \vert \, f \text{ is a.s.\ continuous, and } \int(1+\omega^2)^{\beta/2} \|\calF(f)(\omega)\|^2 dw < M<\infty\}$. Formally, we define the null and alternative class of distributions  as follows: 
     \begin{align}
         &\nullclass = \{ P \text{ with density } p: p \in \Sobolev(M)\}, \quad \text{and}  \label{eq:local-null-class}\\ 
         &\altclass = \{ (P,Q) \text{ with densities } p, q \in \Sobolev(M): \|p - q\|_{L^2} \geq \Delta_{n} \}, \label{eq:local-alt-class}
     \end{align}
     for some sequence $\Delta_{n}$ decaying to zero. In particular, we assume that under $H_0$, $\Pnm=\Qnm$ and $\Pnm\in \nullclass$, while under $H_1$, we assume that $(\Pnm,\Qnm) \in \altclass$. 
     
     Our next result shows that for suitably chosen scale parameter, the $\cmmd$ test $\Psi$ with the Gaussian kernel is minimax rate-optimal for the above  class of local alternatives. For simplicity, we state this result with $n=m$, noting that the result easily extends to the case when there exist constants $0<c\leq C$, such that $c\leq n/m \leq C$. 
     
    \begin{theorem}
        \label{prop:smooth-alternative}
        Consider the case when $n=m$, and let $\{\Delta_n: n \geq 1\}$ be a sequence such that $ \lim_{n \to \infty} \Delta_n n^{2{\beta}/(d+4\beta)} = \infty$.
        On applying the $\cmmd$ test $\Psi$ with the Gaussian kernel $k_{s_n}(x,y) = \exp(-s_n \|x-y\|_2^2)$, if we choose the scale as $s_n \asymp n^{4/(d+4\beta)}$, then we have
        \begin{align}
            \lim_{n \to \infty} \sup_{\Pnm \in \nullclass}  \mathbb{E}_{\Pnm}[\Psi(\Xsample, \Ysample)] \leq \alpha \quad  \text{and} \quad  
            \lim_{n \to \infty} \inf_{(\Pnm,\Qnm)\in \altclass} \mathbb{E}_{\Pnm, \Qnm}[\Psi(\Xsample, \Ysample)] = 1. \label{eq:smooth-power}
        \end{align}
    \end{theorem} 
    The proof of this statement is in~\Cref{appendix:proof-consistency}, and it follows by verifying that the conditions required by~\Cref{prop:general-consistency} are satisfied for the above choices of $\Delta_n$ and $s_n$.

    \begin{remark}
        \label{remark:minimax-optimality} \sloppy \citet[Theorem~5~(ii)]{li2019optimality} showed a converse of the above statement: if $\lim_{n \to \infty}\Delta_n n^{2\beta/(d+4\beta)} < \infty$, then there exists an $\alpha \in(0,1)$ such that any asymptotically level $\alpha$ test $\widetilde{\Psi}$ must have $\lim_{n \to \infty} \inf_{(P,Q) \in \mc{P}(\Delta_n)} \mathbb{E}_{P,Q} [\widetilde{\Psi}(\Xsample, \Ysample)] < 1$. Hence, the sequence of $\{\Delta_n: n \geq 1\}$ used in~\Cref{prop:smooth-alternative} represents the smallest $L_2$-deviations that can be detected by any test, and~\eqref{eq:smooth-power} shows that our $\cmmd$ test $\Psi$ can detect such changes, establishing its minimax rate-optimality. 
    \end{remark}

\section{Experiments}
\label{sec:experiments}

We now present experimental validation of the theoretical claims of the previous section. In particular, our experiments demonstrate that \textbf{(i)} the limiting null distribution of $\csmmd$ is $N(0,1)$ under a wide range of choices of dimension $d$, sample sizes $n,m$ and the kernel $k$, and \textbf{(ii)} the power of our $\cmmd$ test is competitive with the kernel-MMD permutation test. We now describe the experiments in more detail. Additional experimental results are reported in~\Cref{appendix:additional-experiments}.

\noindent \textbf{Limiting null distribution of $\csmmd$.} We showed in~\Cref{theorem:asymptotic-limit} that the statistic $\csmmd$ has a limiting normal distribution under some mild assumptions. We empirically test this result when $\Xsample$ and $\Ysample$ are drawn from $N(\boldsymbol{0}, I_d)$ with $\boldsymbol{0}$ denoting the all-zeros vector in~$\mathbb{R}^d$, and in particular, study the effects of (i) dimension:  $d=10$ versus $d=500$, (ii) skewness of the samples: $n/m =1$ versus $n/m=0.1$, and (iii) choice of kernel: Gaussian versus Quadratic, both with scale parameters selected using \emph{median} heuristic. 

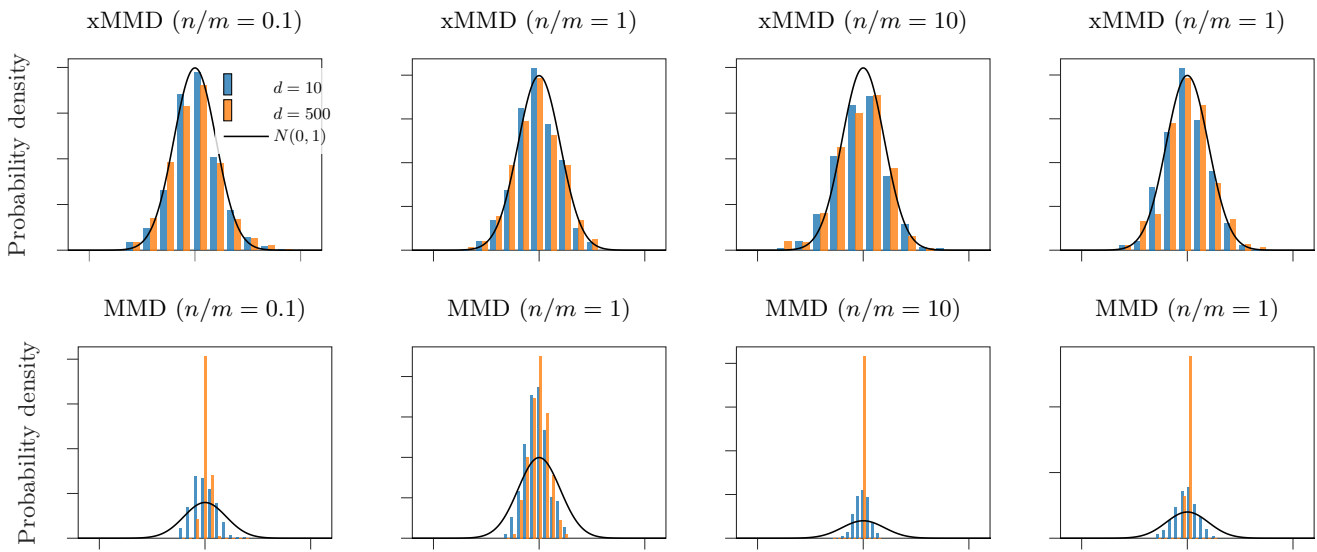
\begin{figure}[htb!]
    \centering
    \def\figwidth{0.30\linewidth}
    \def\figheight{0.25\linewidth} %
    \hspace*{-1cm}
    \begin{tabular}{cccc}
        \input{FinalFigs/Null_Dists_d_10_500_n_20_m_200_kernel__Gaussian_RBF_2022_10_12_22_14_00cross}
    &
        \input{FinalFigs/Null_Dists_d_10_500_n_200_m_200_kernel__Gaussian_RBF_2022_10_12_22_42_59cross}
    &
        \input{FinalFigs/Null_Dists_d_10_500_n_200_m_20_kernel__Gaussian_Poly_2_2022_10_12_23_28_39cross}
    &
        \input{FinalFigs/Null_Dists_d_10_500_n_200_m_200_kernel__Gaussian_Poly_2_2022_10_12_23_29_33cross} \\
        \input{FinalFigs/Null_Dists_d_10_500_n_20_m_200_kernel__Gaussian_RBF_2022_10_12_22_14_00mmd} 
    &
        \input{FinalFigs/Null_Dists_d_10_500_n_200_m_200_kernel__Gaussian_RBF_2022_10_12_22_42_59mmd}
    &
        \input{FinalFigs/Null_Dists_d_10_500_n_200_m_20_kernel__Gaussian_Poly_2_2022_10_12_23_28_39mmd}
    &
        \input{FinalFigs/Null_Dists_d_10_500_n_200_m_200_kernel__Gaussian_Poly_2_2022_10_12_23_29_33mmd}
    \end{tabular}
    \caption{The first two columns show the null distribution of the $\csmmd$ statistic (top row) and the $\mmdhat^2$ statistic scaled by its empirical standard deviation (bottom row) using the Gaussian kernel with scale-parameter chosen using the median heuristic. The last two columns show the null distribution for the two statistics using the Quadratic kernel with scale parameter chosen using the median heuristic. The figures demonstrate that the null distribution of $\mmdhat^2$ changes significantly with dimension~($d$), the ratio $n/m$ and the choice of the kernel, unlike our proposed statistic.}
    \label{fig:null-distributions-1}
   \end{figure}

As shown in the first row of~\Cref{fig:null-distributions-1}, the distribution of $\csmmd$ is robust to all these effects, and is close to $N(0,1)$ in all cases. In contrast, the distribution of the kernel-MMD statistic scaled by its empirical standard deviation~(obtained using $200$ bootstrap samples) in the bottom row of~\Cref{fig:null-distributions-1} shows strong changes with these parameters. We present additional figures and details of the implementation in~\Cref{appendix:additional-experiments}. 

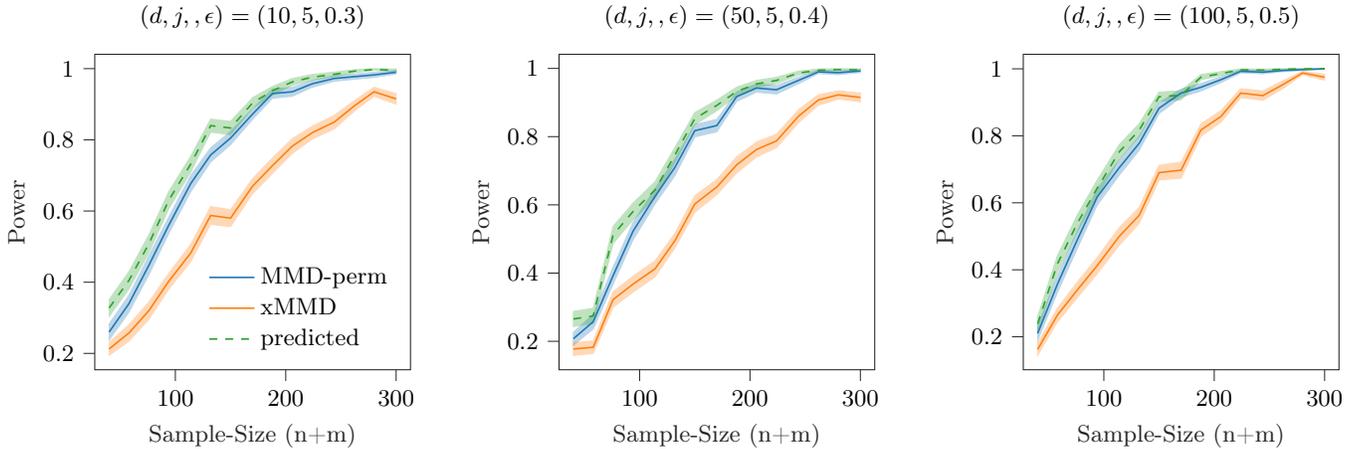
\begin{figure}[htb!]
        \def\figwidth{0.35\linewidth}
        \def\figheight{0.35\linewidth} %
    \centering
    \hspace*{-1cm}
    \begin{tabular}{ccc}
        \input{FinalFigs/PowerCurve_RBF2022_10_13_15_48_24_}
    &
        \input{FinalFigs/PowerCurve_RBF2022_10_13_16_05_01_}
    &
        \input{FinalFigs/PowerCurve_RBF2022_10_13_17_14_52_}
    \end{tabular}
    
    \caption{Curves showing the variation in power versus sample-size for the $\cmmd$ test  and the kernel-MMD permutation test. $\Xsample$ are drawn from $N(\boldsymbol{0}, I_d)$ \iid and $\Ysample$ is drawn from $N(a_{\epsilon,j}, I_d)$ where $a_{\epsilon,j}$ is obtained by perturbing the first $j\leq d$ coordinates of $\boldsymbol{0}$ by $\epsilon$, the kernel used is the Gaussian kernel with scale parameter chosen via the median heuristic. The dashed curve shows the predicted power of the kernel-MMD permutation test using the heuristic defined in~\eqref{eq:power-prediction-heuristic}.}
    \label{fig:predicted-power}
   \end{figure}

\noindent \textbf{Evaluation of the power of $\Psi$.} For $d \geq 1$ and $j \leq d$, let $a_{\epsilon, j}$ denote the element of $\mathbb{R}^d$ with first $j$ coordinates equal to $\epsilon$, and others equal to $0$.  
    We consider the two-sample testing problem with $P=N(\boldsymbol{0}, I_d)$  $Q=N(a_{\epsilon, j}, I_d)$  for different choices of $\epsilon$ and $d$ and $j$. We compare the performance of our proposed test $\Psi$ with the kernel-MMD permutation test, implemented with $B=200$ permutations, and plot the power-curves (using $200$ trials) in~\Cref{fig:predicted-power}. We also propose a heuristic for predicting the power of the permutation test~(denoted by $\rho_{\text{perm}}$) using the power of $\Psi$~(denoted by $\rho_{\Psi}$) as follows~(with $\Phi$ denoting the standard normal cdf, {and $z_{\alpha}$ its $\alpha$-quantile}):
    \begin{align}
        \label{eq:power-prediction-heuristic}
        \widehat{\rho}_{\text{perm}} = \Phi \lp z_{\alpha} + \sqrt{2} \lp \Phi^{-1}(\rho_{\Psi}) - z_{\alpha} \rp  \rp. 
    \end{align}
    This heuristic is motivated by the power expressions derived by~\citet{kim2020dimension} for the problems of one-sample Gaussian mean and covariance testing~(we discuss this further in~\Cref{appendix:additional-experiments}). The term $\sqrt{2}$ in the above expression quantifies the price to pay for sample-splitting.  As shown in~\Cref{fig:predicted-power}, this heuristic gives us an accurate estimate of the power of the kernel-MMD permutation test, without incurring the computational burden.

\begin{figure}[htb!]
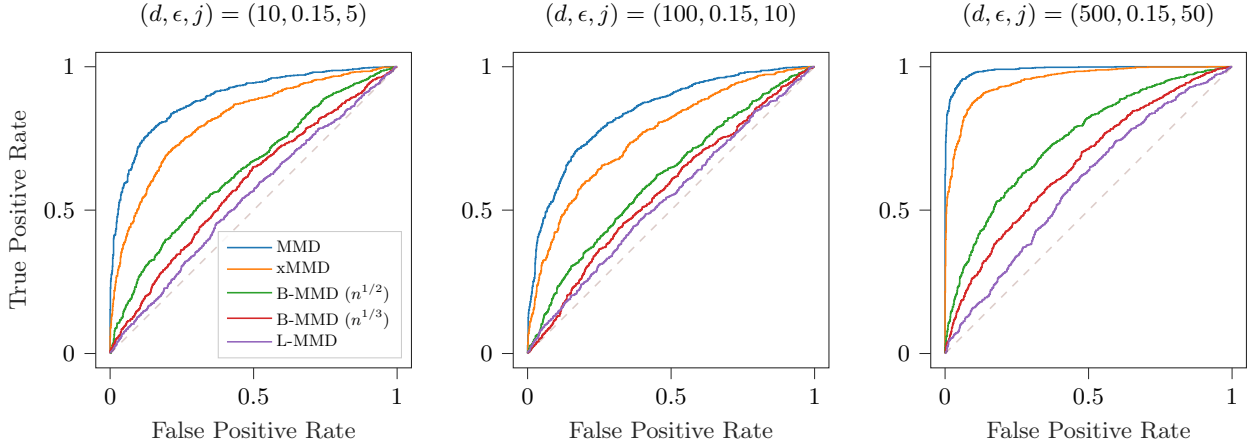

        \def\figwidth{0.35\linewidth}
        \def\figheight{0.35\linewidth} %
        \centering
        \begin{tabular}{ccc}
        \hspace*{-2cm}
        \input{FinalFigs/ROC_curve_n_200_d_10_eps_0.15}
    &
        \input{FinalFigs/ROC_curve_n_200_d_100_eps_0.15}
    &
        \input{FinalFigs/ROC_curve_n_200_d_500_eps_0.15}
    \end{tabular}
    
    \caption{ROC curves highlighting the trade-off  between type-I and type-II errors achieved by the MMD, cross-MMD, batch-MMD with batch sizes $n^{1/2}$ and $n^{1/3}$, and linear-MMD statistics. In all the figures, we use $n=m=200$. }
    \label{fig:ROC}
   \end{figure}

    We now use ROC curves  to compare the tradeoff between type-I and type-II errors for the usual MMD, linear and block-MMDs with our $\csmmd$. We use the same distributions $P=N(\boldsymbol{0}, I_d)$ and $Q=N(a_{\epsilon,j}, I_d)$ as before, and plot results for $(d, j, \epsilon) \in \{(10, 5, 0.2), (100, 20, 0.15), (500, 100, 0.1)\}$ in~\Cref{fig:ROC}. Due to sample splitting, the tradeoff achieved by our proposed statistic is slightly worse than that of $\mmdhat^2$, but  significantly better than other computationally efficient variants of kernel-MMD statistics. More details about the implementation, and additional figures are presented in~\Cref{appendix:additional-experiments}. 
    
\section{Conclusion and future work}
\label{sec:conclusion} 
    We proposed a variant of the kernel-MMD statistic, called cross-MMD, based on the ideas of sample-splitting and studentization, and showed that it has a standard normal limiting null distribution. Using this key result, we introduced a permutation-free (and hence computationally efficient) $\dmmd$ test for the two-sample problem. Experiments indicate that the power achieved by our test is within a $\sqrt{2}$-factor of the power of the kernel-MMD permutation test (that requires recomputing the statistic hundreds of times). In other words, our results achieve the following favorable  tradeoff: \emph{we get a {significant} reduction in computation at the price of a small reduction in power}. 
    
    \cite{sejdinovic2013equivalence} establish in some generality that distance-based two-sample tests (like the energy distance~\citep{szekely2013energy}) can be viewed as kernel-MMD tests with a particular choice of kernel $k$. Hence our results are broadly applicable to distance-based statistics as well. 
    
    Since two-sample testing and independence testing can be reduced to each other, it is an interesting direction for future work to see if the ideas developed in our paper can be used for designing permutation-free versions of kernel-based independence tests like HSIC~\citep{gretton2007kernel} or distance covariance~\citep{szekely2009brownian,lyons2013distance}. 
    
    Our techniques seem to rely on the specific structure of two-sample U-statistics of degree $2$. Extending these to more general U-statistics of higher degrees is another important question for future work.

    A final question is to figure out whether it is possible to achieve minimax optimal power using a sub-quadratic time test statistic. One potential approach would be to work with a kernel approximated by random Fourier features~\citep{rahimi2007random,zhao2015fastmmd}. Depending on the number of random features, our test statistic can be computed in sub-quadratic time and it would be interesting to see whether the resulting test can still be minimax optimal in power. We leave this important question for future work.

\subsection*{Acknowledgements}
The authors thank Anirban Chatterjee for informing them of a bug in the proof of~\Cref{prop:smooth-alternative}. 

\newpage 
\bibliographystyle{abbrvnat}
\bibliography{ref}

\newpage 
\begin{appendix}

\section{Background}
\label{appendix:background}
    To keep the paper self-contained, we collect the relevant definitions and theorems from prior work that are used in proving the main results of our paper. 
    
    \paragraph{Central Limit Theorems.} We first recall a central limit theorem for studentized statistics by~\citet{bentkus1996berry}. 
    \begin{fact}[Berry Esseen CLT]
        \label{fact:lyapunov-clt} For some \iid~$\sim P$ random variables $W_1, \ldots, W_n$, define the statistic $\csmmd = \frac{ \sum_{i=1}^n W_i}{\sqrt{ \frac{1}{n} \sum_{i=1}^n \lp W_i - \bar{W}_n \rp^2 } }$. If $\mathbb{E}_P[W_i]=0$ and $0<\mathbb{E}_P[W_i^2]<\infty$, then there exists a universal constant $C<\infty$ such that 
        \begin{align}
            \sup_{x \in \reals} | \mathbb{P}_P\lp T \leq x \rp - \Phi(x)| \leq C \frac{ \mathbb{E}_P[|W_1^3|]}{ \mathbb{E}_P[W_1^2]^{3/2}\; \sqrt{n}}. 
        \end{align}

    \end{fact}

    \begin{remark}
        \label{remark:sufficient-condition-clt}
        Note that by Cauchy--Schwarz inequality, we have 
        \begin{align}
            \mathbb{E}_P[W_1^3] = \mathbb{E}_P[W_1^2 \times W_1] \leq \sqrt{ \mathbb{E}_P[W_1^4] \mathbb{E}_P[W_1^2] }. 
        \end{align}
        This implies the following 
        \begin{align}
            \frac{\mathbb{E}_P[W_1^3]}{\mathbb{E}_P[W_1^2]^{3/2}} \leq \lp \frac{\mathbb{E}_P[W_1^4]}{\mathbb{E}_P[W_1^2]^2} \rp^{1/2}.  
        \end{align}
        Thus a sufficient condition for applying~\Cref{fact:lyapunov-clt} to show the convergence in distribution to $N(0,1)$ for a triangular sequence $\{W_{i,n}: 1 \leq i \leq n, \; n \geq 1\}$, with $\{W_{i,n}: 1 \leq i \leq n\}$ drawn \iid from some distribution $P_n$ is 
        \begin{align}
            \lim_{n \to \infty}  \frac{\mathbb{E}_{P_n}[W_{1,n}^4]}{\mathbb{E}_{P_n}[n W_{1,n}^2]^2}  = 0. 
        \end{align}
    \end{remark}
    
    We next recall a consequence of Lindeberg's Central Limit Theorem~(CLT), as stated in~\citep[Lemma~11.3.3]{lehmann2005testing}. 
    \begin{fact}
        \label{fact:lindeberg-clt}
        Let $Z_1, Z_2, \ldots$ be a sequence of \iid zero-mean random variables with finite variance $\sigma^2$. Let $c_1, c_2, \ldots$ be a real-valued sequence, satisfying: 
        \begin{align}
            \lim_{n \to \infty} \max_{1\leq i \leq n} \frac{c_i^2}{\sum_{j=1}^n c_j^2} = 0. 
        \end{align}
        Then, we have 
        \begin{align}
            \frac{ \sum_{i=1}^n c_i Y_i }{\sqrt{\sum_{j=1}^nc_j^2}} \convdist N(0, \sigma^2). 
        \end{align}
    \end{fact}

    \paragraph{Null distribution of MMD statistic.}
    Assuming that $(n,m)$ are such that $n/m \to c$ for some $c>0$, and let $\{u_l: l \geq 1\}$ and $\{v_l: l \geq 1\}$ denote two independent sequences of \iid $N(0,1)$ random variables. Furthermore, let $\{\lambda_l: l \geq 1\}$ denote the eigenvalues of the kernel operator $f(\cdot) \mapsto \int_{\mc{X}}f(x)k(\cdot, x)dP(x)$. Using techniques from  the theory of U-statistics, \citet{gretton2012kernel} showed that 
        \begin{align}
            \label{eq:null-dist-mmd2}
            (n+m) \mmdhat^2 \convdist \sum_{l=1}^{\infty} \lambda_l \lp \frac{\lp c^{1/2} u_l - v_l\rp^2 }{1+c} - \frac{(1+c)^2}{c} \rp. 
        \end{align}
    \eqref{eq:null-dist-mmd2} shows that the null distribution of $\mmdhat$ is an infinite combination of chi-squared random variables, weighted by the eigenvalues of the kernel operator. Due to this form, the null distribution has a complex dependence on the kernel and the null distribution $P$.         
        
    \paragraph{Gaussian kernel calculations.}
    Next, we recall some facts derived by\citet{li2019optimality}, about the  the Gaussian kernel $k_s(x,y) \defined \exp\lp - s \|x-y\|_2^2 \rp$, and  probability distributions that admit density functions lying in the Sobolev ball $\in \Sobolev(M)$. 
    \begin{fact}
    \label{fact:li-yuan}
    Consider a  Gaussian kernel that varies with sample size, $k_n(x, y) = \exp(-s_n \|x-y\|_2^2) $. Let $\bark_n$ be as defined in~\eqref{eq:h1-general}, $\calX = \mathbb{R}^d$ and $X_1, X_2, X_3, X_4 \sim P_n$ \iid, $Y_1, Y_2 \sim Q_n$, where $P_n$ and $Q_n$ have densities $p_n$ and $q_n$ in $\Sobolev(M)$ and $\|p_n-q_n\|_{L^2} = \Delta_n$, for some real valued sequence $\{\Delta_n: n \geq 1\}$ converging to $0$.   
    Then, we have the following: 
    \begin{align}
        &\mathbb{E}_{P_n}[\bark_n^2(X_1, X_2)] \asymp s^{-d/2}, \quad \text{and}\quad  \mathbb{E}_{Q_n}[\bark_n^2(Y_1, Y_2)] \asymp s^{-d/2} \label{eq:li-yuan-1} \\
        &\mathbb{E}_{P_n}[\bark_n^4(X_1, X_2)] \lesssim s^{-d/2}, \label{eq:li-yuan-2} \\
        &\mathbb{E}_{P_n}[\bark_n^2(X_1, X_2) \bark_n^2(X_1, X_3)] \lesssim s^{-3d/4}, \label{eq:li-yuan-3} \\
        &\mmdval(P_n, Q_n) = \dmmd(P_n, Q_n) \gtrsim s_n^{-d/2} \Delta_n \label{eq:li-yuan-5}. 
    \end{align}
    \end{fact}
    
    \paragraph{Additional Notation.} We use $U = o_P(u_n)$ and $U = O_P(u_n)$ to denote that $U/u_n \convprob 0$ and that $U/u_n$ is stochastically bounded. For real valued sequences, we use $a_n \lesssim b_n$ if there exists a constant $C$ such that $a_n \leq C b_n$ for all $n$. We use $a_n \asymp b_n$ if $a_n \lesssim b_n$ and $b_n \lesssim a_n$.

\section{Gaussian limiting distribution of \texorpdfstring{$\csmmd$}{c-MMD}}
\label{appendix:proof-asymptotic-limit}

    In this section, we present the results about the limiting null distribution of the statistic $\csmmd$. The general outline of the section is as follows: 
    
    \begin{itemize}
        \item In \Cref{subsec:null-general}, we state the most general version of the result on the limiting distribution of $\csmmd$~(\Cref{theorem:asymptotic-limit}), that we alluded to in~\Cref{subsec:asymptotic-gaussian-1}. We then prove this result in~\Cref{subsec:null-general-proof}. 
        
        \item In~\Cref{subsec:proof-fixed-k-P}, we show how the general result can be used to prove~\Cref{theorem:asymptotic-normality-2}, where the kernel is allowed to change with $n$ while the distribution $P$ is fixed. 
       \item Finally, in~\Cref{subsec:proof-fixed-k-P}, we show how~\Cref{theorem:asymptotic-normality-2} can be used to conclude the result for the case when both the kernel $k$ and null distribution $P$ are fixed with $n$. 
    \end{itemize}
    
    \subsection{Statement of the general result~(both  \texorpdfstring{$k_n$}{k\_n} and \texorpdfstring{$P_n$}{P\_n} changing with \texorpdfstring{$n$}{n})}
    \label{subsec:null-general}

    As stated in~\Cref{remark:notation}, we  assume that $m \equiv m_n$ is some non-decreasing function of $n$. 
    We consider a sequence of positive-definite kernels $\{k_{n}: n\geq 2\}$, and probability distributions $\{P_{n}: n \geq 1, 2\}$, and define 
    \begin{align}
    \label{eq:h1-general}
        \bark(x,y) \equiv \bark_{n}(x, y) = \langle k_{n}(x, \cdot) - \mu_{P_{n}}, k_{n}(y, \cdot) - \mu_{P_{n}} \rangle_k,
    \end{align}
    where $\mu_{P_{n}}$ denotes the embedding of the distribution $P_{n}$ into the RKHS associated with the kernel $k_{n}$. For any fixed values of $n $, we use $\{ (\lambda_{l, n}, \varphi_{l, n}): l \geq 1\}$ to denote the eigenvalue-eigenfunction sequence associated with the integral operator $g \mapsto \int \bark(\cdot, x)g(x) dP_{n}(x)$. If $\bark$ happens to be square-integrable (in addition to being symmetric), 
    it has the following representation:
    \begin{align}
        \label{eq:eigen-decomposition-1}
        \bark_n(x, y) = \sum_{l=1}^{\infty} \lambda_{l,n} \varphi_{l,n}(x) \varphi_{l,n}(y). 
    \end{align}
    We now state the assumption required to prove the limiting normal distribution of the statistic $\csmmd$. As we will see in~\Cref{subsec:proof-fixed-P}, in the special case of fixed $P$, the condition in~\eqref{eq:assump-kernel-1} is a weaker version of that used in~\Cref{theorem:asymptotic-normality-2}. 
    \begin{assumption}
    \label{assump:kernel-1}
    For $\bark$ introduced in~\eqref{eq:h1},  $\{(\lambda_{l,n}, \varphi_{l,n}): l \geq 1\}$ introduced in~\eqref{eq:eigen-decomposition-1} and for a sequence $\{P_n: n \geq 1\}$,  we assume that
    \begin{align}
    &\frac{ \mathbb{E}_{P_{n}}[\bark^4(X_1, X_2)](n^{-1}+m_n^{-1}) + \mathbb{E}_{P_{n}}[\bark^2(X_1, X_3)\bark^2(X_2, X_3)] }{\mathbb{E}_{P_n}[\bark^2(X_1, X_2)]^2 \lp \frac{1}{n^{-1} +m_n^{-1}} \rp}  \rightarrow 0, \quad \text{and} \label{eq:assump-kernel-1} \\ 
    & \lim_{n \to \infty} \frac{\lambda_{1,n}^2}{ \sum_{l=1}^{\infty} \lambda_{l, n}^2} \quad \text{exists}. \label{eq:assump-kernel-2}
    \end{align}
    \end{assumption}
    We now state the main result of this section.
    \begin{theorem}
    \label{theorem:asymptotic-limit}
    Suppose the sequence $\{m_n: n \geq 1\}$ satisfies $\lim_{n \to \infty} n/m_n$ exists and is non-zero.
    Let $\{k_{n}: n \geq 1\}$ be a sequence of positive definite kernels, and  let $\nullclass$ denote a family of distributions such that, for every $n\geq 1$ and  $P_{n} \in \nullclass$,~\Cref{assump:kernel-1} is satisfied by the pair $(\bark_n, P_{n})$ with $\bark_n$ defined in~\eqref{eq:h1-general}.  
    Then,  we have that 
    \begin{align}
    \lim_{n \to \infty} \sup_{P_{n} \in \nullclass} \sup_{x\in \mathbb{R}} \;| \mathbb{P}_{P_{n}}( \csmmd \leq x) - \Phi(x)| = 0.
    \end{align}
    \end{theorem}
    We now present the proof of this result.

    \subsubsection{Proof of the general result with changing \texorpdfstring{$k_n$}{k\_n} and \texorpdfstring{$P_n$}{P\_n}}
        \label{subsec:null-general-proof}  
 
    To simplify the notation, we will drop the subscripts from $k_{n}$, $\bark_{n}$, $P_{n}$, $\lambda_{l,n, m}$ and $\varphi_{l,n, m}$ in this proof outline. Furthermore, note that as mentioned in~\Cref{remark:notation}, we assume that $n_1 = n/2$ and $n_1 = m/2$.

    For any $x \in \calX$, introduce the term  $\tildek(x, \cdot)$ to denote $k(x, \cdot)- \mu$. Next, we define the following terms
    \begin{align}
        S_X &= \langle \muhat_{1} - \mu, \, \overbrace{(\muhat_{2}-\mu) - (\nuhat_{2} - \mu)}^{\defined g_2} \rangle_k, \quad \text{and}
        \quad S_Y = \langle \nuhat_{1} - \mu, \, g_2 \rangle_k ,
    \end{align}
    and note that we can write $\crossmmd = \crossU_X -\crossU_Y = S_X - S_Y$~($S_X$ differs from $\crossU_X$ due to the extra $\mu$ term in the first argument of the inner product). Recall that we use $\mu$ and $\nu$ to denote the kernel embeddings of the distributions $P$ and $Q$. 
    
    We can further rewrite $S_X$ and $S_Y$ in terms of $\{W_i: 1 \leq i \leq n_1\}$ and $\{Z_j: 1 \leq j \leq m_1\}$ as follows: 
    \begin{align}
    \label{eq:W-Z}
        S_X & = \frac{1}{n_1} \sum_{i=1}^{n_1} \overbrace{\langle \tildek(X_i, \cdot), \, g_2 \rangle_k}^{\defined W_i}, \quad \text{and} \quad     S_Y  = \frac{1}{m_1} \sum_{j=1}^{m_1} \overbrace{\langle \tildek(Y_j, \cdot), \, g_2 \rangle_k}^{\defined Z_j}.
    \end{align}

    With these terms defined,  we proceed in the following steps: 
    \begin{itemize}
        \item \textbf{Step 1:} First, we consider the standardized random variables $T_{s,X}$ and $T_{s,Y}$, defined as 
        \begin{align}
            T_{s,X} \defined \frac{\sqrt{n_1}S_X }{ \mathbb{E}_{P_n}[W_i^2|\Xsample_2, \Ysample_2] }, \quad \text{and} \quad 
            T_{s,Y} \defined \frac{\sqrt{m_1}S_Y}{ \mathbb{E}_{P_n}[Z_j^2|\Xsample_2, \Ysample_2] }, 
        \end{align}
        and prove that they converge in distribution to $N(0,1)$ conditioned on $(\Xsample_2, \Ysample_2)$. To prove that the limiting distribution is Gaussian, we verify that $\frac{\mathbb{E}_{P_n}[W_i^4|\Xsample_2, \Ysample_2]}{n_1 \mathbb{E}_{P_n}[W_i^2|\Xsample_2, \Ysample_2]^2} \stackrel{p}{\rightarrow} 0$ and $\frac{\mathbb{E}_{P_n}[Z_j^4|\Xsample_2, \Ysample_2]}{m_1 \mathbb{E}_{P_n}[Z_j^2|\Xsample_2, \Ysample_2]^2} \stackrel{p}{\rightarrow} 0$. This is formally shown in~\Cref{lemma:asymp-normality-1} below. 
        
        \item \sloppy \textbf{Step 2:} Next, building upon the previous result, and using the conditional independence of $T_{s,X}$ and $T_{s,Y}$, we show in~\Cref{lemma:asymp-normality-2} below, that the standardized statistic $T_s = (S_X - S_Y)/\sqrt{ n_1^{-1} \mathbb{E}_{P_n}[W_1^2|\Xsample_2, \Ysample_2] + m_1^{-1} \mathbb{E}_{P_n}[Z_1^2|\Xsample_2, \Ysample_2] }$ also converges in distribution to $N(0, 1)$. 
        
        \item \textbf{Step 3:} We then prove in~\Cref{lemma:asymp-normality-3} below that the ratio $\frac{n_1^{-1} \mathbb{E}_{P_n}[W_1^2|\Xsample_2, \Ysample_2] + m_1^{-1} \mathbb{E}_{P_n}[Z_1^2|\Xsample_2, \Ysample_2]}{n_1^{-1} \sigmahat_X^2 + m_1^{-1} \sigmahat_Y^2}$ converges in probability to $1$. 
    \end{itemize}
    It only remains to state and prove the three lemmas used above, which we do after this proof. Barring that, combining the above three steps completes the proof of the theorem.  
    \qed
    
     Before proceeding, we first introduce the terms $a_i = \langle \tildek(X_i, \cdot), \muhat_{2} - \mu \rangle_k$ and $b_i = \langle \tildek(X_i, \cdot), \nuhat_{2} - \mu \rangle_k$, and note that we can further decompose $W_i$ into $a_i - b_i$ for $1 \leq i \leq n_1$. Similarly, for $1 \leq j \leq m_1$, we can write $Z_j$ as $c_j-d_j$ with $c_j = \langle \tildek(Y_j, \cdot), \muhat_{2} - \mu \rangle_k$ and $d_j =  \langle \tildek(Y_j, \cdot), \nuhat_{2} - \mu \rangle_k$. 
        
         We now state and prove the intermediate results to obtain~\Cref{theorem:asymptotic-limit}. 
        
        \begin{lemma}
        \label{lemma:asymp-normality-1} 
        Under the conditions of~\Cref{theorem:asymptotic-limit}, we have the following: 
        \begin{align}
            \frac{\mathbb{E}_{P_n}[W_i^4|\Xsample_2, \Ysample_2]}{n_1 \mathbb{E}_{P_n}[W_i^2|\Xsample_2, \Ysample_2]^2} \stackrel{p}{\rightarrow} 0, \quad \text{and} \quad  
            \frac{\mathbb{E}_{P_n}[Z_j^4|\Xsample_2, \Ysample_2]}{m_1 \mathbb{E}_{P_n}[Z_j^2|\Xsample_2, \Ysample_2]^2} \stackrel{p}{\rightarrow} 0. 
        \end{align}
        Hence, as a consequence of the Lyapunov form of CLT~(see~\Cref{fact:lyapunov-clt} and~\Cref{remark:sufficient-condition-clt} in~\Cref{appendix:background}), this means that $T_{s,X} \convdist N(0,1)$ and $T_{s,Y} \convdist N(0,1)$ conditioned on $(\Xsample_2, \Ysample_2)$. 
        \end{lemma}
        
        \begin{proof}
        We describe the steps for proving the first statement (involving $W_i$), noting that the other statement follows in an entirely analogous manner. Throughout this proof, we will use the shorthand $\mathbb{E}_2[\cdot]$ to denote the $\mathbb{E}_{P_n}[\cdot | \Xsample_2, \Ysample_2]$. 
        
        By two applications of the AM-GM inequality, we observe that $W_i^4 = (a_i-b_i)^4 \leq 16(a_i^4 + b_i^4)$. Hence, we have the following: 
            \begin{align}
                \frac{\mathbb{E}_2[W_i^4]}{16 n_1 \mathbb{E}_2[W_i^2]^2} & \leq  \frac{ \mathbb{E}_2[a_i^4 + b_i^4]}{n_1 \mathbb{E}_2[(a_i-b_i)^2]} \\
                &=  \frac{n_1\mathbb{E}_2[a_i^4]}{\mathbb{E}_{P_n}[\bark(X_1, X_2)^2]^2} \times \frac{\mathbb{E}_{P_n}[\bark(X_1,X_2)^2]^2 }{n_1^2\mathbb{E}_2[(a_i-b_i)^2]} + \frac{m_1 \mathbb{E}_2[b_i^4]}{\mathbb{E}_{P_n}[\bark(Y_1,Y_2)^2]^2} \times \frac{ \mathbb{E}_{P_n}[\bark(Y_1,Y_2)^2]^2}{m_1^2 \mathbb{E}_2[(a_i-b_i)^2]} \label{eq:proof-lemma-0}\\
                & \defined A_1 \times A_2 + B_1 \times B_2. \label{eq:proof-lemma-1}
            \end{align}
            Thus, to complete the proof, it suffices to show that $A_1 \times A_2$ and $B_1 \times B_2$ converge in probability to $0$. This can be shown in two steps:
            \begin{itemize}
                \item Under the assumptions of~\Cref{theorem:asymptotic-limit}, we have $A_1 \stackrel{p}{\rightarrow} 0$ and $B_1 \stackrel{p}{\rightarrow} 0$. To prove this result, it suffices to show that $\mathbb{E}_{P_n}[A_1] \to 0$ and $\mathbb{E}_{P_n}[B_1] \to 0$. The result then follows by an application of Markov's inequality. 
               
                \item $A_2$ and $B_2$ are bounded in probability. 
            \end{itemize}
            
            We first show that $\mathbb{E}_{P_n}[A_1] \to 0$. The result for $B_1$ follows similarly. 
            \begin{align}
                \mathbb{E}_{P_n} [A_1] &= \frac{n_1}{\mathbb{E}_{P_n}[\bark(X_1, X_2)^2] } \mathbb{E}_{P_n}\lb \mathbb{E}_2 \lb a_i^2 \rb \rb \\
                & \stackrel{(i)}{=} \frac{n_1}{\mathbb{E}_{P_n}[\bark(X_1, X_2)^2] }  \lp \frac{ \mathbb{E}_{P_n}\lb \bark^4(X_1, X_2) \rb }{n_1^3} + \frac{ 3n_1(n_1-1)}{n_1^4} \mathbb{E}[\bark^2(X_1, X_3)\bark^2(X_2, X_3)] \rp \\
                & \leq \frac{3}{\mathbb{E}_{P_n}[\bark(X_1, X_2)^2] }  \lp \frac{ \mathbb{E}_{P_n}\lb \bark^4(X_1, X_2) \rb }{n_1^2} + \frac{1}{n_1} \mathbb{E}[\bark^2(X_1, X_3)\bark^2(X_2, X_3)] \rp, 
            \end{align}
            which goes to $0$ as required, by invoking the condition in~\eqref{eq:assump-kernel-1} of~\Cref{assump:kernel-1}.  For~(i), we used the expression derived by~\citet{kim2020dimension} while proving their~Theorem 6. 
            
            To complete the proof, we show that $A_2$ is bounded in probability~(the result for $B_2$ follows similarly). We consider two cases, depending on whether  $\rho_1 \defined \lim_{n, m \to \infty} \frac{ \lambda_1^2}{\sum_{l} \lambda_l^2}$ is equal to $0$ or greater than $0$~(the existence of this limit is assumed). 
            
            \emph{Case~1: $\rho_1 > 0$.} We first observe that as a consequence of~\eqref{eq:h1-general} and the orthonormality of the eigenfunctions, we have 
            \begin{align}
                \mathbb{E}_{P_n}[\bark(X_1, X_2)^2] &= \mathbb{E}_{P_n}\left[ \sum_{l, l'} \lambda_l \lambda_{l'} \varphi_l(X_1)\varphi_{l'}(X_1) \varphi_l(X_2) \varphi_{l'}(X_2) \right] 
                 = \sum_{l=1}^{\infty}\lambda_l^2. 
            \end{align}
            Using this, we obtain the following: 
            \begin{align}
                \frac{1}{(A_2)^{1/2}} = \frac{n_1 \mathbb{E}_2[ a_i^2 + b_i^2 - 2a_i b_i ]}{\sum_{l=1}^{\infty}\lambda_l^2}. 
            \end{align}
            By repeated use of~\eqref{eq:h1-general}, we can show that the following identities hold: 
            \begin{align}
                \mathbb{E}_2[a_i^2] &= \frac{1}{(n-n_1)^2} \sum_{l=1}^{\infty} \lambda_l^2 \lp \sum_{i'} \varphi_l(X_{i'}) \rp^2, \\ 
                \mathbb{E}_2[b_i^2] &= \frac{1}{(m-m_1)^2} \sum_{l=1}^{\infty} \lambda_l^2 \lp \sum_{j'} \varphi_l(Y_{j'}) \rp^2, \quad \text{and} \\
                \mathbb{E}_2[a_ib_i] &= \frac{1}{(n-n_1)(m-m_1)} \sum_{l=1}^{\infty} \lambda_l^2 \lp \sum_{i'} \varphi_l(X_{i'}) \rp\lp \sum_{j'} \varphi_l(Y_{j'}) \rp. 
            \end{align}
            Plugging these equalities in the expression for $A_2$, and using $\rho_l = \frac{\lambda_l}{\sum_{l'} \lambda_{l'}^2}$, we get 
            \begin{align}
                (A_2)^{1/2} &= \frac{1}{n_1 \sum_{l}\rho_l \lp \frac{1}{n-n_1} \sum_{i'} \varphi_l(X_{i'}) - \frac{1}{m-m_1} \sum_{j'} \varphi_l(Y_{j'}) \rp^2  } \\
                & \leq  \frac{1}{ \rho_1 \lp \frac{\sqrt{n_1}}{n-n_1} \sum_{i'} \varphi_1(X_{i'}) - \frac{\sqrt{n_1}}{m-m_1} \sum_{j'} \varphi_1(Y_{j'}) \rp^2  }  \label{eq:lemma1-proof-1}
            \end{align}
            Since $n_1 = n/2$, $m_1 = m/2$, we have $\sqrt{n_1}/(n-n_1) = \sqrt{2/n}$ and $\sqrt{n_1}/(m-m_1) = \sqrt{2n}/m$. Introduce the notation $u_{i'} = \sqrt{2/n}/\sqrt{1+n/m}$ and $v_{j'} = (\sqrt{2n}/m)/\sqrt{1+n/m}$, and note that 
            \begin{align}
                (A_2)^{1/2} &\leq \frac{1}{\lp 1 + \frac{n}{m} \rp \rho_1 \lp \sum_{i'} u_{i'} \varphi_1(X_{i'}) - \sum_{j'} v_{j'} \varphi_1(Y_{j'}) \rp^2  } \\
                &\leq \frac{1}{ \rho_1 \lp \sum_{i'} u_{i'} \varphi_1(X_{i'}) - \sum_{j'} v_{j'} \varphi_1(Y_{j'}) \rp^2  }. \label{eq:lemma1-proof-20} \\
            \end{align}
            Next, we note that 
            \begin{align}
                \lim_{n \to \infty} \max_{i', j'} \frac{ u_{i'}^2 + v_{j'}^2}{\sum_{i'} u_{i'}^2 + \sum_{j'}v_{j'}^2} &= \lim_{n \to \infty} \frac{2}{n+ n^2/m} + \frac{2}{m + m^2/n}  \\
                & \leq \lim_{n \to \infty} 2 \lp \frac{1}{n} + \frac{1}{m} \rp = 0. 
                \label{eq:lemma1-proof-21}
            \end{align}
            Thus, by an application of Lindeberg's CLT, we observe that the denominator in~\eqref{eq:lemma1-proof-20} converges in distribution to $N(0, \rho_1)^2$. This implies that $A_2 = \mc{O}_P(1)$, as required.

            \emph{Case~2: $\rho_1 =0$.} 
            Again, we observe that 
            \begin{align}
                (A_2)^{-1/2} =   \frac{n_1 \mathbb{E}_2[a_i^2]}{\mathbb{E}_{P_n}[\bark(X_1, X_2)^2]} + \frac{n_1 \mathbb{E}_2[b_i^2]}{\mathbb{E}_{P_n}[\bark(X_1, X_2)^2]} - 2\frac{n_1 \mathbb{E}_2[a_i b_i]}{\mathbb{E}_{P_n}[\bark(X_1, X_2)^2]}. 
            \end{align}
            The first two terms in the display above are $1 + o_P(1)$, as shown in \citep[pg 55, Step 2]{kim2020dimension}. For the last term, we introduce the notation $g(x, y) = \mathbb{E}_{P_n}[\bark(X, x)\bark(X, y)]$, and note the following: 
            \begin{align}
           R \defined \frac{n_1 \mathbb{E}_2[a_i b_i]}{\mathbb{E}_{P_n}[\bark(X_1, X_2)^2]} = \frac{n_1}{(n-n_1)(m-m_1)} \sum_{i', j'} g(X_{i'}, Y_{j'}). 
            \end{align}
            Since $X_{i'}$ and $Y_{j'}$ are independent, we observe that $\mathbb{E}_{P_n}[g(X_{i'}, Y_{j'})]=0$, and hence $\mathbb{E}_{P_n}[R]=0$. Furthermore, the variance of $R$ satisfies 
            \begin{align}
                \mathbb{E}_{P_n}[R^2] &=  \frac{n_1^2}{(n-n_1)(m-m_1)} \frac{ \mathbb{E}_{P_n}[g(X_1, X_2)^2] }{\mathbb{E}_{P_n}[\bark(X_1, X_2)^2]} \\ &=  \frac{n_1^2}{(n-n_1)(m-m_1)} \frac{ \mathbb{E}_{P_n}[\bark(X_1,X_3)^2 \bark(X_2, X_3)^2] }{\mathbb{E}_{P_n}[\bark(X_1, X_2)^2]^2} \\
                & = \frac{ \sum_{l} \lambda_l^4 }{ (\sum_{l} \lambda_l^2)^2} \leq \frac{ \lambda_1^2}{\sum_{l'} \lambda_{l'}^2}  \sum_{l} \frac{\lambda_l^2}{\sum_{l'} \lambda_{l'}^2} = \frac{ \lambda_1^2}{\sum_{l'} \lambda_{l'}^2} \rightarrow \rho_1 = 0. 
            \end{align}
            This implies that the term $R$ is $o_P(1)$, and hence we have 
            \begin{align}
                (A_2)^{1/2} = \frac{1}{2 + o_P(1)} = \calO_P(1),
            \end{align}
            as required. This completes the proof.
        \end{proof}
        
        Next, we show that we can use~\Cref{lemma:asymp-normality-1} to obtain the limiting distribution of the standardized statistic $T_s = \frac{S_X - S_Y}{\sqrt{n_1^{-1} \mathbb{E}[W_1^2|\Xsample_2, \Ysample_2] + m_1^{-1} \mathbb{E}[Z_1^2|\Xsample_2, \Ysample_2]}}$. 
        
        \begin{lemma}
        \label{lemma:asymp-normality-2} 
        Under the conditions of~\Cref{theorem:asymptotic-limit}, the standardized statistic $T_s$ converges in distribution to $N(0,1)$. %
        \end{lemma}
        
        \begin{proof}
        This statement simply follows from the observation that $\mathbb{E}_2[Z_1^2] = \mathbb{E}_2[W_1^2]$ almost surely under the null hypothesis. Then, the term $\alpha_{n} \defined (\sqrt{n_1^{-1} \mathbb{E}_2[W_1^2]})/(\sqrt{n_1^{-1} \mathbb{E}_2[W_1^2] + m_1^{-1} \mathbb{E}_2[Z_1^2]}) = \sqrt{1/(1+n_1m_1^{-1})}$ converges to a constant (say $\alpha \in (0,1)$).  
        
        Using the result of~\Cref{lemma:asymp-normality-1}, we can then conclude that $\alpha_{n} T_{s,X} \convdist N(0, \alpha^2)$ and $\sqrt{1-\alpha_{n}^2} T_{s,Y} \convdist N(0, 1-\alpha^2)$. This implies, due to \levy's continuity theorem~\citep[Theorem~3.3.17.~(i)]{durrett2019probability}, the pointwise convergence of the characteristic functions of these sequences. In particular, let $\psi_{n,X}$ and $\psi_{n,Y}$ denote the characteristic functions of $\alpha_{n}T_{s,X}$ and $\sqrt{1-\alpha_{n}^2} T_{s,Y}$ respectively. Then, due to the conditional independence of $T_{s,X}$ and $T_{s,Y}$ given $(\Xsample_2, \Ysample_2)$, we note that the characteristic function of $T_s = \alpha_{n}T_{s,X} = \sqrt{1-\alpha_{n}^2} T_{s,Y}$, denoted by $\psi_{n}(t)$, satisfies
        \begin{align}
            \psi_{n}(t)  & \defined \mathbb{E}_{P_n}[\exp \lp i t T_s \rp | \Xsample_2, \Ysample_2] \\
            &= \mathbb{E}_{P_n}[\exp \lp i t\, \alpha_{n} T_{s,X} \rp | \Xsample_2, \Ysample_2] \times \mathbb{E}_{P_n}\left[\exp \lp -i t\, \sqrt{1-\alpha_{n}^2} T_{s,Y} \rp \vert \Xsample_2, \Ysample_2\right] \\
            & = \psi_{n,X}(t) \times  \psi_{n, Y}(-t). 
        \end{align}
        Now, taking the limit $n \to \infty$, we get that 
        \begin{align}
            \lim_{n \to \infty} \psi_{n}(t) & = \lim_{n \to \infty} \psi_{n, X}(t) \times  \psi_{n, Y}(-t) \\
            & = \exp \lp -\frac{1}{2}\lp \alpha^2 t^2 \rp \rp \times \exp \lp - \frac{1}{2} \lp (1-\alpha^2)t^2  \rp \rp \\
            & = \exp \lp -\frac{t^2}{2} \rp. 
        \end{align}
        Thus, we have shown that conditioned on $(\Xsample_2, \Ysample_2)$, the characteristic function, $\psi_{n}$ of $T_{s}$ converges pointwise to the characteristic function of a $N(0, 1)$ distribution. Hence, by the other direction of \levy's continuity theorem~\citep[Theorem~3.3.17.~(ii)]{durrett2019probability}, we conclude that $T_s \convdist N(0,1)$. 
        
        Finally, we pass from the conditional statement to the unconditional one by noting that $T_s \convdist N(0,1)$ conditioned on $(\Xsample_2, \Ysample_2)$ implies that $\sup_{x \in \reals} |\Pr_{P_n}(T_s \leq x) - \Phi(x)| \convprob 0$, because the $N(0,1)$ distribution is continuous. This fact, coupled with the boundedness of $\sup_{x \in \reals}|\Pr_{P_n}(T_s\leq x) - \Phi(x)|$ implies that it also converges in expectation, as required. 
        Thus, we have shown that the limiting distribution of the standardized statistic $T_s$ is $N(0, 1)$ unconditionally. 
        \end{proof}
        
        We now prove that the studentized statistic also has the same limiting distribution as the standardized statistic $T_s$ by appealing to Slutsky's theorem and the continuous mapping theorem. 
        \begin{lemma}
        \label{lemma:asymp-normality-3}
        The ratio of $\sigmahat^2$ and the conditional variance $n_1^{-1} \mathbb{E}_2[W_1^2] + m_1^{-1} \mathbb{E}_2[Z_1^2]$ converges in probability to $1$. Stated formally, 
        \begin{align}
            \frac{ n_1^{-1} \sigmahat_X^2 + m_1^{-1} \sigmahat_Y^2}{n_1^{-1} \mathbb{E}_2[W_1^2] + m_1^{-1} \mathbb{E}_2[Z_1^2]} \convprob 1. 
        \end{align}
        Recall that we use the notation $\mathbb{E}_{2}[\cdot]$ to denote the conditional expectation on the second half of the data, i.e., $\mathbb{E}_{P_n}[\cdot|\Xsample_2, \Ysample_2]$. 
        \end{lemma}
        
        \begin{proof}
        Since $\mathbb{E}_2[W_1^2] = \mathbb{E}_2[Z_1^2]$ almost surely, it suffices to show the following two statements to conclude the result: 
        \begin{align}
            \frac{ \sigmahat_X^2}{\mathbb{E}_2[W_1^2]} \convprob 1, \quad \text{and} \quad \frac{ \sigmahat_Y^2}{\mathbb{E}_2[Z_1^2]} \convprob 1. 
        \end{align}
        We provide the details of the first statement, since the second can be obtained similarly.  Consider the following: 
        \begin{align}
            \frac{ (n_1-1)^{-1} \sum_{i=1}^{n_1} (W_i- \bar{U}_X)^2 - \mathbb{E}_2[W_1^2]}{\mathbb{E}_2[W_1^2] }  &= 
            \frac{ \sum_{i=1}^{n_1} (W_i- \bar{U}_X)^2 - (n_1-1)\mathbb{E}_2[W_1^2]}{\mathbb{E}_{P_n}[\bark^2(X_1, X_2)]} \times \frac{\mathbb{E}_{P_n}[\bark^2(X_1, X_2)]}{(n_1-1)\mathbb{E}_2[W_1^2] } \\
            & = C_1 \times C_2. 
        \end{align}
        Note that $C_2 = \frac{n_1}{n_1-1} \sqrt{A_2}$, where $A_2$ was introduced in~\eqref{eq:proof-lemma-1} and shown to be $O_P(1)$ in  the proof of~\Cref{lemma:asymp-normality-1}. Hence, to complete the proof, we will show that $C_1 \convprob 0$. This can be concluded by noting that $\mathbb{E}_{P_n}[C_1]=0$, and that the variance of $C_1$ satisfies: 
        \begin{align}
            \var_{P_n}[C_1] &= \mathbb{E}_{P_n}[ \var_{P_n}[C_1|\Xsample_2, \Ysample_2] ] + \var_{P_n}[ \mathbb{E}_{P_n}[C_1|\Xsample_2, \Ysample_2] ] \\
            & = \frac{(n_1-1)^2}{\mathbb{E}_{P_n}[\bark^2(X_1, X_2)]^2} \mathbb{E}_{P_n}\left[ \var_{P_n}\left[\frac{1}{n_1-1} \sum_{i=1}^{n_1} (W_i - \bar{U}_X)^2 \right] \right] \\
            &  \leq  \frac{(n_1-1)^2}{\mathbb{E}_{P_n}[\bark^2(X_1, X_2)]^2} \frac{ \mathbb{E}_{P_n}[W_1^4]}{n_1} \leq \frac{n_1  \mathbb{E}_{P_n}[W_1^4]}{\mathbb{E}_{P_n}[\bark^2(X_1, X_2)]^2}  \leq 16 \frac{n_1 \mathbb{E}_{P_n}[a_1^4 + b_1^4]}{\mathbb{E}_{P_n}[\bark^2(X_1, X_2)]^2} \\
            & = 16 \lp A_1 + B_1 \rp, 
        \end{align}
        where the terms $A_1$ and $B_1$ were introduced in~\eqref{eq:proof-lemma-0}. As mentioned during the proof of~\Cref{lemma:asymp-normality-1}, both of these terms can be shown to converge in probability  to $0$ as required. 
        \end{proof}   
        
        The previous three lemmas prove that for any sequence $\{P_n: n \geq 1\}$ with $P_n \in \nullclass$, we have $\lim_{n \to \infty} \sup_{x \in \mathbb{R}}|\mathbb{P}_{P_n}\lp \csmmd \leq x \rp - \Phi(x)| = 0$. This is sufficient to conclude the uniform result 
        \begin{align}
            \lim_{n \to \infty} \sup_{P_n \in \nullclass} \sup_{x \in \mathbb{R}} |\mathbb{P}_{P_n} \lp \csmmd \leq x \rp - \Phi(x)| = 0. 
        \end{align}
        This is because we can select a sequence $P_n'$ such that for all $n$, we have 
        \begin{align}
            \sup_{x \in \mathbb{R}} | \mathbb{P}_{P_n'} \lp \csmmd \rp - \Phi(x)| \leq & \sup_{P_n \in \nullclass} \sup_{x \in \mathbb{R}} | \mathbb{P}_{P_n} \lp \csmmd \rp - \Phi(x)| \\
            \leq & \sup_{x \in \mathbb{R}} | \mathbb{P}_{P_n'} \lp \csmmd \rp - \Phi(x)| + \frac{1}{n}. 
        \end{align}
        Since the left and right terms converge to zero, it follows that the middle term does too, as required. This completes the proof of~\Cref{theorem:asymptotic-limit}.

    \subsection{Fixed \texorpdfstring{$P$}{P}, changing \texorpdfstring{$k_n$}{kn}~(Theorem~\ref{theorem:asymptotic-normality-2})}
    \label{subsec:proof-fixed-P}
        We note that the statement of~\Cref{theorem:asymptotic-normality-2} requires an additional technical assumption on the eigenvalues of the kernel operator, introduced in~\eqref{eq:h1-general}. We repeat the statement of~\Cref{theorem:asymptotic-normality-2} with this additional requirement below.             
        \begin{manualtheorem}{5'}
            \label{theorem:repeat}
            Suppose $P$ is fixed, but the kernel $k_n$ changes with $n$. If
            \begin{align}
                \label{eq:kernel-condition-weak-repeat}
                \lim_{n \to \infty} \frac{\mathbb{E}_P[\bark_n(X_1,X_2)^4]}{ \mathbb{E}_P[\bark_n(X_1, X_2)^2]^2} \lp \frac{1}{n} + \frac{1}{m_n} \rp  = 0,   \quad \text{and} \quad \lim_{n \to \infty} \frac{\lambda_{1,n}^2}{\sum_{l=1}^{\infty} \lambda_{l,n}^2} \text{ exists}, 
            \end{align}
            then we have $\csmmd \convdist N(0,1)$. 
        \end{manualtheorem}
        
        \begin{proof}
            The proof of this statement will follow the general outline of the proof of~\Cref{theorem:asymptotic-limit}. However, in this special case when $P$ is fixed, we can remove the condition that $\lim_{n \to \infty} m_n/n$ exists and is non-zero, that is required by~\Cref{theorem:asymptotic-limit}. 
            
            We will carry over the notations used in the proof of~\Cref{theorem:asymptotic-limit}, and in particular, we will use $\crossU_X = \frac{1}{n_1} \sum_{i=1}^{n_1} W_i$ and $\crossU_Y = \frac{1}{m_1} \sum_{j=1}^{m_1} Z_j$. Since $W_i$ and $Z_j$ are identically distributed under the null, we have $\mathbb{E}_{P}[W_i^2|\Xsample_2, \Ysample_2] = \mathbb{E}_P[Z_j^2|\Xsample_2, \Ysample_2]$, and we will use $\sigma_2^2$ to denote this conditional variance. Then, note the following: 
            \begin{align}
                \csmmd &= \frac{\crossU_X - \crossU_Y}{\sigmahat} = \frac{\crossU_X - \crossU_Y}{\sigma_2 \lp \sqrt{n_1^{-1} + m_1^{-1} }\rp} \times \frac{\sigma_2 \lp \sqrt{n_1^{-1} + m_1^{-1} }\rp}{\sigmahat} \\ 
                & \defined T_1 \times T_2. \label{eq:proof-asymp-normal-1}
            \end{align}
            To complete the proof, we will show that $T_1 \convdist N(0,1)$ and $T_2 \convprob 1$. The result then follows by an application of Slutsky's theorem. 
            
            First, we consider the term $T_1$ in~\eqref{eq:proof-asymp-normal-1}. Let $\Wtilde_i \defined W_i/\sigma_2$ and $\Ztilde_j \defined Z_j/\sigma_2$. Then, conditioned on $(\Xsample_2, \Ysample_2)$, the terms $\Wtilde_i$ and $\Ztilde_j$ are independent and identically distributed. Introducing the constants $u_i = \sqrt{ \frac{m_1}{n_1(m_1+n_1)}}$ and $v_j = \sqrt{ \frac{n_1}{m_1(m_1+n_1)}}$, we can write 
            \begin{align}
                T_1 = \sum_{i=1}^{n_1} u_i \Wtilde_i - \sum_{j=1}^{m_1} v_j \Ztilde_j. 
            \end{align}
            We can check that the constants $(u_i)$ and $(v_j)$ satisfy the property:
            \begin{align}
                \lim_{n \to \infty} \max_{i, j} \frac{ u_i^2 + v_j^2}{ \sum_{i'=1}^{n_1} u_{i'}^2  + \sum_{j'=1}^{m_1} v_{j'}^2 } \leq \lim_{n \to \infty} \max_{i, j}  \frac{1}{m_1} + \frac{1}{n_1} = 0. 
            \end{align}
            Thus, by an application of Lindeberg's CLT, we note that $T_1 \convdist N(0,1)$ conditioned on $(\Xsample_2, \Ysample_2)$. Since the limiting distribution~(in this case, standard normal) is continuous, this also means that the $T_1$ converges to $N(0,1)$ in the Kolmogorov-Smirnov metric, that is, $\lim_{n \to \infty} \sup_{x \in \mathbb{R}}|\mathbb{P}_P\lp T_1 \leq x | \Xsample_2, \Ysample_2 \rp - \Phi(x)| \convprob 0$. Since the random variable $\sup_{x \in \mathbb{R}} |\mathbb{P}_P\lp T_1 \leq x | \Xsample_2, \Ysample_2 \rp - \Phi(x)|$ is bounded, convergence in probability implies that $\lim_{n \to \infty} \mathbb{E}_P\lb \sup_{x \in \mathbb{R}}|\mathbb{P}_P\lp T_1 \leq x | \Xsample_2, \Ysample_2 \rp - \Phi(x)| \rb = 0$, which in turn implies that $\lim_{n \to \infty} \sup_{x \in \mathbb{R}} | \mathbb{E}_P\lb \mathbb{P}_P\lp T_1 \leq x | \Xsample_2, \Ysample_2 \rp - \Phi(x)\rb |  = 0$, as required.

        We now consider the second term, $T_2$, in~\eqref{eq:proof-asymp-normal-1}. It remains to show that $T_2 \convprob 1$. We will show that $1/T_2^2 -1 \convprob 0$, and the result will follow by an application of the continuous mapping theorem. 
        \begin{align}
            \left \lvert \frac{1}{T_2^2} -1\right \rvert = \left \lvert \frac{ \frac{ \sigmahat_X^2}{n_1}  + \frac{ \sigmahat_Y^2}{m_1} }{ \sigma_2^2 \lp \frac{1}{n_1} + \frac{1}{m_1} \rp } - 1 \right \rvert \leq \left \lvert \frac{\sigmahat_X^2}{\sigma_2^2} - 1 \right \rvert +  \left \lvert \frac{\sigmahat_Y^2}{\sigma_2^2} - 1 \right \rvert. \label{eq:proof-asymp-normal-3}
        \end{align}
        Thus, it suffices to show that both terms in~\eqref{eq:proof-asymp-normal-3} converge in probability to $0$. This is exactly the result that is proved in~\Cref{lemma:asymp-normality-3} under the two conditions listed in~\Cref{assump:kernel-1}. The condition on eigenvalues is already assumed in the statement of~Theorem~\ref{theorem:repeat}, and thus we will show that the condition on the kernels, stated in~\eqref{eq:kernel-condition-weak-repeat}, implies the condition~\eqref{eq:assump-kernel-1}. To prove this, we first, we note that 
        \begin{align}
            \mathbb{E}_P \lb \bark_n(X_1, X_2)^2 \bark_n(X_1, X_3)^2\rb &\leq \mathbb{E}_P \lb \bark_n(X_1, X_2)^4 \rb^{1/2} \mathbb{E}_P \lb \bark_n(X_1, X_3)^4 \rb^{1/2} \\
            & = \mathbb{E}_P \lb \bark_n(X_1, X_2)^4 \rb. 
        \end{align}
        Thus, the term in~\eqref{eq:assump-kernel-1} is upper bounded by 
        \begin{align}
            \frac{ \mathbb{E}_P \lb \bark_n(X_1, X_2)^4 \rb}{ \mathbb{E}_P \lb \bark_n(X_1, X_2)^2 \rb^2}\lp \frac{1}{n} + \frac{1}{m_n} \rp \lp 1 + \frac{1}{n} + \frac{1}{m_n} \rp. 
        \end{align}
        Since, we have assumed that $\lim_{n \to \infty} m_n \to \infty$, there exists and $n_0$, such that for all $n\geq n_0$, $1+ \frac{1}{n} + \frac{1}{m_n} \leq 2$. This implies that if~\eqref{eq:kernel-condition-weak-repeat} is satisfied, then~\eqref{eq:assump-kernel-1} in~\Cref{assump:kernel-1} is also satisfied, as required. 
        \end{proof}            
        
    \subsection{Fixed \texorpdfstring{$k$}{k}, and fixed \texorpdfstring{$P$}{P}~(Theorem~\ref{prop:simple-asymptotic-normality})}
        \label{subsec:proof-fixed-k-P}
        We prove~\Cref{prop:simple-asymptotic-normality} by showing that under the bounded fourth moment assumption on $\bark$,  both the conditions required by~Theorem~\ref{theorem:repeat} are satisfied. 
        
        Note that since $\mathbb{E}_P[\bark(X_1, X_2)]=0$, the positive and finite fourth moment also implies that the second moment of $\bark(X_1, X_2)$ is also positive and finite. Hence, we have that 
        \begin{align}
            \frac{ \mathbb{E}_P[\bark(X_1, X_2)^4]}{\mathbb{E}_P[\bark(X_1, X_2)^2]^2} < \infty. 
        \end{align}
        This, in turn, implies 
        \begin{align}
            \lim_{n \to \infty} \frac{ \mathbb{E}_P[\bark(X_1, X_2)^4]}{\mathbb{E}_P[\bark(X_1, X_2)^2]^2} \lp \frac{1}{n} + \frac{1}{m_n} \rp = 0, 
        \end{align}
        as required by~\Cref{theorem:asymptotic-normality-2}. 
        
        For the second part of the condition, we note that as kernel $k$ and probability distribution $P$ are fixed, the term $\frac{\lambda_1^2}{\sum_{l}\lambda_l^2}$ doesn't change with $n$, and hence its limit exists. Thus, both the conditions for~Theorem~\ref{theorem:repeat} are satisfied, as required.

\section{Consistency against fixed and local alternatives~(Section~\ref{sec:conclusion})}
\label{appendix:proof-consistency}

    \subsection{Proof of~Theorem~\ref{prop:general-consistency}~(General conditions for consistency)}
     
    \begin{proof}
        
        We begin by noting that 
    \begin{align}
        \label{eq:smooth-power-proof-1}
        \mathbb{E}_{P_n,Q_n}[1-\Psi(\Xsample, \Ysample)] & = \mathbb{P}_{P_n,Q_n}\lp \csmmd \leq z_{1-\alpha} \rp = \mathbb{P}_{P_n,Q_n}\lp \crossmmd \leq z_{1-\alpha} \sigmahat \rp. 
    \end{align}
    Now,  introduce the event $\calE = \{ \sigmahat^2 \leq \mathbb{E}[\sigmahat^2]/\delta_{n} \}$, where $(\delta_{n})$ is a positive sequence converging to zero. By an application of Markov's inequality, we have $\mathbb{P}_{P_n,Q_n}\lp \calE^c \rp \leq \delta_{n}$, which implies that
    \begin{align}
        \mathbb{P}_{P_n,Q_n}\lp \crossmmd \leq z_{1-\alpha} \sqrt{\sigmahat^2} \rp &= \mathbb{P}_{P_n,Q_n}\lp \{ \crossmmd \leq z_{1-\alpha} \sqrt{\sigmahat^2} \} \cap \calE \rp \\
        &\qquad + \mathbb{P}_{P_n,Q_n}\lp \{ \crossmmd \leq z_{1-\alpha} \sqrt{\sigmahat^2} \} \cap \calE^c \rp \\
        & \leq \mathbb{P}_{P_n,Q_n} \lp \crossmmd \leq z_{1-\alpha} \sqrt{ \mathbb{E}_{P_n,Q_n}[\sigmahat^2]/\delta_{n}} \rp + \mathbb{P}_{P_n,Q_n} \lp \calE^c \rp \\
        & \leq \mathbb{P}_{P_n,Q_n} \lp \crossmmd \leq z_{1-\alpha} \sqrt{ \mathbb{E}_{P_n,Q_n}[\sigmahat^2]/\delta_{n}} \rp + \delta_{n}.  \label{eq:proof-consistency-1}
    \end{align}
    By the assumption that $\delta_{n} \to 0$, it suffices to show that the worst-case value of the first term in~\eqref{eq:proof-consistency-1} converges to zero to complete the proof. 
    
    To do this, we observe that~\eqref{eq:general-consistency-conditions} implies that there exists a finite value of $n$, say $n_0$,  such that for all $n \geq n_0$ and $m\geq m_{n_0}$, we have 
    \begin{align}
        \sup_{(\Pnm, \Qnm) \in \altclass} \frac{\mathbb{E}_{P_n,Q_n}[\sigmahat^2]}{\mmdval^4 \delta_{n}} \leq \frac{1}{4 z_{1-\alpha}^2},  
    \end{align}
    which implies that $z_{1-\alpha} \sqrt{\mathbb{E}_{P_n,Q_n}[\sigmahat^2]/\delta_n} \leq \mmdval^2/2$. Furthermore, since $\crossmmd = \langle \muhat_{1}-\nuhat_{1}, \muhat_{2} - \nuhat_{2}\rangle_k$, it follows that $\mathbb{E}_{P_n,Q_n}[\crossmmd] = \mmdval^2$. Combining these two observations, we get  for all $n\geq n_0$: 
    \begin{align}
        \mathbb{P}_{P_n,Q_n}\lp \crossmmd  \leq z_{1-\alpha}\sqrt{\mathbb{E}_{P_n,Q_n}[\sigmahat^2]/\delta_{n}} \rp &\leq \mathbb{P}_{P_n,Q_n} \lp \crossmmd- \mathbb{E}_{P_n,Q_n}[\crossmmd] \leq \frac{\mmdval^2}{2} - \mmdval^2 \rp   \\
        &\stackrel{(i)}{\leq}  4\frac{ \var_{P_n,Q_n}(\crossmmd) }{\mmdval^4}, 
    \end{align}
    where \texttt{(i)} follows from Chebychev's inequality. This implies that 
    \begin{align}
        \sup_{(\Pnm, \Qnm) \in \altclass} \mathbb{P}_{P_n,Q_n} \lp \csmmd < z_{1-\alpha} \rp \leq \sup_{(\Pnm, \Qnm) \in \altclass} 4 \frac{\var_{P_n,Q_n}(\crossU)}{\mmdval^4}. 
    \end{align}
    The required conclusion that $\sup_{(\Pnm, \Qnm) \in \altclass} \mathbb{P}_{P_n,Q_n}(\csmmd \leq z_{1-\alpha}) \to 0$ now follows from the second term in~\eqref{eq:general-consistency-conditions}. 
    \end{proof}
    
    \subsection{Proof of~Theorem~\ref{corollary:fixed-alternative}~(Consistency against fixed alternative)}
        We prove~\Cref{corollary:fixed-alternative} by showing that the sufficient conditions for consistency, as derived in~\Cref{prop:general-consistency}, are satisfied under the assumptions of~\Cref{corollary:fixed-alternative}. 
        
        First, since the kernel is assumed to be characteristic, and $\Pnm=P \neq Q = \Qnm$, it means that the kernel-MMD distance between $P$ and $Q$ must be strictly positive. In other words, we have $\mmdval = \dmmd(P, Q) \defined \gamma >0$ for all $n \geq 1$. Hence, in order to verify the condition~\eqref{eq:general-consistency-conditions}, it suffices to show that the following two properties hold: 
        \begin{align}
            &\lim_{n \to \infty} \mathbb{E}_{P,Q}[\sigmahat^2] = \lim_{n \to \infty} \frac{2}{n} \mathbb{E}_{P,Q}[\sigmahat_X^2] + \frac{2}{m_n} \mathbb{E}_{P,Q}[\sigmahat_Y^2] = 0, \quad \text{and} \label{eq:consistency-simple-condition-1} \\
            &\lim_{n \to \infty} \var_{P,Q} \lp \crossmmd \rp = 0. \label{eq:consistency-simple-condition-2}
        \end{align}
        In the equality in~\eqref{eq:consistency-simple-condition-1}, we used the fact that $n_1 = n/2$ and $m_1 = m_n/2$~(see~\Cref{remark:notation}). 
        
        \paragraph{Verifying~\eqref{eq:consistency-simple-condition-1}.} We begin by noting that it suffices to show that $\mathbb{E}_{P,Q}[\sigmahat_X^2]<\infty$ and $\mathbb{E}_{P,Q}[\sigmahat_Y^2]<\infty$ to conclude~\eqref{eq:consistency-simple-condition-1}~(this is because we have assumed in~\Cref{remark:notation} that $\lim_{n \to \infty} m_n = \infty$). We present the details for $\sigmahat_X^2$ as the same arguments can be used to conclude the result for $\sigmahat_Y^2$. 
        
        Recall that $\sigmahat_X^2 = \frac{1}{n_1} \sum_{i=1}^{n_1} \lp \langle k(X_i, \cdot), g_2 \rangle_k - \crossU_X \rp^2$, where $g_2 = \muhat_2 - \nuhat_2$. Since $X_1, \ldots, X_{n_1}$ are \iid, this implies that 
        \begin{align}
            \mathbb{E}_{P,Q}[\sigmahat_X^2] &= \mathbb{E}_{P,Q}\lb \frac{1}{n_1} \sum_{i=1}^{n_1} \lp \langle k(X_i, \cdot), g_2 \rangle_k - \crossU_X \rp^2 \rb = \mathbb{E}_{P,Q}\lb \lp \langle k(X_1, \cdot), g_2 \rangle_k - \crossU_X \rp^2 \rb \\ 
            & = \mathbb{E}_{P,Q}\lb \langle k(X_1, \cdot) - \muhat_1, g_2 \rangle_k^2 \rb =  \mathbb{E}_{P,Q}\lb \langle k(X_1, \cdot) - \muhat_1, \muhat_2 - \nuhat_2 \rangle_k^2 \rb \label{eq:consistency-simple-temp1} \\
            & \leq \mathbb{E}_{P,Q}\lb \| k(X_1, \cdot) - \muhat_1 \|_k^2 \|\muhat_2 - \nuhat_2 \|_k^2 \rb \label{eq:consistency-simple-temp2}\\
            & \leq \mathbb{E}_{P,Q}\lb \| k(X_1, \cdot) - \muhat_1 \|_k^2 \rb \, \mathbb{E}_{P,Q}\lb \|\muhat_2 - \nuhat_2 \|_k^2 \rb \label{eq:consistency-simple-temp3}\\
            & \leq \lp 2 \mathbb{E}_{P,Q}\lb \|k(X_1, \cdot) \|_k^2 + \|\muhat_1 \|_k^2 \rb \rp \times \lp 2 \mathbb{E}_{P,Q}\lb \|\muhat_2 \|_k^2 + \|\nuhat_2 \|_k^2 \rb \rp \label{eq:consistency-simple-temp4} \\
            & \leq \lp 4 \mathbb{E}_{P,Q}\lb k(X_1, X_1) \rb \rp \times \lp 2 \mathbb{E}_{P,Q}\lb k(X_2, X_2) + k(Y_1, Y_1) \rb \rp < \infty. \label{eq:consistency-simple-temp5} 
        \end{align}
        In the above display: \\
        \eqref{eq:consistency-simple-temp1} uses the fact that $\crossmmd_X = \langle \muhat_1, g_2 \rangle_k =  \langle \muhat_1, \muhat_2 - \nuhat_2 \rangle_k$, and the linearity of inner product, \\
        \eqref{eq:consistency-simple-temp2} uses the Cauchy--Schwarz inequality, \\
        \eqref{eq:consistency-simple-temp3} uses the fact that the two terms inside the expectation are independent, \\
        \eqref{eq:consistency-simple-temp4} uses the fact that $ \|a - b\|_k^2 \leq \lp \|a\|_k + \|b\|_k \rp^2 \leq 2\lp \|a\|_k^2 + \|b\|_k^2 \rp$,  and \\
        \eqref{eq:consistency-simple-temp5} uses the facts that $\|k(X_1, \cdot)\|_k^2 = k(X_1, X_1)$, $\mathbb{E}_{P,Q}[\|\muhat_1\|_k^2] \leq \mathbb{E}_{P,Q}[k(X_1, X_1)]$, $\mathbb{E}_{P,Q}[\|\muhat_2\|_k^2] \leq \mathbb{E}_{P,Q}[k(X_2, X_2)]$ for $X_2 \sim P$ independent of $X_1$ and $\mathbb{E}_{P,Q}[\|\muhat_2\|_k^2]$ and $\mathbb{E}_{P,Q}[\|\nuhat_2\|_k^2] \leq \mathbb{E}_{P,Q}[k(Y_1, Y_1)]$ for $Y_1 \sim Q$. We show the details for the bound for $\mathbb{E}_{P,Q}[\|\muhat_1\|_k^2]$ below: 
        \begin{align}
            \mathbb{E}_{P,Q}[\|\muhat_1\|_k^2] &= \mathbb{E}_{P,Q}\lb \frac{4}{n^2} \sum_{i=1}^{n/2}\sum_{l=1}^{n/2} \langle k(X_i, \cdot), k(X_l, \cdot) \rangle_k \rb  \\
            &\leq \frac{4}{n^2} \sum_{i=1}^{n/2}\sum_{l=1}^{n/2}  \lp \mathbb{E}_{P,Q}[k(X_i, X_i)] \mathbb{E}_{P,Q}[k(X_l, X_l)] \rp^{1/2} \\ &= \mathbb{E}_{P,Q}[k(X_1, X_1)],  
        \end{align}
        where the inequality follows from an application of Cauchy--Schwarz inequality. The bounds for $\mathbb{E}_{P,Q}[\|\muhat_2\|_k^2]$ and $\mathbb{E}_{P,Q}[\|\nuhat_2\|_k^2]$ also follow from the same steps. 
        
        Thus, we have shown that $\mathbb{E}_{P,Q}[\sigmahat_X^2]< \infty$. The result for $\mathbb{E}_{P,Q}[\sigmahat_Y^2]$ follows in an analogous manner.

        \paragraph{Verifying~\eqref{eq:consistency-simple-condition-2}.} We begin by noting that the expected value of $\crossmmd = \langle \muhat_1 - \nuhat_1, \muhat_2 - \nuhat_2 \rangle_k = \crossU_X - \crossU_Y$ is equal to $\dmmd^2(P, Q) = \|\mu-\nu\|_k^2 = \gamma^2$. Thus, we have 
        \begin{align}
            \var_{P,Q}(\crossmmd) &= \mathbb{E}_{P,Q}\lb \lp \crossmmd - \langle \mu - \nu, \mu - \nu \rangle_k \rp^2 \rb \\
            & = \mathbb{E}_{P,Q}\lb \lp \lp  \crossU_X - \langle \mu, \mu- \nu \rangle_k \rp - \lp  \crossU_Y - \langle \nu, \mu- \nu \rangle_k \rp \rp^2 \rb \\
            & = 2\mathbb{E}_{P,Q}\lb  \lp  \crossU_X - \langle \mu, \mu- \nu \rangle_k \rp^2 \rb   + 2\mathbb{E}_{P,Q}\lb \lp  \crossU_Y - \langle \nu, \mu- \nu \rangle_k  \rp^2 \rb. \label{eq:consistency-simple-temp6}
        \end{align}
        We present the details for showing that the first term in~\eqref{eq:consistency-simple-temp6} converges to $0$ with $n$. The result for the second term can be proved similarly. 

        Before proceeding, we introduce some notation: we will use $\mutilde_1$ to denote $\muhat_1 - \mu$, the centered version of $\muhat_1$. Similarly, we will use $\mutilde_2$, $\nutilde_1$, $\nutilde_2$ and $\gtilde_2$ to represent $\muhat_2-\mu$, $\nuhat_1 - \nu$, $\nuhat_2-\nu$ and $g_2 - (\mu - \nu)$ respectively. With these notations, note that we can write 
        \begin{align}
            \mathbb{E}_{P,Q}\lb \lp \crossU_X - \langle \mu, \nu - \mu \rangle_k \rp^2 \rb &= \mathbb{E}_{P,Q} \lb \lp     \langle \mutilde_1, g_2 \rangle_k + \langle \mu, \gtilde_2 \rangle_k \rp^2 \rb\\ 
            &\leq 2 \mathbb{E}_{P,Q} \lb   \langle \mutilde_1, g_2 \rangle_k^2 \rb  + 2\mathbb{E}_{P,Q}\lb \langle \mu, \gtilde_2 \rangle_k^2 \rb.   \label{eq:consistency-simple-temp7}
        \end{align}
        
        We now show that the first term of~\eqref{eq:consistency-simple-temp7} is $\mc{O}(1/n)$. 
        \begin{align}
            \mathbb{E}_{P,Q}\lb \langle \mutilde_1, g_2 \rangle_k^2 \rb &\leq \mathbb{E}_{P,Q}\lb \|\mutilde_1\|_k^2 \|\muhat_2-\nuhat_2\|_k^2 \rb \leq \mathbb{E}_{P,Q}[\|\mutilde_1\|_k^2] \mathbb{E}_{P,Q}\lb 2\lp \|\muhat_2\|_k^2 + \|\nuhat_2\|_k^2 \rp\rb\\ 
            &\leq \mathbb{E}_{P,Q}\lb \|\mutilde_1\|_k^2 \rb \lp 2 \mathbb{E}_{P,Q}[k(X_1, X_1)] + 2 \mathbb{E}_{P,Q}[k(Y_1, Y_1)]  \rp \label{eq:consistency-simple-temp8} \\
            & = \mc{O}\lp \mathbb{E}_{P,Q} \lb \frac{4}{n^2} \sum_{i=1}^{n/2} \sum_{l=1}^{n/2}\langle \tildek(X_i, \cdot), \tildek(X_l, \cdot)  \rangle_k\rb \rp \label{eq:consistency-simple-temp9} \\
            & = \mc{O}\lp \mathbb{E}_{P,Q} \lb \frac{4}{n^2} \sum_{i=1}^{n/2} \langle \tildek(X_i, \cdot), \tildek(X_i, \cdot) \rangle_k \rb \rp \label{eq:consistency-simple-temp10} \\
            & = \mc{O}\lp  \frac{2}{n}  \mathbb{E}_{P,Q} \lb k(X_1, X_1) - \|\mu\|_k^2  \rb \rp = \mc{O}\lp \frac{1}{n} \rp. \label{eq:consistency-simple-temp11}  \\
        \end{align}
        In the above display: \\
        \eqref{eq:consistency-simple-temp8} bounds $\mathbb{E}_{P,Q}[\|\muhat_2\|_k^2]$ with $\mathbb{E}_{P,Q}[k(X_1, X_1)]$ and $\mathbb{E}_{P,Q}[\|\nuhat_2\|_k^2]$ with $\mathbb{E}_{P,Q}[k(Y_1, Y_1)]$ following the same argument as in~\eqref{eq:consistency-simple-temp5}. \\
        \eqref{eq:consistency-simple-temp9} simply expands $\|\mutilde_1\|_k^2$, and \\
        \eqref{eq:consistency-simple-temp10} uses the fact that for $l\neq i$, we have $\mathbb{E}_{P,Q}[\langle \tildek(X_i, \cdot), \tildek(X_l, \cdot) \rangle_k h]= 0$. 
        
        We next show that the second term in~\eqref{eq:consistency-simple-temp7} is  $\mc{O}(1/n + 1/m_n)$. 
        \begin{align}
            \mathbb{E}_{P,Q} \lb \langle \mu, \gtilde_2 \rangle_k^2 \rb  & \leq 2\mathbb{E}_{P,Q}[\|\mu\|_k^2] \lp \mathbb{E}_{P,Q}\lb \|\mutilde_2\|_k^2 + \|\nutilde\|_k^2 \rb \rp \\
            &\leq 2 \mathbb{E}_{P,Q}[\|\mu\|_k^2] \lp  \frac{2}{n} \mathbb{E}_{P,Q}[k(X_1, X_1) - \|\mu\|_k^2] + \frac{2}{m_n} \mathbb{E}_{P,Q}[k(Y_1, Y_2) - \|\nu\|_k^2] \rp \\
            & = \mc{O}\lp \frac{1}{n} + \frac{1}{m_n} \rp. 
        \end{align}
    
    Thus, since $\lim_{n \to \infty} m_n = \infty$, both the terms in~\eqref{eq:consistency-simple-temp7} converge to $0$ as $n$ goes to infinity. This completes the proof that $\lim_{n \to \infty} \mathbb{E}_{P,Q}[(\crossU_X - \langle \mu, \mu - \nu \rangle_k)^2] = 0$. We can use the same arguments to show that $\lim_{n \to \infty}\mathbb{E}_{P,Q}[(\crossU_Y - \langle \nu, \mu - \nu \rangle_k)^2] = 0$. Together, these two statements imply that $\lim_{n \to \infty} \var_{P,Q}(\crossmmd) = 0$ following~\eqref{eq:consistency-simple-temp6}.

    \subsection{Proof of~Theorem~\ref{prop:smooth-alternative}~(Type-I error control and consistency against local alternative)}
            \paragraph{Type-I error bound.}
            To obtain the bound on the type-I error, we verify the conditions required by~\Cref{theorem:asymptotic-limit}, by using the expressions for moments of the Gaussian kernel derived by~\citet{li2019optimality}, and recalled in~\Cref{fact:li-yuan}. 
            
            First, we note that the scale parameters $s_n = n^{4/(d+4\beta)}$, satisfies the property: 
            \begin{align}
                \lim_{n \to \infty} \frac{s_n}{n^{4/d}} = \lim_{n \to \infty} n^{- \frac{4}{d}(1-\frac{d}{d+4\beta})} = 0. 
            \end{align}
            In other words, we have $s_n = o(n^{4/d})$. We now verify the required conditions: 
            \begin{itemize}
                \item Since we have assumed $m_n=n$ in this case, $\lim_{n \to \infty} n/m_n = 1$ exists. 
                \item For checking the condition on the eigenvalues, it suffices to show that 
                \begin{align}
                    \lim_{n \to \infty} \frac{\mathbb{E}_{P_n,Q_n}[\mathbb{E}_{P_n,Q_n}[\bark(X_1, X_2)\bark(X_1, X_3)|X_2, X_3]^2] }{\mathbb{E}_{P_n,Q_n}[\bark(X_1, X_2)^2]^2} = 0, 
                \end{align}
                since this is equivalent to $\lim_{n \to \infty} \frac{\lambda_1^2}{\sum_{l} \lambda_l^2} = 0$. This result follows by a combination of~\eqref{eq:li-yuan-1} and~\eqref{eq:li-yuan-3}. 
                \item We next check the condition~\eqref{eq:assump-kernel-1}. We do this in two steps. First we consider the term, 
                \begin{align}
                    \frac{ \mathbb{E}_{P_n,Q_n}[\bark_n(X_1, X_2)^4] }{ \mathbb{E}_{P_n,Q_n}[\bark_n(X_1, X_2)^2]^2 n^2} \lesssim \frac{ s_n^{-d/2} }{(s_n^{-d/2})^2 } \frac{1}{n^2} = \frac{s_n^{d/2}}{n^2} \to 0, 
                \end{align}
                where the first inequality uses~\eqref{eq:li-yuan-1} and~\eqref{eq:li-yuan-2}, while the last step uses the fact that $s_n = o(n^{4/d})$. Next, we consider the quantity 
                \begin{align}
                    \frac{\mathbb{E}_{P_n,Q_n}[\bark^2_n(X_1, X_2) \bark_n^2(X_1, X_3)]}{n \mathbb{E}_{P_n,Q_n}[\bark_n^2(X_1, X_2)]^2 } \lesssim \frac{1}{n} \frac{s_n^{-3d/4}}{(s_n^{-d/2})^2} = \frac{s_n^{d/4}}{n} = \lp \frac{s_n}{n^{4/d}}\rp^{d/4} \to 0. 
                \end{align}
            \end{itemize}
             Together with~\Cref{theorem:asymptotic-limit}, the above conditions imply that the statistic $\csmmd$ computed using Gaussian kernel with scale parameter $s_n = n^{4/(d+4\beta)}$ has a standard normal null distribution uniformly over the class $\nullclass$. This implies the required result about asymptotic type-I error of the $\cmmd$ test $\Psi$.

            \paragraph{Consistency.} 
                To prove the consistency results, we verify that the sufficient conditions established by the general result,~\Cref{prop:general-consistency}, are satisfied by the Gaussian kernel with scale parameter $s_n = n^{4/(d+4\beta)}$. 
                
                We first check the condition on the variance of $\crossmmd$. Note that we have the following: 
                \begin{align}
                    \crossU_X &= \langle \muhat_1, \muhat_2 - \nuhat_2\rangle_k = \langle \mutilde_1 + \mu, \gtilde_2 + \mu-\nu \rangle_k \\
                    & = \langle \mutilde_1, \gtilde_2 \rangle_k + \langle \mutilde_1, \mu - \nu \rangle_k + \langle \mu, \gtilde_2 \rangle_k + \langle \mu, \mu - \nu \rangle_k \label{eq:local-alt-1}
                \end{align}
                Recall that we use $\mutilde_1$ to denote $\muhat_1 - \mu$, and similarly use $\mutilde_2, \nutilde_1, \nutilde_2$ and  $\gtilde_2$  to denote $\muhat_2-\mu, \nuhat_1-\nu, \nuhat_2-\nu$ and $g_2 -(\mu-\nu)$ respectively.
                Similarly, on expanding the term $\crossU_Y$, we get  
                \begin{align}
                    \crossU_Y &= \langle \nuhat_1, \muhat_2 - \nuhat_2\rangle_k = \langle \nutilde_1 + \nu, \gtilde_2 + \mu-\nu \rangle_k \\
                    & = \langle \nutilde_1, \gtilde_2 \rangle_k + \langle \nutilde_1, \mu - \nu \rangle_k + \langle \nu, \gtilde_2 \rangle_k + \langle \nu, \mu - \nu \rangle_k \label{eq:local-alt-2}
                \end{align}               
                Since $\crossmmd = \crossU_X - \crossU_Y$, we get that 
                \begin{align}
                    \crossmmd = \langle \mutilde_1 - \nutilde_1, \gtilde_2 \rangle_k + \langle \mutilde_1 - \nutilde_1, \mu - \nu \rangle_k + \langle \mu - \nu, \gtilde_2 \rangle_k + \mmdval^2. 
                \end{align}
                Therefore, the variance of $\crossmmd$ is 
                \begin{align}
                    \var\lp \crossmmd \rp = \mathbb{E}_{P_n,Q_n}\lb \langle \mutilde_1 - \nutilde_1, \gtilde_2 \rangle_k^2 + \langle \mutilde_1 - \nutilde_1, \mu - \nu \rangle_k^2 + \langle \mu - \nu, \gtilde_2 \rangle_k^2  \rb, \label{eq:local-alt-3}
                \end{align}
                since all the cross terms are zero in expectation, due to the sample-splitting used in defining $\crossmmd$.
                To establish the consistency of the cross-MMD test, we need to show that 
                \begin{align}
                    \lim_{n \to \infty} \frac{\mathbb{V}(\crossmmd)}{\gamma_n^4} = 0, \quad \text{where} \quad \gamma_n^2 = \dmmd^2(P_n, Q_n). \label{eq:bug-fix-0}
                \end{align}
                We  will first derive an upper bound on $\mathbb{V}(\crossmmd)$ by individually considering the three terms in~\eqref{eq:local-alt-3}, and then show the condition on the scale parameter~(i.e., $s_n \asymp n^{4/(d+4\beta)}$) and the separation term~(i.e., $\lim_{n\to \infty} \Delta_n n^{2\beta/(d+4\beta)} = \infty$) together lead to the required conclusion stated in~\eqref{eq:bug-fix-0}. 

                \begin{description}
                \item[Bounding $\mathbb{E}\lb \langle \mutilde_1-\nutilde_1, \gtilde_2 \rangle_k^2 \rb$.] 
                Observe that 
                \begin{align}
                    \mathbb{E}\lb \langle \mutilde_1 - \nutilde_1, \gtilde_2 \rangle_k^2 \rb &=
                    \mathbb{E}\lb \lp \langle \mutilde_1, \gtilde_2 \rangle_k - \langle \nutilde_1, \gtilde_2 \rangle_k\rp^2 \rb 
                    \stackrel{(i)}{\lesssim} \mathbb{E}\lb \langle \mutilde_1, \gtilde_2 \rangle_k^2 + \langle \nutilde_1, \gtilde_2 \rangle_k^2 \rb.  \label{eq:bug-fix-1}
                \end{align}
                In the above display, $(i)$ uses the fact that $(a-b)^2 \leq 2(a^2 + b^2)$ for any $a, b \in \mathbb{R}$. 
                Now, we expand the first term in~\eqref{eq:bug-fix-1}: 
                \begin{align}
                    \mathbb{E}\lb \langle \mutilde_1, \gtilde_2 \rangle_k^2 \rb &= \mathbb{E}\lb \lp \frac{2}{n} \sum_{i=1}^{n/2} \langle \bark_{s_n}(X_i, \cdot), \gtilde_2 \rangle_k \rp^2 \rb \\
                    & = \frac{4}{n^2} \sum_{i=1}^{n/2} \sum_{j=1}^{n/2} \mathbb{E}\lb \langle \bark_{s_n}(X_i, \cdot), \gtilde_2 \rangle_k \langle \bark_{s_n}(X_j, \cdot), \gtilde_2 \rangle_k \rb \\
                    & = \frac{4}{n^2} \lp \sum_{i=1}^{n/2} \mathbb{E}\lb \langle \bark_{s_n}(X_i, \cdot), \gtilde_2 \rangle_k^2 \rb + \sum_{i \neq j}  \mathbb{E}\lb \langle \bark_{s_n}(X_i, \cdot), \gtilde_2 \rangle_k \langle \bark_{s_n}(X_j, \cdot), \gtilde_2 \rangle_k \rb \rp  \\
                    & = \frac{2}{n} \mathbb{E} \lb \langle \bark_{s_n}(X_1, \cdot), \gtilde_2 \rangle_k^2 \rb   \label{eq:bug-fix-2}
                \end{align}
                To get the last inequality, we used the fact that the cross terms are zero. Now, on expanding~\eqref{eq:bug-fix-2}, we get the following using the fact that $\gtilde_2 = \mutilde_2 - \nutilde_2$. 
                \begin{align}
                    \mathbb{E}\lb \langle \mutilde_1, \gtilde_2 \rangle_k^2 \rb &= \frac{2}{n} \lp \mathbb{E}\lb \langle \bark_{s_n}(X_1, \cdot), \mutilde_2 \rangle_k^2 + \langle \bark_{s_n}(X_1, \cdot), \nutilde_2 \rangle_k^2 - 2 \langle \bark_{s_n}(X_1, \cdot), \mutilde_2\rangle \langle \bark_{s_n}(X_1, \cdot), \nutilde_2 \rangle_k \rb \rp  \\
                    &\stackrel{(i)}{\leq} \frac{4}{n} \lp \mathbb{E}\lb \langle \bark_{s_n}(X_1, \cdot), \mutilde_2 \rangle_k^2 + \langle \bark_{s_n}(X_1, \cdot), \nutilde_2 \rangle_k^2  \rb \rp  \\
                    & \stackrel{(ii)}{=} \frac{8}{n^2} \mathbb{E}\lb \langle \bark_{s_n}(X_1, \cdot),  \bark_{s_n}(X_n, \cdot)\rangle_k^2 + \langle \bark_{s_n}(X_1, \cdot), \bark_{s_n}(Y_n, \cdot) \rangle_k^2 \rb \\ 
                    & \lesssim \frac{8}{n^2} \mathbb{E}\lb k_{s_n}(X_1, X_n)^2 + k_{s_n}(X_1, Y_n)^2 \rb.  \label{eq:bug-fix-3}
                \end{align}
                The equality $(i)$ uses the fact that $(a-b)^2 \leq 2 (a^2 + b^2)$, $(ii)$ uses the fact that the cross-terms in both summations are again all zero. Now, we know from Fact~14, that $\mathbb{E}[k_{s_n}(X_1, X_n)^2] \lesssim s_n^{-d/2}$, a similar bound holds for $\mathbb{E}[k_{s_n}(X_1, Y_n)^2]$ as we show in~\Cref{lemma:bug-fix-1}.

                \item[Bound on $\mathbb{E}\lb \langle \mutilde_1 - \nutilde_1, \mu - \nu\rangle_k^2\rb$ and $\mathbb{E}\lb \langle \mu - \nu, \gtilde_2 \rangle_k^2\rb$.]
                The two terms are identically distributed, so we only consider the first term. 
                Observe that 
                \begin{align}
                    \mathbb{E}\lb  \langle \mutilde_1 - \nutilde_1, \mu - \nu\rangle_k^2\rb &= \frac{2}{n} \mathbb{E}\lb \langle \bark_{s_n}(X_1, \cdot) - \bark_{s_n}(Y_1, \cdot), \mu - \nu\rangle_k^2\rb. 
                \end{align}
                Let us introduce the notation $f=\mu-\nu$, and observe that we have 
                \begin{align}
                    \mathbb{E}\lb  \langle \mutilde_1 - \nutilde_1, \mu - \nu\rangle_k^2\rb &\lesssim \frac{2}{n} \lp \mathbb{E}\lb f(X)^2  + f(Y)^2\rb \rp. 
                \end{align}
                Now, by~\Cref{lemma:bug-fix-2}, we have the following, with $r_n \defined p_n - q_n$, 
                \begin{align}
                \mathbb{E}\lb f(X)^2 \rb \lesssim M \|r_n\|_{L^2}^2 s_n^{-3d/4}, \quad \text{and} \quad  
                \mathbb{E}\lb f(Y)^2 \rb \lesssim M \|r_n\|_{L^2}^2 s_n^{-3d/4}, \label{eq:bug-fix-4}
                \end{align}
                which implies that 
                \begin{align}
                \mathbb{E}\lb \langle \mutilde_1 - \nutilde_1, \mu - \nu \rangle_k^2 \rb \lesssim \frac{2M \Delta_n^2}{n} s_n^{-3d/4} \quad \text{and} \quad 
                \mathbb{E}\lb \langle \mu - \nu, \gtilde_2 \rangle_k^2 \rb \lesssim \frac{2M \Delta_n^2}{n} s_n^{-3d/4}
                \end{align}
            \end{description}

                In the two steps above, we have proved that  
                \begin{align}
                \mathbb{V}\lp \crossmmd \rp \lesssim \frac{s_n^{-d/2}}{n^2} + \frac{s_n^{-3d/4} \Delta_n^2}{n} \quad \implies \quad  
                \frac{\mathbb{V}\lp \crossmmd \rp}{\gamma_n^4} \lesssim \frac{s_n^{-d/2}}{n^2 \gamma_n^4} + \frac{s_n^{-3d/4} \Delta_n^2}{n \gamma_n^4},  
                \end{align}
                where $\gamma_n^2 = \|\mu - \nu\|_k^2$ and $\Delta_n^2 = \|r_n\|_{L^2}^2 = \|p_n - q_n\|_{L^2}^2$. 
                Now, by~\Cref{fact:li-yuan}, we know that  $\gamma_n^2 \gtrsim s_n^{-d/2} \Delta_n^2$, which implies that $\gamma_n^{-4} \lesssim s_n^d \Delta_n^{-4}$. Using this and the fact that $s_n \asymp n^{4/(d+4\beta)}$ in the above expression gives us 
                \begin{align}
                \frac{\mathbb{V}\lp \crossmmd \rp}{\gamma_n^4} &\lesssim \frac{s_n^{d-d/2}}{n^2 \Delta_n^4} + \frac{s_n^{d-3d/4} }{n \Delta_n^2} \asymp \frac{n^{2d/(d+4\beta)}}{n^2 \Delta_n^4} + \frac{n^{d/(d+4\beta)}}{n \Delta_n^2}  \\
                & = \frac{1}{\lp n^{2\beta/(d +4\beta)} \Delta_n\rp^4} + \frac{1}{\lp n^{2\beta/(d + 4\beta)} \Delta_n\rp^2}. 
                \end{align}
                This implies that if the detection boundary is such that $\lim_{n \to \infty} \Delta_n n^{2\beta/(d+4\beta)} = \infty$, then $\mathbb{V}(\crossmmd)/\gamma_n^4 \to 0$ as $n \to \infty$, which verifies the sufficient condition for the minimax optimality of the cross-MMD test.

            \subsubsection{Auxiliary Lemmas}

            \begin{lemma}
                \label{lemma:bug-fix-1} Suppose $X \sim P_n$ and $Y \sim Q_n$ denote two independent $\mathbb{R}^d$-valued random variables with densities $p_n, q_n \in \Sobolev(M)$ respectively. Then, we have 
                \begin{align}
                    \mathbb{E}[k_{s_n}(X, Y)^2] \leq C s_n^{-d/2}, \quad \text{where} \quad  C \equiv C_{p_n,q_n,d} = (\pi/2)^{d/2} \|p_n\|_{L^2} \|q_n\|_{L^2}. 
                \end{align}
            \end{lemma}

            \emph{Proof of~\Cref{lemma:bug-fix-1}.}
            We begin by expanding $\mathbb{E}[k_{s_n}(X,Y)^2]$ as an integral, denoted by $I$, as follows: 
            \begin{align}
            I &= \int \int k_{s_n}^2(x, y) dP_n(x) dQ_n(y) = \int \int e^{ - 2 s_n \|x-y\|_2^2} p_n(x) q_n(y) dx dy \\
            & = \int \int e^{- 2 s_n \|u\|^2} q_n(x-u) p_n(x) du dx && (u\leftarrow x-y) \\
            & = \int \lp k_{2s_n} \star q_n \rp(x) p_n(x) dx && (\text{Definition of convolution}) \\
            &  \leq \|k_{2s_n} \star q_n \|_{L^2} \|p_n\|_{L^2},  && (\text{Cauchy-Schwarz})
            \end{align}
            where $\star$ denotes the convolution operation. By Plancherel's theorem, we know that Fourier transforms preserve the $L^2$ norms, which implies that 
            \begin{align}
                I \leq  \| \mc{F}\lp k_{2s_n} \star q_n \rp \|_{L^2} \|p_n\|_{L^2} = \|\mc{F}(k_{2s_n}) \mc{F}(q_n) \|_{L^2} \|p_n\|_{L^2} \leq \|\mc{F}(k_{2s_n})\|_{L^\infty} \|\mc{F}(q_n) \|_{L^2} \|p_n\|_{L^2}.
            \end{align}
            In the last inequality, we have used the fact that for any two functions $f \in L^{\infty}(\mathbb{R}^d)$ and $g \in L^2(\mathbb{R}^d)$, we have $\|fg\|_{L^2}^2 = \int |f(x)g(x)|^2 dx \leq \|f\|_{L^\infty}^2 \int |g(x)|^2 dx = \|f\|_{L^\infty} \|g\|_{L^2}^2$. 
            The Fourier transform of the Gaussian kernel also has the same Gaussian form (and hence is in $L^{\infty}$):  
            \begin{align}
                \mc{F}(k_{2s_n})(\omega) = \frac{1}{(2\pi)^{d/2}} \int e^{-2s_n \|x\|^2_2} e^{-2 \pi i \omega x} dx = \lp \frac{\pi}{2s_n} \rp^{d/2} \exp \lp - \|\omega \|_2^2/8s_n \rp. 
            \end{align}
            Thus, the infinity norm of the Fourier transform of $k_{2s_n}$ is equal to $(\pi/2s_n)^{d/2}$, which concludes the proof.  
            \hfill \qedsymbol

            \begin{lemma}
                \label{lemma:bug-fix-2} Let $X \sim P_n$ and $Y \sim Q_n$, and let $\mu = \mathbb{E}\lb k_{s_n}(X, \cdot) \rb$ and $\nu = \mathbb{E}\lb k_{s_n}(Y, \cdot) \rb$, where $k_{s_n}(x, y) = \exp(-s_n \|x-y\|_2^2)$. Then, we have the following, with $f = \mu - \nu$ and $r_n = p_n - q_n$: 
                \begin{align}
                    \mathbb{E}\lb f(X)^2 \rb \lesssim M \|r_n\|_{L^2}^2 s_n^{-3d/4}, \quad \text{and} \quad  
                    \mathbb{E}\lb f(Y)^2 \rb \lesssim M \|r_n\|_{L^2}^2 s_n^{-3d/4}. 
                \end{align}
            \end{lemma}
            We include the details of the first bound in the statement of~\Cref{lemma:bug-fix-2}; the exact same argument holds for the second term. Throughout this proof, all the integrands that we work with will be absolutely integrable, and hence we will interchange the order of integrals when needed without further justification. 
            \begin{align}
                \mathbb{E}\lb f(X)^2 \rb &= \mathbb{E}\lb (\mu(X)-\nu(X))^2 \rb = \mathbb{E}\lb\lp \int k_{s_n}(X, a)p_n(a)da - \int k_{s_n}(X, a) q_n(a) da \rp^2 \rb\\ 
                & = \mathbb{E}\lb \lp \int k_{s_n}(X, a) (p_n(a)-q_n(a)) da \rp^2 \rb 
                = \mathbb{E}\lb \lp \int k_{s_n}(X, a) r_n(a) da \rp^2 \rb  && (r_n=p_n-q_n)\\
                &=\mathbb{E}\lb \int k_{s_n}(X, a)r_n(a)da  \int k_{s_n}(X, b) r_n(b) db \rb 
                \\ 
                &= \mathbb{E}\lb \int \int k_{s_n}(X, a)k_{s_n}(X, b) r_n(a) r_n(b)da\, db \rb\label{eq:bug-fix-5}
            \end{align}
            Now, we use the following equality that follows from the parallelogram identity: 
            \begin{align}
                k_{s_n}(X, a) k_{s_n}(X, b) & = \exp\lp - s_n \lp \|X-a\|_2^2 + \|X-b\|_2^2 \rp \rp \\
                &= \exp\lp - s_n \lp 2\|X-u\|_2^2 + \frac{1}{2}\|a-b\|_2^2 \rp \rp && (u \defined (a+b)/2)\\
                &= \exp\lp - 2s_n \|X-u\|_2^2 - \frac{s_n}{2}\|a-b\|_2^2  \rp  \\
                & = k_{2s_n}(X, u) k_{s_n/2}(a, b).  \label{eq:bug-fix-8}
            \end{align}
            Plugging this into~\eqref{eq:bug-fix-5}, we get 
            \begin{align}
                \mathbb{E}[f(X)^2] & = \mathbb{E}\lb \int \int k_{2s_n}(X, u) k_{s_n/2}(a, b) r_n(a) r_n(b) da\, db \rb \\
                & = \int \int \lp \int k_{2s_n}(x, u) p(x) dx\rp k_{s_n/2}(a, b) r_n(a) r_n(b) da\, db.\label{eq:bug-fix-6}
            \end{align}
            Next, we will rewrite the inner integral. Introduce the random variable $Z \sim N(0, 4 s_n I_d)$, and observe that 
            \begin{align}
                k_{2s_n}(x, u) = \exp\lp - 2 s_n \|x-u\|_2^2 \rp = \mathbb{E}\lb e^{-i Z^T(x-u)} \rb. 
            \end{align}
            This implies that 
            \begin{align}
                \int k_{2s_n}(x, u) p_n(x) dx &= \int \mathbb{E}\lb e^{-iZ^T(x-u)} \rb p_n(x) dx = \mathbb{E}\lb \int e^{-iZ^T(x-u)}  p_n(x) dx  \rb \\
                &= \mathbb{E}\lb e^{iZ^T u} \int e^{-iZ^Tx}  p_n(x) dx  \rb = (2\pi)^{d/2} \mathbb{E}\lb e^{iZ^T u} \frac{1}{(2\pi)^{d/2}} \int e^{-iZ^Tx}  p_n(x) dx  \rb\\
                & = (2\pi)^{d/2} \mathbb{E}\lb \calF(p_n)(Z) e^{iZ^T u} \rb \leq (2\pi)^{d/2} \sqrt{\mathbb{E}\lb |\calF(p_n)|^2 \rb \mathbb{E}\lb |e^{iZ^Tu}|^2 \rb} \\
                & = (2\pi)^{d/2} \sqrt{\mathbb{E}\lb |\calF(p_n)|^2 \rb} \lesssim (2\pi)^{d/2}  \|p_n\|_{L^2} s_n^{-d/4}.\label{eq:bug-fix-7}
            \end{align}
            Using~\eqref{eq:bug-fix-7} in~\eqref{eq:bug-fix-6}, and following an argument similar to the proof of~\Cref{lemma:bug-fix-1}, we get
            \begin{align}
                \mathbb{E}\lb f(X)^2 \rb & \lesssim M s_n^{-d/4} \int \lp \int e^{-s_n/2\|a-b\|_2^2} r_n(b) db \rp r_n(a) da = M s_n^{-d/4} \int (k_{s_n/2} \star r_n)(a) r_n(a) da  \\
                & \leq M s_n^{-d/4} \| k_{s_n/2} \star r_n \|_{L^2} \|r_n\|_{L^2}  = M s_n^{-d/4} \| \calF k_{s_n/2} \calF r_n \|_{L^2} \|r_n\|_{L^2}  \\
                & \leq M s_n^{-d/4} \|\calF k_{s_n/2}\|_{L^2} \|r_n \|_{L^2} \leq M \|r_n\|_{L^2} s_n^{-d/4}  \|\calF k_{s_n/2} \|_{L^{\infty}} \lesssim M \|r_n\|_{L^2} s_n^{-3d/4}. 
            \end{align}
            This completes the proof. \hfill \qedsymbol

\section{Gaussian Limit for General Two-Sample U-Statistic}

\label{sec:general-U-statistic}

    We now generalize the asymptotic normality for kernel-MMD statistic stated in~\Cref{theorem:asymptotic-limit} to a larger class of two-sample U-statistics. 
    As before, given $\Xsample = (X_1, \ldots, X_n)$ and $\Ysample = (Y_1, \ldots, Y_m)$, we consider the two-sample  U-statistic with arbitrary kernel $h$ defined as 
    \begin{align}
        \label{eq:U-statistic-def} 
        U = \frac{1}{ \binom{n}{2}  } \frac{1}{\binom{m}{2}} \sum_{i'<i} \sum_{j'<j} h(X_i, X_{i'}, Y_j, Y_{j'}). 
    \end{align} 
    We assume that $h$ is a degenerate kernel, similar to the MMD case, and satisfies 
    \begin{align}
        \mathbb{E}_P[h(X, x', Y, y')] = \mathbb{E}_{P}[h(x, X', y, Y')] = 0, 
    \end{align}
    when $X, X', Y, Y'$ are \iid random variables drawn from any distribution $P$.  
    
    With $\Xsample_1 = (X_1, \ldots, X_{n_1})$ and $\Xsample_2 = (X_{n_1+1}, \ldots, X_n)$ and $\Ysample_1 = (Y_1, \ldots, Y_{m_1})$ and $\Ysample_2 = (Y_{m_1+1}, \ldots, Y_m)$, we introduce the following terms:
    \begin{align}
        &\phi(x, y) \defined \frac{1}{n_2} \frac{1}{m_2} \sum_{X_{i'}\in \Xsample_2} \sum_{Y_{j'} \in \Ysample_2} h(x, X_{i'}, y, Y_{j'}), \quad \text{with } n_2 = n-n_1, \text{ and } m_2 = m-m_1 \label{eq:general-phi} \\ 
         &q(x_1, x_2,   y_2) \defined \mathbb{E}[h(x_1, x_2, Y, y_2)] \quad \text{and} \quad \bar{q}(x) \defined \frac{1}{n_2 m_2} \sum_{X_{i'} \in \Xsample_2, Y_{j'}\in \Ysample_2} q(x, X_{i'}, Y_{j'}), \label{eq:general-q}\\
        &r(x_2, y_1, y_2) \defined \mathbb{E}[h(X, x_2, y_1, y_2)]\quad \text{and} \quad \bar{r}(y) \defined \frac{1}{n_2 m_2} \sum_{X_{i'} \in \Xsample_2, Y_{j'}\in \Ysample_2} r(X_{i'}, y, Y_{j'}). \label{eq:general-r}
    \end{align}

    Using the above terms, we can now define the statistic $T = \crossU/\sigmahat$, with 
    \begin{align}
        &\crossU = \frac{1}{n_1} \frac{1}{m_1} \sum_{X_i \in \Xsample_1} \sum_{Y_j \in \Ysample_1} \phi(X_i, Y_j), \quad \text{and} \quad 
         \sigmahat^2 = \frac{\sigmahat_X^2}{n_1} + \frac{\sigmahat_Y^2}{m_1}, \quad \text{where}  \label{eq:T-statistic-general} \\
        & \sigmahat_X^2 = \frac{1}{n_1}\sum_{i=1}^{n_1} \lp \bar{q}(X_i)-\frac{1}{n_1} \sum_{l=1}^{n_1}\bar{q}(X_{l}) \rp^2, \quad  \sigmahat_Y^2 = \frac{1}{m_1}\sum_{j=1}^{m_1} \lp \bar{r}(Y_j)-\frac{1}{m_1} \sum_{l=1}^{m_1}\bar{r}(Y_{l}) \rp^2.
    \end{align}
    
    \begin{remark}
        Note that the cross U-statistic written above corresponds exactly with the definition of the cross U-statistic for the kernel-MMD case in~\eqref{eq:cross-U}. 
        To motivate the definitions of the empirical variance terms, 
        note that in  the case of kernel-MMD statistic, we have $h(x_1, x_2, y_1, y_2) = \langle k(x_1, \cdot) - k(y_1, \cdot), k(x_2, \cdot) - k(y_2, \cdot)\rangle_k$. We  can check that in this case, we have $q(x_1, x_2, y_2) = \langle \tildek(x_1, \cdot), k(x_1, \cdot) - k(x_2, \cdot) \rangle_k$. This implies that $\barq(X_i)$ equals the term $W_i$ introduced~\eqref{eq:W-Z}, and thus $\frac{1}{n_1}\sum_{i=1}^{n_1} \barq(X_i)$ is a centered analog of $\crossU_X$. Hence, the term $\sigmahat_X^2$ defined above reduces exactly to the $\sigmahat_X^2$ introduced in~\eqref{eq:two-sample-sigma-1}. 
    \end{remark}
    
    We next state the assumptions required to show the limiting Gaussian distribution of the statistic $T$ when $\Xsample$ and $\Ysample$ are drawn independently from the same distribution. 
    \begin{assumption}
        \label{assump:general}
        Let $(h_n, P_n)$ be a sequence of kernel and probability distribution pairs, and let $\Xsample$ and $\Ysample$ be two \iid samples of sizes $n$ and $m_n$ respectively, drawn independently from $P_n$. With $\phi$, $\barq_n$ and $\barr_n$ as defined in~\eqref{eq:general-phi},~\eqref{eq:general-q} and~\eqref{eq:general-r} respectively, we assume the following are true:
        \begin{align}
            &\lim_{n \to \infty} \mathbb{E}_{P_n}\lb \frac{\mathbb{E}_{P_n}[\phi^2(X_1, Y_1)|\Xsample_2, \Ysample_2]}{m_n\mathbb{E}_{P_n}[\barq(X_1)^2|\Xsample_2, \Ysample_2] + n\mathbb{E}_{P_n}[\barr(Y_1)^2|\Xsample_2, \Ysample_2]} \rb = 0, \quad \text{and} \label{eq:assump-general-1}             \\
           &\lim_{n \to \infty} \mathbb{E}_{P_n}\lb \frac{1}{n}\frac{ \mathbb{E}_{P_n}[\barq^4(X_1)| \Xsample_2, \Ysample_2]}{\mathbb{E}_{P_n}[\barq^2(X_1)|\Xsample_2, \Ysample_2]^2} + \frac{1}{m_n}\frac{ \mathbb{E}_{P_n}[\barr^4(Y_1)| \Xsample_2, \Ysample_2]}{\mathbb{E}_{P_n}[\barr^2(Y_1)|\Xsample_2, \Ysample_2]^2} \rb = 0.\label{eq:assump-general-2}
        \end{align}
    \end{assumption}
    \begin{remark}
        Note that in specific the case of kernel-MMD statistic, we can check that $\mathbb{E}_{P_n}[\phi(X_1, Y_1)^2|\Xsample_2, \Ysample_2] = \mathbb{E}_{P_n}[\barq(X_1)^2 + \barr(Y_1)^2 | \Xsample_2, \Ysample_2]$. Hence~\eqref{eq:assump-general-1} always holds. 
        The second condition of~\Cref{assump:general}, stated in~\eqref{eq:assump-general-2}, is a stronger version of the moment conditions used by~\Cref{theorem:asymptotic-normality-2} and~\Cref{theorem:asymptotic-limit}. 
    \end{remark}
    
    We now state the main result of this section.  
    \begin{theorem}
        \label{theorem:general-U-statistic}
        For every $n \geq 1$, let $\Xsample$ and $\Ysample$ denote independent samples of sizes $n$ and $m_n$ respectively, drawn from a distribution $P_n$. Suppose the sample-sizes are such that $\lim_{n \to \infty} m_n/n$ exists and is non-zero. Let $(h_n, P_n)$ denote a sequence satisfying the conditions of~\Cref{assump:general}. 
        Then,  we have that 
        \begin{align}
        \lim_{n,m \to \infty}  \sup_{x\in \mathbb{R}} \;| \mathbb{P}_{P_{n}}(T \leq x) - \Phi(x)| = 0.
        \end{align}
    \end{theorem}

    \subsection{Proof of~Theorem~\ref{theorem:general-U-statistic}}
    \label{appendix:proof-general-U-statistic}
        Before describing the details, we first present the outline of the proof.
        \begin{enumerate}
            \item We first consider the standardized version of the statistic, defined as $T_s = \crossU/\sigma_P$, where $\sigma_P^2 = n_1^{-1} \EPn[\barq(X_1)^2|\Xsample_2, \Ysample_2] + m_1^{-1}\EPn[\barr(Y_1)^2|\Xsample_2, \Ysample_2]$. In~\Cref{lemma:general-1}, we show that the difference between $T_s$ and its projected variant, $T_{P,s} = \crossU_P/\sigma_P = \lp n_1^{-1} \sum_i \barq(X_i) + m_1^{-1}\sum_j \barr(Y_j) \rp/\sigma_P$, converges in probability to $0$. Hence, we can focus on the term $T_{P,s}$. This result uses the condition~\eqref{eq:assump-general-1} of~\Cref{assump:general}. 
            
            \item We then show in~\Cref{lemma:general-2}, that the statistic $T_{P,s}$ converges in distribution to $N(0,1)$. This combined with the previous result implies that $T_s \convdist N(0,1)$. 
            
            \item To complete the proof, we show in~\Cref{lemma:general-3}, that the ratio of the empirical variance $\sigmahat^2$ and the conditional variance $\sigma_P^2$ converge in probability to $1$. This fact combined with the continuous mapping theorem and Slutsky's theorem implies the result. The proof of~\Cref{lemma:general-3} relies on the condition~\eqref{eq:assump-general-2} of~\Cref{assump:general}. 
        \end{enumerate}
        
        We now present the details of the steps outlined above.
       
        Consider the standardized statistic, $T_s$, defined as $\crossU/\sigma_P$, where $\sigma_P^2 = \var_{P_n}(\crossU_P | \Xsample_2, \Ysample_2) = n_1^{-1} \EPn[\barq^2(X_1)|\Xsample_2, \Ysample_2] + m_1^{-1} \EPn[\barr^2(Y_1) | \Xsample_2, \Ysample_2] \defined n_1^{-1} \sigmaPX^2 + m_1^{-1} \sigmaPY^2$. Introduce the term $T_{P,s} = \frac{ \crossU_P}{\sigma_P}$. 
        
        \begin{lemma}
            \label{lemma:general-1}
            Under the conditions of~\Cref{assump:general}, we have $T_p - T_{P,s} \convprob 0$. 
        \end{lemma}        
 
        \begin{proof}
        We  first show that $T_s - T_{P,s} \convprob 0$, conditioned on the second half of the observations, $(\Xsample_2, \Ysample_2)$. As a result of this, the conditional limiting distributions of the two random variables $T_s$ and $T_{P,s}$ are the same.
        Since $\crossU_P$ is the projection of $\crossU$ on the sum on independent~(conditioned on $(\Xsample_2, \Ysample_2)$) random variables, we  have 
        \begin{align}
            \var_{P_n}(T_s - T_{P,S}|\Xsample_, \Ysample_2) &= \var_{P_n}(T_s |\Xsample_, \Ysample_2) + \var_{P_n}(T_{P,s} |\Xsample_, \Ysample_2) - 2 \EPn\lb \lp T_{P,s} + (T_s-T_{P,s}) \rp T_{P,s} | \Xsample_2, \Ysample_2 \rb \\
            & = \var_{P_n}(T_s |\Xsample_, \Ysample_2) - \var_{P_n}(T_{P,s} |\Xsample_, \Ysample_2) = \var_{P_n}(T_s |\Xsample_, \Ysample_2) - 1, 
        \end{align}
        using the fact that $(T_{s} - T_{P,s}) \perp T_{P,s}$ conditioned on $(\Xsample_2, \Ysample_2)$. Next, using the formula for the variance of two-sample U-statistics, we have 
        \begin{align}
            \var_{P_n}(T_s|\Xsample_2, \Ysample_2) &=\lp \frac{ \sigmaPX^2}{n_1} + \frac{\sigmaPY^2}{m_1} + \frac{1}{n_1 m_1} \EPn\lb \phi(X_1, X_2)^2  | \Xsample_2, \Ysample_2 \rb \rp / \sigma_P^2 \\
            &= 1 + \frac{1}{n_1 m_1} \frac{\EPn[\phi^2(X_1, X_2)|\Xsample_2, \Ysample_2]}{\sigma_P^2}. 
        \end{align}
        The result then follows by an application of the condition~\eqref{eq:assump-general-1} of~\Cref{assump:general}, and the fact that $n_1 = n/2$ and $m_1=m_n/2$. 
        \end{proof}
        
        Our next result establishes the limiting distribution of the statistic $T_{P,s}$. 
        \begin{lemma}
            \label{lemma:general-2}
            Under~\Cref{assump:general}, we have $T_{P,s} \convdist N(0,1)$. 
        \end{lemma}
        
        \begin{proof}
            Recall that $T_{P,s} = \crossU_P/\sigma_P$, where $\crossU_P \defined \crossU_{P,X} - \crossU_{P,Y} = \frac{1}{n_1} \sum_{i=1}^{n_1} \barq(X_i) - \frac{1}{m_1} \sum_{j=1}^{m_1}\barr(Y_j)$, and $\sigma_P^2 = n_1^{-1} \sigmaPX^2 + m_1^{-1} \sigmaPY^2$. Introduce the terms $T_X = \crossU_{P,X}/\sqrt{n_1^{-1} \sigmaPX^2}$ and $T_Y = \crossU_{P,Y}/\sqrt{m_1^{-1} \sigmaPY^2}$. The result then follows in the following two steps: 
            \begin{itemize}
                \item We first observe that $T_X$ and $T_Y$ conditioned on $(\Xsample_2, \Ysample_2)$ converge in distribution to $N(0,1)$. The result follows by applying Lindeberg's CLT. 
                
                \item Next, using the assumption that $\lim_{n \to \infty} m_n/n$ exists, and is non-zero, we next observe that $T_{P,s} \convdist N(0,1)$. The proof of this result follows from the same argument used in~\Cref{lemma:asymp-normality-2}. 
            \end{itemize}
        \end{proof}
        Together, the previous two lemmas imply that $T_s \convdist N(0,1)$. To complete the proof, we need to show that the ratio of the conditional variance $\sigma_P^2$, and the empirical variance $\sigmahat^2$ converge in probability to $1$. 
        
        \begin{lemma}
            \label{lemma:general-3}
            Under~\Cref{assump:general}, we have $\frac{\sigmahat^2}{\sigma_P^2} \convprob 1$. 
        \end{lemma}        
        
        \begin{proof}
            We begin by noting the following 
            \begin{align}
                \frac{ \sigmahat^2}{\sigma_P^2} - 1 &= \frac{ n_1^{-1} \lp \sigmahat_X^2 - \sigmaPX^2 \rp m_1^{-1} \lp \sigmahat_Y^2 - \sigmaPY^2 \rp }{\sigma_P^2} \\
                & \leq \lv \frac{\sigmahat_X^2}{\sigmaPX^2} - 1 \rv + \lv \frac{\sigmahat_Y^2}{\sigmaPY^2} - 1 \rv.  \label{eq:proof-general-1}
            \end{align}
            Thus it suffices to show that the two terms in~\eqref{eq:proof-general-1} converge in probability to $0$. Since $n_1/(n_1-1)$ converges to $1$, it suffices to consider 
            \begin{align}
                E \defined \frac{(n_1-1)^{-1} \sum_{i=1}^{n_1} \lp \barq(X_i) - \crossU_{P,X} \rp^2 - \EPn[\barq(X_1)|\Xsample_2, \Ysample_2]}{\EPn[\barq^2(X_1)|\Xsample_2, \Ysample_2]}. 
            \end{align}
            First note that $\EPn[E|\Xsample_2, \Ysample_2]=0$. Hence, its variance can be written as 
            \begin{align}
                \var_{P_n}(E) &= \EPn[\var_{P_n}(E|\Xsample_2, \Ysample_2)] 
                 \leq \frac{1}{n_1}\EPn\lb \frac{ \EPn[\barq^4(X_1)| \Xsample_2, \Ysample_2]}{\EPn[\barq^2(X_1)|\Xsample_2, \Ysample_2]^2} \rb. \label{eq:proof-general-2}
            \end{align}
            The last term in~\eqref{eq:proof-general-2} converges to $0$ by~\Cref{assump:general}, implying that $\frac{\sigmahat_X^2}{\sigmaPX^2}$ converges in the second moment to $1$, which in turn implies their convergence in probability to $1$. Following the same arguments, we can also show that $\frac{\sigmahat_Y^2}{\sigmaPY^2}$ also converge in probability to $1$, as required. 
        \end{proof}

\section{Additional Experiments}
\label{appendix:additional-experiments}

\paragraph{Computing Infrastructure.} All the experiments were performed on a workstation with Intel(R) Core(TM) i7-9700K CPU \@ 3.60GHz and 32 GB of RAM with an NVIDIA GTX 1080 GPU. 

\subsection{Implementation details of experiments reported in the main text}

    \paragraph{Details for~\Cref{fig:general-null-distribution}.}
        For the null distribution, we set $n=500$ and $m=625$ and generated both $\Xsample$ and $\Ysample$ from $N(\boldsymbol{0}, I_d)$ for $d=10$ and $100$. In both cases, we computed the $\csmmd$ statistic $2000$ times to plot the histogram.     
        
        For the second figure,  we obtain the power curves for the $\cmmd$ test and the MMD test with $200$ permutations for testing $P=N(\boldsymbol{0}, I_d)$ againt $Q= N(a_{\epsilon, j}, I_d)$. Here $d=10, j=5$ and $\epsilon=0.2$, and recall that $a_{\epsilon, j}$ is the vector in $\mathbb{R}^d$ obtained by setting the first $j \leq d$ coordinates of $\boldsymbol{0}$ equal to $\epsilon$. 
        We selected $n$ and $m$ from $20$ equally spaced points in the intervals $[10, 400]$ and $[10, 500]$ respectively, and ran $200$ trials of the tests for every $(n,m)$ pair to obtain the power curves. The error regions in the figure correspond to one bootstrap standard deviation with $200$ bootstrap samples.  
        
        For the third figure, we set $d=100$, $j=20$, $\epsilon=0.1$, $P=N(\boldsymbol{0}, I_d)$ and $Q=N(a_{\epsilon, j}, I_d)$. We ran the two tests, $\cmmd$ and MMD with $200$ permutations, for $20$ different $(n,m)$ pairs in the range $[10, 500]$, and repeated the experiment $200$ times for every such pair. The figure plots the wall-clock time, measure by Python's  \texttt{time.time()} function, and plot the power against the average wall-clock time over the $200$ trials. The size of the marker is proportional to the sample size~(i.e., $n+m$).
        
    \paragraph{Details for~\Cref{fig:null-distributions-1}.}  The two kernels used in this figure are the Gaussian and Quadratic kernels. The Gaussian kernel with scale parameter $s>0$ is defined as $k_s(x,y)= \exp(-s \|x-y\|_2^2)$, while the Quadratic kernel with scale $s>0$ is defined as $k_Q(x,y) = \lp 1 + s(x^Ty) \rp^2$. With $w$ denoting the median of the pairwise distance between all the observations, we set $s=1/(2w^2)$ for the Gaussian kernel and $s=1/w$ for the Quadratic kernel. 
    
    \paragraph{Details for~\Cref{fig:predicted-power}.}
    Given observations $X_1, X_2, \ldots, X_n \iid P$, consider the problem of one-sample mean-testing, that is, testing $H_0:\mathbb{E}[X_i]=\boldsymbol{0}$ versus $H_1: \mathbb{E}[X_i]=a \neq \boldsymbol{0}$. When the distribution $P$ is a multivariate Gaussian,   \citet{kim2020dimension} showed that power of their test using a one-sample studentized U-statistic based on a bi-linear kernel is asymptotically $\Phi \lp z_{\alpha} + \frac{ a^Ta}{2 \sqrt{\text{tr}(\Sigma^2)}} \rp$. The power achieved by the test using the full U-statistic is $\Phi \lp z_{\alpha} + \frac{ a^Ta}{ \sqrt{2 \text{tr}(\Sigma^2)}} \rp$, which differs from the previous expression by a factor of $\sqrt{2}$. A similar relation also holds for the problem of Gaussian covariance testing.  Our heuristic in~\eqref{eq:power-prediction-heuristic} is based on these two observations. 
    
    \paragraph{Details for~\Cref{fig:ROC}.} For plotting the ROC curves, we proceed as follows. We fix $n=m=200$, and then compute the MMD, block-MMD, linear-MMD and cross-MMD statistics for $1000$ independent repetitions of `null' and `alternative' trials. For every null trial, we calculate all the statistics on independent samples of sizes $n$ and $m$ drawn from $P = N(\boldsymbol{0}, I_d)$, while for every alternative trial we calculate the statistics on independent samples of size $n$ and $m$ drawn from $P=N(\boldsymbol{0}, I_d)$ and $Q=N(a_{\epsilon,j}, I_d)$ respectively. Recall that $a_{\epsilon,j}$ is obtained by setting the first $j$ coordinates of $\boldsymbol{0}$ equal to $\epsilon$. Having obtained $2000$ values for every statistic, we then plot the tradeoff between false positives~(FP) and true positives~(TP) as the rejection threshold is increased. The ability of a statistic to distinguish between the null and the alternative is quantified by the area under the curve.  In~\Cref{fig:ROC}, we used $(d, j, \epsilon) \in \{ (10, 5, 0.1), (100, 20, 0.1), (500, 100, 0.1)\}$. 

\subsection{Additional Figures}
    
\begin{figure}[htb!]
        \def\figwidth{0.30\linewidth}
        \def\figheight{0.25\linewidth} %
    \centering
    \hspace*{-1cm}
    \begin{tabular}{llll}
        \input{FinalFigs/Null_Dists_d_10_100_n_100_m_20_kernel__Dirichlet_RBF_2022_10_15_22_20_23cross}
    &
        \input{FinalFigs/Null_Dists_d_10_100_n_100_m_100_kernel__Dirichlet_RBF_2022_10_15_22_22_56cross}
    &
        \input{FinalFigs/Null_Dists_d_10_100_n_20_m_100_kernel__Dirichlet_Poly_5_2022_10_15_22_11_41cross}
    &
        \input{FinalFigs/Null_Dists_d_10_100_n_100_m_100_kernel__Dirichlet_Poly_5_2022_10_15_22_14_39cross} \\
        \input{FinalFigs/Null_Dists_d_10_100_n_100_m_20_kernel__Dirichlet_RBF_2022_10_15_22_20_23mmd}
    &
        \input{FinalFigs/Null_Dists_d_10_100_n_100_m_100_kernel__Dirichlet_RBF_2022_10_15_22_22_56mmd}
    &
        \input{FinalFigs/Null_Dists_d_10_100_n_20_m_100_kernel__Dirichlet_Poly_5_2022_10_15_22_11_41mmd}
    &
        \input{FinalFigs/Null_Dists_d_10_100_n_100_m_100_kernel__Dirichlet_Poly_5_2022_10_15_22_14_39mmd}
    \end{tabular}
    \caption{The first two columns show the null distribution of the $\csmmd$ statistic (top row) and the $\mmdhat^2$ statistic scaled by its empirical standard deviation (bottom row) using the Gaussian kernel with scale-parameter chosen using the median heuristic. The last two columns show the null distribution for the two statistics using the Polynomial kernel of degree $5$ with scale parameter chosen using the median heuristic. The figures demonstrate that the null distribution of $\mmdhat^2$ changes significantly with dimension~($d$), the ratio $n/m$ and the choice of the kernel, unlike our proposed statistic.}
    \label{fig:supp-null-1}
   \end{figure}
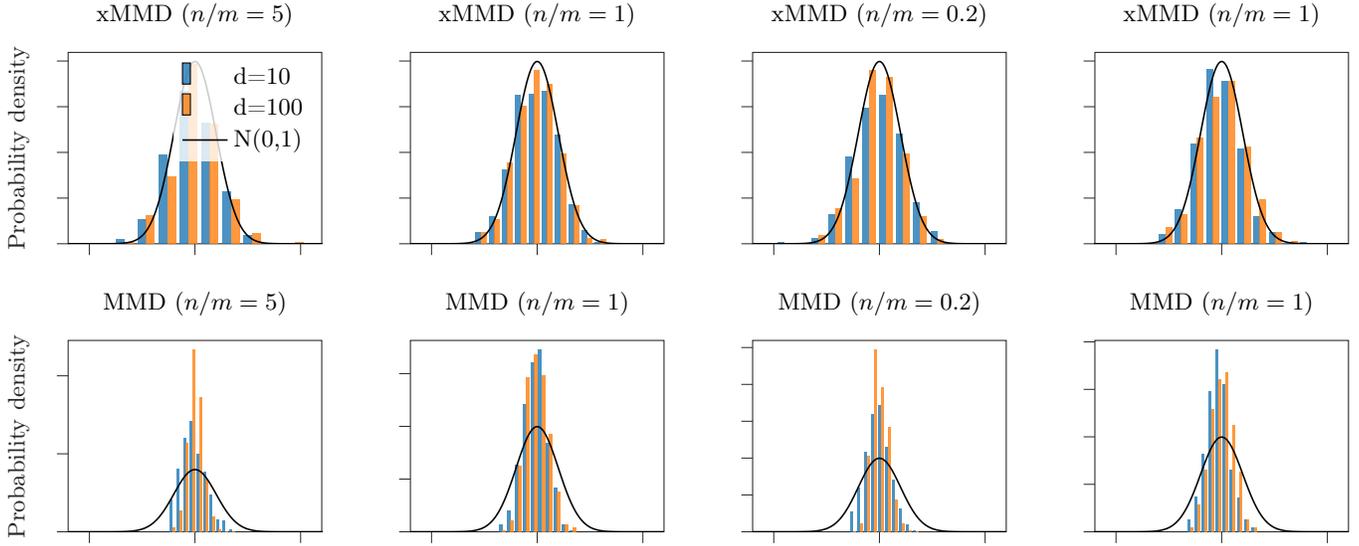

    \paragraph{Null Distribution.} \Cref{fig:supp-null-1} denotes the null distribution of our proposed statistic~($\csmmd$) along with that of the usual MMD normalized by its empirical standard deviation. The null distribution in~\Cref{fig:supp-null-1} is \texttt{Dirichlet} with parameter $2\times \boldsymbol{1} \in \mathbb{R}^d$ for $d \in \{10, 500\}$. 
    
    \paragraph{Power Curves.} In~\Cref{fig:supp-power-1}, we plot the power curves for the different tests using a Gaussian Kernel, and we report the results of the same experiment with a polynomial kernel of degree $5$ in~\Cref{fig:supp-power-2}. Recall that the polynomial kernel of degree $r$ and scale parameter $s>0$ is defined as $k(x,y) = \lp 1 + (x^Ty)/s \rp^r$. In both instances, we selected the scale parameter using the median heuristic.  
    
    From the figure, we can see that the $\cmmd$ test is competitive with the computationally more costly tests, namely the MMD permutation test and the MMD-spectral test of~\citet{gretton2009fast}. Furthermore, the performance of $\cmmd$ test is significantly better than the existing computationally efficient tests, namely block-MMD test~(with block-size $\sqrt{n}$) and linear-MMD test. 

    \begin{figure}[htb!]
        \def\figwidth{0.35\linewidth}
        \def\figheight{0.35\linewidth} %
        \centering
        \hspace*{-2cm}
        \begin{tabular}{lll}
            \input{FinalFigs/PowerCurveAllMethods_RBFd_10_eps_0_2}
            &
            \input{FinalFigs/PowerCurveAllMethods_RBFd_50_eps_0_125}
        & 
            \input{FinalFigs/PowerCurveAllMethods_RBFd_100_eps_0_15}
    \end{tabular}
    
    \caption{Power Curves for the different tests using Gaussian kernel with scale parameter chosen via median heuristic. The two distributions are $P=N(\boldsymbol{0}, I_d)$ and $Q=N(a_{\epsilon, j}, I_d)$ where $a_{\epsilon, j}$ is obtained by setting the first $j \leq d$ coordinates of $\boldsymbol{0}\in \mathbb{R}^d$ equal to $\epsilon$. The figures demonstrate that the $\cmmd$ test is competitive with more computationally expensive tests~(MMD-perm and MMD-spectral), while performing significantly better than the low complexity alternatives (B-MMD and L-MMD). The batch-size used in the B-MMD test was $\sqrt{n}$.}
    \label{fig:supp-power-1}
   \end{figure}
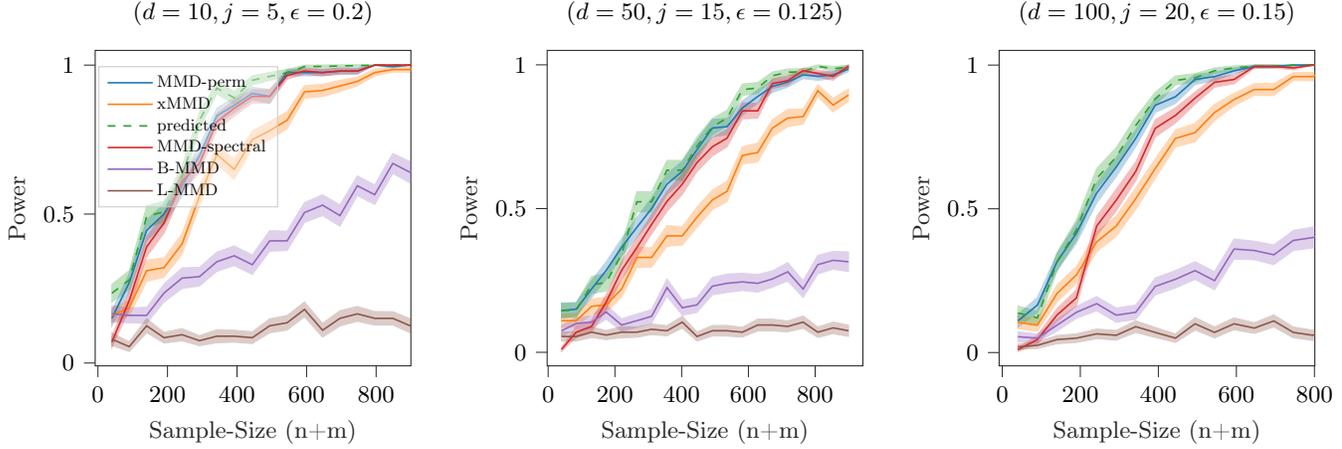  
    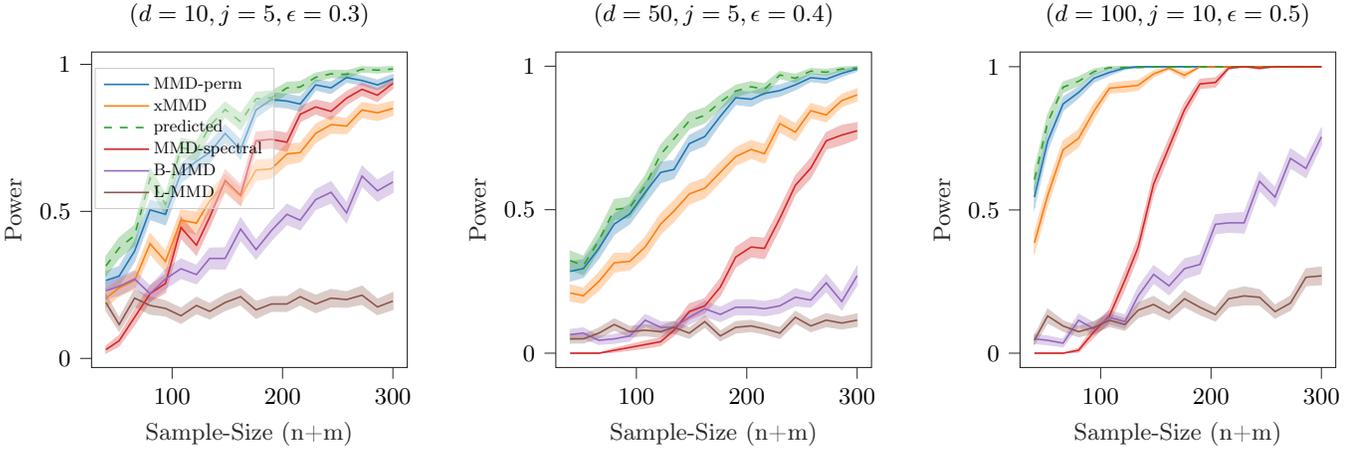
\begin{figure}[htb!]
        \def\figwidth{0.35\linewidth}
        \def\figheight{0.35\linewidth} %
        \centering
        \hspace*{-2cm}
        \begin{tabular}{ccc}
            \input{FinalFigs/PowerCurveAllMethods_Polynomiald_10_eps_0_3}
        &
            \input{FinalFigs/PowerCurveAllMethods_Polynomiald_50_eps_0_4}
        &
            \input{FinalFigs/PowerCurveAllMethods_Polynomiald_100_eps_0_5}
        \end{tabular}
        
        \caption{Power curves of the different kernel-based tests using a polynomial kernel of degree $5$, i.e., $k(x,y) = \lp 1 + (x^Ty)/s \rp^5$ with $s$ chosen via the median heuristic.}
        \label{fig:supp-power-2}
   \end{figure}  

\begin{figure}[htb!]
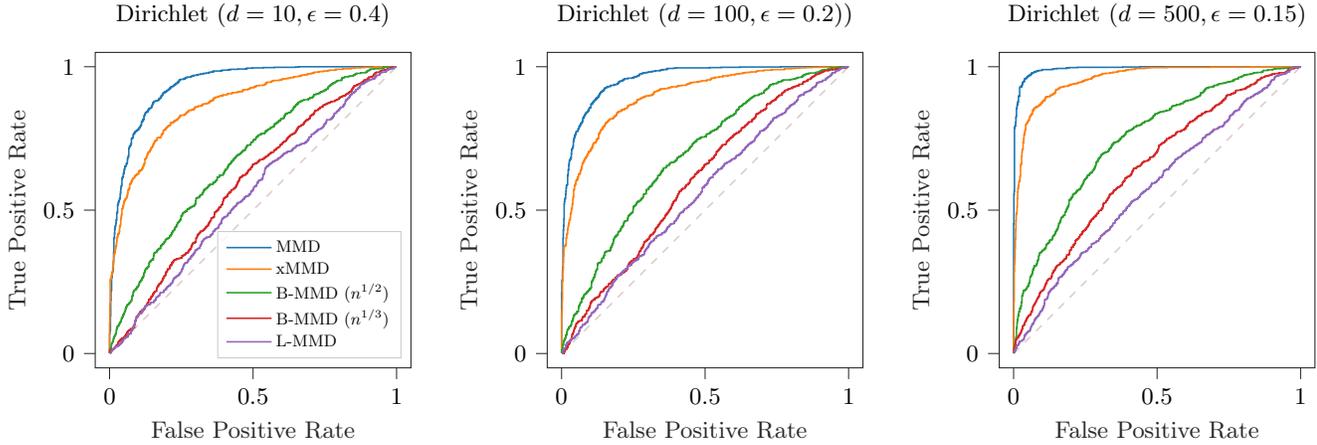

    \def\figwidth{0.35\linewidth}
    \def\figheight{0.35\linewidth} %
    \centering
    \hspace*{-2cm}
    \begin{tabular}{ccc}
        \input{FinalFigs/ROC_curve_n_200_d_10_Dirichlet}   
    &
        \input{FinalFigs/ROC_curve_n_200_d_100_Dirichlet}
    &
        \input{FinalFigs/ROC_curve_n_200_d_500_Dirichlet}
    \end{tabular}
    
    \caption{ROC curves using the different statistics with Gaussian kernel for testing two Dirichlet distributions in dimensions $d \in \{10, 100, 500\}$ with sample-size $n=m=200$. The two distributions are $P=\texttt{Dirichlet}(\boldsymbol{1})$ and $Q= \texttt{Dirichlet}((1+\epsilon)\times \boldsymbol{1})$ where $\boldsymbol{1} \in \mathbb{R}^d$ is the all-ones vector.}
    \label{fig:supp-roc-1}
   \end{figure}

    \paragraph{ROC curves.}  In~\Cref{fig:supp-roc-1}, we plot some additional ROC curves for the different statistics. As before, we used $1000$ `null trials' and another $1000$ 'alternative trials' with sample sizes $n=200$ and $m=200$. The data generating distributions $P$ and $Q$ were both Dirichlet with parameters $\boldsymbol{1} \in \mathbb{R}^d$ and $(1+\epsilon) \times \boldsymbol{1} \in \mathbb{R}^d$ for $(d, \epsilon) \in \{(10, 0.4), (100, 0.2), (500, 0.15)\}$.

        \subsection{Comparison with ME and SCF tests of~\citet{jitkrittum2016interpretable}}
        \label{appendix:linear-time-tests}
            We now present some experimental results comparing the performance of our cross-MDD test with the linear time mean embedding~(MD) and smoothed characteristic function~(SCF) tests of \citet{jitkrittum2016interpretable}.
            These tests proceed in the following steps: 
            \begin{itemize}
                \item Fix $J$, and choose points $\{v_1, \ldots, v_J\}$ from $\mathbb{R}^d$, where $d$ is the dimension of the observation space. 
                \item Using $\Xsample$ and $\Ysample$ with $n=m$, compute $\{z_i: 1 \leq i \leq n\}$, where $z_i = [k(v_J, X_i) - k(v_J, Y_i)]_{j=1}^J \in \mathbb{R}^J$ for ME test, and $z_i = [\hat{l}(X_i)\sin(X_i^Tv_j - \hat{l}(Y_i)\sin(Y_i^Tv_j), \hat{l}(X_i)\cos(X_i^Tv_j) - \hat{l}(Y_i)\cos(Y_i^Tv_j)]_{j=1}^J \in \mathbb{R}^{2J}$ for the SCF test. 
                Define $\bar{z}_n = \frac{1}{n} \sum_{i=1}^n z_i$, and $S_n = \frac{1}{n-1} (z_i - \bar{z}_n)(z_i - \bar{z}_n)^T$. 
                
                \item Using the above, define the test statistic 
                \begin{align}
                    \label{eq:linear-time-test-stat} 
                    \hat{\lambda}_n \defined \bar{z}_n^T \lp S_n + \gamma_n I \rp^{-1} \bar{z}_n,  
                \end{align}
                where $\gamma_n$ is some regularization parameter that converges to $0$ with $n$, and $I$ denotes the identity matrix. 
                For a fixed $d$ and $J$, \citet{jitkrittum2016interpretable} show that the above statistic has a $\chi^2(J)$~(resp. $\chi^2(2J)$) limiting null distribution in the ME~(resp. SCF) case. This result is used to calibrate the test at a given level $\alpha$. 
            \end{itemize}
            
\begin{figure}[htb!]
                    \def\figwidth{0.50\linewidth}
                    \def\figheight{0.45\linewidth} %
                \centering
                \begin{tabular}{cc}
                    \input{FinalFigs/Type_I_comparison_d_10_seed_5541_}
                &
                    \input{FinalFigs/Power_comparison_my_1_0d_10_seed_9876_}
                \end{tabular}
                \caption{The figures plot the variation of the type-I error (left) and the power~(right) with sample-size of the three tests: cross-MMD, and the two linear time tests, ME and SCF, proposed by~\citet{jitkrittum2016interpretable}. }
                \label{fig:supp-linear-time-1}
               \end{figure}
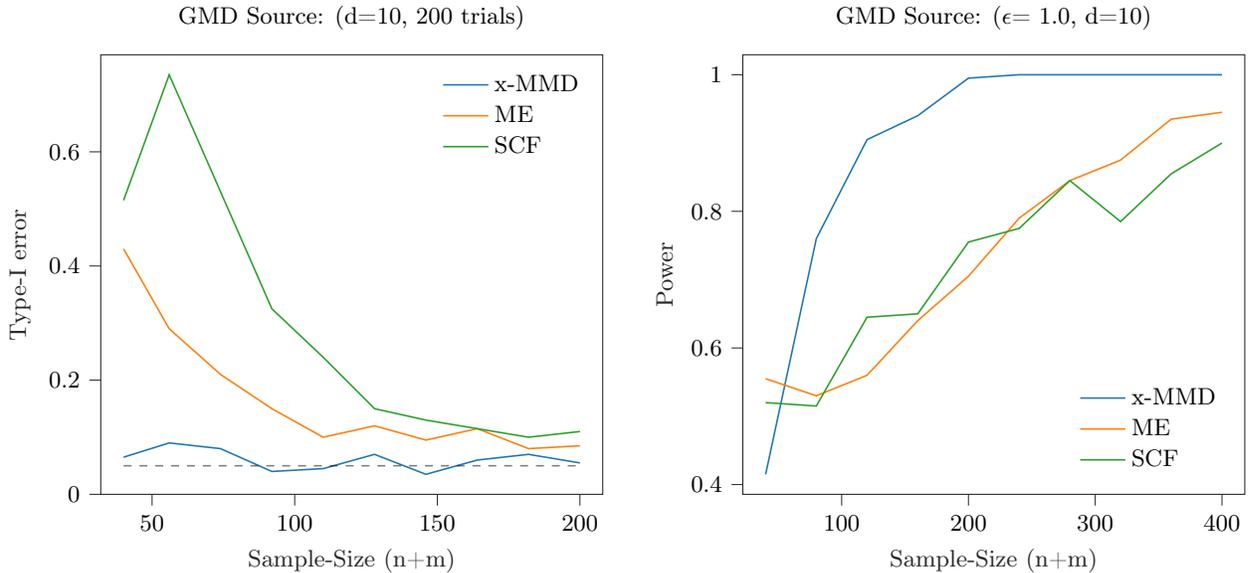

            In~\Cref{fig:supp-linear-time-1}, we plot the variation of type-I error and power with sample-size of the three tests for the Gaussian Mean Difference~(GMD) source with $d=10$. As the figures suggest, the cross-MMD achieves higher power and tighter control over the type-I error than the ME and SCF tests in this regime. 
              
\begin{figure}[htb!]
                   \def\figwidth{0.60\linewidth}
                   \def\figheight{0.50\linewidth} %
                   \centering
                       \input{FinalFigs/Type_I_comparison_d_100_seed_1417_}
                   \caption{The ME and SCF tests provide poor control over the type-I error in the regime when $d/n$ is large, suggesting that the limiting null distribution is different (or the convergence rate is slow) in this regime.}
                   \label{fig:supp-linear-time-2}
               \end{figure}
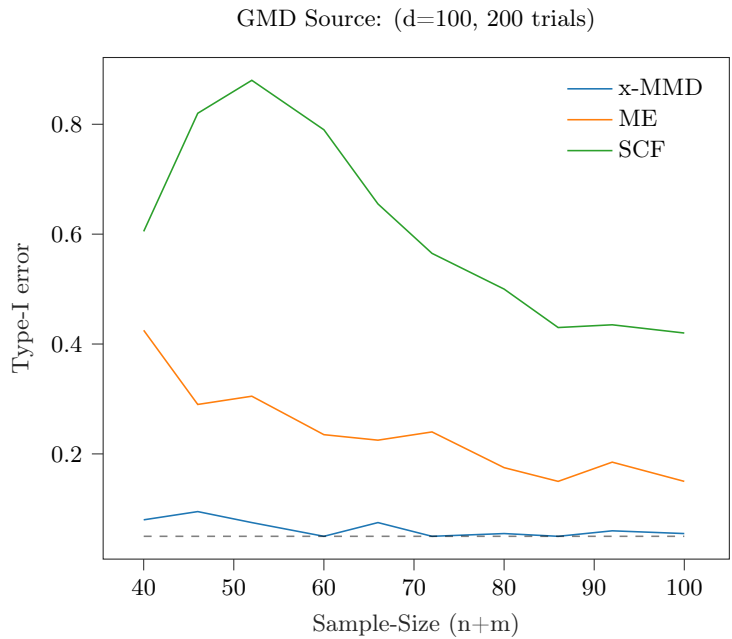

            The ME and SCF tests are calibrated based on the limiting distribution of their statistic in the low dimensional regime: fixed $d$, and $n \to \infty$. However, the high type-I error of these tests for small $n$ values suggests that their limiting distribution may be different in the high dimensional regime, when both $d$ and $n$ go to infinity. We further observe this in~\Cref{fig:supp-linear-time-2} when $d=100$ and $d/n>1$. 
        
            We end this section with a discussion of some key points of difference between the ME and SCF tests, and our proposed cross-MMD test. 
            \begin{itemize}
                \item  The ME and SCF tests require the kernel to be uniformly bounded, whereas our test requires only mild moment conditions that are even satisfied by unbounded kernels if the underlying distributions are not too heavy-tailed~(formally described in~\Cref{assump:kernel-1}). Furthermore, the ME and SCF tests have several tuning parameters: number of features $J$, $\{v_1, \ldots, v_J\}$, bandwidth, step-size for gradient ascent etc. In practice, $J$ is usually set to $5$, and the other parameters are selected by solving a $Jd + 1$ dimensional optimization problem via gradient ascent. While each step of gradient ascent has linear in $n$ complexity, the number of steps needed may be large for higher dimensions, resulting in a higher computational overhead. 

                \item More importantly, the ME and SCF tests are only valid in the `low-dimensional setting': fixed $d$ and $J$, with $n \to \infty$. In the high dimensional setting, when $(d, n) \to \infty$,  the limiting null distribution may no longer be $\chi^2(J)$. This is also suggested by the behavior of type-I error of ME and SCF tests in~\Cref{fig:supp-linear-time-1} and~\Cref{fig:supp-linear-time-2}. This results in the following practical issue: \emph{given a problem with $n=500$ and $d=200$, how should one calibrate the threshold for those tests?}  

                Our proposed test does not suffer from this, because in both high and low dimensional settings, our statistic has the same limiting distribution. This is a significant practical advantage of our cross-MMD test over ME and SCF tests.

                \item In the regime where the number of features, $J$, is allowed to increase with $n$, we expect that the resulting ME and SCF tests may have low power (for small regularization parameter $\gamma_n$). This is because, the test statistic $\hat{\lambda}_n$ used by ME and SCF tests is similar to Hotelling's $T^2$ statistic, for which \citet{bai1996effect} characterized the asymptotic power in this regime. In particular,  their Theorem 2.1 implies that the power of the $T^2$ test grows slowly with $n$, especially when $J/n \approx 1$. 
                
                Finally, we note that our ideas also extend to more general degenerate U-statistics (as discussed in~\Cref{appendix:proof-general-U-statistic}). Hence, they are also applicable in cases beyond MMD distance, where we may not have good linear time alternatives.
            \end{itemize}
        
        \subsection{Type-I Error and goodness-of-fit test of null distribution}
        \label{appendix:type-I-and-gof}
\begin{figure}[htb!]
                \centering
                \begin{tabular}{cc}
                    \includegraphics[width=.48\textwidth]{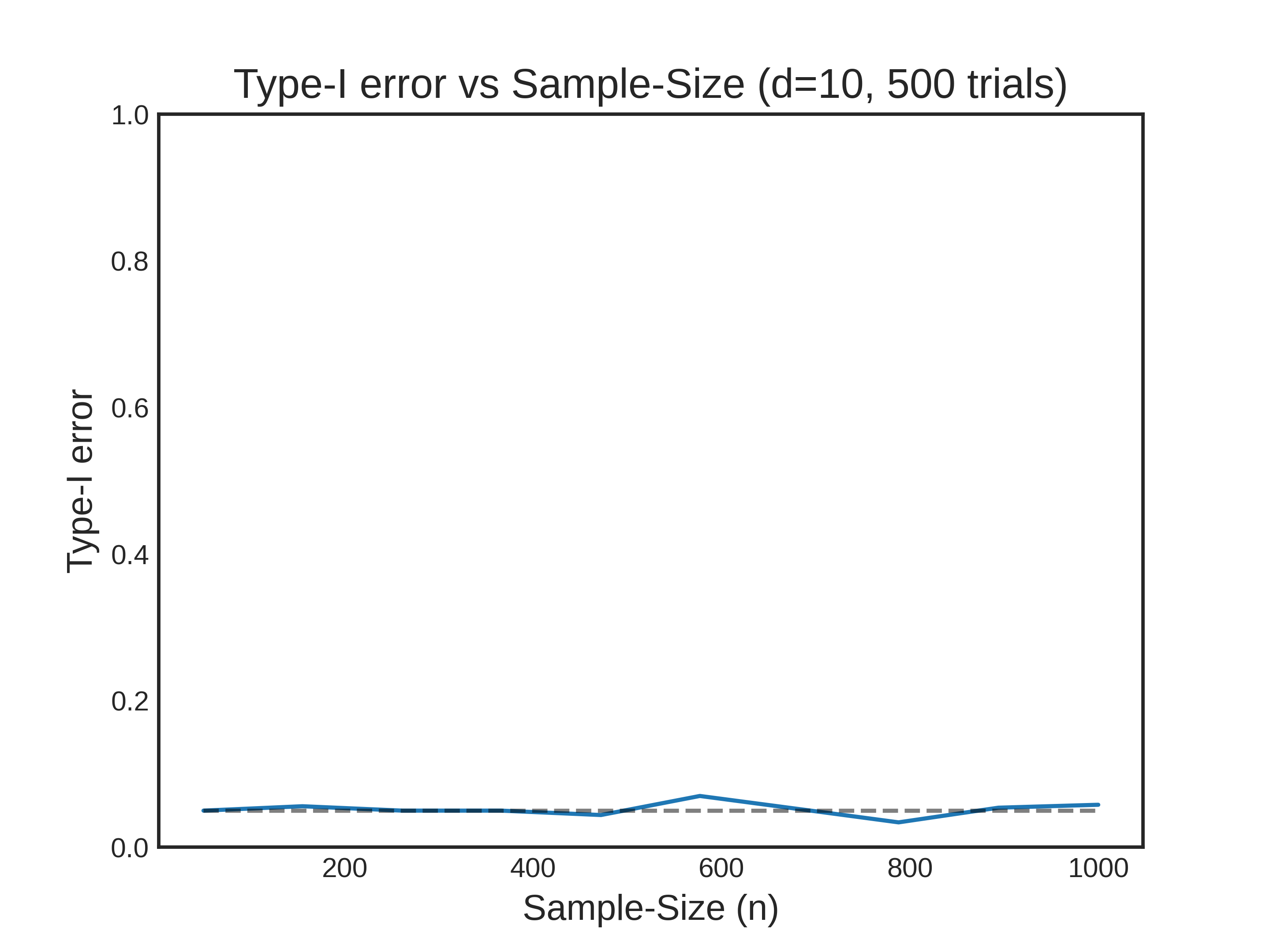}
                &
                     \includegraphics[width=.48\textwidth]{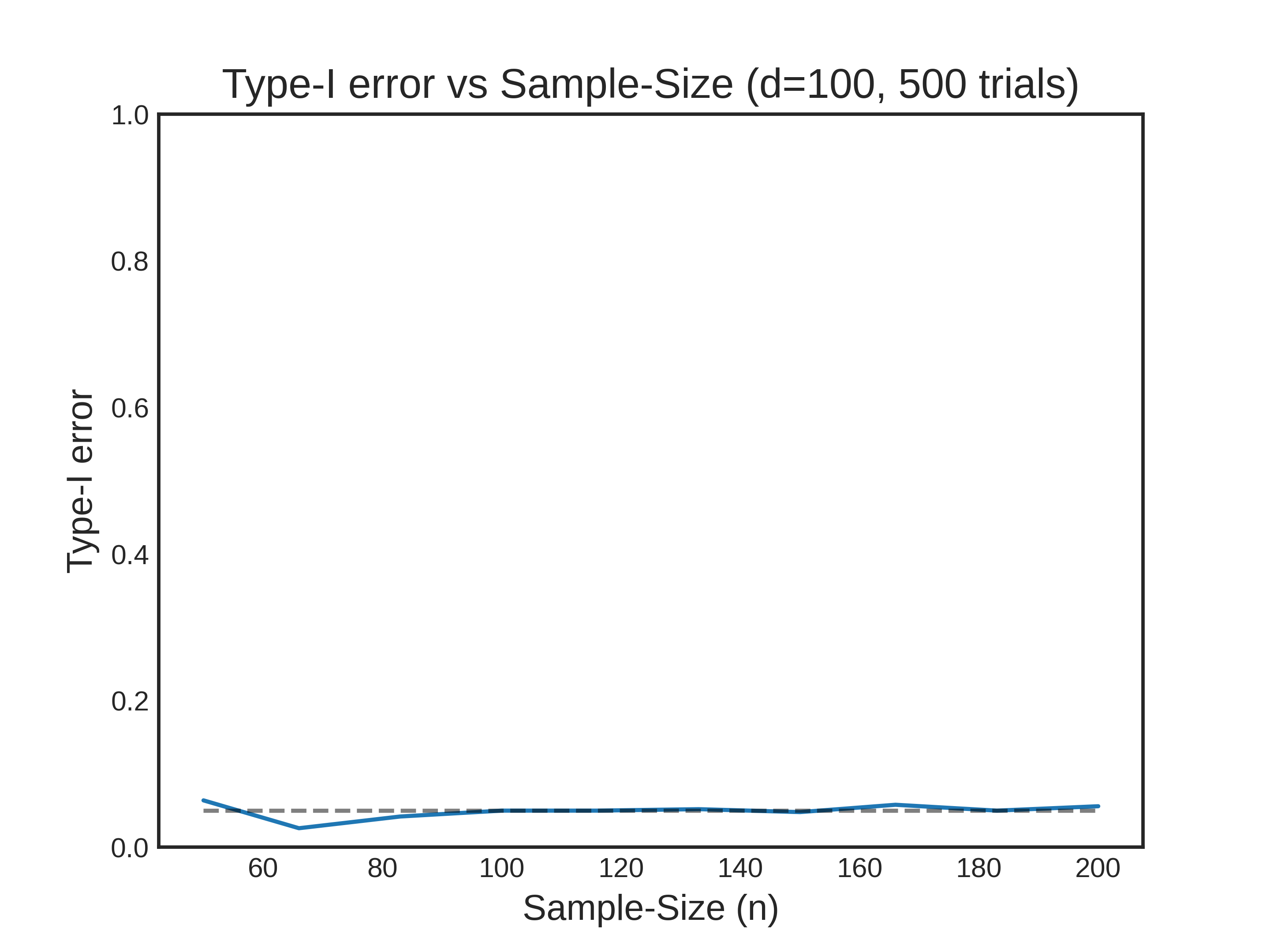} 
                \end{tabular}
                \caption{
                The two figures show the variation of the type-I error of the cross-MMD test with sample-size for dimensions $d \in \{10, 100\}$. The dashed horizontal line denotes the level $\alpha=0.05$. In summary, these tests do not find evidence against the null hypothesis that the null distribution is Gaussian.}
                \label{fig:supp-type-I-errors}
               \end{figure}

            In this section, we experimentally verify the limiting Gaussian distribution of the $\csmmd$ statistic under the null. 
            We first plot the variation of the type-I error of our cross-MMD test with sample size in~\Cref{fig:supp-type-I-errors}. We considered the case when $\Xsample$ and $\Ysample$ are both drawn \iid from a multivariate Gaussian vector in dimension $d \in \{10, 100\}$, and $n=m$.  
            
\begin{figure}[htb!]
                    \centering
                    \begin{tabular}{cc}
                        \includegraphics[width=.48\textwidth]{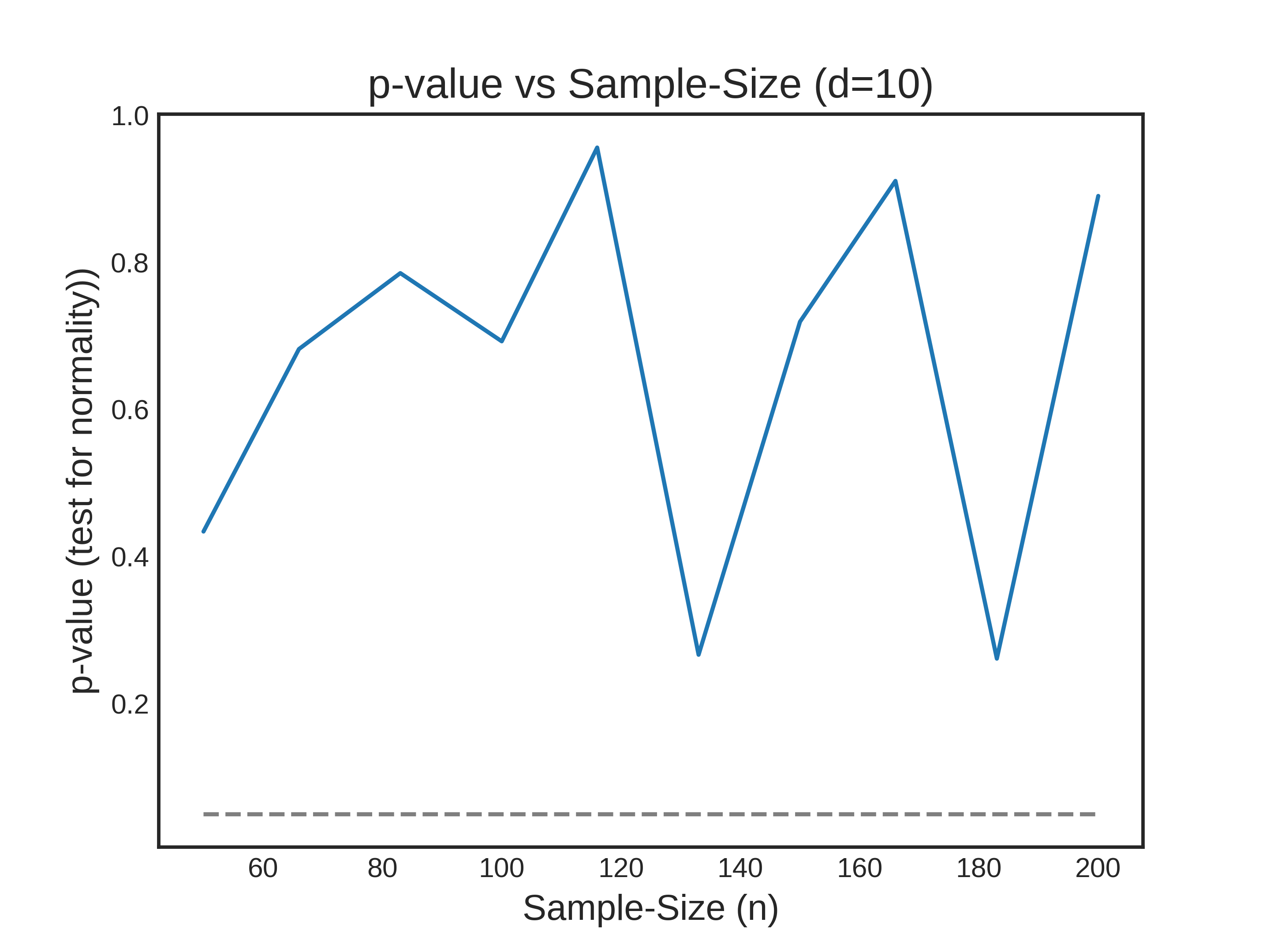}
                    &
                         \includegraphics[width=.48\textwidth]{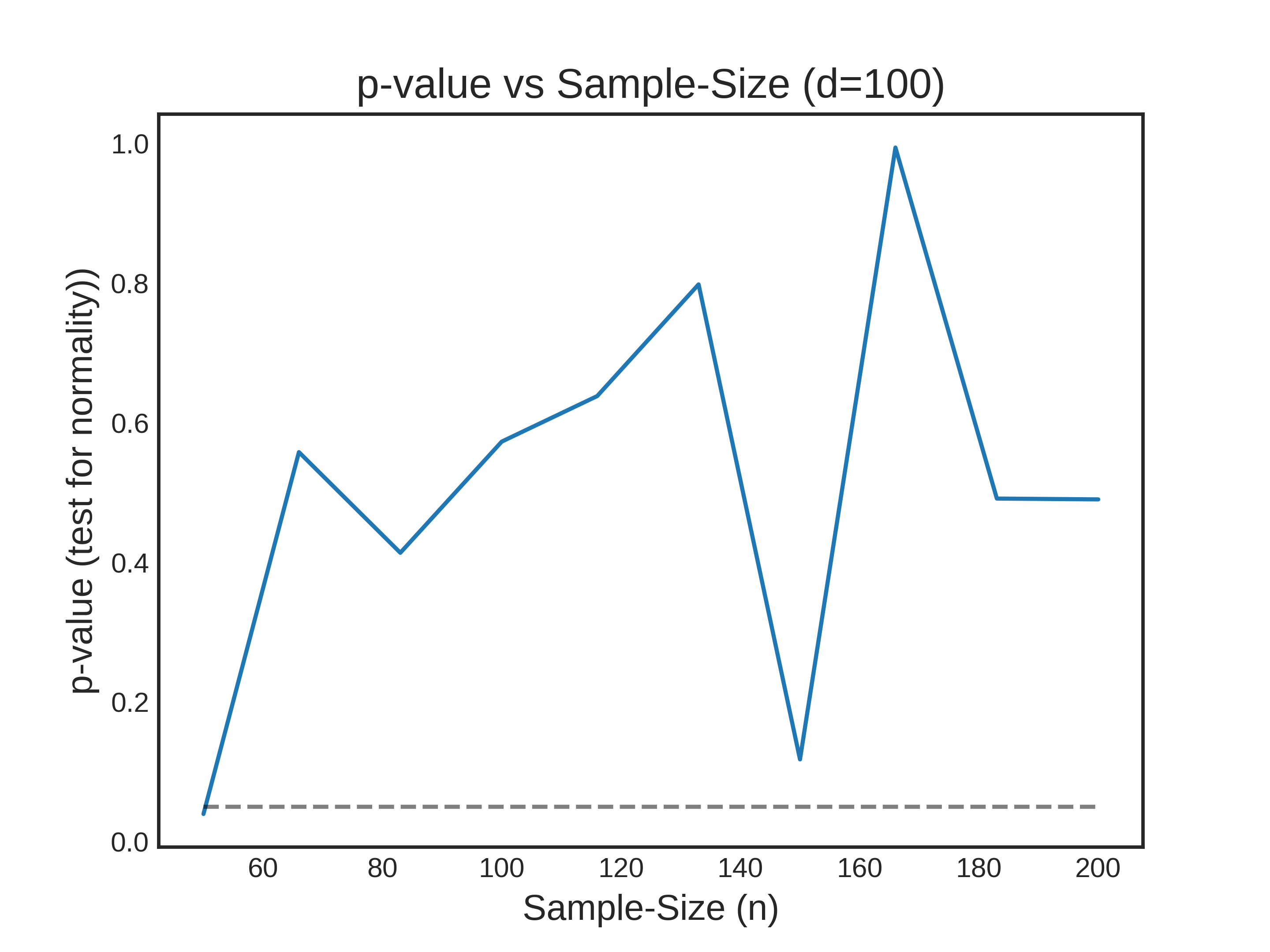} 
                    \end{tabular}
                    \caption{
                    The two figures show p-values for the test for normality proposed by~\citet{d1973tests} (using the implementation \texttt{scipy.stats.normaltest}) of the cross-MMD statistic for dimensions $d \in \{10, 100\}$. In both dimension regimes, the test does not find evidence against the null that the cross-MMD statistic is normally distributed under the null.}
                    \label{fig:supp-gof-test}
                   \end{figure}

            Next, we plot the p-values for the test for normality proposed by~\citet{d1973tests}, and implemented in the function \texttt{scipy.stats.normaltest} in Python.  We performed this test at different sample-sizes~($n$), and for each value of $n$, we calculated the $\csmmd$ statistic on $200$ different indpendent sample pairs. The results are shown in~\Cref{fig:supp-gof-test} 
                 
        \subsection{Comparison with Friedman-Rafsky test}
        \label{subsec:FR-test}
            We now compare the performance of our cross-MMD test with the Friedman-Rafsky two-sample test. This test, proposed by~\citet{friedman1979multivariate}, uses a graph-based statistic that is a multivariate generalization of the Wald-Wolfowitz runs statistic introduced by~\citet{wald1940test}. This statistic, denoted by $R$, is constructed as follows: 
            \begin{itemize}
                \item Pool the samples $\Xsample$ and $\Ysample$ to get $\mathbb{Z}$ of size $N=n+m$. Construct the complete graph with $N$ nodes, and edge weights equal to the euclidean distance between two end points. 
                \item Construct the minimal spanning tree (MST) of the complete graph $G$, and denote the $0$-$1$ valued adjacency matrix of this MST by $M$. 
                \item The statistic $R$ is defined as  one more than the number of edges in $M$ with endpoints from different samples. 
            \end{itemize}
            The statistic $R$ is expected to take a large value under the null when $\Xsample$ and $\Ysample$ are drawn from the same distribution. Hence, the FR test rejects the null for small values of $R$. The rejection threshold can be obtained either by the limiting distribution of $R$ characterized by~\citep[Theorem~1]{henze1999multivariate}, or using the permutation-test. 
            
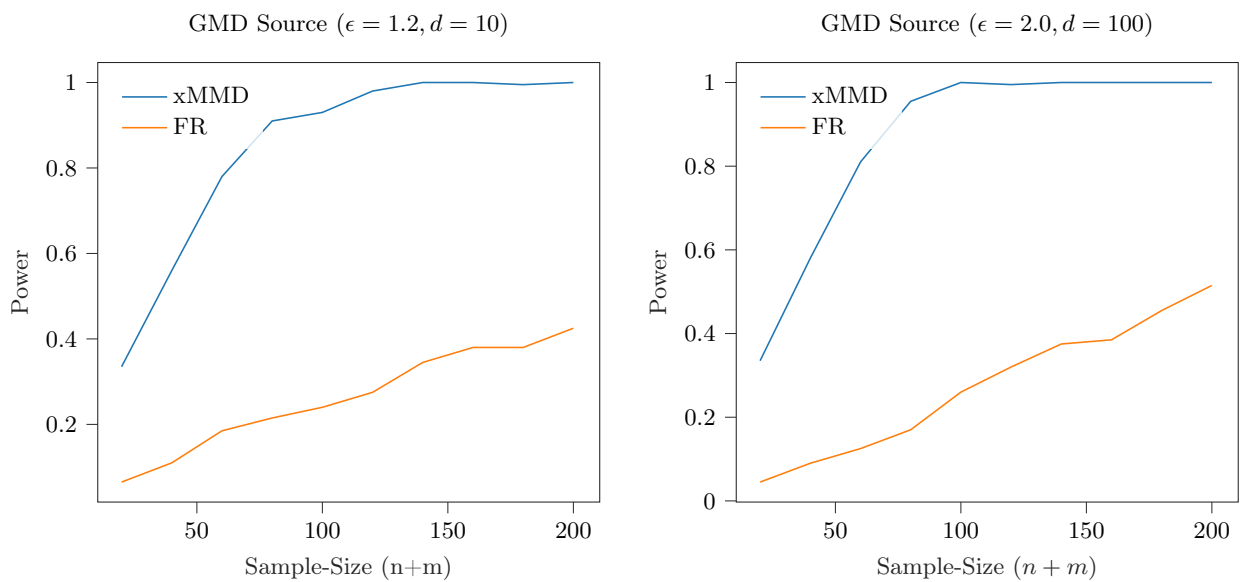
\begin{figure}[htb!]
                    \def\figwidth{0.50\linewidth}
                    \def\figheight{0.45\linewidth} %
                \centering
                \hspace*{-2cm}
                \begin{tabular}{cc}
                    \input{FinalFigs/Power_FR_d_10_seed_1484}
                &
                    \input{FinalFigs/Power_FR_d_100_seed_2404}
                \end{tabular}
                \caption{
                The figures show the power curves for Friedman-Rafsky~(FR) test and our cross-MMD test in the low~($d=10$) and high~($d=100$) dimensional settings with $m=n$ in both plots. The figures indicate that our cross-MMD test is significantly more powerful than the FR test.}
                \label{fig:supp-FR-1}
               \end{figure}

            In~\Cref{fig:supp-FR-1}, we compare the power of the FR permutation-test with our cross-MMD test in a low dimensional~($d/n$ small) and a high dimensional~($d/n$ large) problem. In both cases, it is observed that the power of FR test is significantly smaller than that of cross-MMD test. 
            
\end{appendix}

\end{document}

%% file: preamble.tex
\usepackage{amsmath,amssymb,amsthm,amsfonts,latexsym,bbm,xspace,graphicx,float,mathtools,mathdots}
\usepackage{braket,caption,subcaption,ellipsis,xcolor,textcomp}
\usepackage{combelow} %
\usepackage[nameinlink]{cleveref}
\crefname{ineq}{inequality}{inequalities}
\creflabelformat{ineq}{#2{\upshape(#1)}#3}

\newtheorem{theorem}{Theorem}

\newtheorem*{namedtheorem}{\theoremname}
\newcommand{\theoremname}{testing}

\newtheorem{lemma}[theorem]{Lemma}

\newtheorem{fact}[theorem]{Fact}

\newtheorem{assumption}{Assumption}

\theoremstyle{definition}

\newtheorem{remark}[theorem]{Remark}

\renewcommand{\Pr}{\mathop{\mathbb{P}\/}}

\newcommand{\calE}{\mathcal{E}}
\newcommand{\calF}{\mathcal{F}}

\newcommand{\calO}{\mathcal{O}}

\newcommand{\calX}{\mathcal{X}}

\newcommand{\ignore}[1]{}

\newcommand{\defined}{\coloneqq}
\usepackage{autonum}

\usepackage{natbib}
\setcitestyle{authoryear,open={(},close={)}} %

%% file: newMacros.tex
\newcommand{\lp}{\left(}
\newcommand{\rp}{\right)}
\newcommand{\lb}{\left[}
\newcommand{\rb}{\right]}
\newcommand{\lv}{\left\lvert}
\newcommand{\rv}{\right \rvert}
\newcommand{\mmdhat}{\widehat{\mathrm{MMD}}}

\newcommand{\var}{\mathbb{V}\xspace}
\newcommand{\Xsample}{\mathbb{X}\xspace}
\newcommand{\Ysample}{\mathbb{Y}\xspace}
\newcommand{\muhat}{\widehat{\mu}}
\newcommand{\nuhat}{\widehat{\nu}}

\newcommand{\sigmahat}{\widehat{\sigma}}
\newcommand{\tildek}{\widetilde{k}}

\newcommand{\convprob}{\stackrel{p}{\longrightarrow}}
\newcommand{\convdist}{\stackrel{d}{\longrightarrow}}

\newcommand{\levy}{L\'evy\xspace}

\newcommand{\reals}{\mathbb{R}}
\newcommand{\iid}{\text{i.i.d.}\xspace}

\newcommand{\Sobolev}{\mathcal{W}^{\beta,2}}
\newcommand{\dmmd}{{\mathrm{MMD}}}
\newcommand{\cmmd}{{\mathrm{xMMD}}}
\newcommand{\mmdval}{\gamma_{n}}

\newcommand{\ind}{\mathbbm{1}}
\newcommand{\mc}[1]{\mathcal{#1}}

\newcommand{\crossU}{\bar{U}}
\newcommand{\crossmmd}{\mathrm{x}\mmdhat^2}
\newcommand{\csmmd}{\bar{\mathrm{x}}\mmdhat^2}
\newcommand{\barq}{\bar{q}}
\newcommand{\barr}{\bar{r}}

\newcommand{\nullclass}{\mc{P}_{n}^{(0)}}
\newcommand{\altclass}{\mc{P}_{n}^{(1)}}
\newcommand{\Pnm}{P_{n}}
\newcommand{\Qnm}{Q_{n}}
\newcommand{\bark}{\bar{k}}

\newcommand{\mutilde}{\widetilde{\mu}}
\newcommand{\nutilde}{\widetilde{\nu}}
\newcommand{\gtilde}{\widetilde{g}}
\newcommand{\Wtilde}{\widetilde{W}}
\newcommand{\Ztilde}{\widetilde{Z}}

\newcommand{\sigmaPX}{\sigma_{P,X}}
\newcommand{\sigmaPY}{\sigma_{P,Y}}
\newcommand{\EPn}{\mathbb{E}_{P_n}}

%% file: FinalFigs/Null_Dists_d_10_500_n_100_m_150_kernel__Gaussian_RBF_2022_10_13_18_01_37cross.tex
\begin{tikzpicture}

\definecolor{darkorange25512714}{RGB}{255,127,14}
\definecolor{darkslategray38}{RGB}{38,38,38}
\definecolor{lightgray204}{RGB}{204,204,204}
\definecolor{steelblue31119180}{RGB}{31,119,180}

\begin{axis}[
axis line style={darkslategray38},
height=\figheight,
legend cell align={left},
legend style={fill opacity=0.8, draw opacity=1, text opacity=1, draw=none},
tick align=outside,
tick pos=left,
title={Null distribution of $\cmmd$},
width=\figwidth,
x grid style={lightgray204},
xlabel=\textcolor{darkslategray38}{Statistic Value},
xmin=-6, xmax=6,
xtick style={color=darkslategray38},
y grid style={lightgray204},
ylabel=\textcolor{darkslategray38}{Probability density},
ymin=0, ymax=0.418868408525629,
ytick style={color=darkslategray38}
]
\draw[draw=none,fill=steelblue31119180,fill opacity=0.8] (axis cs:-2.92182064056396,0) rectangle (axis cs:-2.68546175956726,0.0169234208171233);
\addlegendimage{ybar,ybar legend,draw=none,fill=steelblue31119180,fill opacity=0.8}
\addlegendentry{d=10}

\draw[draw=none,fill=steelblue31119180,fill opacity=0.8] (axis cs:-2.33092355728149,0) rectangle (axis cs:-2.09456467628479,0.0592319489606815);
\draw[draw=none,fill=steelblue31119180,fill opacity=0.8] (axis cs:-1.74002647399902,0) rectangle (axis cs:-1.50366759300232,0.105771380107021);
\draw[draw=none,fill=steelblue31119180,fill opacity=0.8] (axis cs:-1.14912927150726,0) rectangle (axis cs:-0.912770390510559,0.275005532797841);
\draw[draw=none,fill=steelblue31119180,fill opacity=0.8] (axis cs:-0.558232128620148,0) rectangle (axis cs:-0.321873247623444,0.338468348058881);
\draw[draw=none,fill=steelblue31119180,fill opacity=0.8] (axis cs:0.0326650738716125,0) rectangle (axis cs:0.269023954868317,0.359622619812561);
\draw[draw=none,fill=steelblue31119180,fill opacity=0.8] (axis cs:0.623562276363373,0) rectangle (axis cs:0.859921157360077,0.304621513252993);
\draw[draw=none,fill=steelblue31119180,fill opacity=0.8] (axis cs:1.21445941925049,0) rectangle (axis cs:1.45081830024719,0.15654164255839);
\draw[draw=none,fill=steelblue31119180,fill opacity=0.8] (axis cs:1.80535662174225,0) rectangle (axis cs:2.04171562194824,0.0507702419662985);
\draw[draw=none,fill=steelblue31119180,fill opacity=0.8] (axis cs:2.39625406265259,0) rectangle (axis cs:2.63261294364929,0.0253851312256849);
\draw[draw=none,fill=darkorange25512714,fill opacity=0.8] (axis cs:-2.68546199798584,0) rectangle (axis cs:-2.44910311698914,0.00846171040856165);
\addlegendimage{ybar,ybar legend,draw=none,fill=darkorange25512714,fill opacity=0.8}
\addlegendentry{d=500}

\draw[draw=none,fill=darkorange25512714,fill opacity=0.8] (axis cs:-2.09456491470337,0) rectangle (axis cs:-1.85820603370667,0.05500109546349);
\draw[draw=none,fill=darkorange25512714,fill opacity=0.8] (axis cs:-1.50366759300232,0) rectangle (axis cs:-1.26730871200562,0.148079932149829);
\draw[draw=none,fill=darkorange25512714,fill opacity=0.8] (axis cs:-0.912770450115204,0) rectangle (axis cs:-0.6764115691185,0.258082115394897);
\draw[draw=none,fill=darkorange25512714,fill opacity=0.8] (axis cs:-0.321873247623444,0) rectangle (axis cs:-0.0855143666267395,0.376546037215505);
\draw[draw=none,fill=darkorange25512714,fill opacity=0.8] (axis cs:0.269023954868317,0) rectangle (axis cs:0.505382835865021,0.393469454618449);
\draw[draw=none,fill=darkorange25512714,fill opacity=0.8] (axis cs:0.859921157360077,0) rectangle (axis cs:1.09628009796143,0.220004426238272);
\draw[draw=none,fill=darkorange25512714,fill opacity=0.8] (axis cs:1.45081830024719,0) rectangle (axis cs:1.6871771812439,0.143849076945548);
\draw[draw=none,fill=darkorange25512714,fill opacity=0.8] (axis cs:2.04171562194824,0) rectangle (axis cs:2.27807450294495,0.0634628024578731);
\draw[draw=none,fill=darkorange25512714,fill opacity=0.8] (axis cs:2.63261270523071,0) rectangle (axis cs:2.86897158622742,0.0253851312256849);
\addplot [semithick, black]
table {%
-10 7.69459862670642e-23
-9.97997997997998 9.39820210218911e-23
-9.95995995995996 1.14743877987917e-22
-9.93993993993994 1.40036162642795e-22
-9.91991991991992 1.70834984871876e-22
-9.8998998998999 2.08324025950642e-22
-9.87987987987988 2.53938085193762e-22
-9.85985985985986 3.09415635142992e-22
-9.83983983983984 3.7686222201397e-22
-9.81981981981982 4.58826916995414e-22
-9.7997997997998 5.58394465795474e-22
-9.77977977977978 6.79296312742742e-22
-9.75975975975976 8.26044308669654e-22
-9.73973973973974 1.00409166885045e-21
-9.71971971971972 1.22002665237565e-21
-9.6996996996997 1.48180551599987e-21
-9.67967967967968 1.79903258756111e-21
-9.65965965965966 2.18329684678274e-21
-9.63963963963964 2.64857624243904e-21
-9.61961961961962 3.21172317125736e-21
-9.5995995995996 3.89304716291232e-21
-9.57957957957958 4.71701393695898e-21
-9.55955955955956 5.71308371631411e-21
-9.53953953953954 6.91671611023779e-21
-9.51951951951952 8.37057415073758e-21
-9.4994994994995 1.01259663374544e-20
-9.47947947947948 1.22445730038445e-20
-9.45945945945946 1.48005121824234e-20
-9.43943943943944 1.78828106797973e-20
-9.41941941941942 2.15983585814365e-20
-9.3993993993994 2.60754402558043e-20
-9.37937937937938 3.14679525475421e-20
-9.35935935935936 3.79604417474457e-20
-9.33933933933934 4.5774115701642e-20
-9.31931931931932 5.51740167796855e-20
-9.2992992992993 6.6477576193283e-20
-9.27927927927928 8.00648113242436e-20
-9.25925925925926 9.63904764362469e-20
-9.23923923923924 1.1599853476858e-19
-9.21921921921922 1.39539388139188e-19
-9.1991991991992 1.67790380698338e-19
-9.17917917917918 2.01680188581445e-19
-9.15915915915916 2.42317819504618e-19
-9.13913913913914 2.91027078875695e-19
-9.11911911911912 3.49387515332417e-19
-9.0990990990991 4.19283042962206e-19
-9.07907907907908 5.0295965472299e-19
-9.05905905905906 6.03093897535385e-19
-9.03903903903904 7.22874080905507e-19
-9.01901901901902 8.66096545670873e-19
-8.998998998999 1.03727973679138e-18
-8.97897897897898 1.24179931485728e-18
-8.95895895895896 1.48604811781133e-18
-8.93893893893894 1.77762546207239e-18
-8.91891891891892 2.12556106807901e-18
-8.8988988988989 2.54057982941549e-18
-8.87887887887888 3.03541474067764e-18
-8.85885885885886 3.62517658454128e-18
-8.83883883883884 4.32779048512422e-18
-8.81881881881882 5.16451119999019e-18
-8.7987987987988 6.16053109048305e-18
-8.77877877877878 7.34569713009337e-18
-8.75875875875876 8.75535614211525e-18
-8.73873873873874 1.04313507694241e-17
-8.71871871871872 1.24231925503958e-17
-8.6986986986987 1.47894429982973e-17
-8.67867867867868 1.75993388643364e-17
-8.65865865865866 2.09347039316773e-17
-8.63863863863864 2.48921968838275e-17
-8.61861861861862 2.95859531836935e-17
-8.5985985985986 3.51506886838449e-17
-8.57857857857858 4.17453440895294e-17
-8.55855855855856 4.95573626747691e-17
-8.53853853853854 5.88077091106193e-17
-8.51851851851852 6.97567552529606e-17
-8.4984984984985 8.27111796592674e-17
-8.47847847847848 9.8032051926925e-17
-8.45845845845846 1.16144301209627e-16
-8.43843843843844 1.37547801096493e-16
-8.41841841841842 1.62830341150177e-16
-8.3983983983984 1.92682799625235e-16
-8.37837837837838 2.27916883183252e-16
-8.35835835835836 2.69485858889318e-16
-8.33833833833834 3.18508772685576e-16
-8.31831831831832 3.76298728354137e-16
-8.2982982982983 4.44395893386326e-16
-8.27827827827828 5.24606005103494e-16
-8.25825825825826 6.19045274051723e-16
-8.23823823823824 7.30192724674871e-16
-8.21821821821822 8.60951178494166e-16
-8.1981981981982 1.01471827585831e-15
-8.17817817817818 1.19546915264237e-15
-8.15815815815816 1.40785264250169e-15
-8.13813813813814 1.65730316851105e-15
-8.11811811811812 1.95017082606148e-15
-8.0980980980981 2.29387254841619e-15
-8.07807807807808 2.69706769497134e-15
-8.05805805805806 3.16986191875719e-15
-8.03803803803804 3.72404376402735e-15
-8.01801801801802 4.3733591283238e-15
-7.997997997998 5.13382950919607e-15
-7.97797797797798 6.02412085866193e-15
-7.95795795795796 7.06597090549303e-15
-7.93793793793794 8.28468399583074e-15
-7.91791791791792 9.70970386854708e-15
-7.8978978978979 1.13752763482777e-14
-7.87787787787788 1.33212157347805e-14
-7.85785785785786 1.55937907247683e-14
-7.83783783783784 1.82467480586655e-14
-7.81781781781782 2.13424947819475e-14
-7.7977977977978 2.49534630966872e-14
-7.77777777777778 2.91636853079909e-14
-7.75775775775776 3.40706104038538e-14
-7.73773773773774 3.9787198415571e-14
-7.71771771771772 4.64443339685918e-14
-7.6976976976977 5.4193606440538e-14
-7.67767767767768 6.32105109959252e-14
-7.65765765765766 7.36981325813117e-14
-7.63763763763764 8.58913838706879e-14
-7.61761761761762 1.00061878296601e-13
-7.5975975975976 1.16523530854699e-13
-7.57757757757758 1.35638992516664e-13
-7.55755755755756 1.57827039041884e-13
-7.53753753753754 1.83571051982057e-13
-7.51751751751752 2.13428748996349e-13
-7.4974974974975 2.48043342543513e-13
-7.47747747747748 2.88156330935935e-13
-7.45745745745746 3.34622154016965e-13
-7.43743743743744 3.88424977793732e-13
-7.41741741741742 4.50697908714384e-13
-7.3973973973974 5.22744979473711e-13
-7.37737737737738 6.06066294885249e-13
-7.35735735735736 7.02386779168738e-13
-7.33733733733734 8.13689025751835e-13
-7.31731731731732 9.42250818252909e-13
-7.2972972972973 1.09068796768221e-12
-7.27727727727728 1.262003197176e-12
-7.25725725725726 1.45964190299847e-12
-7.23723723723724 1.68755573049416e-12
-7.21721721721722 1.95027502769792e-12
-7.1971971971972 2.25299137914218e-12
-7.17717717717718 2.60165157997871e-12
-7.15715715715716 3.00306458801653e-12
-7.13713713713714 3.4650231910834e-12
-7.11711711711712 3.99644235193919e-12
-7.0970970970971 4.60751644580953e-12
-7.07707707707708 5.30989788981514e-12
-7.05705705705706 6.11689998287646e-12
-7.03703703703704 7.0437271332242e-12
-7.01701701701702 8.10773605306645e-12
-6.996996996997 9.32873195138555e-12
-6.97697697697698 1.07293042619726e-11
-6.95695695695696 1.23352070109962e-11
-6.93693693693694 1.41757895636779e-11
-6.91691691691692 1.62844842008237e-11
-6.8968968968969 1.86993577716893e-11
-6.87687687687688 2.14637355595386e-11
-6.85685685685686 2.46269064908967e-11
-6.83683683683684 2.82449199306815e-11
-6.81681681681682 3.2381485546105e-11
-6.7967967967968 3.71089891068559e-11
-6.77677677677678 4.25096386334913e-11
-6.75675675675676 4.86767570277083e-11
-6.73673673673674 5.57162392366004e-11
-6.71671671671672 6.37481941394491e-11
-6.6966966966967 7.29087937236032e-11
-6.67667667667668 8.33523547614402e-11
-6.65665665665666 9.52536811418151e-11
-6.63663663663664 1.08810698278135e-10
-6.61661661661662 1.24247414645768e-10
-6.5965965965966 1.41817249531744e-10
-6.57657657657658 1.61806770551278e-10
-6.55655655655656 1.84539889444164e-10
-6.53653653653654 2.10382570159622e-10
-6.51651651651652 2.39748109325542e-10
-6.4964964964965 2.73103055937438e-10
-6.47647647647648 3.10973844559381e-10
-6.45645645645646 3.53954224575809e-10
-6.43643643643644 4.027135771479e-10
-6.41641641641642 4.58006221596996e-10
-6.3963963963964 5.20681824054169e-10
-6.37637637637638 5.91697033481538e-10
-6.35635635635636 6.72128483698846e-10
-6.33633633633634 7.63187314959523e-10
-6.31631631631632 8.66235385046001e-10
-6.2962962962963 9.8280335793813e-10
-6.27627627627628 1.11461087800766e-09
-6.25625625625626 1.26358905957513e-09
-6.23623623623624 1.43190554571787e-09
-6.21621621621622 1.62199241663862e-09
-6.1961961961962 1.83657725691024e-09
-6.17617617617618 2.0787177227378e-09
-6.15615615615616 2.35183998527873e-09
-6.13613613613614 2.65978146430928e-09
-6.11611611611612 3.00683830841829e-09
-6.0960960960961 3.39781812376754e-09
-6.07607607607608 3.83809850362696e-09
-6.05605605605606 4.33369196574699e-09
-6.03603603603604 4.89131796456919e-09
-6.01601601601602 5.51848271073395e-09
-5.995995995996 6.22356760178439e-09
-5.97597597597598 7.01592714588943e-09
-5.95595595595596 7.90599734535251e-09
-5.93593593593594 8.90541559921139e-09
-5.91591591591592 1.00271532849868e-08
-5.8958958958959 1.12856622892642e-08
-5.87587587587588 1.2697036876002e-08
-5.85585585585586 1.42791924110059e-08
-5.83583583583584 1.60520626017053e-08
-5.81581581581582 1.80378170640688e-08
-5.7957957957958 2.02611011941336e-08
-5.77577577577578 2.27493005011625e-08
-5.75575575575576 2.55328317539416e-08
-5.73573573573574 2.86454635022852e-08
-5.71571571571572 3.2124668763613e-08
-5.6956956956957 3.60120129107462e-08
-5.67567567567568 4.03535800631662e-08
-5.65565565565566 4.5200441571292e-08
-5.63563563563564 5.06091704933412e-08
-5.61561561561562 5.66424062986154e-08
-5.5955955955956 6.33694743912418e-08
-5.57557557557558 7.08670654362614e-08
-5.55555555555556 7.92199798873018e-08
-5.53553553553554 8.85219435638491e-08
-5.51551551551552 9.8876500608364e-08
-5.4954954954955 1.10397990671284e-07
-5.47547547547548 1.23212617727566e-07
-5.45545545545546 1.3745961852414e-07
-5.43543543543544 1.5329253929596e-07
-5.41541541541542 1.70880630071684e-07
-5.3953953953954 1.90410366621162e-07
-5.37537537537538 2.1208711087848e-07
-5.35535535535536 2.36136921509202e-07
-5.33533533533534 2.62808527181656e-07
-5.31531531531532 2.92375476052561e-07
-5.2952952952953 3.25138475990267e-07
-5.27527527527528 3.61427941137511e-07
-5.25525525525526 4.01606761563285e-07
-5.23523523523524 4.46073313973501e-07
-5.21521521521522 4.95264732746229e-07
-5.1951951951952 5.4966046193278e-07
-5.17517517517518 6.09786110324583e-07
-5.15515515515516 6.76217633231267e-07
-5.13513513513514 7.49585866251233e-07
-5.11511511511512 8.30581438046261e-07
-5.0950950950951 9.19960090959837e-07
-5.07507507507508 1.01854844024876e-06
-5.05505505505506 1.12725020473309e-06
-5.03503503503504 1.24705294381396e-06
-5.01501501501502 1.37903533806611e-06
-4.99499499499499 1.5243750529858e-06
-4.97497497497497 1.68435722796805e-06
-4.95495495495495 1.86038363520377e-06
-4.93493493493493 2.05398255592983e-06
-4.91491491491491 2.26681942433715e-06
-4.89489489489489 2.50070829244518e-06
-4.87487487487487 2.75762417238901e-06
-4.85485485485485 3.03971631583941e-06
-4.83483483483483 3.34932249368796e-06
-4.81481481481481 3.68898434268124e-06
-4.79479479479479 4.06146384937932e-06
-4.77477477477477 4.4697610456467e-06
-4.75475475475475 4.91713299385613e-06
-4.73473473473473 5.40711414409909e-06
-4.71471471471471 5.94353814994721e-06
-4.69469469469469 6.53056123369604e-06
-4.67467467467467 7.17268719654363e-06
-4.65465465465465 7.87479417380527e-06
-4.63463463463463 8.64216324004121e-06
-4.61461461461461 9.48050897386715e-06
-4.59459459459459 1.03960120972233e-05
-4.57457457457457 1.13953543089884e-05
-4.55455455455455 1.24857554380297e-05
-4.53453453453453 1.3675013046071e-05
-4.51451451451451 1.49715446161227e-05
-4.49449449449449 1.63844324676437e-05
-4.47447447447447 1.79234715450684e-05
-4.45445445445445 1.95992202318354e-05
-4.43443443443443 2.14230543475548e-05
-4.41441441441441 2.34072244914564e-05
-4.39439439439439 2.55649169007242e-05
-4.37437437437437 2.79103179977393e-05
-4.35435435435435 3.04586828055866e-05
-4.33433433433433 3.32264074164067e-05
-4.31431431431431 3.62311057022691e-05
-4.29429429429429 3.94916904631592e-05
-4.27427427427427 4.3028459211397e-05
-4.25425425425425 4.68631847962824e-05
-4.23423423423423 5.10192110769697e-05
-4.21421421421421 5.55215538554582e-05
-4.19419419419419 6.03970072851107e-05
-4.17417417417417 6.56742559732345e-05
-4.15415415415415 7.13839929989176e-05
-4.13413413413413 7.75590440694795e-05
-4.11411411411411 8.42344980404937e-05
-4.09409409409409 9.14478440253317e-05
-4.07407407407407 9.92391153205018e-05
-4.05405405405405 0.000107651040372646
-4.03403403403403 0.000116729201011866
-4.01401401401401 0.000126522198173995
-3.99399399399399 0.000137081825331481
-3.97397397397397 0.000148463249848567
-3.95395395395395 0.000160725202471485
-3.93393393393393 0.000173930175158222
-3.91391391391391 0.00018814462744512
-3.89389389389389 0.000203439201538965
-3.87387387387387 0.000219888946313312
-3.85385385385385 0.000237573550376443
-3.83383383383383 0.000256577584365551
-3.81381381381381 0.000276990752607344
-3.79379379379379 0.000298908154269281
-3.77377377377377 0.000322430554107926
-3.75375375375375 0.000347664662901427
-3.73373373373373 0.000374723427631836
-3.71371371371371 0.000403726331459719
-3.69369369369369 0.000434799703508357
-3.67367367367367 0.000468077038447599
-3.65365365365365 0.00050369932583812
-3.63363363363363 0.000541815389165405
-3.61361361361361 0.000582582234459159
-3.59359359359359 0.000626165408357979
-3.57357357357357 0.000672739365441021
-3.55355355355355 0.000722487844607978
-3.53353353353353 0.00077560425424601
-3.51351351351351 0.000832292065877155
-3.49349349349349 0.000892765215932443
-3.47347347347347 0.000957248515249216
-3.45345345345345 0.0010259780658361
-3.43343343343343 0.00109920168439588
-3.41341341341341 0.00117717933203981
-3.39339339339339 0.00126018354956833
-3.37337337337337 0.00134849989763212
-3.35335335335335 0.00144242740102448
-3.33333333333333 0.00154227899629111
-3.31331331331331 0.00164838198177652
-3.29329329329329 0.00176107846915772
-3.27327327327327 0.00188072583544552
-3.25325325325325 0.00200769717436226
-3.23323323323323 0.00214238174593163
-3.21321321321321 0.00228518542304204
-3.19319319319319 0.00243653113367012
-3.17317317317317 0.00259685929737497
-3.15315315315315 0.00276662825459747
-3.13313313313313 0.00294631468722261
-3.11311311311311 0.00313641402878609
-3.09309309309309 0.00333744086263052
-3.07307307307307 0.00354992930624086
-3.05305305305305 0.00377443337991422
-3.03303303303303 0.00401152735784579
-3.01301301301301 0.00426180609964128
-2.99299299299299 0.00452588536019618
-2.97297297297297 0.0048044020758154
-2.95295295295295 0.00509801462438215
-2.93293293293293 0.00540740305732385
-2.91291291291291 0.00573326930106519
-2.89289289289289 0.00607633732560526
-2.87287287287287 0.00643735327780636
-2.85285285285285 0.00681708557693873
-2.83283283283283 0.00721632496998623
-2.81281281281281 0.00763588454418632
-2.79279279279279 0.00807659969425075
-2.77277277277277 0.00853932804169477
-2.75275275275275 0.00902494930369032
-2.73273273273273 0.00953436510885489
-2.71271271271271 0.0100684987573917
-2.69269269269269 0.0106282949230102
-2.67267267267267 0.0112147192940778
-2.65265265265265 0.0118287581514852
-2.63263263263263 0.0124714178807513
-2.61261261261261 0.0131437244159435
-2.59259259259259 0.0138467226130541
-2.57257257257257 0.0145814755505492
-2.55255255255255 0.0153490637548887
-2.53253253253253 0.0161505843489182
-2.51251251251251 0.0169871501211409
-2.49249249249249 0.0178598885140022
-2.47247247247247 0.018769940529451
-2.45245245245245 0.0197184595501959
-2.43243243243243 0.0207066100752274
-2.41241241241241 0.0217355663683581
-2.39239239239239 0.0228065110187135
-2.37237237237237 0.0239206334123108
-2.35235235235235 0.0250791281140699
-2.33233233233233 0.0262831931598317
-2.31231231231231 0.0275340282581906
-2.29229229229229 0.0288328329022027
-2.27227227227227 0.0301808043912899
-2.25225225225225 0.0315791357639331
-2.23223223223223 0.033029013642035
-2.21221221221221 0.0345316159881232
-2.19219219219219 0.0360881097768736
-2.17217217217217 0.0376996485827434
-2.15215215215215 0.0393673700858293
-2.13213213213213 0.0410923934983949
-2.11211211211211 0.0428758169148479
-2.09209209209209 0.0447187145882915
-2.07207207207207 0.0466221341371235
-2.05205205205205 0.0485870936855041
-2.03203203203203 0.0506145789418747
-2.01201201201201 0.0527055402200587
-1.99199199199199 0.0548608894078376
-1.97197197197197 0.057081496888248
-1.95195195195195 0.059368188419199
-1.93193193193193 0.0617217419773594
-1.91191191191191 0.0641428845726061
-1.89189189189189 0.0666322890396674
-1.87187187187187 0.0691905708139176
-1.85185185185185 0.0718182846986055
-1.83183183183183 0.0745159216311036
-1.81181181181181 0.077283905456062
-1.79179179179179 0.0801225897136326
-1.77177177177177 0.083032254451193
-1.75175175175175 0.0860131030672496
-1.73173173173173 0.0890652591964251
-1.71171171171171 0.0921887636446459
-1.69169169169169 0.0953835713838294
-1.67167167167167 0.0986495486155338
-1.65165165165165 0.101986469913169
-1.63163163163163 0.10539401545248
-1.61161161161161 0.108871768340093
-1.59159159159159 0.112419212049971
-1.57157157157157 0.116035727977651
-1.55155155155155 0.119720593122119
-1.53153153153153 0.123472977905145
-1.51151151151151 0.127291944137829
-1.49149149149149 0.13117644314399
-1.47147147147147 0.135125314049902
-1.45145145145145 0.139137282249685
-1.43143143143143 0.143210958055468
-1.41141141141141 0.147344835541168
-1.39139139139139 0.151537291588457
-1.37137137137137 0.155786585143159
-1.35135135135135 0.160090856689972
-1.33133133133133 0.164448127952996
-1.31131131131131 0.168856301829129
-1.29129129129129 0.173313162560933
-1.27127127127127 0.17781637615506
-1.25125125125125 0.182363491051798
-1.23123123123123 0.186951939050736
-1.21121121121121 0.191579036496956
-1.19119119119119 0.1962419857315
-1.17117117117117 0.200937876809264
-1.15115115115115 0.205663689486728
-1.13113113113113 0.210416295481265
-1.11111111111111 0.215192461003031
-1.09109109109109 0.219988849559688
-1.07107107107107 0.224802025033432
-1.05105105105105 0.229628455029052
-1.03103103103103 0.234464514490888
-1.01101101101101 0.239306489585817
-0.990990990990991 0.24415058184851
-0.970970970970971 0.24899291258444
-0.950950950950951 0.253829527525259
-0.930930930930931 0.258656401730343
-0.910910910910911 0.2634694447275
-0.890890890890891 0.268264505884996
-0.870870870870871 0.273037380006279
-0.850850850850851 0.277783813137949
-0.830830830830831 0.282499508580786
-0.810810810810811 0.287180133092853
-0.790790790790791 0.291821323272996
-0.77077077077077 0.296418692112302
-0.75075075075075 0.300967835700437
-0.73073073073073 0.305464340073112
-0.71071071071071 0.309903788186304
-0.69069069069069 0.314281767002296
-0.67067067067067 0.318593874672039
-0.65065065065065 0.322835727797843
-0.63063063063063 0.327002968759958
-0.61061061061061 0.331091273090187
-0.59059059059059 0.33509635687531
-0.57057057057057 0.339013984172804
-0.55055055055055 0.34283997442106
-0.53053053053053 0.346570209826128
-0.51051051051051 0.350200642706842
-0.49049049049049 0.353727302780113
-0.47047047047047 0.357146304368113
-0.45045045045045 0.360453853509139
-0.43043043043043 0.363646254953996
-0.41041041041041 0.366719919029892
-0.39039039039039 0.369671368354051
-0.37037037037037 0.372497244379499
-0.35035035035035 0.375194313755802
-0.33033033033033 0.377759474487924
-0.31031031031031 0.38018976187679
-0.29029029029029 0.382482354225654
-0.27027027027027 0.384634578296894
-0.25025025025025 0.386643914504485
-0.23023023023023 0.388508001828027
-0.21021021021021 0.390224642434919
-0.19019019019019 0.391791805998011
-0.17017017017017 0.393207633696876
-0.15015015015015 0.394470441891644
-0.13013013013013 0.395578725459258
-0.11011011011011 0.396531160782876
-0.0900900900900901 0.397326608386124
-0.07007007007007 0.397964115204853
-0.05005005005005 0.398442916490068
-0.03003003003003 0.398762437336696
-0.01001001001001 0.398922293833933
0.01001001001001 0.398922293833933
0.03003003003003 0.398762437336696
0.05005005005005 0.398442916490068
0.07007007007007 0.397964115204853
0.0900900900900901 0.397326608386124
0.11011011011011 0.396531160782876
0.13013013013013 0.395578725459258
0.15015015015015 0.394470441891644
0.17017017017017 0.393207633696876
0.19019019019019 0.391791805998011
0.21021021021021 0.390224642434919
0.23023023023023 0.388508001828027
0.25025025025025 0.386643914504485
0.27027027027027 0.384634578296894
0.29029029029029 0.382482354225654
0.31031031031031 0.38018976187679
0.33033033033033 0.377759474487924
0.35035035035035 0.375194313755802
0.37037037037037 0.372497244379499
0.39039039039039 0.369671368354051
0.41041041041041 0.366719919029892
0.43043043043043 0.363646254953996
0.45045045045045 0.360453853509139
0.47047047047047 0.357146304368113
0.49049049049049 0.353727302780113
0.51051051051051 0.350200642706842
0.53053053053053 0.346570209826128
0.55055055055055 0.34283997442106
0.57057057057057 0.339013984172804
0.59059059059059 0.33509635687531
0.61061061061061 0.331091273090187
0.63063063063063 0.327002968759958
0.65065065065065 0.322835727797843
0.67067067067067 0.318593874672039
0.69069069069069 0.314281767002296
0.71071071071071 0.309903788186304
0.73073073073073 0.305464340073112
0.75075075075075 0.300967835700437
0.77077077077077 0.296418692112302
0.790790790790791 0.291821323272996
0.810810810810811 0.287180133092853
0.830830830830831 0.282499508580786
0.850850850850851 0.277783813137949
0.870870870870871 0.273037380006279
0.890890890890891 0.268264505884996
0.910910910910911 0.2634694447275
0.930930930930931 0.258656401730343
0.950950950950951 0.253829527525259
0.970970970970971 0.24899291258444
0.990990990990991 0.24415058184851
1.01101101101101 0.239306489585817
1.03103103103103 0.234464514490888
1.05105105105105 0.229628455029052
1.07107107107107 0.224802025033432
1.09109109109109 0.219988849559688
1.11111111111111 0.215192461003031
1.13113113113113 0.210416295481265
1.15115115115115 0.205663689486728
1.17117117117117 0.200937876809264
1.19119119119119 0.1962419857315
1.21121121121121 0.191579036496956
1.23123123123123 0.186951939050736
1.25125125125125 0.182363491051798
1.27127127127127 0.17781637615506
1.29129129129129 0.173313162560933
1.31131131131131 0.168856301829129
1.33133133133133 0.164448127952996
1.35135135135135 0.160090856689972
1.37137137137137 0.155786585143159
1.39139139139139 0.151537291588457
1.41141141141141 0.147344835541168
1.43143143143143 0.143210958055468
1.45145145145145 0.139137282249685
1.47147147147147 0.135125314049902
1.49149149149149 0.13117644314399
1.51151151151151 0.127291944137829
1.53153153153153 0.123472977905145
1.55155155155155 0.119720593122119
1.57157157157157 0.116035727977651
1.59159159159159 0.112419212049971
1.61161161161161 0.108871768340093
1.63163163163163 0.10539401545248
1.65165165165165 0.101986469913169
1.67167167167167 0.0986495486155338
1.69169169169169 0.0953835713838294
1.71171171171171 0.0921887636446459
1.73173173173173 0.0890652591964251
1.75175175175175 0.0860131030672496
1.77177177177177 0.083032254451193
1.79179179179179 0.0801225897136326
1.81181181181181 0.077283905456062
1.83183183183183 0.0745159216311036
1.85185185185185 0.0718182846986055
1.87187187187187 0.0691905708139176
1.89189189189189 0.0666322890396674
1.91191191191191 0.0641428845726061
1.93193193193193 0.0617217419773594
1.95195195195195 0.059368188419199
1.97197197197197 0.057081496888248
1.99199199199199 0.0548608894078376
2.01201201201201 0.0527055402200588
2.03203203203203 0.0506145789418748
2.05205205205205 0.0485870936855042
2.07207207207207 0.0466221341371236
2.09209209209209 0.0447187145882916
2.11211211211211 0.0428758169148479
2.13213213213213 0.041092393498395
2.15215215215215 0.0393673700858294
2.17217217217217 0.0376996485827434
2.19219219219219 0.0360881097768737
2.21221221221221 0.0345316159881233
2.23223223223223 0.033029013642035
2.25225225225225 0.0315791357639332
2.27227227227227 0.03018080439129
2.29229229229229 0.0288328329022028
2.31231231231231 0.0275340282581906
2.33233233233233 0.0262831931598317
2.35235235235235 0.02507912811407
2.37237237237237 0.0239206334123108
2.39239239239239 0.0228065110187136
2.41241241241241 0.0217355663683581
2.43243243243243 0.0207066100752275
2.45245245245245 0.0197184595501959
2.47247247247247 0.0187699405294511
2.49249249249249 0.0178598885140022
2.51251251251251 0.016987150121141
2.53253253253253 0.0161505843489182
2.55255255255255 0.0153490637548888
2.57257257257257 0.0145814755505493
2.59259259259259 0.0138467226130541
2.61261261261261 0.0131437244159435
2.63263263263263 0.0124714178807514
2.65265265265265 0.0118287581514852
2.67267267267267 0.0112147192940778
2.69269269269269 0.0106282949230103
2.71271271271271 0.0100684987573917
2.73273273273273 0.00953436510885491
2.75275275275275 0.00902494930369034
2.77277277277277 0.00853932804169479
2.79279279279279 0.00807659969425077
2.81281281281281 0.00763588454418634
2.83283283283283 0.00721632496998621
2.85285285285285 0.00681708557693871
2.87287287287287 0.00643735327780635
2.89289289289289 0.00607633732560524
2.91291291291291 0.00573326930106518
2.93293293293293 0.00540740305732384
2.95295295295295 0.00509801462438214
2.97297297297297 0.00480440207581539
2.99299299299299 0.00452588536019617
3.01301301301301 0.00426180609964127
3.03303303303303 0.00401152735784578
3.05305305305305 0.00377443337991421
3.07307307307307 0.00354992930624085
3.09309309309309 0.00333744086263051
3.11311311311311 0.00313641402878608
3.13313313313313 0.0029463146872226
3.15315315315315 0.00276662825459746
3.17317317317317 0.00259685929737496
3.19319319319319 0.00243653113367011
3.21321321321321 0.00228518542304203
3.23323323323323 0.00214238174593163
3.25325325325325 0.00200769717436225
3.27327327327327 0.00188072583544551
3.29329329329329 0.00176107846915771
3.31331331331331 0.00164838198177652
3.33333333333333 0.0015422789962911
3.35335335335335 0.00144242740102448
3.37337337337337 0.00134849989763212
3.39339339339339 0.00126018354956833
3.41341341341341 0.00117717933203981
3.43343343343343 0.00109920168439588
3.45345345345345 0.0010259780658361
3.47347347347347 0.000957248515249212
3.49349349349349 0.000892765215932441
3.51351351351351 0.000832292065877152
3.53353353353353 0.000775604254246008
3.55355355355355 0.000722487844607976
3.57357357357357 0.000672739365441019
3.59359359359359 0.000626165408357979
3.61361361361361 0.000582582234459159
3.63363363363363 0.000541815389165405
3.65365365365365 0.00050369932583812
3.67367367367367 0.000468077038447599
3.69369369369369 0.000434799703508357
3.71371371371371 0.000403726331459719
3.73373373373373 0.000374723427631836
3.75375375375375 0.000347664662901427
3.77377377377377 0.000322430554107926
3.79379379379379 0.000298908154269281
3.81381381381381 0.000276990752607344
3.83383383383383 0.000256577584365551
3.85385385385385 0.000237573550376443
3.87387387387387 0.000219888946313312
3.89389389389389 0.000203439201538965
3.91391391391391 0.00018814462744512
3.93393393393393 0.000173930175158222
3.95395395395395 0.000160725202471485
3.97397397397397 0.000148463249848567
3.99399399399399 0.000137081825331481
4.01401401401401 0.000126522198173995
4.03403403403403 0.000116729201011866
4.05405405405405 0.000107651040372646
4.07407407407407 9.92391153205018e-05
4.09409409409409 9.14478440253317e-05
4.11411411411411 8.42344980404937e-05
4.13413413413413 7.75590440694795e-05
4.15415415415415 7.13839929989176e-05
4.17417417417417 6.56742559732345e-05
4.19419419419419 6.03970072851107e-05
4.21421421421421 5.55215538554582e-05
4.23423423423423 5.10192110769697e-05
4.25425425425425 4.68631847962824e-05
4.27427427427427 4.3028459211397e-05
4.29429429429429 3.94916904631592e-05
4.31431431431431 3.62311057022691e-05
4.33433433433433 3.32264074164067e-05
4.35435435435435 3.04586828055866e-05
4.37437437437437 2.79103179977393e-05
4.39439439439439 2.55649169007242e-05
4.41441441441441 2.34072244914564e-05
4.43443443443443 2.14230543475548e-05
4.45445445445445 1.95992202318354e-05
4.47447447447447 1.79234715450684e-05
4.49449449449449 1.63844324676437e-05
4.51451451451451 1.49715446161227e-05
4.53453453453453 1.3675013046071e-05
4.55455455455455 1.24857554380297e-05
4.57457457457457 1.13953543089884e-05
4.59459459459459 1.03960120972233e-05
4.61461461461461 9.48050897386715e-06
4.63463463463463 8.64216324004121e-06
4.65465465465465 7.87479417380527e-06
4.67467467467467 7.17268719654363e-06
4.69469469469469 6.53056123369604e-06
4.71471471471471 5.94353814994721e-06
4.73473473473473 5.40711414409909e-06
4.75475475475475 4.91713299385613e-06
4.77477477477477 4.4697610456467e-06
4.79479479479479 4.06146384937932e-06
4.81481481481481 3.68898434268124e-06
4.83483483483483 3.34932249368796e-06
4.85485485485485 3.03971631583941e-06
4.87487487487487 2.75762417238901e-06
4.89489489489489 2.50070829244518e-06
4.91491491491491 2.26681942433715e-06
4.93493493493493 2.05398255592983e-06
4.95495495495495 1.86038363520377e-06
4.97497497497497 1.68435722796805e-06
4.99499499499499 1.5243750529858e-06
5.01501501501502 1.37903533806611e-06
5.03503503503504 1.24705294381396e-06
5.05505505505506 1.12725020473309e-06
5.07507507507508 1.01854844024876e-06
5.0950950950951 9.19960090959837e-07
5.11511511511512 8.30581438046261e-07
5.13513513513514 7.49585866251233e-07
5.15515515515516 6.76217633231267e-07
5.17517517517518 6.09786110324583e-07
5.1951951951952 5.4966046193278e-07
5.21521521521522 4.95264732746229e-07
5.23523523523524 4.46073313973501e-07
5.25525525525526 4.01606761563285e-07
5.27527527527528 3.61427941137511e-07
5.2952952952953 3.25138475990267e-07
5.31531531531532 2.92375476052561e-07
5.33533533533534 2.62808527181656e-07
5.35535535535536 2.36136921509202e-07
5.37537537537538 2.1208711087848e-07
5.3953953953954 1.90410366621162e-07
5.41541541541542 1.70880630071684e-07
5.43543543543544 1.5329253929596e-07
5.45545545545546 1.3745961852414e-07
5.47547547547548 1.23212617727566e-07
5.4954954954955 1.10397990671284e-07
5.51551551551552 9.8876500608364e-08
5.53553553553554 8.85219435638491e-08
5.55555555555556 7.92199798873018e-08
5.57557557557558 7.08670654362614e-08
5.5955955955956 6.33694743912418e-08
5.61561561561562 5.66424062986154e-08
5.63563563563564 5.06091704933412e-08
5.65565565565566 4.5200441571292e-08
5.67567567567568 4.03535800631662e-08
5.6956956956957 3.60120129107462e-08
5.71571571571572 3.2124668763613e-08
5.73573573573574 2.86454635022852e-08
5.75575575575576 2.55328317539416e-08
5.77577577577578 2.27493005011625e-08
5.7957957957958 2.02611011941336e-08
5.81581581581582 1.80378170640688e-08
5.83583583583584 1.60520626017053e-08
5.85585585585586 1.42791924110059e-08
5.87587587587588 1.2697036876002e-08
5.8958958958959 1.12856622892642e-08
5.91591591591592 1.00271532849868e-08
5.93593593593594 8.90541559921139e-09
5.95595595595596 7.90599734535251e-09
5.97597597597598 7.01592714588943e-09
5.995995995996 6.22356760178439e-09
6.01601601601602 5.5184827107339e-09
6.03603603603604 4.89131796456914e-09
6.05605605605606 4.33369196574694e-09
6.07607607607608 3.83809850362692e-09
6.0960960960961 3.3978181237675e-09
6.11611611611612 3.00683830841826e-09
6.13613613613614 2.65978146430924e-09
6.15615615615616 2.35183998527871e-09
6.17617617617618 2.07871772273778e-09
6.1961961961962 1.83657725691022e-09
6.21621621621622 1.6219924166386e-09
6.23623623623624 1.43190554571785e-09
6.25625625625626 1.26358905957511e-09
6.27627627627628 1.11461087800764e-09
6.2962962962963 9.82803357938119e-10
6.31631631631632 8.66235385045992e-10
6.33633633633634 7.63187314959515e-10
6.35635635635636 6.72128483698836e-10
6.37637637637638 5.91697033481532e-10
6.3963963963964 5.20681824054164e-10
6.41641641641642 4.58006221596989e-10
6.43643643643644 4.02713577147895e-10
6.45645645645646 3.53954224575805e-10
6.47647647647648 3.10973844559378e-10
6.4964964964965 2.73103055937435e-10
6.51651651651652 2.39748109325539e-10
6.53653653653654 2.10382570159619e-10
6.55655655655656 1.84539889444162e-10
6.57657657657658 1.61806770551276e-10
6.5965965965966 1.41817249531742e-10
6.61661661661662 1.24247414645767e-10
6.63663663663664 1.08810698278133e-10
6.65665665665666 9.52536811418137e-11
6.67667667667668 8.33523547614393e-11
6.6966966966967 7.29087937236022e-11
6.71671671671672 6.37481941394484e-11
6.73673673673674 5.57162392365998e-11
6.75675675675676 4.86767570277076e-11
6.77677677677678 4.25096386334907e-11
6.7967967967968 3.71089891068556e-11
6.81681681681682 3.23814855461048e-11
6.83683683683684 2.82449199306813e-11
6.85685685685686 2.46269064908966e-11
6.87687687687688 2.14637355595386e-11
6.8968968968969 1.86993577716892e-11
6.91691691691692 1.62844842008236e-11
6.93693693693694 1.41757895636779e-11
6.95695695695696 1.23352070109961e-11
6.97697697697698 1.07293042619725e-11
6.996996996997 9.32873195138548e-12
7.01701701701702 8.10773605306642e-12
7.03703703703704 7.04372713322415e-12
7.05705705705706 6.11689998287643e-12
7.07707707707708 5.30989788981512e-12
7.0970970970971 4.6075164458095e-12
7.11711711711712 3.99644235193917e-12
7.13713713713714 3.46502319108337e-12
7.15715715715716 3.00306458801651e-12
7.17717717717718 2.60165157997869e-12
7.1971971971972 2.25299137914217e-12
7.21721721721722 1.95027502769791e-12
7.23723723723724 1.68755573049414e-12
7.25725725725726 1.45964190299846e-12
7.27727727727728 1.262003197176e-12
7.2972972972973 1.0906879676822e-12
7.31731731731732 9.42250818252902e-13
7.33733733733734 8.13689025751829e-13
7.35735735735736 7.02386779168733e-13
7.37737737737738 6.06066294885245e-13
7.3973973973974 5.22744979473708e-13
7.41741741741742 4.50697908714381e-13
7.43743743743744 3.88424977793729e-13
7.45745745745746 3.34622154016963e-13
7.47747747747748 2.88156330935933e-13
7.4974974974975 2.48043342543511e-13
7.51751751751752 2.13428748996348e-13
7.53753753753754 1.83571051982055e-13
7.55755755755756 1.57827039041883e-13
7.57757757757758 1.35638992516663e-13
7.5975975975976 1.16523530854698e-13
7.61761761761762 1.000618782966e-13
7.63763763763764 8.58913838706873e-14
7.65765765765766 7.36981325813112e-14
7.67767767767768 6.32105109959248e-14
7.6976976976977 5.41936064405376e-14
7.71771771771772 4.64443339685915e-14
7.73773773773774 3.97871984155707e-14
7.75775775775776 3.40706104038536e-14
7.77777777777778 2.91636853079907e-14
7.7977977977978 2.4953463096687e-14
7.81781781781782 2.13424947819474e-14
7.83783783783784 1.82467480586654e-14
7.85785785785786 1.55937907247682e-14
7.87787787787788 1.33212157347804e-14
7.8978978978979 1.13752763482777e-14
7.91791791791792 9.70970386854701e-15
7.93793793793794 8.28468399583068e-15
7.95795795795796 7.06597090549298e-15
7.97797797797798 6.02412085866189e-15
7.997997997998 5.13382950919603e-15
8.01801801801802 4.3733591283238e-15
8.03803803803804 3.72404376402735e-15
8.05805805805806 3.16986191875719e-15
8.07807807807808 2.69706769497134e-15
8.0980980980981 2.29387254841619e-15
8.11811811811812 1.95017082606148e-15
8.13813813813814 1.65730316851105e-15
8.15815815815816 1.40785264250169e-15
8.17817817817818 1.19546915264237e-15
8.1981981981982 1.01471827585831e-15
8.21821821821822 8.60951178494166e-16
8.23823823823824 7.30192724674871e-16
8.25825825825826 6.19045274051723e-16
8.27827827827828 5.24606005103494e-16
8.2982982982983 4.44395893386326e-16
8.31831831831832 3.76298728354137e-16
8.33833833833834 3.18508772685576e-16
8.35835835835836 2.69485858889318e-16
8.37837837837838 2.27916883183252e-16
8.3983983983984 1.92682799625235e-16
8.41841841841842 1.62830341150177e-16
8.43843843843844 1.37547801096493e-16
8.45845845845846 1.16144301209627e-16
8.47847847847848 9.8032051926925e-17
8.4984984984985 8.27111796592674e-17
8.51851851851852 6.97567552529606e-17
8.53853853853854 5.88077091106193e-17
8.55855855855856 4.95573626747691e-17
8.57857857857858 4.17453440895294e-17
8.5985985985986 3.51506886838449e-17
8.61861861861862 2.95859531836935e-17
8.63863863863864 2.48921968838275e-17
8.65865865865866 2.09347039316773e-17
8.67867867867868 1.75993388643364e-17
8.6986986986987 1.47894429982973e-17
8.71871871871872 1.24231925503958e-17
8.73873873873874 1.04313507694241e-17
8.75875875875876 8.75535614211525e-18
8.77877877877878 7.34569713009337e-18
8.7987987987988 6.16053109048305e-18
8.81881881881882 5.16451119999019e-18
8.83883883883884 4.32779048512422e-18
8.85885885885886 3.62517658454128e-18
8.87887887887888 3.03541474067764e-18
8.8988988988989 2.54057982941549e-18
8.91891891891892 2.12556106807901e-18
8.93893893893894 1.77762546207239e-18
8.95895895895896 1.48604811781133e-18
8.97897897897898 1.24179931485728e-18
8.998998998999 1.03727973679138e-18
9.01901901901902 8.66096545670873e-19
9.03903903903904 7.22874080905507e-19
9.05905905905906 6.03093897535385e-19
9.07907907907908 5.0295965472299e-19
9.0990990990991 4.19283042962206e-19
9.11911911911912 3.49387515332417e-19
9.13913913913914 2.91027078875695e-19
9.15915915915916 2.42317819504618e-19
9.17917917917918 2.01680188581445e-19
9.1991991991992 1.67790380698338e-19
9.21921921921922 1.39539388139188e-19
9.23923923923924 1.1599853476858e-19
9.25925925925926 9.63904764362469e-20
9.27927927927928 8.00648113242436e-20
9.2992992992993 6.6477576193283e-20
9.31931931931932 5.51740167796855e-20
9.33933933933934 4.5774115701642e-20
9.35935935935936 3.79604417474457e-20
9.37937937937938 3.14679525475421e-20
9.3993993993994 2.60754402558043e-20
9.41941941941942 2.15983585814365e-20
9.43943943943944 1.78828106797973e-20
9.45945945945946 1.48005121824234e-20
9.47947947947948 1.22445730038445e-20
9.4994994994995 1.01259663374544e-20
9.51951951951952 8.37057415073758e-21
9.53953953953954 6.91671611023779e-21
9.55955955955956 5.71308371631411e-21
9.57957957957958 4.71701393695898e-21
9.5995995995996 3.89304716291232e-21
9.61961961961962 3.21172317125736e-21
9.63963963963964 2.64857624243904e-21
9.65965965965966 2.18329684678274e-21
9.67967967967968 1.79903258756111e-21
9.6996996996997 1.48180551599987e-21
9.71971971971972 1.22002665237565e-21
9.73973973973974 1.00409166885045e-21
9.75975975975976 8.26044308669654e-22
9.77977977977978 6.79296312742742e-22
9.7997997997998 5.58394465795474e-22
9.81981981981982 4.58826916995414e-22
9.83983983983984 3.7686222201397e-22
9.85985985985986 3.09415635142992e-22
9.87987987987988 2.53938085193762e-22
9.8998998998999 2.08324025950642e-22
9.91991991991992 1.70834984871876e-22
9.93993993993994 1.40036162642795e-22
9.95995995995996 1.14743877987917e-22
9.97997997997998 9.39820210218911e-23
10 7.69459862670642e-23
};
\addlegendentry{N(0,1)}
\end{axis}

\end{tikzpicture}

%% file: FinalFigs/PowerCurve_RBFd_10_eps_0_3.tex
\begin{tikzpicture}

\definecolor{darkorange25512714}{RGB}{255,127,14}
\definecolor{darkslategray38}{RGB}{38,38,38}
\definecolor{lightgray204}{RGB}{204,204,204}
\definecolor{steelblue31119180}{RGB}{31,119,180}

\begin{axis}[
axis line style={darkslategray38},
height=\figheight,
legend cell align={left},
legend style={
  fill opacity=0.8,
  draw opacity=1,
  text opacity=1,
  at={(0.20,0.37)},
  anchor=north west,
  draw=none
},
tick align=outside,
tick pos=left,
title={Power vs Sample-Size},
width=\figwidth,
x grid style={lightgray204},
xlabel=\textcolor{darkslategray38}{Sample-Size (n+m)},
xmin=22, xmax=418,
xtick style={color=darkslategray38},
y grid style={lightgray204},
ylabel=\textcolor{darkslategray38}{Power},
ymin=0.146207485526875, ymax=1.04080947825317,
ytick style={color=darkslategray38}
]
\path [draw=steelblue31119180, fill=steelblue31119180, opacity=0.3]
(axis cs:40,0.261180736383563)
--(axis cs:40,0.218819263616437)
--(axis cs:58,0.335989094274684)
--(axis cs:76,0.468377409446745)
--(axis cs:96,0.544650702672461)
--(axis cs:114,0.669502805069961)
--(axis cs:134,0.722502000088897)
--(axis cs:152,0.765747704862475)
--(axis cs:172,0.868029142626255)
--(axis cs:190,0.888396149683508)
--(axis cs:210,0.903773252615787)
--(axis cs:228,0.945551091332858)
--(axis cs:248,0.967136407071192)
--(axis cs:266,0.9784)
--(axis cs:286,0.975588460735842)
--(axis cs:304,0.985260290752588)
--(axis cs:324,0.981624255366339)
--(axis cs:342,0.985070813961717)
--(axis cs:362,1)
--(axis cs:380,0.994854248688935)
--(axis cs:400,1)
--(axis cs:400,1)
--(axis cs:400,1)
--(axis cs:380,1.00014575131106)
--(axis cs:362,1)
--(axis cs:342,0.994929186038283)
--(axis cs:324,0.993375744633661)
--(axis cs:304,0.994739709247412)
--(axis cs:286,0.989411539264158)
--(axis cs:266,0.9916)
--(axis cs:248,0.982863592928808)
--(axis cs:228,0.964448908667142)
--(axis cs:210,0.931226747384213)
--(axis cs:190,0.916603850316492)
--(axis cs:172,0.901970857373745)
--(axis cs:152,0.809252295137525)
--(axis cs:134,0.767497999911103)
--(axis cs:114,0.710497194930038)
--(axis cs:96,0.595349297327539)
--(axis cs:76,0.516622590553255)
--(axis cs:58,0.384010905725316)
--(axis cs:40,0.261180736383563)
--cycle;

\path [draw=darkorange25512714, fill=darkorange25512714, opacity=0.3]
(axis cs:40,0.228128787531021)
--(axis cs:40,0.186871212468979)
--(axis cs:58,0.263837744322445)
--(axis cs:76,0.299858411339966)
--(axis cs:96,0.36777777013889)
--(axis cs:114,0.451560433817236)
--(axis cs:134,0.541842703279574)
--(axis cs:152,0.587663514292754)
--(axis cs:172,0.660802145843425)
--(axis cs:190,0.729262853015054)
--(axis cs:210,0.783122181656595)
--(axis cs:228,0.828191748040022)
--(axis cs:248,0.860770703093076)
--(axis cs:266,0.872860845524758)
--(axis cs:286,0.861094251198132)
--(axis cs:304,0.912422142471788)
--(axis cs:324,0.95629519708724)
--(axis cs:342,0.914487722144067)
--(axis cs:362,0.966536558398027)
--(axis cs:380,0.978507323837584)
--(axis cs:400,0.964578756805274)
--(axis cs:400,0.980421243194726)
--(axis cs:400,0.980421243194726)
--(axis cs:380,0.991492676162416)
--(axis cs:362,0.983463441601973)
--(axis cs:342,0.940512277855933)
--(axis cs:324,0.97370480291276)
--(axis cs:304,0.937577857528212)
--(axis cs:286,0.893905748801868)
--(axis cs:266,0.902139154475242)
--(axis cs:248,0.894229296906923)
--(axis cs:228,0.866808251959978)
--(axis cs:210,0.821877818343405)
--(axis cs:190,0.775737146984946)
--(axis cs:172,0.704197854156575)
--(axis cs:152,0.632336485707246)
--(axis cs:134,0.593157296720426)
--(axis cs:114,0.503439566182764)
--(axis cs:96,0.41222222986111)
--(axis cs:76,0.340141588660034)
--(axis cs:58,0.311162255677555)
--(axis cs:40,0.228128787531021)
--cycle;

\addplot [semithick, steelblue31119180]
table {%
40 0.24
58 0.36
76 0.4925
96 0.57
114 0.69
134 0.745
152 0.7875
172 0.885
190 0.9025
210 0.9175
228 0.955
248 0.975
266 0.985
286 0.9825
304 0.99
324 0.9875
342 0.99
362 1
380 0.9975
400 1
};
\addlegendentry{$\dmmd$-perm}
\addplot [semithick, darkorange25512714]
table {%
40 0.2075
58 0.2875
76 0.32
96 0.39
114 0.4775
134 0.5675
152 0.61
172 0.6825
190 0.7525
210 0.8025
228 0.8475
248 0.8775
266 0.8875
286 0.8775
304 0.925
324 0.965
342 0.9275
362 0.975
380 0.985
400 0.9725
};
\addlegendentry{$\cmmd$}
\end{axis}

\end{tikzpicture}

%% file: FinalFigs/PowerVsTime_RBFd_10_eps_0_2.tex
\begin{tikzpicture}

\definecolor{darkorange25512714}{RGB}{255,127,14}
\definecolor{darkslategray38}{RGB}{38,38,38}
\definecolor{lightgray204}{RGB}{204,204,204}
\definecolor{steelblue31119180}{RGB}{31,119,180}

\begin{axis}[
axis line style={darkslategray38},
height=\figheight,
legend cell align={left},
legend style={
  fill opacity=0.8,
  draw opacity=1,
  text opacity=1,
  at={(0.13,0.47)},
  anchor=north west,
  draw=none
},
log basis x={10},
tick align=outside,
tick pos=left,
title={Power vs Computation},
width=\figwidth,
x grid style={lightgray204},
xlabel=\textcolor{darkslategray38}{Running time (seconds)},
xmin=0.000257061877313564, xmax=0.183328546056211,
xmode=log,
xtick style={color=darkslategray38},
y grid style={lightgray204},
ylabel=\textcolor{darkslategray38}{Power},
ymin=0.0665, ymax=1.0235,
ytick style={color=darkslategray38}
]
\addplot [
  draw=steelblue31119180,
  fill=steelblue31119180,
  mark=*,
  only marks,
  opacity=0.7, 
  scatter, 
  visualization depends on={\thisrow{sizedata} \as\perpointmarksize},
  scatter/@pre marker code/.style={/tikz/mark size=\perpointmarksize},
  scatter/@post marker code/.style={}
]
table{%
x  y  sizedata
0.00511279344558716 0.12 2.5731003930322744
0.00987482786178589 0.16 2.786890959991885
0.0103487944602966 0.27 3.3042441035892036
0.0111070585250854 0.35 3.634537147509613
0.0142297315597534 0.34 3.5949104827605796
0.0176255679130554 0.39 3.7889016702586726
0.0221033596992493 0.49 4.149766841949446
0.0268570113182068 0.6 4.51351666838205
0.0321097445487976 0.65 4.669499673796933
0.0389047718048096 0.67 4.730452938009265
0.0455186533927917 0.7 4.820437914901168
0.0525294017791748 0.79 5.08083967169851
0.0612401103973389 0.88 5.3285309277130155
0.0691829800605774 0.8 5.108953969950291
0.0786550760269165 0.83 5.192383591354463
0.0890532875061035 0.86 5.274493724753429
0.0991631650924683 0.92 5.434993784527796
0.110637104511261 0.91 5.408574538664542
0.122460055351257 0.91 5.408574538664542
0.136000194549561 0.98 5.590888196275101
};
\addlegendentry{$\dmmd$-perm}
\addplot [
  draw=darkorange25512714,
  fill=darkorange25512714,
  mark=triangle*,
  only marks,
  opacity = 0.7, 
  scatter, 
  visualization depends on={\thisrow{sizedata} \as\perpointmarksize},
  scatter/@pre marker code/.style={/tikz/mark size=\perpointmarksize},
  scatter/@post marker code/.style={}
]
table{%
x  y  sizedata
0.000394997596740723 0.13 2.6281789380078857
0.000451819896697998 0.11 2.516816786152189
0.000350978374481201 0.17 2.837823130579125
0.000346519947052002 0.25 3.2163754912903433
0.000372955799102783 0.27 3.3042441035892036
0.000386786460876465 0.25 3.2163754912903433
0.000418229103088379 0.35 3.634537147509613
0.000450630187988281 0.38 3.7509061530946823
0.000480389595031738 0.48 4.115104609238709
0.00052215576171875 0.5 4.1841419359420025
0.000569009780883789 0.48 4.115104609238709
0.000617935657501221 0.6 4.51351666838205
0.000664288997650146 0.66 4.700075116543891
0.000797414779663086 0.59 4.481668642168583
0.000893704891204834 0.66 4.700075116543891
0.000978443622589111 0.62 4.576547881416016
0.00104090213775635 0.76 4.995547525227759
0.00113027811050415 0.7 4.820437914901168
0.00121775150299072 0.75 4.9667913363905045
0.00135158538818359 0.84 5.219897150072258
};
\addlegendentry{$\cmmd$}
\end{axis}

\end{tikzpicture}

%% file: FinalFigs/Null_Dists_d_10_500_n_20_m_200_kernel__Gaussian_RBF_2022_10_12_22_14_00cross.tex
\begin{tikzpicture}

\definecolor{darkorange25512714}{RGB}{255,127,14}
\definecolor{darkslategray38}{RGB}{38,38,38}
\definecolor{lightgray204}{RGB}{204,204,204}
\definecolor{steelblue31119180}{RGB}{31,119,180}

\begin{axis}[
axis line style={darkslategray38},
height=\figheight,
legend cell align={left},
legend style={fill opacity=0.8, draw opacity=1, text opacity=1, draw=none, 
nodes={scale=0.7, transform shape}, 
at={(axis cs:7,0.4)},anchor=north east
},
tick align=outside,
tick pos=left,
title={$\cmmd~(n/m=0.1)$},
width=\figwidth,
x grid style={lightgray204},
xmin=-6, xmax=6,
xtick style={},
y grid style={lightgray204},
ylabel=\textcolor{darkslategray38}{Probability density},
ymin=0, ymax=0.418868408525629,
ytick style={color=darkslategray38}, 
xticklabels=empty,
yticklabels=empty
]
\draw[draw=none,fill=steelblue31119180,fill opacity=0.8] (axis cs:-3.25350737571716,0) rectangle (axis cs:-2.93425297737122,0.0175408703185094);
\addlegendimage{ybar,ybar legend,draw=none,fill=steelblue31119180,fill opacity=0.8}
\addlegendentry{$d=10$}

\draw[draw=none,fill=steelblue31119180,fill opacity=0.8] (axis cs:-2.45537137985229,0) rectangle (axis cs:-2.13611698150635,0.0476109479439746);
\draw[draw=none,fill=steelblue31119180,fill opacity=0.8] (axis cs:-1.65723550319672,0) rectangle (axis cs:-1.33798110485077,0.130303627542456);
\draw[draw=none,fill=steelblue31119180,fill opacity=0.8] (axis cs:-0.859099686145782,0) rectangle (axis cs:-0.539845287799835,0.340794102803346);
\draw[draw=none,fill=steelblue31119180,fill opacity=0.8] (axis cs:-0.0609637945890427,0) rectangle (axis cs:0.258290588855743,0.388405043636166);
\draw[draw=none,fill=steelblue31119180,fill opacity=0.8] (axis cs:0.737172067165375,0) rectangle (axis cs:1.05642652511597,0.202972958287287);
\draw[draw=none,fill=steelblue31119180,fill opacity=0.8] (axis cs:1.53530788421631,0) rectangle (axis cs:1.85456228256226,0.0877043777915321);
\draw[draw=none,fill=steelblue31119180,fill opacity=0.8] (axis cs:2.33344388008118,0) rectangle (axis cs:2.65269827842712,0.027564224786229);
\draw[draw=none,fill=steelblue31119180,fill opacity=0.8] (axis cs:3.13157963752747,0) rectangle (axis cs:3.45083403587341,0.00751751585078974);
\draw[draw=none,fill=steelblue31119180,fill opacity=0.8] (axis cs:3.92971587181091,0) rectangle (axis cs:4.24897003173828,0.00250583936547235);
\draw[draw=none,fill=darkorange25512714,fill opacity=0.8] (axis cs:-2.93425297737122,0) rectangle (axis cs:-2.61499857902527,0.0175408703185094);
\addlegendimage{ybar,ybar legend,draw=none,fill=darkorange25512714,fill opacity=0.8}
\addlegendentry{$d=500$}

\draw[draw=none,fill=darkorange25512714,fill opacity=0.8] (axis cs:-2.13611698150635,0) rectangle (axis cs:-1.8168625831604,0.0701635022332257);
\draw[draw=none,fill=darkorange25512714,fill opacity=0.8] (axis cs:-1.33798122406006,0) rectangle (axis cs:-1.01872682571411,0.192949602322483);
\draw[draw=none,fill=darkorange25512714,fill opacity=0.8] (axis cs:-0.539845287799835,0) rectangle (axis cs:-0.220590889453888,0.315735712891335);
\draw[draw=none,fill=darkorange25512714,fill opacity=0.8] (axis cs:0.258290588855743,0) rectangle (axis cs:0.577544987201691,0.360840814732955);
\draw[draw=none,fill=darkorange25512714,fill opacity=0.8] (axis cs:1.05642652511597,0) rectangle (axis cs:1.37568092346191,0.190443763331282);
\draw[draw=none,fill=darkorange25512714,fill opacity=0.8] (axis cs:1.85456228256226,0) rectangle (axis cs:2.1738166809082,0.0676576628677534);
\draw[draw=none,fill=darkorange25512714,fill opacity=0.8] (axis cs:2.65269827842712,0) rectangle (axis cs:2.97195267677307,0.0250583861692991);
\draw[draw=none,fill=darkorange25512714,fill opacity=0.8] (axis cs:3.45083403587341,0) rectangle (axis cs:3.77008843421936,0.0100233544677197);
\draw[draw=none,fill=darkorange25512714,fill opacity=0.8] (axis cs:4.24897003173828,0) rectangle (axis cs:4.56822443008423,0.00250583936547235);
\addplot [semithick, black]
table {%
-10 7.69459862670642e-23
-9.97997997997998 9.39820210218911e-23
-9.95995995995996 1.14743877987917e-22
-9.93993993993994 1.40036162642795e-22
-9.91991991991992 1.70834984871876e-22
-9.8998998998999 2.08324025950642e-22
-9.87987987987988 2.53938085193762e-22
-9.85985985985986 3.09415635142992e-22
-9.83983983983984 3.7686222201397e-22
-9.81981981981982 4.58826916995414e-22
-9.7997997997998 5.58394465795474e-22
-9.77977977977978 6.79296312742742e-22
-9.75975975975976 8.26044308669654e-22
-9.73973973973974 1.00409166885045e-21
-9.71971971971972 1.22002665237565e-21
-9.6996996996997 1.48180551599987e-21
-9.67967967967968 1.79903258756111e-21
-9.65965965965966 2.18329684678274e-21
-9.63963963963964 2.64857624243904e-21
-9.61961961961962 3.21172317125736e-21
-9.5995995995996 3.89304716291232e-21
-9.57957957957958 4.71701393695898e-21
-9.55955955955956 5.71308371631411e-21
-9.53953953953954 6.91671611023779e-21
-9.51951951951952 8.37057415073758e-21
-9.4994994994995 1.01259663374544e-20
-9.47947947947948 1.22445730038445e-20
-9.45945945945946 1.48005121824234e-20
-9.43943943943944 1.78828106797973e-20
-9.41941941941942 2.15983585814365e-20
-9.3993993993994 2.60754402558043e-20
-9.37937937937938 3.14679525475421e-20
-9.35935935935936 3.79604417474457e-20
-9.33933933933934 4.5774115701642e-20
-9.31931931931932 5.51740167796855e-20
-9.2992992992993 6.6477576193283e-20
-9.27927927927928 8.00648113242436e-20
-9.25925925925926 9.63904764362469e-20
-9.23923923923924 1.1599853476858e-19
-9.21921921921922 1.39539388139188e-19
-9.1991991991992 1.67790380698338e-19
-9.17917917917918 2.01680188581445e-19
-9.15915915915916 2.42317819504618e-19
-9.13913913913914 2.91027078875695e-19
-9.11911911911912 3.49387515332417e-19
-9.0990990990991 4.19283042962206e-19
-9.07907907907908 5.0295965472299e-19
-9.05905905905906 6.03093897535385e-19
-9.03903903903904 7.22874080905507e-19
-9.01901901901902 8.66096545670873e-19
-8.998998998999 1.03727973679138e-18
-8.97897897897898 1.24179931485728e-18
-8.95895895895896 1.48604811781133e-18
-8.93893893893894 1.77762546207239e-18
-8.91891891891892 2.12556106807901e-18
-8.8988988988989 2.54057982941549e-18
-8.87887887887888 3.03541474067764e-18
-8.85885885885886 3.62517658454128e-18
-8.83883883883884 4.32779048512422e-18
-8.81881881881882 5.16451119999019e-18
-8.7987987987988 6.16053109048305e-18
-8.77877877877878 7.34569713009337e-18
-8.75875875875876 8.75535614211525e-18
-8.73873873873874 1.04313507694241e-17
-8.71871871871872 1.24231925503958e-17
-8.6986986986987 1.47894429982973e-17
-8.67867867867868 1.75993388643364e-17
-8.65865865865866 2.09347039316773e-17
-8.63863863863864 2.48921968838275e-17
-8.61861861861862 2.95859531836935e-17
-8.5985985985986 3.51506886838449e-17
-8.57857857857858 4.17453440895294e-17
-8.55855855855856 4.95573626747691e-17
-8.53853853853854 5.88077091106193e-17
-8.51851851851852 6.97567552529606e-17
-8.4984984984985 8.27111796592674e-17
-8.47847847847848 9.8032051926925e-17
-8.45845845845846 1.16144301209627e-16
-8.43843843843844 1.37547801096493e-16
-8.41841841841842 1.62830341150177e-16
-8.3983983983984 1.92682799625235e-16
-8.37837837837838 2.27916883183252e-16
-8.35835835835836 2.69485858889318e-16
-8.33833833833834 3.18508772685576e-16
-8.31831831831832 3.76298728354137e-16
-8.2982982982983 4.44395893386326e-16
-8.27827827827828 5.24606005103494e-16
-8.25825825825826 6.19045274051723e-16
-8.23823823823824 7.30192724674871e-16
-8.21821821821822 8.60951178494166e-16
-8.1981981981982 1.01471827585831e-15
-8.17817817817818 1.19546915264237e-15
-8.15815815815816 1.40785264250169e-15
-8.13813813813814 1.65730316851105e-15
-8.11811811811812 1.95017082606148e-15
-8.0980980980981 2.29387254841619e-15
-8.07807807807808 2.69706769497134e-15
-8.05805805805806 3.16986191875719e-15
-8.03803803803804 3.72404376402735e-15
-8.01801801801802 4.3733591283238e-15
-7.997997997998 5.13382950919607e-15
-7.97797797797798 6.02412085866193e-15
-7.95795795795796 7.06597090549303e-15
-7.93793793793794 8.28468399583074e-15
-7.91791791791792 9.70970386854708e-15
-7.8978978978979 1.13752763482777e-14
-7.87787787787788 1.33212157347805e-14
-7.85785785785786 1.55937907247683e-14
-7.83783783783784 1.82467480586655e-14
-7.81781781781782 2.13424947819475e-14
-7.7977977977978 2.49534630966872e-14
-7.77777777777778 2.91636853079909e-14
-7.75775775775776 3.40706104038538e-14
-7.73773773773774 3.9787198415571e-14
-7.71771771771772 4.64443339685918e-14
-7.6976976976977 5.4193606440538e-14
-7.67767767767768 6.32105109959252e-14
-7.65765765765766 7.36981325813117e-14
-7.63763763763764 8.58913838706879e-14
-7.61761761761762 1.00061878296601e-13
-7.5975975975976 1.16523530854699e-13
-7.57757757757758 1.35638992516664e-13
-7.55755755755756 1.57827039041884e-13
-7.53753753753754 1.83571051982057e-13
-7.51751751751752 2.13428748996349e-13
-7.4974974974975 2.48043342543513e-13
-7.47747747747748 2.88156330935935e-13
-7.45745745745746 3.34622154016965e-13
-7.43743743743744 3.88424977793732e-13
-7.41741741741742 4.50697908714384e-13
-7.3973973973974 5.22744979473711e-13
-7.37737737737738 6.06066294885249e-13
-7.35735735735736 7.02386779168738e-13
-7.33733733733734 8.13689025751835e-13
-7.31731731731732 9.42250818252909e-13
-7.2972972972973 1.09068796768221e-12
-7.27727727727728 1.262003197176e-12
-7.25725725725726 1.45964190299847e-12
-7.23723723723724 1.68755573049416e-12
-7.21721721721722 1.95027502769792e-12
-7.1971971971972 2.25299137914218e-12
-7.17717717717718 2.60165157997871e-12
-7.15715715715716 3.00306458801653e-12
-7.13713713713714 3.4650231910834e-12
-7.11711711711712 3.99644235193919e-12
-7.0970970970971 4.60751644580953e-12
-7.07707707707708 5.30989788981514e-12
-7.05705705705706 6.11689998287646e-12
-7.03703703703704 7.0437271332242e-12
-7.01701701701702 8.10773605306645e-12
-6.996996996997 9.32873195138555e-12
-6.97697697697698 1.07293042619726e-11
-6.95695695695696 1.23352070109962e-11
-6.93693693693694 1.41757895636779e-11
-6.91691691691692 1.62844842008237e-11
-6.8968968968969 1.86993577716893e-11
-6.87687687687688 2.14637355595386e-11
-6.85685685685686 2.46269064908967e-11
-6.83683683683684 2.82449199306815e-11
-6.81681681681682 3.2381485546105e-11
-6.7967967967968 3.71089891068559e-11
-6.77677677677678 4.25096386334913e-11
-6.75675675675676 4.86767570277083e-11
-6.73673673673674 5.57162392366004e-11
-6.71671671671672 6.37481941394491e-11
-6.6966966966967 7.29087937236032e-11
-6.67667667667668 8.33523547614402e-11
-6.65665665665666 9.52536811418151e-11
-6.63663663663664 1.08810698278135e-10
-6.61661661661662 1.24247414645768e-10
-6.5965965965966 1.41817249531744e-10
-6.57657657657658 1.61806770551278e-10
-6.55655655655656 1.84539889444164e-10
-6.53653653653654 2.10382570159622e-10
-6.51651651651652 2.39748109325542e-10
-6.4964964964965 2.73103055937438e-10
-6.47647647647648 3.10973844559381e-10
-6.45645645645646 3.53954224575809e-10
-6.43643643643644 4.027135771479e-10
-6.41641641641642 4.58006221596996e-10
-6.3963963963964 5.20681824054169e-10
-6.37637637637638 5.91697033481538e-10
-6.35635635635636 6.72128483698846e-10
-6.33633633633634 7.63187314959523e-10
-6.31631631631632 8.66235385046001e-10
-6.2962962962963 9.8280335793813e-10
-6.27627627627628 1.11461087800766e-09
-6.25625625625626 1.26358905957513e-09
-6.23623623623624 1.43190554571787e-09
-6.21621621621622 1.62199241663862e-09
-6.1961961961962 1.83657725691024e-09
-6.17617617617618 2.0787177227378e-09
-6.15615615615616 2.35183998527873e-09
-6.13613613613614 2.65978146430928e-09
-6.11611611611612 3.00683830841829e-09
-6.0960960960961 3.39781812376754e-09
-6.07607607607608 3.83809850362696e-09
-6.05605605605606 4.33369196574699e-09
-6.03603603603604 4.89131796456919e-09
-6.01601601601602 5.51848271073395e-09
-5.995995995996 6.22356760178439e-09
-5.97597597597598 7.01592714588943e-09
-5.95595595595596 7.90599734535251e-09
-5.93593593593594 8.90541559921139e-09
-5.91591591591592 1.00271532849868e-08
-5.8958958958959 1.12856622892642e-08
-5.87587587587588 1.2697036876002e-08
-5.85585585585586 1.42791924110059e-08
-5.83583583583584 1.60520626017053e-08
-5.81581581581582 1.80378170640688e-08
-5.7957957957958 2.02611011941336e-08
-5.77577577577578 2.27493005011625e-08
-5.75575575575576 2.55328317539416e-08
-5.73573573573574 2.86454635022852e-08
-5.71571571571572 3.2124668763613e-08
-5.6956956956957 3.60120129107462e-08
-5.67567567567568 4.03535800631662e-08
-5.65565565565566 4.5200441571292e-08
-5.63563563563564 5.06091704933412e-08
-5.61561561561562 5.66424062986154e-08
-5.5955955955956 6.33694743912418e-08
-5.57557557557558 7.08670654362614e-08
-5.55555555555556 7.92199798873018e-08
-5.53553553553554 8.85219435638491e-08
-5.51551551551552 9.8876500608364e-08
-5.4954954954955 1.10397990671284e-07
-5.47547547547548 1.23212617727566e-07
-5.45545545545546 1.3745961852414e-07
-5.43543543543544 1.5329253929596e-07
-5.41541541541542 1.70880630071684e-07
-5.3953953953954 1.90410366621162e-07
-5.37537537537538 2.1208711087848e-07
-5.35535535535536 2.36136921509202e-07
-5.33533533533534 2.62808527181656e-07
-5.31531531531532 2.92375476052561e-07
-5.2952952952953 3.25138475990267e-07
-5.27527527527528 3.61427941137511e-07
-5.25525525525526 4.01606761563285e-07
-5.23523523523524 4.46073313973501e-07
-5.21521521521522 4.95264732746229e-07
-5.1951951951952 5.4966046193278e-07
-5.17517517517518 6.09786110324583e-07
-5.15515515515516 6.76217633231267e-07
-5.13513513513514 7.49585866251233e-07
-5.11511511511512 8.30581438046261e-07
-5.0950950950951 9.19960090959837e-07
-5.07507507507508 1.01854844024876e-06
-5.05505505505506 1.12725020473309e-06
-5.03503503503504 1.24705294381396e-06
-5.01501501501502 1.37903533806611e-06
-4.99499499499499 1.5243750529858e-06
-4.97497497497497 1.68435722796805e-06
-4.95495495495495 1.86038363520377e-06
-4.93493493493493 2.05398255592983e-06
-4.91491491491491 2.26681942433715e-06
-4.89489489489489 2.50070829244518e-06
-4.87487487487487 2.75762417238901e-06
-4.85485485485485 3.03971631583941e-06
-4.83483483483483 3.34932249368796e-06
-4.81481481481481 3.68898434268124e-06
-4.79479479479479 4.06146384937932e-06
-4.77477477477477 4.4697610456467e-06
-4.75475475475475 4.91713299385613e-06
-4.73473473473473 5.40711414409909e-06
-4.71471471471471 5.94353814994721e-06
-4.69469469469469 6.53056123369604e-06
-4.67467467467467 7.17268719654363e-06
-4.65465465465465 7.87479417380527e-06
-4.63463463463463 8.64216324004121e-06
-4.61461461461461 9.48050897386715e-06
-4.59459459459459 1.03960120972233e-05
-4.57457457457457 1.13953543089884e-05
-4.55455455455455 1.24857554380297e-05
-4.53453453453453 1.3675013046071e-05
-4.51451451451451 1.49715446161227e-05
-4.49449449449449 1.63844324676437e-05
-4.47447447447447 1.79234715450684e-05
-4.45445445445445 1.95992202318354e-05
-4.43443443443443 2.14230543475548e-05
-4.41441441441441 2.34072244914564e-05
-4.39439439439439 2.55649169007242e-05
-4.37437437437437 2.79103179977393e-05
-4.35435435435435 3.04586828055866e-05
-4.33433433433433 3.32264074164067e-05
-4.31431431431431 3.62311057022691e-05
-4.29429429429429 3.94916904631592e-05
-4.27427427427427 4.3028459211397e-05
-4.25425425425425 4.68631847962824e-05
-4.23423423423423 5.10192110769697e-05
-4.21421421421421 5.55215538554582e-05
-4.19419419419419 6.03970072851107e-05
-4.17417417417417 6.56742559732345e-05
-4.15415415415415 7.13839929989176e-05
-4.13413413413413 7.75590440694795e-05
-4.11411411411411 8.42344980404937e-05
-4.09409409409409 9.14478440253317e-05
-4.07407407407407 9.92391153205018e-05
-4.05405405405405 0.000107651040372646
-4.03403403403403 0.000116729201011866
-4.01401401401401 0.000126522198173995
-3.99399399399399 0.000137081825331481
-3.97397397397397 0.000148463249848567
-3.95395395395395 0.000160725202471485
-3.93393393393393 0.000173930175158222
-3.91391391391391 0.00018814462744512
-3.89389389389389 0.000203439201538965
-3.87387387387387 0.000219888946313312
-3.85385385385385 0.000237573550376443
-3.83383383383383 0.000256577584365551
-3.81381381381381 0.000276990752607344
-3.79379379379379 0.000298908154269281
-3.77377377377377 0.000322430554107926
-3.75375375375375 0.000347664662901427
-3.73373373373373 0.000374723427631836
-3.71371371371371 0.000403726331459719
-3.69369369369369 0.000434799703508357
-3.67367367367367 0.000468077038447599
-3.65365365365365 0.00050369932583812
-3.63363363363363 0.000541815389165405
-3.61361361361361 0.000582582234459159
-3.59359359359359 0.000626165408357979
-3.57357357357357 0.000672739365441021
-3.55355355355355 0.000722487844607978
-3.53353353353353 0.00077560425424601
-3.51351351351351 0.000832292065877155
-3.49349349349349 0.000892765215932443
-3.47347347347347 0.000957248515249216
-3.45345345345345 0.0010259780658361
-3.43343343343343 0.00109920168439588
-3.41341341341341 0.00117717933203981
-3.39339339339339 0.00126018354956833
-3.37337337337337 0.00134849989763212
-3.35335335335335 0.00144242740102448
-3.33333333333333 0.00154227899629111
-3.31331331331331 0.00164838198177652
-3.29329329329329 0.00176107846915772
-3.27327327327327 0.00188072583544552
-3.25325325325325 0.00200769717436226
-3.23323323323323 0.00214238174593163
-3.21321321321321 0.00228518542304204
-3.19319319319319 0.00243653113367012
-3.17317317317317 0.00259685929737497
-3.15315315315315 0.00276662825459747
-3.13313313313313 0.00294631468722261
-3.11311311311311 0.00313641402878609
-3.09309309309309 0.00333744086263052
-3.07307307307307 0.00354992930624086
-3.05305305305305 0.00377443337991422
-3.03303303303303 0.00401152735784579
-3.01301301301301 0.00426180609964128
-2.99299299299299 0.00452588536019618
-2.97297297297297 0.0048044020758154
-2.95295295295295 0.00509801462438215
-2.93293293293293 0.00540740305732385
-2.91291291291291 0.00573326930106519
-2.89289289289289 0.00607633732560526
-2.87287287287287 0.00643735327780636
-2.85285285285285 0.00681708557693873
-2.83283283283283 0.00721632496998623
-2.81281281281281 0.00763588454418632
-2.79279279279279 0.00807659969425075
-2.77277277277277 0.00853932804169477
-2.75275275275275 0.00902494930369032
-2.73273273273273 0.00953436510885489
-2.71271271271271 0.0100684987573917
-2.69269269269269 0.0106282949230102
-2.67267267267267 0.0112147192940778
-2.65265265265265 0.0118287581514852
-2.63263263263263 0.0124714178807513
-2.61261261261261 0.0131437244159435
-2.59259259259259 0.0138467226130541
-2.57257257257257 0.0145814755505492
-2.55255255255255 0.0153490637548887
-2.53253253253253 0.0161505843489182
-2.51251251251251 0.0169871501211409
-2.49249249249249 0.0178598885140022
-2.47247247247247 0.018769940529451
-2.45245245245245 0.0197184595501959
-2.43243243243243 0.0207066100752274
-2.41241241241241 0.0217355663683581
-2.39239239239239 0.0228065110187135
-2.37237237237237 0.0239206334123108
-2.35235235235235 0.0250791281140699
-2.33233233233233 0.0262831931598317
-2.31231231231231 0.0275340282581906
-2.29229229229229 0.0288328329022027
-2.27227227227227 0.0301808043912899
-2.25225225225225 0.0315791357639331
-2.23223223223223 0.033029013642035
-2.21221221221221 0.0345316159881232
-2.19219219219219 0.0360881097768736
-2.17217217217217 0.0376996485827434
-2.15215215215215 0.0393673700858293
-2.13213213213213 0.0410923934983949
-2.11211211211211 0.0428758169148479
-2.09209209209209 0.0447187145882915
-2.07207207207207 0.0466221341371235
-2.05205205205205 0.0485870936855041
-2.03203203203203 0.0506145789418747
-2.01201201201201 0.0527055402200587
-1.99199199199199 0.0548608894078376
-1.97197197197197 0.057081496888248
-1.95195195195195 0.059368188419199
-1.93193193193193 0.0617217419773594
-1.91191191191191 0.0641428845726061
-1.89189189189189 0.0666322890396674
-1.87187187187187 0.0691905708139176
-1.85185185185185 0.0718182846986055
-1.83183183183183 0.0745159216311036
-1.81181181181181 0.077283905456062
-1.79179179179179 0.0801225897136326
-1.77177177177177 0.083032254451193
-1.75175175175175 0.0860131030672496
-1.73173173173173 0.0890652591964251
-1.71171171171171 0.0921887636446459
-1.69169169169169 0.0953835713838294
-1.67167167167167 0.0986495486155338
-1.65165165165165 0.101986469913169
-1.63163163163163 0.10539401545248
-1.61161161161161 0.108871768340093
-1.59159159159159 0.112419212049971
-1.57157157157157 0.116035727977651
-1.55155155155155 0.119720593122119
-1.53153153153153 0.123472977905145
-1.51151151151151 0.127291944137829
-1.49149149149149 0.13117644314399
-1.47147147147147 0.135125314049902
-1.45145145145145 0.139137282249685
-1.43143143143143 0.143210958055468
-1.41141141141141 0.147344835541168
-1.39139139139139 0.151537291588457
-1.37137137137137 0.155786585143159
-1.35135135135135 0.160090856689972
-1.33133133133133 0.164448127952996
-1.31131131131131 0.168856301829129
-1.29129129129129 0.173313162560933
-1.27127127127127 0.17781637615506
-1.25125125125125 0.182363491051798
-1.23123123123123 0.186951939050736
-1.21121121121121 0.191579036496956
-1.19119119119119 0.1962419857315
-1.17117117117117 0.200937876809264
-1.15115115115115 0.205663689486728
-1.13113113113113 0.210416295481265
-1.11111111111111 0.215192461003031
-1.09109109109109 0.219988849559688
-1.07107107107107 0.224802025033432
-1.05105105105105 0.229628455029052
-1.03103103103103 0.234464514490888
-1.01101101101101 0.239306489585817
-0.990990990990991 0.24415058184851
-0.970970970970971 0.24899291258444
-0.950950950950951 0.253829527525259
-0.930930930930931 0.258656401730343
-0.910910910910911 0.2634694447275
-0.890890890890891 0.268264505884996
-0.870870870870871 0.273037380006279
-0.850850850850851 0.277783813137949
-0.830830830830831 0.282499508580786
-0.810810810810811 0.287180133092853
-0.790790790790791 0.291821323272996
-0.77077077077077 0.296418692112302
-0.75075075075075 0.300967835700437
-0.73073073073073 0.305464340073112
-0.71071071071071 0.309903788186304
-0.69069069069069 0.314281767002296
-0.67067067067067 0.318593874672039
-0.65065065065065 0.322835727797843
-0.63063063063063 0.327002968759958
-0.61061061061061 0.331091273090187
-0.59059059059059 0.33509635687531
-0.57057057057057 0.339013984172804
-0.55055055055055 0.34283997442106
-0.53053053053053 0.346570209826128
-0.51051051051051 0.350200642706842
-0.49049049049049 0.353727302780113
-0.47047047047047 0.357146304368113
-0.45045045045045 0.360453853509139
-0.43043043043043 0.363646254953996
-0.41041041041041 0.366719919029892
-0.39039039039039 0.369671368354051
-0.37037037037037 0.372497244379499
-0.35035035035035 0.375194313755802
-0.33033033033033 0.377759474487924
-0.31031031031031 0.38018976187679
-0.29029029029029 0.382482354225654
-0.27027027027027 0.384634578296894
-0.25025025025025 0.386643914504485
-0.23023023023023 0.388508001828027
-0.21021021021021 0.390224642434919
-0.19019019019019 0.391791805998011
-0.17017017017017 0.393207633696876
-0.15015015015015 0.394470441891644
-0.13013013013013 0.395578725459258
-0.11011011011011 0.396531160782876
-0.0900900900900901 0.397326608386124
-0.07007007007007 0.397964115204853
-0.05005005005005 0.398442916490068
-0.03003003003003 0.398762437336696
-0.01001001001001 0.398922293833933
0.01001001001001 0.398922293833933
0.03003003003003 0.398762437336696
0.05005005005005 0.398442916490068
0.07007007007007 0.397964115204853
0.0900900900900901 0.397326608386124
0.11011011011011 0.396531160782876
0.13013013013013 0.395578725459258
0.15015015015015 0.394470441891644
0.17017017017017 0.393207633696876
0.19019019019019 0.391791805998011
0.21021021021021 0.390224642434919
0.23023023023023 0.388508001828027
0.25025025025025 0.386643914504485
0.27027027027027 0.384634578296894
0.29029029029029 0.382482354225654
0.31031031031031 0.38018976187679
0.33033033033033 0.377759474487924
0.35035035035035 0.375194313755802
0.37037037037037 0.372497244379499
0.39039039039039 0.369671368354051
0.41041041041041 0.366719919029892
0.43043043043043 0.363646254953996
0.45045045045045 0.360453853509139
0.47047047047047 0.357146304368113
0.49049049049049 0.353727302780113
0.51051051051051 0.350200642706842
0.53053053053053 0.346570209826128
0.55055055055055 0.34283997442106
0.57057057057057 0.339013984172804
0.59059059059059 0.33509635687531
0.61061061061061 0.331091273090187
0.63063063063063 0.327002968759958
0.65065065065065 0.322835727797843
0.67067067067067 0.318593874672039
0.69069069069069 0.314281767002296
0.71071071071071 0.309903788186304
0.73073073073073 0.305464340073112
0.75075075075075 0.300967835700437
0.77077077077077 0.296418692112302
0.790790790790791 0.291821323272996
0.810810810810811 0.287180133092853
0.830830830830831 0.282499508580786
0.850850850850851 0.277783813137949
0.870870870870871 0.273037380006279
0.890890890890891 0.268264505884996
0.910910910910911 0.2634694447275
0.930930930930931 0.258656401730343
0.950950950950951 0.253829527525259
0.970970970970971 0.24899291258444
0.990990990990991 0.24415058184851
1.01101101101101 0.239306489585817
1.03103103103103 0.234464514490888
1.05105105105105 0.229628455029052
1.07107107107107 0.224802025033432
1.09109109109109 0.219988849559688
1.11111111111111 0.215192461003031
1.13113113113113 0.210416295481265
1.15115115115115 0.205663689486728
1.17117117117117 0.200937876809264
1.19119119119119 0.1962419857315
1.21121121121121 0.191579036496956
1.23123123123123 0.186951939050736
1.25125125125125 0.182363491051798
1.27127127127127 0.17781637615506
1.29129129129129 0.173313162560933
1.31131131131131 0.168856301829129
1.33133133133133 0.164448127952996
1.35135135135135 0.160090856689972
1.37137137137137 0.155786585143159
1.39139139139139 0.151537291588457
1.41141141141141 0.147344835541168
1.43143143143143 0.143210958055468
1.45145145145145 0.139137282249685
1.47147147147147 0.135125314049902
1.49149149149149 0.13117644314399
1.51151151151151 0.127291944137829
1.53153153153153 0.123472977905145
1.55155155155155 0.119720593122119
1.57157157157157 0.116035727977651
1.59159159159159 0.112419212049971
1.61161161161161 0.108871768340093
1.63163163163163 0.10539401545248
1.65165165165165 0.101986469913169
1.67167167167167 0.0986495486155338
1.69169169169169 0.0953835713838294
1.71171171171171 0.0921887636446459
1.73173173173173 0.0890652591964251
1.75175175175175 0.0860131030672496
1.77177177177177 0.083032254451193
1.79179179179179 0.0801225897136326
1.81181181181181 0.077283905456062
1.83183183183183 0.0745159216311036
1.85185185185185 0.0718182846986055
1.87187187187187 0.0691905708139176
1.89189189189189 0.0666322890396674
1.91191191191191 0.0641428845726061
1.93193193193193 0.0617217419773594
1.95195195195195 0.059368188419199
1.97197197197197 0.057081496888248
1.99199199199199 0.0548608894078376
2.01201201201201 0.0527055402200588
2.03203203203203 0.0506145789418748
2.05205205205205 0.0485870936855042
2.07207207207207 0.0466221341371236
2.09209209209209 0.0447187145882916
2.11211211211211 0.0428758169148479
2.13213213213213 0.041092393498395
2.15215215215215 0.0393673700858294
2.17217217217217 0.0376996485827434
2.19219219219219 0.0360881097768737
2.21221221221221 0.0345316159881233
2.23223223223223 0.033029013642035
2.25225225225225 0.0315791357639332
2.27227227227227 0.03018080439129
2.29229229229229 0.0288328329022028
2.31231231231231 0.0275340282581906
2.33233233233233 0.0262831931598317
2.35235235235235 0.02507912811407
2.37237237237237 0.0239206334123108
2.39239239239239 0.0228065110187136
2.41241241241241 0.0217355663683581
2.43243243243243 0.0207066100752275
2.45245245245245 0.0197184595501959
2.47247247247247 0.0187699405294511
2.49249249249249 0.0178598885140022
2.51251251251251 0.016987150121141
2.53253253253253 0.0161505843489182
2.55255255255255 0.0153490637548888
2.57257257257257 0.0145814755505493
2.59259259259259 0.0138467226130541
2.61261261261261 0.0131437244159435
2.63263263263263 0.0124714178807514
2.65265265265265 0.0118287581514852
2.67267267267267 0.0112147192940778
2.69269269269269 0.0106282949230103
2.71271271271271 0.0100684987573917
2.73273273273273 0.00953436510885491
2.75275275275275 0.00902494930369034
2.77277277277277 0.00853932804169479
2.79279279279279 0.00807659969425077
2.81281281281281 0.00763588454418634
2.83283283283283 0.00721632496998621
2.85285285285285 0.00681708557693871
2.87287287287287 0.00643735327780635
2.89289289289289 0.00607633732560524
2.91291291291291 0.00573326930106518
2.93293293293293 0.00540740305732384
2.95295295295295 0.00509801462438214
2.97297297297297 0.00480440207581539
2.99299299299299 0.00452588536019617
3.01301301301301 0.00426180609964127
3.03303303303303 0.00401152735784578
3.05305305305305 0.00377443337991421
3.07307307307307 0.00354992930624085
3.09309309309309 0.00333744086263051
3.11311311311311 0.00313641402878608
3.13313313313313 0.0029463146872226
3.15315315315315 0.00276662825459746
3.17317317317317 0.00259685929737496
3.19319319319319 0.00243653113367011
3.21321321321321 0.00228518542304203
3.23323323323323 0.00214238174593163
3.25325325325325 0.00200769717436225
3.27327327327327 0.00188072583544551
3.29329329329329 0.00176107846915771
3.31331331331331 0.00164838198177652
3.33333333333333 0.0015422789962911
3.35335335335335 0.00144242740102448
3.37337337337337 0.00134849989763212
3.39339339339339 0.00126018354956833
3.41341341341341 0.00117717933203981
3.43343343343343 0.00109920168439588
3.45345345345345 0.0010259780658361
3.47347347347347 0.000957248515249212
3.49349349349349 0.000892765215932441
3.51351351351351 0.000832292065877152
3.53353353353353 0.000775604254246008
3.55355355355355 0.000722487844607976
3.57357357357357 0.000672739365441019
3.59359359359359 0.000626165408357979
3.61361361361361 0.000582582234459159
3.63363363363363 0.000541815389165405
3.65365365365365 0.00050369932583812
3.67367367367367 0.000468077038447599
3.69369369369369 0.000434799703508357
3.71371371371371 0.000403726331459719
3.73373373373373 0.000374723427631836
3.75375375375375 0.000347664662901427
3.77377377377377 0.000322430554107926
3.79379379379379 0.000298908154269281
3.81381381381381 0.000276990752607344
3.83383383383383 0.000256577584365551
3.85385385385385 0.000237573550376443
3.87387387387387 0.000219888946313312
3.89389389389389 0.000203439201538965
3.91391391391391 0.00018814462744512
3.93393393393393 0.000173930175158222
3.95395395395395 0.000160725202471485
3.97397397397397 0.000148463249848567
3.99399399399399 0.000137081825331481
4.01401401401401 0.000126522198173995
4.03403403403403 0.000116729201011866
4.05405405405405 0.000107651040372646
4.07407407407407 9.92391153205018e-05
4.09409409409409 9.14478440253317e-05
4.11411411411411 8.42344980404937e-05
4.13413413413413 7.75590440694795e-05
4.15415415415415 7.13839929989176e-05
4.17417417417417 6.56742559732345e-05
4.19419419419419 6.03970072851107e-05
4.21421421421421 5.55215538554582e-05
4.23423423423423 5.10192110769697e-05
4.25425425425425 4.68631847962824e-05
4.27427427427427 4.3028459211397e-05
4.29429429429429 3.94916904631592e-05
4.31431431431431 3.62311057022691e-05
4.33433433433433 3.32264074164067e-05
4.35435435435435 3.04586828055866e-05
4.37437437437437 2.79103179977393e-05
4.39439439439439 2.55649169007242e-05
4.41441441441441 2.34072244914564e-05
4.43443443443443 2.14230543475548e-05
4.45445445445445 1.95992202318354e-05
4.47447447447447 1.79234715450684e-05
4.49449449449449 1.63844324676437e-05
4.51451451451451 1.49715446161227e-05
4.53453453453453 1.3675013046071e-05
4.55455455455455 1.24857554380297e-05
4.57457457457457 1.13953543089884e-05
4.59459459459459 1.03960120972233e-05
4.61461461461461 9.48050897386715e-06
4.63463463463463 8.64216324004121e-06
4.65465465465465 7.87479417380527e-06
4.67467467467467 7.17268719654363e-06
4.69469469469469 6.53056123369604e-06
4.71471471471471 5.94353814994721e-06
4.73473473473473 5.40711414409909e-06
4.75475475475475 4.91713299385613e-06
4.77477477477477 4.4697610456467e-06
4.79479479479479 4.06146384937932e-06
4.81481481481481 3.68898434268124e-06
4.83483483483483 3.34932249368796e-06
4.85485485485485 3.03971631583941e-06
4.87487487487487 2.75762417238901e-06
4.89489489489489 2.50070829244518e-06
4.91491491491491 2.26681942433715e-06
4.93493493493493 2.05398255592983e-06
4.95495495495495 1.86038363520377e-06
4.97497497497497 1.68435722796805e-06
4.99499499499499 1.5243750529858e-06
5.01501501501502 1.37903533806611e-06
5.03503503503504 1.24705294381396e-06
5.05505505505506 1.12725020473309e-06
5.07507507507508 1.01854844024876e-06
5.0950950950951 9.19960090959837e-07
5.11511511511512 8.30581438046261e-07
5.13513513513514 7.49585866251233e-07
5.15515515515516 6.76217633231267e-07
5.17517517517518 6.09786110324583e-07
5.1951951951952 5.4966046193278e-07
5.21521521521522 4.95264732746229e-07
5.23523523523524 4.46073313973501e-07
5.25525525525526 4.01606761563285e-07
5.27527527527528 3.61427941137511e-07
5.2952952952953 3.25138475990267e-07
5.31531531531532 2.92375476052561e-07
5.33533533533534 2.62808527181656e-07
5.35535535535536 2.36136921509202e-07
5.37537537537538 2.1208711087848e-07
5.3953953953954 1.90410366621162e-07
5.41541541541542 1.70880630071684e-07
5.43543543543544 1.5329253929596e-07
5.45545545545546 1.3745961852414e-07
5.47547547547548 1.23212617727566e-07
5.4954954954955 1.10397990671284e-07
5.51551551551552 9.8876500608364e-08
5.53553553553554 8.85219435638491e-08
5.55555555555556 7.92199798873018e-08
5.57557557557558 7.08670654362614e-08
5.5955955955956 6.33694743912418e-08
5.61561561561562 5.66424062986154e-08
5.63563563563564 5.06091704933412e-08
5.65565565565566 4.5200441571292e-08
5.67567567567568 4.03535800631662e-08
5.6956956956957 3.60120129107462e-08
5.71571571571572 3.2124668763613e-08
5.73573573573574 2.86454635022852e-08
5.75575575575576 2.55328317539416e-08
5.77577577577578 2.27493005011625e-08
5.7957957957958 2.02611011941336e-08
5.81581581581582 1.80378170640688e-08
5.83583583583584 1.60520626017053e-08
5.85585585585586 1.42791924110059e-08
5.87587587587588 1.2697036876002e-08
5.8958958958959 1.12856622892642e-08
5.91591591591592 1.00271532849868e-08
5.93593593593594 8.90541559921139e-09
5.95595595595596 7.90599734535251e-09
5.97597597597598 7.01592714588943e-09
5.995995995996 6.22356760178439e-09
6.01601601601602 5.5184827107339e-09
6.03603603603604 4.89131796456914e-09
6.05605605605606 4.33369196574694e-09
6.07607607607608 3.83809850362692e-09
6.0960960960961 3.3978181237675e-09
6.11611611611612 3.00683830841826e-09
6.13613613613614 2.65978146430924e-09
6.15615615615616 2.35183998527871e-09
6.17617617617618 2.07871772273778e-09
6.1961961961962 1.83657725691022e-09
6.21621621621622 1.6219924166386e-09
6.23623623623624 1.43190554571785e-09
6.25625625625626 1.26358905957511e-09
6.27627627627628 1.11461087800764e-09
6.2962962962963 9.82803357938119e-10
6.31631631631632 8.66235385045992e-10
6.33633633633634 7.63187314959515e-10
6.35635635635636 6.72128483698836e-10
6.37637637637638 5.91697033481532e-10
6.3963963963964 5.20681824054164e-10
6.41641641641642 4.58006221596989e-10
6.43643643643644 4.02713577147895e-10
6.45645645645646 3.53954224575805e-10
6.47647647647648 3.10973844559378e-10
6.4964964964965 2.73103055937435e-10
6.51651651651652 2.39748109325539e-10
6.53653653653654 2.10382570159619e-10
6.55655655655656 1.84539889444162e-10
6.57657657657658 1.61806770551276e-10
6.5965965965966 1.41817249531742e-10
6.61661661661662 1.24247414645767e-10
6.63663663663664 1.08810698278133e-10
6.65665665665666 9.52536811418137e-11
6.67667667667668 8.33523547614393e-11
6.6966966966967 7.29087937236022e-11
6.71671671671672 6.37481941394484e-11
6.73673673673674 5.57162392365998e-11
6.75675675675676 4.86767570277076e-11
6.77677677677678 4.25096386334907e-11
6.7967967967968 3.71089891068556e-11
6.81681681681682 3.23814855461048e-11
6.83683683683684 2.82449199306813e-11
6.85685685685686 2.46269064908966e-11
6.87687687687688 2.14637355595386e-11
6.8968968968969 1.86993577716892e-11
6.91691691691692 1.62844842008236e-11
6.93693693693694 1.41757895636779e-11
6.95695695695696 1.23352070109961e-11
6.97697697697698 1.07293042619725e-11
6.996996996997 9.32873195138548e-12
7.01701701701702 8.10773605306642e-12
7.03703703703704 7.04372713322415e-12
7.05705705705706 6.11689998287643e-12
7.07707707707708 5.30989788981512e-12
7.0970970970971 4.6075164458095e-12
7.11711711711712 3.99644235193917e-12
7.13713713713714 3.46502319108337e-12
7.15715715715716 3.00306458801651e-12
7.17717717717718 2.60165157997869e-12
7.1971971971972 2.25299137914217e-12
7.21721721721722 1.95027502769791e-12
7.23723723723724 1.68755573049414e-12
7.25725725725726 1.45964190299846e-12
7.27727727727728 1.262003197176e-12
7.2972972972973 1.0906879676822e-12
7.31731731731732 9.42250818252902e-13
7.33733733733734 8.13689025751829e-13
7.35735735735736 7.02386779168733e-13
7.37737737737738 6.06066294885245e-13
7.3973973973974 5.22744979473708e-13
7.41741741741742 4.50697908714381e-13
7.43743743743744 3.88424977793729e-13
7.45745745745746 3.34622154016963e-13
7.47747747747748 2.88156330935933e-13
7.4974974974975 2.48043342543511e-13
7.51751751751752 2.13428748996348e-13
7.53753753753754 1.83571051982055e-13
7.55755755755756 1.57827039041883e-13
7.57757757757758 1.35638992516663e-13
7.5975975975976 1.16523530854698e-13
7.61761761761762 1.000618782966e-13
7.63763763763764 8.58913838706873e-14
7.65765765765766 7.36981325813112e-14
7.67767767767768 6.32105109959248e-14
7.6976976976977 5.41936064405376e-14
7.71771771771772 4.64443339685915e-14
7.73773773773774 3.97871984155707e-14
7.75775775775776 3.40706104038536e-14
7.77777777777778 2.91636853079907e-14
7.7977977977978 2.4953463096687e-14
7.81781781781782 2.13424947819474e-14
7.83783783783784 1.82467480586654e-14
7.85785785785786 1.55937907247682e-14
7.87787787787788 1.33212157347804e-14
7.8978978978979 1.13752763482777e-14
7.91791791791792 9.70970386854701e-15
7.93793793793794 8.28468399583068e-15
7.95795795795796 7.06597090549298e-15
7.97797797797798 6.02412085866189e-15
7.997997997998 5.13382950919603e-15
8.01801801801802 4.3733591283238e-15
8.03803803803804 3.72404376402735e-15
8.05805805805806 3.16986191875719e-15
8.07807807807808 2.69706769497134e-15
8.0980980980981 2.29387254841619e-15
8.11811811811812 1.95017082606148e-15
8.13813813813814 1.65730316851105e-15
8.15815815815816 1.40785264250169e-15
8.17817817817818 1.19546915264237e-15
8.1981981981982 1.01471827585831e-15
8.21821821821822 8.60951178494166e-16
8.23823823823824 7.30192724674871e-16
8.25825825825826 6.19045274051723e-16
8.27827827827828 5.24606005103494e-16
8.2982982982983 4.44395893386326e-16
8.31831831831832 3.76298728354137e-16
8.33833833833834 3.18508772685576e-16
8.35835835835836 2.69485858889318e-16
8.37837837837838 2.27916883183252e-16
8.3983983983984 1.92682799625235e-16
8.41841841841842 1.62830341150177e-16
8.43843843843844 1.37547801096493e-16
8.45845845845846 1.16144301209627e-16
8.47847847847848 9.8032051926925e-17
8.4984984984985 8.27111796592674e-17
8.51851851851852 6.97567552529606e-17
8.53853853853854 5.88077091106193e-17
8.55855855855856 4.95573626747691e-17
8.57857857857858 4.17453440895294e-17
8.5985985985986 3.51506886838449e-17
8.61861861861862 2.95859531836935e-17
8.63863863863864 2.48921968838275e-17
8.65865865865866 2.09347039316773e-17
8.67867867867868 1.75993388643364e-17
8.6986986986987 1.47894429982973e-17
8.71871871871872 1.24231925503958e-17
8.73873873873874 1.04313507694241e-17
8.75875875875876 8.75535614211525e-18
8.77877877877878 7.34569713009337e-18
8.7987987987988 6.16053109048305e-18
8.81881881881882 5.16451119999019e-18
8.83883883883884 4.32779048512422e-18
8.85885885885886 3.62517658454128e-18
8.87887887887888 3.03541474067764e-18
8.8988988988989 2.54057982941549e-18
8.91891891891892 2.12556106807901e-18
8.93893893893894 1.77762546207239e-18
8.95895895895896 1.48604811781133e-18
8.97897897897898 1.24179931485728e-18
8.998998998999 1.03727973679138e-18
9.01901901901902 8.66096545670873e-19
9.03903903903904 7.22874080905507e-19
9.05905905905906 6.03093897535385e-19
9.07907907907908 5.0295965472299e-19
9.0990990990991 4.19283042962206e-19
9.11911911911912 3.49387515332417e-19
9.13913913913914 2.91027078875695e-19
9.15915915915916 2.42317819504618e-19
9.17917917917918 2.01680188581445e-19
9.1991991991992 1.67790380698338e-19
9.21921921921922 1.39539388139188e-19
9.23923923923924 1.1599853476858e-19
9.25925925925926 9.63904764362469e-20
9.27927927927928 8.00648113242436e-20
9.2992992992993 6.6477576193283e-20
9.31931931931932 5.51740167796855e-20
9.33933933933934 4.5774115701642e-20
9.35935935935936 3.79604417474457e-20
9.37937937937938 3.14679525475421e-20
9.3993993993994 2.60754402558043e-20
9.41941941941942 2.15983585814365e-20
9.43943943943944 1.78828106797973e-20
9.45945945945946 1.48005121824234e-20
9.47947947947948 1.22445730038445e-20
9.4994994994995 1.01259663374544e-20
9.51951951951952 8.37057415073758e-21
9.53953953953954 6.91671611023779e-21
9.55955955955956 5.71308371631411e-21
9.57957957957958 4.71701393695898e-21
9.5995995995996 3.89304716291232e-21
9.61961961961962 3.21172317125736e-21
9.63963963963964 2.64857624243904e-21
9.65965965965966 2.18329684678274e-21
9.67967967967968 1.79903258756111e-21
9.6996996996997 1.48180551599987e-21
9.71971971971972 1.22002665237565e-21
9.73973973973974 1.00409166885045e-21
9.75975975975976 8.26044308669654e-22
9.77977977977978 6.79296312742742e-22
9.7997997997998 5.58394465795474e-22
9.81981981981982 4.58826916995414e-22
9.83983983983984 3.7686222201397e-22
9.85985985985986 3.09415635142992e-22
9.87987987987988 2.53938085193762e-22
9.8998998998999 2.08324025950642e-22
9.91991991991992 1.70834984871876e-22
9.93993993993994 1.40036162642795e-22
9.95995995995996 1.14743877987917e-22
9.97997997997998 9.39820210218911e-23
10 7.69459862670642e-23
};
\addlegendentry{$N(0,1)$}
\end{axis}

\end{tikzpicture}

%% file: FinalFigs/Null_Dists_d_10_500_n_200_m_200_kernel__Gaussian_RBF_2022_10_12_22_42_59cross.tex
\begin{tikzpicture}

\definecolor{darkorange25512714}{RGB}{255,127,14}
\definecolor{darkslategray38}{RGB}{38,38,38}
\definecolor{lightgray204}{RGB}{204,204,204}
\definecolor{steelblue31119180}{RGB}{31,119,180}

\begin{axis}[
axis line style={darkslategray38},
height=\figheight,
legend cell align={left},
legend style={fill opacity=0.8, draw opacity=1, text opacity=1, draw=none},
tick align=outside,
tick pos=left,
title={$\cmmd~(n/m=1)$},
width=\figwidth,
x grid style={lightgray204},
xmin=-6, xmax=6,
xtick style={color=darkslategray38},
y grid style={lightgray204},
ylabel={},
ymin=0, ymax=0.437345126786171,
ytick style={color=darkslategray38}, 
xticklabels=empty,
yticklabels=empty
]
\draw[draw=none,fill=steelblue31119180,fill opacity=0.8] (axis cs:-3.61791682243347,0) rectangle (axis cs:-3.35670447349548,0.00306264122341969);
\addlegendimage{ybar,ybar legend,draw=none,fill=steelblue31119180,fill opacity=0.8}
\addlegendentry{d=10}

\draw[draw=none,fill=steelblue31119180,fill opacity=0.8] (axis cs:-2.96488571166992,0) rectangle (axis cs:-2.70367336273193,0.0214384885639379);
\draw[draw=none,fill=steelblue31119180,fill opacity=0.8] (axis cs:-2.31185483932495,0) rectangle (axis cs:-2.05064225196838,0.067378094615521);
\draw[draw=none,fill=steelblue31119180,fill opacity=0.8] (axis cs:-1.65882337093353,0) rectangle (axis cs:-1.39761078357697,0.137818829895384);
\draw[draw=none,fill=steelblue31119180,fill opacity=0.8] (axis cs:-1.00579214096069,0) rectangle (axis cs:-0.744579792022705,0.324639969682488);
\draw[draw=none,fill=steelblue31119180,fill opacity=0.8] (axis cs:-0.352761059999466,0) rectangle (axis cs:-0.0915484726428986,0.416519168367782);
\draw[draw=none,fill=steelblue31119180,fill opacity=0.8] (axis cs:0.300270080566406,0) rectangle (axis cs:0.561482667922974,0.287888248724791);
\draw[draw=none,fill=steelblue31119180,fill opacity=0.8] (axis cs:0.953301429748535,0) rectangle (axis cs:1.21451377868652,0.20519696196912);
\draw[draw=none,fill=steelblue31119180,fill opacity=0.8] (axis cs:1.6063324213028,0) rectangle (axis cs:1.86754500865936,0.0490022506294698);
\draw[draw=none,fill=steelblue31119180,fill opacity=0.8] (axis cs:2.25936365127563,0) rectangle (axis cs:2.52057600021362,0.0183758473405182);
\draw[draw=none,fill=darkorange25512714,fill opacity=0.8] (axis cs:-3.3567042350769,0) rectangle (axis cs:-3.09549188613892,0.00612528244683939);
\addlegendimage{ybar,ybar legend,draw=none,fill=darkorange25512714,fill opacity=0.8}
\addlegendentry{d=500}

\draw[draw=none,fill=darkorange25512714,fill opacity=0.8] (axis cs:-2.70367312431335,0) rectangle (axis cs:-2.44246077537537,0.0214384885639379);
\draw[draw=none,fill=darkorange25512714,fill opacity=0.8] (axis cs:-2.0506420135498,0) rectangle (axis cs:-1.78942942619324,0.0765660166085465);
\draw[draw=none,fill=darkorange25512714,fill opacity=0.8] (axis cs:-1.39761090278625,0) rectangle (axis cs:-1.13639831542969,0.192946361853537);
\draw[draw=none,fill=darkorange25512714,fill opacity=0.8] (axis cs:-0.744579613208771,0) rectangle (axis cs:-0.483367264270782,0.294013557448291);
\draw[draw=none,fill=darkorange25512714,fill opacity=0.8] (axis cs:-0.0915484726428986,0) rectangle (axis cs:0.16966387629509,0.392018040816736);
\draw[draw=none,fill=darkorange25512714,fill opacity=0.8] (axis cs:0.561482667922974,0) rectangle (axis cs:0.822695016860962,0.263387121173745);
\draw[draw=none,fill=darkorange25512714,fill opacity=0.8] (axis cs:1.21451377868652,0) rectangle (axis cs:1.47572612762451,0.192946397075441);
\draw[draw=none,fill=darkorange25512714,fill opacity=0.8] (axis cs:1.86754488945007,0) rectangle (axis cs:2.12875747680664,0.067378094615521);
\draw[draw=none,fill=darkorange25512714,fill opacity=0.8] (axis cs:2.5205762386322,0) rectangle (axis cs:2.78178858757019,0.0245011297873576);
\addplot [semithick, black]
table {%
-10 7.69459862670642e-23
-9.97997997997998 9.39820210218911e-23
-9.95995995995996 1.14743877987917e-22
-9.93993993993994 1.40036162642795e-22
-9.91991991991992 1.70834984871876e-22
-9.8998998998999 2.08324025950642e-22
-9.87987987987988 2.53938085193762e-22
-9.85985985985986 3.09415635142992e-22
-9.83983983983984 3.7686222201397e-22
-9.81981981981982 4.58826916995414e-22
-9.7997997997998 5.58394465795474e-22
-9.77977977977978 6.79296312742742e-22
-9.75975975975976 8.26044308669654e-22
-9.73973973973974 1.00409166885045e-21
-9.71971971971972 1.22002665237565e-21
-9.6996996996997 1.48180551599987e-21
-9.67967967967968 1.79903258756111e-21
-9.65965965965966 2.18329684678274e-21
-9.63963963963964 2.64857624243904e-21
-9.61961961961962 3.21172317125736e-21
-9.5995995995996 3.89304716291232e-21
-9.57957957957958 4.71701393695898e-21
-9.55955955955956 5.71308371631411e-21
-9.53953953953954 6.91671611023779e-21
-9.51951951951952 8.37057415073758e-21
-9.4994994994995 1.01259663374544e-20
-9.47947947947948 1.22445730038445e-20
-9.45945945945946 1.48005121824234e-20
-9.43943943943944 1.78828106797973e-20
-9.41941941941942 2.15983585814365e-20
-9.3993993993994 2.60754402558043e-20
-9.37937937937938 3.14679525475421e-20
-9.35935935935936 3.79604417474457e-20
-9.33933933933934 4.5774115701642e-20
-9.31931931931932 5.51740167796855e-20
-9.2992992992993 6.6477576193283e-20
-9.27927927927928 8.00648113242436e-20
-9.25925925925926 9.63904764362469e-20
-9.23923923923924 1.1599853476858e-19
-9.21921921921922 1.39539388139188e-19
-9.1991991991992 1.67790380698338e-19
-9.17917917917918 2.01680188581445e-19
-9.15915915915916 2.42317819504618e-19
-9.13913913913914 2.91027078875695e-19
-9.11911911911912 3.49387515332417e-19
-9.0990990990991 4.19283042962206e-19
-9.07907907907908 5.0295965472299e-19
-9.05905905905906 6.03093897535385e-19
-9.03903903903904 7.22874080905507e-19
-9.01901901901902 8.66096545670873e-19
-8.998998998999 1.03727973679138e-18
-8.97897897897898 1.24179931485728e-18
-8.95895895895896 1.48604811781133e-18
-8.93893893893894 1.77762546207239e-18
-8.91891891891892 2.12556106807901e-18
-8.8988988988989 2.54057982941549e-18
-8.87887887887888 3.03541474067764e-18
-8.85885885885886 3.62517658454128e-18
-8.83883883883884 4.32779048512422e-18
-8.81881881881882 5.16451119999019e-18
-8.7987987987988 6.16053109048305e-18
-8.77877877877878 7.34569713009337e-18
-8.75875875875876 8.75535614211525e-18
-8.73873873873874 1.04313507694241e-17
-8.71871871871872 1.24231925503958e-17
-8.6986986986987 1.47894429982973e-17
-8.67867867867868 1.75993388643364e-17
-8.65865865865866 2.09347039316773e-17
-8.63863863863864 2.48921968838275e-17
-8.61861861861862 2.95859531836935e-17
-8.5985985985986 3.51506886838449e-17
-8.57857857857858 4.17453440895294e-17
-8.55855855855856 4.95573626747691e-17
-8.53853853853854 5.88077091106193e-17
-8.51851851851852 6.97567552529606e-17
-8.4984984984985 8.27111796592674e-17
-8.47847847847848 9.8032051926925e-17
-8.45845845845846 1.16144301209627e-16
-8.43843843843844 1.37547801096493e-16
-8.41841841841842 1.62830341150177e-16
-8.3983983983984 1.92682799625235e-16
-8.37837837837838 2.27916883183252e-16
-8.35835835835836 2.69485858889318e-16
-8.33833833833834 3.18508772685576e-16
-8.31831831831832 3.76298728354137e-16
-8.2982982982983 4.44395893386326e-16
-8.27827827827828 5.24606005103494e-16
-8.25825825825826 6.19045274051723e-16
-8.23823823823824 7.30192724674871e-16
-8.21821821821822 8.60951178494166e-16
-8.1981981981982 1.01471827585831e-15
-8.17817817817818 1.19546915264237e-15
-8.15815815815816 1.40785264250169e-15
-8.13813813813814 1.65730316851105e-15
-8.11811811811812 1.95017082606148e-15
-8.0980980980981 2.29387254841619e-15
-8.07807807807808 2.69706769497134e-15
-8.05805805805806 3.16986191875719e-15
-8.03803803803804 3.72404376402735e-15
-8.01801801801802 4.3733591283238e-15
-7.997997997998 5.13382950919607e-15
-7.97797797797798 6.02412085866193e-15
-7.95795795795796 7.06597090549303e-15
-7.93793793793794 8.28468399583074e-15
-7.91791791791792 9.70970386854708e-15
-7.8978978978979 1.13752763482777e-14
-7.87787787787788 1.33212157347805e-14
-7.85785785785786 1.55937907247683e-14
-7.83783783783784 1.82467480586655e-14
-7.81781781781782 2.13424947819475e-14
-7.7977977977978 2.49534630966872e-14
-7.77777777777778 2.91636853079909e-14
-7.75775775775776 3.40706104038538e-14
-7.73773773773774 3.9787198415571e-14
-7.71771771771772 4.64443339685918e-14
-7.6976976976977 5.4193606440538e-14
-7.67767767767768 6.32105109959252e-14
-7.65765765765766 7.36981325813117e-14
-7.63763763763764 8.58913838706879e-14
-7.61761761761762 1.00061878296601e-13
-7.5975975975976 1.16523530854699e-13
-7.57757757757758 1.35638992516664e-13
-7.55755755755756 1.57827039041884e-13
-7.53753753753754 1.83571051982057e-13
-7.51751751751752 2.13428748996349e-13
-7.4974974974975 2.48043342543513e-13
-7.47747747747748 2.88156330935935e-13
-7.45745745745746 3.34622154016965e-13
-7.43743743743744 3.88424977793732e-13
-7.41741741741742 4.50697908714384e-13
-7.3973973973974 5.22744979473711e-13
-7.37737737737738 6.06066294885249e-13
-7.35735735735736 7.02386779168738e-13
-7.33733733733734 8.13689025751835e-13
-7.31731731731732 9.42250818252909e-13
-7.2972972972973 1.09068796768221e-12
-7.27727727727728 1.262003197176e-12
-7.25725725725726 1.45964190299847e-12
-7.23723723723724 1.68755573049416e-12
-7.21721721721722 1.95027502769792e-12
-7.1971971971972 2.25299137914218e-12
-7.17717717717718 2.60165157997871e-12
-7.15715715715716 3.00306458801653e-12
-7.13713713713714 3.4650231910834e-12
-7.11711711711712 3.99644235193919e-12
-7.0970970970971 4.60751644580953e-12
-7.07707707707708 5.30989788981514e-12
-7.05705705705706 6.11689998287646e-12
-7.03703703703704 7.0437271332242e-12
-7.01701701701702 8.10773605306645e-12
-6.996996996997 9.32873195138555e-12
-6.97697697697698 1.07293042619726e-11
-6.95695695695696 1.23352070109962e-11
-6.93693693693694 1.41757895636779e-11
-6.91691691691692 1.62844842008237e-11
-6.8968968968969 1.86993577716893e-11
-6.87687687687688 2.14637355595386e-11
-6.85685685685686 2.46269064908967e-11
-6.83683683683684 2.82449199306815e-11
-6.81681681681682 3.2381485546105e-11
-6.7967967967968 3.71089891068559e-11
-6.77677677677678 4.25096386334913e-11
-6.75675675675676 4.86767570277083e-11
-6.73673673673674 5.57162392366004e-11
-6.71671671671672 6.37481941394491e-11
-6.6966966966967 7.29087937236032e-11
-6.67667667667668 8.33523547614402e-11
-6.65665665665666 9.52536811418151e-11
-6.63663663663664 1.08810698278135e-10
-6.61661661661662 1.24247414645768e-10
-6.5965965965966 1.41817249531744e-10
-6.57657657657658 1.61806770551278e-10
-6.55655655655656 1.84539889444164e-10
-6.53653653653654 2.10382570159622e-10
-6.51651651651652 2.39748109325542e-10
-6.4964964964965 2.73103055937438e-10
-6.47647647647648 3.10973844559381e-10
-6.45645645645646 3.53954224575809e-10
-6.43643643643644 4.027135771479e-10
-6.41641641641642 4.58006221596996e-10
-6.3963963963964 5.20681824054169e-10
-6.37637637637638 5.91697033481538e-10
-6.35635635635636 6.72128483698846e-10
-6.33633633633634 7.63187314959523e-10
-6.31631631631632 8.66235385046001e-10
-6.2962962962963 9.8280335793813e-10
-6.27627627627628 1.11461087800766e-09
-6.25625625625626 1.26358905957513e-09
-6.23623623623624 1.43190554571787e-09
-6.21621621621622 1.62199241663862e-09
-6.1961961961962 1.83657725691024e-09
-6.17617617617618 2.0787177227378e-09
-6.15615615615616 2.35183998527873e-09
-6.13613613613614 2.65978146430928e-09
-6.11611611611612 3.00683830841829e-09
-6.0960960960961 3.39781812376754e-09
-6.07607607607608 3.83809850362696e-09
-6.05605605605606 4.33369196574699e-09
-6.03603603603604 4.89131796456919e-09
-6.01601601601602 5.51848271073395e-09
-5.995995995996 6.22356760178439e-09
-5.97597597597598 7.01592714588943e-09
-5.95595595595596 7.90599734535251e-09
-5.93593593593594 8.90541559921139e-09
-5.91591591591592 1.00271532849868e-08
-5.8958958958959 1.12856622892642e-08
-5.87587587587588 1.2697036876002e-08
-5.85585585585586 1.42791924110059e-08
-5.83583583583584 1.60520626017053e-08
-5.81581581581582 1.80378170640688e-08
-5.7957957957958 2.02611011941336e-08
-5.77577577577578 2.27493005011625e-08
-5.75575575575576 2.55328317539416e-08
-5.73573573573574 2.86454635022852e-08
-5.71571571571572 3.2124668763613e-08
-5.6956956956957 3.60120129107462e-08
-5.67567567567568 4.03535800631662e-08
-5.65565565565566 4.5200441571292e-08
-5.63563563563564 5.06091704933412e-08
-5.61561561561562 5.66424062986154e-08
-5.5955955955956 6.33694743912418e-08
-5.57557557557558 7.08670654362614e-08
-5.55555555555556 7.92199798873018e-08
-5.53553553553554 8.85219435638491e-08
-5.51551551551552 9.8876500608364e-08
-5.4954954954955 1.10397990671284e-07
-5.47547547547548 1.23212617727566e-07
-5.45545545545546 1.3745961852414e-07
-5.43543543543544 1.5329253929596e-07
-5.41541541541542 1.70880630071684e-07
-5.3953953953954 1.90410366621162e-07
-5.37537537537538 2.1208711087848e-07
-5.35535535535536 2.36136921509202e-07
-5.33533533533534 2.62808527181656e-07
-5.31531531531532 2.92375476052561e-07
-5.2952952952953 3.25138475990267e-07
-5.27527527527528 3.61427941137511e-07
-5.25525525525526 4.01606761563285e-07
-5.23523523523524 4.46073313973501e-07
-5.21521521521522 4.95264732746229e-07
-5.1951951951952 5.4966046193278e-07
-5.17517517517518 6.09786110324583e-07
-5.15515515515516 6.76217633231267e-07
-5.13513513513514 7.49585866251233e-07
-5.11511511511512 8.30581438046261e-07
-5.0950950950951 9.19960090959837e-07
-5.07507507507508 1.01854844024876e-06
-5.05505505505506 1.12725020473309e-06
-5.03503503503504 1.24705294381396e-06
-5.01501501501502 1.37903533806611e-06
-4.99499499499499 1.5243750529858e-06
-4.97497497497497 1.68435722796805e-06
-4.95495495495495 1.86038363520377e-06
-4.93493493493493 2.05398255592983e-06
-4.91491491491491 2.26681942433715e-06
-4.89489489489489 2.50070829244518e-06
-4.87487487487487 2.75762417238901e-06
-4.85485485485485 3.03971631583941e-06
-4.83483483483483 3.34932249368796e-06
-4.81481481481481 3.68898434268124e-06
-4.79479479479479 4.06146384937932e-06
-4.77477477477477 4.4697610456467e-06
-4.75475475475475 4.91713299385613e-06
-4.73473473473473 5.40711414409909e-06
-4.71471471471471 5.94353814994721e-06
-4.69469469469469 6.53056123369604e-06
-4.67467467467467 7.17268719654363e-06
-4.65465465465465 7.87479417380527e-06
-4.63463463463463 8.64216324004121e-06
-4.61461461461461 9.48050897386715e-06
-4.59459459459459 1.03960120972233e-05
-4.57457457457457 1.13953543089884e-05
-4.55455455455455 1.24857554380297e-05
-4.53453453453453 1.3675013046071e-05
-4.51451451451451 1.49715446161227e-05
-4.49449449449449 1.63844324676437e-05
-4.47447447447447 1.79234715450684e-05
-4.45445445445445 1.95992202318354e-05
-4.43443443443443 2.14230543475548e-05
-4.41441441441441 2.34072244914564e-05
-4.39439439439439 2.55649169007242e-05
-4.37437437437437 2.79103179977393e-05
-4.35435435435435 3.04586828055866e-05
-4.33433433433433 3.32264074164067e-05
-4.31431431431431 3.62311057022691e-05
-4.29429429429429 3.94916904631592e-05
-4.27427427427427 4.3028459211397e-05
-4.25425425425425 4.68631847962824e-05
-4.23423423423423 5.10192110769697e-05
-4.21421421421421 5.55215538554582e-05
-4.19419419419419 6.03970072851107e-05
-4.17417417417417 6.56742559732345e-05
-4.15415415415415 7.13839929989176e-05
-4.13413413413413 7.75590440694795e-05
-4.11411411411411 8.42344980404937e-05
-4.09409409409409 9.14478440253317e-05
-4.07407407407407 9.92391153205018e-05
-4.05405405405405 0.000107651040372646
-4.03403403403403 0.000116729201011866
-4.01401401401401 0.000126522198173995
-3.99399399399399 0.000137081825331481
-3.97397397397397 0.000148463249848567
-3.95395395395395 0.000160725202471485
-3.93393393393393 0.000173930175158222
-3.91391391391391 0.00018814462744512
-3.89389389389389 0.000203439201538965
-3.87387387387387 0.000219888946313312
-3.85385385385385 0.000237573550376443
-3.83383383383383 0.000256577584365551
-3.81381381381381 0.000276990752607344
-3.79379379379379 0.000298908154269281
-3.77377377377377 0.000322430554107926
-3.75375375375375 0.000347664662901427
-3.73373373373373 0.000374723427631836
-3.71371371371371 0.000403726331459719
-3.69369369369369 0.000434799703508357
-3.67367367367367 0.000468077038447599
-3.65365365365365 0.00050369932583812
-3.63363363363363 0.000541815389165405
-3.61361361361361 0.000582582234459159
-3.59359359359359 0.000626165408357979
-3.57357357357357 0.000672739365441021
-3.55355355355355 0.000722487844607978
-3.53353353353353 0.00077560425424601
-3.51351351351351 0.000832292065877155
-3.49349349349349 0.000892765215932443
-3.47347347347347 0.000957248515249216
-3.45345345345345 0.0010259780658361
-3.43343343343343 0.00109920168439588
-3.41341341341341 0.00117717933203981
-3.39339339339339 0.00126018354956833
-3.37337337337337 0.00134849989763212
-3.35335335335335 0.00144242740102448
-3.33333333333333 0.00154227899629111
-3.31331331331331 0.00164838198177652
-3.29329329329329 0.00176107846915772
-3.27327327327327 0.00188072583544552
-3.25325325325325 0.00200769717436226
-3.23323323323323 0.00214238174593163
-3.21321321321321 0.00228518542304204
-3.19319319319319 0.00243653113367012
-3.17317317317317 0.00259685929737497
-3.15315315315315 0.00276662825459747
-3.13313313313313 0.00294631468722261
-3.11311311311311 0.00313641402878609
-3.09309309309309 0.00333744086263052
-3.07307307307307 0.00354992930624086
-3.05305305305305 0.00377443337991422
-3.03303303303303 0.00401152735784579
-3.01301301301301 0.00426180609964128
-2.99299299299299 0.00452588536019618
-2.97297297297297 0.0048044020758154
-2.95295295295295 0.00509801462438215
-2.93293293293293 0.00540740305732385
-2.91291291291291 0.00573326930106519
-2.89289289289289 0.00607633732560526
-2.87287287287287 0.00643735327780636
-2.85285285285285 0.00681708557693873
-2.83283283283283 0.00721632496998623
-2.81281281281281 0.00763588454418632
-2.79279279279279 0.00807659969425075
-2.77277277277277 0.00853932804169477
-2.75275275275275 0.00902494930369032
-2.73273273273273 0.00953436510885489
-2.71271271271271 0.0100684987573917
-2.69269269269269 0.0106282949230102
-2.67267267267267 0.0112147192940778
-2.65265265265265 0.0118287581514852
-2.63263263263263 0.0124714178807513
-2.61261261261261 0.0131437244159435
-2.59259259259259 0.0138467226130541
-2.57257257257257 0.0145814755505492
-2.55255255255255 0.0153490637548887
-2.53253253253253 0.0161505843489182
-2.51251251251251 0.0169871501211409
-2.49249249249249 0.0178598885140022
-2.47247247247247 0.018769940529451
-2.45245245245245 0.0197184595501959
-2.43243243243243 0.0207066100752274
-2.41241241241241 0.0217355663683581
-2.39239239239239 0.0228065110187135
-2.37237237237237 0.0239206334123108
-2.35235235235235 0.0250791281140699
-2.33233233233233 0.0262831931598317
-2.31231231231231 0.0275340282581906
-2.29229229229229 0.0288328329022027
-2.27227227227227 0.0301808043912899
-2.25225225225225 0.0315791357639331
-2.23223223223223 0.033029013642035
-2.21221221221221 0.0345316159881232
-2.19219219219219 0.0360881097768736
-2.17217217217217 0.0376996485827434
-2.15215215215215 0.0393673700858293
-2.13213213213213 0.0410923934983949
-2.11211211211211 0.0428758169148479
-2.09209209209209 0.0447187145882915
-2.07207207207207 0.0466221341371235
-2.05205205205205 0.0485870936855041
-2.03203203203203 0.0506145789418747
-2.01201201201201 0.0527055402200587
-1.99199199199199 0.0548608894078376
-1.97197197197197 0.057081496888248
-1.95195195195195 0.059368188419199
-1.93193193193193 0.0617217419773594
-1.91191191191191 0.0641428845726061
-1.89189189189189 0.0666322890396674
-1.87187187187187 0.0691905708139176
-1.85185185185185 0.0718182846986055
-1.83183183183183 0.0745159216311036
-1.81181181181181 0.077283905456062
-1.79179179179179 0.0801225897136326
-1.77177177177177 0.083032254451193
-1.75175175175175 0.0860131030672496
-1.73173173173173 0.0890652591964251
-1.71171171171171 0.0921887636446459
-1.69169169169169 0.0953835713838294
-1.67167167167167 0.0986495486155338
-1.65165165165165 0.101986469913169
-1.63163163163163 0.10539401545248
-1.61161161161161 0.108871768340093
-1.59159159159159 0.112419212049971
-1.57157157157157 0.116035727977651
-1.55155155155155 0.119720593122119
-1.53153153153153 0.123472977905145
-1.51151151151151 0.127291944137829
-1.49149149149149 0.13117644314399
-1.47147147147147 0.135125314049902
-1.45145145145145 0.139137282249685
-1.43143143143143 0.143210958055468
-1.41141141141141 0.147344835541168
-1.39139139139139 0.151537291588457
-1.37137137137137 0.155786585143159
-1.35135135135135 0.160090856689972
-1.33133133133133 0.164448127952996
-1.31131131131131 0.168856301829129
-1.29129129129129 0.173313162560933
-1.27127127127127 0.17781637615506
-1.25125125125125 0.182363491051798
-1.23123123123123 0.186951939050736
-1.21121121121121 0.191579036496956
-1.19119119119119 0.1962419857315
-1.17117117117117 0.200937876809264
-1.15115115115115 0.205663689486728
-1.13113113113113 0.210416295481265
-1.11111111111111 0.215192461003031
-1.09109109109109 0.219988849559688
-1.07107107107107 0.224802025033432
-1.05105105105105 0.229628455029052
-1.03103103103103 0.234464514490888
-1.01101101101101 0.239306489585817
-0.990990990990991 0.24415058184851
-0.970970970970971 0.24899291258444
-0.950950950950951 0.253829527525259
-0.930930930930931 0.258656401730343
-0.910910910910911 0.2634694447275
-0.890890890890891 0.268264505884996
-0.870870870870871 0.273037380006279
-0.850850850850851 0.277783813137949
-0.830830830830831 0.282499508580786
-0.810810810810811 0.287180133092853
-0.790790790790791 0.291821323272996
-0.77077077077077 0.296418692112302
-0.75075075075075 0.300967835700437
-0.73073073073073 0.305464340073112
-0.71071071071071 0.309903788186304
-0.69069069069069 0.314281767002296
-0.67067067067067 0.318593874672039
-0.65065065065065 0.322835727797843
-0.63063063063063 0.327002968759958
-0.61061061061061 0.331091273090187
-0.59059059059059 0.33509635687531
-0.57057057057057 0.339013984172804
-0.55055055055055 0.34283997442106
-0.53053053053053 0.346570209826128
-0.51051051051051 0.350200642706842
-0.49049049049049 0.353727302780113
-0.47047047047047 0.357146304368113
-0.45045045045045 0.360453853509139
-0.43043043043043 0.363646254953996
-0.41041041041041 0.366719919029892
-0.39039039039039 0.369671368354051
-0.37037037037037 0.372497244379499
-0.35035035035035 0.375194313755802
-0.33033033033033 0.377759474487924
-0.31031031031031 0.38018976187679
-0.29029029029029 0.382482354225654
-0.27027027027027 0.384634578296894
-0.25025025025025 0.386643914504485
-0.23023023023023 0.388508001828027
-0.21021021021021 0.390224642434919
-0.19019019019019 0.391791805998011
-0.17017017017017 0.393207633696876
-0.15015015015015 0.394470441891644
-0.13013013013013 0.395578725459258
-0.11011011011011 0.396531160782876
-0.0900900900900901 0.397326608386124
-0.07007007007007 0.397964115204853
-0.05005005005005 0.398442916490068
-0.03003003003003 0.398762437336696
-0.01001001001001 0.398922293833933
0.01001001001001 0.398922293833933
0.03003003003003 0.398762437336696
0.05005005005005 0.398442916490068
0.07007007007007 0.397964115204853
0.0900900900900901 0.397326608386124
0.11011011011011 0.396531160782876
0.13013013013013 0.395578725459258
0.15015015015015 0.394470441891644
0.17017017017017 0.393207633696876
0.19019019019019 0.391791805998011
0.21021021021021 0.390224642434919
0.23023023023023 0.388508001828027
0.25025025025025 0.386643914504485
0.27027027027027 0.384634578296894
0.29029029029029 0.382482354225654
0.31031031031031 0.38018976187679
0.33033033033033 0.377759474487924
0.35035035035035 0.375194313755802
0.37037037037037 0.372497244379499
0.39039039039039 0.369671368354051
0.41041041041041 0.366719919029892
0.43043043043043 0.363646254953996
0.45045045045045 0.360453853509139
0.47047047047047 0.357146304368113
0.49049049049049 0.353727302780113
0.51051051051051 0.350200642706842
0.53053053053053 0.346570209826128
0.55055055055055 0.34283997442106
0.57057057057057 0.339013984172804
0.59059059059059 0.33509635687531
0.61061061061061 0.331091273090187
0.63063063063063 0.327002968759958
0.65065065065065 0.322835727797843
0.67067067067067 0.318593874672039
0.69069069069069 0.314281767002296
0.71071071071071 0.309903788186304
0.73073073073073 0.305464340073112
0.75075075075075 0.300967835700437
0.77077077077077 0.296418692112302
0.790790790790791 0.291821323272996
0.810810810810811 0.287180133092853
0.830830830830831 0.282499508580786
0.850850850850851 0.277783813137949
0.870870870870871 0.273037380006279
0.890890890890891 0.268264505884996
0.910910910910911 0.2634694447275
0.930930930930931 0.258656401730343
0.950950950950951 0.253829527525259
0.970970970970971 0.24899291258444
0.990990990990991 0.24415058184851
1.01101101101101 0.239306489585817
1.03103103103103 0.234464514490888
1.05105105105105 0.229628455029052
1.07107107107107 0.224802025033432
1.09109109109109 0.219988849559688
1.11111111111111 0.215192461003031
1.13113113113113 0.210416295481265
1.15115115115115 0.205663689486728
1.17117117117117 0.200937876809264
1.19119119119119 0.1962419857315
1.21121121121121 0.191579036496956
1.23123123123123 0.186951939050736
1.25125125125125 0.182363491051798
1.27127127127127 0.17781637615506
1.29129129129129 0.173313162560933
1.31131131131131 0.168856301829129
1.33133133133133 0.164448127952996
1.35135135135135 0.160090856689972
1.37137137137137 0.155786585143159
1.39139139139139 0.151537291588457
1.41141141141141 0.147344835541168
1.43143143143143 0.143210958055468
1.45145145145145 0.139137282249685
1.47147147147147 0.135125314049902
1.49149149149149 0.13117644314399
1.51151151151151 0.127291944137829
1.53153153153153 0.123472977905145
1.55155155155155 0.119720593122119
1.57157157157157 0.116035727977651
1.59159159159159 0.112419212049971
1.61161161161161 0.108871768340093
1.63163163163163 0.10539401545248
1.65165165165165 0.101986469913169
1.67167167167167 0.0986495486155338
1.69169169169169 0.0953835713838294
1.71171171171171 0.0921887636446459
1.73173173173173 0.0890652591964251
1.75175175175175 0.0860131030672496
1.77177177177177 0.083032254451193
1.79179179179179 0.0801225897136326
1.81181181181181 0.077283905456062
1.83183183183183 0.0745159216311036
1.85185185185185 0.0718182846986055
1.87187187187187 0.0691905708139176
1.89189189189189 0.0666322890396674
1.91191191191191 0.0641428845726061
1.93193193193193 0.0617217419773594
1.95195195195195 0.059368188419199
1.97197197197197 0.057081496888248
1.99199199199199 0.0548608894078376
2.01201201201201 0.0527055402200588
2.03203203203203 0.0506145789418748
2.05205205205205 0.0485870936855042
2.07207207207207 0.0466221341371236
2.09209209209209 0.0447187145882916
2.11211211211211 0.0428758169148479
2.13213213213213 0.041092393498395
2.15215215215215 0.0393673700858294
2.17217217217217 0.0376996485827434
2.19219219219219 0.0360881097768737
2.21221221221221 0.0345316159881233
2.23223223223223 0.033029013642035
2.25225225225225 0.0315791357639332
2.27227227227227 0.03018080439129
2.29229229229229 0.0288328329022028
2.31231231231231 0.0275340282581906
2.33233233233233 0.0262831931598317
2.35235235235235 0.02507912811407
2.37237237237237 0.0239206334123108
2.39239239239239 0.0228065110187136
2.41241241241241 0.0217355663683581
2.43243243243243 0.0207066100752275
2.45245245245245 0.0197184595501959
2.47247247247247 0.0187699405294511
2.49249249249249 0.0178598885140022
2.51251251251251 0.016987150121141
2.53253253253253 0.0161505843489182
2.55255255255255 0.0153490637548888
2.57257257257257 0.0145814755505493
2.59259259259259 0.0138467226130541
2.61261261261261 0.0131437244159435
2.63263263263263 0.0124714178807514
2.65265265265265 0.0118287581514852
2.67267267267267 0.0112147192940778
2.69269269269269 0.0106282949230103
2.71271271271271 0.0100684987573917
2.73273273273273 0.00953436510885491
2.75275275275275 0.00902494930369034
2.77277277277277 0.00853932804169479
2.79279279279279 0.00807659969425077
2.81281281281281 0.00763588454418634
2.83283283283283 0.00721632496998621
2.85285285285285 0.00681708557693871
2.87287287287287 0.00643735327780635
2.89289289289289 0.00607633732560524
2.91291291291291 0.00573326930106518
2.93293293293293 0.00540740305732384
2.95295295295295 0.00509801462438214
2.97297297297297 0.00480440207581539
2.99299299299299 0.00452588536019617
3.01301301301301 0.00426180609964127
3.03303303303303 0.00401152735784578
3.05305305305305 0.00377443337991421
3.07307307307307 0.00354992930624085
3.09309309309309 0.00333744086263051
3.11311311311311 0.00313641402878608
3.13313313313313 0.0029463146872226
3.15315315315315 0.00276662825459746
3.17317317317317 0.00259685929737496
3.19319319319319 0.00243653113367011
3.21321321321321 0.00228518542304203
3.23323323323323 0.00214238174593163
3.25325325325325 0.00200769717436225
3.27327327327327 0.00188072583544551
3.29329329329329 0.00176107846915771
3.31331331331331 0.00164838198177652
3.33333333333333 0.0015422789962911
3.35335335335335 0.00144242740102448
3.37337337337337 0.00134849989763212
3.39339339339339 0.00126018354956833
3.41341341341341 0.00117717933203981
3.43343343343343 0.00109920168439588
3.45345345345345 0.0010259780658361
3.47347347347347 0.000957248515249212
3.49349349349349 0.000892765215932441
3.51351351351351 0.000832292065877152
3.53353353353353 0.000775604254246008
3.55355355355355 0.000722487844607976
3.57357357357357 0.000672739365441019
3.59359359359359 0.000626165408357979
3.61361361361361 0.000582582234459159
3.63363363363363 0.000541815389165405
3.65365365365365 0.00050369932583812
3.67367367367367 0.000468077038447599
3.69369369369369 0.000434799703508357
3.71371371371371 0.000403726331459719
3.73373373373373 0.000374723427631836
3.75375375375375 0.000347664662901427
3.77377377377377 0.000322430554107926
3.79379379379379 0.000298908154269281
3.81381381381381 0.000276990752607344
3.83383383383383 0.000256577584365551
3.85385385385385 0.000237573550376443
3.87387387387387 0.000219888946313312
3.89389389389389 0.000203439201538965
3.91391391391391 0.00018814462744512
3.93393393393393 0.000173930175158222
3.95395395395395 0.000160725202471485
3.97397397397397 0.000148463249848567
3.99399399399399 0.000137081825331481
4.01401401401401 0.000126522198173995
4.03403403403403 0.000116729201011866
4.05405405405405 0.000107651040372646
4.07407407407407 9.92391153205018e-05
4.09409409409409 9.14478440253317e-05
4.11411411411411 8.42344980404937e-05
4.13413413413413 7.75590440694795e-05
4.15415415415415 7.13839929989176e-05
4.17417417417417 6.56742559732345e-05
4.19419419419419 6.03970072851107e-05
4.21421421421421 5.55215538554582e-05
4.23423423423423 5.10192110769697e-05
4.25425425425425 4.68631847962824e-05
4.27427427427427 4.3028459211397e-05
4.29429429429429 3.94916904631592e-05
4.31431431431431 3.62311057022691e-05
4.33433433433433 3.32264074164067e-05
4.35435435435435 3.04586828055866e-05
4.37437437437437 2.79103179977393e-05
4.39439439439439 2.55649169007242e-05
4.41441441441441 2.34072244914564e-05
4.43443443443443 2.14230543475548e-05
4.45445445445445 1.95992202318354e-05
4.47447447447447 1.79234715450684e-05
4.49449449449449 1.63844324676437e-05
4.51451451451451 1.49715446161227e-05
4.53453453453453 1.3675013046071e-05
4.55455455455455 1.24857554380297e-05
4.57457457457457 1.13953543089884e-05
4.59459459459459 1.03960120972233e-05
4.61461461461461 9.48050897386715e-06
4.63463463463463 8.64216324004121e-06
4.65465465465465 7.87479417380527e-06
4.67467467467467 7.17268719654363e-06
4.69469469469469 6.53056123369604e-06
4.71471471471471 5.94353814994721e-06
4.73473473473473 5.40711414409909e-06
4.75475475475475 4.91713299385613e-06
4.77477477477477 4.4697610456467e-06
4.79479479479479 4.06146384937932e-06
4.81481481481481 3.68898434268124e-06
4.83483483483483 3.34932249368796e-06
4.85485485485485 3.03971631583941e-06
4.87487487487487 2.75762417238901e-06
4.89489489489489 2.50070829244518e-06
4.91491491491491 2.26681942433715e-06
4.93493493493493 2.05398255592983e-06
4.95495495495495 1.86038363520377e-06
4.97497497497497 1.68435722796805e-06
4.99499499499499 1.5243750529858e-06
5.01501501501502 1.37903533806611e-06
5.03503503503504 1.24705294381396e-06
5.05505505505506 1.12725020473309e-06
5.07507507507508 1.01854844024876e-06
5.0950950950951 9.19960090959837e-07
5.11511511511512 8.30581438046261e-07
5.13513513513514 7.49585866251233e-07
5.15515515515516 6.76217633231267e-07
5.17517517517518 6.09786110324583e-07
5.1951951951952 5.4966046193278e-07
5.21521521521522 4.95264732746229e-07
5.23523523523524 4.46073313973501e-07
5.25525525525526 4.01606761563285e-07
5.27527527527528 3.61427941137511e-07
5.2952952952953 3.25138475990267e-07
5.31531531531532 2.92375476052561e-07
5.33533533533534 2.62808527181656e-07
5.35535535535536 2.36136921509202e-07
5.37537537537538 2.1208711087848e-07
5.3953953953954 1.90410366621162e-07
5.41541541541542 1.70880630071684e-07
5.43543543543544 1.5329253929596e-07
5.45545545545546 1.3745961852414e-07
5.47547547547548 1.23212617727566e-07
5.4954954954955 1.10397990671284e-07
5.51551551551552 9.8876500608364e-08
5.53553553553554 8.85219435638491e-08
5.55555555555556 7.92199798873018e-08
5.57557557557558 7.08670654362614e-08
5.5955955955956 6.33694743912418e-08
5.61561561561562 5.66424062986154e-08
5.63563563563564 5.06091704933412e-08
5.65565565565566 4.5200441571292e-08
5.67567567567568 4.03535800631662e-08
5.6956956956957 3.60120129107462e-08
5.71571571571572 3.2124668763613e-08
5.73573573573574 2.86454635022852e-08
5.75575575575576 2.55328317539416e-08
5.77577577577578 2.27493005011625e-08
5.7957957957958 2.02611011941336e-08
5.81581581581582 1.80378170640688e-08
5.83583583583584 1.60520626017053e-08
5.85585585585586 1.42791924110059e-08
5.87587587587588 1.2697036876002e-08
5.8958958958959 1.12856622892642e-08
5.91591591591592 1.00271532849868e-08
5.93593593593594 8.90541559921139e-09
5.95595595595596 7.90599734535251e-09
5.97597597597598 7.01592714588943e-09
5.995995995996 6.22356760178439e-09
6.01601601601602 5.5184827107339e-09
6.03603603603604 4.89131796456914e-09
6.05605605605606 4.33369196574694e-09
6.07607607607608 3.83809850362692e-09
6.0960960960961 3.3978181237675e-09
6.11611611611612 3.00683830841826e-09
6.13613613613614 2.65978146430924e-09
6.15615615615616 2.35183998527871e-09
6.17617617617618 2.07871772273778e-09
6.1961961961962 1.83657725691022e-09
6.21621621621622 1.6219924166386e-09
6.23623623623624 1.43190554571785e-09
6.25625625625626 1.26358905957511e-09
6.27627627627628 1.11461087800764e-09
6.2962962962963 9.82803357938119e-10
6.31631631631632 8.66235385045992e-10
6.33633633633634 7.63187314959515e-10
6.35635635635636 6.72128483698836e-10
6.37637637637638 5.91697033481532e-10
6.3963963963964 5.20681824054164e-10
6.41641641641642 4.58006221596989e-10
6.43643643643644 4.02713577147895e-10
6.45645645645646 3.53954224575805e-10
6.47647647647648 3.10973844559378e-10
6.4964964964965 2.73103055937435e-10
6.51651651651652 2.39748109325539e-10
6.53653653653654 2.10382570159619e-10
6.55655655655656 1.84539889444162e-10
6.57657657657658 1.61806770551276e-10
6.5965965965966 1.41817249531742e-10
6.61661661661662 1.24247414645767e-10
6.63663663663664 1.08810698278133e-10
6.65665665665666 9.52536811418137e-11
6.67667667667668 8.33523547614393e-11
6.6966966966967 7.29087937236022e-11
6.71671671671672 6.37481941394484e-11
6.73673673673674 5.57162392365998e-11
6.75675675675676 4.86767570277076e-11
6.77677677677678 4.25096386334907e-11
6.7967967967968 3.71089891068556e-11
6.81681681681682 3.23814855461048e-11
6.83683683683684 2.82449199306813e-11
6.85685685685686 2.46269064908966e-11
6.87687687687688 2.14637355595386e-11
6.8968968968969 1.86993577716892e-11
6.91691691691692 1.62844842008236e-11
6.93693693693694 1.41757895636779e-11
6.95695695695696 1.23352070109961e-11
6.97697697697698 1.07293042619725e-11
6.996996996997 9.32873195138548e-12
7.01701701701702 8.10773605306642e-12
7.03703703703704 7.04372713322415e-12
7.05705705705706 6.11689998287643e-12
7.07707707707708 5.30989788981512e-12
7.0970970970971 4.6075164458095e-12
7.11711711711712 3.99644235193917e-12
7.13713713713714 3.46502319108337e-12
7.15715715715716 3.00306458801651e-12
7.17717717717718 2.60165157997869e-12
7.1971971971972 2.25299137914217e-12
7.21721721721722 1.95027502769791e-12
7.23723723723724 1.68755573049414e-12
7.25725725725726 1.45964190299846e-12
7.27727727727728 1.262003197176e-12
7.2972972972973 1.0906879676822e-12
7.31731731731732 9.42250818252902e-13
7.33733733733734 8.13689025751829e-13
7.35735735735736 7.02386779168733e-13
7.37737737737738 6.06066294885245e-13
7.3973973973974 5.22744979473708e-13
7.41741741741742 4.50697908714381e-13
7.43743743743744 3.88424977793729e-13
7.45745745745746 3.34622154016963e-13
7.47747747747748 2.88156330935933e-13
7.4974974974975 2.48043342543511e-13
7.51751751751752 2.13428748996348e-13
7.53753753753754 1.83571051982055e-13
7.55755755755756 1.57827039041883e-13
7.57757757757758 1.35638992516663e-13
7.5975975975976 1.16523530854698e-13
7.61761761761762 1.000618782966e-13
7.63763763763764 8.58913838706873e-14
7.65765765765766 7.36981325813112e-14
7.67767767767768 6.32105109959248e-14
7.6976976976977 5.41936064405376e-14
7.71771771771772 4.64443339685915e-14
7.73773773773774 3.97871984155707e-14
7.75775775775776 3.40706104038536e-14
7.77777777777778 2.91636853079907e-14
7.7977977977978 2.4953463096687e-14
7.81781781781782 2.13424947819474e-14
7.83783783783784 1.82467480586654e-14
7.85785785785786 1.55937907247682e-14
7.87787787787788 1.33212157347804e-14
7.8978978978979 1.13752763482777e-14
7.91791791791792 9.70970386854701e-15
7.93793793793794 8.28468399583068e-15
7.95795795795796 7.06597090549298e-15
7.97797797797798 6.02412085866189e-15
7.997997997998 5.13382950919603e-15
8.01801801801802 4.3733591283238e-15
8.03803803803804 3.72404376402735e-15
8.05805805805806 3.16986191875719e-15
8.07807807807808 2.69706769497134e-15
8.0980980980981 2.29387254841619e-15
8.11811811811812 1.95017082606148e-15
8.13813813813814 1.65730316851105e-15
8.15815815815816 1.40785264250169e-15
8.17817817817818 1.19546915264237e-15
8.1981981981982 1.01471827585831e-15
8.21821821821822 8.60951178494166e-16
8.23823823823824 7.30192724674871e-16
8.25825825825826 6.19045274051723e-16
8.27827827827828 5.24606005103494e-16
8.2982982982983 4.44395893386326e-16
8.31831831831832 3.76298728354137e-16
8.33833833833834 3.18508772685576e-16
8.35835835835836 2.69485858889318e-16
8.37837837837838 2.27916883183252e-16
8.3983983983984 1.92682799625235e-16
8.41841841841842 1.62830341150177e-16
8.43843843843844 1.37547801096493e-16
8.45845845845846 1.16144301209627e-16
8.47847847847848 9.8032051926925e-17
8.4984984984985 8.27111796592674e-17
8.51851851851852 6.97567552529606e-17
8.53853853853854 5.88077091106193e-17
8.55855855855856 4.95573626747691e-17
8.57857857857858 4.17453440895294e-17
8.5985985985986 3.51506886838449e-17
8.61861861861862 2.95859531836935e-17
8.63863863863864 2.48921968838275e-17
8.65865865865866 2.09347039316773e-17
8.67867867867868 1.75993388643364e-17
8.6986986986987 1.47894429982973e-17
8.71871871871872 1.24231925503958e-17
8.73873873873874 1.04313507694241e-17
8.75875875875876 8.75535614211525e-18
8.77877877877878 7.34569713009337e-18
8.7987987987988 6.16053109048305e-18
8.81881881881882 5.16451119999019e-18
8.83883883883884 4.32779048512422e-18
8.85885885885886 3.62517658454128e-18
8.87887887887888 3.03541474067764e-18
8.8988988988989 2.54057982941549e-18
8.91891891891892 2.12556106807901e-18
8.93893893893894 1.77762546207239e-18
8.95895895895896 1.48604811781133e-18
8.97897897897898 1.24179931485728e-18
8.998998998999 1.03727973679138e-18
9.01901901901902 8.66096545670873e-19
9.03903903903904 7.22874080905507e-19
9.05905905905906 6.03093897535385e-19
9.07907907907908 5.0295965472299e-19
9.0990990990991 4.19283042962206e-19
9.11911911911912 3.49387515332417e-19
9.13913913913914 2.91027078875695e-19
9.15915915915916 2.42317819504618e-19
9.17917917917918 2.01680188581445e-19
9.1991991991992 1.67790380698338e-19
9.21921921921922 1.39539388139188e-19
9.23923923923924 1.1599853476858e-19
9.25925925925926 9.63904764362469e-20
9.27927927927928 8.00648113242436e-20
9.2992992992993 6.6477576193283e-20
9.31931931931932 5.51740167796855e-20
9.33933933933934 4.5774115701642e-20
9.35935935935936 3.79604417474457e-20
9.37937937937938 3.14679525475421e-20
9.3993993993994 2.60754402558043e-20
9.41941941941942 2.15983585814365e-20
9.43943943943944 1.78828106797973e-20
9.45945945945946 1.48005121824234e-20
9.47947947947948 1.22445730038445e-20
9.4994994994995 1.01259663374544e-20
9.51951951951952 8.37057415073758e-21
9.53953953953954 6.91671611023779e-21
9.55955955955956 5.71308371631411e-21
9.57957957957958 4.71701393695898e-21
9.5995995995996 3.89304716291232e-21
9.61961961961962 3.21172317125736e-21
9.63963963963964 2.64857624243904e-21
9.65965965965966 2.18329684678274e-21
9.67967967967968 1.79903258756111e-21
9.6996996996997 1.48180551599987e-21
9.71971971971972 1.22002665237565e-21
9.73973973973974 1.00409166885045e-21
9.75975975975976 8.26044308669654e-22
9.77977977977978 6.79296312742742e-22
9.7997997997998 5.58394465795474e-22
9.81981981981982 4.58826916995414e-22
9.83983983983984 3.7686222201397e-22
9.85985985985986 3.09415635142992e-22
9.87987987987988 2.53938085193762e-22
9.8998998998999 2.08324025950642e-22
9.91991991991992 1.70834984871876e-22
9.93993993993994 1.40036162642795e-22
9.95995995995996 1.14743877987917e-22
9.97997997997998 9.39820210218911e-23
10 7.69459862670642e-23
};
\addlegendentry{N(0,1)}; 
\legend{}; 
\end{axis}

\end{tikzpicture}

%% file: FinalFigs/Null_Dists_d_10_500_n_200_m_20_kernel__Gaussian_Poly_2_2022_10_12_23_28_39cross.tex
\begin{tikzpicture}

\definecolor{darkorange25512714}{RGB}{255,127,14}
\definecolor{darkslategray38}{RGB}{38,38,38}
\definecolor{lightgray204}{RGB}{204,204,204}
\definecolor{steelblue31119180}{RGB}{31,119,180}

\begin{axis}[
axis line style={darkslategray38},
height=\figheight,
legend cell align={left},
legend style={fill opacity=0.8, draw opacity=1, text opacity=1, draw=none},
tick align=outside,
tick pos=left,
title={$\cmmd~(n/m=10)$},
width=\figwidth,
x grid style={lightgray204},
xmin=-6, xmax=6,
xtick style={color=darkslategray38},
y grid style={lightgray204},
ylabel = {}, 
ymin=0, ymax=0.418868408525629,
ytick style={color=darkslategray38}, 
xticklabels=empty,
yticklabels=empty
]
\draw[draw=none,fill=steelblue31119180,fill opacity=0.8] (axis cs:-4.04229593276978,0) rectangle (axis cs:-3.70769119262695,0.00478176267120867);
\addlegendimage{ybar,ybar legend,draw=none,fill=steelblue31119180,fill opacity=0.8}
\addlegendentry{d=10}

\draw[draw=none,fill=steelblue31119180,fill opacity=0.8] (axis cs:-3.2057843208313,0) rectangle (axis cs:-2.87117958068848,0.0191270506848347);
\draw[draw=none,fill=steelblue31119180,fill opacity=0.8] (axis cs:-2.36927270889282,0) rectangle (axis cs:-2.03466820716858,0.0788990953186664);
\draw[draw=none,fill=steelblue31119180,fill opacity=0.8] (axis cs:-1.53276109695435,0) rectangle (axis cs:-1.1981565952301,0.205615809512884);
\draw[draw=none,fill=steelblue31119180,fill opacity=0.8] (axis cs:-0.696249485015869,0) rectangle (axis cs:-0.361644983291626,0.317987240293181);
\draw[draw=none,fill=steelblue31119180,fill opacity=0.8] (axis cs:0.14026203751564,0) rectangle (axis cs:0.474866539239883,0.337114292340891);
\draw[draw=none,fill=steelblue31119180,fill opacity=0.8] (axis cs:0.976773619651794,0) rectangle (axis cs:1.31137812137604,0.162579942405536);
\draw[draw=none,fill=steelblue31119180,fill opacity=0.8] (axis cs:1.81328511238098,0) rectangle (axis cs:2.14788961410522,0.0573811602317573);
\draw[draw=none,fill=steelblue31119180,fill opacity=0.8] (axis cs:2.64979648590088,0) rectangle (axis cs:2.9844012260437,0.00717264400681301);
\draw[draw=none,fill=steelblue31119180,fill opacity=0.8] (axis cs:3.48630809783936,0) rectangle (axis cs:3.82091283798218,0.00478176267120867);
\draw[draw=none,fill=darkorange25512714,fill opacity=0.8] (axis cs:-3.70769119262695,0) rectangle (axis cs:-3.37308645248413,0.0191270506848347);
\addlegendimage{ybar,ybar legend,draw=none,fill=darkorange25512714,fill opacity=0.8}
\addlegendentry{d=500}

\draw[draw=none,fill=darkorange25512714,fill opacity=0.8] (axis cs:-2.87117958068848,0) rectangle (axis cs:-2.53657484054565,0.0167361693492303);
\draw[draw=none,fill=darkorange25512714,fill opacity=0.8] (axis cs:-2.03466796875,0) rectangle (axis cs:-1.70006346702576,0.0812899769949896);
\draw[draw=none,fill=darkorange25512714,fill opacity=0.8] (axis cs:-1.19815647602081,0) rectangle (axis cs:-0.86355197429657,0.224742861560594);
\draw[draw=none,fill=darkorange25512714,fill opacity=0.8] (axis cs:-0.361644923686981,0) rectangle (axis cs:-0.027040421962738,0.298860188245471);
\draw[draw=none,fill=darkorange25512714,fill opacity=0.8] (axis cs:0.474866628646851,0) rectangle (axis cs:0.809471130371094,0.339505173846855);
\draw[draw=none,fill=darkorange25512714,fill opacity=0.8] (axis cs:1.31137812137604,0) rectangle (axis cs:1.64598262310028,0.179316112947283);
\draw[draw=none,fill=darkorange25512714,fill opacity=0.8] (axis cs:2.14788961410522,0) rectangle (axis cs:2.48249411582947,0.0310814617922019);
\draw[draw=none,fill=darkorange25512714,fill opacity=0.8] (axis cs:2.9844012260437,0) rectangle (axis cs:3.31900596618652,0.00478176267120867);
\draw[draw=none,fill=darkorange25512714,fill opacity=0.8] (axis cs:3.82091283798218,0) rectangle (axis cs:4.155517578125,0);
\addplot [semithick, black]
table {%
-10 7.69459862670642e-23
-9.97997997997998 9.39820210218911e-23
-9.95995995995996 1.14743877987917e-22
-9.93993993993994 1.40036162642795e-22
-9.91991991991992 1.70834984871876e-22
-9.8998998998999 2.08324025950642e-22
-9.87987987987988 2.53938085193762e-22
-9.85985985985986 3.09415635142992e-22
-9.83983983983984 3.7686222201397e-22
-9.81981981981982 4.58826916995414e-22
-9.7997997997998 5.58394465795474e-22
-9.77977977977978 6.79296312742742e-22
-9.75975975975976 8.26044308669654e-22
-9.73973973973974 1.00409166885045e-21
-9.71971971971972 1.22002665237565e-21
-9.6996996996997 1.48180551599987e-21
-9.67967967967968 1.79903258756111e-21
-9.65965965965966 2.18329684678274e-21
-9.63963963963964 2.64857624243904e-21
-9.61961961961962 3.21172317125736e-21
-9.5995995995996 3.89304716291232e-21
-9.57957957957958 4.71701393695898e-21
-9.55955955955956 5.71308371631411e-21
-9.53953953953954 6.91671611023779e-21
-9.51951951951952 8.37057415073758e-21
-9.4994994994995 1.01259663374544e-20
-9.47947947947948 1.22445730038445e-20
-9.45945945945946 1.48005121824234e-20
-9.43943943943944 1.78828106797973e-20
-9.41941941941942 2.15983585814365e-20
-9.3993993993994 2.60754402558043e-20
-9.37937937937938 3.14679525475421e-20
-9.35935935935936 3.79604417474457e-20
-9.33933933933934 4.5774115701642e-20
-9.31931931931932 5.51740167796855e-20
-9.2992992992993 6.6477576193283e-20
-9.27927927927928 8.00648113242436e-20
-9.25925925925926 9.63904764362469e-20
-9.23923923923924 1.1599853476858e-19
-9.21921921921922 1.39539388139188e-19
-9.1991991991992 1.67790380698338e-19
-9.17917917917918 2.01680188581445e-19
-9.15915915915916 2.42317819504618e-19
-9.13913913913914 2.91027078875695e-19
-9.11911911911912 3.49387515332417e-19
-9.0990990990991 4.19283042962206e-19
-9.07907907907908 5.0295965472299e-19
-9.05905905905906 6.03093897535385e-19
-9.03903903903904 7.22874080905507e-19
-9.01901901901902 8.66096545670873e-19
-8.998998998999 1.03727973679138e-18
-8.97897897897898 1.24179931485728e-18
-8.95895895895896 1.48604811781133e-18
-8.93893893893894 1.77762546207239e-18
-8.91891891891892 2.12556106807901e-18
-8.8988988988989 2.54057982941549e-18
-8.87887887887888 3.03541474067764e-18
-8.85885885885886 3.62517658454128e-18
-8.83883883883884 4.32779048512422e-18
-8.81881881881882 5.16451119999019e-18
-8.7987987987988 6.16053109048305e-18
-8.77877877877878 7.34569713009337e-18
-8.75875875875876 8.75535614211525e-18
-8.73873873873874 1.04313507694241e-17
-8.71871871871872 1.24231925503958e-17
-8.6986986986987 1.47894429982973e-17
-8.67867867867868 1.75993388643364e-17
-8.65865865865866 2.09347039316773e-17
-8.63863863863864 2.48921968838275e-17
-8.61861861861862 2.95859531836935e-17
-8.5985985985986 3.51506886838449e-17
-8.57857857857858 4.17453440895294e-17
-8.55855855855856 4.95573626747691e-17
-8.53853853853854 5.88077091106193e-17
-8.51851851851852 6.97567552529606e-17
-8.4984984984985 8.27111796592674e-17
-8.47847847847848 9.8032051926925e-17
-8.45845845845846 1.16144301209627e-16
-8.43843843843844 1.37547801096493e-16
-8.41841841841842 1.62830341150177e-16
-8.3983983983984 1.92682799625235e-16
-8.37837837837838 2.27916883183252e-16
-8.35835835835836 2.69485858889318e-16
-8.33833833833834 3.18508772685576e-16
-8.31831831831832 3.76298728354137e-16
-8.2982982982983 4.44395893386326e-16
-8.27827827827828 5.24606005103494e-16
-8.25825825825826 6.19045274051723e-16
-8.23823823823824 7.30192724674871e-16
-8.21821821821822 8.60951178494166e-16
-8.1981981981982 1.01471827585831e-15
-8.17817817817818 1.19546915264237e-15
-8.15815815815816 1.40785264250169e-15
-8.13813813813814 1.65730316851105e-15
-8.11811811811812 1.95017082606148e-15
-8.0980980980981 2.29387254841619e-15
-8.07807807807808 2.69706769497134e-15
-8.05805805805806 3.16986191875719e-15
-8.03803803803804 3.72404376402735e-15
-8.01801801801802 4.3733591283238e-15
-7.997997997998 5.13382950919607e-15
-7.97797797797798 6.02412085866193e-15
-7.95795795795796 7.06597090549303e-15
-7.93793793793794 8.28468399583074e-15
-7.91791791791792 9.70970386854708e-15
-7.8978978978979 1.13752763482777e-14
-7.87787787787788 1.33212157347805e-14
-7.85785785785786 1.55937907247683e-14
-7.83783783783784 1.82467480586655e-14
-7.81781781781782 2.13424947819475e-14
-7.7977977977978 2.49534630966872e-14
-7.77777777777778 2.91636853079909e-14
-7.75775775775776 3.40706104038538e-14
-7.73773773773774 3.9787198415571e-14
-7.71771771771772 4.64443339685918e-14
-7.6976976976977 5.4193606440538e-14
-7.67767767767768 6.32105109959252e-14
-7.65765765765766 7.36981325813117e-14
-7.63763763763764 8.58913838706879e-14
-7.61761761761762 1.00061878296601e-13
-7.5975975975976 1.16523530854699e-13
-7.57757757757758 1.35638992516664e-13
-7.55755755755756 1.57827039041884e-13
-7.53753753753754 1.83571051982057e-13
-7.51751751751752 2.13428748996349e-13
-7.4974974974975 2.48043342543513e-13
-7.47747747747748 2.88156330935935e-13
-7.45745745745746 3.34622154016965e-13
-7.43743743743744 3.88424977793732e-13
-7.41741741741742 4.50697908714384e-13
-7.3973973973974 5.22744979473711e-13
-7.37737737737738 6.06066294885249e-13
-7.35735735735736 7.02386779168738e-13
-7.33733733733734 8.13689025751835e-13
-7.31731731731732 9.42250818252909e-13
-7.2972972972973 1.09068796768221e-12
-7.27727727727728 1.262003197176e-12
-7.25725725725726 1.45964190299847e-12
-7.23723723723724 1.68755573049416e-12
-7.21721721721722 1.95027502769792e-12
-7.1971971971972 2.25299137914218e-12
-7.17717717717718 2.60165157997871e-12
-7.15715715715716 3.00306458801653e-12
-7.13713713713714 3.4650231910834e-12
-7.11711711711712 3.99644235193919e-12
-7.0970970970971 4.60751644580953e-12
-7.07707707707708 5.30989788981514e-12
-7.05705705705706 6.11689998287646e-12
-7.03703703703704 7.0437271332242e-12
-7.01701701701702 8.10773605306645e-12
-6.996996996997 9.32873195138555e-12
-6.97697697697698 1.07293042619726e-11
-6.95695695695696 1.23352070109962e-11
-6.93693693693694 1.41757895636779e-11
-6.91691691691692 1.62844842008237e-11
-6.8968968968969 1.86993577716893e-11
-6.87687687687688 2.14637355595386e-11
-6.85685685685686 2.46269064908967e-11
-6.83683683683684 2.82449199306815e-11
-6.81681681681682 3.2381485546105e-11
-6.7967967967968 3.71089891068559e-11
-6.77677677677678 4.25096386334913e-11
-6.75675675675676 4.86767570277083e-11
-6.73673673673674 5.57162392366004e-11
-6.71671671671672 6.37481941394491e-11
-6.6966966966967 7.29087937236032e-11
-6.67667667667668 8.33523547614402e-11
-6.65665665665666 9.52536811418151e-11
-6.63663663663664 1.08810698278135e-10
-6.61661661661662 1.24247414645768e-10
-6.5965965965966 1.41817249531744e-10
-6.57657657657658 1.61806770551278e-10
-6.55655655655656 1.84539889444164e-10
-6.53653653653654 2.10382570159622e-10
-6.51651651651652 2.39748109325542e-10
-6.4964964964965 2.73103055937438e-10
-6.47647647647648 3.10973844559381e-10
-6.45645645645646 3.53954224575809e-10
-6.43643643643644 4.027135771479e-10
-6.41641641641642 4.58006221596996e-10
-6.3963963963964 5.20681824054169e-10
-6.37637637637638 5.91697033481538e-10
-6.35635635635636 6.72128483698846e-10
-6.33633633633634 7.63187314959523e-10
-6.31631631631632 8.66235385046001e-10
-6.2962962962963 9.8280335793813e-10
-6.27627627627628 1.11461087800766e-09
-6.25625625625626 1.26358905957513e-09
-6.23623623623624 1.43190554571787e-09
-6.21621621621622 1.62199241663862e-09
-6.1961961961962 1.83657725691024e-09
-6.17617617617618 2.0787177227378e-09
-6.15615615615616 2.35183998527873e-09
-6.13613613613614 2.65978146430928e-09
-6.11611611611612 3.00683830841829e-09
-6.0960960960961 3.39781812376754e-09
-6.07607607607608 3.83809850362696e-09
-6.05605605605606 4.33369196574699e-09
-6.03603603603604 4.89131796456919e-09
-6.01601601601602 5.51848271073395e-09
-5.995995995996 6.22356760178439e-09
-5.97597597597598 7.01592714588943e-09
-5.95595595595596 7.90599734535251e-09
-5.93593593593594 8.90541559921139e-09
-5.91591591591592 1.00271532849868e-08
-5.8958958958959 1.12856622892642e-08
-5.87587587587588 1.2697036876002e-08
-5.85585585585586 1.42791924110059e-08
-5.83583583583584 1.60520626017053e-08
-5.81581581581582 1.80378170640688e-08
-5.7957957957958 2.02611011941336e-08
-5.77577577577578 2.27493005011625e-08
-5.75575575575576 2.55328317539416e-08
-5.73573573573574 2.86454635022852e-08
-5.71571571571572 3.2124668763613e-08
-5.6956956956957 3.60120129107462e-08
-5.67567567567568 4.03535800631662e-08
-5.65565565565566 4.5200441571292e-08
-5.63563563563564 5.06091704933412e-08
-5.61561561561562 5.66424062986154e-08
-5.5955955955956 6.33694743912418e-08
-5.57557557557558 7.08670654362614e-08
-5.55555555555556 7.92199798873018e-08
-5.53553553553554 8.85219435638491e-08
-5.51551551551552 9.8876500608364e-08
-5.4954954954955 1.10397990671284e-07
-5.47547547547548 1.23212617727566e-07
-5.45545545545546 1.3745961852414e-07
-5.43543543543544 1.5329253929596e-07
-5.41541541541542 1.70880630071684e-07
-5.3953953953954 1.90410366621162e-07
-5.37537537537538 2.1208711087848e-07
-5.35535535535536 2.36136921509202e-07
-5.33533533533534 2.62808527181656e-07
-5.31531531531532 2.92375476052561e-07
-5.2952952952953 3.25138475990267e-07
-5.27527527527528 3.61427941137511e-07
-5.25525525525526 4.01606761563285e-07
-5.23523523523524 4.46073313973501e-07
-5.21521521521522 4.95264732746229e-07
-5.1951951951952 5.4966046193278e-07
-5.17517517517518 6.09786110324583e-07
-5.15515515515516 6.76217633231267e-07
-5.13513513513514 7.49585866251233e-07
-5.11511511511512 8.30581438046261e-07
-5.0950950950951 9.19960090959837e-07
-5.07507507507508 1.01854844024876e-06
-5.05505505505506 1.12725020473309e-06
-5.03503503503504 1.24705294381396e-06
-5.01501501501502 1.37903533806611e-06
-4.99499499499499 1.5243750529858e-06
-4.97497497497497 1.68435722796805e-06
-4.95495495495495 1.86038363520377e-06
-4.93493493493493 2.05398255592983e-06
-4.91491491491491 2.26681942433715e-06
-4.89489489489489 2.50070829244518e-06
-4.87487487487487 2.75762417238901e-06
-4.85485485485485 3.03971631583941e-06
-4.83483483483483 3.34932249368796e-06
-4.81481481481481 3.68898434268124e-06
-4.79479479479479 4.06146384937932e-06
-4.77477477477477 4.4697610456467e-06
-4.75475475475475 4.91713299385613e-06
-4.73473473473473 5.40711414409909e-06
-4.71471471471471 5.94353814994721e-06
-4.69469469469469 6.53056123369604e-06
-4.67467467467467 7.17268719654363e-06
-4.65465465465465 7.87479417380527e-06
-4.63463463463463 8.64216324004121e-06
-4.61461461461461 9.48050897386715e-06
-4.59459459459459 1.03960120972233e-05
-4.57457457457457 1.13953543089884e-05
-4.55455455455455 1.24857554380297e-05
-4.53453453453453 1.3675013046071e-05
-4.51451451451451 1.49715446161227e-05
-4.49449449449449 1.63844324676437e-05
-4.47447447447447 1.79234715450684e-05
-4.45445445445445 1.95992202318354e-05
-4.43443443443443 2.14230543475548e-05
-4.41441441441441 2.34072244914564e-05
-4.39439439439439 2.55649169007242e-05
-4.37437437437437 2.79103179977393e-05
-4.35435435435435 3.04586828055866e-05
-4.33433433433433 3.32264074164067e-05
-4.31431431431431 3.62311057022691e-05
-4.29429429429429 3.94916904631592e-05
-4.27427427427427 4.3028459211397e-05
-4.25425425425425 4.68631847962824e-05
-4.23423423423423 5.10192110769697e-05
-4.21421421421421 5.55215538554582e-05
-4.19419419419419 6.03970072851107e-05
-4.17417417417417 6.56742559732345e-05
-4.15415415415415 7.13839929989176e-05
-4.13413413413413 7.75590440694795e-05
-4.11411411411411 8.42344980404937e-05
-4.09409409409409 9.14478440253317e-05
-4.07407407407407 9.92391153205018e-05
-4.05405405405405 0.000107651040372646
-4.03403403403403 0.000116729201011866
-4.01401401401401 0.000126522198173995
-3.99399399399399 0.000137081825331481
-3.97397397397397 0.000148463249848567
-3.95395395395395 0.000160725202471485
-3.93393393393393 0.000173930175158222
-3.91391391391391 0.00018814462744512
-3.89389389389389 0.000203439201538965
-3.87387387387387 0.000219888946313312
-3.85385385385385 0.000237573550376443
-3.83383383383383 0.000256577584365551
-3.81381381381381 0.000276990752607344
-3.79379379379379 0.000298908154269281
-3.77377377377377 0.000322430554107926
-3.75375375375375 0.000347664662901427
-3.73373373373373 0.000374723427631836
-3.71371371371371 0.000403726331459719
-3.69369369369369 0.000434799703508357
-3.67367367367367 0.000468077038447599
-3.65365365365365 0.00050369932583812
-3.63363363363363 0.000541815389165405
-3.61361361361361 0.000582582234459159
-3.59359359359359 0.000626165408357979
-3.57357357357357 0.000672739365441021
-3.55355355355355 0.000722487844607978
-3.53353353353353 0.00077560425424601
-3.51351351351351 0.000832292065877155
-3.49349349349349 0.000892765215932443
-3.47347347347347 0.000957248515249216
-3.45345345345345 0.0010259780658361
-3.43343343343343 0.00109920168439588
-3.41341341341341 0.00117717933203981
-3.39339339339339 0.00126018354956833
-3.37337337337337 0.00134849989763212
-3.35335335335335 0.00144242740102448
-3.33333333333333 0.00154227899629111
-3.31331331331331 0.00164838198177652
-3.29329329329329 0.00176107846915772
-3.27327327327327 0.00188072583544552
-3.25325325325325 0.00200769717436226
-3.23323323323323 0.00214238174593163
-3.21321321321321 0.00228518542304204
-3.19319319319319 0.00243653113367012
-3.17317317317317 0.00259685929737497
-3.15315315315315 0.00276662825459747
-3.13313313313313 0.00294631468722261
-3.11311311311311 0.00313641402878609
-3.09309309309309 0.00333744086263052
-3.07307307307307 0.00354992930624086
-3.05305305305305 0.00377443337991422
-3.03303303303303 0.00401152735784579
-3.01301301301301 0.00426180609964128
-2.99299299299299 0.00452588536019618
-2.97297297297297 0.0048044020758154
-2.95295295295295 0.00509801462438215
-2.93293293293293 0.00540740305732385
-2.91291291291291 0.00573326930106519
-2.89289289289289 0.00607633732560526
-2.87287287287287 0.00643735327780636
-2.85285285285285 0.00681708557693873
-2.83283283283283 0.00721632496998623
-2.81281281281281 0.00763588454418632
-2.79279279279279 0.00807659969425075
-2.77277277277277 0.00853932804169477
-2.75275275275275 0.00902494930369032
-2.73273273273273 0.00953436510885489
-2.71271271271271 0.0100684987573917
-2.69269269269269 0.0106282949230102
-2.67267267267267 0.0112147192940778
-2.65265265265265 0.0118287581514852
-2.63263263263263 0.0124714178807513
-2.61261261261261 0.0131437244159435
-2.59259259259259 0.0138467226130541
-2.57257257257257 0.0145814755505492
-2.55255255255255 0.0153490637548887
-2.53253253253253 0.0161505843489182
-2.51251251251251 0.0169871501211409
-2.49249249249249 0.0178598885140022
-2.47247247247247 0.018769940529451
-2.45245245245245 0.0197184595501959
-2.43243243243243 0.0207066100752274
-2.41241241241241 0.0217355663683581
-2.39239239239239 0.0228065110187135
-2.37237237237237 0.0239206334123108
-2.35235235235235 0.0250791281140699
-2.33233233233233 0.0262831931598317
-2.31231231231231 0.0275340282581906
-2.29229229229229 0.0288328329022027
-2.27227227227227 0.0301808043912899
-2.25225225225225 0.0315791357639331
-2.23223223223223 0.033029013642035
-2.21221221221221 0.0345316159881232
-2.19219219219219 0.0360881097768736
-2.17217217217217 0.0376996485827434
-2.15215215215215 0.0393673700858293
-2.13213213213213 0.0410923934983949
-2.11211211211211 0.0428758169148479
-2.09209209209209 0.0447187145882915
-2.07207207207207 0.0466221341371235
-2.05205205205205 0.0485870936855041
-2.03203203203203 0.0506145789418747
-2.01201201201201 0.0527055402200587
-1.99199199199199 0.0548608894078376
-1.97197197197197 0.057081496888248
-1.95195195195195 0.059368188419199
-1.93193193193193 0.0617217419773594
-1.91191191191191 0.0641428845726061
-1.89189189189189 0.0666322890396674
-1.87187187187187 0.0691905708139176
-1.85185185185185 0.0718182846986055
-1.83183183183183 0.0745159216311036
-1.81181181181181 0.077283905456062
-1.79179179179179 0.0801225897136326
-1.77177177177177 0.083032254451193
-1.75175175175175 0.0860131030672496
-1.73173173173173 0.0890652591964251
-1.71171171171171 0.0921887636446459
-1.69169169169169 0.0953835713838294
-1.67167167167167 0.0986495486155338
-1.65165165165165 0.101986469913169
-1.63163163163163 0.10539401545248
-1.61161161161161 0.108871768340093
-1.59159159159159 0.112419212049971
-1.57157157157157 0.116035727977651
-1.55155155155155 0.119720593122119
-1.53153153153153 0.123472977905145
-1.51151151151151 0.127291944137829
-1.49149149149149 0.13117644314399
-1.47147147147147 0.135125314049902
-1.45145145145145 0.139137282249685
-1.43143143143143 0.143210958055468
-1.41141141141141 0.147344835541168
-1.39139139139139 0.151537291588457
-1.37137137137137 0.155786585143159
-1.35135135135135 0.160090856689972
-1.33133133133133 0.164448127952996
-1.31131131131131 0.168856301829129
-1.29129129129129 0.173313162560933
-1.27127127127127 0.17781637615506
-1.25125125125125 0.182363491051798
-1.23123123123123 0.186951939050736
-1.21121121121121 0.191579036496956
-1.19119119119119 0.1962419857315
-1.17117117117117 0.200937876809264
-1.15115115115115 0.205663689486728
-1.13113113113113 0.210416295481265
-1.11111111111111 0.215192461003031
-1.09109109109109 0.219988849559688
-1.07107107107107 0.224802025033432
-1.05105105105105 0.229628455029052
-1.03103103103103 0.234464514490888
-1.01101101101101 0.239306489585817
-0.990990990990991 0.24415058184851
-0.970970970970971 0.24899291258444
-0.950950950950951 0.253829527525259
-0.930930930930931 0.258656401730343
-0.910910910910911 0.2634694447275
-0.890890890890891 0.268264505884996
-0.870870870870871 0.273037380006279
-0.850850850850851 0.277783813137949
-0.830830830830831 0.282499508580786
-0.810810810810811 0.287180133092853
-0.790790790790791 0.291821323272996
-0.77077077077077 0.296418692112302
-0.75075075075075 0.300967835700437
-0.73073073073073 0.305464340073112
-0.71071071071071 0.309903788186304
-0.69069069069069 0.314281767002296
-0.67067067067067 0.318593874672039
-0.65065065065065 0.322835727797843
-0.63063063063063 0.327002968759958
-0.61061061061061 0.331091273090187
-0.59059059059059 0.33509635687531
-0.57057057057057 0.339013984172804
-0.55055055055055 0.34283997442106
-0.53053053053053 0.346570209826128
-0.51051051051051 0.350200642706842
-0.49049049049049 0.353727302780113
-0.47047047047047 0.357146304368113
-0.45045045045045 0.360453853509139
-0.43043043043043 0.363646254953996
-0.41041041041041 0.366719919029892
-0.39039039039039 0.369671368354051
-0.37037037037037 0.372497244379499
-0.35035035035035 0.375194313755802
-0.33033033033033 0.377759474487924
-0.31031031031031 0.38018976187679
-0.29029029029029 0.382482354225654
-0.27027027027027 0.384634578296894
-0.25025025025025 0.386643914504485
-0.23023023023023 0.388508001828027
-0.21021021021021 0.390224642434919
-0.19019019019019 0.391791805998011
-0.17017017017017 0.393207633696876
-0.15015015015015 0.394470441891644
-0.13013013013013 0.395578725459258
-0.11011011011011 0.396531160782876
-0.0900900900900901 0.397326608386124
-0.07007007007007 0.397964115204853
-0.05005005005005 0.398442916490068
-0.03003003003003 0.398762437336696
-0.01001001001001 0.398922293833933
0.01001001001001 0.398922293833933
0.03003003003003 0.398762437336696
0.05005005005005 0.398442916490068
0.07007007007007 0.397964115204853
0.0900900900900901 0.397326608386124
0.11011011011011 0.396531160782876
0.13013013013013 0.395578725459258
0.15015015015015 0.394470441891644
0.17017017017017 0.393207633696876
0.19019019019019 0.391791805998011
0.21021021021021 0.390224642434919
0.23023023023023 0.388508001828027
0.25025025025025 0.386643914504485
0.27027027027027 0.384634578296894
0.29029029029029 0.382482354225654
0.31031031031031 0.38018976187679
0.33033033033033 0.377759474487924
0.35035035035035 0.375194313755802
0.37037037037037 0.372497244379499
0.39039039039039 0.369671368354051
0.41041041041041 0.366719919029892
0.43043043043043 0.363646254953996
0.45045045045045 0.360453853509139
0.47047047047047 0.357146304368113
0.49049049049049 0.353727302780113
0.51051051051051 0.350200642706842
0.53053053053053 0.346570209826128
0.55055055055055 0.34283997442106
0.57057057057057 0.339013984172804
0.59059059059059 0.33509635687531
0.61061061061061 0.331091273090187
0.63063063063063 0.327002968759958
0.65065065065065 0.322835727797843
0.67067067067067 0.318593874672039
0.69069069069069 0.314281767002296
0.71071071071071 0.309903788186304
0.73073073073073 0.305464340073112
0.75075075075075 0.300967835700437
0.77077077077077 0.296418692112302
0.790790790790791 0.291821323272996
0.810810810810811 0.287180133092853
0.830830830830831 0.282499508580786
0.850850850850851 0.277783813137949
0.870870870870871 0.273037380006279
0.890890890890891 0.268264505884996
0.910910910910911 0.2634694447275
0.930930930930931 0.258656401730343
0.950950950950951 0.253829527525259
0.970970970970971 0.24899291258444
0.990990990990991 0.24415058184851
1.01101101101101 0.239306489585817
1.03103103103103 0.234464514490888
1.05105105105105 0.229628455029052
1.07107107107107 0.224802025033432
1.09109109109109 0.219988849559688
1.11111111111111 0.215192461003031
1.13113113113113 0.210416295481265
1.15115115115115 0.205663689486728
1.17117117117117 0.200937876809264
1.19119119119119 0.1962419857315
1.21121121121121 0.191579036496956
1.23123123123123 0.186951939050736
1.25125125125125 0.182363491051798
1.27127127127127 0.17781637615506
1.29129129129129 0.173313162560933
1.31131131131131 0.168856301829129
1.33133133133133 0.164448127952996
1.35135135135135 0.160090856689972
1.37137137137137 0.155786585143159
1.39139139139139 0.151537291588457
1.41141141141141 0.147344835541168
1.43143143143143 0.143210958055468
1.45145145145145 0.139137282249685
1.47147147147147 0.135125314049902
1.49149149149149 0.13117644314399
1.51151151151151 0.127291944137829
1.53153153153153 0.123472977905145
1.55155155155155 0.119720593122119
1.57157157157157 0.116035727977651
1.59159159159159 0.112419212049971
1.61161161161161 0.108871768340093
1.63163163163163 0.10539401545248
1.65165165165165 0.101986469913169
1.67167167167167 0.0986495486155338
1.69169169169169 0.0953835713838294
1.71171171171171 0.0921887636446459
1.73173173173173 0.0890652591964251
1.75175175175175 0.0860131030672496
1.77177177177177 0.083032254451193
1.79179179179179 0.0801225897136326
1.81181181181181 0.077283905456062
1.83183183183183 0.0745159216311036
1.85185185185185 0.0718182846986055
1.87187187187187 0.0691905708139176
1.89189189189189 0.0666322890396674
1.91191191191191 0.0641428845726061
1.93193193193193 0.0617217419773594
1.95195195195195 0.059368188419199
1.97197197197197 0.057081496888248
1.99199199199199 0.0548608894078376
2.01201201201201 0.0527055402200588
2.03203203203203 0.0506145789418748
2.05205205205205 0.0485870936855042
2.07207207207207 0.0466221341371236
2.09209209209209 0.0447187145882916
2.11211211211211 0.0428758169148479
2.13213213213213 0.041092393498395
2.15215215215215 0.0393673700858294
2.17217217217217 0.0376996485827434
2.19219219219219 0.0360881097768737
2.21221221221221 0.0345316159881233
2.23223223223223 0.033029013642035
2.25225225225225 0.0315791357639332
2.27227227227227 0.03018080439129
2.29229229229229 0.0288328329022028
2.31231231231231 0.0275340282581906
2.33233233233233 0.0262831931598317
2.35235235235235 0.02507912811407
2.37237237237237 0.0239206334123108
2.39239239239239 0.0228065110187136
2.41241241241241 0.0217355663683581
2.43243243243243 0.0207066100752275
2.45245245245245 0.0197184595501959
2.47247247247247 0.0187699405294511
2.49249249249249 0.0178598885140022
2.51251251251251 0.016987150121141
2.53253253253253 0.0161505843489182
2.55255255255255 0.0153490637548888
2.57257257257257 0.0145814755505493
2.59259259259259 0.0138467226130541
2.61261261261261 0.0131437244159435
2.63263263263263 0.0124714178807514
2.65265265265265 0.0118287581514852
2.67267267267267 0.0112147192940778
2.69269269269269 0.0106282949230103
2.71271271271271 0.0100684987573917
2.73273273273273 0.00953436510885491
2.75275275275275 0.00902494930369034
2.77277277277277 0.00853932804169479
2.79279279279279 0.00807659969425077
2.81281281281281 0.00763588454418634
2.83283283283283 0.00721632496998621
2.85285285285285 0.00681708557693871
2.87287287287287 0.00643735327780635
2.89289289289289 0.00607633732560524
2.91291291291291 0.00573326930106518
2.93293293293293 0.00540740305732384
2.95295295295295 0.00509801462438214
2.97297297297297 0.00480440207581539
2.99299299299299 0.00452588536019617
3.01301301301301 0.00426180609964127
3.03303303303303 0.00401152735784578
3.05305305305305 0.00377443337991421
3.07307307307307 0.00354992930624085
3.09309309309309 0.00333744086263051
3.11311311311311 0.00313641402878608
3.13313313313313 0.0029463146872226
3.15315315315315 0.00276662825459746
3.17317317317317 0.00259685929737496
3.19319319319319 0.00243653113367011
3.21321321321321 0.00228518542304203
3.23323323323323 0.00214238174593163
3.25325325325325 0.00200769717436225
3.27327327327327 0.00188072583544551
3.29329329329329 0.00176107846915771
3.31331331331331 0.00164838198177652
3.33333333333333 0.0015422789962911
3.35335335335335 0.00144242740102448
3.37337337337337 0.00134849989763212
3.39339339339339 0.00126018354956833
3.41341341341341 0.00117717933203981
3.43343343343343 0.00109920168439588
3.45345345345345 0.0010259780658361
3.47347347347347 0.000957248515249212
3.49349349349349 0.000892765215932441
3.51351351351351 0.000832292065877152
3.53353353353353 0.000775604254246008
3.55355355355355 0.000722487844607976
3.57357357357357 0.000672739365441019
3.59359359359359 0.000626165408357979
3.61361361361361 0.000582582234459159
3.63363363363363 0.000541815389165405
3.65365365365365 0.00050369932583812
3.67367367367367 0.000468077038447599
3.69369369369369 0.000434799703508357
3.71371371371371 0.000403726331459719
3.73373373373373 0.000374723427631836
3.75375375375375 0.000347664662901427
3.77377377377377 0.000322430554107926
3.79379379379379 0.000298908154269281
3.81381381381381 0.000276990752607344
3.83383383383383 0.000256577584365551
3.85385385385385 0.000237573550376443
3.87387387387387 0.000219888946313312
3.89389389389389 0.000203439201538965
3.91391391391391 0.00018814462744512
3.93393393393393 0.000173930175158222
3.95395395395395 0.000160725202471485
3.97397397397397 0.000148463249848567
3.99399399399399 0.000137081825331481
4.01401401401401 0.000126522198173995
4.03403403403403 0.000116729201011866
4.05405405405405 0.000107651040372646
4.07407407407407 9.92391153205018e-05
4.09409409409409 9.14478440253317e-05
4.11411411411411 8.42344980404937e-05
4.13413413413413 7.75590440694795e-05
4.15415415415415 7.13839929989176e-05
4.17417417417417 6.56742559732345e-05
4.19419419419419 6.03970072851107e-05
4.21421421421421 5.55215538554582e-05
4.23423423423423 5.10192110769697e-05
4.25425425425425 4.68631847962824e-05
4.27427427427427 4.3028459211397e-05
4.29429429429429 3.94916904631592e-05
4.31431431431431 3.62311057022691e-05
4.33433433433433 3.32264074164067e-05
4.35435435435435 3.04586828055866e-05
4.37437437437437 2.79103179977393e-05
4.39439439439439 2.55649169007242e-05
4.41441441441441 2.34072244914564e-05
4.43443443443443 2.14230543475548e-05
4.45445445445445 1.95992202318354e-05
4.47447447447447 1.79234715450684e-05
4.49449449449449 1.63844324676437e-05
4.51451451451451 1.49715446161227e-05
4.53453453453453 1.3675013046071e-05
4.55455455455455 1.24857554380297e-05
4.57457457457457 1.13953543089884e-05
4.59459459459459 1.03960120972233e-05
4.61461461461461 9.48050897386715e-06
4.63463463463463 8.64216324004121e-06
4.65465465465465 7.87479417380527e-06
4.67467467467467 7.17268719654363e-06
4.69469469469469 6.53056123369604e-06
4.71471471471471 5.94353814994721e-06
4.73473473473473 5.40711414409909e-06
4.75475475475475 4.91713299385613e-06
4.77477477477477 4.4697610456467e-06
4.79479479479479 4.06146384937932e-06
4.81481481481481 3.68898434268124e-06
4.83483483483483 3.34932249368796e-06
4.85485485485485 3.03971631583941e-06
4.87487487487487 2.75762417238901e-06
4.89489489489489 2.50070829244518e-06
4.91491491491491 2.26681942433715e-06
4.93493493493493 2.05398255592983e-06
4.95495495495495 1.86038363520377e-06
4.97497497497497 1.68435722796805e-06
4.99499499499499 1.5243750529858e-06
5.01501501501502 1.37903533806611e-06
5.03503503503504 1.24705294381396e-06
5.05505505505506 1.12725020473309e-06
5.07507507507508 1.01854844024876e-06
5.0950950950951 9.19960090959837e-07
5.11511511511512 8.30581438046261e-07
5.13513513513514 7.49585866251233e-07
5.15515515515516 6.76217633231267e-07
5.17517517517518 6.09786110324583e-07
5.1951951951952 5.4966046193278e-07
5.21521521521522 4.95264732746229e-07
5.23523523523524 4.46073313973501e-07
5.25525525525526 4.01606761563285e-07
5.27527527527528 3.61427941137511e-07
5.2952952952953 3.25138475990267e-07
5.31531531531532 2.92375476052561e-07
5.33533533533534 2.62808527181656e-07
5.35535535535536 2.36136921509202e-07
5.37537537537538 2.1208711087848e-07
5.3953953953954 1.90410366621162e-07
5.41541541541542 1.70880630071684e-07
5.43543543543544 1.5329253929596e-07
5.45545545545546 1.3745961852414e-07
5.47547547547548 1.23212617727566e-07
5.4954954954955 1.10397990671284e-07
5.51551551551552 9.8876500608364e-08
5.53553553553554 8.85219435638491e-08
5.55555555555556 7.92199798873018e-08
5.57557557557558 7.08670654362614e-08
5.5955955955956 6.33694743912418e-08
5.61561561561562 5.66424062986154e-08
5.63563563563564 5.06091704933412e-08
5.65565565565566 4.5200441571292e-08
5.67567567567568 4.03535800631662e-08
5.6956956956957 3.60120129107462e-08
5.71571571571572 3.2124668763613e-08
5.73573573573574 2.86454635022852e-08
5.75575575575576 2.55328317539416e-08
5.77577577577578 2.27493005011625e-08
5.7957957957958 2.02611011941336e-08
5.81581581581582 1.80378170640688e-08
5.83583583583584 1.60520626017053e-08
5.85585585585586 1.42791924110059e-08
5.87587587587588 1.2697036876002e-08
5.8958958958959 1.12856622892642e-08
5.91591591591592 1.00271532849868e-08
5.93593593593594 8.90541559921139e-09
5.95595595595596 7.90599734535251e-09
5.97597597597598 7.01592714588943e-09
5.995995995996 6.22356760178439e-09
6.01601601601602 5.5184827107339e-09
6.03603603603604 4.89131796456914e-09
6.05605605605606 4.33369196574694e-09
6.07607607607608 3.83809850362692e-09
6.0960960960961 3.3978181237675e-09
6.11611611611612 3.00683830841826e-09
6.13613613613614 2.65978146430924e-09
6.15615615615616 2.35183998527871e-09
6.17617617617618 2.07871772273778e-09
6.1961961961962 1.83657725691022e-09
6.21621621621622 1.6219924166386e-09
6.23623623623624 1.43190554571785e-09
6.25625625625626 1.26358905957511e-09
6.27627627627628 1.11461087800764e-09
6.2962962962963 9.82803357938119e-10
6.31631631631632 8.66235385045992e-10
6.33633633633634 7.63187314959515e-10
6.35635635635636 6.72128483698836e-10
6.37637637637638 5.91697033481532e-10
6.3963963963964 5.20681824054164e-10
6.41641641641642 4.58006221596989e-10
6.43643643643644 4.02713577147895e-10
6.45645645645646 3.53954224575805e-10
6.47647647647648 3.10973844559378e-10
6.4964964964965 2.73103055937435e-10
6.51651651651652 2.39748109325539e-10
6.53653653653654 2.10382570159619e-10
6.55655655655656 1.84539889444162e-10
6.57657657657658 1.61806770551276e-10
6.5965965965966 1.41817249531742e-10
6.61661661661662 1.24247414645767e-10
6.63663663663664 1.08810698278133e-10
6.65665665665666 9.52536811418137e-11
6.67667667667668 8.33523547614393e-11
6.6966966966967 7.29087937236022e-11
6.71671671671672 6.37481941394484e-11
6.73673673673674 5.57162392365998e-11
6.75675675675676 4.86767570277076e-11
6.77677677677678 4.25096386334907e-11
6.7967967967968 3.71089891068556e-11
6.81681681681682 3.23814855461048e-11
6.83683683683684 2.82449199306813e-11
6.85685685685686 2.46269064908966e-11
6.87687687687688 2.14637355595386e-11
6.8968968968969 1.86993577716892e-11
6.91691691691692 1.62844842008236e-11
6.93693693693694 1.41757895636779e-11
6.95695695695696 1.23352070109961e-11
6.97697697697698 1.07293042619725e-11
6.996996996997 9.32873195138548e-12
7.01701701701702 8.10773605306642e-12
7.03703703703704 7.04372713322415e-12
7.05705705705706 6.11689998287643e-12
7.07707707707708 5.30989788981512e-12
7.0970970970971 4.6075164458095e-12
7.11711711711712 3.99644235193917e-12
7.13713713713714 3.46502319108337e-12
7.15715715715716 3.00306458801651e-12
7.17717717717718 2.60165157997869e-12
7.1971971971972 2.25299137914217e-12
7.21721721721722 1.95027502769791e-12
7.23723723723724 1.68755573049414e-12
7.25725725725726 1.45964190299846e-12
7.27727727727728 1.262003197176e-12
7.2972972972973 1.0906879676822e-12
7.31731731731732 9.42250818252902e-13
7.33733733733734 8.13689025751829e-13
7.35735735735736 7.02386779168733e-13
7.37737737737738 6.06066294885245e-13
7.3973973973974 5.22744979473708e-13
7.41741741741742 4.50697908714381e-13
7.43743743743744 3.88424977793729e-13
7.45745745745746 3.34622154016963e-13
7.47747747747748 2.88156330935933e-13
7.4974974974975 2.48043342543511e-13
7.51751751751752 2.13428748996348e-13
7.53753753753754 1.83571051982055e-13
7.55755755755756 1.57827039041883e-13
7.57757757757758 1.35638992516663e-13
7.5975975975976 1.16523530854698e-13
7.61761761761762 1.000618782966e-13
7.63763763763764 8.58913838706873e-14
7.65765765765766 7.36981325813112e-14
7.67767767767768 6.32105109959248e-14
7.6976976976977 5.41936064405376e-14
7.71771771771772 4.64443339685915e-14
7.73773773773774 3.97871984155707e-14
7.75775775775776 3.40706104038536e-14
7.77777777777778 2.91636853079907e-14
7.7977977977978 2.4953463096687e-14
7.81781781781782 2.13424947819474e-14
7.83783783783784 1.82467480586654e-14
7.85785785785786 1.55937907247682e-14
7.87787787787788 1.33212157347804e-14
7.8978978978979 1.13752763482777e-14
7.91791791791792 9.70970386854701e-15
7.93793793793794 8.28468399583068e-15
7.95795795795796 7.06597090549298e-15
7.97797797797798 6.02412085866189e-15
7.997997997998 5.13382950919603e-15
8.01801801801802 4.3733591283238e-15
8.03803803803804 3.72404376402735e-15
8.05805805805806 3.16986191875719e-15
8.07807807807808 2.69706769497134e-15
8.0980980980981 2.29387254841619e-15
8.11811811811812 1.95017082606148e-15
8.13813813813814 1.65730316851105e-15
8.15815815815816 1.40785264250169e-15
8.17817817817818 1.19546915264237e-15
8.1981981981982 1.01471827585831e-15
8.21821821821822 8.60951178494166e-16
8.23823823823824 7.30192724674871e-16
8.25825825825826 6.19045274051723e-16
8.27827827827828 5.24606005103494e-16
8.2982982982983 4.44395893386326e-16
8.31831831831832 3.76298728354137e-16
8.33833833833834 3.18508772685576e-16
8.35835835835836 2.69485858889318e-16
8.37837837837838 2.27916883183252e-16
8.3983983983984 1.92682799625235e-16
8.41841841841842 1.62830341150177e-16
8.43843843843844 1.37547801096493e-16
8.45845845845846 1.16144301209627e-16
8.47847847847848 9.8032051926925e-17
8.4984984984985 8.27111796592674e-17
8.51851851851852 6.97567552529606e-17
8.53853853853854 5.88077091106193e-17
8.55855855855856 4.95573626747691e-17
8.57857857857858 4.17453440895294e-17
8.5985985985986 3.51506886838449e-17
8.61861861861862 2.95859531836935e-17
8.63863863863864 2.48921968838275e-17
8.65865865865866 2.09347039316773e-17
8.67867867867868 1.75993388643364e-17
8.6986986986987 1.47894429982973e-17
8.71871871871872 1.24231925503958e-17
8.73873873873874 1.04313507694241e-17
8.75875875875876 8.75535614211525e-18
8.77877877877878 7.34569713009337e-18
8.7987987987988 6.16053109048305e-18
8.81881881881882 5.16451119999019e-18
8.83883883883884 4.32779048512422e-18
8.85885885885886 3.62517658454128e-18
8.87887887887888 3.03541474067764e-18
8.8988988988989 2.54057982941549e-18
8.91891891891892 2.12556106807901e-18
8.93893893893894 1.77762546207239e-18
8.95895895895896 1.48604811781133e-18
8.97897897897898 1.24179931485728e-18
8.998998998999 1.03727973679138e-18
9.01901901901902 8.66096545670873e-19
9.03903903903904 7.22874080905507e-19
9.05905905905906 6.03093897535385e-19
9.07907907907908 5.0295965472299e-19
9.0990990990991 4.19283042962206e-19
9.11911911911912 3.49387515332417e-19
9.13913913913914 2.91027078875695e-19
9.15915915915916 2.42317819504618e-19
9.17917917917918 2.01680188581445e-19
9.1991991991992 1.67790380698338e-19
9.21921921921922 1.39539388139188e-19
9.23923923923924 1.1599853476858e-19
9.25925925925926 9.63904764362469e-20
9.27927927927928 8.00648113242436e-20
9.2992992992993 6.6477576193283e-20
9.31931931931932 5.51740167796855e-20
9.33933933933934 4.5774115701642e-20
9.35935935935936 3.79604417474457e-20
9.37937937937938 3.14679525475421e-20
9.3993993993994 2.60754402558043e-20
9.41941941941942 2.15983585814365e-20
9.43943943943944 1.78828106797973e-20
9.45945945945946 1.48005121824234e-20
9.47947947947948 1.22445730038445e-20
9.4994994994995 1.01259663374544e-20
9.51951951951952 8.37057415073758e-21
9.53953953953954 6.91671611023779e-21
9.55955955955956 5.71308371631411e-21
9.57957957957958 4.71701393695898e-21
9.5995995995996 3.89304716291232e-21
9.61961961961962 3.21172317125736e-21
9.63963963963964 2.64857624243904e-21
9.65965965965966 2.18329684678274e-21
9.67967967967968 1.79903258756111e-21
9.6996996996997 1.48180551599987e-21
9.71971971971972 1.22002665237565e-21
9.73973973973974 1.00409166885045e-21
9.75975975975976 8.26044308669654e-22
9.77977977977978 6.79296312742742e-22
9.7997997997998 5.58394465795474e-22
9.81981981981982 4.58826916995414e-22
9.83983983983984 3.7686222201397e-22
9.85985985985986 3.09415635142992e-22
9.87987987987988 2.53938085193762e-22
9.8998998998999 2.08324025950642e-22
9.91991991991992 1.70834984871876e-22
9.93993993993994 1.40036162642795e-22
9.95995995995996 1.14743877987917e-22
9.97997997997998 9.39820210218911e-23
10 7.69459862670642e-23
};
\addlegendentry{N(0,1)}; 
\legend{}; 
\end{axis}

\end{tikzpicture}

%% file: FinalFigs/Null_Dists_d_10_500_n_200_m_200_kernel__Gaussian_Poly_2_2022_10_12_23_29_33cross.tex
\begin{tikzpicture}

\definecolor{darkorange25512714}{RGB}{255,127,14}
\definecolor{darkslategray38}{RGB}{38,38,38}
\definecolor{lightgray204}{RGB}{204,204,204}
\definecolor{steelblue31119180}{RGB}{31,119,180}

\begin{axis}[
axis line style={darkslategray38},
height=\figheight,
legend cell align={left},
legend style={fill opacity=0.8, draw opacity=1, text opacity=1, draw=none},
tick align=outside,
tick pos=left,
title={$\cmmd~(n/m=1)$},
width=\figwidth,
x grid style={lightgray204},
xmin=-6, xmax=6,
xtick style={color=darkslategray38},
y grid style={lightgray204},
ylabel = {}, 
ymin=0, ymax=0.436886580486519,
ytick style={color=darkslategray38}, 
xticklabels=empty,
yticklabels=empty
]
\draw[draw=none,fill=steelblue31119180,fill opacity=0.8] (axis cs:-3.21416425704956,0) rectangle (axis cs:-2.93152809143066,0.0113219725906111);
\addlegendimage{ybar,ybar legend,draw=none,fill=steelblue31119180,fill opacity=0.8}
\addlegendentry{d=10}

\draw[draw=none,fill=steelblue31119180,fill opacity=0.8] (axis cs:-2.50757360458374,0) rectangle (axis cs:-2.22493720054626,0.0198134486908317);
\draw[draw=none,fill=steelblue31119180,fill opacity=0.8] (axis cs:-1.80098283290863,0) rectangle (axis cs:-1.51834666728973,0.144355150530291);
\draw[draw=none,fill=steelblue31119180,fill opacity=0.8] (axis cs:-1.0943922996521,0) rectangle (axis cs:-0.811755895614624,0.268896826344148);
\draw[draw=none,fill=steelblue31119180,fill opacity=0.8] (axis cs:-0.387801617383957,0) rectangle (axis cs:-0.105165213346481,0.416082457606209);
\draw[draw=none,fill=steelblue31119180,fill opacity=0.8] (axis cs:0.318789094686508,0) rectangle (axis cs:0.601425528526306,0.297201755433006);
\draw[draw=none,fill=steelblue31119180,fill opacity=0.8] (axis cs:1.02537977695465,0) rectangle (axis cs:1.30801618099213,0.181151546168689);
\draw[draw=none,fill=steelblue31119180,fill opacity=0.8] (axis cs:1.73197054862976,0) rectangle (axis cs:2.01460695266724,0.0622708387426141);
\draw[draw=none,fill=steelblue31119180,fill opacity=0.8] (axis cs:2.43856143951416,0) rectangle (axis cs:2.72119760513306,0.0113219725906111);
\draw[draw=none,fill=steelblue31119180,fill opacity=0.8] (axis cs:3.14515209197998,0) rectangle (axis cs:3.42778825759888,0.00283049314765276);
\draw[draw=none,fill=darkorange25512714,fill opacity=0.8] (axis cs:-2.93152809143066,0) rectangle (axis cs:-2.64889192581177,0.0113219725906111);
\addlegendimage{ybar,ybar legend,draw=none,fill=darkorange25512714,fill opacity=0.8}
\addlegendentry{d=500}

\draw[draw=none,fill=darkorange25512714,fill opacity=0.8] (axis cs:-2.22493743896484,0) rectangle (axis cs:-1.94230103492737,0.0651013314127329);
\draw[draw=none,fill=darkorange25512714,fill opacity=0.8] (axis cs:-1.51834654808044,0) rectangle (axis cs:-1.23571038246155,0.0820843012819302);
\draw[draw=none,fill=darkorange25512714,fill opacity=0.8] (axis cs:-0.811756014823914,0) rectangle (axis cs:-0.529119610786438,0.288710276706349);
\draw[draw=none,fill=darkorange25512714,fill opacity=0.8] (axis cs:-0.105165332555771,0) rectangle (axis cs:0.177471071481705,0.393438514335122);
\draw[draw=none,fill=darkorange25512714,fill opacity=0.8] (axis cs:0.601425409317017,0) rectangle (axis cs:0.884061813354492,0.331167670339635);
\draw[draw=none,fill=darkorange25512714,fill opacity=0.8] (axis cs:1.30801606178284,0) rectangle (axis cs:1.59065246582031,0.152846617079832);
\draw[draw=none,fill=darkorange25512714,fill opacity=0.8] (axis cs:2.01460695266724,0) rectangle (axis cs:2.29724335670471,0.0707623167529705);
\draw[draw=none,fill=darkorange25512714,fill opacity=0.8] (axis cs:2.72119760513306,0) rectangle (axis cs:3.00383377075195,0.0141524657382638);
\draw[draw=none,fill=darkorange25512714,fill opacity=0.8] (axis cs:3.42778825759888,0) rectangle (axis cs:3.71042442321777,0.00566098629530553);
\addplot [semithick, black]
table {%
-10 7.69459862670642e-23
-9.97997997997998 9.39820210218911e-23
-9.95995995995996 1.14743877987917e-22
-9.93993993993994 1.40036162642795e-22
-9.91991991991992 1.70834984871876e-22
-9.8998998998999 2.08324025950642e-22
-9.87987987987988 2.53938085193762e-22
-9.85985985985986 3.09415635142992e-22
-9.83983983983984 3.7686222201397e-22
-9.81981981981982 4.58826916995414e-22
-9.7997997997998 5.58394465795474e-22
-9.77977977977978 6.79296312742742e-22
-9.75975975975976 8.26044308669654e-22
-9.73973973973974 1.00409166885045e-21
-9.71971971971972 1.22002665237565e-21
-9.6996996996997 1.48180551599987e-21
-9.67967967967968 1.79903258756111e-21
-9.65965965965966 2.18329684678274e-21
-9.63963963963964 2.64857624243904e-21
-9.61961961961962 3.21172317125736e-21
-9.5995995995996 3.89304716291232e-21
-9.57957957957958 4.71701393695898e-21
-9.55955955955956 5.71308371631411e-21
-9.53953953953954 6.91671611023779e-21
-9.51951951951952 8.37057415073758e-21
-9.4994994994995 1.01259663374544e-20
-9.47947947947948 1.22445730038445e-20
-9.45945945945946 1.48005121824234e-20
-9.43943943943944 1.78828106797973e-20
-9.41941941941942 2.15983585814365e-20
-9.3993993993994 2.60754402558043e-20
-9.37937937937938 3.14679525475421e-20
-9.35935935935936 3.79604417474457e-20
-9.33933933933934 4.5774115701642e-20
-9.31931931931932 5.51740167796855e-20
-9.2992992992993 6.6477576193283e-20
-9.27927927927928 8.00648113242436e-20
-9.25925925925926 9.63904764362469e-20
-9.23923923923924 1.1599853476858e-19
-9.21921921921922 1.39539388139188e-19
-9.1991991991992 1.67790380698338e-19
-9.17917917917918 2.01680188581445e-19
-9.15915915915916 2.42317819504618e-19
-9.13913913913914 2.91027078875695e-19
-9.11911911911912 3.49387515332417e-19
-9.0990990990991 4.19283042962206e-19
-9.07907907907908 5.0295965472299e-19
-9.05905905905906 6.03093897535385e-19
-9.03903903903904 7.22874080905507e-19
-9.01901901901902 8.66096545670873e-19
-8.998998998999 1.03727973679138e-18
-8.97897897897898 1.24179931485728e-18
-8.95895895895896 1.48604811781133e-18
-8.93893893893894 1.77762546207239e-18
-8.91891891891892 2.12556106807901e-18
-8.8988988988989 2.54057982941549e-18
-8.87887887887888 3.03541474067764e-18
-8.85885885885886 3.62517658454128e-18
-8.83883883883884 4.32779048512422e-18
-8.81881881881882 5.16451119999019e-18
-8.7987987987988 6.16053109048305e-18
-8.77877877877878 7.34569713009337e-18
-8.75875875875876 8.75535614211525e-18
-8.73873873873874 1.04313507694241e-17
-8.71871871871872 1.24231925503958e-17
-8.6986986986987 1.47894429982973e-17
-8.67867867867868 1.75993388643364e-17
-8.65865865865866 2.09347039316773e-17
-8.63863863863864 2.48921968838275e-17
-8.61861861861862 2.95859531836935e-17
-8.5985985985986 3.51506886838449e-17
-8.57857857857858 4.17453440895294e-17
-8.55855855855856 4.95573626747691e-17
-8.53853853853854 5.88077091106193e-17
-8.51851851851852 6.97567552529606e-17
-8.4984984984985 8.27111796592674e-17
-8.47847847847848 9.8032051926925e-17
-8.45845845845846 1.16144301209627e-16
-8.43843843843844 1.37547801096493e-16
-8.41841841841842 1.62830341150177e-16
-8.3983983983984 1.92682799625235e-16
-8.37837837837838 2.27916883183252e-16
-8.35835835835836 2.69485858889318e-16
-8.33833833833834 3.18508772685576e-16
-8.31831831831832 3.76298728354137e-16
-8.2982982982983 4.44395893386326e-16
-8.27827827827828 5.24606005103494e-16
-8.25825825825826 6.19045274051723e-16
-8.23823823823824 7.30192724674871e-16
-8.21821821821822 8.60951178494166e-16
-8.1981981981982 1.01471827585831e-15
-8.17817817817818 1.19546915264237e-15
-8.15815815815816 1.40785264250169e-15
-8.13813813813814 1.65730316851105e-15
-8.11811811811812 1.95017082606148e-15
-8.0980980980981 2.29387254841619e-15
-8.07807807807808 2.69706769497134e-15
-8.05805805805806 3.16986191875719e-15
-8.03803803803804 3.72404376402735e-15
-8.01801801801802 4.3733591283238e-15
-7.997997997998 5.13382950919607e-15
-7.97797797797798 6.02412085866193e-15
-7.95795795795796 7.06597090549303e-15
-7.93793793793794 8.28468399583074e-15
-7.91791791791792 9.70970386854708e-15
-7.8978978978979 1.13752763482777e-14
-7.87787787787788 1.33212157347805e-14
-7.85785785785786 1.55937907247683e-14
-7.83783783783784 1.82467480586655e-14
-7.81781781781782 2.13424947819475e-14
-7.7977977977978 2.49534630966872e-14
-7.77777777777778 2.91636853079909e-14
-7.75775775775776 3.40706104038538e-14
-7.73773773773774 3.9787198415571e-14
-7.71771771771772 4.64443339685918e-14
-7.6976976976977 5.4193606440538e-14
-7.67767767767768 6.32105109959252e-14
-7.65765765765766 7.36981325813117e-14
-7.63763763763764 8.58913838706879e-14
-7.61761761761762 1.00061878296601e-13
-7.5975975975976 1.16523530854699e-13
-7.57757757757758 1.35638992516664e-13
-7.55755755755756 1.57827039041884e-13
-7.53753753753754 1.83571051982057e-13
-7.51751751751752 2.13428748996349e-13
-7.4974974974975 2.48043342543513e-13
-7.47747747747748 2.88156330935935e-13
-7.45745745745746 3.34622154016965e-13
-7.43743743743744 3.88424977793732e-13
-7.41741741741742 4.50697908714384e-13
-7.3973973973974 5.22744979473711e-13
-7.37737737737738 6.06066294885249e-13
-7.35735735735736 7.02386779168738e-13
-7.33733733733734 8.13689025751835e-13
-7.31731731731732 9.42250818252909e-13
-7.2972972972973 1.09068796768221e-12
-7.27727727727728 1.262003197176e-12
-7.25725725725726 1.45964190299847e-12
-7.23723723723724 1.68755573049416e-12
-7.21721721721722 1.95027502769792e-12
-7.1971971971972 2.25299137914218e-12
-7.17717717717718 2.60165157997871e-12
-7.15715715715716 3.00306458801653e-12
-7.13713713713714 3.4650231910834e-12
-7.11711711711712 3.99644235193919e-12
-7.0970970970971 4.60751644580953e-12
-7.07707707707708 5.30989788981514e-12
-7.05705705705706 6.11689998287646e-12
-7.03703703703704 7.0437271332242e-12
-7.01701701701702 8.10773605306645e-12
-6.996996996997 9.32873195138555e-12
-6.97697697697698 1.07293042619726e-11
-6.95695695695696 1.23352070109962e-11
-6.93693693693694 1.41757895636779e-11
-6.91691691691692 1.62844842008237e-11
-6.8968968968969 1.86993577716893e-11
-6.87687687687688 2.14637355595386e-11
-6.85685685685686 2.46269064908967e-11
-6.83683683683684 2.82449199306815e-11
-6.81681681681682 3.2381485546105e-11
-6.7967967967968 3.71089891068559e-11
-6.77677677677678 4.25096386334913e-11
-6.75675675675676 4.86767570277083e-11
-6.73673673673674 5.57162392366004e-11
-6.71671671671672 6.37481941394491e-11
-6.6966966966967 7.29087937236032e-11
-6.67667667667668 8.33523547614402e-11
-6.65665665665666 9.52536811418151e-11
-6.63663663663664 1.08810698278135e-10
-6.61661661661662 1.24247414645768e-10
-6.5965965965966 1.41817249531744e-10
-6.57657657657658 1.61806770551278e-10
-6.55655655655656 1.84539889444164e-10
-6.53653653653654 2.10382570159622e-10
-6.51651651651652 2.39748109325542e-10
-6.4964964964965 2.73103055937438e-10
-6.47647647647648 3.10973844559381e-10
-6.45645645645646 3.53954224575809e-10
-6.43643643643644 4.027135771479e-10
-6.41641641641642 4.58006221596996e-10
-6.3963963963964 5.20681824054169e-10
-6.37637637637638 5.91697033481538e-10
-6.35635635635636 6.72128483698846e-10
-6.33633633633634 7.63187314959523e-10
-6.31631631631632 8.66235385046001e-10
-6.2962962962963 9.8280335793813e-10
-6.27627627627628 1.11461087800766e-09
-6.25625625625626 1.26358905957513e-09
-6.23623623623624 1.43190554571787e-09
-6.21621621621622 1.62199241663862e-09
-6.1961961961962 1.83657725691024e-09
-6.17617617617618 2.0787177227378e-09
-6.15615615615616 2.35183998527873e-09
-6.13613613613614 2.65978146430928e-09
-6.11611611611612 3.00683830841829e-09
-6.0960960960961 3.39781812376754e-09
-6.07607607607608 3.83809850362696e-09
-6.05605605605606 4.33369196574699e-09
-6.03603603603604 4.89131796456919e-09
-6.01601601601602 5.51848271073395e-09
-5.995995995996 6.22356760178439e-09
-5.97597597597598 7.01592714588943e-09
-5.95595595595596 7.90599734535251e-09
-5.93593593593594 8.90541559921139e-09
-5.91591591591592 1.00271532849868e-08
-5.8958958958959 1.12856622892642e-08
-5.87587587587588 1.2697036876002e-08
-5.85585585585586 1.42791924110059e-08
-5.83583583583584 1.60520626017053e-08
-5.81581581581582 1.80378170640688e-08
-5.7957957957958 2.02611011941336e-08
-5.77577577577578 2.27493005011625e-08
-5.75575575575576 2.55328317539416e-08
-5.73573573573574 2.86454635022852e-08
-5.71571571571572 3.2124668763613e-08
-5.6956956956957 3.60120129107462e-08
-5.67567567567568 4.03535800631662e-08
-5.65565565565566 4.5200441571292e-08
-5.63563563563564 5.06091704933412e-08
-5.61561561561562 5.66424062986154e-08
-5.5955955955956 6.33694743912418e-08
-5.57557557557558 7.08670654362614e-08
-5.55555555555556 7.92199798873018e-08
-5.53553553553554 8.85219435638491e-08
-5.51551551551552 9.8876500608364e-08
-5.4954954954955 1.10397990671284e-07
-5.47547547547548 1.23212617727566e-07
-5.45545545545546 1.3745961852414e-07
-5.43543543543544 1.5329253929596e-07
-5.41541541541542 1.70880630071684e-07
-5.3953953953954 1.90410366621162e-07
-5.37537537537538 2.1208711087848e-07
-5.35535535535536 2.36136921509202e-07
-5.33533533533534 2.62808527181656e-07
-5.31531531531532 2.92375476052561e-07
-5.2952952952953 3.25138475990267e-07
-5.27527527527528 3.61427941137511e-07
-5.25525525525526 4.01606761563285e-07
-5.23523523523524 4.46073313973501e-07
-5.21521521521522 4.95264732746229e-07
-5.1951951951952 5.4966046193278e-07
-5.17517517517518 6.09786110324583e-07
-5.15515515515516 6.76217633231267e-07
-5.13513513513514 7.49585866251233e-07
-5.11511511511512 8.30581438046261e-07
-5.0950950950951 9.19960090959837e-07
-5.07507507507508 1.01854844024876e-06
-5.05505505505506 1.12725020473309e-06
-5.03503503503504 1.24705294381396e-06
-5.01501501501502 1.37903533806611e-06
-4.99499499499499 1.5243750529858e-06
-4.97497497497497 1.68435722796805e-06
-4.95495495495495 1.86038363520377e-06
-4.93493493493493 2.05398255592983e-06
-4.91491491491491 2.26681942433715e-06
-4.89489489489489 2.50070829244518e-06
-4.87487487487487 2.75762417238901e-06
-4.85485485485485 3.03971631583941e-06
-4.83483483483483 3.34932249368796e-06
-4.81481481481481 3.68898434268124e-06
-4.79479479479479 4.06146384937932e-06
-4.77477477477477 4.4697610456467e-06
-4.75475475475475 4.91713299385613e-06
-4.73473473473473 5.40711414409909e-06
-4.71471471471471 5.94353814994721e-06
-4.69469469469469 6.53056123369604e-06
-4.67467467467467 7.17268719654363e-06
-4.65465465465465 7.87479417380527e-06
-4.63463463463463 8.64216324004121e-06
-4.61461461461461 9.48050897386715e-06
-4.59459459459459 1.03960120972233e-05
-4.57457457457457 1.13953543089884e-05
-4.55455455455455 1.24857554380297e-05
-4.53453453453453 1.3675013046071e-05
-4.51451451451451 1.49715446161227e-05
-4.49449449449449 1.63844324676437e-05
-4.47447447447447 1.79234715450684e-05
-4.45445445445445 1.95992202318354e-05
-4.43443443443443 2.14230543475548e-05
-4.41441441441441 2.34072244914564e-05
-4.39439439439439 2.55649169007242e-05
-4.37437437437437 2.79103179977393e-05
-4.35435435435435 3.04586828055866e-05
-4.33433433433433 3.32264074164067e-05
-4.31431431431431 3.62311057022691e-05
-4.29429429429429 3.94916904631592e-05
-4.27427427427427 4.3028459211397e-05
-4.25425425425425 4.68631847962824e-05
-4.23423423423423 5.10192110769697e-05
-4.21421421421421 5.55215538554582e-05
-4.19419419419419 6.03970072851107e-05
-4.17417417417417 6.56742559732345e-05
-4.15415415415415 7.13839929989176e-05
-4.13413413413413 7.75590440694795e-05
-4.11411411411411 8.42344980404937e-05
-4.09409409409409 9.14478440253317e-05
-4.07407407407407 9.92391153205018e-05
-4.05405405405405 0.000107651040372646
-4.03403403403403 0.000116729201011866
-4.01401401401401 0.000126522198173995
-3.99399399399399 0.000137081825331481
-3.97397397397397 0.000148463249848567
-3.95395395395395 0.000160725202471485
-3.93393393393393 0.000173930175158222
-3.91391391391391 0.00018814462744512
-3.89389389389389 0.000203439201538965
-3.87387387387387 0.000219888946313312
-3.85385385385385 0.000237573550376443
-3.83383383383383 0.000256577584365551
-3.81381381381381 0.000276990752607344
-3.79379379379379 0.000298908154269281
-3.77377377377377 0.000322430554107926
-3.75375375375375 0.000347664662901427
-3.73373373373373 0.000374723427631836
-3.71371371371371 0.000403726331459719
-3.69369369369369 0.000434799703508357
-3.67367367367367 0.000468077038447599
-3.65365365365365 0.00050369932583812
-3.63363363363363 0.000541815389165405
-3.61361361361361 0.000582582234459159
-3.59359359359359 0.000626165408357979
-3.57357357357357 0.000672739365441021
-3.55355355355355 0.000722487844607978
-3.53353353353353 0.00077560425424601
-3.51351351351351 0.000832292065877155
-3.49349349349349 0.000892765215932443
-3.47347347347347 0.000957248515249216
-3.45345345345345 0.0010259780658361
-3.43343343343343 0.00109920168439588
-3.41341341341341 0.00117717933203981
-3.39339339339339 0.00126018354956833
-3.37337337337337 0.00134849989763212
-3.35335335335335 0.00144242740102448
-3.33333333333333 0.00154227899629111
-3.31331331331331 0.00164838198177652
-3.29329329329329 0.00176107846915772
-3.27327327327327 0.00188072583544552
-3.25325325325325 0.00200769717436226
-3.23323323323323 0.00214238174593163
-3.21321321321321 0.00228518542304204
-3.19319319319319 0.00243653113367012
-3.17317317317317 0.00259685929737497
-3.15315315315315 0.00276662825459747
-3.13313313313313 0.00294631468722261
-3.11311311311311 0.00313641402878609
-3.09309309309309 0.00333744086263052
-3.07307307307307 0.00354992930624086
-3.05305305305305 0.00377443337991422
-3.03303303303303 0.00401152735784579
-3.01301301301301 0.00426180609964128
-2.99299299299299 0.00452588536019618
-2.97297297297297 0.0048044020758154
-2.95295295295295 0.00509801462438215
-2.93293293293293 0.00540740305732385
-2.91291291291291 0.00573326930106519
-2.89289289289289 0.00607633732560526
-2.87287287287287 0.00643735327780636
-2.85285285285285 0.00681708557693873
-2.83283283283283 0.00721632496998623
-2.81281281281281 0.00763588454418632
-2.79279279279279 0.00807659969425075
-2.77277277277277 0.00853932804169477
-2.75275275275275 0.00902494930369032
-2.73273273273273 0.00953436510885489
-2.71271271271271 0.0100684987573917
-2.69269269269269 0.0106282949230102
-2.67267267267267 0.0112147192940778
-2.65265265265265 0.0118287581514852
-2.63263263263263 0.0124714178807513
-2.61261261261261 0.0131437244159435
-2.59259259259259 0.0138467226130541
-2.57257257257257 0.0145814755505492
-2.55255255255255 0.0153490637548887
-2.53253253253253 0.0161505843489182
-2.51251251251251 0.0169871501211409
-2.49249249249249 0.0178598885140022
-2.47247247247247 0.018769940529451
-2.45245245245245 0.0197184595501959
-2.43243243243243 0.0207066100752274
-2.41241241241241 0.0217355663683581
-2.39239239239239 0.0228065110187135
-2.37237237237237 0.0239206334123108
-2.35235235235235 0.0250791281140699
-2.33233233233233 0.0262831931598317
-2.31231231231231 0.0275340282581906
-2.29229229229229 0.0288328329022027
-2.27227227227227 0.0301808043912899
-2.25225225225225 0.0315791357639331
-2.23223223223223 0.033029013642035
-2.21221221221221 0.0345316159881232
-2.19219219219219 0.0360881097768736
-2.17217217217217 0.0376996485827434
-2.15215215215215 0.0393673700858293
-2.13213213213213 0.0410923934983949
-2.11211211211211 0.0428758169148479
-2.09209209209209 0.0447187145882915
-2.07207207207207 0.0466221341371235
-2.05205205205205 0.0485870936855041
-2.03203203203203 0.0506145789418747
-2.01201201201201 0.0527055402200587
-1.99199199199199 0.0548608894078376
-1.97197197197197 0.057081496888248
-1.95195195195195 0.059368188419199
-1.93193193193193 0.0617217419773594
-1.91191191191191 0.0641428845726061
-1.89189189189189 0.0666322890396674
-1.87187187187187 0.0691905708139176
-1.85185185185185 0.0718182846986055
-1.83183183183183 0.0745159216311036
-1.81181181181181 0.077283905456062
-1.79179179179179 0.0801225897136326
-1.77177177177177 0.083032254451193
-1.75175175175175 0.0860131030672496
-1.73173173173173 0.0890652591964251
-1.71171171171171 0.0921887636446459
-1.69169169169169 0.0953835713838294
-1.67167167167167 0.0986495486155338
-1.65165165165165 0.101986469913169
-1.63163163163163 0.10539401545248
-1.61161161161161 0.108871768340093
-1.59159159159159 0.112419212049971
-1.57157157157157 0.116035727977651
-1.55155155155155 0.119720593122119
-1.53153153153153 0.123472977905145
-1.51151151151151 0.127291944137829
-1.49149149149149 0.13117644314399
-1.47147147147147 0.135125314049902
-1.45145145145145 0.139137282249685
-1.43143143143143 0.143210958055468
-1.41141141141141 0.147344835541168
-1.39139139139139 0.151537291588457
-1.37137137137137 0.155786585143159
-1.35135135135135 0.160090856689972
-1.33133133133133 0.164448127952996
-1.31131131131131 0.168856301829129
-1.29129129129129 0.173313162560933
-1.27127127127127 0.17781637615506
-1.25125125125125 0.182363491051798
-1.23123123123123 0.186951939050736
-1.21121121121121 0.191579036496956
-1.19119119119119 0.1962419857315
-1.17117117117117 0.200937876809264
-1.15115115115115 0.205663689486728
-1.13113113113113 0.210416295481265
-1.11111111111111 0.215192461003031
-1.09109109109109 0.219988849559688
-1.07107107107107 0.224802025033432
-1.05105105105105 0.229628455029052
-1.03103103103103 0.234464514490888
-1.01101101101101 0.239306489585817
-0.990990990990991 0.24415058184851
-0.970970970970971 0.24899291258444
-0.950950950950951 0.253829527525259
-0.930930930930931 0.258656401730343
-0.910910910910911 0.2634694447275
-0.890890890890891 0.268264505884996
-0.870870870870871 0.273037380006279
-0.850850850850851 0.277783813137949
-0.830830830830831 0.282499508580786
-0.810810810810811 0.287180133092853
-0.790790790790791 0.291821323272996
-0.77077077077077 0.296418692112302
-0.75075075075075 0.300967835700437
-0.73073073073073 0.305464340073112
-0.71071071071071 0.309903788186304
-0.69069069069069 0.314281767002296
-0.67067067067067 0.318593874672039
-0.65065065065065 0.322835727797843
-0.63063063063063 0.327002968759958
-0.61061061061061 0.331091273090187
-0.59059059059059 0.33509635687531
-0.57057057057057 0.339013984172804
-0.55055055055055 0.34283997442106
-0.53053053053053 0.346570209826128
-0.51051051051051 0.350200642706842
-0.49049049049049 0.353727302780113
-0.47047047047047 0.357146304368113
-0.45045045045045 0.360453853509139
-0.43043043043043 0.363646254953996
-0.41041041041041 0.366719919029892
-0.39039039039039 0.369671368354051
-0.37037037037037 0.372497244379499
-0.35035035035035 0.375194313755802
-0.33033033033033 0.377759474487924
-0.31031031031031 0.38018976187679
-0.29029029029029 0.382482354225654
-0.27027027027027 0.384634578296894
-0.25025025025025 0.386643914504485
-0.23023023023023 0.388508001828027
-0.21021021021021 0.390224642434919
-0.19019019019019 0.391791805998011
-0.17017017017017 0.393207633696876
-0.15015015015015 0.394470441891644
-0.13013013013013 0.395578725459258
-0.11011011011011 0.396531160782876
-0.0900900900900901 0.397326608386124
-0.07007007007007 0.397964115204853
-0.05005005005005 0.398442916490068
-0.03003003003003 0.398762437336696
-0.01001001001001 0.398922293833933
0.01001001001001 0.398922293833933
0.03003003003003 0.398762437336696
0.05005005005005 0.398442916490068
0.07007007007007 0.397964115204853
0.0900900900900901 0.397326608386124
0.11011011011011 0.396531160782876
0.13013013013013 0.395578725459258
0.15015015015015 0.394470441891644
0.17017017017017 0.393207633696876
0.19019019019019 0.391791805998011
0.21021021021021 0.390224642434919
0.23023023023023 0.388508001828027
0.25025025025025 0.386643914504485
0.27027027027027 0.384634578296894
0.29029029029029 0.382482354225654
0.31031031031031 0.38018976187679
0.33033033033033 0.377759474487924
0.35035035035035 0.375194313755802
0.37037037037037 0.372497244379499
0.39039039039039 0.369671368354051
0.41041041041041 0.366719919029892
0.43043043043043 0.363646254953996
0.45045045045045 0.360453853509139
0.47047047047047 0.357146304368113
0.49049049049049 0.353727302780113
0.51051051051051 0.350200642706842
0.53053053053053 0.346570209826128
0.55055055055055 0.34283997442106
0.57057057057057 0.339013984172804
0.59059059059059 0.33509635687531
0.61061061061061 0.331091273090187
0.63063063063063 0.327002968759958
0.65065065065065 0.322835727797843
0.67067067067067 0.318593874672039
0.69069069069069 0.314281767002296
0.71071071071071 0.309903788186304
0.73073073073073 0.305464340073112
0.75075075075075 0.300967835700437
0.77077077077077 0.296418692112302
0.790790790790791 0.291821323272996
0.810810810810811 0.287180133092853
0.830830830830831 0.282499508580786
0.850850850850851 0.277783813137949
0.870870870870871 0.273037380006279
0.890890890890891 0.268264505884996
0.910910910910911 0.2634694447275
0.930930930930931 0.258656401730343
0.950950950950951 0.253829527525259
0.970970970970971 0.24899291258444
0.990990990990991 0.24415058184851
1.01101101101101 0.239306489585817
1.03103103103103 0.234464514490888
1.05105105105105 0.229628455029052
1.07107107107107 0.224802025033432
1.09109109109109 0.219988849559688
1.11111111111111 0.215192461003031
1.13113113113113 0.210416295481265
1.15115115115115 0.205663689486728
1.17117117117117 0.200937876809264
1.19119119119119 0.1962419857315
1.21121121121121 0.191579036496956
1.23123123123123 0.186951939050736
1.25125125125125 0.182363491051798
1.27127127127127 0.17781637615506
1.29129129129129 0.173313162560933
1.31131131131131 0.168856301829129
1.33133133133133 0.164448127952996
1.35135135135135 0.160090856689972
1.37137137137137 0.155786585143159
1.39139139139139 0.151537291588457
1.41141141141141 0.147344835541168
1.43143143143143 0.143210958055468
1.45145145145145 0.139137282249685
1.47147147147147 0.135125314049902
1.49149149149149 0.13117644314399
1.51151151151151 0.127291944137829
1.53153153153153 0.123472977905145
1.55155155155155 0.119720593122119
1.57157157157157 0.116035727977651
1.59159159159159 0.112419212049971
1.61161161161161 0.108871768340093
1.63163163163163 0.10539401545248
1.65165165165165 0.101986469913169
1.67167167167167 0.0986495486155338
1.69169169169169 0.0953835713838294
1.71171171171171 0.0921887636446459
1.73173173173173 0.0890652591964251
1.75175175175175 0.0860131030672496
1.77177177177177 0.083032254451193
1.79179179179179 0.0801225897136326
1.81181181181181 0.077283905456062
1.83183183183183 0.0745159216311036
1.85185185185185 0.0718182846986055
1.87187187187187 0.0691905708139176
1.89189189189189 0.0666322890396674
1.91191191191191 0.0641428845726061
1.93193193193193 0.0617217419773594
1.95195195195195 0.059368188419199
1.97197197197197 0.057081496888248
1.99199199199199 0.0548608894078376
2.01201201201201 0.0527055402200588
2.03203203203203 0.0506145789418748
2.05205205205205 0.0485870936855042
2.07207207207207 0.0466221341371236
2.09209209209209 0.0447187145882916
2.11211211211211 0.0428758169148479
2.13213213213213 0.041092393498395
2.15215215215215 0.0393673700858294
2.17217217217217 0.0376996485827434
2.19219219219219 0.0360881097768737
2.21221221221221 0.0345316159881233
2.23223223223223 0.033029013642035
2.25225225225225 0.0315791357639332
2.27227227227227 0.03018080439129
2.29229229229229 0.0288328329022028
2.31231231231231 0.0275340282581906
2.33233233233233 0.0262831931598317
2.35235235235235 0.02507912811407
2.37237237237237 0.0239206334123108
2.39239239239239 0.0228065110187136
2.41241241241241 0.0217355663683581
2.43243243243243 0.0207066100752275
2.45245245245245 0.0197184595501959
2.47247247247247 0.0187699405294511
2.49249249249249 0.0178598885140022
2.51251251251251 0.016987150121141
2.53253253253253 0.0161505843489182
2.55255255255255 0.0153490637548888
2.57257257257257 0.0145814755505493
2.59259259259259 0.0138467226130541
2.61261261261261 0.0131437244159435
2.63263263263263 0.0124714178807514
2.65265265265265 0.0118287581514852
2.67267267267267 0.0112147192940778
2.69269269269269 0.0106282949230103
2.71271271271271 0.0100684987573917
2.73273273273273 0.00953436510885491
2.75275275275275 0.00902494930369034
2.77277277277277 0.00853932804169479
2.79279279279279 0.00807659969425077
2.81281281281281 0.00763588454418634
2.83283283283283 0.00721632496998621
2.85285285285285 0.00681708557693871
2.87287287287287 0.00643735327780635
2.89289289289289 0.00607633732560524
2.91291291291291 0.00573326930106518
2.93293293293293 0.00540740305732384
2.95295295295295 0.00509801462438214
2.97297297297297 0.00480440207581539
2.99299299299299 0.00452588536019617
3.01301301301301 0.00426180609964127
3.03303303303303 0.00401152735784578
3.05305305305305 0.00377443337991421
3.07307307307307 0.00354992930624085
3.09309309309309 0.00333744086263051
3.11311311311311 0.00313641402878608
3.13313313313313 0.0029463146872226
3.15315315315315 0.00276662825459746
3.17317317317317 0.00259685929737496
3.19319319319319 0.00243653113367011
3.21321321321321 0.00228518542304203
3.23323323323323 0.00214238174593163
3.25325325325325 0.00200769717436225
3.27327327327327 0.00188072583544551
3.29329329329329 0.00176107846915771
3.31331331331331 0.00164838198177652
3.33333333333333 0.0015422789962911
3.35335335335335 0.00144242740102448
3.37337337337337 0.00134849989763212
3.39339339339339 0.00126018354956833
3.41341341341341 0.00117717933203981
3.43343343343343 0.00109920168439588
3.45345345345345 0.0010259780658361
3.47347347347347 0.000957248515249212
3.49349349349349 0.000892765215932441
3.51351351351351 0.000832292065877152
3.53353353353353 0.000775604254246008
3.55355355355355 0.000722487844607976
3.57357357357357 0.000672739365441019
3.59359359359359 0.000626165408357979
3.61361361361361 0.000582582234459159
3.63363363363363 0.000541815389165405
3.65365365365365 0.00050369932583812
3.67367367367367 0.000468077038447599
3.69369369369369 0.000434799703508357
3.71371371371371 0.000403726331459719
3.73373373373373 0.000374723427631836
3.75375375375375 0.000347664662901427
3.77377377377377 0.000322430554107926
3.79379379379379 0.000298908154269281
3.81381381381381 0.000276990752607344
3.83383383383383 0.000256577584365551
3.85385385385385 0.000237573550376443
3.87387387387387 0.000219888946313312
3.89389389389389 0.000203439201538965
3.91391391391391 0.00018814462744512
3.93393393393393 0.000173930175158222
3.95395395395395 0.000160725202471485
3.97397397397397 0.000148463249848567
3.99399399399399 0.000137081825331481
4.01401401401401 0.000126522198173995
4.03403403403403 0.000116729201011866
4.05405405405405 0.000107651040372646
4.07407407407407 9.92391153205018e-05
4.09409409409409 9.14478440253317e-05
4.11411411411411 8.42344980404937e-05
4.13413413413413 7.75590440694795e-05
4.15415415415415 7.13839929989176e-05
4.17417417417417 6.56742559732345e-05
4.19419419419419 6.03970072851107e-05
4.21421421421421 5.55215538554582e-05
4.23423423423423 5.10192110769697e-05
4.25425425425425 4.68631847962824e-05
4.27427427427427 4.3028459211397e-05
4.29429429429429 3.94916904631592e-05
4.31431431431431 3.62311057022691e-05
4.33433433433433 3.32264074164067e-05
4.35435435435435 3.04586828055866e-05
4.37437437437437 2.79103179977393e-05
4.39439439439439 2.55649169007242e-05
4.41441441441441 2.34072244914564e-05
4.43443443443443 2.14230543475548e-05
4.45445445445445 1.95992202318354e-05
4.47447447447447 1.79234715450684e-05
4.49449449449449 1.63844324676437e-05
4.51451451451451 1.49715446161227e-05
4.53453453453453 1.3675013046071e-05
4.55455455455455 1.24857554380297e-05
4.57457457457457 1.13953543089884e-05
4.59459459459459 1.03960120972233e-05
4.61461461461461 9.48050897386715e-06
4.63463463463463 8.64216324004121e-06
4.65465465465465 7.87479417380527e-06
4.67467467467467 7.17268719654363e-06
4.69469469469469 6.53056123369604e-06
4.71471471471471 5.94353814994721e-06
4.73473473473473 5.40711414409909e-06
4.75475475475475 4.91713299385613e-06
4.77477477477477 4.4697610456467e-06
4.79479479479479 4.06146384937932e-06
4.81481481481481 3.68898434268124e-06
4.83483483483483 3.34932249368796e-06
4.85485485485485 3.03971631583941e-06
4.87487487487487 2.75762417238901e-06
4.89489489489489 2.50070829244518e-06
4.91491491491491 2.26681942433715e-06
4.93493493493493 2.05398255592983e-06
4.95495495495495 1.86038363520377e-06
4.97497497497497 1.68435722796805e-06
4.99499499499499 1.5243750529858e-06
5.01501501501502 1.37903533806611e-06
5.03503503503504 1.24705294381396e-06
5.05505505505506 1.12725020473309e-06
5.07507507507508 1.01854844024876e-06
5.0950950950951 9.19960090959837e-07
5.11511511511512 8.30581438046261e-07
5.13513513513514 7.49585866251233e-07
5.15515515515516 6.76217633231267e-07
5.17517517517518 6.09786110324583e-07
5.1951951951952 5.4966046193278e-07
5.21521521521522 4.95264732746229e-07
5.23523523523524 4.46073313973501e-07
5.25525525525526 4.01606761563285e-07
5.27527527527528 3.61427941137511e-07
5.2952952952953 3.25138475990267e-07
5.31531531531532 2.92375476052561e-07
5.33533533533534 2.62808527181656e-07
5.35535535535536 2.36136921509202e-07
5.37537537537538 2.1208711087848e-07
5.3953953953954 1.90410366621162e-07
5.41541541541542 1.70880630071684e-07
5.43543543543544 1.5329253929596e-07
5.45545545545546 1.3745961852414e-07
5.47547547547548 1.23212617727566e-07
5.4954954954955 1.10397990671284e-07
5.51551551551552 9.8876500608364e-08
5.53553553553554 8.85219435638491e-08
5.55555555555556 7.92199798873018e-08
5.57557557557558 7.08670654362614e-08
5.5955955955956 6.33694743912418e-08
5.61561561561562 5.66424062986154e-08
5.63563563563564 5.06091704933412e-08
5.65565565565566 4.5200441571292e-08
5.67567567567568 4.03535800631662e-08
5.6956956956957 3.60120129107462e-08
5.71571571571572 3.2124668763613e-08
5.73573573573574 2.86454635022852e-08
5.75575575575576 2.55328317539416e-08
5.77577577577578 2.27493005011625e-08
5.7957957957958 2.02611011941336e-08
5.81581581581582 1.80378170640688e-08
5.83583583583584 1.60520626017053e-08
5.85585585585586 1.42791924110059e-08
5.87587587587588 1.2697036876002e-08
5.8958958958959 1.12856622892642e-08
5.91591591591592 1.00271532849868e-08
5.93593593593594 8.90541559921139e-09
5.95595595595596 7.90599734535251e-09
5.97597597597598 7.01592714588943e-09
5.995995995996 6.22356760178439e-09
6.01601601601602 5.5184827107339e-09
6.03603603603604 4.89131796456914e-09
6.05605605605606 4.33369196574694e-09
6.07607607607608 3.83809850362692e-09
6.0960960960961 3.3978181237675e-09
6.11611611611612 3.00683830841826e-09
6.13613613613614 2.65978146430924e-09
6.15615615615616 2.35183998527871e-09
6.17617617617618 2.07871772273778e-09
6.1961961961962 1.83657725691022e-09
6.21621621621622 1.6219924166386e-09
6.23623623623624 1.43190554571785e-09
6.25625625625626 1.26358905957511e-09
6.27627627627628 1.11461087800764e-09
6.2962962962963 9.82803357938119e-10
6.31631631631632 8.66235385045992e-10
6.33633633633634 7.63187314959515e-10
6.35635635635636 6.72128483698836e-10
6.37637637637638 5.91697033481532e-10
6.3963963963964 5.20681824054164e-10
6.41641641641642 4.58006221596989e-10
6.43643643643644 4.02713577147895e-10
6.45645645645646 3.53954224575805e-10
6.47647647647648 3.10973844559378e-10
6.4964964964965 2.73103055937435e-10
6.51651651651652 2.39748109325539e-10
6.53653653653654 2.10382570159619e-10
6.55655655655656 1.84539889444162e-10
6.57657657657658 1.61806770551276e-10
6.5965965965966 1.41817249531742e-10
6.61661661661662 1.24247414645767e-10
6.63663663663664 1.08810698278133e-10
6.65665665665666 9.52536811418137e-11
6.67667667667668 8.33523547614393e-11
6.6966966966967 7.29087937236022e-11
6.71671671671672 6.37481941394484e-11
6.73673673673674 5.57162392365998e-11
6.75675675675676 4.86767570277076e-11
6.77677677677678 4.25096386334907e-11
6.7967967967968 3.71089891068556e-11
6.81681681681682 3.23814855461048e-11
6.83683683683684 2.82449199306813e-11
6.85685685685686 2.46269064908966e-11
6.87687687687688 2.14637355595386e-11
6.8968968968969 1.86993577716892e-11
6.91691691691692 1.62844842008236e-11
6.93693693693694 1.41757895636779e-11
6.95695695695696 1.23352070109961e-11
6.97697697697698 1.07293042619725e-11
6.996996996997 9.32873195138548e-12
7.01701701701702 8.10773605306642e-12
7.03703703703704 7.04372713322415e-12
7.05705705705706 6.11689998287643e-12
7.07707707707708 5.30989788981512e-12
7.0970970970971 4.6075164458095e-12
7.11711711711712 3.99644235193917e-12
7.13713713713714 3.46502319108337e-12
7.15715715715716 3.00306458801651e-12
7.17717717717718 2.60165157997869e-12
7.1971971971972 2.25299137914217e-12
7.21721721721722 1.95027502769791e-12
7.23723723723724 1.68755573049414e-12
7.25725725725726 1.45964190299846e-12
7.27727727727728 1.262003197176e-12
7.2972972972973 1.0906879676822e-12
7.31731731731732 9.42250818252902e-13
7.33733733733734 8.13689025751829e-13
7.35735735735736 7.02386779168733e-13
7.37737737737738 6.06066294885245e-13
7.3973973973974 5.22744979473708e-13
7.41741741741742 4.50697908714381e-13
7.43743743743744 3.88424977793729e-13
7.45745745745746 3.34622154016963e-13
7.47747747747748 2.88156330935933e-13
7.4974974974975 2.48043342543511e-13
7.51751751751752 2.13428748996348e-13
7.53753753753754 1.83571051982055e-13
7.55755755755756 1.57827039041883e-13
7.57757757757758 1.35638992516663e-13
7.5975975975976 1.16523530854698e-13
7.61761761761762 1.000618782966e-13
7.63763763763764 8.58913838706873e-14
7.65765765765766 7.36981325813112e-14
7.67767767767768 6.32105109959248e-14
7.6976976976977 5.41936064405376e-14
7.71771771771772 4.64443339685915e-14
7.73773773773774 3.97871984155707e-14
7.75775775775776 3.40706104038536e-14
7.77777777777778 2.91636853079907e-14
7.7977977977978 2.4953463096687e-14
7.81781781781782 2.13424947819474e-14
7.83783783783784 1.82467480586654e-14
7.85785785785786 1.55937907247682e-14
7.87787787787788 1.33212157347804e-14
7.8978978978979 1.13752763482777e-14
7.91791791791792 9.70970386854701e-15
7.93793793793794 8.28468399583068e-15
7.95795795795796 7.06597090549298e-15
7.97797797797798 6.02412085866189e-15
7.997997997998 5.13382950919603e-15
8.01801801801802 4.3733591283238e-15
8.03803803803804 3.72404376402735e-15
8.05805805805806 3.16986191875719e-15
8.07807807807808 2.69706769497134e-15
8.0980980980981 2.29387254841619e-15
8.11811811811812 1.95017082606148e-15
8.13813813813814 1.65730316851105e-15
8.15815815815816 1.40785264250169e-15
8.17817817817818 1.19546915264237e-15
8.1981981981982 1.01471827585831e-15
8.21821821821822 8.60951178494166e-16
8.23823823823824 7.30192724674871e-16
8.25825825825826 6.19045274051723e-16
8.27827827827828 5.24606005103494e-16
8.2982982982983 4.44395893386326e-16
8.31831831831832 3.76298728354137e-16
8.33833833833834 3.18508772685576e-16
8.35835835835836 2.69485858889318e-16
8.37837837837838 2.27916883183252e-16
8.3983983983984 1.92682799625235e-16
8.41841841841842 1.62830341150177e-16
8.43843843843844 1.37547801096493e-16
8.45845845845846 1.16144301209627e-16
8.47847847847848 9.8032051926925e-17
8.4984984984985 8.27111796592674e-17
8.51851851851852 6.97567552529606e-17
8.53853853853854 5.88077091106193e-17
8.55855855855856 4.95573626747691e-17
8.57857857857858 4.17453440895294e-17
8.5985985985986 3.51506886838449e-17
8.61861861861862 2.95859531836935e-17
8.63863863863864 2.48921968838275e-17
8.65865865865866 2.09347039316773e-17
8.67867867867868 1.75993388643364e-17
8.6986986986987 1.47894429982973e-17
8.71871871871872 1.24231925503958e-17
8.73873873873874 1.04313507694241e-17
8.75875875875876 8.75535614211525e-18
8.77877877877878 7.34569713009337e-18
8.7987987987988 6.16053109048305e-18
8.81881881881882 5.16451119999019e-18
8.83883883883884 4.32779048512422e-18
8.85885885885886 3.62517658454128e-18
8.87887887887888 3.03541474067764e-18
8.8988988988989 2.54057982941549e-18
8.91891891891892 2.12556106807901e-18
8.93893893893894 1.77762546207239e-18
8.95895895895896 1.48604811781133e-18
8.97897897897898 1.24179931485728e-18
8.998998998999 1.03727973679138e-18
9.01901901901902 8.66096545670873e-19
9.03903903903904 7.22874080905507e-19
9.05905905905906 6.03093897535385e-19
9.07907907907908 5.0295965472299e-19
9.0990990990991 4.19283042962206e-19
9.11911911911912 3.49387515332417e-19
9.13913913913914 2.91027078875695e-19
9.15915915915916 2.42317819504618e-19
9.17917917917918 2.01680188581445e-19
9.1991991991992 1.67790380698338e-19
9.21921921921922 1.39539388139188e-19
9.23923923923924 1.1599853476858e-19
9.25925925925926 9.63904764362469e-20
9.27927927927928 8.00648113242436e-20
9.2992992992993 6.6477576193283e-20
9.31931931931932 5.51740167796855e-20
9.33933933933934 4.5774115701642e-20
9.35935935935936 3.79604417474457e-20
9.37937937937938 3.14679525475421e-20
9.3993993993994 2.60754402558043e-20
9.41941941941942 2.15983585814365e-20
9.43943943943944 1.78828106797973e-20
9.45945945945946 1.48005121824234e-20
9.47947947947948 1.22445730038445e-20
9.4994994994995 1.01259663374544e-20
9.51951951951952 8.37057415073758e-21
9.53953953953954 6.91671611023779e-21
9.55955955955956 5.71308371631411e-21
9.57957957957958 4.71701393695898e-21
9.5995995995996 3.89304716291232e-21
9.61961961961962 3.21172317125736e-21
9.63963963963964 2.64857624243904e-21
9.65965965965966 2.18329684678274e-21
9.67967967967968 1.79903258756111e-21
9.6996996996997 1.48180551599987e-21
9.71971971971972 1.22002665237565e-21
9.73973973973974 1.00409166885045e-21
9.75975975975976 8.26044308669654e-22
9.77977977977978 6.79296312742742e-22
9.7997997997998 5.58394465795474e-22
9.81981981981982 4.58826916995414e-22
9.83983983983984 3.7686222201397e-22
9.85985985985986 3.09415635142992e-22
9.87987987987988 2.53938085193762e-22
9.8998998998999 2.08324025950642e-22
9.91991991991992 1.70834984871876e-22
9.93993993993994 1.40036162642795e-22
9.95995995995996 1.14743877987917e-22
9.97997997997998 9.39820210218911e-23
10 7.69459862670642e-23
};
\addlegendentry{N(0,1)}; 
\legend{}; 
\end{axis}

\end{tikzpicture}

%% file: FinalFigs/Null_Dists_d_10_500_n_20_m_200_kernel__Gaussian_RBF_2022_10_12_22_14_00mmd.tex
\begin{tikzpicture}

\definecolor{darkorange25512714}{RGB}{255,127,14}
\definecolor{darkslategray38}{RGB}{38,38,38}
\definecolor{lightgray204}{RGB}{204,204,204}
\definecolor{steelblue31119180}{RGB}{31,119,180}

\begin{axis}[
axis line style={darkslategray38},
height=\figheight,
legend cell align={left},
legend style={fill opacity=0.8, draw opacity=1, text opacity=1, draw=none},
tick align=outside,
tick pos=left,
title={$\dmmd~(n/m=0.1)$},
width=\figwidth,
x grid style={lightgray204},
xmin=-6, xmax=6,
xtick style={color=darkslategray38},
y grid style={lightgray204},
ylabel=\textcolor{darkslategray38}{Probability density},
ymin=0, ymax=2.1377055648682,
ytick style={color=darkslategray38}, 
xticklabels=empty,
yticklabels=empty
]
\draw[draw=none,fill=steelblue31119180,fill opacity=0.8] (axis cs:-1.2091383934021,0) rectangle (axis cs:-1.07475125789642,0.11310610451011);
\addlegendimage{ybar,ybar legend,draw=none,fill=steelblue31119180,fill opacity=0.8}
\addlegendentry{mmd (d=10)}

\draw[draw=none,fill=steelblue31119180,fill opacity=0.8] (axis cs:-0.873170495033264,0) rectangle (axis cs:-0.738783478736877,0.351224281579576);
\draw[draw=none,fill=steelblue31119180,fill opacity=0.8] (axis cs:-0.537202894687653,0) rectangle (axis cs:-0.402815759181976,0.69649548566752);
\draw[draw=none,fill=steelblue31119180,fill opacity=0.8] (axis cs:-0.201235204935074,0) rectangle (axis cs:-0.0668480694293976,0.666730780465715);
\draw[draw=none,fill=steelblue31119180,fill opacity=0.8] (axis cs:0.134732484817505,0) rectangle (axis cs:0.269119620323181,0.541718759128393);
\draw[draw=none,fill=steelblue31119180,fill opacity=0.8] (axis cs:0.470700204372406,0) rectangle (axis cs:0.605087339878082,0.386941936481956);
\draw[draw=none,fill=steelblue31119180,fill opacity=0.8] (axis cs:0.806667983531952,0) rectangle (axis cs:0.941054999828339,0.172635663827249);
\draw[draw=none,fill=steelblue31119180,fill opacity=0.8] (axis cs:1.14263558387756,0) rectangle (axis cs:1.27702260017395,0.0297647696253878);
\draw[draw=none,fill=steelblue31119180,fill opacity=0.8] (axis cs:1.47860336303711,0) rectangle (axis cs:1.61299049854279,0.0119059036256583);
\draw[draw=none,fill=steelblue31119180,fill opacity=0.8] (axis cs:1.81457102298737,0) rectangle (axis cs:1.94895803928375,0.00595295392507756);
\draw[draw=none,fill=darkorange25512714,fill opacity=0.8] (axis cs:-1.07475113868713,0) rectangle (axis cs:-0.940364062786102,0);
\addlegendimage{ybar,ybar legend,draw=none,fill=darkorange25512714,fill opacity=0.8}
\addlegendentry{mmd (d=500)}

\draw[draw=none,fill=darkorange25512714,fill opacity=0.8] (axis cs:-0.738783478736877,0) rectangle (axis cs:-0.604396402835846,0);
\draw[draw=none,fill=darkorange25512714,fill opacity=0.8] (axis cs:-0.402815788984299,0) rectangle (axis cs:-0.268428713083267,0.214306303282314);
\draw[draw=none,fill=darkorange25512714,fill opacity=0.8] (axis cs:-0.06684809923172,0) rectangle (axis cs:0.0675389766693115,2.03591006177924);
\draw[draw=none,fill=darkorange25512714,fill opacity=0.8] (axis cs:0.269119590520859,0) rectangle (axis cs:0.40350666642189,0.702448500847807);
\draw[draw=none,fill=darkorange25512714,fill opacity=0.8] (axis cs:0.605087280273438,0) rectangle (axis cs:0.739474356174469,0.0238118114758126);
\draw[draw=none,fill=darkorange25512714,fill opacity=0.8] (axis cs:0.941055059432983,0) rectangle (axis cs:1.07544207572937,0);
\draw[draw=none,fill=darkorange25512714,fill opacity=0.8] (axis cs:1.27702271938324,0) rectangle (axis cs:1.41140985488892,0);
\draw[draw=none,fill=darkorange25512714,fill opacity=0.8] (axis cs:1.6129903793335,0) rectangle (axis cs:1.74737751483917,0);
\draw[draw=none,fill=darkorange25512714,fill opacity=0.8] (axis cs:1.94895815849304,0) rectangle (axis cs:2.08334517478943,0);
\addplot [semithick, black]
table {%
-10 7.69459862670642e-23
-9.97997997997998 9.39820210218911e-23
-9.95995995995996 1.14743877987917e-22
-9.93993993993994 1.40036162642795e-22
-9.91991991991992 1.70834984871876e-22
-9.8998998998999 2.08324025950642e-22
-9.87987987987988 2.53938085193762e-22
-9.85985985985986 3.09415635142992e-22
-9.83983983983984 3.7686222201397e-22
-9.81981981981982 4.58826916995414e-22
-9.7997997997998 5.58394465795474e-22
-9.77977977977978 6.79296312742742e-22
-9.75975975975976 8.26044308669654e-22
-9.73973973973974 1.00409166885045e-21
-9.71971971971972 1.22002665237565e-21
-9.6996996996997 1.48180551599987e-21
-9.67967967967968 1.79903258756111e-21
-9.65965965965966 2.18329684678274e-21
-9.63963963963964 2.64857624243904e-21
-9.61961961961962 3.21172317125736e-21
-9.5995995995996 3.89304716291232e-21
-9.57957957957958 4.71701393695898e-21
-9.55955955955956 5.71308371631411e-21
-9.53953953953954 6.91671611023779e-21
-9.51951951951952 8.37057415073758e-21
-9.4994994994995 1.01259663374544e-20
-9.47947947947948 1.22445730038445e-20
-9.45945945945946 1.48005121824234e-20
-9.43943943943944 1.78828106797973e-20
-9.41941941941942 2.15983585814365e-20
-9.3993993993994 2.60754402558043e-20
-9.37937937937938 3.14679525475421e-20
-9.35935935935936 3.79604417474457e-20
-9.33933933933934 4.5774115701642e-20
-9.31931931931932 5.51740167796855e-20
-9.2992992992993 6.6477576193283e-20
-9.27927927927928 8.00648113242436e-20
-9.25925925925926 9.63904764362469e-20
-9.23923923923924 1.1599853476858e-19
-9.21921921921922 1.39539388139188e-19
-9.1991991991992 1.67790380698338e-19
-9.17917917917918 2.01680188581445e-19
-9.15915915915916 2.42317819504618e-19
-9.13913913913914 2.91027078875695e-19
-9.11911911911912 3.49387515332417e-19
-9.0990990990991 4.19283042962206e-19
-9.07907907907908 5.0295965472299e-19
-9.05905905905906 6.03093897535385e-19
-9.03903903903904 7.22874080905507e-19
-9.01901901901902 8.66096545670873e-19
-8.998998998999 1.03727973679138e-18
-8.97897897897898 1.24179931485728e-18
-8.95895895895896 1.48604811781133e-18
-8.93893893893894 1.77762546207239e-18
-8.91891891891892 2.12556106807901e-18
-8.8988988988989 2.54057982941549e-18
-8.87887887887888 3.03541474067764e-18
-8.85885885885886 3.62517658454128e-18
-8.83883883883884 4.32779048512422e-18
-8.81881881881882 5.16451119999019e-18
-8.7987987987988 6.16053109048305e-18
-8.77877877877878 7.34569713009337e-18
-8.75875875875876 8.75535614211525e-18
-8.73873873873874 1.04313507694241e-17
-8.71871871871872 1.24231925503958e-17
-8.6986986986987 1.47894429982973e-17
-8.67867867867868 1.75993388643364e-17
-8.65865865865866 2.09347039316773e-17
-8.63863863863864 2.48921968838275e-17
-8.61861861861862 2.95859531836935e-17
-8.5985985985986 3.51506886838449e-17
-8.57857857857858 4.17453440895294e-17
-8.55855855855856 4.95573626747691e-17
-8.53853853853854 5.88077091106193e-17
-8.51851851851852 6.97567552529606e-17
-8.4984984984985 8.27111796592674e-17
-8.47847847847848 9.8032051926925e-17
-8.45845845845846 1.16144301209627e-16
-8.43843843843844 1.37547801096493e-16
-8.41841841841842 1.62830341150177e-16
-8.3983983983984 1.92682799625235e-16
-8.37837837837838 2.27916883183252e-16
-8.35835835835836 2.69485858889318e-16
-8.33833833833834 3.18508772685576e-16
-8.31831831831832 3.76298728354137e-16
-8.2982982982983 4.44395893386326e-16
-8.27827827827828 5.24606005103494e-16
-8.25825825825826 6.19045274051723e-16
-8.23823823823824 7.30192724674871e-16
-8.21821821821822 8.60951178494166e-16
-8.1981981981982 1.01471827585831e-15
-8.17817817817818 1.19546915264237e-15
-8.15815815815816 1.40785264250169e-15
-8.13813813813814 1.65730316851105e-15
-8.11811811811812 1.95017082606148e-15
-8.0980980980981 2.29387254841619e-15
-8.07807807807808 2.69706769497134e-15
-8.05805805805806 3.16986191875719e-15
-8.03803803803804 3.72404376402735e-15
-8.01801801801802 4.3733591283238e-15
-7.997997997998 5.13382950919607e-15
-7.97797797797798 6.02412085866193e-15
-7.95795795795796 7.06597090549303e-15
-7.93793793793794 8.28468399583074e-15
-7.91791791791792 9.70970386854708e-15
-7.8978978978979 1.13752763482777e-14
-7.87787787787788 1.33212157347805e-14
-7.85785785785786 1.55937907247683e-14
-7.83783783783784 1.82467480586655e-14
-7.81781781781782 2.13424947819475e-14
-7.7977977977978 2.49534630966872e-14
-7.77777777777778 2.91636853079909e-14
-7.75775775775776 3.40706104038538e-14
-7.73773773773774 3.9787198415571e-14
-7.71771771771772 4.64443339685918e-14
-7.6976976976977 5.4193606440538e-14
-7.67767767767768 6.32105109959252e-14
-7.65765765765766 7.36981325813117e-14
-7.63763763763764 8.58913838706879e-14
-7.61761761761762 1.00061878296601e-13
-7.5975975975976 1.16523530854699e-13
-7.57757757757758 1.35638992516664e-13
-7.55755755755756 1.57827039041884e-13
-7.53753753753754 1.83571051982057e-13
-7.51751751751752 2.13428748996349e-13
-7.4974974974975 2.48043342543513e-13
-7.47747747747748 2.88156330935935e-13
-7.45745745745746 3.34622154016965e-13
-7.43743743743744 3.88424977793732e-13
-7.41741741741742 4.50697908714384e-13
-7.3973973973974 5.22744979473711e-13
-7.37737737737738 6.06066294885249e-13
-7.35735735735736 7.02386779168738e-13
-7.33733733733734 8.13689025751835e-13
-7.31731731731732 9.42250818252909e-13
-7.2972972972973 1.09068796768221e-12
-7.27727727727728 1.262003197176e-12
-7.25725725725726 1.45964190299847e-12
-7.23723723723724 1.68755573049416e-12
-7.21721721721722 1.95027502769792e-12
-7.1971971971972 2.25299137914218e-12
-7.17717717717718 2.60165157997871e-12
-7.15715715715716 3.00306458801653e-12
-7.13713713713714 3.4650231910834e-12
-7.11711711711712 3.99644235193919e-12
-7.0970970970971 4.60751644580953e-12
-7.07707707707708 5.30989788981514e-12
-7.05705705705706 6.11689998287646e-12
-7.03703703703704 7.0437271332242e-12
-7.01701701701702 8.10773605306645e-12
-6.996996996997 9.32873195138555e-12
-6.97697697697698 1.07293042619726e-11
-6.95695695695696 1.23352070109962e-11
-6.93693693693694 1.41757895636779e-11
-6.91691691691692 1.62844842008237e-11
-6.8968968968969 1.86993577716893e-11
-6.87687687687688 2.14637355595386e-11
-6.85685685685686 2.46269064908967e-11
-6.83683683683684 2.82449199306815e-11
-6.81681681681682 3.2381485546105e-11
-6.7967967967968 3.71089891068559e-11
-6.77677677677678 4.25096386334913e-11
-6.75675675675676 4.86767570277083e-11
-6.73673673673674 5.57162392366004e-11
-6.71671671671672 6.37481941394491e-11
-6.6966966966967 7.29087937236032e-11
-6.67667667667668 8.33523547614402e-11
-6.65665665665666 9.52536811418151e-11
-6.63663663663664 1.08810698278135e-10
-6.61661661661662 1.24247414645768e-10
-6.5965965965966 1.41817249531744e-10
-6.57657657657658 1.61806770551278e-10
-6.55655655655656 1.84539889444164e-10
-6.53653653653654 2.10382570159622e-10
-6.51651651651652 2.39748109325542e-10
-6.4964964964965 2.73103055937438e-10
-6.47647647647648 3.10973844559381e-10
-6.45645645645646 3.53954224575809e-10
-6.43643643643644 4.027135771479e-10
-6.41641641641642 4.58006221596996e-10
-6.3963963963964 5.20681824054169e-10
-6.37637637637638 5.91697033481538e-10
-6.35635635635636 6.72128483698846e-10
-6.33633633633634 7.63187314959523e-10
-6.31631631631632 8.66235385046001e-10
-6.2962962962963 9.8280335793813e-10
-6.27627627627628 1.11461087800766e-09
-6.25625625625626 1.26358905957513e-09
-6.23623623623624 1.43190554571787e-09
-6.21621621621622 1.62199241663862e-09
-6.1961961961962 1.83657725691024e-09
-6.17617617617618 2.0787177227378e-09
-6.15615615615616 2.35183998527873e-09
-6.13613613613614 2.65978146430928e-09
-6.11611611611612 3.00683830841829e-09
-6.0960960960961 3.39781812376754e-09
-6.07607607607608 3.83809850362696e-09
-6.05605605605606 4.33369196574699e-09
-6.03603603603604 4.89131796456919e-09
-6.01601601601602 5.51848271073395e-09
-5.995995995996 6.22356760178439e-09
-5.97597597597598 7.01592714588943e-09
-5.95595595595596 7.90599734535251e-09
-5.93593593593594 8.90541559921139e-09
-5.91591591591592 1.00271532849868e-08
-5.8958958958959 1.12856622892642e-08
-5.87587587587588 1.2697036876002e-08
-5.85585585585586 1.42791924110059e-08
-5.83583583583584 1.60520626017053e-08
-5.81581581581582 1.80378170640688e-08
-5.7957957957958 2.02611011941336e-08
-5.77577577577578 2.27493005011625e-08
-5.75575575575576 2.55328317539416e-08
-5.73573573573574 2.86454635022852e-08
-5.71571571571572 3.2124668763613e-08
-5.6956956956957 3.60120129107462e-08
-5.67567567567568 4.03535800631662e-08
-5.65565565565566 4.5200441571292e-08
-5.63563563563564 5.06091704933412e-08
-5.61561561561562 5.66424062986154e-08
-5.5955955955956 6.33694743912418e-08
-5.57557557557558 7.08670654362614e-08
-5.55555555555556 7.92199798873018e-08
-5.53553553553554 8.85219435638491e-08
-5.51551551551552 9.8876500608364e-08
-5.4954954954955 1.10397990671284e-07
-5.47547547547548 1.23212617727566e-07
-5.45545545545546 1.3745961852414e-07
-5.43543543543544 1.5329253929596e-07
-5.41541541541542 1.70880630071684e-07
-5.3953953953954 1.90410366621162e-07
-5.37537537537538 2.1208711087848e-07
-5.35535535535536 2.36136921509202e-07
-5.33533533533534 2.62808527181656e-07
-5.31531531531532 2.92375476052561e-07
-5.2952952952953 3.25138475990267e-07
-5.27527527527528 3.61427941137511e-07
-5.25525525525526 4.01606761563285e-07
-5.23523523523524 4.46073313973501e-07
-5.21521521521522 4.95264732746229e-07
-5.1951951951952 5.4966046193278e-07
-5.17517517517518 6.09786110324583e-07
-5.15515515515516 6.76217633231267e-07
-5.13513513513514 7.49585866251233e-07
-5.11511511511512 8.30581438046261e-07
-5.0950950950951 9.19960090959837e-07
-5.07507507507508 1.01854844024876e-06
-5.05505505505506 1.12725020473309e-06
-5.03503503503504 1.24705294381396e-06
-5.01501501501502 1.37903533806611e-06
-4.99499499499499 1.5243750529858e-06
-4.97497497497497 1.68435722796805e-06
-4.95495495495495 1.86038363520377e-06
-4.93493493493493 2.05398255592983e-06
-4.91491491491491 2.26681942433715e-06
-4.89489489489489 2.50070829244518e-06
-4.87487487487487 2.75762417238901e-06
-4.85485485485485 3.03971631583941e-06
-4.83483483483483 3.34932249368796e-06
-4.81481481481481 3.68898434268124e-06
-4.79479479479479 4.06146384937932e-06
-4.77477477477477 4.4697610456467e-06
-4.75475475475475 4.91713299385613e-06
-4.73473473473473 5.40711414409909e-06
-4.71471471471471 5.94353814994721e-06
-4.69469469469469 6.53056123369604e-06
-4.67467467467467 7.17268719654363e-06
-4.65465465465465 7.87479417380527e-06
-4.63463463463463 8.64216324004121e-06
-4.61461461461461 9.48050897386715e-06
-4.59459459459459 1.03960120972233e-05
-4.57457457457457 1.13953543089884e-05
-4.55455455455455 1.24857554380297e-05
-4.53453453453453 1.3675013046071e-05
-4.51451451451451 1.49715446161227e-05
-4.49449449449449 1.63844324676437e-05
-4.47447447447447 1.79234715450684e-05
-4.45445445445445 1.95992202318354e-05
-4.43443443443443 2.14230543475548e-05
-4.41441441441441 2.34072244914564e-05
-4.39439439439439 2.55649169007242e-05
-4.37437437437437 2.79103179977393e-05
-4.35435435435435 3.04586828055866e-05
-4.33433433433433 3.32264074164067e-05
-4.31431431431431 3.62311057022691e-05
-4.29429429429429 3.94916904631592e-05
-4.27427427427427 4.3028459211397e-05
-4.25425425425425 4.68631847962824e-05
-4.23423423423423 5.10192110769697e-05
-4.21421421421421 5.55215538554582e-05
-4.19419419419419 6.03970072851107e-05
-4.17417417417417 6.56742559732345e-05
-4.15415415415415 7.13839929989176e-05
-4.13413413413413 7.75590440694795e-05
-4.11411411411411 8.42344980404937e-05
-4.09409409409409 9.14478440253317e-05
-4.07407407407407 9.92391153205018e-05
-4.05405405405405 0.000107651040372646
-4.03403403403403 0.000116729201011866
-4.01401401401401 0.000126522198173995
-3.99399399399399 0.000137081825331481
-3.97397397397397 0.000148463249848567
-3.95395395395395 0.000160725202471485
-3.93393393393393 0.000173930175158222
-3.91391391391391 0.00018814462744512
-3.89389389389389 0.000203439201538965
-3.87387387387387 0.000219888946313312
-3.85385385385385 0.000237573550376443
-3.83383383383383 0.000256577584365551
-3.81381381381381 0.000276990752607344
-3.79379379379379 0.000298908154269281
-3.77377377377377 0.000322430554107926
-3.75375375375375 0.000347664662901427
-3.73373373373373 0.000374723427631836
-3.71371371371371 0.000403726331459719
-3.69369369369369 0.000434799703508357
-3.67367367367367 0.000468077038447599
-3.65365365365365 0.00050369932583812
-3.63363363363363 0.000541815389165405
-3.61361361361361 0.000582582234459159
-3.59359359359359 0.000626165408357979
-3.57357357357357 0.000672739365441021
-3.55355355355355 0.000722487844607978
-3.53353353353353 0.00077560425424601
-3.51351351351351 0.000832292065877155
-3.49349349349349 0.000892765215932443
-3.47347347347347 0.000957248515249216
-3.45345345345345 0.0010259780658361
-3.43343343343343 0.00109920168439588
-3.41341341341341 0.00117717933203981
-3.39339339339339 0.00126018354956833
-3.37337337337337 0.00134849989763212
-3.35335335335335 0.00144242740102448
-3.33333333333333 0.00154227899629111
-3.31331331331331 0.00164838198177652
-3.29329329329329 0.00176107846915772
-3.27327327327327 0.00188072583544552
-3.25325325325325 0.00200769717436226
-3.23323323323323 0.00214238174593163
-3.21321321321321 0.00228518542304204
-3.19319319319319 0.00243653113367012
-3.17317317317317 0.00259685929737497
-3.15315315315315 0.00276662825459747
-3.13313313313313 0.00294631468722261
-3.11311311311311 0.00313641402878609
-3.09309309309309 0.00333744086263052
-3.07307307307307 0.00354992930624086
-3.05305305305305 0.00377443337991422
-3.03303303303303 0.00401152735784579
-3.01301301301301 0.00426180609964128
-2.99299299299299 0.00452588536019618
-2.97297297297297 0.0048044020758154
-2.95295295295295 0.00509801462438215
-2.93293293293293 0.00540740305732385
-2.91291291291291 0.00573326930106519
-2.89289289289289 0.00607633732560526
-2.87287287287287 0.00643735327780636
-2.85285285285285 0.00681708557693873
-2.83283283283283 0.00721632496998623
-2.81281281281281 0.00763588454418632
-2.79279279279279 0.00807659969425075
-2.77277277277277 0.00853932804169477
-2.75275275275275 0.00902494930369032
-2.73273273273273 0.00953436510885489
-2.71271271271271 0.0100684987573917
-2.69269269269269 0.0106282949230102
-2.67267267267267 0.0112147192940778
-2.65265265265265 0.0118287581514852
-2.63263263263263 0.0124714178807513
-2.61261261261261 0.0131437244159435
-2.59259259259259 0.0138467226130541
-2.57257257257257 0.0145814755505492
-2.55255255255255 0.0153490637548887
-2.53253253253253 0.0161505843489182
-2.51251251251251 0.0169871501211409
-2.49249249249249 0.0178598885140022
-2.47247247247247 0.018769940529451
-2.45245245245245 0.0197184595501959
-2.43243243243243 0.0207066100752274
-2.41241241241241 0.0217355663683581
-2.39239239239239 0.0228065110187135
-2.37237237237237 0.0239206334123108
-2.35235235235235 0.0250791281140699
-2.33233233233233 0.0262831931598317
-2.31231231231231 0.0275340282581906
-2.29229229229229 0.0288328329022027
-2.27227227227227 0.0301808043912899
-2.25225225225225 0.0315791357639331
-2.23223223223223 0.033029013642035
-2.21221221221221 0.0345316159881232
-2.19219219219219 0.0360881097768736
-2.17217217217217 0.0376996485827434
-2.15215215215215 0.0393673700858293
-2.13213213213213 0.0410923934983949
-2.11211211211211 0.0428758169148479
-2.09209209209209 0.0447187145882915
-2.07207207207207 0.0466221341371235
-2.05205205205205 0.0485870936855041
-2.03203203203203 0.0506145789418747
-2.01201201201201 0.0527055402200587
-1.99199199199199 0.0548608894078376
-1.97197197197197 0.057081496888248
-1.95195195195195 0.059368188419199
-1.93193193193193 0.0617217419773594
-1.91191191191191 0.0641428845726061
-1.89189189189189 0.0666322890396674
-1.87187187187187 0.0691905708139176
-1.85185185185185 0.0718182846986055
-1.83183183183183 0.0745159216311036
-1.81181181181181 0.077283905456062
-1.79179179179179 0.0801225897136326
-1.77177177177177 0.083032254451193
-1.75175175175175 0.0860131030672496
-1.73173173173173 0.0890652591964251
-1.71171171171171 0.0921887636446459
-1.69169169169169 0.0953835713838294
-1.67167167167167 0.0986495486155338
-1.65165165165165 0.101986469913169
-1.63163163163163 0.10539401545248
-1.61161161161161 0.108871768340093
-1.59159159159159 0.112419212049971
-1.57157157157157 0.116035727977651
-1.55155155155155 0.119720593122119
-1.53153153153153 0.123472977905145
-1.51151151151151 0.127291944137829
-1.49149149149149 0.13117644314399
-1.47147147147147 0.135125314049902
-1.45145145145145 0.139137282249685
-1.43143143143143 0.143210958055468
-1.41141141141141 0.147344835541168
-1.39139139139139 0.151537291588457
-1.37137137137137 0.155786585143159
-1.35135135135135 0.160090856689972
-1.33133133133133 0.164448127952996
-1.31131131131131 0.168856301829129
-1.29129129129129 0.173313162560933
-1.27127127127127 0.17781637615506
-1.25125125125125 0.182363491051798
-1.23123123123123 0.186951939050736
-1.21121121121121 0.191579036496956
-1.19119119119119 0.1962419857315
-1.17117117117117 0.200937876809264
-1.15115115115115 0.205663689486728
-1.13113113113113 0.210416295481265
-1.11111111111111 0.215192461003031
-1.09109109109109 0.219988849559688
-1.07107107107107 0.224802025033432
-1.05105105105105 0.229628455029052
-1.03103103103103 0.234464514490888
-1.01101101101101 0.239306489585817
-0.990990990990991 0.24415058184851
-0.970970970970971 0.24899291258444
-0.950950950950951 0.253829527525259
-0.930930930930931 0.258656401730343
-0.910910910910911 0.2634694447275
-0.890890890890891 0.268264505884996
-0.870870870870871 0.273037380006279
-0.850850850850851 0.277783813137949
-0.830830830830831 0.282499508580786
-0.810810810810811 0.287180133092853
-0.790790790790791 0.291821323272996
-0.77077077077077 0.296418692112302
-0.75075075075075 0.300967835700437
-0.73073073073073 0.305464340073112
-0.71071071071071 0.309903788186304
-0.69069069069069 0.314281767002296
-0.67067067067067 0.318593874672039
-0.65065065065065 0.322835727797843
-0.63063063063063 0.327002968759958
-0.61061061061061 0.331091273090187
-0.59059059059059 0.33509635687531
-0.57057057057057 0.339013984172804
-0.55055055055055 0.34283997442106
-0.53053053053053 0.346570209826128
-0.51051051051051 0.350200642706842
-0.49049049049049 0.353727302780113
-0.47047047047047 0.357146304368113
-0.45045045045045 0.360453853509139
-0.43043043043043 0.363646254953996
-0.41041041041041 0.366719919029892
-0.39039039039039 0.369671368354051
-0.37037037037037 0.372497244379499
-0.35035035035035 0.375194313755802
-0.33033033033033 0.377759474487924
-0.31031031031031 0.38018976187679
-0.29029029029029 0.382482354225654
-0.27027027027027 0.384634578296894
-0.25025025025025 0.386643914504485
-0.23023023023023 0.388508001828027
-0.21021021021021 0.390224642434919
-0.19019019019019 0.391791805998011
-0.17017017017017 0.393207633696876
-0.15015015015015 0.394470441891644
-0.13013013013013 0.395578725459258
-0.11011011011011 0.396531160782876
-0.0900900900900901 0.397326608386124
-0.07007007007007 0.397964115204853
-0.05005005005005 0.398442916490068
-0.03003003003003 0.398762437336696
-0.01001001001001 0.398922293833933
0.01001001001001 0.398922293833933
0.03003003003003 0.398762437336696
0.05005005005005 0.398442916490068
0.07007007007007 0.397964115204853
0.0900900900900901 0.397326608386124
0.11011011011011 0.396531160782876
0.13013013013013 0.395578725459258
0.15015015015015 0.394470441891644
0.17017017017017 0.393207633696876
0.19019019019019 0.391791805998011
0.21021021021021 0.390224642434919
0.23023023023023 0.388508001828027
0.25025025025025 0.386643914504485
0.27027027027027 0.384634578296894
0.29029029029029 0.382482354225654
0.31031031031031 0.38018976187679
0.33033033033033 0.377759474487924
0.35035035035035 0.375194313755802
0.37037037037037 0.372497244379499
0.39039039039039 0.369671368354051
0.41041041041041 0.366719919029892
0.43043043043043 0.363646254953996
0.45045045045045 0.360453853509139
0.47047047047047 0.357146304368113
0.49049049049049 0.353727302780113
0.51051051051051 0.350200642706842
0.53053053053053 0.346570209826128
0.55055055055055 0.34283997442106
0.57057057057057 0.339013984172804
0.59059059059059 0.33509635687531
0.61061061061061 0.331091273090187
0.63063063063063 0.327002968759958
0.65065065065065 0.322835727797843
0.67067067067067 0.318593874672039
0.69069069069069 0.314281767002296
0.71071071071071 0.309903788186304
0.73073073073073 0.305464340073112
0.75075075075075 0.300967835700437
0.77077077077077 0.296418692112302
0.790790790790791 0.291821323272996
0.810810810810811 0.287180133092853
0.830830830830831 0.282499508580786
0.850850850850851 0.277783813137949
0.870870870870871 0.273037380006279
0.890890890890891 0.268264505884996
0.910910910910911 0.2634694447275
0.930930930930931 0.258656401730343
0.950950950950951 0.253829527525259
0.970970970970971 0.24899291258444
0.990990990990991 0.24415058184851
1.01101101101101 0.239306489585817
1.03103103103103 0.234464514490888
1.05105105105105 0.229628455029052
1.07107107107107 0.224802025033432
1.09109109109109 0.219988849559688
1.11111111111111 0.215192461003031
1.13113113113113 0.210416295481265
1.15115115115115 0.205663689486728
1.17117117117117 0.200937876809264
1.19119119119119 0.1962419857315
1.21121121121121 0.191579036496956
1.23123123123123 0.186951939050736
1.25125125125125 0.182363491051798
1.27127127127127 0.17781637615506
1.29129129129129 0.173313162560933
1.31131131131131 0.168856301829129
1.33133133133133 0.164448127952996
1.35135135135135 0.160090856689972
1.37137137137137 0.155786585143159
1.39139139139139 0.151537291588457
1.41141141141141 0.147344835541168
1.43143143143143 0.143210958055468
1.45145145145145 0.139137282249685
1.47147147147147 0.135125314049902
1.49149149149149 0.13117644314399
1.51151151151151 0.127291944137829
1.53153153153153 0.123472977905145
1.55155155155155 0.119720593122119
1.57157157157157 0.116035727977651
1.59159159159159 0.112419212049971
1.61161161161161 0.108871768340093
1.63163163163163 0.10539401545248
1.65165165165165 0.101986469913169
1.67167167167167 0.0986495486155338
1.69169169169169 0.0953835713838294
1.71171171171171 0.0921887636446459
1.73173173173173 0.0890652591964251
1.75175175175175 0.0860131030672496
1.77177177177177 0.083032254451193
1.79179179179179 0.0801225897136326
1.81181181181181 0.077283905456062
1.83183183183183 0.0745159216311036
1.85185185185185 0.0718182846986055
1.87187187187187 0.0691905708139176
1.89189189189189 0.0666322890396674
1.91191191191191 0.0641428845726061
1.93193193193193 0.0617217419773594
1.95195195195195 0.059368188419199
1.97197197197197 0.057081496888248
1.99199199199199 0.0548608894078376
2.01201201201201 0.0527055402200588
2.03203203203203 0.0506145789418748
2.05205205205205 0.0485870936855042
2.07207207207207 0.0466221341371236
2.09209209209209 0.0447187145882916
2.11211211211211 0.0428758169148479
2.13213213213213 0.041092393498395
2.15215215215215 0.0393673700858294
2.17217217217217 0.0376996485827434
2.19219219219219 0.0360881097768737
2.21221221221221 0.0345316159881233
2.23223223223223 0.033029013642035
2.25225225225225 0.0315791357639332
2.27227227227227 0.03018080439129
2.29229229229229 0.0288328329022028
2.31231231231231 0.0275340282581906
2.33233233233233 0.0262831931598317
2.35235235235235 0.02507912811407
2.37237237237237 0.0239206334123108
2.39239239239239 0.0228065110187136
2.41241241241241 0.0217355663683581
2.43243243243243 0.0207066100752275
2.45245245245245 0.0197184595501959
2.47247247247247 0.0187699405294511
2.49249249249249 0.0178598885140022
2.51251251251251 0.016987150121141
2.53253253253253 0.0161505843489182
2.55255255255255 0.0153490637548888
2.57257257257257 0.0145814755505493
2.59259259259259 0.0138467226130541
2.61261261261261 0.0131437244159435
2.63263263263263 0.0124714178807514
2.65265265265265 0.0118287581514852
2.67267267267267 0.0112147192940778
2.69269269269269 0.0106282949230103
2.71271271271271 0.0100684987573917
2.73273273273273 0.00953436510885491
2.75275275275275 0.00902494930369034
2.77277277277277 0.00853932804169479
2.79279279279279 0.00807659969425077
2.81281281281281 0.00763588454418634
2.83283283283283 0.00721632496998621
2.85285285285285 0.00681708557693871
2.87287287287287 0.00643735327780635
2.89289289289289 0.00607633732560524
2.91291291291291 0.00573326930106518
2.93293293293293 0.00540740305732384
2.95295295295295 0.00509801462438214
2.97297297297297 0.00480440207581539
2.99299299299299 0.00452588536019617
3.01301301301301 0.00426180609964127
3.03303303303303 0.00401152735784578
3.05305305305305 0.00377443337991421
3.07307307307307 0.00354992930624085
3.09309309309309 0.00333744086263051
3.11311311311311 0.00313641402878608
3.13313313313313 0.0029463146872226
3.15315315315315 0.00276662825459746
3.17317317317317 0.00259685929737496
3.19319319319319 0.00243653113367011
3.21321321321321 0.00228518542304203
3.23323323323323 0.00214238174593163
3.25325325325325 0.00200769717436225
3.27327327327327 0.00188072583544551
3.29329329329329 0.00176107846915771
3.31331331331331 0.00164838198177652
3.33333333333333 0.0015422789962911
3.35335335335335 0.00144242740102448
3.37337337337337 0.00134849989763212
3.39339339339339 0.00126018354956833
3.41341341341341 0.00117717933203981
3.43343343343343 0.00109920168439588
3.45345345345345 0.0010259780658361
3.47347347347347 0.000957248515249212
3.49349349349349 0.000892765215932441
3.51351351351351 0.000832292065877152
3.53353353353353 0.000775604254246008
3.55355355355355 0.000722487844607976
3.57357357357357 0.000672739365441019
3.59359359359359 0.000626165408357979
3.61361361361361 0.000582582234459159
3.63363363363363 0.000541815389165405
3.65365365365365 0.00050369932583812
3.67367367367367 0.000468077038447599
3.69369369369369 0.000434799703508357
3.71371371371371 0.000403726331459719
3.73373373373373 0.000374723427631836
3.75375375375375 0.000347664662901427
3.77377377377377 0.000322430554107926
3.79379379379379 0.000298908154269281
3.81381381381381 0.000276990752607344
3.83383383383383 0.000256577584365551
3.85385385385385 0.000237573550376443
3.87387387387387 0.000219888946313312
3.89389389389389 0.000203439201538965
3.91391391391391 0.00018814462744512
3.93393393393393 0.000173930175158222
3.95395395395395 0.000160725202471485
3.97397397397397 0.000148463249848567
3.99399399399399 0.000137081825331481
4.01401401401401 0.000126522198173995
4.03403403403403 0.000116729201011866
4.05405405405405 0.000107651040372646
4.07407407407407 9.92391153205018e-05
4.09409409409409 9.14478440253317e-05
4.11411411411411 8.42344980404937e-05
4.13413413413413 7.75590440694795e-05
4.15415415415415 7.13839929989176e-05
4.17417417417417 6.56742559732345e-05
4.19419419419419 6.03970072851107e-05
4.21421421421421 5.55215538554582e-05
4.23423423423423 5.10192110769697e-05
4.25425425425425 4.68631847962824e-05
4.27427427427427 4.3028459211397e-05
4.29429429429429 3.94916904631592e-05
4.31431431431431 3.62311057022691e-05
4.33433433433433 3.32264074164067e-05
4.35435435435435 3.04586828055866e-05
4.37437437437437 2.79103179977393e-05
4.39439439439439 2.55649169007242e-05
4.41441441441441 2.34072244914564e-05
4.43443443443443 2.14230543475548e-05
4.45445445445445 1.95992202318354e-05
4.47447447447447 1.79234715450684e-05
4.49449449449449 1.63844324676437e-05
4.51451451451451 1.49715446161227e-05
4.53453453453453 1.3675013046071e-05
4.55455455455455 1.24857554380297e-05
4.57457457457457 1.13953543089884e-05
4.59459459459459 1.03960120972233e-05
4.61461461461461 9.48050897386715e-06
4.63463463463463 8.64216324004121e-06
4.65465465465465 7.87479417380527e-06
4.67467467467467 7.17268719654363e-06
4.69469469469469 6.53056123369604e-06
4.71471471471471 5.94353814994721e-06
4.73473473473473 5.40711414409909e-06
4.75475475475475 4.91713299385613e-06
4.77477477477477 4.4697610456467e-06
4.79479479479479 4.06146384937932e-06
4.81481481481481 3.68898434268124e-06
4.83483483483483 3.34932249368796e-06
4.85485485485485 3.03971631583941e-06
4.87487487487487 2.75762417238901e-06
4.89489489489489 2.50070829244518e-06
4.91491491491491 2.26681942433715e-06
4.93493493493493 2.05398255592983e-06
4.95495495495495 1.86038363520377e-06
4.97497497497497 1.68435722796805e-06
4.99499499499499 1.5243750529858e-06
5.01501501501502 1.37903533806611e-06
5.03503503503504 1.24705294381396e-06
5.05505505505506 1.12725020473309e-06
5.07507507507508 1.01854844024876e-06
5.0950950950951 9.19960090959837e-07
5.11511511511512 8.30581438046261e-07
5.13513513513514 7.49585866251233e-07
5.15515515515516 6.76217633231267e-07
5.17517517517518 6.09786110324583e-07
5.1951951951952 5.4966046193278e-07
5.21521521521522 4.95264732746229e-07
5.23523523523524 4.46073313973501e-07
5.25525525525526 4.01606761563285e-07
5.27527527527528 3.61427941137511e-07
5.2952952952953 3.25138475990267e-07
5.31531531531532 2.92375476052561e-07
5.33533533533534 2.62808527181656e-07
5.35535535535536 2.36136921509202e-07
5.37537537537538 2.1208711087848e-07
5.3953953953954 1.90410366621162e-07
5.41541541541542 1.70880630071684e-07
5.43543543543544 1.5329253929596e-07
5.45545545545546 1.3745961852414e-07
5.47547547547548 1.23212617727566e-07
5.4954954954955 1.10397990671284e-07
5.51551551551552 9.8876500608364e-08
5.53553553553554 8.85219435638491e-08
5.55555555555556 7.92199798873018e-08
5.57557557557558 7.08670654362614e-08
5.5955955955956 6.33694743912418e-08
5.61561561561562 5.66424062986154e-08
5.63563563563564 5.06091704933412e-08
5.65565565565566 4.5200441571292e-08
5.67567567567568 4.03535800631662e-08
5.6956956956957 3.60120129107462e-08
5.71571571571572 3.2124668763613e-08
5.73573573573574 2.86454635022852e-08
5.75575575575576 2.55328317539416e-08
5.77577577577578 2.27493005011625e-08
5.7957957957958 2.02611011941336e-08
5.81581581581582 1.80378170640688e-08
5.83583583583584 1.60520626017053e-08
5.85585585585586 1.42791924110059e-08
5.87587587587588 1.2697036876002e-08
5.8958958958959 1.12856622892642e-08
5.91591591591592 1.00271532849868e-08
5.93593593593594 8.90541559921139e-09
5.95595595595596 7.90599734535251e-09
5.97597597597598 7.01592714588943e-09
5.995995995996 6.22356760178439e-09
6.01601601601602 5.5184827107339e-09
6.03603603603604 4.89131796456914e-09
6.05605605605606 4.33369196574694e-09
6.07607607607608 3.83809850362692e-09
6.0960960960961 3.3978181237675e-09
6.11611611611612 3.00683830841826e-09
6.13613613613614 2.65978146430924e-09
6.15615615615616 2.35183998527871e-09
6.17617617617618 2.07871772273778e-09
6.1961961961962 1.83657725691022e-09
6.21621621621622 1.6219924166386e-09
6.23623623623624 1.43190554571785e-09
6.25625625625626 1.26358905957511e-09
6.27627627627628 1.11461087800764e-09
6.2962962962963 9.82803357938119e-10
6.31631631631632 8.66235385045992e-10
6.33633633633634 7.63187314959515e-10
6.35635635635636 6.72128483698836e-10
6.37637637637638 5.91697033481532e-10
6.3963963963964 5.20681824054164e-10
6.41641641641642 4.58006221596989e-10
6.43643643643644 4.02713577147895e-10
6.45645645645646 3.53954224575805e-10
6.47647647647648 3.10973844559378e-10
6.4964964964965 2.73103055937435e-10
6.51651651651652 2.39748109325539e-10
6.53653653653654 2.10382570159619e-10
6.55655655655656 1.84539889444162e-10
6.57657657657658 1.61806770551276e-10
6.5965965965966 1.41817249531742e-10
6.61661661661662 1.24247414645767e-10
6.63663663663664 1.08810698278133e-10
6.65665665665666 9.52536811418137e-11
6.67667667667668 8.33523547614393e-11
6.6966966966967 7.29087937236022e-11
6.71671671671672 6.37481941394484e-11
6.73673673673674 5.57162392365998e-11
6.75675675675676 4.86767570277076e-11
6.77677677677678 4.25096386334907e-11
6.7967967967968 3.71089891068556e-11
6.81681681681682 3.23814855461048e-11
6.83683683683684 2.82449199306813e-11
6.85685685685686 2.46269064908966e-11
6.87687687687688 2.14637355595386e-11
6.8968968968969 1.86993577716892e-11
6.91691691691692 1.62844842008236e-11
6.93693693693694 1.41757895636779e-11
6.95695695695696 1.23352070109961e-11
6.97697697697698 1.07293042619725e-11
6.996996996997 9.32873195138548e-12
7.01701701701702 8.10773605306642e-12
7.03703703703704 7.04372713322415e-12
7.05705705705706 6.11689998287643e-12
7.07707707707708 5.30989788981512e-12
7.0970970970971 4.6075164458095e-12
7.11711711711712 3.99644235193917e-12
7.13713713713714 3.46502319108337e-12
7.15715715715716 3.00306458801651e-12
7.17717717717718 2.60165157997869e-12
7.1971971971972 2.25299137914217e-12
7.21721721721722 1.95027502769791e-12
7.23723723723724 1.68755573049414e-12
7.25725725725726 1.45964190299846e-12
7.27727727727728 1.262003197176e-12
7.2972972972973 1.0906879676822e-12
7.31731731731732 9.42250818252902e-13
7.33733733733734 8.13689025751829e-13
7.35735735735736 7.02386779168733e-13
7.37737737737738 6.06066294885245e-13
7.3973973973974 5.22744979473708e-13
7.41741741741742 4.50697908714381e-13
7.43743743743744 3.88424977793729e-13
7.45745745745746 3.34622154016963e-13
7.47747747747748 2.88156330935933e-13
7.4974974974975 2.48043342543511e-13
7.51751751751752 2.13428748996348e-13
7.53753753753754 1.83571051982055e-13
7.55755755755756 1.57827039041883e-13
7.57757757757758 1.35638992516663e-13
7.5975975975976 1.16523530854698e-13
7.61761761761762 1.000618782966e-13
7.63763763763764 8.58913838706873e-14
7.65765765765766 7.36981325813112e-14
7.67767767767768 6.32105109959248e-14
7.6976976976977 5.41936064405376e-14
7.71771771771772 4.64443339685915e-14
7.73773773773774 3.97871984155707e-14
7.75775775775776 3.40706104038536e-14
7.77777777777778 2.91636853079907e-14
7.7977977977978 2.4953463096687e-14
7.81781781781782 2.13424947819474e-14
7.83783783783784 1.82467480586654e-14
7.85785785785786 1.55937907247682e-14
7.87787787787788 1.33212157347804e-14
7.8978978978979 1.13752763482777e-14
7.91791791791792 9.70970386854701e-15
7.93793793793794 8.28468399583068e-15
7.95795795795796 7.06597090549298e-15
7.97797797797798 6.02412085866189e-15
7.997997997998 5.13382950919603e-15
8.01801801801802 4.3733591283238e-15
8.03803803803804 3.72404376402735e-15
8.05805805805806 3.16986191875719e-15
8.07807807807808 2.69706769497134e-15
8.0980980980981 2.29387254841619e-15
8.11811811811812 1.95017082606148e-15
8.13813813813814 1.65730316851105e-15
8.15815815815816 1.40785264250169e-15
8.17817817817818 1.19546915264237e-15
8.1981981981982 1.01471827585831e-15
8.21821821821822 8.60951178494166e-16
8.23823823823824 7.30192724674871e-16
8.25825825825826 6.19045274051723e-16
8.27827827827828 5.24606005103494e-16
8.2982982982983 4.44395893386326e-16
8.31831831831832 3.76298728354137e-16
8.33833833833834 3.18508772685576e-16
8.35835835835836 2.69485858889318e-16
8.37837837837838 2.27916883183252e-16
8.3983983983984 1.92682799625235e-16
8.41841841841842 1.62830341150177e-16
8.43843843843844 1.37547801096493e-16
8.45845845845846 1.16144301209627e-16
8.47847847847848 9.8032051926925e-17
8.4984984984985 8.27111796592674e-17
8.51851851851852 6.97567552529606e-17
8.53853853853854 5.88077091106193e-17
8.55855855855856 4.95573626747691e-17
8.57857857857858 4.17453440895294e-17
8.5985985985986 3.51506886838449e-17
8.61861861861862 2.95859531836935e-17
8.63863863863864 2.48921968838275e-17
8.65865865865866 2.09347039316773e-17
8.67867867867868 1.75993388643364e-17
8.6986986986987 1.47894429982973e-17
8.71871871871872 1.24231925503958e-17
8.73873873873874 1.04313507694241e-17
8.75875875875876 8.75535614211525e-18
8.77877877877878 7.34569713009337e-18
8.7987987987988 6.16053109048305e-18
8.81881881881882 5.16451119999019e-18
8.83883883883884 4.32779048512422e-18
8.85885885885886 3.62517658454128e-18
8.87887887887888 3.03541474067764e-18
8.8988988988989 2.54057982941549e-18
8.91891891891892 2.12556106807901e-18
8.93893893893894 1.77762546207239e-18
8.95895895895896 1.48604811781133e-18
8.97897897897898 1.24179931485728e-18
8.998998998999 1.03727973679138e-18
9.01901901901902 8.66096545670873e-19
9.03903903903904 7.22874080905507e-19
9.05905905905906 6.03093897535385e-19
9.07907907907908 5.0295965472299e-19
9.0990990990991 4.19283042962206e-19
9.11911911911912 3.49387515332417e-19
9.13913913913914 2.91027078875695e-19
9.15915915915916 2.42317819504618e-19
9.17917917917918 2.01680188581445e-19
9.1991991991992 1.67790380698338e-19
9.21921921921922 1.39539388139188e-19
9.23923923923924 1.1599853476858e-19
9.25925925925926 9.63904764362469e-20
9.27927927927928 8.00648113242436e-20
9.2992992992993 6.6477576193283e-20
9.31931931931932 5.51740167796855e-20
9.33933933933934 4.5774115701642e-20
9.35935935935936 3.79604417474457e-20
9.37937937937938 3.14679525475421e-20
9.3993993993994 2.60754402558043e-20
9.41941941941942 2.15983585814365e-20
9.43943943943944 1.78828106797973e-20
9.45945945945946 1.48005121824234e-20
9.47947947947948 1.22445730038445e-20
9.4994994994995 1.01259663374544e-20
9.51951951951952 8.37057415073758e-21
9.53953953953954 6.91671611023779e-21
9.55955955955956 5.71308371631411e-21
9.57957957957958 4.71701393695898e-21
9.5995995995996 3.89304716291232e-21
9.61961961961962 3.21172317125736e-21
9.63963963963964 2.64857624243904e-21
9.65965965965966 2.18329684678274e-21
9.67967967967968 1.79903258756111e-21
9.6996996996997 1.48180551599987e-21
9.71971971971972 1.22002665237565e-21
9.73973973973974 1.00409166885045e-21
9.75975975975976 8.26044308669654e-22
9.77977977977978 6.79296312742742e-22
9.7997997997998 5.58394465795474e-22
9.81981981981982 4.58826916995414e-22
9.83983983983984 3.7686222201397e-22
9.85985985985986 3.09415635142992e-22
9.87987987987988 2.53938085193762e-22
9.8998998998999 2.08324025950642e-22
9.91991991991992 1.70834984871876e-22
9.93993993993994 1.40036162642795e-22
9.95995995995996 1.14743877987917e-22
9.97997997997998 9.39820210218911e-23
10 7.69459862670642e-23
};
\addlegendentry{N(0,1)}; 
\legend{}; 
\end{axis}

\end{tikzpicture}

%% file: FinalFigs/Null_Dists_d_10_500_n_200_m_200_kernel__Gaussian_RBF_2022_10_12_22_42_59mmd.tex
\begin{tikzpicture}

\definecolor{darkorange25512714}{RGB}{255,127,14}
\definecolor{darkslategray38}{RGB}{38,38,38}
\definecolor{lightgray204}{RGB}{204,204,204}
\definecolor{steelblue31119180}{RGB}{31,119,180}

\begin{axis}[
axis line style={darkslategray38},
height=\figheight,
legend cell align={left},
legend style={fill opacity=0.8, draw opacity=1, text opacity=1, draw=none},
tick align=outside,
tick pos=left,
title={$\dmmd~(n/m=1)$},
width=\figwidth,
x grid style={lightgray204},
xmin=-6, xmax=6,
xtick style={color=darkslategray38},
y grid style={lightgray204},
ylabel = {}, 
ymin=0, ymax=0.946759975888624,
ytick style={color=darkslategray38}, 
xticklabels=empty,
yticklabels=empty
]
\draw[draw=none,fill=steelblue31119180,fill opacity=0.8] (axis cs:-1.65599834918976,0) rectangle (axis cs:-1.53267252445221,0.0194606367089131);
\addlegendimage{ybar,ybar legend,draw=none,fill=steelblue31119180,fill opacity=0.8}
\addlegendentry{mmd (d=10)}

\draw[draw=none,fill=steelblue31119180,fill opacity=0.8] (axis cs:-1.3476836681366,0) rectangle (axis cs:-1.22435784339905,0.103790062447537);
\draw[draw=none,fill=steelblue31119180,fill opacity=0.8] (axis cs:-1.03936898708344,0) rectangle (axis cs:-0.916043162345886,0.233527685653477);
\draw[draw=none,fill=steelblue31119180,fill opacity=0.8] (axis cs:-0.731054306030273,0) rectangle (axis cs:-0.607728481292725,0.467055281013915);
\draw[draw=none,fill=steelblue31119180,fill opacity=0.8] (axis cs:-0.422739654779434,0) rectangle (axis cs:-0.299413830041885,0.707069800423844);
\draw[draw=none,fill=steelblue31119180,fill opacity=0.8] (axis cs:-0.114424973726273,0) rectangle (axis cs:0.00890085101127625,0.74599107384167);
\draw[draw=none,fill=steelblue31119180,fill opacity=0.8] (axis cs:0.193889677524567,0) rectangle (axis cs:0.317215502262115,0.531924070043626);
\draw[draw=none,fill=steelblue31119180,fill opacity=0.8] (axis cs:0.502204358577728,0) rectangle (axis cs:0.625530183315277,0.201093265430185);
\draw[draw=none,fill=steelblue31119180,fill opacity=0.8] (axis cs:0.810518980026245,0) rectangle (axis cs:0.933844804763794,0.181632609283189);
\draw[draw=none,fill=steelblue31119180,fill opacity=0.8] (axis cs:1.11883366107941,0) rectangle (axis cs:1.24215948581696,0.0518950312237683);
\draw[draw=none,fill=darkorange25512714,fill opacity=0.8] (axis cs:-1.53267240524292,0) rectangle (axis cs:-1.40934658050537,0);
\addlegendimage{ybar,ybar legend,draw=none,fill=darkorange25512714,fill opacity=0.8}
\addlegendentry{mmd (d=500)}

\draw[draw=none,fill=darkorange25512714,fill opacity=0.8] (axis cs:-1.22435772418976,0) rectangle (axis cs:-1.10103189945221,0.0194606367089131);
\draw[draw=none,fill=darkorange25512714,fill opacity=0.8] (axis cs:-0.916043162345886,0) rectangle (axis cs:-0.792717337608337,0.18811952455419);
\draw[draw=none,fill=darkorange25512714,fill opacity=0.8] (axis cs:-0.607728481292725,0) rectangle (axis cs:-0.484402656555176,0.402186491984205);
\draw[draw=none,fill=darkorange25512714,fill opacity=0.8] (axis cs:-0.299413770437241,0) rectangle (axis cs:-0.176087945699692,0.694096042617902);
\draw[draw=none,fill=darkorange25512714,fill opacity=0.8] (axis cs:0.00890091061592102,0) rectangle (axis cs:0.13222673535347,0.901676167512975);
\draw[draw=none,fill=darkorange25512714,fill opacity=0.8] (axis cs:0.31721556186676,0) rectangle (axis cs:0.440541386604309,0.616253495782249);
\draw[draw=none,fill=darkorange25512714,fill opacity=0.8] (axis cs:0.625530242919922,0) rectangle (axis cs:0.748856067657471,0.311370217440287);
\draw[draw=none,fill=darkorange25512714,fill opacity=0.8] (axis cs:0.933844923973083,0) rectangle (axis cs:1.05717074871063,0.0908163046415946);
\draw[draw=none,fill=darkorange25512714,fill opacity=0.8] (axis cs:1.24215960502625,0) rectangle (axis cs:1.36548542976379,0.0194606367089131);
\addplot [semithick, black]
table {%
-10 7.69459862670642e-23
-9.97997997997998 9.39820210218911e-23
-9.95995995995996 1.14743877987917e-22
-9.93993993993994 1.40036162642795e-22
-9.91991991991992 1.70834984871876e-22
-9.8998998998999 2.08324025950642e-22
-9.87987987987988 2.53938085193762e-22
-9.85985985985986 3.09415635142992e-22
-9.83983983983984 3.7686222201397e-22
-9.81981981981982 4.58826916995414e-22
-9.7997997997998 5.58394465795474e-22
-9.77977977977978 6.79296312742742e-22
-9.75975975975976 8.26044308669654e-22
-9.73973973973974 1.00409166885045e-21
-9.71971971971972 1.22002665237565e-21
-9.6996996996997 1.48180551599987e-21
-9.67967967967968 1.79903258756111e-21
-9.65965965965966 2.18329684678274e-21
-9.63963963963964 2.64857624243904e-21
-9.61961961961962 3.21172317125736e-21
-9.5995995995996 3.89304716291232e-21
-9.57957957957958 4.71701393695898e-21
-9.55955955955956 5.71308371631411e-21
-9.53953953953954 6.91671611023779e-21
-9.51951951951952 8.37057415073758e-21
-9.4994994994995 1.01259663374544e-20
-9.47947947947948 1.22445730038445e-20
-9.45945945945946 1.48005121824234e-20
-9.43943943943944 1.78828106797973e-20
-9.41941941941942 2.15983585814365e-20
-9.3993993993994 2.60754402558043e-20
-9.37937937937938 3.14679525475421e-20
-9.35935935935936 3.79604417474457e-20
-9.33933933933934 4.5774115701642e-20
-9.31931931931932 5.51740167796855e-20
-9.2992992992993 6.6477576193283e-20
-9.27927927927928 8.00648113242436e-20
-9.25925925925926 9.63904764362469e-20
-9.23923923923924 1.1599853476858e-19
-9.21921921921922 1.39539388139188e-19
-9.1991991991992 1.67790380698338e-19
-9.17917917917918 2.01680188581445e-19
-9.15915915915916 2.42317819504618e-19
-9.13913913913914 2.91027078875695e-19
-9.11911911911912 3.49387515332417e-19
-9.0990990990991 4.19283042962206e-19
-9.07907907907908 5.0295965472299e-19
-9.05905905905906 6.03093897535385e-19
-9.03903903903904 7.22874080905507e-19
-9.01901901901902 8.66096545670873e-19
-8.998998998999 1.03727973679138e-18
-8.97897897897898 1.24179931485728e-18
-8.95895895895896 1.48604811781133e-18
-8.93893893893894 1.77762546207239e-18
-8.91891891891892 2.12556106807901e-18
-8.8988988988989 2.54057982941549e-18
-8.87887887887888 3.03541474067764e-18
-8.85885885885886 3.62517658454128e-18
-8.83883883883884 4.32779048512422e-18
-8.81881881881882 5.16451119999019e-18
-8.7987987987988 6.16053109048305e-18
-8.77877877877878 7.34569713009337e-18
-8.75875875875876 8.75535614211525e-18
-8.73873873873874 1.04313507694241e-17
-8.71871871871872 1.24231925503958e-17
-8.6986986986987 1.47894429982973e-17
-8.67867867867868 1.75993388643364e-17
-8.65865865865866 2.09347039316773e-17
-8.63863863863864 2.48921968838275e-17
-8.61861861861862 2.95859531836935e-17
-8.5985985985986 3.51506886838449e-17
-8.57857857857858 4.17453440895294e-17
-8.55855855855856 4.95573626747691e-17
-8.53853853853854 5.88077091106193e-17
-8.51851851851852 6.97567552529606e-17
-8.4984984984985 8.27111796592674e-17
-8.47847847847848 9.8032051926925e-17
-8.45845845845846 1.16144301209627e-16
-8.43843843843844 1.37547801096493e-16
-8.41841841841842 1.62830341150177e-16
-8.3983983983984 1.92682799625235e-16
-8.37837837837838 2.27916883183252e-16
-8.35835835835836 2.69485858889318e-16
-8.33833833833834 3.18508772685576e-16
-8.31831831831832 3.76298728354137e-16
-8.2982982982983 4.44395893386326e-16
-8.27827827827828 5.24606005103494e-16
-8.25825825825826 6.19045274051723e-16
-8.23823823823824 7.30192724674871e-16
-8.21821821821822 8.60951178494166e-16
-8.1981981981982 1.01471827585831e-15
-8.17817817817818 1.19546915264237e-15
-8.15815815815816 1.40785264250169e-15
-8.13813813813814 1.65730316851105e-15
-8.11811811811812 1.95017082606148e-15
-8.0980980980981 2.29387254841619e-15
-8.07807807807808 2.69706769497134e-15
-8.05805805805806 3.16986191875719e-15
-8.03803803803804 3.72404376402735e-15
-8.01801801801802 4.3733591283238e-15
-7.997997997998 5.13382950919607e-15
-7.97797797797798 6.02412085866193e-15
-7.95795795795796 7.06597090549303e-15
-7.93793793793794 8.28468399583074e-15
-7.91791791791792 9.70970386854708e-15
-7.8978978978979 1.13752763482777e-14
-7.87787787787788 1.33212157347805e-14
-7.85785785785786 1.55937907247683e-14
-7.83783783783784 1.82467480586655e-14
-7.81781781781782 2.13424947819475e-14
-7.7977977977978 2.49534630966872e-14
-7.77777777777778 2.91636853079909e-14
-7.75775775775776 3.40706104038538e-14
-7.73773773773774 3.9787198415571e-14
-7.71771771771772 4.64443339685918e-14
-7.6976976976977 5.4193606440538e-14
-7.67767767767768 6.32105109959252e-14
-7.65765765765766 7.36981325813117e-14
-7.63763763763764 8.58913838706879e-14
-7.61761761761762 1.00061878296601e-13
-7.5975975975976 1.16523530854699e-13
-7.57757757757758 1.35638992516664e-13
-7.55755755755756 1.57827039041884e-13
-7.53753753753754 1.83571051982057e-13
-7.51751751751752 2.13428748996349e-13
-7.4974974974975 2.48043342543513e-13
-7.47747747747748 2.88156330935935e-13
-7.45745745745746 3.34622154016965e-13
-7.43743743743744 3.88424977793732e-13
-7.41741741741742 4.50697908714384e-13
-7.3973973973974 5.22744979473711e-13
-7.37737737737738 6.06066294885249e-13
-7.35735735735736 7.02386779168738e-13
-7.33733733733734 8.13689025751835e-13
-7.31731731731732 9.42250818252909e-13
-7.2972972972973 1.09068796768221e-12
-7.27727727727728 1.262003197176e-12
-7.25725725725726 1.45964190299847e-12
-7.23723723723724 1.68755573049416e-12
-7.21721721721722 1.95027502769792e-12
-7.1971971971972 2.25299137914218e-12
-7.17717717717718 2.60165157997871e-12
-7.15715715715716 3.00306458801653e-12
-7.13713713713714 3.4650231910834e-12
-7.11711711711712 3.99644235193919e-12
-7.0970970970971 4.60751644580953e-12
-7.07707707707708 5.30989788981514e-12
-7.05705705705706 6.11689998287646e-12
-7.03703703703704 7.0437271332242e-12
-7.01701701701702 8.10773605306645e-12
-6.996996996997 9.32873195138555e-12
-6.97697697697698 1.07293042619726e-11
-6.95695695695696 1.23352070109962e-11
-6.93693693693694 1.41757895636779e-11
-6.91691691691692 1.62844842008237e-11
-6.8968968968969 1.86993577716893e-11
-6.87687687687688 2.14637355595386e-11
-6.85685685685686 2.46269064908967e-11
-6.83683683683684 2.82449199306815e-11
-6.81681681681682 3.2381485546105e-11
-6.7967967967968 3.71089891068559e-11
-6.77677677677678 4.25096386334913e-11
-6.75675675675676 4.86767570277083e-11
-6.73673673673674 5.57162392366004e-11
-6.71671671671672 6.37481941394491e-11
-6.6966966966967 7.29087937236032e-11
-6.67667667667668 8.33523547614402e-11
-6.65665665665666 9.52536811418151e-11
-6.63663663663664 1.08810698278135e-10
-6.61661661661662 1.24247414645768e-10
-6.5965965965966 1.41817249531744e-10
-6.57657657657658 1.61806770551278e-10
-6.55655655655656 1.84539889444164e-10
-6.53653653653654 2.10382570159622e-10
-6.51651651651652 2.39748109325542e-10
-6.4964964964965 2.73103055937438e-10
-6.47647647647648 3.10973844559381e-10
-6.45645645645646 3.53954224575809e-10
-6.43643643643644 4.027135771479e-10
-6.41641641641642 4.58006221596996e-10
-6.3963963963964 5.20681824054169e-10
-6.37637637637638 5.91697033481538e-10
-6.35635635635636 6.72128483698846e-10
-6.33633633633634 7.63187314959523e-10
-6.31631631631632 8.66235385046001e-10
-6.2962962962963 9.8280335793813e-10
-6.27627627627628 1.11461087800766e-09
-6.25625625625626 1.26358905957513e-09
-6.23623623623624 1.43190554571787e-09
-6.21621621621622 1.62199241663862e-09
-6.1961961961962 1.83657725691024e-09
-6.17617617617618 2.0787177227378e-09
-6.15615615615616 2.35183998527873e-09
-6.13613613613614 2.65978146430928e-09
-6.11611611611612 3.00683830841829e-09
-6.0960960960961 3.39781812376754e-09
-6.07607607607608 3.83809850362696e-09
-6.05605605605606 4.33369196574699e-09
-6.03603603603604 4.89131796456919e-09
-6.01601601601602 5.51848271073395e-09
-5.995995995996 6.22356760178439e-09
-5.97597597597598 7.01592714588943e-09
-5.95595595595596 7.90599734535251e-09
-5.93593593593594 8.90541559921139e-09
-5.91591591591592 1.00271532849868e-08
-5.8958958958959 1.12856622892642e-08
-5.87587587587588 1.2697036876002e-08
-5.85585585585586 1.42791924110059e-08
-5.83583583583584 1.60520626017053e-08
-5.81581581581582 1.80378170640688e-08
-5.7957957957958 2.02611011941336e-08
-5.77577577577578 2.27493005011625e-08
-5.75575575575576 2.55328317539416e-08
-5.73573573573574 2.86454635022852e-08
-5.71571571571572 3.2124668763613e-08
-5.6956956956957 3.60120129107462e-08
-5.67567567567568 4.03535800631662e-08
-5.65565565565566 4.5200441571292e-08
-5.63563563563564 5.06091704933412e-08
-5.61561561561562 5.66424062986154e-08
-5.5955955955956 6.33694743912418e-08
-5.57557557557558 7.08670654362614e-08
-5.55555555555556 7.92199798873018e-08
-5.53553553553554 8.85219435638491e-08
-5.51551551551552 9.8876500608364e-08
-5.4954954954955 1.10397990671284e-07
-5.47547547547548 1.23212617727566e-07
-5.45545545545546 1.3745961852414e-07
-5.43543543543544 1.5329253929596e-07
-5.41541541541542 1.70880630071684e-07
-5.3953953953954 1.90410366621162e-07
-5.37537537537538 2.1208711087848e-07
-5.35535535535536 2.36136921509202e-07
-5.33533533533534 2.62808527181656e-07
-5.31531531531532 2.92375476052561e-07
-5.2952952952953 3.25138475990267e-07
-5.27527527527528 3.61427941137511e-07
-5.25525525525526 4.01606761563285e-07
-5.23523523523524 4.46073313973501e-07
-5.21521521521522 4.95264732746229e-07
-5.1951951951952 5.4966046193278e-07
-5.17517517517518 6.09786110324583e-07
-5.15515515515516 6.76217633231267e-07
-5.13513513513514 7.49585866251233e-07
-5.11511511511512 8.30581438046261e-07
-5.0950950950951 9.19960090959837e-07
-5.07507507507508 1.01854844024876e-06
-5.05505505505506 1.12725020473309e-06
-5.03503503503504 1.24705294381396e-06
-5.01501501501502 1.37903533806611e-06
-4.99499499499499 1.5243750529858e-06
-4.97497497497497 1.68435722796805e-06
-4.95495495495495 1.86038363520377e-06
-4.93493493493493 2.05398255592983e-06
-4.91491491491491 2.26681942433715e-06
-4.89489489489489 2.50070829244518e-06
-4.87487487487487 2.75762417238901e-06
-4.85485485485485 3.03971631583941e-06
-4.83483483483483 3.34932249368796e-06
-4.81481481481481 3.68898434268124e-06
-4.79479479479479 4.06146384937932e-06
-4.77477477477477 4.4697610456467e-06
-4.75475475475475 4.91713299385613e-06
-4.73473473473473 5.40711414409909e-06
-4.71471471471471 5.94353814994721e-06
-4.69469469469469 6.53056123369604e-06
-4.67467467467467 7.17268719654363e-06
-4.65465465465465 7.87479417380527e-06
-4.63463463463463 8.64216324004121e-06
-4.61461461461461 9.48050897386715e-06
-4.59459459459459 1.03960120972233e-05
-4.57457457457457 1.13953543089884e-05
-4.55455455455455 1.24857554380297e-05
-4.53453453453453 1.3675013046071e-05
-4.51451451451451 1.49715446161227e-05
-4.49449449449449 1.63844324676437e-05
-4.47447447447447 1.79234715450684e-05
-4.45445445445445 1.95992202318354e-05
-4.43443443443443 2.14230543475548e-05
-4.41441441441441 2.34072244914564e-05
-4.39439439439439 2.55649169007242e-05
-4.37437437437437 2.79103179977393e-05
-4.35435435435435 3.04586828055866e-05
-4.33433433433433 3.32264074164067e-05
-4.31431431431431 3.62311057022691e-05
-4.29429429429429 3.94916904631592e-05
-4.27427427427427 4.3028459211397e-05
-4.25425425425425 4.68631847962824e-05
-4.23423423423423 5.10192110769697e-05
-4.21421421421421 5.55215538554582e-05
-4.19419419419419 6.03970072851107e-05
-4.17417417417417 6.56742559732345e-05
-4.15415415415415 7.13839929989176e-05
-4.13413413413413 7.75590440694795e-05
-4.11411411411411 8.42344980404937e-05
-4.09409409409409 9.14478440253317e-05
-4.07407407407407 9.92391153205018e-05
-4.05405405405405 0.000107651040372646
-4.03403403403403 0.000116729201011866
-4.01401401401401 0.000126522198173995
-3.99399399399399 0.000137081825331481
-3.97397397397397 0.000148463249848567
-3.95395395395395 0.000160725202471485
-3.93393393393393 0.000173930175158222
-3.91391391391391 0.00018814462744512
-3.89389389389389 0.000203439201538965
-3.87387387387387 0.000219888946313312
-3.85385385385385 0.000237573550376443
-3.83383383383383 0.000256577584365551
-3.81381381381381 0.000276990752607344
-3.79379379379379 0.000298908154269281
-3.77377377377377 0.000322430554107926
-3.75375375375375 0.000347664662901427
-3.73373373373373 0.000374723427631836
-3.71371371371371 0.000403726331459719
-3.69369369369369 0.000434799703508357
-3.67367367367367 0.000468077038447599
-3.65365365365365 0.00050369932583812
-3.63363363363363 0.000541815389165405
-3.61361361361361 0.000582582234459159
-3.59359359359359 0.000626165408357979
-3.57357357357357 0.000672739365441021
-3.55355355355355 0.000722487844607978
-3.53353353353353 0.00077560425424601
-3.51351351351351 0.000832292065877155
-3.49349349349349 0.000892765215932443
-3.47347347347347 0.000957248515249216
-3.45345345345345 0.0010259780658361
-3.43343343343343 0.00109920168439588
-3.41341341341341 0.00117717933203981
-3.39339339339339 0.00126018354956833
-3.37337337337337 0.00134849989763212
-3.35335335335335 0.00144242740102448
-3.33333333333333 0.00154227899629111
-3.31331331331331 0.00164838198177652
-3.29329329329329 0.00176107846915772
-3.27327327327327 0.00188072583544552
-3.25325325325325 0.00200769717436226
-3.23323323323323 0.00214238174593163
-3.21321321321321 0.00228518542304204
-3.19319319319319 0.00243653113367012
-3.17317317317317 0.00259685929737497
-3.15315315315315 0.00276662825459747
-3.13313313313313 0.00294631468722261
-3.11311311311311 0.00313641402878609
-3.09309309309309 0.00333744086263052
-3.07307307307307 0.00354992930624086
-3.05305305305305 0.00377443337991422
-3.03303303303303 0.00401152735784579
-3.01301301301301 0.00426180609964128
-2.99299299299299 0.00452588536019618
-2.97297297297297 0.0048044020758154
-2.95295295295295 0.00509801462438215
-2.93293293293293 0.00540740305732385
-2.91291291291291 0.00573326930106519
-2.89289289289289 0.00607633732560526
-2.87287287287287 0.00643735327780636
-2.85285285285285 0.00681708557693873
-2.83283283283283 0.00721632496998623
-2.81281281281281 0.00763588454418632
-2.79279279279279 0.00807659969425075
-2.77277277277277 0.00853932804169477
-2.75275275275275 0.00902494930369032
-2.73273273273273 0.00953436510885489
-2.71271271271271 0.0100684987573917
-2.69269269269269 0.0106282949230102
-2.67267267267267 0.0112147192940778
-2.65265265265265 0.0118287581514852
-2.63263263263263 0.0124714178807513
-2.61261261261261 0.0131437244159435
-2.59259259259259 0.0138467226130541
-2.57257257257257 0.0145814755505492
-2.55255255255255 0.0153490637548887
-2.53253253253253 0.0161505843489182
-2.51251251251251 0.0169871501211409
-2.49249249249249 0.0178598885140022
-2.47247247247247 0.018769940529451
-2.45245245245245 0.0197184595501959
-2.43243243243243 0.0207066100752274
-2.41241241241241 0.0217355663683581
-2.39239239239239 0.0228065110187135
-2.37237237237237 0.0239206334123108
-2.35235235235235 0.0250791281140699
-2.33233233233233 0.0262831931598317
-2.31231231231231 0.0275340282581906
-2.29229229229229 0.0288328329022027
-2.27227227227227 0.0301808043912899
-2.25225225225225 0.0315791357639331
-2.23223223223223 0.033029013642035
-2.21221221221221 0.0345316159881232
-2.19219219219219 0.0360881097768736
-2.17217217217217 0.0376996485827434
-2.15215215215215 0.0393673700858293
-2.13213213213213 0.0410923934983949
-2.11211211211211 0.0428758169148479
-2.09209209209209 0.0447187145882915
-2.07207207207207 0.0466221341371235
-2.05205205205205 0.0485870936855041
-2.03203203203203 0.0506145789418747
-2.01201201201201 0.0527055402200587
-1.99199199199199 0.0548608894078376
-1.97197197197197 0.057081496888248
-1.95195195195195 0.059368188419199
-1.93193193193193 0.0617217419773594
-1.91191191191191 0.0641428845726061
-1.89189189189189 0.0666322890396674
-1.87187187187187 0.0691905708139176
-1.85185185185185 0.0718182846986055
-1.83183183183183 0.0745159216311036
-1.81181181181181 0.077283905456062
-1.79179179179179 0.0801225897136326
-1.77177177177177 0.083032254451193
-1.75175175175175 0.0860131030672496
-1.73173173173173 0.0890652591964251
-1.71171171171171 0.0921887636446459
-1.69169169169169 0.0953835713838294
-1.67167167167167 0.0986495486155338
-1.65165165165165 0.101986469913169
-1.63163163163163 0.10539401545248
-1.61161161161161 0.108871768340093
-1.59159159159159 0.112419212049971
-1.57157157157157 0.116035727977651
-1.55155155155155 0.119720593122119
-1.53153153153153 0.123472977905145
-1.51151151151151 0.127291944137829
-1.49149149149149 0.13117644314399
-1.47147147147147 0.135125314049902
-1.45145145145145 0.139137282249685
-1.43143143143143 0.143210958055468
-1.41141141141141 0.147344835541168
-1.39139139139139 0.151537291588457
-1.37137137137137 0.155786585143159
-1.35135135135135 0.160090856689972
-1.33133133133133 0.164448127952996
-1.31131131131131 0.168856301829129
-1.29129129129129 0.173313162560933
-1.27127127127127 0.17781637615506
-1.25125125125125 0.182363491051798
-1.23123123123123 0.186951939050736
-1.21121121121121 0.191579036496956
-1.19119119119119 0.1962419857315
-1.17117117117117 0.200937876809264
-1.15115115115115 0.205663689486728
-1.13113113113113 0.210416295481265
-1.11111111111111 0.215192461003031
-1.09109109109109 0.219988849559688
-1.07107107107107 0.224802025033432
-1.05105105105105 0.229628455029052
-1.03103103103103 0.234464514490888
-1.01101101101101 0.239306489585817
-0.990990990990991 0.24415058184851
-0.970970970970971 0.24899291258444
-0.950950950950951 0.253829527525259
-0.930930930930931 0.258656401730343
-0.910910910910911 0.2634694447275
-0.890890890890891 0.268264505884996
-0.870870870870871 0.273037380006279
-0.850850850850851 0.277783813137949
-0.830830830830831 0.282499508580786
-0.810810810810811 0.287180133092853
-0.790790790790791 0.291821323272996
-0.77077077077077 0.296418692112302
-0.75075075075075 0.300967835700437
-0.73073073073073 0.305464340073112
-0.71071071071071 0.309903788186304
-0.69069069069069 0.314281767002296
-0.67067067067067 0.318593874672039
-0.65065065065065 0.322835727797843
-0.63063063063063 0.327002968759958
-0.61061061061061 0.331091273090187
-0.59059059059059 0.33509635687531
-0.57057057057057 0.339013984172804
-0.55055055055055 0.34283997442106
-0.53053053053053 0.346570209826128
-0.51051051051051 0.350200642706842
-0.49049049049049 0.353727302780113
-0.47047047047047 0.357146304368113
-0.45045045045045 0.360453853509139
-0.43043043043043 0.363646254953996
-0.41041041041041 0.366719919029892
-0.39039039039039 0.369671368354051
-0.37037037037037 0.372497244379499
-0.35035035035035 0.375194313755802
-0.33033033033033 0.377759474487924
-0.31031031031031 0.38018976187679
-0.29029029029029 0.382482354225654
-0.27027027027027 0.384634578296894
-0.25025025025025 0.386643914504485
-0.23023023023023 0.388508001828027
-0.21021021021021 0.390224642434919
-0.19019019019019 0.391791805998011
-0.17017017017017 0.393207633696876
-0.15015015015015 0.394470441891644
-0.13013013013013 0.395578725459258
-0.11011011011011 0.396531160782876
-0.0900900900900901 0.397326608386124
-0.07007007007007 0.397964115204853
-0.05005005005005 0.398442916490068
-0.03003003003003 0.398762437336696
-0.01001001001001 0.398922293833933
0.01001001001001 0.398922293833933
0.03003003003003 0.398762437336696
0.05005005005005 0.398442916490068
0.07007007007007 0.397964115204853
0.0900900900900901 0.397326608386124
0.11011011011011 0.396531160782876
0.13013013013013 0.395578725459258
0.15015015015015 0.394470441891644
0.17017017017017 0.393207633696876
0.19019019019019 0.391791805998011
0.21021021021021 0.390224642434919
0.23023023023023 0.388508001828027
0.25025025025025 0.386643914504485
0.27027027027027 0.384634578296894
0.29029029029029 0.382482354225654
0.31031031031031 0.38018976187679
0.33033033033033 0.377759474487924
0.35035035035035 0.375194313755802
0.37037037037037 0.372497244379499
0.39039039039039 0.369671368354051
0.41041041041041 0.366719919029892
0.43043043043043 0.363646254953996
0.45045045045045 0.360453853509139
0.47047047047047 0.357146304368113
0.49049049049049 0.353727302780113
0.51051051051051 0.350200642706842
0.53053053053053 0.346570209826128
0.55055055055055 0.34283997442106
0.57057057057057 0.339013984172804
0.59059059059059 0.33509635687531
0.61061061061061 0.331091273090187
0.63063063063063 0.327002968759958
0.65065065065065 0.322835727797843
0.67067067067067 0.318593874672039
0.69069069069069 0.314281767002296
0.71071071071071 0.309903788186304
0.73073073073073 0.305464340073112
0.75075075075075 0.300967835700437
0.77077077077077 0.296418692112302
0.790790790790791 0.291821323272996
0.810810810810811 0.287180133092853
0.830830830830831 0.282499508580786
0.850850850850851 0.277783813137949
0.870870870870871 0.273037380006279
0.890890890890891 0.268264505884996
0.910910910910911 0.2634694447275
0.930930930930931 0.258656401730343
0.950950950950951 0.253829527525259
0.970970970970971 0.24899291258444
0.990990990990991 0.24415058184851
1.01101101101101 0.239306489585817
1.03103103103103 0.234464514490888
1.05105105105105 0.229628455029052
1.07107107107107 0.224802025033432
1.09109109109109 0.219988849559688
1.11111111111111 0.215192461003031
1.13113113113113 0.210416295481265
1.15115115115115 0.205663689486728
1.17117117117117 0.200937876809264
1.19119119119119 0.1962419857315
1.21121121121121 0.191579036496956
1.23123123123123 0.186951939050736
1.25125125125125 0.182363491051798
1.27127127127127 0.17781637615506
1.29129129129129 0.173313162560933
1.31131131131131 0.168856301829129
1.33133133133133 0.164448127952996
1.35135135135135 0.160090856689972
1.37137137137137 0.155786585143159
1.39139139139139 0.151537291588457
1.41141141141141 0.147344835541168
1.43143143143143 0.143210958055468
1.45145145145145 0.139137282249685
1.47147147147147 0.135125314049902
1.49149149149149 0.13117644314399
1.51151151151151 0.127291944137829
1.53153153153153 0.123472977905145
1.55155155155155 0.119720593122119
1.57157157157157 0.116035727977651
1.59159159159159 0.112419212049971
1.61161161161161 0.108871768340093
1.63163163163163 0.10539401545248
1.65165165165165 0.101986469913169
1.67167167167167 0.0986495486155338
1.69169169169169 0.0953835713838294
1.71171171171171 0.0921887636446459
1.73173173173173 0.0890652591964251
1.75175175175175 0.0860131030672496
1.77177177177177 0.083032254451193
1.79179179179179 0.0801225897136326
1.81181181181181 0.077283905456062
1.83183183183183 0.0745159216311036
1.85185185185185 0.0718182846986055
1.87187187187187 0.0691905708139176
1.89189189189189 0.0666322890396674
1.91191191191191 0.0641428845726061
1.93193193193193 0.0617217419773594
1.95195195195195 0.059368188419199
1.97197197197197 0.057081496888248
1.99199199199199 0.0548608894078376
2.01201201201201 0.0527055402200588
2.03203203203203 0.0506145789418748
2.05205205205205 0.0485870936855042
2.07207207207207 0.0466221341371236
2.09209209209209 0.0447187145882916
2.11211211211211 0.0428758169148479
2.13213213213213 0.041092393498395
2.15215215215215 0.0393673700858294
2.17217217217217 0.0376996485827434
2.19219219219219 0.0360881097768737
2.21221221221221 0.0345316159881233
2.23223223223223 0.033029013642035
2.25225225225225 0.0315791357639332
2.27227227227227 0.03018080439129
2.29229229229229 0.0288328329022028
2.31231231231231 0.0275340282581906
2.33233233233233 0.0262831931598317
2.35235235235235 0.02507912811407
2.37237237237237 0.0239206334123108
2.39239239239239 0.0228065110187136
2.41241241241241 0.0217355663683581
2.43243243243243 0.0207066100752275
2.45245245245245 0.0197184595501959
2.47247247247247 0.0187699405294511
2.49249249249249 0.0178598885140022
2.51251251251251 0.016987150121141
2.53253253253253 0.0161505843489182
2.55255255255255 0.0153490637548888
2.57257257257257 0.0145814755505493
2.59259259259259 0.0138467226130541
2.61261261261261 0.0131437244159435
2.63263263263263 0.0124714178807514
2.65265265265265 0.0118287581514852
2.67267267267267 0.0112147192940778
2.69269269269269 0.0106282949230103
2.71271271271271 0.0100684987573917
2.73273273273273 0.00953436510885491
2.75275275275275 0.00902494930369034
2.77277277277277 0.00853932804169479
2.79279279279279 0.00807659969425077
2.81281281281281 0.00763588454418634
2.83283283283283 0.00721632496998621
2.85285285285285 0.00681708557693871
2.87287287287287 0.00643735327780635
2.89289289289289 0.00607633732560524
2.91291291291291 0.00573326930106518
2.93293293293293 0.00540740305732384
2.95295295295295 0.00509801462438214
2.97297297297297 0.00480440207581539
2.99299299299299 0.00452588536019617
3.01301301301301 0.00426180609964127
3.03303303303303 0.00401152735784578
3.05305305305305 0.00377443337991421
3.07307307307307 0.00354992930624085
3.09309309309309 0.00333744086263051
3.11311311311311 0.00313641402878608
3.13313313313313 0.0029463146872226
3.15315315315315 0.00276662825459746
3.17317317317317 0.00259685929737496
3.19319319319319 0.00243653113367011
3.21321321321321 0.00228518542304203
3.23323323323323 0.00214238174593163
3.25325325325325 0.00200769717436225
3.27327327327327 0.00188072583544551
3.29329329329329 0.00176107846915771
3.31331331331331 0.00164838198177652
3.33333333333333 0.0015422789962911
3.35335335335335 0.00144242740102448
3.37337337337337 0.00134849989763212
3.39339339339339 0.00126018354956833
3.41341341341341 0.00117717933203981
3.43343343343343 0.00109920168439588
3.45345345345345 0.0010259780658361
3.47347347347347 0.000957248515249212
3.49349349349349 0.000892765215932441
3.51351351351351 0.000832292065877152
3.53353353353353 0.000775604254246008
3.55355355355355 0.000722487844607976
3.57357357357357 0.000672739365441019
3.59359359359359 0.000626165408357979
3.61361361361361 0.000582582234459159
3.63363363363363 0.000541815389165405
3.65365365365365 0.00050369932583812
3.67367367367367 0.000468077038447599
3.69369369369369 0.000434799703508357
3.71371371371371 0.000403726331459719
3.73373373373373 0.000374723427631836
3.75375375375375 0.000347664662901427
3.77377377377377 0.000322430554107926
3.79379379379379 0.000298908154269281
3.81381381381381 0.000276990752607344
3.83383383383383 0.000256577584365551
3.85385385385385 0.000237573550376443
3.87387387387387 0.000219888946313312
3.89389389389389 0.000203439201538965
3.91391391391391 0.00018814462744512
3.93393393393393 0.000173930175158222
3.95395395395395 0.000160725202471485
3.97397397397397 0.000148463249848567
3.99399399399399 0.000137081825331481
4.01401401401401 0.000126522198173995
4.03403403403403 0.000116729201011866
4.05405405405405 0.000107651040372646
4.07407407407407 9.92391153205018e-05
4.09409409409409 9.14478440253317e-05
4.11411411411411 8.42344980404937e-05
4.13413413413413 7.75590440694795e-05
4.15415415415415 7.13839929989176e-05
4.17417417417417 6.56742559732345e-05
4.19419419419419 6.03970072851107e-05
4.21421421421421 5.55215538554582e-05
4.23423423423423 5.10192110769697e-05
4.25425425425425 4.68631847962824e-05
4.27427427427427 4.3028459211397e-05
4.29429429429429 3.94916904631592e-05
4.31431431431431 3.62311057022691e-05
4.33433433433433 3.32264074164067e-05
4.35435435435435 3.04586828055866e-05
4.37437437437437 2.79103179977393e-05
4.39439439439439 2.55649169007242e-05
4.41441441441441 2.34072244914564e-05
4.43443443443443 2.14230543475548e-05
4.45445445445445 1.95992202318354e-05
4.47447447447447 1.79234715450684e-05
4.49449449449449 1.63844324676437e-05
4.51451451451451 1.49715446161227e-05
4.53453453453453 1.3675013046071e-05
4.55455455455455 1.24857554380297e-05
4.57457457457457 1.13953543089884e-05
4.59459459459459 1.03960120972233e-05
4.61461461461461 9.48050897386715e-06
4.63463463463463 8.64216324004121e-06
4.65465465465465 7.87479417380527e-06
4.67467467467467 7.17268719654363e-06
4.69469469469469 6.53056123369604e-06
4.71471471471471 5.94353814994721e-06
4.73473473473473 5.40711414409909e-06
4.75475475475475 4.91713299385613e-06
4.77477477477477 4.4697610456467e-06
4.79479479479479 4.06146384937932e-06
4.81481481481481 3.68898434268124e-06
4.83483483483483 3.34932249368796e-06
4.85485485485485 3.03971631583941e-06
4.87487487487487 2.75762417238901e-06
4.89489489489489 2.50070829244518e-06
4.91491491491491 2.26681942433715e-06
4.93493493493493 2.05398255592983e-06
4.95495495495495 1.86038363520377e-06
4.97497497497497 1.68435722796805e-06
4.99499499499499 1.5243750529858e-06
5.01501501501502 1.37903533806611e-06
5.03503503503504 1.24705294381396e-06
5.05505505505506 1.12725020473309e-06
5.07507507507508 1.01854844024876e-06
5.0950950950951 9.19960090959837e-07
5.11511511511512 8.30581438046261e-07
5.13513513513514 7.49585866251233e-07
5.15515515515516 6.76217633231267e-07
5.17517517517518 6.09786110324583e-07
5.1951951951952 5.4966046193278e-07
5.21521521521522 4.95264732746229e-07
5.23523523523524 4.46073313973501e-07
5.25525525525526 4.01606761563285e-07
5.27527527527528 3.61427941137511e-07
5.2952952952953 3.25138475990267e-07
5.31531531531532 2.92375476052561e-07
5.33533533533534 2.62808527181656e-07
5.35535535535536 2.36136921509202e-07
5.37537537537538 2.1208711087848e-07
5.3953953953954 1.90410366621162e-07
5.41541541541542 1.70880630071684e-07
5.43543543543544 1.5329253929596e-07
5.45545545545546 1.3745961852414e-07
5.47547547547548 1.23212617727566e-07
5.4954954954955 1.10397990671284e-07
5.51551551551552 9.8876500608364e-08
5.53553553553554 8.85219435638491e-08
5.55555555555556 7.92199798873018e-08
5.57557557557558 7.08670654362614e-08
5.5955955955956 6.33694743912418e-08
5.61561561561562 5.66424062986154e-08
5.63563563563564 5.06091704933412e-08
5.65565565565566 4.5200441571292e-08
5.67567567567568 4.03535800631662e-08
5.6956956956957 3.60120129107462e-08
5.71571571571572 3.2124668763613e-08
5.73573573573574 2.86454635022852e-08
5.75575575575576 2.55328317539416e-08
5.77577577577578 2.27493005011625e-08
5.7957957957958 2.02611011941336e-08
5.81581581581582 1.80378170640688e-08
5.83583583583584 1.60520626017053e-08
5.85585585585586 1.42791924110059e-08
5.87587587587588 1.2697036876002e-08
5.8958958958959 1.12856622892642e-08
5.91591591591592 1.00271532849868e-08
5.93593593593594 8.90541559921139e-09
5.95595595595596 7.90599734535251e-09
5.97597597597598 7.01592714588943e-09
5.995995995996 6.22356760178439e-09
6.01601601601602 5.5184827107339e-09
6.03603603603604 4.89131796456914e-09
6.05605605605606 4.33369196574694e-09
6.07607607607608 3.83809850362692e-09
6.0960960960961 3.3978181237675e-09
6.11611611611612 3.00683830841826e-09
6.13613613613614 2.65978146430924e-09
6.15615615615616 2.35183998527871e-09
6.17617617617618 2.07871772273778e-09
6.1961961961962 1.83657725691022e-09
6.21621621621622 1.6219924166386e-09
6.23623623623624 1.43190554571785e-09
6.25625625625626 1.26358905957511e-09
6.27627627627628 1.11461087800764e-09
6.2962962962963 9.82803357938119e-10
6.31631631631632 8.66235385045992e-10
6.33633633633634 7.63187314959515e-10
6.35635635635636 6.72128483698836e-10
6.37637637637638 5.91697033481532e-10
6.3963963963964 5.20681824054164e-10
6.41641641641642 4.58006221596989e-10
6.43643643643644 4.02713577147895e-10
6.45645645645646 3.53954224575805e-10
6.47647647647648 3.10973844559378e-10
6.4964964964965 2.73103055937435e-10
6.51651651651652 2.39748109325539e-10
6.53653653653654 2.10382570159619e-10
6.55655655655656 1.84539889444162e-10
6.57657657657658 1.61806770551276e-10
6.5965965965966 1.41817249531742e-10
6.61661661661662 1.24247414645767e-10
6.63663663663664 1.08810698278133e-10
6.65665665665666 9.52536811418137e-11
6.67667667667668 8.33523547614393e-11
6.6966966966967 7.29087937236022e-11
6.71671671671672 6.37481941394484e-11
6.73673673673674 5.57162392365998e-11
6.75675675675676 4.86767570277076e-11
6.77677677677678 4.25096386334907e-11
6.7967967967968 3.71089891068556e-11
6.81681681681682 3.23814855461048e-11
6.83683683683684 2.82449199306813e-11
6.85685685685686 2.46269064908966e-11
6.87687687687688 2.14637355595386e-11
6.8968968968969 1.86993577716892e-11
6.91691691691692 1.62844842008236e-11
6.93693693693694 1.41757895636779e-11
6.95695695695696 1.23352070109961e-11
6.97697697697698 1.07293042619725e-11
6.996996996997 9.32873195138548e-12
7.01701701701702 8.10773605306642e-12
7.03703703703704 7.04372713322415e-12
7.05705705705706 6.11689998287643e-12
7.07707707707708 5.30989788981512e-12
7.0970970970971 4.6075164458095e-12
7.11711711711712 3.99644235193917e-12
7.13713713713714 3.46502319108337e-12
7.15715715715716 3.00306458801651e-12
7.17717717717718 2.60165157997869e-12
7.1971971971972 2.25299137914217e-12
7.21721721721722 1.95027502769791e-12
7.23723723723724 1.68755573049414e-12
7.25725725725726 1.45964190299846e-12
7.27727727727728 1.262003197176e-12
7.2972972972973 1.0906879676822e-12
7.31731731731732 9.42250818252902e-13
7.33733733733734 8.13689025751829e-13
7.35735735735736 7.02386779168733e-13
7.37737737737738 6.06066294885245e-13
7.3973973973974 5.22744979473708e-13
7.41741741741742 4.50697908714381e-13
7.43743743743744 3.88424977793729e-13
7.45745745745746 3.34622154016963e-13
7.47747747747748 2.88156330935933e-13
7.4974974974975 2.48043342543511e-13
7.51751751751752 2.13428748996348e-13
7.53753753753754 1.83571051982055e-13
7.55755755755756 1.57827039041883e-13
7.57757757757758 1.35638992516663e-13
7.5975975975976 1.16523530854698e-13
7.61761761761762 1.000618782966e-13
7.63763763763764 8.58913838706873e-14
7.65765765765766 7.36981325813112e-14
7.67767767767768 6.32105109959248e-14
7.6976976976977 5.41936064405376e-14
7.71771771771772 4.64443339685915e-14
7.73773773773774 3.97871984155707e-14
7.75775775775776 3.40706104038536e-14
7.77777777777778 2.91636853079907e-14
7.7977977977978 2.4953463096687e-14
7.81781781781782 2.13424947819474e-14
7.83783783783784 1.82467480586654e-14
7.85785785785786 1.55937907247682e-14
7.87787787787788 1.33212157347804e-14
7.8978978978979 1.13752763482777e-14
7.91791791791792 9.70970386854701e-15
7.93793793793794 8.28468399583068e-15
7.95795795795796 7.06597090549298e-15
7.97797797797798 6.02412085866189e-15
7.997997997998 5.13382950919603e-15
8.01801801801802 4.3733591283238e-15
8.03803803803804 3.72404376402735e-15
8.05805805805806 3.16986191875719e-15
8.07807807807808 2.69706769497134e-15
8.0980980980981 2.29387254841619e-15
8.11811811811812 1.95017082606148e-15
8.13813813813814 1.65730316851105e-15
8.15815815815816 1.40785264250169e-15
8.17817817817818 1.19546915264237e-15
8.1981981981982 1.01471827585831e-15
8.21821821821822 8.60951178494166e-16
8.23823823823824 7.30192724674871e-16
8.25825825825826 6.19045274051723e-16
8.27827827827828 5.24606005103494e-16
8.2982982982983 4.44395893386326e-16
8.31831831831832 3.76298728354137e-16
8.33833833833834 3.18508772685576e-16
8.35835835835836 2.69485858889318e-16
8.37837837837838 2.27916883183252e-16
8.3983983983984 1.92682799625235e-16
8.41841841841842 1.62830341150177e-16
8.43843843843844 1.37547801096493e-16
8.45845845845846 1.16144301209627e-16
8.47847847847848 9.8032051926925e-17
8.4984984984985 8.27111796592674e-17
8.51851851851852 6.97567552529606e-17
8.53853853853854 5.88077091106193e-17
8.55855855855856 4.95573626747691e-17
8.57857857857858 4.17453440895294e-17
8.5985985985986 3.51506886838449e-17
8.61861861861862 2.95859531836935e-17
8.63863863863864 2.48921968838275e-17
8.65865865865866 2.09347039316773e-17
8.67867867867868 1.75993388643364e-17
8.6986986986987 1.47894429982973e-17
8.71871871871872 1.24231925503958e-17
8.73873873873874 1.04313507694241e-17
8.75875875875876 8.75535614211525e-18
8.77877877877878 7.34569713009337e-18
8.7987987987988 6.16053109048305e-18
8.81881881881882 5.16451119999019e-18
8.83883883883884 4.32779048512422e-18
8.85885885885886 3.62517658454128e-18
8.87887887887888 3.03541474067764e-18
8.8988988988989 2.54057982941549e-18
8.91891891891892 2.12556106807901e-18
8.93893893893894 1.77762546207239e-18
8.95895895895896 1.48604811781133e-18
8.97897897897898 1.24179931485728e-18
8.998998998999 1.03727973679138e-18
9.01901901901902 8.66096545670873e-19
9.03903903903904 7.22874080905507e-19
9.05905905905906 6.03093897535385e-19
9.07907907907908 5.0295965472299e-19
9.0990990990991 4.19283042962206e-19
9.11911911911912 3.49387515332417e-19
9.13913913913914 2.91027078875695e-19
9.15915915915916 2.42317819504618e-19
9.17917917917918 2.01680188581445e-19
9.1991991991992 1.67790380698338e-19
9.21921921921922 1.39539388139188e-19
9.23923923923924 1.1599853476858e-19
9.25925925925926 9.63904764362469e-20
9.27927927927928 8.00648113242436e-20
9.2992992992993 6.6477576193283e-20
9.31931931931932 5.51740167796855e-20
9.33933933933934 4.5774115701642e-20
9.35935935935936 3.79604417474457e-20
9.37937937937938 3.14679525475421e-20
9.3993993993994 2.60754402558043e-20
9.41941941941942 2.15983585814365e-20
9.43943943943944 1.78828106797973e-20
9.45945945945946 1.48005121824234e-20
9.47947947947948 1.22445730038445e-20
9.4994994994995 1.01259663374544e-20
9.51951951951952 8.37057415073758e-21
9.53953953953954 6.91671611023779e-21
9.55955955955956 5.71308371631411e-21
9.57957957957958 4.71701393695898e-21
9.5995995995996 3.89304716291232e-21
9.61961961961962 3.21172317125736e-21
9.63963963963964 2.64857624243904e-21
9.65965965965966 2.18329684678274e-21
9.67967967967968 1.79903258756111e-21
9.6996996996997 1.48180551599987e-21
9.71971971971972 1.22002665237565e-21
9.73973973973974 1.00409166885045e-21
9.75975975975976 8.26044308669654e-22
9.77977977977978 6.79296312742742e-22
9.7997997997998 5.58394465795474e-22
9.81981981981982 4.58826916995414e-22
9.83983983983984 3.7686222201397e-22
9.85985985985986 3.09415635142992e-22
9.87987987987988 2.53938085193762e-22
9.8998998998999 2.08324025950642e-22
9.91991991991992 1.70834984871876e-22
9.93993993993994 1.40036162642795e-22
9.95995995995996 1.14743877987917e-22
9.97997997997998 9.39820210218911e-23
10 7.69459862670642e-23
};
\addlegendentry{N(0,1)}; 
\legend{}; 
\end{axis}

\end{tikzpicture}

%% file: FinalFigs/Null_Dists_d_10_500_n_200_m_20_kernel__Gaussian_Poly_2_2022_10_12_23_28_39mmd.tex
\begin{tikzpicture}

\definecolor{darkorange25512714}{RGB}{255,127,14}
\definecolor{darkslategray38}{RGB}{38,38,38}
\definecolor{lightgray204}{RGB}{204,204,204}
\definecolor{steelblue31119180}{RGB}{31,119,180}

\begin{axis}[
axis line style={darkslategray38},
height=\figheight,
legend cell align={left},
legend style={fill opacity=0.8, draw opacity=1, text opacity=1, draw=none},
tick align=outside,
tick pos=left,
title={$\dmmd~(n/m=10)$},
width=\figwidth,
x grid style={lightgray204},
xmin=-6, xmax=6,
xtick style={color=darkslategray38},
y grid style={lightgray204},
ylabel = {}, 
ymin=0, ymax=4.37727296784821,
ytick style={color=darkslategray38}, 
xticklabels=empty,
yticklabels=empty
]
\draw[draw=none,fill=steelblue31119180,fill opacity=0.8] (axis cs:-1.50988662242889,0) rectangle (axis cs:-1.41393649578094,0.00833766072414511);
\addlegendimage{ybar,ybar legend,draw=none,fill=steelblue31119180,fill opacity=0.8}
\addlegendentry{mmd (d=10)}

\draw[draw=none,fill=steelblue31119180,fill opacity=0.8] (axis cs:-1.27001118659973,0) rectangle (axis cs:-1.17406105995178,0);
\draw[draw=none,fill=steelblue31119180,fill opacity=0.8] (axis cs:-1.03013586997986,0) rectangle (axis cs:-0.934185743331909,0.0416883139795067);
\draw[draw=none,fill=steelblue31119180,fill opacity=0.8] (axis cs:-0.790260493755341,0) rectangle (axis cs:-0.694310367107391,0.12506494193852);
\draw[draw=none,fill=steelblue31119180,fill opacity=0.8] (axis cs:-0.550385117530823,0) rectangle (axis cs:-0.454434990882874,0.550285744529489);
\draw[draw=none,fill=steelblue31119180,fill opacity=0.8] (axis cs:-0.310509741306305,0) rectangle (axis cs:-0.214559614658356,0.967168884324556);
\draw[draw=none,fill=steelblue31119180,fill opacity=0.8] (axis cs:-0.0706343501806259,0) rectangle (axis cs:0.0253157764673233,1.09223382626308);
\draw[draw=none,fill=steelblue31119180,fill opacity=0.8] (axis cs:0.169241026043892,0) rectangle (axis cs:0.26519113779068,0.925480570345049);
\draw[draw=none,fill=steelblue31119180,fill opacity=0.8] (axis cs:0.409116387367249,0) rectangle (axis cs:0.505066514015198,0.350181837427856);
\draw[draw=none,fill=steelblue31119180,fill opacity=0.8] (axis cs:0.648991763591766,0) rectangle (axis cs:0.744941890239716,0.108389616346717);
\draw[draw=none,fill=darkorange25512714,fill opacity=0.8] (axis cs:-1.41393637657166,0) rectangle (axis cs:-1.31798624992371,0);
\addlegendimage{ybar,ybar legend,draw=none,fill=darkorange25512714,fill opacity=0.8}
\addlegendentry{mmd (d=500)}

\draw[draw=none,fill=darkorange25512714,fill opacity=0.8] (axis cs:-1.17406105995178,0) rectangle (axis cs:-1.07811093330383,0);
\draw[draw=none,fill=darkorange25512714,fill opacity=0.8] (axis cs:-0.934185743331909,0) rectangle (axis cs:-0.83823561668396,0);
\draw[draw=none,fill=darkorange25512714,fill opacity=0.8] (axis cs:-0.694310307502747,0) rectangle (axis cs:-0.598360180854797,0);
\draw[draw=none,fill=darkorange25512714,fill opacity=0.8] (axis cs:-0.454434961080551,0) rectangle (axis cs:-0.358484834432602,0);
\draw[draw=none,fill=darkorange25512714,fill opacity=0.8] (axis cs:-0.214559584856033,0) rectangle (axis cs:-0.118609458208084,0);
\draw[draw=none,fill=darkorange25512714,fill opacity=0.8] (axis cs:0.0253157913684845,0) rectangle (axis cs:0.121265918016434,4.16883139795067);
\draw[draw=none,fill=darkorange25512714,fill opacity=0.8] (axis cs:0.265191167593002,0) rectangle (axis cs:0.361141294240952,0);
\draw[draw=none,fill=darkorange25512714,fill opacity=0.8] (axis cs:0.505066514015198,0) rectangle (axis cs:0.601016640663147,0);
\draw[draw=none,fill=darkorange25512714,fill opacity=0.8] (axis cs:0.74494194984436,0) rectangle (axis cs:0.84089207649231,0);
\addplot [semithick, black]
table {%
-10 7.69459862670642e-23
-9.97997997997998 9.39820210218911e-23
-9.95995995995996 1.14743877987917e-22
-9.93993993993994 1.40036162642795e-22
-9.91991991991992 1.70834984871876e-22
-9.8998998998999 2.08324025950642e-22
-9.87987987987988 2.53938085193762e-22
-9.85985985985986 3.09415635142992e-22
-9.83983983983984 3.7686222201397e-22
-9.81981981981982 4.58826916995414e-22
-9.7997997997998 5.58394465795474e-22
-9.77977977977978 6.79296312742742e-22
-9.75975975975976 8.26044308669654e-22
-9.73973973973974 1.00409166885045e-21
-9.71971971971972 1.22002665237565e-21
-9.6996996996997 1.48180551599987e-21
-9.67967967967968 1.79903258756111e-21
-9.65965965965966 2.18329684678274e-21
-9.63963963963964 2.64857624243904e-21
-9.61961961961962 3.21172317125736e-21
-9.5995995995996 3.89304716291232e-21
-9.57957957957958 4.71701393695898e-21
-9.55955955955956 5.71308371631411e-21
-9.53953953953954 6.91671611023779e-21
-9.51951951951952 8.37057415073758e-21
-9.4994994994995 1.01259663374544e-20
-9.47947947947948 1.22445730038445e-20
-9.45945945945946 1.48005121824234e-20
-9.43943943943944 1.78828106797973e-20
-9.41941941941942 2.15983585814365e-20
-9.3993993993994 2.60754402558043e-20
-9.37937937937938 3.14679525475421e-20
-9.35935935935936 3.79604417474457e-20
-9.33933933933934 4.5774115701642e-20
-9.31931931931932 5.51740167796855e-20
-9.2992992992993 6.6477576193283e-20
-9.27927927927928 8.00648113242436e-20
-9.25925925925926 9.63904764362469e-20
-9.23923923923924 1.1599853476858e-19
-9.21921921921922 1.39539388139188e-19
-9.1991991991992 1.67790380698338e-19
-9.17917917917918 2.01680188581445e-19
-9.15915915915916 2.42317819504618e-19
-9.13913913913914 2.91027078875695e-19
-9.11911911911912 3.49387515332417e-19
-9.0990990990991 4.19283042962206e-19
-9.07907907907908 5.0295965472299e-19
-9.05905905905906 6.03093897535385e-19
-9.03903903903904 7.22874080905507e-19
-9.01901901901902 8.66096545670873e-19
-8.998998998999 1.03727973679138e-18
-8.97897897897898 1.24179931485728e-18
-8.95895895895896 1.48604811781133e-18
-8.93893893893894 1.77762546207239e-18
-8.91891891891892 2.12556106807901e-18
-8.8988988988989 2.54057982941549e-18
-8.87887887887888 3.03541474067764e-18
-8.85885885885886 3.62517658454128e-18
-8.83883883883884 4.32779048512422e-18
-8.81881881881882 5.16451119999019e-18
-8.7987987987988 6.16053109048305e-18
-8.77877877877878 7.34569713009337e-18
-8.75875875875876 8.75535614211525e-18
-8.73873873873874 1.04313507694241e-17
-8.71871871871872 1.24231925503958e-17
-8.6986986986987 1.47894429982973e-17
-8.67867867867868 1.75993388643364e-17
-8.65865865865866 2.09347039316773e-17
-8.63863863863864 2.48921968838275e-17
-8.61861861861862 2.95859531836935e-17
-8.5985985985986 3.51506886838449e-17
-8.57857857857858 4.17453440895294e-17
-8.55855855855856 4.95573626747691e-17
-8.53853853853854 5.88077091106193e-17
-8.51851851851852 6.97567552529606e-17
-8.4984984984985 8.27111796592674e-17
-8.47847847847848 9.8032051926925e-17
-8.45845845845846 1.16144301209627e-16
-8.43843843843844 1.37547801096493e-16
-8.41841841841842 1.62830341150177e-16
-8.3983983983984 1.92682799625235e-16
-8.37837837837838 2.27916883183252e-16
-8.35835835835836 2.69485858889318e-16
-8.33833833833834 3.18508772685576e-16
-8.31831831831832 3.76298728354137e-16
-8.2982982982983 4.44395893386326e-16
-8.27827827827828 5.24606005103494e-16
-8.25825825825826 6.19045274051723e-16
-8.23823823823824 7.30192724674871e-16
-8.21821821821822 8.60951178494166e-16
-8.1981981981982 1.01471827585831e-15
-8.17817817817818 1.19546915264237e-15
-8.15815815815816 1.40785264250169e-15
-8.13813813813814 1.65730316851105e-15
-8.11811811811812 1.95017082606148e-15
-8.0980980980981 2.29387254841619e-15
-8.07807807807808 2.69706769497134e-15
-8.05805805805806 3.16986191875719e-15
-8.03803803803804 3.72404376402735e-15
-8.01801801801802 4.3733591283238e-15
-7.997997997998 5.13382950919607e-15
-7.97797797797798 6.02412085866193e-15
-7.95795795795796 7.06597090549303e-15
-7.93793793793794 8.28468399583074e-15
-7.91791791791792 9.70970386854708e-15
-7.8978978978979 1.13752763482777e-14
-7.87787787787788 1.33212157347805e-14
-7.85785785785786 1.55937907247683e-14
-7.83783783783784 1.82467480586655e-14
-7.81781781781782 2.13424947819475e-14
-7.7977977977978 2.49534630966872e-14
-7.77777777777778 2.91636853079909e-14
-7.75775775775776 3.40706104038538e-14
-7.73773773773774 3.9787198415571e-14
-7.71771771771772 4.64443339685918e-14
-7.6976976976977 5.4193606440538e-14
-7.67767767767768 6.32105109959252e-14
-7.65765765765766 7.36981325813117e-14
-7.63763763763764 8.58913838706879e-14
-7.61761761761762 1.00061878296601e-13
-7.5975975975976 1.16523530854699e-13
-7.57757757757758 1.35638992516664e-13
-7.55755755755756 1.57827039041884e-13
-7.53753753753754 1.83571051982057e-13
-7.51751751751752 2.13428748996349e-13
-7.4974974974975 2.48043342543513e-13
-7.47747747747748 2.88156330935935e-13
-7.45745745745746 3.34622154016965e-13
-7.43743743743744 3.88424977793732e-13
-7.41741741741742 4.50697908714384e-13
-7.3973973973974 5.22744979473711e-13
-7.37737737737738 6.06066294885249e-13
-7.35735735735736 7.02386779168738e-13
-7.33733733733734 8.13689025751835e-13
-7.31731731731732 9.42250818252909e-13
-7.2972972972973 1.09068796768221e-12
-7.27727727727728 1.262003197176e-12
-7.25725725725726 1.45964190299847e-12
-7.23723723723724 1.68755573049416e-12
-7.21721721721722 1.95027502769792e-12
-7.1971971971972 2.25299137914218e-12
-7.17717717717718 2.60165157997871e-12
-7.15715715715716 3.00306458801653e-12
-7.13713713713714 3.4650231910834e-12
-7.11711711711712 3.99644235193919e-12
-7.0970970970971 4.60751644580953e-12
-7.07707707707708 5.30989788981514e-12
-7.05705705705706 6.11689998287646e-12
-7.03703703703704 7.0437271332242e-12
-7.01701701701702 8.10773605306645e-12
-6.996996996997 9.32873195138555e-12
-6.97697697697698 1.07293042619726e-11
-6.95695695695696 1.23352070109962e-11
-6.93693693693694 1.41757895636779e-11
-6.91691691691692 1.62844842008237e-11
-6.8968968968969 1.86993577716893e-11
-6.87687687687688 2.14637355595386e-11
-6.85685685685686 2.46269064908967e-11
-6.83683683683684 2.82449199306815e-11
-6.81681681681682 3.2381485546105e-11
-6.7967967967968 3.71089891068559e-11
-6.77677677677678 4.25096386334913e-11
-6.75675675675676 4.86767570277083e-11
-6.73673673673674 5.57162392366004e-11
-6.71671671671672 6.37481941394491e-11
-6.6966966966967 7.29087937236032e-11
-6.67667667667668 8.33523547614402e-11
-6.65665665665666 9.52536811418151e-11
-6.63663663663664 1.08810698278135e-10
-6.61661661661662 1.24247414645768e-10
-6.5965965965966 1.41817249531744e-10
-6.57657657657658 1.61806770551278e-10
-6.55655655655656 1.84539889444164e-10
-6.53653653653654 2.10382570159622e-10
-6.51651651651652 2.39748109325542e-10
-6.4964964964965 2.73103055937438e-10
-6.47647647647648 3.10973844559381e-10
-6.45645645645646 3.53954224575809e-10
-6.43643643643644 4.027135771479e-10
-6.41641641641642 4.58006221596996e-10
-6.3963963963964 5.20681824054169e-10
-6.37637637637638 5.91697033481538e-10
-6.35635635635636 6.72128483698846e-10
-6.33633633633634 7.63187314959523e-10
-6.31631631631632 8.66235385046001e-10
-6.2962962962963 9.8280335793813e-10
-6.27627627627628 1.11461087800766e-09
-6.25625625625626 1.26358905957513e-09
-6.23623623623624 1.43190554571787e-09
-6.21621621621622 1.62199241663862e-09
-6.1961961961962 1.83657725691024e-09
-6.17617617617618 2.0787177227378e-09
-6.15615615615616 2.35183998527873e-09
-6.13613613613614 2.65978146430928e-09
-6.11611611611612 3.00683830841829e-09
-6.0960960960961 3.39781812376754e-09
-6.07607607607608 3.83809850362696e-09
-6.05605605605606 4.33369196574699e-09
-6.03603603603604 4.89131796456919e-09
-6.01601601601602 5.51848271073395e-09
-5.995995995996 6.22356760178439e-09
-5.97597597597598 7.01592714588943e-09
-5.95595595595596 7.90599734535251e-09
-5.93593593593594 8.90541559921139e-09
-5.91591591591592 1.00271532849868e-08
-5.8958958958959 1.12856622892642e-08
-5.87587587587588 1.2697036876002e-08
-5.85585585585586 1.42791924110059e-08
-5.83583583583584 1.60520626017053e-08
-5.81581581581582 1.80378170640688e-08
-5.7957957957958 2.02611011941336e-08
-5.77577577577578 2.27493005011625e-08
-5.75575575575576 2.55328317539416e-08
-5.73573573573574 2.86454635022852e-08
-5.71571571571572 3.2124668763613e-08
-5.6956956956957 3.60120129107462e-08
-5.67567567567568 4.03535800631662e-08
-5.65565565565566 4.5200441571292e-08
-5.63563563563564 5.06091704933412e-08
-5.61561561561562 5.66424062986154e-08
-5.5955955955956 6.33694743912418e-08
-5.57557557557558 7.08670654362614e-08
-5.55555555555556 7.92199798873018e-08
-5.53553553553554 8.85219435638491e-08
-5.51551551551552 9.8876500608364e-08
-5.4954954954955 1.10397990671284e-07
-5.47547547547548 1.23212617727566e-07
-5.45545545545546 1.3745961852414e-07
-5.43543543543544 1.5329253929596e-07
-5.41541541541542 1.70880630071684e-07
-5.3953953953954 1.90410366621162e-07
-5.37537537537538 2.1208711087848e-07
-5.35535535535536 2.36136921509202e-07
-5.33533533533534 2.62808527181656e-07
-5.31531531531532 2.92375476052561e-07
-5.2952952952953 3.25138475990267e-07
-5.27527527527528 3.61427941137511e-07
-5.25525525525526 4.01606761563285e-07
-5.23523523523524 4.46073313973501e-07
-5.21521521521522 4.95264732746229e-07
-5.1951951951952 5.4966046193278e-07
-5.17517517517518 6.09786110324583e-07
-5.15515515515516 6.76217633231267e-07
-5.13513513513514 7.49585866251233e-07
-5.11511511511512 8.30581438046261e-07
-5.0950950950951 9.19960090959837e-07
-5.07507507507508 1.01854844024876e-06
-5.05505505505506 1.12725020473309e-06
-5.03503503503504 1.24705294381396e-06
-5.01501501501502 1.37903533806611e-06
-4.99499499499499 1.5243750529858e-06
-4.97497497497497 1.68435722796805e-06
-4.95495495495495 1.86038363520377e-06
-4.93493493493493 2.05398255592983e-06
-4.91491491491491 2.26681942433715e-06
-4.89489489489489 2.50070829244518e-06
-4.87487487487487 2.75762417238901e-06
-4.85485485485485 3.03971631583941e-06
-4.83483483483483 3.34932249368796e-06
-4.81481481481481 3.68898434268124e-06
-4.79479479479479 4.06146384937932e-06
-4.77477477477477 4.4697610456467e-06
-4.75475475475475 4.91713299385613e-06
-4.73473473473473 5.40711414409909e-06
-4.71471471471471 5.94353814994721e-06
-4.69469469469469 6.53056123369604e-06
-4.67467467467467 7.17268719654363e-06
-4.65465465465465 7.87479417380527e-06
-4.63463463463463 8.64216324004121e-06
-4.61461461461461 9.48050897386715e-06
-4.59459459459459 1.03960120972233e-05
-4.57457457457457 1.13953543089884e-05
-4.55455455455455 1.24857554380297e-05
-4.53453453453453 1.3675013046071e-05
-4.51451451451451 1.49715446161227e-05
-4.49449449449449 1.63844324676437e-05
-4.47447447447447 1.79234715450684e-05
-4.45445445445445 1.95992202318354e-05
-4.43443443443443 2.14230543475548e-05
-4.41441441441441 2.34072244914564e-05
-4.39439439439439 2.55649169007242e-05
-4.37437437437437 2.79103179977393e-05
-4.35435435435435 3.04586828055866e-05
-4.33433433433433 3.32264074164067e-05
-4.31431431431431 3.62311057022691e-05
-4.29429429429429 3.94916904631592e-05
-4.27427427427427 4.3028459211397e-05
-4.25425425425425 4.68631847962824e-05
-4.23423423423423 5.10192110769697e-05
-4.21421421421421 5.55215538554582e-05
-4.19419419419419 6.03970072851107e-05
-4.17417417417417 6.56742559732345e-05
-4.15415415415415 7.13839929989176e-05
-4.13413413413413 7.75590440694795e-05
-4.11411411411411 8.42344980404937e-05
-4.09409409409409 9.14478440253317e-05
-4.07407407407407 9.92391153205018e-05
-4.05405405405405 0.000107651040372646
-4.03403403403403 0.000116729201011866
-4.01401401401401 0.000126522198173995
-3.99399399399399 0.000137081825331481
-3.97397397397397 0.000148463249848567
-3.95395395395395 0.000160725202471485
-3.93393393393393 0.000173930175158222
-3.91391391391391 0.00018814462744512
-3.89389389389389 0.000203439201538965
-3.87387387387387 0.000219888946313312
-3.85385385385385 0.000237573550376443
-3.83383383383383 0.000256577584365551
-3.81381381381381 0.000276990752607344
-3.79379379379379 0.000298908154269281
-3.77377377377377 0.000322430554107926
-3.75375375375375 0.000347664662901427
-3.73373373373373 0.000374723427631836
-3.71371371371371 0.000403726331459719
-3.69369369369369 0.000434799703508357
-3.67367367367367 0.000468077038447599
-3.65365365365365 0.00050369932583812
-3.63363363363363 0.000541815389165405
-3.61361361361361 0.000582582234459159
-3.59359359359359 0.000626165408357979
-3.57357357357357 0.000672739365441021
-3.55355355355355 0.000722487844607978
-3.53353353353353 0.00077560425424601
-3.51351351351351 0.000832292065877155
-3.49349349349349 0.000892765215932443
-3.47347347347347 0.000957248515249216
-3.45345345345345 0.0010259780658361
-3.43343343343343 0.00109920168439588
-3.41341341341341 0.00117717933203981
-3.39339339339339 0.00126018354956833
-3.37337337337337 0.00134849989763212
-3.35335335335335 0.00144242740102448
-3.33333333333333 0.00154227899629111
-3.31331331331331 0.00164838198177652
-3.29329329329329 0.00176107846915772
-3.27327327327327 0.00188072583544552
-3.25325325325325 0.00200769717436226
-3.23323323323323 0.00214238174593163
-3.21321321321321 0.00228518542304204
-3.19319319319319 0.00243653113367012
-3.17317317317317 0.00259685929737497
-3.15315315315315 0.00276662825459747
-3.13313313313313 0.00294631468722261
-3.11311311311311 0.00313641402878609
-3.09309309309309 0.00333744086263052
-3.07307307307307 0.00354992930624086
-3.05305305305305 0.00377443337991422
-3.03303303303303 0.00401152735784579
-3.01301301301301 0.00426180609964128
-2.99299299299299 0.00452588536019618
-2.97297297297297 0.0048044020758154
-2.95295295295295 0.00509801462438215
-2.93293293293293 0.00540740305732385
-2.91291291291291 0.00573326930106519
-2.89289289289289 0.00607633732560526
-2.87287287287287 0.00643735327780636
-2.85285285285285 0.00681708557693873
-2.83283283283283 0.00721632496998623
-2.81281281281281 0.00763588454418632
-2.79279279279279 0.00807659969425075
-2.77277277277277 0.00853932804169477
-2.75275275275275 0.00902494930369032
-2.73273273273273 0.00953436510885489
-2.71271271271271 0.0100684987573917
-2.69269269269269 0.0106282949230102
-2.67267267267267 0.0112147192940778
-2.65265265265265 0.0118287581514852
-2.63263263263263 0.0124714178807513
-2.61261261261261 0.0131437244159435
-2.59259259259259 0.0138467226130541
-2.57257257257257 0.0145814755505492
-2.55255255255255 0.0153490637548887
-2.53253253253253 0.0161505843489182
-2.51251251251251 0.0169871501211409
-2.49249249249249 0.0178598885140022
-2.47247247247247 0.018769940529451
-2.45245245245245 0.0197184595501959
-2.43243243243243 0.0207066100752274
-2.41241241241241 0.0217355663683581
-2.39239239239239 0.0228065110187135
-2.37237237237237 0.0239206334123108
-2.35235235235235 0.0250791281140699
-2.33233233233233 0.0262831931598317
-2.31231231231231 0.0275340282581906
-2.29229229229229 0.0288328329022027
-2.27227227227227 0.0301808043912899
-2.25225225225225 0.0315791357639331
-2.23223223223223 0.033029013642035
-2.21221221221221 0.0345316159881232
-2.19219219219219 0.0360881097768736
-2.17217217217217 0.0376996485827434
-2.15215215215215 0.0393673700858293
-2.13213213213213 0.0410923934983949
-2.11211211211211 0.0428758169148479
-2.09209209209209 0.0447187145882915
-2.07207207207207 0.0466221341371235
-2.05205205205205 0.0485870936855041
-2.03203203203203 0.0506145789418747
-2.01201201201201 0.0527055402200587
-1.99199199199199 0.0548608894078376
-1.97197197197197 0.057081496888248
-1.95195195195195 0.059368188419199
-1.93193193193193 0.0617217419773594
-1.91191191191191 0.0641428845726061
-1.89189189189189 0.0666322890396674
-1.87187187187187 0.0691905708139176
-1.85185185185185 0.0718182846986055
-1.83183183183183 0.0745159216311036
-1.81181181181181 0.077283905456062
-1.79179179179179 0.0801225897136326
-1.77177177177177 0.083032254451193
-1.75175175175175 0.0860131030672496
-1.73173173173173 0.0890652591964251
-1.71171171171171 0.0921887636446459
-1.69169169169169 0.0953835713838294
-1.67167167167167 0.0986495486155338
-1.65165165165165 0.101986469913169
-1.63163163163163 0.10539401545248
-1.61161161161161 0.108871768340093
-1.59159159159159 0.112419212049971
-1.57157157157157 0.116035727977651
-1.55155155155155 0.119720593122119
-1.53153153153153 0.123472977905145
-1.51151151151151 0.127291944137829
-1.49149149149149 0.13117644314399
-1.47147147147147 0.135125314049902
-1.45145145145145 0.139137282249685
-1.43143143143143 0.143210958055468
-1.41141141141141 0.147344835541168
-1.39139139139139 0.151537291588457
-1.37137137137137 0.155786585143159
-1.35135135135135 0.160090856689972
-1.33133133133133 0.164448127952996
-1.31131131131131 0.168856301829129
-1.29129129129129 0.173313162560933
-1.27127127127127 0.17781637615506
-1.25125125125125 0.182363491051798
-1.23123123123123 0.186951939050736
-1.21121121121121 0.191579036496956
-1.19119119119119 0.1962419857315
-1.17117117117117 0.200937876809264
-1.15115115115115 0.205663689486728
-1.13113113113113 0.210416295481265
-1.11111111111111 0.215192461003031
-1.09109109109109 0.219988849559688
-1.07107107107107 0.224802025033432
-1.05105105105105 0.229628455029052
-1.03103103103103 0.234464514490888
-1.01101101101101 0.239306489585817
-0.990990990990991 0.24415058184851
-0.970970970970971 0.24899291258444
-0.950950950950951 0.253829527525259
-0.930930930930931 0.258656401730343
-0.910910910910911 0.2634694447275
-0.890890890890891 0.268264505884996
-0.870870870870871 0.273037380006279
-0.850850850850851 0.277783813137949
-0.830830830830831 0.282499508580786
-0.810810810810811 0.287180133092853
-0.790790790790791 0.291821323272996
-0.77077077077077 0.296418692112302
-0.75075075075075 0.300967835700437
-0.73073073073073 0.305464340073112
-0.71071071071071 0.309903788186304
-0.69069069069069 0.314281767002296
-0.67067067067067 0.318593874672039
-0.65065065065065 0.322835727797843
-0.63063063063063 0.327002968759958
-0.61061061061061 0.331091273090187
-0.59059059059059 0.33509635687531
-0.57057057057057 0.339013984172804
-0.55055055055055 0.34283997442106
-0.53053053053053 0.346570209826128
-0.51051051051051 0.350200642706842
-0.49049049049049 0.353727302780113
-0.47047047047047 0.357146304368113
-0.45045045045045 0.360453853509139
-0.43043043043043 0.363646254953996
-0.41041041041041 0.366719919029892
-0.39039039039039 0.369671368354051
-0.37037037037037 0.372497244379499
-0.35035035035035 0.375194313755802
-0.33033033033033 0.377759474487924
-0.31031031031031 0.38018976187679
-0.29029029029029 0.382482354225654
-0.27027027027027 0.384634578296894
-0.25025025025025 0.386643914504485
-0.23023023023023 0.388508001828027
-0.21021021021021 0.390224642434919
-0.19019019019019 0.391791805998011
-0.17017017017017 0.393207633696876
-0.15015015015015 0.394470441891644
-0.13013013013013 0.395578725459258
-0.11011011011011 0.396531160782876
-0.0900900900900901 0.397326608386124
-0.07007007007007 0.397964115204853
-0.05005005005005 0.398442916490068
-0.03003003003003 0.398762437336696
-0.01001001001001 0.398922293833933
0.01001001001001 0.398922293833933
0.03003003003003 0.398762437336696
0.05005005005005 0.398442916490068
0.07007007007007 0.397964115204853
0.0900900900900901 0.397326608386124
0.11011011011011 0.396531160782876
0.13013013013013 0.395578725459258
0.15015015015015 0.394470441891644
0.17017017017017 0.393207633696876
0.19019019019019 0.391791805998011
0.21021021021021 0.390224642434919
0.23023023023023 0.388508001828027
0.25025025025025 0.386643914504485
0.27027027027027 0.384634578296894
0.29029029029029 0.382482354225654
0.31031031031031 0.38018976187679
0.33033033033033 0.377759474487924
0.35035035035035 0.375194313755802
0.37037037037037 0.372497244379499
0.39039039039039 0.369671368354051
0.41041041041041 0.366719919029892
0.43043043043043 0.363646254953996
0.45045045045045 0.360453853509139
0.47047047047047 0.357146304368113
0.49049049049049 0.353727302780113
0.51051051051051 0.350200642706842
0.53053053053053 0.346570209826128
0.55055055055055 0.34283997442106
0.57057057057057 0.339013984172804
0.59059059059059 0.33509635687531
0.61061061061061 0.331091273090187
0.63063063063063 0.327002968759958
0.65065065065065 0.322835727797843
0.67067067067067 0.318593874672039
0.69069069069069 0.314281767002296
0.71071071071071 0.309903788186304
0.73073073073073 0.305464340073112
0.75075075075075 0.300967835700437
0.77077077077077 0.296418692112302
0.790790790790791 0.291821323272996
0.810810810810811 0.287180133092853
0.830830830830831 0.282499508580786
0.850850850850851 0.277783813137949
0.870870870870871 0.273037380006279
0.890890890890891 0.268264505884996
0.910910910910911 0.2634694447275
0.930930930930931 0.258656401730343
0.950950950950951 0.253829527525259
0.970970970970971 0.24899291258444
0.990990990990991 0.24415058184851
1.01101101101101 0.239306489585817
1.03103103103103 0.234464514490888
1.05105105105105 0.229628455029052
1.07107107107107 0.224802025033432
1.09109109109109 0.219988849559688
1.11111111111111 0.215192461003031
1.13113113113113 0.210416295481265
1.15115115115115 0.205663689486728
1.17117117117117 0.200937876809264
1.19119119119119 0.1962419857315
1.21121121121121 0.191579036496956
1.23123123123123 0.186951939050736
1.25125125125125 0.182363491051798
1.27127127127127 0.17781637615506
1.29129129129129 0.173313162560933
1.31131131131131 0.168856301829129
1.33133133133133 0.164448127952996
1.35135135135135 0.160090856689972
1.37137137137137 0.155786585143159
1.39139139139139 0.151537291588457
1.41141141141141 0.147344835541168
1.43143143143143 0.143210958055468
1.45145145145145 0.139137282249685
1.47147147147147 0.135125314049902
1.49149149149149 0.13117644314399
1.51151151151151 0.127291944137829
1.53153153153153 0.123472977905145
1.55155155155155 0.119720593122119
1.57157157157157 0.116035727977651
1.59159159159159 0.112419212049971
1.61161161161161 0.108871768340093
1.63163163163163 0.10539401545248
1.65165165165165 0.101986469913169
1.67167167167167 0.0986495486155338
1.69169169169169 0.0953835713838294
1.71171171171171 0.0921887636446459
1.73173173173173 0.0890652591964251
1.75175175175175 0.0860131030672496
1.77177177177177 0.083032254451193
1.79179179179179 0.0801225897136326
1.81181181181181 0.077283905456062
1.83183183183183 0.0745159216311036
1.85185185185185 0.0718182846986055
1.87187187187187 0.0691905708139176
1.89189189189189 0.0666322890396674
1.91191191191191 0.0641428845726061
1.93193193193193 0.0617217419773594
1.95195195195195 0.059368188419199
1.97197197197197 0.057081496888248
1.99199199199199 0.0548608894078376
2.01201201201201 0.0527055402200588
2.03203203203203 0.0506145789418748
2.05205205205205 0.0485870936855042
2.07207207207207 0.0466221341371236
2.09209209209209 0.0447187145882916
2.11211211211211 0.0428758169148479
2.13213213213213 0.041092393498395
2.15215215215215 0.0393673700858294
2.17217217217217 0.0376996485827434
2.19219219219219 0.0360881097768737
2.21221221221221 0.0345316159881233
2.23223223223223 0.033029013642035
2.25225225225225 0.0315791357639332
2.27227227227227 0.03018080439129
2.29229229229229 0.0288328329022028
2.31231231231231 0.0275340282581906
2.33233233233233 0.0262831931598317
2.35235235235235 0.02507912811407
2.37237237237237 0.0239206334123108
2.39239239239239 0.0228065110187136
2.41241241241241 0.0217355663683581
2.43243243243243 0.0207066100752275
2.45245245245245 0.0197184595501959
2.47247247247247 0.0187699405294511
2.49249249249249 0.0178598885140022
2.51251251251251 0.016987150121141
2.53253253253253 0.0161505843489182
2.55255255255255 0.0153490637548888
2.57257257257257 0.0145814755505493
2.59259259259259 0.0138467226130541
2.61261261261261 0.0131437244159435
2.63263263263263 0.0124714178807514
2.65265265265265 0.0118287581514852
2.67267267267267 0.0112147192940778
2.69269269269269 0.0106282949230103
2.71271271271271 0.0100684987573917
2.73273273273273 0.00953436510885491
2.75275275275275 0.00902494930369034
2.77277277277277 0.00853932804169479
2.79279279279279 0.00807659969425077
2.81281281281281 0.00763588454418634
2.83283283283283 0.00721632496998621
2.85285285285285 0.00681708557693871
2.87287287287287 0.00643735327780635
2.89289289289289 0.00607633732560524
2.91291291291291 0.00573326930106518
2.93293293293293 0.00540740305732384
2.95295295295295 0.00509801462438214
2.97297297297297 0.00480440207581539
2.99299299299299 0.00452588536019617
3.01301301301301 0.00426180609964127
3.03303303303303 0.00401152735784578
3.05305305305305 0.00377443337991421
3.07307307307307 0.00354992930624085
3.09309309309309 0.00333744086263051
3.11311311311311 0.00313641402878608
3.13313313313313 0.0029463146872226
3.15315315315315 0.00276662825459746
3.17317317317317 0.00259685929737496
3.19319319319319 0.00243653113367011
3.21321321321321 0.00228518542304203
3.23323323323323 0.00214238174593163
3.25325325325325 0.00200769717436225
3.27327327327327 0.00188072583544551
3.29329329329329 0.00176107846915771
3.31331331331331 0.00164838198177652
3.33333333333333 0.0015422789962911
3.35335335335335 0.00144242740102448
3.37337337337337 0.00134849989763212
3.39339339339339 0.00126018354956833
3.41341341341341 0.00117717933203981
3.43343343343343 0.00109920168439588
3.45345345345345 0.0010259780658361
3.47347347347347 0.000957248515249212
3.49349349349349 0.000892765215932441
3.51351351351351 0.000832292065877152
3.53353353353353 0.000775604254246008
3.55355355355355 0.000722487844607976
3.57357357357357 0.000672739365441019
3.59359359359359 0.000626165408357979
3.61361361361361 0.000582582234459159
3.63363363363363 0.000541815389165405
3.65365365365365 0.00050369932583812
3.67367367367367 0.000468077038447599
3.69369369369369 0.000434799703508357
3.71371371371371 0.000403726331459719
3.73373373373373 0.000374723427631836
3.75375375375375 0.000347664662901427
3.77377377377377 0.000322430554107926
3.79379379379379 0.000298908154269281
3.81381381381381 0.000276990752607344
3.83383383383383 0.000256577584365551
3.85385385385385 0.000237573550376443
3.87387387387387 0.000219888946313312
3.89389389389389 0.000203439201538965
3.91391391391391 0.00018814462744512
3.93393393393393 0.000173930175158222
3.95395395395395 0.000160725202471485
3.97397397397397 0.000148463249848567
3.99399399399399 0.000137081825331481
4.01401401401401 0.000126522198173995
4.03403403403403 0.000116729201011866
4.05405405405405 0.000107651040372646
4.07407407407407 9.92391153205018e-05
4.09409409409409 9.14478440253317e-05
4.11411411411411 8.42344980404937e-05
4.13413413413413 7.75590440694795e-05
4.15415415415415 7.13839929989176e-05
4.17417417417417 6.56742559732345e-05
4.19419419419419 6.03970072851107e-05
4.21421421421421 5.55215538554582e-05
4.23423423423423 5.10192110769697e-05
4.25425425425425 4.68631847962824e-05
4.27427427427427 4.3028459211397e-05
4.29429429429429 3.94916904631592e-05
4.31431431431431 3.62311057022691e-05
4.33433433433433 3.32264074164067e-05
4.35435435435435 3.04586828055866e-05
4.37437437437437 2.79103179977393e-05
4.39439439439439 2.55649169007242e-05
4.41441441441441 2.34072244914564e-05
4.43443443443443 2.14230543475548e-05
4.45445445445445 1.95992202318354e-05
4.47447447447447 1.79234715450684e-05
4.49449449449449 1.63844324676437e-05
4.51451451451451 1.49715446161227e-05
4.53453453453453 1.3675013046071e-05
4.55455455455455 1.24857554380297e-05
4.57457457457457 1.13953543089884e-05
4.59459459459459 1.03960120972233e-05
4.61461461461461 9.48050897386715e-06
4.63463463463463 8.64216324004121e-06
4.65465465465465 7.87479417380527e-06
4.67467467467467 7.17268719654363e-06
4.69469469469469 6.53056123369604e-06
4.71471471471471 5.94353814994721e-06
4.73473473473473 5.40711414409909e-06
4.75475475475475 4.91713299385613e-06
4.77477477477477 4.4697610456467e-06
4.79479479479479 4.06146384937932e-06
4.81481481481481 3.68898434268124e-06
4.83483483483483 3.34932249368796e-06
4.85485485485485 3.03971631583941e-06
4.87487487487487 2.75762417238901e-06
4.89489489489489 2.50070829244518e-06
4.91491491491491 2.26681942433715e-06
4.93493493493493 2.05398255592983e-06
4.95495495495495 1.86038363520377e-06
4.97497497497497 1.68435722796805e-06
4.99499499499499 1.5243750529858e-06
5.01501501501502 1.37903533806611e-06
5.03503503503504 1.24705294381396e-06
5.05505505505506 1.12725020473309e-06
5.07507507507508 1.01854844024876e-06
5.0950950950951 9.19960090959837e-07
5.11511511511512 8.30581438046261e-07
5.13513513513514 7.49585866251233e-07
5.15515515515516 6.76217633231267e-07
5.17517517517518 6.09786110324583e-07
5.1951951951952 5.4966046193278e-07
5.21521521521522 4.95264732746229e-07
5.23523523523524 4.46073313973501e-07
5.25525525525526 4.01606761563285e-07
5.27527527527528 3.61427941137511e-07
5.2952952952953 3.25138475990267e-07
5.31531531531532 2.92375476052561e-07
5.33533533533534 2.62808527181656e-07
5.35535535535536 2.36136921509202e-07
5.37537537537538 2.1208711087848e-07
5.3953953953954 1.90410366621162e-07
5.41541541541542 1.70880630071684e-07
5.43543543543544 1.5329253929596e-07
5.45545545545546 1.3745961852414e-07
5.47547547547548 1.23212617727566e-07
5.4954954954955 1.10397990671284e-07
5.51551551551552 9.8876500608364e-08
5.53553553553554 8.85219435638491e-08
5.55555555555556 7.92199798873018e-08
5.57557557557558 7.08670654362614e-08
5.5955955955956 6.33694743912418e-08
5.61561561561562 5.66424062986154e-08
5.63563563563564 5.06091704933412e-08
5.65565565565566 4.5200441571292e-08
5.67567567567568 4.03535800631662e-08
5.6956956956957 3.60120129107462e-08
5.71571571571572 3.2124668763613e-08
5.73573573573574 2.86454635022852e-08
5.75575575575576 2.55328317539416e-08
5.77577577577578 2.27493005011625e-08
5.7957957957958 2.02611011941336e-08
5.81581581581582 1.80378170640688e-08
5.83583583583584 1.60520626017053e-08
5.85585585585586 1.42791924110059e-08
5.87587587587588 1.2697036876002e-08
5.8958958958959 1.12856622892642e-08
5.91591591591592 1.00271532849868e-08
5.93593593593594 8.90541559921139e-09
5.95595595595596 7.90599734535251e-09
5.97597597597598 7.01592714588943e-09
5.995995995996 6.22356760178439e-09
6.01601601601602 5.5184827107339e-09
6.03603603603604 4.89131796456914e-09
6.05605605605606 4.33369196574694e-09
6.07607607607608 3.83809850362692e-09
6.0960960960961 3.3978181237675e-09
6.11611611611612 3.00683830841826e-09
6.13613613613614 2.65978146430924e-09
6.15615615615616 2.35183998527871e-09
6.17617617617618 2.07871772273778e-09
6.1961961961962 1.83657725691022e-09
6.21621621621622 1.6219924166386e-09
6.23623623623624 1.43190554571785e-09
6.25625625625626 1.26358905957511e-09
6.27627627627628 1.11461087800764e-09
6.2962962962963 9.82803357938119e-10
6.31631631631632 8.66235385045992e-10
6.33633633633634 7.63187314959515e-10
6.35635635635636 6.72128483698836e-10
6.37637637637638 5.91697033481532e-10
6.3963963963964 5.20681824054164e-10
6.41641641641642 4.58006221596989e-10
6.43643643643644 4.02713577147895e-10
6.45645645645646 3.53954224575805e-10
6.47647647647648 3.10973844559378e-10
6.4964964964965 2.73103055937435e-10
6.51651651651652 2.39748109325539e-10
6.53653653653654 2.10382570159619e-10
6.55655655655656 1.84539889444162e-10
6.57657657657658 1.61806770551276e-10
6.5965965965966 1.41817249531742e-10
6.61661661661662 1.24247414645767e-10
6.63663663663664 1.08810698278133e-10
6.65665665665666 9.52536811418137e-11
6.67667667667668 8.33523547614393e-11
6.6966966966967 7.29087937236022e-11
6.71671671671672 6.37481941394484e-11
6.73673673673674 5.57162392365998e-11
6.75675675675676 4.86767570277076e-11
6.77677677677678 4.25096386334907e-11
6.7967967967968 3.71089891068556e-11
6.81681681681682 3.23814855461048e-11
6.83683683683684 2.82449199306813e-11
6.85685685685686 2.46269064908966e-11
6.87687687687688 2.14637355595386e-11
6.8968968968969 1.86993577716892e-11
6.91691691691692 1.62844842008236e-11
6.93693693693694 1.41757895636779e-11
6.95695695695696 1.23352070109961e-11
6.97697697697698 1.07293042619725e-11
6.996996996997 9.32873195138548e-12
7.01701701701702 8.10773605306642e-12
7.03703703703704 7.04372713322415e-12
7.05705705705706 6.11689998287643e-12
7.07707707707708 5.30989788981512e-12
7.0970970970971 4.6075164458095e-12
7.11711711711712 3.99644235193917e-12
7.13713713713714 3.46502319108337e-12
7.15715715715716 3.00306458801651e-12
7.17717717717718 2.60165157997869e-12
7.1971971971972 2.25299137914217e-12
7.21721721721722 1.95027502769791e-12
7.23723723723724 1.68755573049414e-12
7.25725725725726 1.45964190299846e-12
7.27727727727728 1.262003197176e-12
7.2972972972973 1.0906879676822e-12
7.31731731731732 9.42250818252902e-13
7.33733733733734 8.13689025751829e-13
7.35735735735736 7.02386779168733e-13
7.37737737737738 6.06066294885245e-13
7.3973973973974 5.22744979473708e-13
7.41741741741742 4.50697908714381e-13
7.43743743743744 3.88424977793729e-13
7.45745745745746 3.34622154016963e-13
7.47747747747748 2.88156330935933e-13
7.4974974974975 2.48043342543511e-13
7.51751751751752 2.13428748996348e-13
7.53753753753754 1.83571051982055e-13
7.55755755755756 1.57827039041883e-13
7.57757757757758 1.35638992516663e-13
7.5975975975976 1.16523530854698e-13
7.61761761761762 1.000618782966e-13
7.63763763763764 8.58913838706873e-14
7.65765765765766 7.36981325813112e-14
7.67767767767768 6.32105109959248e-14
7.6976976976977 5.41936064405376e-14
7.71771771771772 4.64443339685915e-14
7.73773773773774 3.97871984155707e-14
7.75775775775776 3.40706104038536e-14
7.77777777777778 2.91636853079907e-14
7.7977977977978 2.4953463096687e-14
7.81781781781782 2.13424947819474e-14
7.83783783783784 1.82467480586654e-14
7.85785785785786 1.55937907247682e-14
7.87787787787788 1.33212157347804e-14
7.8978978978979 1.13752763482777e-14
7.91791791791792 9.70970386854701e-15
7.93793793793794 8.28468399583068e-15
7.95795795795796 7.06597090549298e-15
7.97797797797798 6.02412085866189e-15
7.997997997998 5.13382950919603e-15
8.01801801801802 4.3733591283238e-15
8.03803803803804 3.72404376402735e-15
8.05805805805806 3.16986191875719e-15
8.07807807807808 2.69706769497134e-15
8.0980980980981 2.29387254841619e-15
8.11811811811812 1.95017082606148e-15
8.13813813813814 1.65730316851105e-15
8.15815815815816 1.40785264250169e-15
8.17817817817818 1.19546915264237e-15
8.1981981981982 1.01471827585831e-15
8.21821821821822 8.60951178494166e-16
8.23823823823824 7.30192724674871e-16
8.25825825825826 6.19045274051723e-16
8.27827827827828 5.24606005103494e-16
8.2982982982983 4.44395893386326e-16
8.31831831831832 3.76298728354137e-16
8.33833833833834 3.18508772685576e-16
8.35835835835836 2.69485858889318e-16
8.37837837837838 2.27916883183252e-16
8.3983983983984 1.92682799625235e-16
8.41841841841842 1.62830341150177e-16
8.43843843843844 1.37547801096493e-16
8.45845845845846 1.16144301209627e-16
8.47847847847848 9.8032051926925e-17
8.4984984984985 8.27111796592674e-17
8.51851851851852 6.97567552529606e-17
8.53853853853854 5.88077091106193e-17
8.55855855855856 4.95573626747691e-17
8.57857857857858 4.17453440895294e-17
8.5985985985986 3.51506886838449e-17
8.61861861861862 2.95859531836935e-17
8.63863863863864 2.48921968838275e-17
8.65865865865866 2.09347039316773e-17
8.67867867867868 1.75993388643364e-17
8.6986986986987 1.47894429982973e-17
8.71871871871872 1.24231925503958e-17
8.73873873873874 1.04313507694241e-17
8.75875875875876 8.75535614211525e-18
8.77877877877878 7.34569713009337e-18
8.7987987987988 6.16053109048305e-18
8.81881881881882 5.16451119999019e-18
8.83883883883884 4.32779048512422e-18
8.85885885885886 3.62517658454128e-18
8.87887887887888 3.03541474067764e-18
8.8988988988989 2.54057982941549e-18
8.91891891891892 2.12556106807901e-18
8.93893893893894 1.77762546207239e-18
8.95895895895896 1.48604811781133e-18
8.97897897897898 1.24179931485728e-18
8.998998998999 1.03727973679138e-18
9.01901901901902 8.66096545670873e-19
9.03903903903904 7.22874080905507e-19
9.05905905905906 6.03093897535385e-19
9.07907907907908 5.0295965472299e-19
9.0990990990991 4.19283042962206e-19
9.11911911911912 3.49387515332417e-19
9.13913913913914 2.91027078875695e-19
9.15915915915916 2.42317819504618e-19
9.17917917917918 2.01680188581445e-19
9.1991991991992 1.67790380698338e-19
9.21921921921922 1.39539388139188e-19
9.23923923923924 1.1599853476858e-19
9.25925925925926 9.63904764362469e-20
9.27927927927928 8.00648113242436e-20
9.2992992992993 6.6477576193283e-20
9.31931931931932 5.51740167796855e-20
9.33933933933934 4.5774115701642e-20
9.35935935935936 3.79604417474457e-20
9.37937937937938 3.14679525475421e-20
9.3993993993994 2.60754402558043e-20
9.41941941941942 2.15983585814365e-20
9.43943943943944 1.78828106797973e-20
9.45945945945946 1.48005121824234e-20
9.47947947947948 1.22445730038445e-20
9.4994994994995 1.01259663374544e-20
9.51951951951952 8.37057415073758e-21
9.53953953953954 6.91671611023779e-21
9.55955955955956 5.71308371631411e-21
9.57957957957958 4.71701393695898e-21
9.5995995995996 3.89304716291232e-21
9.61961961961962 3.21172317125736e-21
9.63963963963964 2.64857624243904e-21
9.65965965965966 2.18329684678274e-21
9.67967967967968 1.79903258756111e-21
9.6996996996997 1.48180551599987e-21
9.71971971971972 1.22002665237565e-21
9.73973973973974 1.00409166885045e-21
9.75975975975976 8.26044308669654e-22
9.77977977977978 6.79296312742742e-22
9.7997997997998 5.58394465795474e-22
9.81981981981982 4.58826916995414e-22
9.83983983983984 3.7686222201397e-22
9.85985985985986 3.09415635142992e-22
9.87987987987988 2.53938085193762e-22
9.8998998998999 2.08324025950642e-22
9.91991991991992 1.70834984871876e-22
9.93993993993994 1.40036162642795e-22
9.95995995995996 1.14743877987917e-22
9.97997997997998 9.39820210218911e-23
10 7.69459862670642e-23
};
\addlegendentry{N(0,1)}; 
\legend{}; 
\end{axis}

\end{tikzpicture}

%% file: FinalFigs/Null_Dists_d_10_500_n_200_m_200_kernel__Gaussian_Poly_2_2022_10_12_23_29_33mmd.tex
\begin{tikzpicture}

\definecolor{darkorange25512714}{RGB}{255,127,14}
\definecolor{darkslategray38}{RGB}{38,38,38}
\definecolor{lightgray204}{RGB}{204,204,204}
\definecolor{steelblue31119180}{RGB}{31,119,180}

\begin{axis}[
axis line style={darkslategray38},
height=\figheight,
legend cell align={left},
legend style={fill opacity=0.8, draw opacity=1, text opacity=1, draw=none},
tick align=outside,
tick pos=left,
title={$\dmmd~(n/m=1)$},
width=\figwidth,
x grid style={lightgray204},
xmin=-6, xmax=6,
xtick style={color=darkslategray38},
y grid style={lightgray204},
ylabel = {}, 
ymin=0, ymax=2.91388451992186,
ytick style={color=darkslategray38}, 
xticklabels=empty,
yticklabels=empty
]
\draw[draw=none,fill=steelblue31119180,fill opacity=0.8] (axis cs:-1.48239099979401,0) rectangle (axis cs:-1.36506307125092,0.0613664631209752);
\addlegendimage{ybar,ybar legend,draw=none,fill=steelblue31119180,fill opacity=0.8}
\addlegendentry{mmd (d=10)}

\draw[draw=none,fill=steelblue31119180,fill opacity=0.8] (axis cs:-1.18907117843628,0) rectangle (axis cs:-1.07174324989319,0.11591445389425);
\draw[draw=none,fill=steelblue31119180,fill opacity=0.8] (axis cs:-0.895751357078552,0) rectangle (axis cs:-0.778423428535461,0.279558331995554);
\draw[draw=none,fill=steelblue31119180,fill opacity=0.8] (axis cs:-0.602431535720825,0) rectangle (axis cs:-0.485103607177734,0.470476265062848);
\draw[draw=none,fill=steelblue31119180,fill opacity=0.8] (axis cs:-0.309111773967743,0) rectangle (axis cs:-0.191783845424652,0.70912364589183);
\draw[draw=none,fill=steelblue31119180,fill opacity=0.8] (axis cs:-0.0157919675111771,0) rectangle (axis cs:0.101535961031914,0.777308611842967);
\draw[draw=none,fill=steelblue31119180,fill opacity=0.8] (axis cs:0.277527809143066,0) rectangle (axis cs:0.394855737686157,0.518205741228645);
\draw[draw=none,fill=steelblue31119180,fill opacity=0.8] (axis cs:0.570847630500793,0) rectangle (axis cs:0.688175559043884,0.320469307409537);
\draw[draw=none,fill=steelblue31119180,fill opacity=0.8] (axis cs:0.864167451858521,0) rectangle (axis cs:0.981495380401611,0.129551448470044);
\draw[draw=none,fill=steelblue31119180,fill opacity=0.8] (axis cs:1.15748727321625,0) rectangle (axis cs:1.27481520175934,0.0272739836093223);
\draw[draw=none,fill=darkorange25512714,fill opacity=0.8] (axis cs:-1.36506307125092,0) rectangle (axis cs:-1.24773514270782,0);
\addlegendimage{ybar,ybar legend,draw=none,fill=darkorange25512714,fill opacity=0.8}
\addlegendentry{mmd (d=500)}

\draw[draw=none,fill=darkorange25512714,fill opacity=0.8] (axis cs:-1.07174324989319,0) rectangle (axis cs:-0.954415321350098,0);
\draw[draw=none,fill=darkorange25512714,fill opacity=0.8] (axis cs:-0.778423428535461,0) rectangle (axis cs:-0.661095499992371,0);
\draw[draw=none,fill=darkorange25512714,fill opacity=0.8] (axis cs:-0.485103636980057,0) rectangle (axis cs:-0.367775708436966,0);
\draw[draw=none,fill=darkorange25512714,fill opacity=0.8] (axis cs:-0.191783845424652,0) rectangle (axis cs:-0.0744559168815613,0.634120183345578);
\draw[draw=none,fill=darkorange25512714,fill opacity=0.8] (axis cs:0.101535946130753,0) rectangle (axis cs:0.218863874673843,2.77512811421129);
\draw[draw=none,fill=darkorange25512714,fill opacity=0.8] (axis cs:0.394855737686157,0) rectangle (axis cs:0.512183666229248,0);
\draw[draw=none,fill=darkorange25512714,fill opacity=0.8] (axis cs:0.688175559043884,0) rectangle (axis cs:0.805503487586975,0);
\draw[draw=none,fill=darkorange25512714,fill opacity=0.8] (axis cs:0.981495380401611,0) rectangle (axis cs:1.0988233089447,0);
\draw[draw=none,fill=darkorange25512714,fill opacity=0.8] (axis cs:1.27481520175934,0) rectangle (axis cs:1.39214313030243,0);
\addplot [semithick, black]
table {%
-10 7.69459862670642e-23
-9.97997997997998 9.39820210218911e-23
-9.95995995995996 1.14743877987917e-22
-9.93993993993994 1.40036162642795e-22
-9.91991991991992 1.70834984871876e-22
-9.8998998998999 2.08324025950642e-22
-9.87987987987988 2.53938085193762e-22
-9.85985985985986 3.09415635142992e-22
-9.83983983983984 3.7686222201397e-22
-9.81981981981982 4.58826916995414e-22
-9.7997997997998 5.58394465795474e-22
-9.77977977977978 6.79296312742742e-22
-9.75975975975976 8.26044308669654e-22
-9.73973973973974 1.00409166885045e-21
-9.71971971971972 1.22002665237565e-21
-9.6996996996997 1.48180551599987e-21
-9.67967967967968 1.79903258756111e-21
-9.65965965965966 2.18329684678274e-21
-9.63963963963964 2.64857624243904e-21
-9.61961961961962 3.21172317125736e-21
-9.5995995995996 3.89304716291232e-21
-9.57957957957958 4.71701393695898e-21
-9.55955955955956 5.71308371631411e-21
-9.53953953953954 6.91671611023779e-21
-9.51951951951952 8.37057415073758e-21
-9.4994994994995 1.01259663374544e-20
-9.47947947947948 1.22445730038445e-20
-9.45945945945946 1.48005121824234e-20
-9.43943943943944 1.78828106797973e-20
-9.41941941941942 2.15983585814365e-20
-9.3993993993994 2.60754402558043e-20
-9.37937937937938 3.14679525475421e-20
-9.35935935935936 3.79604417474457e-20
-9.33933933933934 4.5774115701642e-20
-9.31931931931932 5.51740167796855e-20
-9.2992992992993 6.6477576193283e-20
-9.27927927927928 8.00648113242436e-20
-9.25925925925926 9.63904764362469e-20
-9.23923923923924 1.1599853476858e-19
-9.21921921921922 1.39539388139188e-19
-9.1991991991992 1.67790380698338e-19
-9.17917917917918 2.01680188581445e-19
-9.15915915915916 2.42317819504618e-19
-9.13913913913914 2.91027078875695e-19
-9.11911911911912 3.49387515332417e-19
-9.0990990990991 4.19283042962206e-19
-9.07907907907908 5.0295965472299e-19
-9.05905905905906 6.03093897535385e-19
-9.03903903903904 7.22874080905507e-19
-9.01901901901902 8.66096545670873e-19
-8.998998998999 1.03727973679138e-18
-8.97897897897898 1.24179931485728e-18
-8.95895895895896 1.48604811781133e-18
-8.93893893893894 1.77762546207239e-18
-8.91891891891892 2.12556106807901e-18
-8.8988988988989 2.54057982941549e-18
-8.87887887887888 3.03541474067764e-18
-8.85885885885886 3.62517658454128e-18
-8.83883883883884 4.32779048512422e-18
-8.81881881881882 5.16451119999019e-18
-8.7987987987988 6.16053109048305e-18
-8.77877877877878 7.34569713009337e-18
-8.75875875875876 8.75535614211525e-18
-8.73873873873874 1.04313507694241e-17
-8.71871871871872 1.24231925503958e-17
-8.6986986986987 1.47894429982973e-17
-8.67867867867868 1.75993388643364e-17
-8.65865865865866 2.09347039316773e-17
-8.63863863863864 2.48921968838275e-17
-8.61861861861862 2.95859531836935e-17
-8.5985985985986 3.51506886838449e-17
-8.57857857857858 4.17453440895294e-17
-8.55855855855856 4.95573626747691e-17
-8.53853853853854 5.88077091106193e-17
-8.51851851851852 6.97567552529606e-17
-8.4984984984985 8.27111796592674e-17
-8.47847847847848 9.8032051926925e-17
-8.45845845845846 1.16144301209627e-16
-8.43843843843844 1.37547801096493e-16
-8.41841841841842 1.62830341150177e-16
-8.3983983983984 1.92682799625235e-16
-8.37837837837838 2.27916883183252e-16
-8.35835835835836 2.69485858889318e-16
-8.33833833833834 3.18508772685576e-16
-8.31831831831832 3.76298728354137e-16
-8.2982982982983 4.44395893386326e-16
-8.27827827827828 5.24606005103494e-16
-8.25825825825826 6.19045274051723e-16
-8.23823823823824 7.30192724674871e-16
-8.21821821821822 8.60951178494166e-16
-8.1981981981982 1.01471827585831e-15
-8.17817817817818 1.19546915264237e-15
-8.15815815815816 1.40785264250169e-15
-8.13813813813814 1.65730316851105e-15
-8.11811811811812 1.95017082606148e-15
-8.0980980980981 2.29387254841619e-15
-8.07807807807808 2.69706769497134e-15
-8.05805805805806 3.16986191875719e-15
-8.03803803803804 3.72404376402735e-15
-8.01801801801802 4.3733591283238e-15
-7.997997997998 5.13382950919607e-15
-7.97797797797798 6.02412085866193e-15
-7.95795795795796 7.06597090549303e-15
-7.93793793793794 8.28468399583074e-15
-7.91791791791792 9.70970386854708e-15
-7.8978978978979 1.13752763482777e-14
-7.87787787787788 1.33212157347805e-14
-7.85785785785786 1.55937907247683e-14
-7.83783783783784 1.82467480586655e-14
-7.81781781781782 2.13424947819475e-14
-7.7977977977978 2.49534630966872e-14
-7.77777777777778 2.91636853079909e-14
-7.75775775775776 3.40706104038538e-14
-7.73773773773774 3.9787198415571e-14
-7.71771771771772 4.64443339685918e-14
-7.6976976976977 5.4193606440538e-14
-7.67767767767768 6.32105109959252e-14
-7.65765765765766 7.36981325813117e-14
-7.63763763763764 8.58913838706879e-14
-7.61761761761762 1.00061878296601e-13
-7.5975975975976 1.16523530854699e-13
-7.57757757757758 1.35638992516664e-13
-7.55755755755756 1.57827039041884e-13
-7.53753753753754 1.83571051982057e-13
-7.51751751751752 2.13428748996349e-13
-7.4974974974975 2.48043342543513e-13
-7.47747747747748 2.88156330935935e-13
-7.45745745745746 3.34622154016965e-13
-7.43743743743744 3.88424977793732e-13
-7.41741741741742 4.50697908714384e-13
-7.3973973973974 5.22744979473711e-13
-7.37737737737738 6.06066294885249e-13
-7.35735735735736 7.02386779168738e-13
-7.33733733733734 8.13689025751835e-13
-7.31731731731732 9.42250818252909e-13
-7.2972972972973 1.09068796768221e-12
-7.27727727727728 1.262003197176e-12
-7.25725725725726 1.45964190299847e-12
-7.23723723723724 1.68755573049416e-12
-7.21721721721722 1.95027502769792e-12
-7.1971971971972 2.25299137914218e-12
-7.17717717717718 2.60165157997871e-12
-7.15715715715716 3.00306458801653e-12
-7.13713713713714 3.4650231910834e-12
-7.11711711711712 3.99644235193919e-12
-7.0970970970971 4.60751644580953e-12
-7.07707707707708 5.30989788981514e-12
-7.05705705705706 6.11689998287646e-12
-7.03703703703704 7.0437271332242e-12
-7.01701701701702 8.10773605306645e-12
-6.996996996997 9.32873195138555e-12
-6.97697697697698 1.07293042619726e-11
-6.95695695695696 1.23352070109962e-11
-6.93693693693694 1.41757895636779e-11
-6.91691691691692 1.62844842008237e-11
-6.8968968968969 1.86993577716893e-11
-6.87687687687688 2.14637355595386e-11
-6.85685685685686 2.46269064908967e-11
-6.83683683683684 2.82449199306815e-11
-6.81681681681682 3.2381485546105e-11
-6.7967967967968 3.71089891068559e-11
-6.77677677677678 4.25096386334913e-11
-6.75675675675676 4.86767570277083e-11
-6.73673673673674 5.57162392366004e-11
-6.71671671671672 6.37481941394491e-11
-6.6966966966967 7.29087937236032e-11
-6.67667667667668 8.33523547614402e-11
-6.65665665665666 9.52536811418151e-11
-6.63663663663664 1.08810698278135e-10
-6.61661661661662 1.24247414645768e-10
-6.5965965965966 1.41817249531744e-10
-6.57657657657658 1.61806770551278e-10
-6.55655655655656 1.84539889444164e-10
-6.53653653653654 2.10382570159622e-10
-6.51651651651652 2.39748109325542e-10
-6.4964964964965 2.73103055937438e-10
-6.47647647647648 3.10973844559381e-10
-6.45645645645646 3.53954224575809e-10
-6.43643643643644 4.027135771479e-10
-6.41641641641642 4.58006221596996e-10
-6.3963963963964 5.20681824054169e-10
-6.37637637637638 5.91697033481538e-10
-6.35635635635636 6.72128483698846e-10
-6.33633633633634 7.63187314959523e-10
-6.31631631631632 8.66235385046001e-10
-6.2962962962963 9.8280335793813e-10
-6.27627627627628 1.11461087800766e-09
-6.25625625625626 1.26358905957513e-09
-6.23623623623624 1.43190554571787e-09
-6.21621621621622 1.62199241663862e-09
-6.1961961961962 1.83657725691024e-09
-6.17617617617618 2.0787177227378e-09
-6.15615615615616 2.35183998527873e-09
-6.13613613613614 2.65978146430928e-09
-6.11611611611612 3.00683830841829e-09
-6.0960960960961 3.39781812376754e-09
-6.07607607607608 3.83809850362696e-09
-6.05605605605606 4.33369196574699e-09
-6.03603603603604 4.89131796456919e-09
-6.01601601601602 5.51848271073395e-09
-5.995995995996 6.22356760178439e-09
-5.97597597597598 7.01592714588943e-09
-5.95595595595596 7.90599734535251e-09
-5.93593593593594 8.90541559921139e-09
-5.91591591591592 1.00271532849868e-08
-5.8958958958959 1.12856622892642e-08
-5.87587587587588 1.2697036876002e-08
-5.85585585585586 1.42791924110059e-08
-5.83583583583584 1.60520626017053e-08
-5.81581581581582 1.80378170640688e-08
-5.7957957957958 2.02611011941336e-08
-5.77577577577578 2.27493005011625e-08
-5.75575575575576 2.55328317539416e-08
-5.73573573573574 2.86454635022852e-08
-5.71571571571572 3.2124668763613e-08
-5.6956956956957 3.60120129107462e-08
-5.67567567567568 4.03535800631662e-08
-5.65565565565566 4.5200441571292e-08
-5.63563563563564 5.06091704933412e-08
-5.61561561561562 5.66424062986154e-08
-5.5955955955956 6.33694743912418e-08
-5.57557557557558 7.08670654362614e-08
-5.55555555555556 7.92199798873018e-08
-5.53553553553554 8.85219435638491e-08
-5.51551551551552 9.8876500608364e-08
-5.4954954954955 1.10397990671284e-07
-5.47547547547548 1.23212617727566e-07
-5.45545545545546 1.3745961852414e-07
-5.43543543543544 1.5329253929596e-07
-5.41541541541542 1.70880630071684e-07
-5.3953953953954 1.90410366621162e-07
-5.37537537537538 2.1208711087848e-07
-5.35535535535536 2.36136921509202e-07
-5.33533533533534 2.62808527181656e-07
-5.31531531531532 2.92375476052561e-07
-5.2952952952953 3.25138475990267e-07
-5.27527527527528 3.61427941137511e-07
-5.25525525525526 4.01606761563285e-07
-5.23523523523524 4.46073313973501e-07
-5.21521521521522 4.95264732746229e-07
-5.1951951951952 5.4966046193278e-07
-5.17517517517518 6.09786110324583e-07
-5.15515515515516 6.76217633231267e-07
-5.13513513513514 7.49585866251233e-07
-5.11511511511512 8.30581438046261e-07
-5.0950950950951 9.19960090959837e-07
-5.07507507507508 1.01854844024876e-06
-5.05505505505506 1.12725020473309e-06
-5.03503503503504 1.24705294381396e-06
-5.01501501501502 1.37903533806611e-06
-4.99499499499499 1.5243750529858e-06
-4.97497497497497 1.68435722796805e-06
-4.95495495495495 1.86038363520377e-06
-4.93493493493493 2.05398255592983e-06
-4.91491491491491 2.26681942433715e-06
-4.89489489489489 2.50070829244518e-06
-4.87487487487487 2.75762417238901e-06
-4.85485485485485 3.03971631583941e-06
-4.83483483483483 3.34932249368796e-06
-4.81481481481481 3.68898434268124e-06
-4.79479479479479 4.06146384937932e-06
-4.77477477477477 4.4697610456467e-06
-4.75475475475475 4.91713299385613e-06
-4.73473473473473 5.40711414409909e-06
-4.71471471471471 5.94353814994721e-06
-4.69469469469469 6.53056123369604e-06
-4.67467467467467 7.17268719654363e-06
-4.65465465465465 7.87479417380527e-06
-4.63463463463463 8.64216324004121e-06
-4.61461461461461 9.48050897386715e-06
-4.59459459459459 1.03960120972233e-05
-4.57457457457457 1.13953543089884e-05
-4.55455455455455 1.24857554380297e-05
-4.53453453453453 1.3675013046071e-05
-4.51451451451451 1.49715446161227e-05
-4.49449449449449 1.63844324676437e-05
-4.47447447447447 1.79234715450684e-05
-4.45445445445445 1.95992202318354e-05
-4.43443443443443 2.14230543475548e-05
-4.41441441441441 2.34072244914564e-05
-4.39439439439439 2.55649169007242e-05
-4.37437437437437 2.79103179977393e-05
-4.35435435435435 3.04586828055866e-05
-4.33433433433433 3.32264074164067e-05
-4.31431431431431 3.62311057022691e-05
-4.29429429429429 3.94916904631592e-05
-4.27427427427427 4.3028459211397e-05
-4.25425425425425 4.68631847962824e-05
-4.23423423423423 5.10192110769697e-05
-4.21421421421421 5.55215538554582e-05
-4.19419419419419 6.03970072851107e-05
-4.17417417417417 6.56742559732345e-05
-4.15415415415415 7.13839929989176e-05
-4.13413413413413 7.75590440694795e-05
-4.11411411411411 8.42344980404937e-05
-4.09409409409409 9.14478440253317e-05
-4.07407407407407 9.92391153205018e-05
-4.05405405405405 0.000107651040372646
-4.03403403403403 0.000116729201011866
-4.01401401401401 0.000126522198173995
-3.99399399399399 0.000137081825331481
-3.97397397397397 0.000148463249848567
-3.95395395395395 0.000160725202471485
-3.93393393393393 0.000173930175158222
-3.91391391391391 0.00018814462744512
-3.89389389389389 0.000203439201538965
-3.87387387387387 0.000219888946313312
-3.85385385385385 0.000237573550376443
-3.83383383383383 0.000256577584365551
-3.81381381381381 0.000276990752607344
-3.79379379379379 0.000298908154269281
-3.77377377377377 0.000322430554107926
-3.75375375375375 0.000347664662901427
-3.73373373373373 0.000374723427631836
-3.71371371371371 0.000403726331459719
-3.69369369369369 0.000434799703508357
-3.67367367367367 0.000468077038447599
-3.65365365365365 0.00050369932583812
-3.63363363363363 0.000541815389165405
-3.61361361361361 0.000582582234459159
-3.59359359359359 0.000626165408357979
-3.57357357357357 0.000672739365441021
-3.55355355355355 0.000722487844607978
-3.53353353353353 0.00077560425424601
-3.51351351351351 0.000832292065877155
-3.49349349349349 0.000892765215932443
-3.47347347347347 0.000957248515249216
-3.45345345345345 0.0010259780658361
-3.43343343343343 0.00109920168439588
-3.41341341341341 0.00117717933203981
-3.39339339339339 0.00126018354956833
-3.37337337337337 0.00134849989763212
-3.35335335335335 0.00144242740102448
-3.33333333333333 0.00154227899629111
-3.31331331331331 0.00164838198177652
-3.29329329329329 0.00176107846915772
-3.27327327327327 0.00188072583544552
-3.25325325325325 0.00200769717436226
-3.23323323323323 0.00214238174593163
-3.21321321321321 0.00228518542304204
-3.19319319319319 0.00243653113367012
-3.17317317317317 0.00259685929737497
-3.15315315315315 0.00276662825459747
-3.13313313313313 0.00294631468722261
-3.11311311311311 0.00313641402878609
-3.09309309309309 0.00333744086263052
-3.07307307307307 0.00354992930624086
-3.05305305305305 0.00377443337991422
-3.03303303303303 0.00401152735784579
-3.01301301301301 0.00426180609964128
-2.99299299299299 0.00452588536019618
-2.97297297297297 0.0048044020758154
-2.95295295295295 0.00509801462438215
-2.93293293293293 0.00540740305732385
-2.91291291291291 0.00573326930106519
-2.89289289289289 0.00607633732560526
-2.87287287287287 0.00643735327780636
-2.85285285285285 0.00681708557693873
-2.83283283283283 0.00721632496998623
-2.81281281281281 0.00763588454418632
-2.79279279279279 0.00807659969425075
-2.77277277277277 0.00853932804169477
-2.75275275275275 0.00902494930369032
-2.73273273273273 0.00953436510885489
-2.71271271271271 0.0100684987573917
-2.69269269269269 0.0106282949230102
-2.67267267267267 0.0112147192940778
-2.65265265265265 0.0118287581514852
-2.63263263263263 0.0124714178807513
-2.61261261261261 0.0131437244159435
-2.59259259259259 0.0138467226130541
-2.57257257257257 0.0145814755505492
-2.55255255255255 0.0153490637548887
-2.53253253253253 0.0161505843489182
-2.51251251251251 0.0169871501211409
-2.49249249249249 0.0178598885140022
-2.47247247247247 0.018769940529451
-2.45245245245245 0.0197184595501959
-2.43243243243243 0.0207066100752274
-2.41241241241241 0.0217355663683581
-2.39239239239239 0.0228065110187135
-2.37237237237237 0.0239206334123108
-2.35235235235235 0.0250791281140699
-2.33233233233233 0.0262831931598317
-2.31231231231231 0.0275340282581906
-2.29229229229229 0.0288328329022027
-2.27227227227227 0.0301808043912899
-2.25225225225225 0.0315791357639331
-2.23223223223223 0.033029013642035
-2.21221221221221 0.0345316159881232
-2.19219219219219 0.0360881097768736
-2.17217217217217 0.0376996485827434
-2.15215215215215 0.0393673700858293
-2.13213213213213 0.0410923934983949
-2.11211211211211 0.0428758169148479
-2.09209209209209 0.0447187145882915
-2.07207207207207 0.0466221341371235
-2.05205205205205 0.0485870936855041
-2.03203203203203 0.0506145789418747
-2.01201201201201 0.0527055402200587
-1.99199199199199 0.0548608894078376
-1.97197197197197 0.057081496888248
-1.95195195195195 0.059368188419199
-1.93193193193193 0.0617217419773594
-1.91191191191191 0.0641428845726061
-1.89189189189189 0.0666322890396674
-1.87187187187187 0.0691905708139176
-1.85185185185185 0.0718182846986055
-1.83183183183183 0.0745159216311036
-1.81181181181181 0.077283905456062
-1.79179179179179 0.0801225897136326
-1.77177177177177 0.083032254451193
-1.75175175175175 0.0860131030672496
-1.73173173173173 0.0890652591964251
-1.71171171171171 0.0921887636446459
-1.69169169169169 0.0953835713838294
-1.67167167167167 0.0986495486155338
-1.65165165165165 0.101986469913169
-1.63163163163163 0.10539401545248
-1.61161161161161 0.108871768340093
-1.59159159159159 0.112419212049971
-1.57157157157157 0.116035727977651
-1.55155155155155 0.119720593122119
-1.53153153153153 0.123472977905145
-1.51151151151151 0.127291944137829
-1.49149149149149 0.13117644314399
-1.47147147147147 0.135125314049902
-1.45145145145145 0.139137282249685
-1.43143143143143 0.143210958055468
-1.41141141141141 0.147344835541168
-1.39139139139139 0.151537291588457
-1.37137137137137 0.155786585143159
-1.35135135135135 0.160090856689972
-1.33133133133133 0.164448127952996
-1.31131131131131 0.168856301829129
-1.29129129129129 0.173313162560933
-1.27127127127127 0.17781637615506
-1.25125125125125 0.182363491051798
-1.23123123123123 0.186951939050736
-1.21121121121121 0.191579036496956
-1.19119119119119 0.1962419857315
-1.17117117117117 0.200937876809264
-1.15115115115115 0.205663689486728
-1.13113113113113 0.210416295481265
-1.11111111111111 0.215192461003031
-1.09109109109109 0.219988849559688
-1.07107107107107 0.224802025033432
-1.05105105105105 0.229628455029052
-1.03103103103103 0.234464514490888
-1.01101101101101 0.239306489585817
-0.990990990990991 0.24415058184851
-0.970970970970971 0.24899291258444
-0.950950950950951 0.253829527525259
-0.930930930930931 0.258656401730343
-0.910910910910911 0.2634694447275
-0.890890890890891 0.268264505884996
-0.870870870870871 0.273037380006279
-0.850850850850851 0.277783813137949
-0.830830830830831 0.282499508580786
-0.810810810810811 0.287180133092853
-0.790790790790791 0.291821323272996
-0.77077077077077 0.296418692112302
-0.75075075075075 0.300967835700437
-0.73073073073073 0.305464340073112
-0.71071071071071 0.309903788186304
-0.69069069069069 0.314281767002296
-0.67067067067067 0.318593874672039
-0.65065065065065 0.322835727797843
-0.63063063063063 0.327002968759958
-0.61061061061061 0.331091273090187
-0.59059059059059 0.33509635687531
-0.57057057057057 0.339013984172804
-0.55055055055055 0.34283997442106
-0.53053053053053 0.346570209826128
-0.51051051051051 0.350200642706842
-0.49049049049049 0.353727302780113
-0.47047047047047 0.357146304368113
-0.45045045045045 0.360453853509139
-0.43043043043043 0.363646254953996
-0.41041041041041 0.366719919029892
-0.39039039039039 0.369671368354051
-0.37037037037037 0.372497244379499
-0.35035035035035 0.375194313755802
-0.33033033033033 0.377759474487924
-0.31031031031031 0.38018976187679
-0.29029029029029 0.382482354225654
-0.27027027027027 0.384634578296894
-0.25025025025025 0.386643914504485
-0.23023023023023 0.388508001828027
-0.21021021021021 0.390224642434919
-0.19019019019019 0.391791805998011
-0.17017017017017 0.393207633696876
-0.15015015015015 0.394470441891644
-0.13013013013013 0.395578725459258
-0.11011011011011 0.396531160782876
-0.0900900900900901 0.397326608386124
-0.07007007007007 0.397964115204853
-0.05005005005005 0.398442916490068
-0.03003003003003 0.398762437336696
-0.01001001001001 0.398922293833933
0.01001001001001 0.398922293833933
0.03003003003003 0.398762437336696
0.05005005005005 0.398442916490068
0.07007007007007 0.397964115204853
0.0900900900900901 0.397326608386124
0.11011011011011 0.396531160782876
0.13013013013013 0.395578725459258
0.15015015015015 0.394470441891644
0.17017017017017 0.393207633696876
0.19019019019019 0.391791805998011
0.21021021021021 0.390224642434919
0.23023023023023 0.388508001828027
0.25025025025025 0.386643914504485
0.27027027027027 0.384634578296894
0.29029029029029 0.382482354225654
0.31031031031031 0.38018976187679
0.33033033033033 0.377759474487924
0.35035035035035 0.375194313755802
0.37037037037037 0.372497244379499
0.39039039039039 0.369671368354051
0.41041041041041 0.366719919029892
0.43043043043043 0.363646254953996
0.45045045045045 0.360453853509139
0.47047047047047 0.357146304368113
0.49049049049049 0.353727302780113
0.51051051051051 0.350200642706842
0.53053053053053 0.346570209826128
0.55055055055055 0.34283997442106
0.57057057057057 0.339013984172804
0.59059059059059 0.33509635687531
0.61061061061061 0.331091273090187
0.63063063063063 0.327002968759958
0.65065065065065 0.322835727797843
0.67067067067067 0.318593874672039
0.69069069069069 0.314281767002296
0.71071071071071 0.309903788186304
0.73073073073073 0.305464340073112
0.75075075075075 0.300967835700437
0.77077077077077 0.296418692112302
0.790790790790791 0.291821323272996
0.810810810810811 0.287180133092853
0.830830830830831 0.282499508580786
0.850850850850851 0.277783813137949
0.870870870870871 0.273037380006279
0.890890890890891 0.268264505884996
0.910910910910911 0.2634694447275
0.930930930930931 0.258656401730343
0.950950950950951 0.253829527525259
0.970970970970971 0.24899291258444
0.990990990990991 0.24415058184851
1.01101101101101 0.239306489585817
1.03103103103103 0.234464514490888
1.05105105105105 0.229628455029052
1.07107107107107 0.224802025033432
1.09109109109109 0.219988849559688
1.11111111111111 0.215192461003031
1.13113113113113 0.210416295481265
1.15115115115115 0.205663689486728
1.17117117117117 0.200937876809264
1.19119119119119 0.1962419857315
1.21121121121121 0.191579036496956
1.23123123123123 0.186951939050736
1.25125125125125 0.182363491051798
1.27127127127127 0.17781637615506
1.29129129129129 0.173313162560933
1.31131131131131 0.168856301829129
1.33133133133133 0.164448127952996
1.35135135135135 0.160090856689972
1.37137137137137 0.155786585143159
1.39139139139139 0.151537291588457
1.41141141141141 0.147344835541168
1.43143143143143 0.143210958055468
1.45145145145145 0.139137282249685
1.47147147147147 0.135125314049902
1.49149149149149 0.13117644314399
1.51151151151151 0.127291944137829
1.53153153153153 0.123472977905145
1.55155155155155 0.119720593122119
1.57157157157157 0.116035727977651
1.59159159159159 0.112419212049971
1.61161161161161 0.108871768340093
1.63163163163163 0.10539401545248
1.65165165165165 0.101986469913169
1.67167167167167 0.0986495486155338
1.69169169169169 0.0953835713838294
1.71171171171171 0.0921887636446459
1.73173173173173 0.0890652591964251
1.75175175175175 0.0860131030672496
1.77177177177177 0.083032254451193
1.79179179179179 0.0801225897136326
1.81181181181181 0.077283905456062
1.83183183183183 0.0745159216311036
1.85185185185185 0.0718182846986055
1.87187187187187 0.0691905708139176
1.89189189189189 0.0666322890396674
1.91191191191191 0.0641428845726061
1.93193193193193 0.0617217419773594
1.95195195195195 0.059368188419199
1.97197197197197 0.057081496888248
1.99199199199199 0.0548608894078376
2.01201201201201 0.0527055402200588
2.03203203203203 0.0506145789418748
2.05205205205205 0.0485870936855042
2.07207207207207 0.0466221341371236
2.09209209209209 0.0447187145882916
2.11211211211211 0.0428758169148479
2.13213213213213 0.041092393498395
2.15215215215215 0.0393673700858294
2.17217217217217 0.0376996485827434
2.19219219219219 0.0360881097768737
2.21221221221221 0.0345316159881233
2.23223223223223 0.033029013642035
2.25225225225225 0.0315791357639332
2.27227227227227 0.03018080439129
2.29229229229229 0.0288328329022028
2.31231231231231 0.0275340282581906
2.33233233233233 0.0262831931598317
2.35235235235235 0.02507912811407
2.37237237237237 0.0239206334123108
2.39239239239239 0.0228065110187136
2.41241241241241 0.0217355663683581
2.43243243243243 0.0207066100752275
2.45245245245245 0.0197184595501959
2.47247247247247 0.0187699405294511
2.49249249249249 0.0178598885140022
2.51251251251251 0.016987150121141
2.53253253253253 0.0161505843489182
2.55255255255255 0.0153490637548888
2.57257257257257 0.0145814755505493
2.59259259259259 0.0138467226130541
2.61261261261261 0.0131437244159435
2.63263263263263 0.0124714178807514
2.65265265265265 0.0118287581514852
2.67267267267267 0.0112147192940778
2.69269269269269 0.0106282949230103
2.71271271271271 0.0100684987573917
2.73273273273273 0.00953436510885491
2.75275275275275 0.00902494930369034
2.77277277277277 0.00853932804169479
2.79279279279279 0.00807659969425077
2.81281281281281 0.00763588454418634
2.83283283283283 0.00721632496998621
2.85285285285285 0.00681708557693871
2.87287287287287 0.00643735327780635
2.89289289289289 0.00607633732560524
2.91291291291291 0.00573326930106518
2.93293293293293 0.00540740305732384
2.95295295295295 0.00509801462438214
2.97297297297297 0.00480440207581539
2.99299299299299 0.00452588536019617
3.01301301301301 0.00426180609964127
3.03303303303303 0.00401152735784578
3.05305305305305 0.00377443337991421
3.07307307307307 0.00354992930624085
3.09309309309309 0.00333744086263051
3.11311311311311 0.00313641402878608
3.13313313313313 0.0029463146872226
3.15315315315315 0.00276662825459746
3.17317317317317 0.00259685929737496
3.19319319319319 0.00243653113367011
3.21321321321321 0.00228518542304203
3.23323323323323 0.00214238174593163
3.25325325325325 0.00200769717436225
3.27327327327327 0.00188072583544551
3.29329329329329 0.00176107846915771
3.31331331331331 0.00164838198177652
3.33333333333333 0.0015422789962911
3.35335335335335 0.00144242740102448
3.37337337337337 0.00134849989763212
3.39339339339339 0.00126018354956833
3.41341341341341 0.00117717933203981
3.43343343343343 0.00109920168439588
3.45345345345345 0.0010259780658361
3.47347347347347 0.000957248515249212
3.49349349349349 0.000892765215932441
3.51351351351351 0.000832292065877152
3.53353353353353 0.000775604254246008
3.55355355355355 0.000722487844607976
3.57357357357357 0.000672739365441019
3.59359359359359 0.000626165408357979
3.61361361361361 0.000582582234459159
3.63363363363363 0.000541815389165405
3.65365365365365 0.00050369932583812
3.67367367367367 0.000468077038447599
3.69369369369369 0.000434799703508357
3.71371371371371 0.000403726331459719
3.73373373373373 0.000374723427631836
3.75375375375375 0.000347664662901427
3.77377377377377 0.000322430554107926
3.79379379379379 0.000298908154269281
3.81381381381381 0.000276990752607344
3.83383383383383 0.000256577584365551
3.85385385385385 0.000237573550376443
3.87387387387387 0.000219888946313312
3.89389389389389 0.000203439201538965
3.91391391391391 0.00018814462744512
3.93393393393393 0.000173930175158222
3.95395395395395 0.000160725202471485
3.97397397397397 0.000148463249848567
3.99399399399399 0.000137081825331481
4.01401401401401 0.000126522198173995
4.03403403403403 0.000116729201011866
4.05405405405405 0.000107651040372646
4.07407407407407 9.92391153205018e-05
4.09409409409409 9.14478440253317e-05
4.11411411411411 8.42344980404937e-05
4.13413413413413 7.75590440694795e-05
4.15415415415415 7.13839929989176e-05
4.17417417417417 6.56742559732345e-05
4.19419419419419 6.03970072851107e-05
4.21421421421421 5.55215538554582e-05
4.23423423423423 5.10192110769697e-05
4.25425425425425 4.68631847962824e-05
4.27427427427427 4.3028459211397e-05
4.29429429429429 3.94916904631592e-05
4.31431431431431 3.62311057022691e-05
4.33433433433433 3.32264074164067e-05
4.35435435435435 3.04586828055866e-05
4.37437437437437 2.79103179977393e-05
4.39439439439439 2.55649169007242e-05
4.41441441441441 2.34072244914564e-05
4.43443443443443 2.14230543475548e-05
4.45445445445445 1.95992202318354e-05
4.47447447447447 1.79234715450684e-05
4.49449449449449 1.63844324676437e-05
4.51451451451451 1.49715446161227e-05
4.53453453453453 1.3675013046071e-05
4.55455455455455 1.24857554380297e-05
4.57457457457457 1.13953543089884e-05
4.59459459459459 1.03960120972233e-05
4.61461461461461 9.48050897386715e-06
4.63463463463463 8.64216324004121e-06
4.65465465465465 7.87479417380527e-06
4.67467467467467 7.17268719654363e-06
4.69469469469469 6.53056123369604e-06
4.71471471471471 5.94353814994721e-06
4.73473473473473 5.40711414409909e-06
4.75475475475475 4.91713299385613e-06
4.77477477477477 4.4697610456467e-06
4.79479479479479 4.06146384937932e-06
4.81481481481481 3.68898434268124e-06
4.83483483483483 3.34932249368796e-06
4.85485485485485 3.03971631583941e-06
4.87487487487487 2.75762417238901e-06
4.89489489489489 2.50070829244518e-06
4.91491491491491 2.26681942433715e-06
4.93493493493493 2.05398255592983e-06
4.95495495495495 1.86038363520377e-06
4.97497497497497 1.68435722796805e-06
4.99499499499499 1.5243750529858e-06
5.01501501501502 1.37903533806611e-06
5.03503503503504 1.24705294381396e-06
5.05505505505506 1.12725020473309e-06
5.07507507507508 1.01854844024876e-06
5.0950950950951 9.19960090959837e-07
5.11511511511512 8.30581438046261e-07
5.13513513513514 7.49585866251233e-07
5.15515515515516 6.76217633231267e-07
5.17517517517518 6.09786110324583e-07
5.1951951951952 5.4966046193278e-07
5.21521521521522 4.95264732746229e-07
5.23523523523524 4.46073313973501e-07
5.25525525525526 4.01606761563285e-07
5.27527527527528 3.61427941137511e-07
5.2952952952953 3.25138475990267e-07
5.31531531531532 2.92375476052561e-07
5.33533533533534 2.62808527181656e-07
5.35535535535536 2.36136921509202e-07
5.37537537537538 2.1208711087848e-07
5.3953953953954 1.90410366621162e-07
5.41541541541542 1.70880630071684e-07
5.43543543543544 1.5329253929596e-07
5.45545545545546 1.3745961852414e-07
5.47547547547548 1.23212617727566e-07
5.4954954954955 1.10397990671284e-07
5.51551551551552 9.8876500608364e-08
5.53553553553554 8.85219435638491e-08
5.55555555555556 7.92199798873018e-08
5.57557557557558 7.08670654362614e-08
5.5955955955956 6.33694743912418e-08
5.61561561561562 5.66424062986154e-08
5.63563563563564 5.06091704933412e-08
5.65565565565566 4.5200441571292e-08
5.67567567567568 4.03535800631662e-08
5.6956956956957 3.60120129107462e-08
5.71571571571572 3.2124668763613e-08
5.73573573573574 2.86454635022852e-08
5.75575575575576 2.55328317539416e-08
5.77577577577578 2.27493005011625e-08
5.7957957957958 2.02611011941336e-08
5.81581581581582 1.80378170640688e-08
5.83583583583584 1.60520626017053e-08
5.85585585585586 1.42791924110059e-08
5.87587587587588 1.2697036876002e-08
5.8958958958959 1.12856622892642e-08
5.91591591591592 1.00271532849868e-08
5.93593593593594 8.90541559921139e-09
5.95595595595596 7.90599734535251e-09
5.97597597597598 7.01592714588943e-09
5.995995995996 6.22356760178439e-09
6.01601601601602 5.5184827107339e-09
6.03603603603604 4.89131796456914e-09
6.05605605605606 4.33369196574694e-09
6.07607607607608 3.83809850362692e-09
6.0960960960961 3.3978181237675e-09
6.11611611611612 3.00683830841826e-09
6.13613613613614 2.65978146430924e-09
6.15615615615616 2.35183998527871e-09
6.17617617617618 2.07871772273778e-09
6.1961961961962 1.83657725691022e-09
6.21621621621622 1.6219924166386e-09
6.23623623623624 1.43190554571785e-09
6.25625625625626 1.26358905957511e-09
6.27627627627628 1.11461087800764e-09
6.2962962962963 9.82803357938119e-10
6.31631631631632 8.66235385045992e-10
6.33633633633634 7.63187314959515e-10
6.35635635635636 6.72128483698836e-10
6.37637637637638 5.91697033481532e-10
6.3963963963964 5.20681824054164e-10
6.41641641641642 4.58006221596989e-10
6.43643643643644 4.02713577147895e-10
6.45645645645646 3.53954224575805e-10
6.47647647647648 3.10973844559378e-10
6.4964964964965 2.73103055937435e-10
6.51651651651652 2.39748109325539e-10
6.53653653653654 2.10382570159619e-10
6.55655655655656 1.84539889444162e-10
6.57657657657658 1.61806770551276e-10
6.5965965965966 1.41817249531742e-10
6.61661661661662 1.24247414645767e-10
6.63663663663664 1.08810698278133e-10
6.65665665665666 9.52536811418137e-11
6.67667667667668 8.33523547614393e-11
6.6966966966967 7.29087937236022e-11
6.71671671671672 6.37481941394484e-11
6.73673673673674 5.57162392365998e-11
6.75675675675676 4.86767570277076e-11
6.77677677677678 4.25096386334907e-11
6.7967967967968 3.71089891068556e-11
6.81681681681682 3.23814855461048e-11
6.83683683683684 2.82449199306813e-11
6.85685685685686 2.46269064908966e-11
6.87687687687688 2.14637355595386e-11
6.8968968968969 1.86993577716892e-11
6.91691691691692 1.62844842008236e-11
6.93693693693694 1.41757895636779e-11
6.95695695695696 1.23352070109961e-11
6.97697697697698 1.07293042619725e-11
6.996996996997 9.32873195138548e-12
7.01701701701702 8.10773605306642e-12
7.03703703703704 7.04372713322415e-12
7.05705705705706 6.11689998287643e-12
7.07707707707708 5.30989788981512e-12
7.0970970970971 4.6075164458095e-12
7.11711711711712 3.99644235193917e-12
7.13713713713714 3.46502319108337e-12
7.15715715715716 3.00306458801651e-12
7.17717717717718 2.60165157997869e-12
7.1971971971972 2.25299137914217e-12
7.21721721721722 1.95027502769791e-12
7.23723723723724 1.68755573049414e-12
7.25725725725726 1.45964190299846e-12
7.27727727727728 1.262003197176e-12
7.2972972972973 1.0906879676822e-12
7.31731731731732 9.42250818252902e-13
7.33733733733734 8.13689025751829e-13
7.35735735735736 7.02386779168733e-13
7.37737737737738 6.06066294885245e-13
7.3973973973974 5.22744979473708e-13
7.41741741741742 4.50697908714381e-13
7.43743743743744 3.88424977793729e-13
7.45745745745746 3.34622154016963e-13
7.47747747747748 2.88156330935933e-13
7.4974974974975 2.48043342543511e-13
7.51751751751752 2.13428748996348e-13
7.53753753753754 1.83571051982055e-13
7.55755755755756 1.57827039041883e-13
7.57757757757758 1.35638992516663e-13
7.5975975975976 1.16523530854698e-13
7.61761761761762 1.000618782966e-13
7.63763763763764 8.58913838706873e-14
7.65765765765766 7.36981325813112e-14
7.67767767767768 6.32105109959248e-14
7.6976976976977 5.41936064405376e-14
7.71771771771772 4.64443339685915e-14
7.73773773773774 3.97871984155707e-14
7.75775775775776 3.40706104038536e-14
7.77777777777778 2.91636853079907e-14
7.7977977977978 2.4953463096687e-14
7.81781781781782 2.13424947819474e-14
7.83783783783784 1.82467480586654e-14
7.85785785785786 1.55937907247682e-14
7.87787787787788 1.33212157347804e-14
7.8978978978979 1.13752763482777e-14
7.91791791791792 9.70970386854701e-15
7.93793793793794 8.28468399583068e-15
7.95795795795796 7.06597090549298e-15
7.97797797797798 6.02412085866189e-15
7.997997997998 5.13382950919603e-15
8.01801801801802 4.3733591283238e-15
8.03803803803804 3.72404376402735e-15
8.05805805805806 3.16986191875719e-15
8.07807807807808 2.69706769497134e-15
8.0980980980981 2.29387254841619e-15
8.11811811811812 1.95017082606148e-15
8.13813813813814 1.65730316851105e-15
8.15815815815816 1.40785264250169e-15
8.17817817817818 1.19546915264237e-15
8.1981981981982 1.01471827585831e-15
8.21821821821822 8.60951178494166e-16
8.23823823823824 7.30192724674871e-16
8.25825825825826 6.19045274051723e-16
8.27827827827828 5.24606005103494e-16
8.2982982982983 4.44395893386326e-16
8.31831831831832 3.76298728354137e-16
8.33833833833834 3.18508772685576e-16
8.35835835835836 2.69485858889318e-16
8.37837837837838 2.27916883183252e-16
8.3983983983984 1.92682799625235e-16
8.41841841841842 1.62830341150177e-16
8.43843843843844 1.37547801096493e-16
8.45845845845846 1.16144301209627e-16
8.47847847847848 9.8032051926925e-17
8.4984984984985 8.27111796592674e-17
8.51851851851852 6.97567552529606e-17
8.53853853853854 5.88077091106193e-17
8.55855855855856 4.95573626747691e-17
8.57857857857858 4.17453440895294e-17
8.5985985985986 3.51506886838449e-17
8.61861861861862 2.95859531836935e-17
8.63863863863864 2.48921968838275e-17
8.65865865865866 2.09347039316773e-17
8.67867867867868 1.75993388643364e-17
8.6986986986987 1.47894429982973e-17
8.71871871871872 1.24231925503958e-17
8.73873873873874 1.04313507694241e-17
8.75875875875876 8.75535614211525e-18
8.77877877877878 7.34569713009337e-18
8.7987987987988 6.16053109048305e-18
8.81881881881882 5.16451119999019e-18
8.83883883883884 4.32779048512422e-18
8.85885885885886 3.62517658454128e-18
8.87887887887888 3.03541474067764e-18
8.8988988988989 2.54057982941549e-18
8.91891891891892 2.12556106807901e-18
8.93893893893894 1.77762546207239e-18
8.95895895895896 1.48604811781133e-18
8.97897897897898 1.24179931485728e-18
8.998998998999 1.03727973679138e-18
9.01901901901902 8.66096545670873e-19
9.03903903903904 7.22874080905507e-19
9.05905905905906 6.03093897535385e-19
9.07907907907908 5.0295965472299e-19
9.0990990990991 4.19283042962206e-19
9.11911911911912 3.49387515332417e-19
9.13913913913914 2.91027078875695e-19
9.15915915915916 2.42317819504618e-19
9.17917917917918 2.01680188581445e-19
9.1991991991992 1.67790380698338e-19
9.21921921921922 1.39539388139188e-19
9.23923923923924 1.1599853476858e-19
9.25925925925926 9.63904764362469e-20
9.27927927927928 8.00648113242436e-20
9.2992992992993 6.6477576193283e-20
9.31931931931932 5.51740167796855e-20
9.33933933933934 4.5774115701642e-20
9.35935935935936 3.79604417474457e-20
9.37937937937938 3.14679525475421e-20
9.3993993993994 2.60754402558043e-20
9.41941941941942 2.15983585814365e-20
9.43943943943944 1.78828106797973e-20
9.45945945945946 1.48005121824234e-20
9.47947947947948 1.22445730038445e-20
9.4994994994995 1.01259663374544e-20
9.51951951951952 8.37057415073758e-21
9.53953953953954 6.91671611023779e-21
9.55955955955956 5.71308371631411e-21
9.57957957957958 4.71701393695898e-21
9.5995995995996 3.89304716291232e-21
9.61961961961962 3.21172317125736e-21
9.63963963963964 2.64857624243904e-21
9.65965965965966 2.18329684678274e-21
9.67967967967968 1.79903258756111e-21
9.6996996996997 1.48180551599987e-21
9.71971971971972 1.22002665237565e-21
9.73973973973974 1.00409166885045e-21
9.75975975975976 8.26044308669654e-22
9.77977977977978 6.79296312742742e-22
9.7997997997998 5.58394465795474e-22
9.81981981981982 4.58826916995414e-22
9.83983983983984 3.7686222201397e-22
9.85985985985986 3.09415635142992e-22
9.87987987987988 2.53938085193762e-22
9.8998998998999 2.08324025950642e-22
9.91991991991992 1.70834984871876e-22
9.93993993993994 1.40036162642795e-22
9.95995995995996 1.14743877987917e-22
9.97997997997998 9.39820210218911e-23
10 7.69459862670642e-23
};
\addlegendentry{N(0,1)}; 
\legend{}; 
\end{axis}

\end{tikzpicture}

%% file: FinalFigs/PowerCurve_RBF2022_10_13_15_48_24_.tex
\begin{tikzpicture}

\definecolor{darkorange25512714}{RGB}{255,127,14}
\definecolor{darkslategray38}{RGB}{38,38,38}
\definecolor{forestgreen4416044}{RGB}{44,160,44}
\definecolor{lightgray204}{RGB}{204,204,204}
\definecolor{steelblue31119180}{RGB}{31,119,180}

\begin{axis}[
axis line style={darkslategray38},
height=\figheight,
legend cell align={left},
legend style={
  fill opacity=0.8,
  draw opacity=1,
  text opacity=1,
  at={(0.97,0.03)},
  anchor=south east,
  draw=none
},
tick align=outside,
tick pos=left,
title={$(d, j, , \epsilon) = (10, 5, 0.3)$},
width=\figwidth,
x grid style={lightgray204},
xlabel=\textcolor{darkslategray38}{Sample-Size (n+m)},
xmin=27, xmax=313,
xtick style={color=darkslategray38},
y grid style={lightgray204},
ylabel=\textcolor{darkslategray38}{Power},
ymin=0.154013184564948, ymax=1.04034314524104,
ytick style={color=darkslategray38}
]
\path [draw=steelblue31119180, fill=steelblue31119180, opacity=0.3]
(axis cs:40,0.282723332584592)
--(axis cs:40,0.237276667415408)
--(axis cs:58,0.316032261266444)
--(axis cs:76,0.420989636460433)
--(axis cs:94,0.536790357607236)
--(axis cs:114,0.654424583562804)
--(axis cs:132,0.737608988367858)
--(axis cs:150,0.786130584435389)
--(axis cs:170,0.857540810725511)
--(axis cs:188,0.916759059701064)
--(axis cs:206,0.922540220959022)
--(axis cs:224,0.947877443491982)
--(axis cs:244,0.964754527532164)
--(axis cs:262,0.970173720248312)
--(axis cs:280,0.975420098870747)
--(axis cs:300,0.985210754234746)
--(axis cs:300,0.994789245765254)
--(axis cs:300,0.994789245765254)
--(axis cs:280,0.989579901129253)
--(axis cs:262,0.984826279751688)
--(axis cs:244,0.980245472467836)
--(axis cs:224,0.967122556508018)
--(axis cs:206,0.947459779040978)
--(axis cs:188,0.943240940298937)
--(axis cs:170,0.88745918927449)
--(axis cs:150,0.823869415564611)
--(axis cs:132,0.777391011632142)
--(axis cs:114,0.700575416437196)
--(axis cs:94,0.583209642392764)
--(axis cs:76,0.474010363539567)
--(axis cs:58,0.363967738733556)
--(axis cs:40,0.282723332584592)
--cycle;

\path [draw=darkorange25512714, fill=darkorange25512714, opacity=0.3]
(axis cs:40,0.230699089949775)
--(axis cs:40,0.194300910050225)
--(axis cs:58,0.234372851451162)
--(axis cs:76,0.29572295322738)
--(axis cs:94,0.380221661305205)
--(axis cs:114,0.458280383080032)
--(axis cs:132,0.56252737749054)
--(axis cs:150,0.556212345638966)
--(axis cs:170,0.648640462464042)
--(axis cs:188,0.70554607452527)
--(axis cs:206,0.761676575689863)
--(axis cs:224,0.801999670141911)
--(axis cs:244,0.830976860307773)
--(axis cs:262,0.880128130581531)
--(axis cs:280,0.922132793620991)
--(axis cs:300,0.900108727388165)
--(axis cs:300,0.929891272611835)
--(axis cs:300,0.929891272611835)
--(axis cs:280,0.947867206379009)
--(axis cs:262,0.909871869418469)
--(axis cs:244,0.869023139692227)
--(axis cs:224,0.838000329858089)
--(axis cs:206,0.803323424310137)
--(axis cs:188,0.74945392547473)
--(axis cs:170,0.691359537535958)
--(axis cs:150,0.603787654361034)
--(axis cs:132,0.61247262250946)
--(axis cs:114,0.506719616919968)
--(axis cs:94,0.424778338694795)
--(axis cs:76,0.34427704677262)
--(axis cs:58,0.280627148548838)
--(axis cs:40,0.230699089949775)
--cycle;

\path [draw=forestgreen4416044, fill=forestgreen4416044, opacity=0.3]
(axis cs:40,0.350935951445562)
--(axis cs:40,0.304006868184874)
--(axis cs:58,0.380836144249499)
--(axis cs:76,0.482939461142481)
--(axis cs:94,0.60598386857594)
--(axis cs:114,0.709986565405858)
--(axis cs:132,0.821560678431874)
--(axis cs:150,0.814547897625608)
--(axis cs:170,0.889045757255082)
--(axis cs:188,0.925822433540332)
--(axis cs:206,0.95344678987824)
--(axis cs:224,0.968247825284858)
--(axis cs:244,0.977852367920829)
--(axis cs:262,0.988751987026468)
--(axis cs:280,0.995180747737994)
--(axis cs:300,0.992329280362275)
--(axis cs:300,0.998926865697017)
--(axis cs:300,0.998926865697017)
--(axis cs:280,1.00005541975576)
--(axis cs:262,0.997125418432602)
--(axis cs:244,0.990359034219746)
--(axis cs:224,0.983579545381617)
--(axis cs:206,0.972348100715726)
--(axis cs:188,0.949957924259024)
--(axis cs:170,0.918533696559823)
--(axis cs:150,0.851828665974854)
--(axis cs:132,0.858230943978391)
--(axis cs:114,0.754271562599627)
--(axis cs:94,0.654261001046582)
--(axis cs:76,0.532933162241393)
--(axis cs:58,0.429932773397319)
--(axis cs:40,0.350935951445562)
--cycle;

\addplot [semithick, steelblue31119180]
table {%
40 0.26
58 0.34
76 0.4475
94 0.56
114 0.6775
132 0.7575
150 0.805
170 0.8725
188 0.93
206 0.935
224 0.9575
244 0.9725
262 0.9775
280 0.9825
300 0.99
};
\addlegendentry{$\dmmd$-perm}
\addplot [semithick, darkorange25512714]
table {%
40 0.2125
58 0.2575
76 0.32
94 0.4025
114 0.4825
132 0.5875
150 0.58
170 0.67
188 0.7275
206 0.7825
224 0.82
244 0.85
262 0.895
280 0.935
300 0.915
};
\addlegendentry{$\cmmd$}
\addplot [semithick, forestgreen4416044, dashed]
table {%
40 0.327471409815218
58 0.405384458823409
76 0.507936311691937
94 0.630122434811261
114 0.732129064002742
132 0.839895811205133
150 0.833188281800231
170 0.903789726907452
188 0.937890178899678
206 0.962897445296983
224 0.975913685333238
244 0.984105701070288
262 0.992938702729535
280 0.99761808374688
300 0.995628073029646
};
\addlegendentry{predicted}
\end{axis}

\end{tikzpicture}

%% file: FinalFigs/PowerCurve_RBF2022_10_13_16_05_01_.tex
\begin{tikzpicture}

\definecolor{darkorange25512714}{RGB}{255,127,14}
\definecolor{darkslategray38}{RGB}{38,38,38}
\definecolor{forestgreen4416044}{RGB}{44,160,44}
\definecolor{lightgray204}{RGB}{204,204,204}
\definecolor{steelblue31119180}{RGB}{31,119,180}

\begin{axis}[
axis line style={darkslategray38},
height=\figheight,
legend cell align={left},
legend style={
  fill opacity=0.8,
  draw opacity=1,
  text opacity=1,
  at={(0.03,0.97)},
  anchor=north west,
  draw=none
},
tick align=outside,
tick pos=left,
title={$(d, j, , \epsilon) = (50, 5, 0.4)$},
width=\figwidth,
x grid style={lightgray204},
xlabel=\textcolor{darkslategray38}{Sample-Size (n+m)},
xmin=27, xmax=313,
xtick style={color=darkslategray38},
y grid style={lightgray204},
ylabel=\textcolor{darkslategray38}{Power},
ymin=0.116582537365147, ymax=1.04146608927892,
ytick style={color=darkslategray38}
]
\path [draw=steelblue31119180, fill=steelblue31119180, opacity=0.3]
(axis cs:40,0.226611367690461)
--(axis cs:40,0.188388632309539)
--(axis cs:58,0.236098203930511)
--(axis cs:76,0.367421386102896)
--(axis cs:94,0.49673612327793)
--(axis cs:114,0.602442143829699)
--(axis cs:132,0.685673124429964)
--(axis cs:150,0.79931295533216)
--(axis cs:170,0.814222490767339)
--(axis cs:188,0.903444535093068)
--(axis cs:206,0.930541902586532)
--(axis cs:224,0.92495012450261)
--(axis cs:244,0.955403198775113)
--(axis cs:262,0.985024450407241)
--(axis cs:280,0.981767635740813)
--(axis cs:300,0.988110399363268)
--(axis cs:300,0.996889600636732)
--(axis cs:300,0.996889600636732)
--(axis cs:280,0.993232364259187)
--(axis cs:262,0.994975549592759)
--(axis cs:244,0.974596801224887)
--(axis cs:224,0.95004987549739)
--(axis cs:206,0.954458097413468)
--(axis cs:188,0.931555464906932)
--(axis cs:170,0.850777509232661)
--(axis cs:150,0.83568704466784)
--(axis cs:132,0.734326875570036)
--(axis cs:114,0.647557856170301)
--(axis cs:94,0.54826387672207)
--(axis cs:76,0.417578613897104)
--(axis cs:58,0.278901796069489)
--(axis cs:40,0.226611367690461)
--cycle;

\path [draw=darkorange25512714, fill=darkorange25512714, opacity=0.3]
(axis cs:40,0.196377301184226)
--(axis cs:40,0.158622698815773)
--(axis cs:58,0.164391477317296)
--(axis cs:76,0.299664477370772)
--(axis cs:94,0.34415220834854)
--(axis cs:114,0.388725920101926)
--(axis cs:132,0.473191400778592)
--(axis cs:150,0.577918760329267)
--(axis cs:170,0.630325140924912)
--(axis cs:188,0.693497946051431)
--(axis cs:206,0.740700232942758)
--(axis cs:224,0.767783354906323)
--(axis cs:244,0.840981296212675)
--(axis cs:262,0.892185974565778)
--(axis cs:280,0.91035133752218)
--(axis cs:300,0.90215)
--(axis cs:300,0.92785)
--(axis cs:300,0.92785)
--(axis cs:280,0.93464866247782)
--(axis cs:262,0.922814025434222)
--(axis cs:244,0.879018703787325)
--(axis cs:224,0.807216645093677)
--(axis cs:206,0.784299767057242)
--(axis cs:188,0.741502053948569)
--(axis cs:170,0.674674859075088)
--(axis cs:150,0.627081239670733)
--(axis cs:132,0.516808599221408)
--(axis cs:114,0.436274079898074)
--(axis cs:94,0.39084779165146)
--(axis cs:76,0.345335522629228)
--(axis cs:58,0.200608522682704)
--(axis cs:40,0.196377301184226)
--cycle;

\path [draw=forestgreen4416044, fill=forestgreen4416044, opacity=0.3]
(axis cs:40,0.287494491301437)
--(axis cs:40,0.243339004900705)
--(axis cs:58,0.252015959431782)
--(axis cs:76,0.486879886935067)
--(axis cs:94,0.555614209169116)
--(axis cs:114,0.619847480239983)
--(axis cs:132,0.724775425837818)
--(axis cs:150,0.835144779264165)
--(axis cs:170,0.876198465296363)
--(axis cs:188,0.920016723235861)
--(axis cs:206,0.944232935259161)
--(axis cs:224,0.955604902625768)
--(axis cs:244,0.980629350200146)
--(axis cs:262,0.991075065956534)
--(axis cs:280,0.993481373317986)
--(axis cs:300,0.992329280362275)
--(axis cs:300,0.998926865697017)
--(axis cs:300,0.998926865697017)
--(axis cs:280,0.999425927828294)
--(axis cs:262,0.998333081743404)
--(axis cs:244,0.992204559143343)
--(axis cs:224,0.97402911214607)
--(axis cs:206,0.965042661860158)
--(axis cs:188,0.945095629563397)
--(axis cs:170,0.907270039976658)
--(axis cs:150,0.870569601520384)
--(axis cs:132,0.768275449804153)
--(axis cs:114,0.667735269962057)
--(axis cs:94,0.604965354136798)
--(axis cs:76,0.536865788519843)
--(axis cs:58,0.296633290722828)
--(axis cs:40,0.287494491301437)
--cycle;

\addplot [semithick, steelblue31119180]
table {%
40 0.2075
58 0.2575
76 0.3925
94 0.5225
114 0.625
132 0.71
150 0.8175
170 0.8325
188 0.9175
206 0.9425
224 0.9375
244 0.965
262 0.99
280 0.9875
300 0.9925
};
\addlegendentry{mmd-perm}
\addplot [semithick, darkorange25512714]
table {%
40 0.1775
58 0.1825
76 0.3225
94 0.3675
114 0.4125
132 0.495
150 0.6025
170 0.6525
188 0.7175
206 0.7625
224 0.7875
244 0.86
262 0.9075
280 0.9225
300 0.915
};
\addlegendentry{c-mmd}
\addplot [semithick, forestgreen4416044, dashed]
table {%
40 0.265416748101071
58 0.274324625077305
76 0.511872837727455
94 0.580289781652957
114 0.64379137510102
132 0.746525437820986
150 0.852857190392275
170 0.89173425263651
188 0.932556176399629
206 0.95463779855966
224 0.964817007385919
244 0.986416954671745
262 0.994704073849969
280 0.99645365057314
300 0.995628073029646
};
\addlegendentry{predicted}; 
\legend{}; 
\end{axis}

\end{tikzpicture}

%% file: FinalFigs/PowerCurve_RBF2022_10_13_17_14_52_.tex
\begin{tikzpicture}

\definecolor{darkorange25512714}{RGB}{255,127,14}
\definecolor{darkslategray38}{RGB}{38,38,38}
\definecolor{forestgreen4416044}{RGB}{44,160,44}
\definecolor{lightgray204}{RGB}{204,204,204}
\definecolor{steelblue31119180}{RGB}{31,119,180}

\begin{axis}[
axis line style={darkslategray38},
height=\figheight,
legend cell align={left},
legend style={
  fill opacity=0.8,
  draw opacity=1,
  text opacity=1,
  at={(0.97,0.03)},
  anchor=south east,
  draw=none
},
tick align=outside,
tick pos=left,
title={$(d, j, , \epsilon) = (100, 5, 0.5)$},
width=\figwidth,
x grid style={lightgray204},
xlabel=\textcolor{darkslategray38}{Sample-Size (n+m)},
xmin=27, xmax=313,
xtick style={color=darkslategray38},
y grid style={lightgray204},
ylabel=\textcolor{darkslategray38}{Power},
ymin=0.10159618018133, ymax=1.04336264875752,
ytick style={color=darkslategray38}
]
\path [draw=steelblue31119180, fill=steelblue31119180, opacity=0.3]
(axis cs:40,0.231235197049945)
--(axis cs:40,0.188764802950055)
--(axis cs:58,0.332319529516905)
--(axis cs:76,0.463839512738674)
--(axis cs:94,0.59308435760112)
--(axis cs:114,0.682440360292106)
--(axis cs:132,0.756629719844957)
--(axis cs:150,0.866209133118523)
--(axis cs:170,0.915394113002345)
--(axis cs:188,0.933732818675463)
--(axis cs:206,0.958850578053997)
--(axis cs:224,0.987924675557952)
--(axis cs:244,0.985155479899928)
--(axis cs:262,0.99143473791286)
--(axis cs:280,0.995037785549551)
--(axis cs:300,1)
--(axis cs:300,1)
--(axis cs:300,1)
--(axis cs:280,0.999962214450449)
--(axis cs:262,0.99856526208714)
--(axis cs:244,0.994844520100072)
--(axis cs:224,0.997075324442048)
--(axis cs:206,0.976149421946003)
--(axis cs:188,0.956267181324537)
--(axis cs:170,0.939605886997655)
--(axis cs:150,0.898790866881477)
--(axis cs:132,0.798370280155043)
--(axis cs:114,0.727559639707894)
--(axis cs:94,0.64191564239888)
--(axis cs:76,0.516160487261326)
--(axis cs:58,0.382680470483095)
--(axis cs:40,0.231235197049945)
--cycle;

\path [draw=darkorange25512714, fill=darkorange25512714, opacity=0.3]
(axis cs:40,0.180596253065207)
--(axis cs:40,0.144403746934793)
--(axis cs:58,0.244603630837818)
--(axis cs:76,0.316543660904779)
--(axis cs:94,0.388120734347606)
--(axis cs:114,0.473550640957672)
--(axis cs:132,0.538368074180663)
--(axis cs:150,0.667800633454983)
--(axis cs:170,0.673697964897934)
--(axis cs:188,0.798919133665246)
--(axis cs:206,0.840032905937163)
--(axis cs:224,0.91405169620361)
--(axis cs:244,0.905979679925194)
--(axis cs:262,0.940719083227524)
--(axis cs:280,0.98243177792515)
--(axis cs:300,0.966377799367331)
--(axis cs:300,0.983622200632669)
--(axis cs:300,0.983622200632669)
--(axis cs:280,0.99256822207485)
--(axis cs:262,0.964280916772476)
--(axis cs:244,0.934020320074806)
--(axis cs:224,0.94094830379639)
--(axis cs:206,0.874967094062837)
--(axis cs:188,0.836080866334754)
--(axis cs:170,0.721302035102066)
--(axis cs:150,0.712199366545016)
--(axis cs:132,0.586631925819337)
--(axis cs:114,0.526449359042328)
--(axis cs:94,0.436879265652394)
--(axis cs:76,0.363456339095221)
--(axis cs:58,0.285396369162182)
--(axis cs:40,0.180596253065207)
--cycle;

\path [draw=forestgreen4416044, fill=forestgreen4416044, opacity=0.3]
(axis cs:40,0.259980200484672)
--(axis cs:40,0.217353337775451)
--(axis cs:58,0.393415430446018)
--(axis cs:76,0.514114029576054)
--(axis cs:94,0.619847480239983)
--(axis cs:114,0.730577945159302)
--(axis cs:132,0.797611540569048)
--(axis cs:150,0.902775631857634)
--(axis cs:170,0.907664634222838)
--(axis cs:188,0.967359221262882)
--(axis cs:206,0.979954579568221)
--(axis cs:224,0.994193806770588)
--(axis cs:244,0.993108552603373)
--(axis cs:262,0.997115234665635)
--(axis cs:280,0.99955787383945)
--(axis cs:300,0.998890792673032)
--(axis cs:300,1.00055508200406)
--(axis cs:300,1.00055508200406)
--(axis cs:280,1.00032455825181)
--(axis cs:262,1.00053840018503)
--(axis cs:244,0.999269918240475)
--(axis cs:224,0.999707479244144)
--(axis cs:206,0.991762071829178)
--(axis cs:188,0.982927961586952)
--(axis cs:170,0.934616554348278)
--(axis cs:150,0.930424229937154)
--(axis cs:132,0.836282534281065)
--(axis cs:114,0.773753454403969)
--(axis cs:94,0.667735269962057)
--(axis cs:76,0.563961402298636)
--(axis cs:58,0.442739736591965)
--(axis cs:40,0.259980200484672)
--cycle;

\addplot [semithick, steelblue31119180]
table {%
40 0.21
58 0.3575
76 0.49
94 0.6175
114 0.705
132 0.7775
150 0.8825
170 0.9275
188 0.945
206 0.9675
224 0.9925
244 0.99
262 0.995
280 0.9975
300 1
};
\addlegendentry{mmd-perm}
\addplot [semithick, darkorange25512714]
table {%
40 0.1625
58 0.265
76 0.34
94 0.4125
114 0.5
132 0.5625
150 0.69
170 0.6975
188 0.8175
206 0.8575
224 0.9275
244 0.92
262 0.9525
280 0.9875
300 0.975
};
\addlegendentry{c-mmd}
\addplot [semithick, forestgreen4416044, dashed]
table {%
40 0.238666769130062
58 0.418077583518991
76 0.539037715937345
94 0.64379137510102
114 0.752165699781636
132 0.816947037425056
150 0.916599930897394
170 0.921140594285558
188 0.975143591424917
206 0.9858583256987
224 0.996950643007366
244 0.996189235421924
262 0.998826817425331
280 0.999941216045631
300 0.999722937338545
};
\addlegendentry{predicted}; 
\legend{}; 
\end{axis}

\end{tikzpicture}

%% file: FinalFigs/Null_Dists_d_10_100_n_100_m_20_kernel__Dirichlet_RBF_2022_10_15_22_20_23cross.tex
\begin{tikzpicture}

\definecolor{darkorange25512714}{RGB}{255,127,14}
\definecolor{darkslategray38}{RGB}{38,38,38}
\definecolor{lightgray204}{RGB}{204,204,204}
\definecolor{steelblue31119180}{RGB}{31,119,180}

\begin{axis}[
axis line style={darkslategray38},
height=\figheight,
legend cell align={left},
legend style={fill opacity=0.8, draw opacity=1, text opacity=1, draw=none},
tick align=outside,
tick pos=left,
title={$\cmmd$~$(n/m=5)$},
width=\figwidth,
x grid style={lightgray204},
xmin=-6, xmax=6,
xtick style={color=darkslategray38},
y grid style={lightgray204},
ylabel=\textcolor{darkslategray38}{Probability density},
ymin=0, ymax=0.418868408525629,
ytick style={color=darkslategray38}, 
xticklabels=empty,
yticklabels=empty
]
\draw[draw=none,fill=steelblue31119180,fill opacity=0.8] (axis cs:-4.73563480377197,0) rectangle (axis cs:-4.332923412323,0.00198653479001993);
\addlegendimage{ybar,ybar legend,draw=none,fill=steelblue31119180,fill opacity=0.8}
\addlegendentry{d=10}

\draw[draw=none,fill=steelblue31119180,fill opacity=0.8] (axis cs:-3.72885680198669,0) rectangle (axis cs:-3.32614541053772,0.00993267159790989);
\draw[draw=none,fill=steelblue31119180,fill opacity=0.8] (axis cs:-2.72207832336426,0) rectangle (axis cs:-2.31936693191528,0.053636439330538);
\draw[draw=none,fill=steelblue31119180,fill opacity=0.8] (axis cs:-1.7153000831604,0) rectangle (axis cs:-1.31258869171143,0.194680386370491);
\draw[draw=none,fill=steelblue31119180,fill opacity=0.8] (axis cs:-0.708521664142609,0) rectangle (axis cs:-0.305810272693634,0.333737805206555);
\draw[draw=none,fill=steelblue31119180,fill opacity=0.8] (axis cs:0.298256695270538,0) rectangle (axis cs:0.700968086719513,0.264209095788523);
\draw[draw=none,fill=steelblue31119180,fill opacity=0.8] (axis cs:1.30503511428833,0) rectangle (axis cs:1.7077465057373,0.113232469623653);
\draw[draw=none,fill=steelblue31119180,fill opacity=0.8] (axis cs:2.31181335449219,0) rectangle (axis cs:2.71452474594116,0.0178788131101793);
\draw[draw=none,fill=steelblue31119180,fill opacity=0.8] (axis cs:3.31859183311462,0) rectangle (axis cs:3.7213032245636,0.00198653431958198);
\draw[draw=none,fill=steelblue31119180,fill opacity=0.8] (axis cs:4.3253698348999,0) rectangle (axis cs:4.72808122634888,0.00198653479001993);
\draw[draw=none,fill=darkorange25512714,fill opacity=0.8] (axis cs:-4.33292388916016,0) rectangle (axis cs:-3.93021249771118,0);
\addlegendimage{ybar,ybar legend,draw=none,fill=darkorange25512714,fill opacity=0.8}
\addlegendentry{d=100}

\draw[draw=none,fill=darkorange25512714,fill opacity=0.8] (axis cs:-3.32614541053772,0) rectangle (axis cs:-2.92343401908875,0.00397306863916395);
\draw[draw=none,fill=darkorange25512714,fill opacity=0.8] (axis cs:-2.31936693191528,0) rectangle (axis cs:-1.91665554046631,0.0615825784906177);
\draw[draw=none,fill=darkorange25512714,fill opacity=0.8] (axis cs:-1.31258869171143,0) rectangle (axis cs:-0.909877300262451,0.147003557055268);
\draw[draw=none,fill=darkorange25512714,fill opacity=0.8] (axis cs:-0.305810332298279,0) rectangle (axis cs:0.0969010591506958,0.393333841850583);
\draw[draw=none,fill=darkorange25512714,fill opacity=0.8] (axis cs:0.700968027114868,0) rectangle (axis cs:1.10367941856384,0.262222561233722);
\draw[draw=none,fill=darkorange25512714,fill opacity=0.8] (axis cs:1.70774638652802,0) rectangle (axis cs:2.11045789718628,0.0973401931852453);
\draw[draw=none,fill=darkorange25512714,fill opacity=0.8] (axis cs:2.71452474594116,0) rectangle (axis cs:3.11723613739014,0.0218518826902192);
\draw[draw=none,fill=darkorange25512714,fill opacity=0.8] (axis cs:3.7213032245636,0) rectangle (axis cs:4.12401485443115,0.00198653431958198);
\draw[draw=none,fill=darkorange25512714,fill opacity=0.8] (axis cs:4.72808170318604,0) rectangle (axis cs:5.13079309463501,0.00397306958003985);
\addplot [semithick, black]
table {%
-10 7.69459862670642e-23
-9.97997997997998 9.39820210218911e-23
-9.95995995995996 1.14743877987917e-22
-9.93993993993994 1.40036162642795e-22
-9.91991991991992 1.70834984871876e-22
-9.8998998998999 2.08324025950642e-22
-9.87987987987988 2.53938085193762e-22
-9.85985985985986 3.09415635142992e-22
-9.83983983983984 3.7686222201397e-22
-9.81981981981982 4.58826916995414e-22
-9.7997997997998 5.58394465795474e-22
-9.77977977977978 6.79296312742742e-22
-9.75975975975976 8.26044308669654e-22
-9.73973973973974 1.00409166885045e-21
-9.71971971971972 1.22002665237565e-21
-9.6996996996997 1.48180551599987e-21
-9.67967967967968 1.79903258756111e-21
-9.65965965965966 2.18329684678274e-21
-9.63963963963964 2.64857624243904e-21
-9.61961961961962 3.21172317125736e-21
-9.5995995995996 3.89304716291232e-21
-9.57957957957958 4.71701393695898e-21
-9.55955955955956 5.71308371631411e-21
-9.53953953953954 6.91671611023779e-21
-9.51951951951952 8.37057415073758e-21
-9.4994994994995 1.01259663374544e-20
-9.47947947947948 1.22445730038445e-20
-9.45945945945946 1.48005121824234e-20
-9.43943943943944 1.78828106797973e-20
-9.41941941941942 2.15983585814365e-20
-9.3993993993994 2.60754402558043e-20
-9.37937937937938 3.14679525475421e-20
-9.35935935935936 3.79604417474457e-20
-9.33933933933934 4.5774115701642e-20
-9.31931931931932 5.51740167796855e-20
-9.2992992992993 6.6477576193283e-20
-9.27927927927928 8.00648113242436e-20
-9.25925925925926 9.63904764362469e-20
-9.23923923923924 1.1599853476858e-19
-9.21921921921922 1.39539388139188e-19
-9.1991991991992 1.67790380698338e-19
-9.17917917917918 2.01680188581445e-19
-9.15915915915916 2.42317819504618e-19
-9.13913913913914 2.91027078875695e-19
-9.11911911911912 3.49387515332417e-19
-9.0990990990991 4.19283042962206e-19
-9.07907907907908 5.0295965472299e-19
-9.05905905905906 6.03093897535385e-19
-9.03903903903904 7.22874080905507e-19
-9.01901901901902 8.66096545670873e-19
-8.998998998999 1.03727973679138e-18
-8.97897897897898 1.24179931485728e-18
-8.95895895895896 1.48604811781133e-18
-8.93893893893894 1.77762546207239e-18
-8.91891891891892 2.12556106807901e-18
-8.8988988988989 2.54057982941549e-18
-8.87887887887888 3.03541474067764e-18
-8.85885885885886 3.62517658454128e-18
-8.83883883883884 4.32779048512422e-18
-8.81881881881882 5.16451119999019e-18
-8.7987987987988 6.16053109048305e-18
-8.77877877877878 7.34569713009337e-18
-8.75875875875876 8.75535614211525e-18
-8.73873873873874 1.04313507694241e-17
-8.71871871871872 1.24231925503958e-17
-8.6986986986987 1.47894429982973e-17
-8.67867867867868 1.75993388643364e-17
-8.65865865865866 2.09347039316773e-17
-8.63863863863864 2.48921968838275e-17
-8.61861861861862 2.95859531836935e-17
-8.5985985985986 3.51506886838449e-17
-8.57857857857858 4.17453440895294e-17
-8.55855855855856 4.95573626747691e-17
-8.53853853853854 5.88077091106193e-17
-8.51851851851852 6.97567552529606e-17
-8.4984984984985 8.27111796592674e-17
-8.47847847847848 9.8032051926925e-17
-8.45845845845846 1.16144301209627e-16
-8.43843843843844 1.37547801096493e-16
-8.41841841841842 1.62830341150177e-16
-8.3983983983984 1.92682799625235e-16
-8.37837837837838 2.27916883183252e-16
-8.35835835835836 2.69485858889318e-16
-8.33833833833834 3.18508772685576e-16
-8.31831831831832 3.76298728354137e-16
-8.2982982982983 4.44395893386326e-16
-8.27827827827828 5.24606005103494e-16
-8.25825825825826 6.19045274051723e-16
-8.23823823823824 7.30192724674871e-16
-8.21821821821822 8.60951178494166e-16
-8.1981981981982 1.01471827585831e-15
-8.17817817817818 1.19546915264237e-15
-8.15815815815816 1.40785264250169e-15
-8.13813813813814 1.65730316851105e-15
-8.11811811811812 1.95017082606148e-15
-8.0980980980981 2.29387254841619e-15
-8.07807807807808 2.69706769497134e-15
-8.05805805805806 3.16986191875719e-15
-8.03803803803804 3.72404376402735e-15
-8.01801801801802 4.3733591283238e-15
-7.997997997998 5.13382950919607e-15
-7.97797797797798 6.02412085866193e-15
-7.95795795795796 7.06597090549303e-15
-7.93793793793794 8.28468399583074e-15
-7.91791791791792 9.70970386854708e-15
-7.8978978978979 1.13752763482777e-14
-7.87787787787788 1.33212157347805e-14
-7.85785785785786 1.55937907247683e-14
-7.83783783783784 1.82467480586655e-14
-7.81781781781782 2.13424947819475e-14
-7.7977977977978 2.49534630966872e-14
-7.77777777777778 2.91636853079909e-14
-7.75775775775776 3.40706104038538e-14
-7.73773773773774 3.9787198415571e-14
-7.71771771771772 4.64443339685918e-14
-7.6976976976977 5.4193606440538e-14
-7.67767767767768 6.32105109959252e-14
-7.65765765765766 7.36981325813117e-14
-7.63763763763764 8.58913838706879e-14
-7.61761761761762 1.00061878296601e-13
-7.5975975975976 1.16523530854699e-13
-7.57757757757758 1.35638992516664e-13
-7.55755755755756 1.57827039041884e-13
-7.53753753753754 1.83571051982057e-13
-7.51751751751752 2.13428748996349e-13
-7.4974974974975 2.48043342543513e-13
-7.47747747747748 2.88156330935935e-13
-7.45745745745746 3.34622154016965e-13
-7.43743743743744 3.88424977793732e-13
-7.41741741741742 4.50697908714384e-13
-7.3973973973974 5.22744979473711e-13
-7.37737737737738 6.06066294885249e-13
-7.35735735735736 7.02386779168738e-13
-7.33733733733734 8.13689025751835e-13
-7.31731731731732 9.42250818252909e-13
-7.2972972972973 1.09068796768221e-12
-7.27727727727728 1.262003197176e-12
-7.25725725725726 1.45964190299847e-12
-7.23723723723724 1.68755573049416e-12
-7.21721721721722 1.95027502769792e-12
-7.1971971971972 2.25299137914218e-12
-7.17717717717718 2.60165157997871e-12
-7.15715715715716 3.00306458801653e-12
-7.13713713713714 3.4650231910834e-12
-7.11711711711712 3.99644235193919e-12
-7.0970970970971 4.60751644580953e-12
-7.07707707707708 5.30989788981514e-12
-7.05705705705706 6.11689998287646e-12
-7.03703703703704 7.0437271332242e-12
-7.01701701701702 8.10773605306645e-12
-6.996996996997 9.32873195138555e-12
-6.97697697697698 1.07293042619726e-11
-6.95695695695696 1.23352070109962e-11
-6.93693693693694 1.41757895636779e-11
-6.91691691691692 1.62844842008237e-11
-6.8968968968969 1.86993577716893e-11
-6.87687687687688 2.14637355595386e-11
-6.85685685685686 2.46269064908967e-11
-6.83683683683684 2.82449199306815e-11
-6.81681681681682 3.2381485546105e-11
-6.7967967967968 3.71089891068559e-11
-6.77677677677678 4.25096386334913e-11
-6.75675675675676 4.86767570277083e-11
-6.73673673673674 5.57162392366004e-11
-6.71671671671672 6.37481941394491e-11
-6.6966966966967 7.29087937236032e-11
-6.67667667667668 8.33523547614402e-11
-6.65665665665666 9.52536811418151e-11
-6.63663663663664 1.08810698278135e-10
-6.61661661661662 1.24247414645768e-10
-6.5965965965966 1.41817249531744e-10
-6.57657657657658 1.61806770551278e-10
-6.55655655655656 1.84539889444164e-10
-6.53653653653654 2.10382570159622e-10
-6.51651651651652 2.39748109325542e-10
-6.4964964964965 2.73103055937438e-10
-6.47647647647648 3.10973844559381e-10
-6.45645645645646 3.53954224575809e-10
-6.43643643643644 4.027135771479e-10
-6.41641641641642 4.58006221596996e-10
-6.3963963963964 5.20681824054169e-10
-6.37637637637638 5.91697033481538e-10
-6.35635635635636 6.72128483698846e-10
-6.33633633633634 7.63187314959523e-10
-6.31631631631632 8.66235385046001e-10
-6.2962962962963 9.8280335793813e-10
-6.27627627627628 1.11461087800766e-09
-6.25625625625626 1.26358905957513e-09
-6.23623623623624 1.43190554571787e-09
-6.21621621621622 1.62199241663862e-09
-6.1961961961962 1.83657725691024e-09
-6.17617617617618 2.0787177227378e-09
-6.15615615615616 2.35183998527873e-09
-6.13613613613614 2.65978146430928e-09
-6.11611611611612 3.00683830841829e-09
-6.0960960960961 3.39781812376754e-09
-6.07607607607608 3.83809850362696e-09
-6.05605605605606 4.33369196574699e-09
-6.03603603603604 4.89131796456919e-09
-6.01601601601602 5.51848271073395e-09
-5.995995995996 6.22356760178439e-09
-5.97597597597598 7.01592714588943e-09
-5.95595595595596 7.90599734535251e-09
-5.93593593593594 8.90541559921139e-09
-5.91591591591592 1.00271532849868e-08
-5.8958958958959 1.12856622892642e-08
-5.87587587587588 1.2697036876002e-08
-5.85585585585586 1.42791924110059e-08
-5.83583583583584 1.60520626017053e-08
-5.81581581581582 1.80378170640688e-08
-5.7957957957958 2.02611011941336e-08
-5.77577577577578 2.27493005011625e-08
-5.75575575575576 2.55328317539416e-08
-5.73573573573574 2.86454635022852e-08
-5.71571571571572 3.2124668763613e-08
-5.6956956956957 3.60120129107462e-08
-5.67567567567568 4.03535800631662e-08
-5.65565565565566 4.5200441571292e-08
-5.63563563563564 5.06091704933412e-08
-5.61561561561562 5.66424062986154e-08
-5.5955955955956 6.33694743912418e-08
-5.57557557557558 7.08670654362614e-08
-5.55555555555556 7.92199798873018e-08
-5.53553553553554 8.85219435638491e-08
-5.51551551551552 9.8876500608364e-08
-5.4954954954955 1.10397990671284e-07
-5.47547547547548 1.23212617727566e-07
-5.45545545545546 1.3745961852414e-07
-5.43543543543544 1.5329253929596e-07
-5.41541541541542 1.70880630071684e-07
-5.3953953953954 1.90410366621162e-07
-5.37537537537538 2.1208711087848e-07
-5.35535535535536 2.36136921509202e-07
-5.33533533533534 2.62808527181656e-07
-5.31531531531532 2.92375476052561e-07
-5.2952952952953 3.25138475990267e-07
-5.27527527527528 3.61427941137511e-07
-5.25525525525526 4.01606761563285e-07
-5.23523523523524 4.46073313973501e-07
-5.21521521521522 4.95264732746229e-07
-5.1951951951952 5.4966046193278e-07
-5.17517517517518 6.09786110324583e-07
-5.15515515515516 6.76217633231267e-07
-5.13513513513514 7.49585866251233e-07
-5.11511511511512 8.30581438046261e-07
-5.0950950950951 9.19960090959837e-07
-5.07507507507508 1.01854844024876e-06
-5.05505505505506 1.12725020473309e-06
-5.03503503503504 1.24705294381396e-06
-5.01501501501502 1.37903533806611e-06
-4.99499499499499 1.5243750529858e-06
-4.97497497497497 1.68435722796805e-06
-4.95495495495495 1.86038363520377e-06
-4.93493493493493 2.05398255592983e-06
-4.91491491491491 2.26681942433715e-06
-4.89489489489489 2.50070829244518e-06
-4.87487487487487 2.75762417238901e-06
-4.85485485485485 3.03971631583941e-06
-4.83483483483483 3.34932249368796e-06
-4.81481481481481 3.68898434268124e-06
-4.79479479479479 4.06146384937932e-06
-4.77477477477477 4.4697610456467e-06
-4.75475475475475 4.91713299385613e-06
-4.73473473473473 5.40711414409909e-06
-4.71471471471471 5.94353814994721e-06
-4.69469469469469 6.53056123369604e-06
-4.67467467467467 7.17268719654363e-06
-4.65465465465465 7.87479417380527e-06
-4.63463463463463 8.64216324004121e-06
-4.61461461461461 9.48050897386715e-06
-4.59459459459459 1.03960120972233e-05
-4.57457457457457 1.13953543089884e-05
-4.55455455455455 1.24857554380297e-05
-4.53453453453453 1.3675013046071e-05
-4.51451451451451 1.49715446161227e-05
-4.49449449449449 1.63844324676437e-05
-4.47447447447447 1.79234715450684e-05
-4.45445445445445 1.95992202318354e-05
-4.43443443443443 2.14230543475548e-05
-4.41441441441441 2.34072244914564e-05
-4.39439439439439 2.55649169007242e-05
-4.37437437437437 2.79103179977393e-05
-4.35435435435435 3.04586828055866e-05
-4.33433433433433 3.32264074164067e-05
-4.31431431431431 3.62311057022691e-05
-4.29429429429429 3.94916904631592e-05
-4.27427427427427 4.3028459211397e-05
-4.25425425425425 4.68631847962824e-05
-4.23423423423423 5.10192110769697e-05
-4.21421421421421 5.55215538554582e-05
-4.19419419419419 6.03970072851107e-05
-4.17417417417417 6.56742559732345e-05
-4.15415415415415 7.13839929989176e-05
-4.13413413413413 7.75590440694795e-05
-4.11411411411411 8.42344980404937e-05
-4.09409409409409 9.14478440253317e-05
-4.07407407407407 9.92391153205018e-05
-4.05405405405405 0.000107651040372646
-4.03403403403403 0.000116729201011866
-4.01401401401401 0.000126522198173995
-3.99399399399399 0.000137081825331481
-3.97397397397397 0.000148463249848567
-3.95395395395395 0.000160725202471485
-3.93393393393393 0.000173930175158222
-3.91391391391391 0.00018814462744512
-3.89389389389389 0.000203439201538965
-3.87387387387387 0.000219888946313312
-3.85385385385385 0.000237573550376443
-3.83383383383383 0.000256577584365551
-3.81381381381381 0.000276990752607344
-3.79379379379379 0.000298908154269281
-3.77377377377377 0.000322430554107926
-3.75375375375375 0.000347664662901427
-3.73373373373373 0.000374723427631836
-3.71371371371371 0.000403726331459719
-3.69369369369369 0.000434799703508357
-3.67367367367367 0.000468077038447599
-3.65365365365365 0.00050369932583812
-3.63363363363363 0.000541815389165405
-3.61361361361361 0.000582582234459159
-3.59359359359359 0.000626165408357979
-3.57357357357357 0.000672739365441021
-3.55355355355355 0.000722487844607978
-3.53353353353353 0.00077560425424601
-3.51351351351351 0.000832292065877155
-3.49349349349349 0.000892765215932443
-3.47347347347347 0.000957248515249216
-3.45345345345345 0.0010259780658361
-3.43343343343343 0.00109920168439588
-3.41341341341341 0.00117717933203981
-3.39339339339339 0.00126018354956833
-3.37337337337337 0.00134849989763212
-3.35335335335335 0.00144242740102448
-3.33333333333333 0.00154227899629111
-3.31331331331331 0.00164838198177652
-3.29329329329329 0.00176107846915772
-3.27327327327327 0.00188072583544552
-3.25325325325325 0.00200769717436226
-3.23323323323323 0.00214238174593163
-3.21321321321321 0.00228518542304204
-3.19319319319319 0.00243653113367012
-3.17317317317317 0.00259685929737497
-3.15315315315315 0.00276662825459747
-3.13313313313313 0.00294631468722261
-3.11311311311311 0.00313641402878609
-3.09309309309309 0.00333744086263052
-3.07307307307307 0.00354992930624086
-3.05305305305305 0.00377443337991422
-3.03303303303303 0.00401152735784579
-3.01301301301301 0.00426180609964128
-2.99299299299299 0.00452588536019618
-2.97297297297297 0.0048044020758154
-2.95295295295295 0.00509801462438215
-2.93293293293293 0.00540740305732385
-2.91291291291291 0.00573326930106519
-2.89289289289289 0.00607633732560526
-2.87287287287287 0.00643735327780636
-2.85285285285285 0.00681708557693873
-2.83283283283283 0.00721632496998623
-2.81281281281281 0.00763588454418632
-2.79279279279279 0.00807659969425075
-2.77277277277277 0.00853932804169477
-2.75275275275275 0.00902494930369032
-2.73273273273273 0.00953436510885489
-2.71271271271271 0.0100684987573917
-2.69269269269269 0.0106282949230102
-2.67267267267267 0.0112147192940778
-2.65265265265265 0.0118287581514852
-2.63263263263263 0.0124714178807513
-2.61261261261261 0.0131437244159435
-2.59259259259259 0.0138467226130541
-2.57257257257257 0.0145814755505492
-2.55255255255255 0.0153490637548887
-2.53253253253253 0.0161505843489182
-2.51251251251251 0.0169871501211409
-2.49249249249249 0.0178598885140022
-2.47247247247247 0.018769940529451
-2.45245245245245 0.0197184595501959
-2.43243243243243 0.0207066100752274
-2.41241241241241 0.0217355663683581
-2.39239239239239 0.0228065110187135
-2.37237237237237 0.0239206334123108
-2.35235235235235 0.0250791281140699
-2.33233233233233 0.0262831931598317
-2.31231231231231 0.0275340282581906
-2.29229229229229 0.0288328329022027
-2.27227227227227 0.0301808043912899
-2.25225225225225 0.0315791357639331
-2.23223223223223 0.033029013642035
-2.21221221221221 0.0345316159881232
-2.19219219219219 0.0360881097768736
-2.17217217217217 0.0376996485827434
-2.15215215215215 0.0393673700858293
-2.13213213213213 0.0410923934983949
-2.11211211211211 0.0428758169148479
-2.09209209209209 0.0447187145882915
-2.07207207207207 0.0466221341371235
-2.05205205205205 0.0485870936855041
-2.03203203203203 0.0506145789418747
-2.01201201201201 0.0527055402200587
-1.99199199199199 0.0548608894078376
-1.97197197197197 0.057081496888248
-1.95195195195195 0.059368188419199
-1.93193193193193 0.0617217419773594
-1.91191191191191 0.0641428845726061
-1.89189189189189 0.0666322890396674
-1.87187187187187 0.0691905708139176
-1.85185185185185 0.0718182846986055
-1.83183183183183 0.0745159216311036
-1.81181181181181 0.077283905456062
-1.79179179179179 0.0801225897136326
-1.77177177177177 0.083032254451193
-1.75175175175175 0.0860131030672496
-1.73173173173173 0.0890652591964251
-1.71171171171171 0.0921887636446459
-1.69169169169169 0.0953835713838294
-1.67167167167167 0.0986495486155338
-1.65165165165165 0.101986469913169
-1.63163163163163 0.10539401545248
-1.61161161161161 0.108871768340093
-1.59159159159159 0.112419212049971
-1.57157157157157 0.116035727977651
-1.55155155155155 0.119720593122119
-1.53153153153153 0.123472977905145
-1.51151151151151 0.127291944137829
-1.49149149149149 0.13117644314399
-1.47147147147147 0.135125314049902
-1.45145145145145 0.139137282249685
-1.43143143143143 0.143210958055468
-1.41141141141141 0.147344835541168
-1.39139139139139 0.151537291588457
-1.37137137137137 0.155786585143159
-1.35135135135135 0.160090856689972
-1.33133133133133 0.164448127952996
-1.31131131131131 0.168856301829129
-1.29129129129129 0.173313162560933
-1.27127127127127 0.17781637615506
-1.25125125125125 0.182363491051798
-1.23123123123123 0.186951939050736
-1.21121121121121 0.191579036496956
-1.19119119119119 0.1962419857315
-1.17117117117117 0.200937876809264
-1.15115115115115 0.205663689486728
-1.13113113113113 0.210416295481265
-1.11111111111111 0.215192461003031
-1.09109109109109 0.219988849559688
-1.07107107107107 0.224802025033432
-1.05105105105105 0.229628455029052
-1.03103103103103 0.234464514490888
-1.01101101101101 0.239306489585817
-0.990990990990991 0.24415058184851
-0.970970970970971 0.24899291258444
-0.950950950950951 0.253829527525259
-0.930930930930931 0.258656401730343
-0.910910910910911 0.2634694447275
-0.890890890890891 0.268264505884996
-0.870870870870871 0.273037380006279
-0.850850850850851 0.277783813137949
-0.830830830830831 0.282499508580786
-0.810810810810811 0.287180133092853
-0.790790790790791 0.291821323272996
-0.77077077077077 0.296418692112302
-0.75075075075075 0.300967835700437
-0.73073073073073 0.305464340073112
-0.71071071071071 0.309903788186304
-0.69069069069069 0.314281767002296
-0.67067067067067 0.318593874672039
-0.65065065065065 0.322835727797843
-0.63063063063063 0.327002968759958
-0.61061061061061 0.331091273090187
-0.59059059059059 0.33509635687531
-0.57057057057057 0.339013984172804
-0.55055055055055 0.34283997442106
-0.53053053053053 0.346570209826128
-0.51051051051051 0.350200642706842
-0.49049049049049 0.353727302780113
-0.47047047047047 0.357146304368113
-0.45045045045045 0.360453853509139
-0.43043043043043 0.363646254953996
-0.41041041041041 0.366719919029892
-0.39039039039039 0.369671368354051
-0.37037037037037 0.372497244379499
-0.35035035035035 0.375194313755802
-0.33033033033033 0.377759474487924
-0.31031031031031 0.38018976187679
-0.29029029029029 0.382482354225654
-0.27027027027027 0.384634578296894
-0.25025025025025 0.386643914504485
-0.23023023023023 0.388508001828027
-0.21021021021021 0.390224642434919
-0.19019019019019 0.391791805998011
-0.17017017017017 0.393207633696876
-0.15015015015015 0.394470441891644
-0.13013013013013 0.395578725459258
-0.11011011011011 0.396531160782876
-0.0900900900900901 0.397326608386124
-0.07007007007007 0.397964115204853
-0.05005005005005 0.398442916490068
-0.03003003003003 0.398762437336696
-0.01001001001001 0.398922293833933
0.01001001001001 0.398922293833933
0.03003003003003 0.398762437336696
0.05005005005005 0.398442916490068
0.07007007007007 0.397964115204853
0.0900900900900901 0.397326608386124
0.11011011011011 0.396531160782876
0.13013013013013 0.395578725459258
0.15015015015015 0.394470441891644
0.17017017017017 0.393207633696876
0.19019019019019 0.391791805998011
0.21021021021021 0.390224642434919
0.23023023023023 0.388508001828027
0.25025025025025 0.386643914504485
0.27027027027027 0.384634578296894
0.29029029029029 0.382482354225654
0.31031031031031 0.38018976187679
0.33033033033033 0.377759474487924
0.35035035035035 0.375194313755802
0.37037037037037 0.372497244379499
0.39039039039039 0.369671368354051
0.41041041041041 0.366719919029892
0.43043043043043 0.363646254953996
0.45045045045045 0.360453853509139
0.47047047047047 0.357146304368113
0.49049049049049 0.353727302780113
0.51051051051051 0.350200642706842
0.53053053053053 0.346570209826128
0.55055055055055 0.34283997442106
0.57057057057057 0.339013984172804
0.59059059059059 0.33509635687531
0.61061061061061 0.331091273090187
0.63063063063063 0.327002968759958
0.65065065065065 0.322835727797843
0.67067067067067 0.318593874672039
0.69069069069069 0.314281767002296
0.71071071071071 0.309903788186304
0.73073073073073 0.305464340073112
0.75075075075075 0.300967835700437
0.77077077077077 0.296418692112302
0.790790790790791 0.291821323272996
0.810810810810811 0.287180133092853
0.830830830830831 0.282499508580786
0.850850850850851 0.277783813137949
0.870870870870871 0.273037380006279
0.890890890890891 0.268264505884996
0.910910910910911 0.2634694447275
0.930930930930931 0.258656401730343
0.950950950950951 0.253829527525259
0.970970970970971 0.24899291258444
0.990990990990991 0.24415058184851
1.01101101101101 0.239306489585817
1.03103103103103 0.234464514490888
1.05105105105105 0.229628455029052
1.07107107107107 0.224802025033432
1.09109109109109 0.219988849559688
1.11111111111111 0.215192461003031
1.13113113113113 0.210416295481265
1.15115115115115 0.205663689486728
1.17117117117117 0.200937876809264
1.19119119119119 0.1962419857315
1.21121121121121 0.191579036496956
1.23123123123123 0.186951939050736
1.25125125125125 0.182363491051798
1.27127127127127 0.17781637615506
1.29129129129129 0.173313162560933
1.31131131131131 0.168856301829129
1.33133133133133 0.164448127952996
1.35135135135135 0.160090856689972
1.37137137137137 0.155786585143159
1.39139139139139 0.151537291588457
1.41141141141141 0.147344835541168
1.43143143143143 0.143210958055468
1.45145145145145 0.139137282249685
1.47147147147147 0.135125314049902
1.49149149149149 0.13117644314399
1.51151151151151 0.127291944137829
1.53153153153153 0.123472977905145
1.55155155155155 0.119720593122119
1.57157157157157 0.116035727977651
1.59159159159159 0.112419212049971
1.61161161161161 0.108871768340093
1.63163163163163 0.10539401545248
1.65165165165165 0.101986469913169
1.67167167167167 0.0986495486155338
1.69169169169169 0.0953835713838294
1.71171171171171 0.0921887636446459
1.73173173173173 0.0890652591964251
1.75175175175175 0.0860131030672496
1.77177177177177 0.083032254451193
1.79179179179179 0.0801225897136326
1.81181181181181 0.077283905456062
1.83183183183183 0.0745159216311036
1.85185185185185 0.0718182846986055
1.87187187187187 0.0691905708139176
1.89189189189189 0.0666322890396674
1.91191191191191 0.0641428845726061
1.93193193193193 0.0617217419773594
1.95195195195195 0.059368188419199
1.97197197197197 0.057081496888248
1.99199199199199 0.0548608894078376
2.01201201201201 0.0527055402200588
2.03203203203203 0.0506145789418748
2.05205205205205 0.0485870936855042
2.07207207207207 0.0466221341371236
2.09209209209209 0.0447187145882916
2.11211211211211 0.0428758169148479
2.13213213213213 0.041092393498395
2.15215215215215 0.0393673700858294
2.17217217217217 0.0376996485827434
2.19219219219219 0.0360881097768737
2.21221221221221 0.0345316159881233
2.23223223223223 0.033029013642035
2.25225225225225 0.0315791357639332
2.27227227227227 0.03018080439129
2.29229229229229 0.0288328329022028
2.31231231231231 0.0275340282581906
2.33233233233233 0.0262831931598317
2.35235235235235 0.02507912811407
2.37237237237237 0.0239206334123108
2.39239239239239 0.0228065110187136
2.41241241241241 0.0217355663683581
2.43243243243243 0.0207066100752275
2.45245245245245 0.0197184595501959
2.47247247247247 0.0187699405294511
2.49249249249249 0.0178598885140022
2.51251251251251 0.016987150121141
2.53253253253253 0.0161505843489182
2.55255255255255 0.0153490637548888
2.57257257257257 0.0145814755505493
2.59259259259259 0.0138467226130541
2.61261261261261 0.0131437244159435
2.63263263263263 0.0124714178807514
2.65265265265265 0.0118287581514852
2.67267267267267 0.0112147192940778
2.69269269269269 0.0106282949230103
2.71271271271271 0.0100684987573917
2.73273273273273 0.00953436510885491
2.75275275275275 0.00902494930369034
2.77277277277277 0.00853932804169479
2.79279279279279 0.00807659969425077
2.81281281281281 0.00763588454418634
2.83283283283283 0.00721632496998621
2.85285285285285 0.00681708557693871
2.87287287287287 0.00643735327780635
2.89289289289289 0.00607633732560524
2.91291291291291 0.00573326930106518
2.93293293293293 0.00540740305732384
2.95295295295295 0.00509801462438214
2.97297297297297 0.00480440207581539
2.99299299299299 0.00452588536019617
3.01301301301301 0.00426180609964127
3.03303303303303 0.00401152735784578
3.05305305305305 0.00377443337991421
3.07307307307307 0.00354992930624085
3.09309309309309 0.00333744086263051
3.11311311311311 0.00313641402878608
3.13313313313313 0.0029463146872226
3.15315315315315 0.00276662825459746
3.17317317317317 0.00259685929737496
3.19319319319319 0.00243653113367011
3.21321321321321 0.00228518542304203
3.23323323323323 0.00214238174593163
3.25325325325325 0.00200769717436225
3.27327327327327 0.00188072583544551
3.29329329329329 0.00176107846915771
3.31331331331331 0.00164838198177652
3.33333333333333 0.0015422789962911
3.35335335335335 0.00144242740102448
3.37337337337337 0.00134849989763212
3.39339339339339 0.00126018354956833
3.41341341341341 0.00117717933203981
3.43343343343343 0.00109920168439588
3.45345345345345 0.0010259780658361
3.47347347347347 0.000957248515249212
3.49349349349349 0.000892765215932441
3.51351351351351 0.000832292065877152
3.53353353353353 0.000775604254246008
3.55355355355355 0.000722487844607976
3.57357357357357 0.000672739365441019
3.59359359359359 0.000626165408357979
3.61361361361361 0.000582582234459159
3.63363363363363 0.000541815389165405
3.65365365365365 0.00050369932583812
3.67367367367367 0.000468077038447599
3.69369369369369 0.000434799703508357
3.71371371371371 0.000403726331459719
3.73373373373373 0.000374723427631836
3.75375375375375 0.000347664662901427
3.77377377377377 0.000322430554107926
3.79379379379379 0.000298908154269281
3.81381381381381 0.000276990752607344
3.83383383383383 0.000256577584365551
3.85385385385385 0.000237573550376443
3.87387387387387 0.000219888946313312
3.89389389389389 0.000203439201538965
3.91391391391391 0.00018814462744512
3.93393393393393 0.000173930175158222
3.95395395395395 0.000160725202471485
3.97397397397397 0.000148463249848567
3.99399399399399 0.000137081825331481
4.01401401401401 0.000126522198173995
4.03403403403403 0.000116729201011866
4.05405405405405 0.000107651040372646
4.07407407407407 9.92391153205018e-05
4.09409409409409 9.14478440253317e-05
4.11411411411411 8.42344980404937e-05
4.13413413413413 7.75590440694795e-05
4.15415415415415 7.13839929989176e-05
4.17417417417417 6.56742559732345e-05
4.19419419419419 6.03970072851107e-05
4.21421421421421 5.55215538554582e-05
4.23423423423423 5.10192110769697e-05
4.25425425425425 4.68631847962824e-05
4.27427427427427 4.3028459211397e-05
4.29429429429429 3.94916904631592e-05
4.31431431431431 3.62311057022691e-05
4.33433433433433 3.32264074164067e-05
4.35435435435435 3.04586828055866e-05
4.37437437437437 2.79103179977393e-05
4.39439439439439 2.55649169007242e-05
4.41441441441441 2.34072244914564e-05
4.43443443443443 2.14230543475548e-05
4.45445445445445 1.95992202318354e-05
4.47447447447447 1.79234715450684e-05
4.49449449449449 1.63844324676437e-05
4.51451451451451 1.49715446161227e-05
4.53453453453453 1.3675013046071e-05
4.55455455455455 1.24857554380297e-05
4.57457457457457 1.13953543089884e-05
4.59459459459459 1.03960120972233e-05
4.61461461461461 9.48050897386715e-06
4.63463463463463 8.64216324004121e-06
4.65465465465465 7.87479417380527e-06
4.67467467467467 7.17268719654363e-06
4.69469469469469 6.53056123369604e-06
4.71471471471471 5.94353814994721e-06
4.73473473473473 5.40711414409909e-06
4.75475475475475 4.91713299385613e-06
4.77477477477477 4.4697610456467e-06
4.79479479479479 4.06146384937932e-06
4.81481481481481 3.68898434268124e-06
4.83483483483483 3.34932249368796e-06
4.85485485485485 3.03971631583941e-06
4.87487487487487 2.75762417238901e-06
4.89489489489489 2.50070829244518e-06
4.91491491491491 2.26681942433715e-06
4.93493493493493 2.05398255592983e-06
4.95495495495495 1.86038363520377e-06
4.97497497497497 1.68435722796805e-06
4.99499499499499 1.5243750529858e-06
5.01501501501502 1.37903533806611e-06
5.03503503503504 1.24705294381396e-06
5.05505505505506 1.12725020473309e-06
5.07507507507508 1.01854844024876e-06
5.0950950950951 9.19960090959837e-07
5.11511511511512 8.30581438046261e-07
5.13513513513514 7.49585866251233e-07
5.15515515515516 6.76217633231267e-07
5.17517517517518 6.09786110324583e-07
5.1951951951952 5.4966046193278e-07
5.21521521521522 4.95264732746229e-07
5.23523523523524 4.46073313973501e-07
5.25525525525526 4.01606761563285e-07
5.27527527527528 3.61427941137511e-07
5.2952952952953 3.25138475990267e-07
5.31531531531532 2.92375476052561e-07
5.33533533533534 2.62808527181656e-07
5.35535535535536 2.36136921509202e-07
5.37537537537538 2.1208711087848e-07
5.3953953953954 1.90410366621162e-07
5.41541541541542 1.70880630071684e-07
5.43543543543544 1.5329253929596e-07
5.45545545545546 1.3745961852414e-07
5.47547547547548 1.23212617727566e-07
5.4954954954955 1.10397990671284e-07
5.51551551551552 9.8876500608364e-08
5.53553553553554 8.85219435638491e-08
5.55555555555556 7.92199798873018e-08
5.57557557557558 7.08670654362614e-08
5.5955955955956 6.33694743912418e-08
5.61561561561562 5.66424062986154e-08
5.63563563563564 5.06091704933412e-08
5.65565565565566 4.5200441571292e-08
5.67567567567568 4.03535800631662e-08
5.6956956956957 3.60120129107462e-08
5.71571571571572 3.2124668763613e-08
5.73573573573574 2.86454635022852e-08
5.75575575575576 2.55328317539416e-08
5.77577577577578 2.27493005011625e-08
5.7957957957958 2.02611011941336e-08
5.81581581581582 1.80378170640688e-08
5.83583583583584 1.60520626017053e-08
5.85585585585586 1.42791924110059e-08
5.87587587587588 1.2697036876002e-08
5.8958958958959 1.12856622892642e-08
5.91591591591592 1.00271532849868e-08
5.93593593593594 8.90541559921139e-09
5.95595595595596 7.90599734535251e-09
5.97597597597598 7.01592714588943e-09
5.995995995996 6.22356760178439e-09
6.01601601601602 5.5184827107339e-09
6.03603603603604 4.89131796456914e-09
6.05605605605606 4.33369196574694e-09
6.07607607607608 3.83809850362692e-09
6.0960960960961 3.3978181237675e-09
6.11611611611612 3.00683830841826e-09
6.13613613613614 2.65978146430924e-09
6.15615615615616 2.35183998527871e-09
6.17617617617618 2.07871772273778e-09
6.1961961961962 1.83657725691022e-09
6.21621621621622 1.6219924166386e-09
6.23623623623624 1.43190554571785e-09
6.25625625625626 1.26358905957511e-09
6.27627627627628 1.11461087800764e-09
6.2962962962963 9.82803357938119e-10
6.31631631631632 8.66235385045992e-10
6.33633633633634 7.63187314959515e-10
6.35635635635636 6.72128483698836e-10
6.37637637637638 5.91697033481532e-10
6.3963963963964 5.20681824054164e-10
6.41641641641642 4.58006221596989e-10
6.43643643643644 4.02713577147895e-10
6.45645645645646 3.53954224575805e-10
6.47647647647648 3.10973844559378e-10
6.4964964964965 2.73103055937435e-10
6.51651651651652 2.39748109325539e-10
6.53653653653654 2.10382570159619e-10
6.55655655655656 1.84539889444162e-10
6.57657657657658 1.61806770551276e-10
6.5965965965966 1.41817249531742e-10
6.61661661661662 1.24247414645767e-10
6.63663663663664 1.08810698278133e-10
6.65665665665666 9.52536811418137e-11
6.67667667667668 8.33523547614393e-11
6.6966966966967 7.29087937236022e-11
6.71671671671672 6.37481941394484e-11
6.73673673673674 5.57162392365998e-11
6.75675675675676 4.86767570277076e-11
6.77677677677678 4.25096386334907e-11
6.7967967967968 3.71089891068556e-11
6.81681681681682 3.23814855461048e-11
6.83683683683684 2.82449199306813e-11
6.85685685685686 2.46269064908966e-11
6.87687687687688 2.14637355595386e-11
6.8968968968969 1.86993577716892e-11
6.91691691691692 1.62844842008236e-11
6.93693693693694 1.41757895636779e-11
6.95695695695696 1.23352070109961e-11
6.97697697697698 1.07293042619725e-11
6.996996996997 9.32873195138548e-12
7.01701701701702 8.10773605306642e-12
7.03703703703704 7.04372713322415e-12
7.05705705705706 6.11689998287643e-12
7.07707707707708 5.30989788981512e-12
7.0970970970971 4.6075164458095e-12
7.11711711711712 3.99644235193917e-12
7.13713713713714 3.46502319108337e-12
7.15715715715716 3.00306458801651e-12
7.17717717717718 2.60165157997869e-12
7.1971971971972 2.25299137914217e-12
7.21721721721722 1.95027502769791e-12
7.23723723723724 1.68755573049414e-12
7.25725725725726 1.45964190299846e-12
7.27727727727728 1.262003197176e-12
7.2972972972973 1.0906879676822e-12
7.31731731731732 9.42250818252902e-13
7.33733733733734 8.13689025751829e-13
7.35735735735736 7.02386779168733e-13
7.37737737737738 6.06066294885245e-13
7.3973973973974 5.22744979473708e-13
7.41741741741742 4.50697908714381e-13
7.43743743743744 3.88424977793729e-13
7.45745745745746 3.34622154016963e-13
7.47747747747748 2.88156330935933e-13
7.4974974974975 2.48043342543511e-13
7.51751751751752 2.13428748996348e-13
7.53753753753754 1.83571051982055e-13
7.55755755755756 1.57827039041883e-13
7.57757757757758 1.35638992516663e-13
7.5975975975976 1.16523530854698e-13
7.61761761761762 1.000618782966e-13
7.63763763763764 8.58913838706873e-14
7.65765765765766 7.36981325813112e-14
7.67767767767768 6.32105109959248e-14
7.6976976976977 5.41936064405376e-14
7.71771771771772 4.64443339685915e-14
7.73773773773774 3.97871984155707e-14
7.75775775775776 3.40706104038536e-14
7.77777777777778 2.91636853079907e-14
7.7977977977978 2.4953463096687e-14
7.81781781781782 2.13424947819474e-14
7.83783783783784 1.82467480586654e-14
7.85785785785786 1.55937907247682e-14
7.87787787787788 1.33212157347804e-14
7.8978978978979 1.13752763482777e-14
7.91791791791792 9.70970386854701e-15
7.93793793793794 8.28468399583068e-15
7.95795795795796 7.06597090549298e-15
7.97797797797798 6.02412085866189e-15
7.997997997998 5.13382950919603e-15
8.01801801801802 4.3733591283238e-15
8.03803803803804 3.72404376402735e-15
8.05805805805806 3.16986191875719e-15
8.07807807807808 2.69706769497134e-15
8.0980980980981 2.29387254841619e-15
8.11811811811812 1.95017082606148e-15
8.13813813813814 1.65730316851105e-15
8.15815815815816 1.40785264250169e-15
8.17817817817818 1.19546915264237e-15
8.1981981981982 1.01471827585831e-15
8.21821821821822 8.60951178494166e-16
8.23823823823824 7.30192724674871e-16
8.25825825825826 6.19045274051723e-16
8.27827827827828 5.24606005103494e-16
8.2982982982983 4.44395893386326e-16
8.31831831831832 3.76298728354137e-16
8.33833833833834 3.18508772685576e-16
8.35835835835836 2.69485858889318e-16
8.37837837837838 2.27916883183252e-16
8.3983983983984 1.92682799625235e-16
8.41841841841842 1.62830341150177e-16
8.43843843843844 1.37547801096493e-16
8.45845845845846 1.16144301209627e-16
8.47847847847848 9.8032051926925e-17
8.4984984984985 8.27111796592674e-17
8.51851851851852 6.97567552529606e-17
8.53853853853854 5.88077091106193e-17
8.55855855855856 4.95573626747691e-17
8.57857857857858 4.17453440895294e-17
8.5985985985986 3.51506886838449e-17
8.61861861861862 2.95859531836935e-17
8.63863863863864 2.48921968838275e-17
8.65865865865866 2.09347039316773e-17
8.67867867867868 1.75993388643364e-17
8.6986986986987 1.47894429982973e-17
8.71871871871872 1.24231925503958e-17
8.73873873873874 1.04313507694241e-17
8.75875875875876 8.75535614211525e-18
8.77877877877878 7.34569713009337e-18
8.7987987987988 6.16053109048305e-18
8.81881881881882 5.16451119999019e-18
8.83883883883884 4.32779048512422e-18
8.85885885885886 3.62517658454128e-18
8.87887887887888 3.03541474067764e-18
8.8988988988989 2.54057982941549e-18
8.91891891891892 2.12556106807901e-18
8.93893893893894 1.77762546207239e-18
8.95895895895896 1.48604811781133e-18
8.97897897897898 1.24179931485728e-18
8.998998998999 1.03727973679138e-18
9.01901901901902 8.66096545670873e-19
9.03903903903904 7.22874080905507e-19
9.05905905905906 6.03093897535385e-19
9.07907907907908 5.0295965472299e-19
9.0990990990991 4.19283042962206e-19
9.11911911911912 3.49387515332417e-19
9.13913913913914 2.91027078875695e-19
9.15915915915916 2.42317819504618e-19
9.17917917917918 2.01680188581445e-19
9.1991991991992 1.67790380698338e-19
9.21921921921922 1.39539388139188e-19
9.23923923923924 1.1599853476858e-19
9.25925925925926 9.63904764362469e-20
9.27927927927928 8.00648113242436e-20
9.2992992992993 6.6477576193283e-20
9.31931931931932 5.51740167796855e-20
9.33933933933934 4.5774115701642e-20
9.35935935935936 3.79604417474457e-20
9.37937937937938 3.14679525475421e-20
9.3993993993994 2.60754402558043e-20
9.41941941941942 2.15983585814365e-20
9.43943943943944 1.78828106797973e-20
9.45945945945946 1.48005121824234e-20
9.47947947947948 1.22445730038445e-20
9.4994994994995 1.01259663374544e-20
9.51951951951952 8.37057415073758e-21
9.53953953953954 6.91671611023779e-21
9.55955955955956 5.71308371631411e-21
9.57957957957958 4.71701393695898e-21
9.5995995995996 3.89304716291232e-21
9.61961961961962 3.21172317125736e-21
9.63963963963964 2.64857624243904e-21
9.65965965965966 2.18329684678274e-21
9.67967967967968 1.79903258756111e-21
9.6996996996997 1.48180551599987e-21
9.71971971971972 1.22002665237565e-21
9.73973973973974 1.00409166885045e-21
9.75975975975976 8.26044308669654e-22
9.77977977977978 6.79296312742742e-22
9.7997997997998 5.58394465795474e-22
9.81981981981982 4.58826916995414e-22
9.83983983983984 3.7686222201397e-22
9.85985985985986 3.09415635142992e-22
9.87987987987988 2.53938085193762e-22
9.8998998998999 2.08324025950642e-22
9.91991991991992 1.70834984871876e-22
9.93993993993994 1.40036162642795e-22
9.95995995995996 1.14743877987917e-22
9.97997997997998 9.39820210218911e-23
10 7.69459862670642e-23
};
\addlegendentry{N(0,1)}
\end{axis}

\end{tikzpicture}

%% file: FinalFigs/Null_Dists_d_10_100_n_100_m_100_kernel__Dirichlet_RBF_2022_10_15_22_22_56cross.tex
\begin{tikzpicture}

\definecolor{darkorange25512714}{RGB}{255,127,14}
\definecolor{darkslategray38}{RGB}{38,38,38}
\definecolor{lightgray204}{RGB}{204,204,204}
\definecolor{steelblue31119180}{RGB}{31,119,180}

\begin{axis}[
axis line style={darkslategray38},
height=\figheight,
legend cell align={left},
legend style={fill opacity=0.8, draw opacity=1, text opacity=1, draw=none},
tick align=outside,
tick pos=left,
title={$\cmmd$~$(n/m=1)$},
width=\figwidth,
x grid style={lightgray204},
xmin=-6, xmax=6,
xtick style={color=darkslategray38},
y grid style={lightgray204},
ylabel=\textcolor{darkslategray38}{},
ymin=0, ymax=0.418868408525629,
ytick style={color=darkslategray38}, 
xticklabels=empty,
yticklabels=empty
]
\draw[draw=none,fill=steelblue31119180,fill opacity=0.8] (axis cs:-2.91104698181152,0) rectangle (axis cs:-2.65976428985596,0.0254693228180304);
\addlegendimage{ybar,ybar legend,draw=none,fill=steelblue31119180,fill opacity=0.8}
\addlegendentry{d=10}

\draw[draw=none,fill=steelblue31119180,fill opacity=0.8] (axis cs:-2.28284025192261,0) rectangle (axis cs:-2.03155755996704,0.0604896531714138);
\draw[draw=none,fill=steelblue31119180,fill opacity=0.8] (axis cs:-1.65463364124298,0) rectangle (axis cs:-1.40335094928741,0.162366932964944);
\draw[draw=none,fill=steelblue31119180,fill opacity=0.8] (axis cs:-1.02642691135406,0) rectangle (axis cs:-0.775144219398499,0.324733896740841);
\draw[draw=none,fill=steelblue31119180,fill opacity=0.8] (axis cs:-0.398220241069794,0) rectangle (axis cs:-0.146937549114227,0.327917562395163);
\draw[draw=none,fill=steelblue31119180,fill opacity=0.8] (axis cs:0.229986429214478,0) rectangle (axis cs:0.481269121170044,0.334284893703807);
\draw[draw=none,fill=steelblue31119180,fill opacity=0.8] (axis cs:0.858193099498749,0) rectangle (axis cs:1.10947585105896,0.238774924074148);
\draw[draw=none,fill=steelblue31119180,fill opacity=0.8] (axis cs:1.48639976978302,0) rectangle (axis cs:1.73768246173859,0.0859589808225354);
\draw[draw=none,fill=steelblue31119180,fill opacity=0.8] (axis cs:2.11460638046265,0) rectangle (axis cs:2.36588907241821,0.0286529881702841);
\draw[draw=none,fill=steelblue31119180,fill opacity=0.8] (axis cs:2.74281311035156,0) rectangle (axis cs:2.99409580230713,0.00318366535225379);
\draw[draw=none,fill=darkorange25512714,fill opacity=0.8] (axis cs:-2.65976428985596,0) rectangle (axis cs:-2.40848159790039,0.0254693228180304);
\addlegendimage{ybar,ybar legend,draw=none,fill=darkorange25512714,fill opacity=0.8}
\addlegendentry{d=100}

\draw[draw=none,fill=darkorange25512714,fill opacity=0.8] (axis cs:-2.03155755996704,0) rectangle (axis cs:-1.78027486801147,0.0541223212586334);
\draw[draw=none,fill=darkorange25512714,fill opacity=0.8] (axis cs:-1.40335094928741,0) rectangle (axis cs:-1.15206825733185,0.178285259726212);
\draw[draw=none,fill=darkorange25512714,fill opacity=0.8] (axis cs:-0.775144279003143,0) rectangle (axis cs:-0.523861587047577,0.302448237160587);
\draw[draw=none,fill=darkorange25512714,fill opacity=0.8] (axis cs:-0.146937608718872,0) rectangle (axis cs:0.104345083236694,0.378856212864314);
\draw[draw=none,fill=darkorange25512714,fill opacity=0.8] (axis cs:0.481269061565399,0) rectangle (axis cs:0.732551753520966,0.350203221975417);
\draw[draw=none,fill=darkorange25512714,fill opacity=0.8] (axis cs:1.10947573184967,0) rectangle (axis cs:1.36075842380524,0.197387270567962);
\draw[draw=none,fill=darkorange25512714,fill opacity=0.8] (axis cs:1.7376823425293,0) rectangle (axis cs:1.98896503448486,0.0827753148661452);
\draw[draw=none,fill=darkorange25512714,fill opacity=0.8] (axis cs:2.36588907241821,0) rectangle (axis cs:2.61717176437378,0.0127346614090152);
\draw[draw=none,fill=darkorange25512714,fill opacity=0.8] (axis cs:2.99409580230713,0) rectangle (axis cs:3.2453784942627,0.00955099605676138);
\addplot [semithick, black]
table {%
-10 7.69459862670642e-23
-9.97997997997998 9.39820210218911e-23
-9.95995995995996 1.14743877987917e-22
-9.93993993993994 1.40036162642795e-22
-9.91991991991992 1.70834984871876e-22
-9.8998998998999 2.08324025950642e-22
-9.87987987987988 2.53938085193762e-22
-9.85985985985986 3.09415635142992e-22
-9.83983983983984 3.7686222201397e-22
-9.81981981981982 4.58826916995414e-22
-9.7997997997998 5.58394465795474e-22
-9.77977977977978 6.79296312742742e-22
-9.75975975975976 8.26044308669654e-22
-9.73973973973974 1.00409166885045e-21
-9.71971971971972 1.22002665237565e-21
-9.6996996996997 1.48180551599987e-21
-9.67967967967968 1.79903258756111e-21
-9.65965965965966 2.18329684678274e-21
-9.63963963963964 2.64857624243904e-21
-9.61961961961962 3.21172317125736e-21
-9.5995995995996 3.89304716291232e-21
-9.57957957957958 4.71701393695898e-21
-9.55955955955956 5.71308371631411e-21
-9.53953953953954 6.91671611023779e-21
-9.51951951951952 8.37057415073758e-21
-9.4994994994995 1.01259663374544e-20
-9.47947947947948 1.22445730038445e-20
-9.45945945945946 1.48005121824234e-20
-9.43943943943944 1.78828106797973e-20
-9.41941941941942 2.15983585814365e-20
-9.3993993993994 2.60754402558043e-20
-9.37937937937938 3.14679525475421e-20
-9.35935935935936 3.79604417474457e-20
-9.33933933933934 4.5774115701642e-20
-9.31931931931932 5.51740167796855e-20
-9.2992992992993 6.6477576193283e-20
-9.27927927927928 8.00648113242436e-20
-9.25925925925926 9.63904764362469e-20
-9.23923923923924 1.1599853476858e-19
-9.21921921921922 1.39539388139188e-19
-9.1991991991992 1.67790380698338e-19
-9.17917917917918 2.01680188581445e-19
-9.15915915915916 2.42317819504618e-19
-9.13913913913914 2.91027078875695e-19
-9.11911911911912 3.49387515332417e-19
-9.0990990990991 4.19283042962206e-19
-9.07907907907908 5.0295965472299e-19
-9.05905905905906 6.03093897535385e-19
-9.03903903903904 7.22874080905507e-19
-9.01901901901902 8.66096545670873e-19
-8.998998998999 1.03727973679138e-18
-8.97897897897898 1.24179931485728e-18
-8.95895895895896 1.48604811781133e-18
-8.93893893893894 1.77762546207239e-18
-8.91891891891892 2.12556106807901e-18
-8.8988988988989 2.54057982941549e-18
-8.87887887887888 3.03541474067764e-18
-8.85885885885886 3.62517658454128e-18
-8.83883883883884 4.32779048512422e-18
-8.81881881881882 5.16451119999019e-18
-8.7987987987988 6.16053109048305e-18
-8.77877877877878 7.34569713009337e-18
-8.75875875875876 8.75535614211525e-18
-8.73873873873874 1.04313507694241e-17
-8.71871871871872 1.24231925503958e-17
-8.6986986986987 1.47894429982973e-17
-8.67867867867868 1.75993388643364e-17
-8.65865865865866 2.09347039316773e-17
-8.63863863863864 2.48921968838275e-17
-8.61861861861862 2.95859531836935e-17
-8.5985985985986 3.51506886838449e-17
-8.57857857857858 4.17453440895294e-17
-8.55855855855856 4.95573626747691e-17
-8.53853853853854 5.88077091106193e-17
-8.51851851851852 6.97567552529606e-17
-8.4984984984985 8.27111796592674e-17
-8.47847847847848 9.8032051926925e-17
-8.45845845845846 1.16144301209627e-16
-8.43843843843844 1.37547801096493e-16
-8.41841841841842 1.62830341150177e-16
-8.3983983983984 1.92682799625235e-16
-8.37837837837838 2.27916883183252e-16
-8.35835835835836 2.69485858889318e-16
-8.33833833833834 3.18508772685576e-16
-8.31831831831832 3.76298728354137e-16
-8.2982982982983 4.44395893386326e-16
-8.27827827827828 5.24606005103494e-16
-8.25825825825826 6.19045274051723e-16
-8.23823823823824 7.30192724674871e-16
-8.21821821821822 8.60951178494166e-16
-8.1981981981982 1.01471827585831e-15
-8.17817817817818 1.19546915264237e-15
-8.15815815815816 1.40785264250169e-15
-8.13813813813814 1.65730316851105e-15
-8.11811811811812 1.95017082606148e-15
-8.0980980980981 2.29387254841619e-15
-8.07807807807808 2.69706769497134e-15
-8.05805805805806 3.16986191875719e-15
-8.03803803803804 3.72404376402735e-15
-8.01801801801802 4.3733591283238e-15
-7.997997997998 5.13382950919607e-15
-7.97797797797798 6.02412085866193e-15
-7.95795795795796 7.06597090549303e-15
-7.93793793793794 8.28468399583074e-15
-7.91791791791792 9.70970386854708e-15
-7.8978978978979 1.13752763482777e-14
-7.87787787787788 1.33212157347805e-14
-7.85785785785786 1.55937907247683e-14
-7.83783783783784 1.82467480586655e-14
-7.81781781781782 2.13424947819475e-14
-7.7977977977978 2.49534630966872e-14
-7.77777777777778 2.91636853079909e-14
-7.75775775775776 3.40706104038538e-14
-7.73773773773774 3.9787198415571e-14
-7.71771771771772 4.64443339685918e-14
-7.6976976976977 5.4193606440538e-14
-7.67767767767768 6.32105109959252e-14
-7.65765765765766 7.36981325813117e-14
-7.63763763763764 8.58913838706879e-14
-7.61761761761762 1.00061878296601e-13
-7.5975975975976 1.16523530854699e-13
-7.57757757757758 1.35638992516664e-13
-7.55755755755756 1.57827039041884e-13
-7.53753753753754 1.83571051982057e-13
-7.51751751751752 2.13428748996349e-13
-7.4974974974975 2.48043342543513e-13
-7.47747747747748 2.88156330935935e-13
-7.45745745745746 3.34622154016965e-13
-7.43743743743744 3.88424977793732e-13
-7.41741741741742 4.50697908714384e-13
-7.3973973973974 5.22744979473711e-13
-7.37737737737738 6.06066294885249e-13
-7.35735735735736 7.02386779168738e-13
-7.33733733733734 8.13689025751835e-13
-7.31731731731732 9.42250818252909e-13
-7.2972972972973 1.09068796768221e-12
-7.27727727727728 1.262003197176e-12
-7.25725725725726 1.45964190299847e-12
-7.23723723723724 1.68755573049416e-12
-7.21721721721722 1.95027502769792e-12
-7.1971971971972 2.25299137914218e-12
-7.17717717717718 2.60165157997871e-12
-7.15715715715716 3.00306458801653e-12
-7.13713713713714 3.4650231910834e-12
-7.11711711711712 3.99644235193919e-12
-7.0970970970971 4.60751644580953e-12
-7.07707707707708 5.30989788981514e-12
-7.05705705705706 6.11689998287646e-12
-7.03703703703704 7.0437271332242e-12
-7.01701701701702 8.10773605306645e-12
-6.996996996997 9.32873195138555e-12
-6.97697697697698 1.07293042619726e-11
-6.95695695695696 1.23352070109962e-11
-6.93693693693694 1.41757895636779e-11
-6.91691691691692 1.62844842008237e-11
-6.8968968968969 1.86993577716893e-11
-6.87687687687688 2.14637355595386e-11
-6.85685685685686 2.46269064908967e-11
-6.83683683683684 2.82449199306815e-11
-6.81681681681682 3.2381485546105e-11
-6.7967967967968 3.71089891068559e-11
-6.77677677677678 4.25096386334913e-11
-6.75675675675676 4.86767570277083e-11
-6.73673673673674 5.57162392366004e-11
-6.71671671671672 6.37481941394491e-11
-6.6966966966967 7.29087937236032e-11
-6.67667667667668 8.33523547614402e-11
-6.65665665665666 9.52536811418151e-11
-6.63663663663664 1.08810698278135e-10
-6.61661661661662 1.24247414645768e-10
-6.5965965965966 1.41817249531744e-10
-6.57657657657658 1.61806770551278e-10
-6.55655655655656 1.84539889444164e-10
-6.53653653653654 2.10382570159622e-10
-6.51651651651652 2.39748109325542e-10
-6.4964964964965 2.73103055937438e-10
-6.47647647647648 3.10973844559381e-10
-6.45645645645646 3.53954224575809e-10
-6.43643643643644 4.027135771479e-10
-6.41641641641642 4.58006221596996e-10
-6.3963963963964 5.20681824054169e-10
-6.37637637637638 5.91697033481538e-10
-6.35635635635636 6.72128483698846e-10
-6.33633633633634 7.63187314959523e-10
-6.31631631631632 8.66235385046001e-10
-6.2962962962963 9.8280335793813e-10
-6.27627627627628 1.11461087800766e-09
-6.25625625625626 1.26358905957513e-09
-6.23623623623624 1.43190554571787e-09
-6.21621621621622 1.62199241663862e-09
-6.1961961961962 1.83657725691024e-09
-6.17617617617618 2.0787177227378e-09
-6.15615615615616 2.35183998527873e-09
-6.13613613613614 2.65978146430928e-09
-6.11611611611612 3.00683830841829e-09
-6.0960960960961 3.39781812376754e-09
-6.07607607607608 3.83809850362696e-09
-6.05605605605606 4.33369196574699e-09
-6.03603603603604 4.89131796456919e-09
-6.01601601601602 5.51848271073395e-09
-5.995995995996 6.22356760178439e-09
-5.97597597597598 7.01592714588943e-09
-5.95595595595596 7.90599734535251e-09
-5.93593593593594 8.90541559921139e-09
-5.91591591591592 1.00271532849868e-08
-5.8958958958959 1.12856622892642e-08
-5.87587587587588 1.2697036876002e-08
-5.85585585585586 1.42791924110059e-08
-5.83583583583584 1.60520626017053e-08
-5.81581581581582 1.80378170640688e-08
-5.7957957957958 2.02611011941336e-08
-5.77577577577578 2.27493005011625e-08
-5.75575575575576 2.55328317539416e-08
-5.73573573573574 2.86454635022852e-08
-5.71571571571572 3.2124668763613e-08
-5.6956956956957 3.60120129107462e-08
-5.67567567567568 4.03535800631662e-08
-5.65565565565566 4.5200441571292e-08
-5.63563563563564 5.06091704933412e-08
-5.61561561561562 5.66424062986154e-08
-5.5955955955956 6.33694743912418e-08
-5.57557557557558 7.08670654362614e-08
-5.55555555555556 7.92199798873018e-08
-5.53553553553554 8.85219435638491e-08
-5.51551551551552 9.8876500608364e-08
-5.4954954954955 1.10397990671284e-07
-5.47547547547548 1.23212617727566e-07
-5.45545545545546 1.3745961852414e-07
-5.43543543543544 1.5329253929596e-07
-5.41541541541542 1.70880630071684e-07
-5.3953953953954 1.90410366621162e-07
-5.37537537537538 2.1208711087848e-07
-5.35535535535536 2.36136921509202e-07
-5.33533533533534 2.62808527181656e-07
-5.31531531531532 2.92375476052561e-07
-5.2952952952953 3.25138475990267e-07
-5.27527527527528 3.61427941137511e-07
-5.25525525525526 4.01606761563285e-07
-5.23523523523524 4.46073313973501e-07
-5.21521521521522 4.95264732746229e-07
-5.1951951951952 5.4966046193278e-07
-5.17517517517518 6.09786110324583e-07
-5.15515515515516 6.76217633231267e-07
-5.13513513513514 7.49585866251233e-07
-5.11511511511512 8.30581438046261e-07
-5.0950950950951 9.19960090959837e-07
-5.07507507507508 1.01854844024876e-06
-5.05505505505506 1.12725020473309e-06
-5.03503503503504 1.24705294381396e-06
-5.01501501501502 1.37903533806611e-06
-4.99499499499499 1.5243750529858e-06
-4.97497497497497 1.68435722796805e-06
-4.95495495495495 1.86038363520377e-06
-4.93493493493493 2.05398255592983e-06
-4.91491491491491 2.26681942433715e-06
-4.89489489489489 2.50070829244518e-06
-4.87487487487487 2.75762417238901e-06
-4.85485485485485 3.03971631583941e-06
-4.83483483483483 3.34932249368796e-06
-4.81481481481481 3.68898434268124e-06
-4.79479479479479 4.06146384937932e-06
-4.77477477477477 4.4697610456467e-06
-4.75475475475475 4.91713299385613e-06
-4.73473473473473 5.40711414409909e-06
-4.71471471471471 5.94353814994721e-06
-4.69469469469469 6.53056123369604e-06
-4.67467467467467 7.17268719654363e-06
-4.65465465465465 7.87479417380527e-06
-4.63463463463463 8.64216324004121e-06
-4.61461461461461 9.48050897386715e-06
-4.59459459459459 1.03960120972233e-05
-4.57457457457457 1.13953543089884e-05
-4.55455455455455 1.24857554380297e-05
-4.53453453453453 1.3675013046071e-05
-4.51451451451451 1.49715446161227e-05
-4.49449449449449 1.63844324676437e-05
-4.47447447447447 1.79234715450684e-05
-4.45445445445445 1.95992202318354e-05
-4.43443443443443 2.14230543475548e-05
-4.41441441441441 2.34072244914564e-05
-4.39439439439439 2.55649169007242e-05
-4.37437437437437 2.79103179977393e-05
-4.35435435435435 3.04586828055866e-05
-4.33433433433433 3.32264074164067e-05
-4.31431431431431 3.62311057022691e-05
-4.29429429429429 3.94916904631592e-05
-4.27427427427427 4.3028459211397e-05
-4.25425425425425 4.68631847962824e-05
-4.23423423423423 5.10192110769697e-05
-4.21421421421421 5.55215538554582e-05
-4.19419419419419 6.03970072851107e-05
-4.17417417417417 6.56742559732345e-05
-4.15415415415415 7.13839929989176e-05
-4.13413413413413 7.75590440694795e-05
-4.11411411411411 8.42344980404937e-05
-4.09409409409409 9.14478440253317e-05
-4.07407407407407 9.92391153205018e-05
-4.05405405405405 0.000107651040372646
-4.03403403403403 0.000116729201011866
-4.01401401401401 0.000126522198173995
-3.99399399399399 0.000137081825331481
-3.97397397397397 0.000148463249848567
-3.95395395395395 0.000160725202471485
-3.93393393393393 0.000173930175158222
-3.91391391391391 0.00018814462744512
-3.89389389389389 0.000203439201538965
-3.87387387387387 0.000219888946313312
-3.85385385385385 0.000237573550376443
-3.83383383383383 0.000256577584365551
-3.81381381381381 0.000276990752607344
-3.79379379379379 0.000298908154269281
-3.77377377377377 0.000322430554107926
-3.75375375375375 0.000347664662901427
-3.73373373373373 0.000374723427631836
-3.71371371371371 0.000403726331459719
-3.69369369369369 0.000434799703508357
-3.67367367367367 0.000468077038447599
-3.65365365365365 0.00050369932583812
-3.63363363363363 0.000541815389165405
-3.61361361361361 0.000582582234459159
-3.59359359359359 0.000626165408357979
-3.57357357357357 0.000672739365441021
-3.55355355355355 0.000722487844607978
-3.53353353353353 0.00077560425424601
-3.51351351351351 0.000832292065877155
-3.49349349349349 0.000892765215932443
-3.47347347347347 0.000957248515249216
-3.45345345345345 0.0010259780658361
-3.43343343343343 0.00109920168439588
-3.41341341341341 0.00117717933203981
-3.39339339339339 0.00126018354956833
-3.37337337337337 0.00134849989763212
-3.35335335335335 0.00144242740102448
-3.33333333333333 0.00154227899629111
-3.31331331331331 0.00164838198177652
-3.29329329329329 0.00176107846915772
-3.27327327327327 0.00188072583544552
-3.25325325325325 0.00200769717436226
-3.23323323323323 0.00214238174593163
-3.21321321321321 0.00228518542304204
-3.19319319319319 0.00243653113367012
-3.17317317317317 0.00259685929737497
-3.15315315315315 0.00276662825459747
-3.13313313313313 0.00294631468722261
-3.11311311311311 0.00313641402878609
-3.09309309309309 0.00333744086263052
-3.07307307307307 0.00354992930624086
-3.05305305305305 0.00377443337991422
-3.03303303303303 0.00401152735784579
-3.01301301301301 0.00426180609964128
-2.99299299299299 0.00452588536019618
-2.97297297297297 0.0048044020758154
-2.95295295295295 0.00509801462438215
-2.93293293293293 0.00540740305732385
-2.91291291291291 0.00573326930106519
-2.89289289289289 0.00607633732560526
-2.87287287287287 0.00643735327780636
-2.85285285285285 0.00681708557693873
-2.83283283283283 0.00721632496998623
-2.81281281281281 0.00763588454418632
-2.79279279279279 0.00807659969425075
-2.77277277277277 0.00853932804169477
-2.75275275275275 0.00902494930369032
-2.73273273273273 0.00953436510885489
-2.71271271271271 0.0100684987573917
-2.69269269269269 0.0106282949230102
-2.67267267267267 0.0112147192940778
-2.65265265265265 0.0118287581514852
-2.63263263263263 0.0124714178807513
-2.61261261261261 0.0131437244159435
-2.59259259259259 0.0138467226130541
-2.57257257257257 0.0145814755505492
-2.55255255255255 0.0153490637548887
-2.53253253253253 0.0161505843489182
-2.51251251251251 0.0169871501211409
-2.49249249249249 0.0178598885140022
-2.47247247247247 0.018769940529451
-2.45245245245245 0.0197184595501959
-2.43243243243243 0.0207066100752274
-2.41241241241241 0.0217355663683581
-2.39239239239239 0.0228065110187135
-2.37237237237237 0.0239206334123108
-2.35235235235235 0.0250791281140699
-2.33233233233233 0.0262831931598317
-2.31231231231231 0.0275340282581906
-2.29229229229229 0.0288328329022027
-2.27227227227227 0.0301808043912899
-2.25225225225225 0.0315791357639331
-2.23223223223223 0.033029013642035
-2.21221221221221 0.0345316159881232
-2.19219219219219 0.0360881097768736
-2.17217217217217 0.0376996485827434
-2.15215215215215 0.0393673700858293
-2.13213213213213 0.0410923934983949
-2.11211211211211 0.0428758169148479
-2.09209209209209 0.0447187145882915
-2.07207207207207 0.0466221341371235
-2.05205205205205 0.0485870936855041
-2.03203203203203 0.0506145789418747
-2.01201201201201 0.0527055402200587
-1.99199199199199 0.0548608894078376
-1.97197197197197 0.057081496888248
-1.95195195195195 0.059368188419199
-1.93193193193193 0.0617217419773594
-1.91191191191191 0.0641428845726061
-1.89189189189189 0.0666322890396674
-1.87187187187187 0.0691905708139176
-1.85185185185185 0.0718182846986055
-1.83183183183183 0.0745159216311036
-1.81181181181181 0.077283905456062
-1.79179179179179 0.0801225897136326
-1.77177177177177 0.083032254451193
-1.75175175175175 0.0860131030672496
-1.73173173173173 0.0890652591964251
-1.71171171171171 0.0921887636446459
-1.69169169169169 0.0953835713838294
-1.67167167167167 0.0986495486155338
-1.65165165165165 0.101986469913169
-1.63163163163163 0.10539401545248
-1.61161161161161 0.108871768340093
-1.59159159159159 0.112419212049971
-1.57157157157157 0.116035727977651
-1.55155155155155 0.119720593122119
-1.53153153153153 0.123472977905145
-1.51151151151151 0.127291944137829
-1.49149149149149 0.13117644314399
-1.47147147147147 0.135125314049902
-1.45145145145145 0.139137282249685
-1.43143143143143 0.143210958055468
-1.41141141141141 0.147344835541168
-1.39139139139139 0.151537291588457
-1.37137137137137 0.155786585143159
-1.35135135135135 0.160090856689972
-1.33133133133133 0.164448127952996
-1.31131131131131 0.168856301829129
-1.29129129129129 0.173313162560933
-1.27127127127127 0.17781637615506
-1.25125125125125 0.182363491051798
-1.23123123123123 0.186951939050736
-1.21121121121121 0.191579036496956
-1.19119119119119 0.1962419857315
-1.17117117117117 0.200937876809264
-1.15115115115115 0.205663689486728
-1.13113113113113 0.210416295481265
-1.11111111111111 0.215192461003031
-1.09109109109109 0.219988849559688
-1.07107107107107 0.224802025033432
-1.05105105105105 0.229628455029052
-1.03103103103103 0.234464514490888
-1.01101101101101 0.239306489585817
-0.990990990990991 0.24415058184851
-0.970970970970971 0.24899291258444
-0.950950950950951 0.253829527525259
-0.930930930930931 0.258656401730343
-0.910910910910911 0.2634694447275
-0.890890890890891 0.268264505884996
-0.870870870870871 0.273037380006279
-0.850850850850851 0.277783813137949
-0.830830830830831 0.282499508580786
-0.810810810810811 0.287180133092853
-0.790790790790791 0.291821323272996
-0.77077077077077 0.296418692112302
-0.75075075075075 0.300967835700437
-0.73073073073073 0.305464340073112
-0.71071071071071 0.309903788186304
-0.69069069069069 0.314281767002296
-0.67067067067067 0.318593874672039
-0.65065065065065 0.322835727797843
-0.63063063063063 0.327002968759958
-0.61061061061061 0.331091273090187
-0.59059059059059 0.33509635687531
-0.57057057057057 0.339013984172804
-0.55055055055055 0.34283997442106
-0.53053053053053 0.346570209826128
-0.51051051051051 0.350200642706842
-0.49049049049049 0.353727302780113
-0.47047047047047 0.357146304368113
-0.45045045045045 0.360453853509139
-0.43043043043043 0.363646254953996
-0.41041041041041 0.366719919029892
-0.39039039039039 0.369671368354051
-0.37037037037037 0.372497244379499
-0.35035035035035 0.375194313755802
-0.33033033033033 0.377759474487924
-0.31031031031031 0.38018976187679
-0.29029029029029 0.382482354225654
-0.27027027027027 0.384634578296894
-0.25025025025025 0.386643914504485
-0.23023023023023 0.388508001828027
-0.21021021021021 0.390224642434919
-0.19019019019019 0.391791805998011
-0.17017017017017 0.393207633696876
-0.15015015015015 0.394470441891644
-0.13013013013013 0.395578725459258
-0.11011011011011 0.396531160782876
-0.0900900900900901 0.397326608386124
-0.07007007007007 0.397964115204853
-0.05005005005005 0.398442916490068
-0.03003003003003 0.398762437336696
-0.01001001001001 0.398922293833933
0.01001001001001 0.398922293833933
0.03003003003003 0.398762437336696
0.05005005005005 0.398442916490068
0.07007007007007 0.397964115204853
0.0900900900900901 0.397326608386124
0.11011011011011 0.396531160782876
0.13013013013013 0.395578725459258
0.15015015015015 0.394470441891644
0.17017017017017 0.393207633696876
0.19019019019019 0.391791805998011
0.21021021021021 0.390224642434919
0.23023023023023 0.388508001828027
0.25025025025025 0.386643914504485
0.27027027027027 0.384634578296894
0.29029029029029 0.382482354225654
0.31031031031031 0.38018976187679
0.33033033033033 0.377759474487924
0.35035035035035 0.375194313755802
0.37037037037037 0.372497244379499
0.39039039039039 0.369671368354051
0.41041041041041 0.366719919029892
0.43043043043043 0.363646254953996
0.45045045045045 0.360453853509139
0.47047047047047 0.357146304368113
0.49049049049049 0.353727302780113
0.51051051051051 0.350200642706842
0.53053053053053 0.346570209826128
0.55055055055055 0.34283997442106
0.57057057057057 0.339013984172804
0.59059059059059 0.33509635687531
0.61061061061061 0.331091273090187
0.63063063063063 0.327002968759958
0.65065065065065 0.322835727797843
0.67067067067067 0.318593874672039
0.69069069069069 0.314281767002296
0.71071071071071 0.309903788186304
0.73073073073073 0.305464340073112
0.75075075075075 0.300967835700437
0.77077077077077 0.296418692112302
0.790790790790791 0.291821323272996
0.810810810810811 0.287180133092853
0.830830830830831 0.282499508580786
0.850850850850851 0.277783813137949
0.870870870870871 0.273037380006279
0.890890890890891 0.268264505884996
0.910910910910911 0.2634694447275
0.930930930930931 0.258656401730343
0.950950950950951 0.253829527525259
0.970970970970971 0.24899291258444
0.990990990990991 0.24415058184851
1.01101101101101 0.239306489585817
1.03103103103103 0.234464514490888
1.05105105105105 0.229628455029052
1.07107107107107 0.224802025033432
1.09109109109109 0.219988849559688
1.11111111111111 0.215192461003031
1.13113113113113 0.210416295481265
1.15115115115115 0.205663689486728
1.17117117117117 0.200937876809264
1.19119119119119 0.1962419857315
1.21121121121121 0.191579036496956
1.23123123123123 0.186951939050736
1.25125125125125 0.182363491051798
1.27127127127127 0.17781637615506
1.29129129129129 0.173313162560933
1.31131131131131 0.168856301829129
1.33133133133133 0.164448127952996
1.35135135135135 0.160090856689972
1.37137137137137 0.155786585143159
1.39139139139139 0.151537291588457
1.41141141141141 0.147344835541168
1.43143143143143 0.143210958055468
1.45145145145145 0.139137282249685
1.47147147147147 0.135125314049902
1.49149149149149 0.13117644314399
1.51151151151151 0.127291944137829
1.53153153153153 0.123472977905145
1.55155155155155 0.119720593122119
1.57157157157157 0.116035727977651
1.59159159159159 0.112419212049971
1.61161161161161 0.108871768340093
1.63163163163163 0.10539401545248
1.65165165165165 0.101986469913169
1.67167167167167 0.0986495486155338
1.69169169169169 0.0953835713838294
1.71171171171171 0.0921887636446459
1.73173173173173 0.0890652591964251
1.75175175175175 0.0860131030672496
1.77177177177177 0.083032254451193
1.79179179179179 0.0801225897136326
1.81181181181181 0.077283905456062
1.83183183183183 0.0745159216311036
1.85185185185185 0.0718182846986055
1.87187187187187 0.0691905708139176
1.89189189189189 0.0666322890396674
1.91191191191191 0.0641428845726061
1.93193193193193 0.0617217419773594
1.95195195195195 0.059368188419199
1.97197197197197 0.057081496888248
1.99199199199199 0.0548608894078376
2.01201201201201 0.0527055402200588
2.03203203203203 0.0506145789418748
2.05205205205205 0.0485870936855042
2.07207207207207 0.0466221341371236
2.09209209209209 0.0447187145882916
2.11211211211211 0.0428758169148479
2.13213213213213 0.041092393498395
2.15215215215215 0.0393673700858294
2.17217217217217 0.0376996485827434
2.19219219219219 0.0360881097768737
2.21221221221221 0.0345316159881233
2.23223223223223 0.033029013642035
2.25225225225225 0.0315791357639332
2.27227227227227 0.03018080439129
2.29229229229229 0.0288328329022028
2.31231231231231 0.0275340282581906
2.33233233233233 0.0262831931598317
2.35235235235235 0.02507912811407
2.37237237237237 0.0239206334123108
2.39239239239239 0.0228065110187136
2.41241241241241 0.0217355663683581
2.43243243243243 0.0207066100752275
2.45245245245245 0.0197184595501959
2.47247247247247 0.0187699405294511
2.49249249249249 0.0178598885140022
2.51251251251251 0.016987150121141
2.53253253253253 0.0161505843489182
2.55255255255255 0.0153490637548888
2.57257257257257 0.0145814755505493
2.59259259259259 0.0138467226130541
2.61261261261261 0.0131437244159435
2.63263263263263 0.0124714178807514
2.65265265265265 0.0118287581514852
2.67267267267267 0.0112147192940778
2.69269269269269 0.0106282949230103
2.71271271271271 0.0100684987573917
2.73273273273273 0.00953436510885491
2.75275275275275 0.00902494930369034
2.77277277277277 0.00853932804169479
2.79279279279279 0.00807659969425077
2.81281281281281 0.00763588454418634
2.83283283283283 0.00721632496998621
2.85285285285285 0.00681708557693871
2.87287287287287 0.00643735327780635
2.89289289289289 0.00607633732560524
2.91291291291291 0.00573326930106518
2.93293293293293 0.00540740305732384
2.95295295295295 0.00509801462438214
2.97297297297297 0.00480440207581539
2.99299299299299 0.00452588536019617
3.01301301301301 0.00426180609964127
3.03303303303303 0.00401152735784578
3.05305305305305 0.00377443337991421
3.07307307307307 0.00354992930624085
3.09309309309309 0.00333744086263051
3.11311311311311 0.00313641402878608
3.13313313313313 0.0029463146872226
3.15315315315315 0.00276662825459746
3.17317317317317 0.00259685929737496
3.19319319319319 0.00243653113367011
3.21321321321321 0.00228518542304203
3.23323323323323 0.00214238174593163
3.25325325325325 0.00200769717436225
3.27327327327327 0.00188072583544551
3.29329329329329 0.00176107846915771
3.31331331331331 0.00164838198177652
3.33333333333333 0.0015422789962911
3.35335335335335 0.00144242740102448
3.37337337337337 0.00134849989763212
3.39339339339339 0.00126018354956833
3.41341341341341 0.00117717933203981
3.43343343343343 0.00109920168439588
3.45345345345345 0.0010259780658361
3.47347347347347 0.000957248515249212
3.49349349349349 0.000892765215932441
3.51351351351351 0.000832292065877152
3.53353353353353 0.000775604254246008
3.55355355355355 0.000722487844607976
3.57357357357357 0.000672739365441019
3.59359359359359 0.000626165408357979
3.61361361361361 0.000582582234459159
3.63363363363363 0.000541815389165405
3.65365365365365 0.00050369932583812
3.67367367367367 0.000468077038447599
3.69369369369369 0.000434799703508357
3.71371371371371 0.000403726331459719
3.73373373373373 0.000374723427631836
3.75375375375375 0.000347664662901427
3.77377377377377 0.000322430554107926
3.79379379379379 0.000298908154269281
3.81381381381381 0.000276990752607344
3.83383383383383 0.000256577584365551
3.85385385385385 0.000237573550376443
3.87387387387387 0.000219888946313312
3.89389389389389 0.000203439201538965
3.91391391391391 0.00018814462744512
3.93393393393393 0.000173930175158222
3.95395395395395 0.000160725202471485
3.97397397397397 0.000148463249848567
3.99399399399399 0.000137081825331481
4.01401401401401 0.000126522198173995
4.03403403403403 0.000116729201011866
4.05405405405405 0.000107651040372646
4.07407407407407 9.92391153205018e-05
4.09409409409409 9.14478440253317e-05
4.11411411411411 8.42344980404937e-05
4.13413413413413 7.75590440694795e-05
4.15415415415415 7.13839929989176e-05
4.17417417417417 6.56742559732345e-05
4.19419419419419 6.03970072851107e-05
4.21421421421421 5.55215538554582e-05
4.23423423423423 5.10192110769697e-05
4.25425425425425 4.68631847962824e-05
4.27427427427427 4.3028459211397e-05
4.29429429429429 3.94916904631592e-05
4.31431431431431 3.62311057022691e-05
4.33433433433433 3.32264074164067e-05
4.35435435435435 3.04586828055866e-05
4.37437437437437 2.79103179977393e-05
4.39439439439439 2.55649169007242e-05
4.41441441441441 2.34072244914564e-05
4.43443443443443 2.14230543475548e-05
4.45445445445445 1.95992202318354e-05
4.47447447447447 1.79234715450684e-05
4.49449449449449 1.63844324676437e-05
4.51451451451451 1.49715446161227e-05
4.53453453453453 1.3675013046071e-05
4.55455455455455 1.24857554380297e-05
4.57457457457457 1.13953543089884e-05
4.59459459459459 1.03960120972233e-05
4.61461461461461 9.48050897386715e-06
4.63463463463463 8.64216324004121e-06
4.65465465465465 7.87479417380527e-06
4.67467467467467 7.17268719654363e-06
4.69469469469469 6.53056123369604e-06
4.71471471471471 5.94353814994721e-06
4.73473473473473 5.40711414409909e-06
4.75475475475475 4.91713299385613e-06
4.77477477477477 4.4697610456467e-06
4.79479479479479 4.06146384937932e-06
4.81481481481481 3.68898434268124e-06
4.83483483483483 3.34932249368796e-06
4.85485485485485 3.03971631583941e-06
4.87487487487487 2.75762417238901e-06
4.89489489489489 2.50070829244518e-06
4.91491491491491 2.26681942433715e-06
4.93493493493493 2.05398255592983e-06
4.95495495495495 1.86038363520377e-06
4.97497497497497 1.68435722796805e-06
4.99499499499499 1.5243750529858e-06
5.01501501501502 1.37903533806611e-06
5.03503503503504 1.24705294381396e-06
5.05505505505506 1.12725020473309e-06
5.07507507507508 1.01854844024876e-06
5.0950950950951 9.19960090959837e-07
5.11511511511512 8.30581438046261e-07
5.13513513513514 7.49585866251233e-07
5.15515515515516 6.76217633231267e-07
5.17517517517518 6.09786110324583e-07
5.1951951951952 5.4966046193278e-07
5.21521521521522 4.95264732746229e-07
5.23523523523524 4.46073313973501e-07
5.25525525525526 4.01606761563285e-07
5.27527527527528 3.61427941137511e-07
5.2952952952953 3.25138475990267e-07
5.31531531531532 2.92375476052561e-07
5.33533533533534 2.62808527181656e-07
5.35535535535536 2.36136921509202e-07
5.37537537537538 2.1208711087848e-07
5.3953953953954 1.90410366621162e-07
5.41541541541542 1.70880630071684e-07
5.43543543543544 1.5329253929596e-07
5.45545545545546 1.3745961852414e-07
5.47547547547548 1.23212617727566e-07
5.4954954954955 1.10397990671284e-07
5.51551551551552 9.8876500608364e-08
5.53553553553554 8.85219435638491e-08
5.55555555555556 7.92199798873018e-08
5.57557557557558 7.08670654362614e-08
5.5955955955956 6.33694743912418e-08
5.61561561561562 5.66424062986154e-08
5.63563563563564 5.06091704933412e-08
5.65565565565566 4.5200441571292e-08
5.67567567567568 4.03535800631662e-08
5.6956956956957 3.60120129107462e-08
5.71571571571572 3.2124668763613e-08
5.73573573573574 2.86454635022852e-08
5.75575575575576 2.55328317539416e-08
5.77577577577578 2.27493005011625e-08
5.7957957957958 2.02611011941336e-08
5.81581581581582 1.80378170640688e-08
5.83583583583584 1.60520626017053e-08
5.85585585585586 1.42791924110059e-08
5.87587587587588 1.2697036876002e-08
5.8958958958959 1.12856622892642e-08
5.91591591591592 1.00271532849868e-08
5.93593593593594 8.90541559921139e-09
5.95595595595596 7.90599734535251e-09
5.97597597597598 7.01592714588943e-09
5.995995995996 6.22356760178439e-09
6.01601601601602 5.5184827107339e-09
6.03603603603604 4.89131796456914e-09
6.05605605605606 4.33369196574694e-09
6.07607607607608 3.83809850362692e-09
6.0960960960961 3.3978181237675e-09
6.11611611611612 3.00683830841826e-09
6.13613613613614 2.65978146430924e-09
6.15615615615616 2.35183998527871e-09
6.17617617617618 2.07871772273778e-09
6.1961961961962 1.83657725691022e-09
6.21621621621622 1.6219924166386e-09
6.23623623623624 1.43190554571785e-09
6.25625625625626 1.26358905957511e-09
6.27627627627628 1.11461087800764e-09
6.2962962962963 9.82803357938119e-10
6.31631631631632 8.66235385045992e-10
6.33633633633634 7.63187314959515e-10
6.35635635635636 6.72128483698836e-10
6.37637637637638 5.91697033481532e-10
6.3963963963964 5.20681824054164e-10
6.41641641641642 4.58006221596989e-10
6.43643643643644 4.02713577147895e-10
6.45645645645646 3.53954224575805e-10
6.47647647647648 3.10973844559378e-10
6.4964964964965 2.73103055937435e-10
6.51651651651652 2.39748109325539e-10
6.53653653653654 2.10382570159619e-10
6.55655655655656 1.84539889444162e-10
6.57657657657658 1.61806770551276e-10
6.5965965965966 1.41817249531742e-10
6.61661661661662 1.24247414645767e-10
6.63663663663664 1.08810698278133e-10
6.65665665665666 9.52536811418137e-11
6.67667667667668 8.33523547614393e-11
6.6966966966967 7.29087937236022e-11
6.71671671671672 6.37481941394484e-11
6.73673673673674 5.57162392365998e-11
6.75675675675676 4.86767570277076e-11
6.77677677677678 4.25096386334907e-11
6.7967967967968 3.71089891068556e-11
6.81681681681682 3.23814855461048e-11
6.83683683683684 2.82449199306813e-11
6.85685685685686 2.46269064908966e-11
6.87687687687688 2.14637355595386e-11
6.8968968968969 1.86993577716892e-11
6.91691691691692 1.62844842008236e-11
6.93693693693694 1.41757895636779e-11
6.95695695695696 1.23352070109961e-11
6.97697697697698 1.07293042619725e-11
6.996996996997 9.32873195138548e-12
7.01701701701702 8.10773605306642e-12
7.03703703703704 7.04372713322415e-12
7.05705705705706 6.11689998287643e-12
7.07707707707708 5.30989788981512e-12
7.0970970970971 4.6075164458095e-12
7.11711711711712 3.99644235193917e-12
7.13713713713714 3.46502319108337e-12
7.15715715715716 3.00306458801651e-12
7.17717717717718 2.60165157997869e-12
7.1971971971972 2.25299137914217e-12
7.21721721721722 1.95027502769791e-12
7.23723723723724 1.68755573049414e-12
7.25725725725726 1.45964190299846e-12
7.27727727727728 1.262003197176e-12
7.2972972972973 1.0906879676822e-12
7.31731731731732 9.42250818252902e-13
7.33733733733734 8.13689025751829e-13
7.35735735735736 7.02386779168733e-13
7.37737737737738 6.06066294885245e-13
7.3973973973974 5.22744979473708e-13
7.41741741741742 4.50697908714381e-13
7.43743743743744 3.88424977793729e-13
7.45745745745746 3.34622154016963e-13
7.47747747747748 2.88156330935933e-13
7.4974974974975 2.48043342543511e-13
7.51751751751752 2.13428748996348e-13
7.53753753753754 1.83571051982055e-13
7.55755755755756 1.57827039041883e-13
7.57757757757758 1.35638992516663e-13
7.5975975975976 1.16523530854698e-13
7.61761761761762 1.000618782966e-13
7.63763763763764 8.58913838706873e-14
7.65765765765766 7.36981325813112e-14
7.67767767767768 6.32105109959248e-14
7.6976976976977 5.41936064405376e-14
7.71771771771772 4.64443339685915e-14
7.73773773773774 3.97871984155707e-14
7.75775775775776 3.40706104038536e-14
7.77777777777778 2.91636853079907e-14
7.7977977977978 2.4953463096687e-14
7.81781781781782 2.13424947819474e-14
7.83783783783784 1.82467480586654e-14
7.85785785785786 1.55937907247682e-14
7.87787787787788 1.33212157347804e-14
7.8978978978979 1.13752763482777e-14
7.91791791791792 9.70970386854701e-15
7.93793793793794 8.28468399583068e-15
7.95795795795796 7.06597090549298e-15
7.97797797797798 6.02412085866189e-15
7.997997997998 5.13382950919603e-15
8.01801801801802 4.3733591283238e-15
8.03803803803804 3.72404376402735e-15
8.05805805805806 3.16986191875719e-15
8.07807807807808 2.69706769497134e-15
8.0980980980981 2.29387254841619e-15
8.11811811811812 1.95017082606148e-15
8.13813813813814 1.65730316851105e-15
8.15815815815816 1.40785264250169e-15
8.17817817817818 1.19546915264237e-15
8.1981981981982 1.01471827585831e-15
8.21821821821822 8.60951178494166e-16
8.23823823823824 7.30192724674871e-16
8.25825825825826 6.19045274051723e-16
8.27827827827828 5.24606005103494e-16
8.2982982982983 4.44395893386326e-16
8.31831831831832 3.76298728354137e-16
8.33833833833834 3.18508772685576e-16
8.35835835835836 2.69485858889318e-16
8.37837837837838 2.27916883183252e-16
8.3983983983984 1.92682799625235e-16
8.41841841841842 1.62830341150177e-16
8.43843843843844 1.37547801096493e-16
8.45845845845846 1.16144301209627e-16
8.47847847847848 9.8032051926925e-17
8.4984984984985 8.27111796592674e-17
8.51851851851852 6.97567552529606e-17
8.53853853853854 5.88077091106193e-17
8.55855855855856 4.95573626747691e-17
8.57857857857858 4.17453440895294e-17
8.5985985985986 3.51506886838449e-17
8.61861861861862 2.95859531836935e-17
8.63863863863864 2.48921968838275e-17
8.65865865865866 2.09347039316773e-17
8.67867867867868 1.75993388643364e-17
8.6986986986987 1.47894429982973e-17
8.71871871871872 1.24231925503958e-17
8.73873873873874 1.04313507694241e-17
8.75875875875876 8.75535614211525e-18
8.77877877877878 7.34569713009337e-18
8.7987987987988 6.16053109048305e-18
8.81881881881882 5.16451119999019e-18
8.83883883883884 4.32779048512422e-18
8.85885885885886 3.62517658454128e-18
8.87887887887888 3.03541474067764e-18
8.8988988988989 2.54057982941549e-18
8.91891891891892 2.12556106807901e-18
8.93893893893894 1.77762546207239e-18
8.95895895895896 1.48604811781133e-18
8.97897897897898 1.24179931485728e-18
8.998998998999 1.03727973679138e-18
9.01901901901902 8.66096545670873e-19
9.03903903903904 7.22874080905507e-19
9.05905905905906 6.03093897535385e-19
9.07907907907908 5.0295965472299e-19
9.0990990990991 4.19283042962206e-19
9.11911911911912 3.49387515332417e-19
9.13913913913914 2.91027078875695e-19
9.15915915915916 2.42317819504618e-19
9.17917917917918 2.01680188581445e-19
9.1991991991992 1.67790380698338e-19
9.21921921921922 1.39539388139188e-19
9.23923923923924 1.1599853476858e-19
9.25925925925926 9.63904764362469e-20
9.27927927927928 8.00648113242436e-20
9.2992992992993 6.6477576193283e-20
9.31931931931932 5.51740167796855e-20
9.33933933933934 4.5774115701642e-20
9.35935935935936 3.79604417474457e-20
9.37937937937938 3.14679525475421e-20
9.3993993993994 2.60754402558043e-20
9.41941941941942 2.15983585814365e-20
9.43943943943944 1.78828106797973e-20
9.45945945945946 1.48005121824234e-20
9.47947947947948 1.22445730038445e-20
9.4994994994995 1.01259663374544e-20
9.51951951951952 8.37057415073758e-21
9.53953953953954 6.91671611023779e-21
9.55955955955956 5.71308371631411e-21
9.57957957957958 4.71701393695898e-21
9.5995995995996 3.89304716291232e-21
9.61961961961962 3.21172317125736e-21
9.63963963963964 2.64857624243904e-21
9.65965965965966 2.18329684678274e-21
9.67967967967968 1.79903258756111e-21
9.6996996996997 1.48180551599987e-21
9.71971971971972 1.22002665237565e-21
9.73973973973974 1.00409166885045e-21
9.75975975975976 8.26044308669654e-22
9.77977977977978 6.79296312742742e-22
9.7997997997998 5.58394465795474e-22
9.81981981981982 4.58826916995414e-22
9.83983983983984 3.7686222201397e-22
9.85985985985986 3.09415635142992e-22
9.87987987987988 2.53938085193762e-22
9.8998998998999 2.08324025950642e-22
9.91991991991992 1.70834984871876e-22
9.93993993993994 1.40036162642795e-22
9.95995995995996 1.14743877987917e-22
9.97997997997998 9.39820210218911e-23
10 7.69459862670642e-23
};
\addlegendentry{N(0,1)}; 
\legend{}; 
\end{axis}

\end{tikzpicture}

%% file: FinalFigs/Null_Dists_d_10_100_n_20_m_100_kernel__Dirichlet_Poly_5_2022_10_15_22_11_41cross.tex
\begin{tikzpicture}

\definecolor{darkorange25512714}{RGB}{255,127,14}
\definecolor{darkslategray38}{RGB}{38,38,38}
\definecolor{lightgray204}{RGB}{204,204,204}
\definecolor{steelblue31119180}{RGB}{31,119,180}

\begin{axis}[
axis line style={darkslategray38},
height=\figheight,
legend cell align={left},
legend style={fill opacity=0.8, draw opacity=1, text opacity=1, draw=none},
tick align=outside,
tick pos=left,
title={$\cmmd$ $(n/m=0.2)$},
width=\figwidth,
x grid style={lightgray204},
xmin=-6, xmax=6,
xtick style={color=darkslategray38},
y grid style={lightgray204},
ylabel=\textcolor{darkslategray38}{},
ymin=0, ymax=0.418868408525629,
ytick style={color=darkslategray38}, 
xticklabels=empty,
yticklabels=empty
]
\draw[draw=none,fill=steelblue31119180,fill opacity=0.8] (axis cs:-4.80234956741333,0) rectangle (axis cs:-4.48186922073364,0.00249625143059835);
\addlegendimage{ybar,ybar legend,draw=none,fill=steelblue31119180,fill opacity=0.8}
\addlegendentry{d=10}

\draw[draw=none,fill=steelblue31119180,fill opacity=0.8] (axis cs:-4.00114822387695,0) rectangle (axis cs:-3.68066787719727,0);
\draw[draw=none,fill=steelblue31119180,fill opacity=0.8] (axis cs:-3.199946641922,0) rectangle (axis cs:-2.87946629524231,0.0124812571529917);
\draw[draw=none,fill=steelblue31119180,fill opacity=0.8] (axis cs:-2.39874529838562,0) rectangle (axis cs:-2.07826495170593,0.0649025371955571);
\draw[draw=none,fill=steelblue31119180,fill opacity=0.8] (axis cs:-1.59754407405853,0) rectangle (axis cs:-1.27706372737885,0.189715108725475);
\draw[draw=none,fill=steelblue31119180,fill opacity=0.8] (axis cs:-0.796342730522156,0) rectangle (axis cs:-0.475862383842468,0.297053920241204);
\draw[draw=none,fill=steelblue31119180,fill opacity=0.8] (axis cs:0.00485862791538239,0) rectangle (axis cs:0.325338959693909,0.324512685977785);
\draw[draw=none,fill=steelblue31119180,fill opacity=0.8] (axis cs:0.806059956550598,0) rectangle (axis cs:1.12654030323029,0.239640137337441);
\draw[draw=none,fill=steelblue31119180,fill opacity=0.8] (axis cs:1.60726130008698,0) rectangle (axis cs:1.92774164676666,0.0898650515015405);
\draw[draw=none,fill=steelblue31119180,fill opacity=0.8] (axis cs:2.40846276283264,0) rectangle (axis cs:2.72894310951233,0.0274587657365818);
\draw[draw=none,fill=darkorange25512714,fill opacity=0.8] (axis cs:-4.48186874389648,0) rectangle (axis cs:-4.1613883972168,0);
\addlegendimage{ybar,ybar legend,draw=none,fill=darkorange25512714,fill opacity=0.8}
\addlegendentry{d=100}

\draw[draw=none,fill=darkorange25512714,fill opacity=0.8] (axis cs:-3.68066740036011,0) rectangle (axis cs:-3.36018705368042,0);
\draw[draw=none,fill=darkorange25512714,fill opacity=0.8] (axis cs:-2.87946605682373,0) rectangle (axis cs:-2.55898571014404,0.0174737600141884);
\draw[draw=none,fill=darkorange25512714,fill opacity=0.8] (axis cs:-2.07826471328735,0) rectangle (axis cs:-1.75778436660767,0.0773837943485488);
\draw[draw=none,fill=darkorange25512714,fill opacity=0.8] (axis cs:-1.27706348896027,0) rectangle (axis cs:-0.956583142280579,0.142286331544106);
\draw[draw=none,fill=darkorange25512714,fill opacity=0.8] (axis cs:-0.475862145423889,0) rectangle (axis cs:-0.155381798744202,0.379430217450949);
\draw[draw=none,fill=darkorange25512714,fill opacity=0.8] (axis cs:0.325339198112488,0) rectangle (axis cs:0.645819544792175,0.364452708867359);
\draw[draw=none,fill=darkorange25512714,fill opacity=0.8] (axis cs:1.12654054164886,0) rectangle (axis cs:1.44702088832855,0.19720386301727);
\draw[draw=none,fill=darkorange25512714,fill opacity=0.8] (axis cs:1.92774200439453,0) rectangle (axis cs:2.24822235107422,0.0599100343343604);
\draw[draw=none,fill=darkorange25512714,fill opacity=0.8] (axis cs:2.72894334793091,0) rectangle (axis cs:3.0494236946106,0.00998500572239339);
\addplot [semithick, black]
table {%
-10 7.69459862670642e-23
-9.97997997997998 9.39820210218911e-23
-9.95995995995996 1.14743877987917e-22
-9.93993993993994 1.40036162642795e-22
-9.91991991991992 1.70834984871876e-22
-9.8998998998999 2.08324025950642e-22
-9.87987987987988 2.53938085193762e-22
-9.85985985985986 3.09415635142992e-22
-9.83983983983984 3.7686222201397e-22
-9.81981981981982 4.58826916995414e-22
-9.7997997997998 5.58394465795474e-22
-9.77977977977978 6.79296312742742e-22
-9.75975975975976 8.26044308669654e-22
-9.73973973973974 1.00409166885045e-21
-9.71971971971972 1.22002665237565e-21
-9.6996996996997 1.48180551599987e-21
-9.67967967967968 1.79903258756111e-21
-9.65965965965966 2.18329684678274e-21
-9.63963963963964 2.64857624243904e-21
-9.61961961961962 3.21172317125736e-21
-9.5995995995996 3.89304716291232e-21
-9.57957957957958 4.71701393695898e-21
-9.55955955955956 5.71308371631411e-21
-9.53953953953954 6.91671611023779e-21
-9.51951951951952 8.37057415073758e-21
-9.4994994994995 1.01259663374544e-20
-9.47947947947948 1.22445730038445e-20
-9.45945945945946 1.48005121824234e-20
-9.43943943943944 1.78828106797973e-20
-9.41941941941942 2.15983585814365e-20
-9.3993993993994 2.60754402558043e-20
-9.37937937937938 3.14679525475421e-20
-9.35935935935936 3.79604417474457e-20
-9.33933933933934 4.5774115701642e-20
-9.31931931931932 5.51740167796855e-20
-9.2992992992993 6.6477576193283e-20
-9.27927927927928 8.00648113242436e-20
-9.25925925925926 9.63904764362469e-20
-9.23923923923924 1.1599853476858e-19
-9.21921921921922 1.39539388139188e-19
-9.1991991991992 1.67790380698338e-19
-9.17917917917918 2.01680188581445e-19
-9.15915915915916 2.42317819504618e-19
-9.13913913913914 2.91027078875695e-19
-9.11911911911912 3.49387515332417e-19
-9.0990990990991 4.19283042962206e-19
-9.07907907907908 5.0295965472299e-19
-9.05905905905906 6.03093897535385e-19
-9.03903903903904 7.22874080905507e-19
-9.01901901901902 8.66096545670873e-19
-8.998998998999 1.03727973679138e-18
-8.97897897897898 1.24179931485728e-18
-8.95895895895896 1.48604811781133e-18
-8.93893893893894 1.77762546207239e-18
-8.91891891891892 2.12556106807901e-18
-8.8988988988989 2.54057982941549e-18
-8.87887887887888 3.03541474067764e-18
-8.85885885885886 3.62517658454128e-18
-8.83883883883884 4.32779048512422e-18
-8.81881881881882 5.16451119999019e-18
-8.7987987987988 6.16053109048305e-18
-8.77877877877878 7.34569713009337e-18
-8.75875875875876 8.75535614211525e-18
-8.73873873873874 1.04313507694241e-17
-8.71871871871872 1.24231925503958e-17
-8.6986986986987 1.47894429982973e-17
-8.67867867867868 1.75993388643364e-17
-8.65865865865866 2.09347039316773e-17
-8.63863863863864 2.48921968838275e-17
-8.61861861861862 2.95859531836935e-17
-8.5985985985986 3.51506886838449e-17
-8.57857857857858 4.17453440895294e-17
-8.55855855855856 4.95573626747691e-17
-8.53853853853854 5.88077091106193e-17
-8.51851851851852 6.97567552529606e-17
-8.4984984984985 8.27111796592674e-17
-8.47847847847848 9.8032051926925e-17
-8.45845845845846 1.16144301209627e-16
-8.43843843843844 1.37547801096493e-16
-8.41841841841842 1.62830341150177e-16
-8.3983983983984 1.92682799625235e-16
-8.37837837837838 2.27916883183252e-16
-8.35835835835836 2.69485858889318e-16
-8.33833833833834 3.18508772685576e-16
-8.31831831831832 3.76298728354137e-16
-8.2982982982983 4.44395893386326e-16
-8.27827827827828 5.24606005103494e-16
-8.25825825825826 6.19045274051723e-16
-8.23823823823824 7.30192724674871e-16
-8.21821821821822 8.60951178494166e-16
-8.1981981981982 1.01471827585831e-15
-8.17817817817818 1.19546915264237e-15
-8.15815815815816 1.40785264250169e-15
-8.13813813813814 1.65730316851105e-15
-8.11811811811812 1.95017082606148e-15
-8.0980980980981 2.29387254841619e-15
-8.07807807807808 2.69706769497134e-15
-8.05805805805806 3.16986191875719e-15
-8.03803803803804 3.72404376402735e-15
-8.01801801801802 4.3733591283238e-15
-7.997997997998 5.13382950919607e-15
-7.97797797797798 6.02412085866193e-15
-7.95795795795796 7.06597090549303e-15
-7.93793793793794 8.28468399583074e-15
-7.91791791791792 9.70970386854708e-15
-7.8978978978979 1.13752763482777e-14
-7.87787787787788 1.33212157347805e-14
-7.85785785785786 1.55937907247683e-14
-7.83783783783784 1.82467480586655e-14
-7.81781781781782 2.13424947819475e-14
-7.7977977977978 2.49534630966872e-14
-7.77777777777778 2.91636853079909e-14
-7.75775775775776 3.40706104038538e-14
-7.73773773773774 3.9787198415571e-14
-7.71771771771772 4.64443339685918e-14
-7.6976976976977 5.4193606440538e-14
-7.67767767767768 6.32105109959252e-14
-7.65765765765766 7.36981325813117e-14
-7.63763763763764 8.58913838706879e-14
-7.61761761761762 1.00061878296601e-13
-7.5975975975976 1.16523530854699e-13
-7.57757757757758 1.35638992516664e-13
-7.55755755755756 1.57827039041884e-13
-7.53753753753754 1.83571051982057e-13
-7.51751751751752 2.13428748996349e-13
-7.4974974974975 2.48043342543513e-13
-7.47747747747748 2.88156330935935e-13
-7.45745745745746 3.34622154016965e-13
-7.43743743743744 3.88424977793732e-13
-7.41741741741742 4.50697908714384e-13
-7.3973973973974 5.22744979473711e-13
-7.37737737737738 6.06066294885249e-13
-7.35735735735736 7.02386779168738e-13
-7.33733733733734 8.13689025751835e-13
-7.31731731731732 9.42250818252909e-13
-7.2972972972973 1.09068796768221e-12
-7.27727727727728 1.262003197176e-12
-7.25725725725726 1.45964190299847e-12
-7.23723723723724 1.68755573049416e-12
-7.21721721721722 1.95027502769792e-12
-7.1971971971972 2.25299137914218e-12
-7.17717717717718 2.60165157997871e-12
-7.15715715715716 3.00306458801653e-12
-7.13713713713714 3.4650231910834e-12
-7.11711711711712 3.99644235193919e-12
-7.0970970970971 4.60751644580953e-12
-7.07707707707708 5.30989788981514e-12
-7.05705705705706 6.11689998287646e-12
-7.03703703703704 7.0437271332242e-12
-7.01701701701702 8.10773605306645e-12
-6.996996996997 9.32873195138555e-12
-6.97697697697698 1.07293042619726e-11
-6.95695695695696 1.23352070109962e-11
-6.93693693693694 1.41757895636779e-11
-6.91691691691692 1.62844842008237e-11
-6.8968968968969 1.86993577716893e-11
-6.87687687687688 2.14637355595386e-11
-6.85685685685686 2.46269064908967e-11
-6.83683683683684 2.82449199306815e-11
-6.81681681681682 3.2381485546105e-11
-6.7967967967968 3.71089891068559e-11
-6.77677677677678 4.25096386334913e-11
-6.75675675675676 4.86767570277083e-11
-6.73673673673674 5.57162392366004e-11
-6.71671671671672 6.37481941394491e-11
-6.6966966966967 7.29087937236032e-11
-6.67667667667668 8.33523547614402e-11
-6.65665665665666 9.52536811418151e-11
-6.63663663663664 1.08810698278135e-10
-6.61661661661662 1.24247414645768e-10
-6.5965965965966 1.41817249531744e-10
-6.57657657657658 1.61806770551278e-10
-6.55655655655656 1.84539889444164e-10
-6.53653653653654 2.10382570159622e-10
-6.51651651651652 2.39748109325542e-10
-6.4964964964965 2.73103055937438e-10
-6.47647647647648 3.10973844559381e-10
-6.45645645645646 3.53954224575809e-10
-6.43643643643644 4.027135771479e-10
-6.41641641641642 4.58006221596996e-10
-6.3963963963964 5.20681824054169e-10
-6.37637637637638 5.91697033481538e-10
-6.35635635635636 6.72128483698846e-10
-6.33633633633634 7.63187314959523e-10
-6.31631631631632 8.66235385046001e-10
-6.2962962962963 9.8280335793813e-10
-6.27627627627628 1.11461087800766e-09
-6.25625625625626 1.26358905957513e-09
-6.23623623623624 1.43190554571787e-09
-6.21621621621622 1.62199241663862e-09
-6.1961961961962 1.83657725691024e-09
-6.17617617617618 2.0787177227378e-09
-6.15615615615616 2.35183998527873e-09
-6.13613613613614 2.65978146430928e-09
-6.11611611611612 3.00683830841829e-09
-6.0960960960961 3.39781812376754e-09
-6.07607607607608 3.83809850362696e-09
-6.05605605605606 4.33369196574699e-09
-6.03603603603604 4.89131796456919e-09
-6.01601601601602 5.51848271073395e-09
-5.995995995996 6.22356760178439e-09
-5.97597597597598 7.01592714588943e-09
-5.95595595595596 7.90599734535251e-09
-5.93593593593594 8.90541559921139e-09
-5.91591591591592 1.00271532849868e-08
-5.8958958958959 1.12856622892642e-08
-5.87587587587588 1.2697036876002e-08
-5.85585585585586 1.42791924110059e-08
-5.83583583583584 1.60520626017053e-08
-5.81581581581582 1.80378170640688e-08
-5.7957957957958 2.02611011941336e-08
-5.77577577577578 2.27493005011625e-08
-5.75575575575576 2.55328317539416e-08
-5.73573573573574 2.86454635022852e-08
-5.71571571571572 3.2124668763613e-08
-5.6956956956957 3.60120129107462e-08
-5.67567567567568 4.03535800631662e-08
-5.65565565565566 4.5200441571292e-08
-5.63563563563564 5.06091704933412e-08
-5.61561561561562 5.66424062986154e-08
-5.5955955955956 6.33694743912418e-08
-5.57557557557558 7.08670654362614e-08
-5.55555555555556 7.92199798873018e-08
-5.53553553553554 8.85219435638491e-08
-5.51551551551552 9.8876500608364e-08
-5.4954954954955 1.10397990671284e-07
-5.47547547547548 1.23212617727566e-07
-5.45545545545546 1.3745961852414e-07
-5.43543543543544 1.5329253929596e-07
-5.41541541541542 1.70880630071684e-07
-5.3953953953954 1.90410366621162e-07
-5.37537537537538 2.1208711087848e-07
-5.35535535535536 2.36136921509202e-07
-5.33533533533534 2.62808527181656e-07
-5.31531531531532 2.92375476052561e-07
-5.2952952952953 3.25138475990267e-07
-5.27527527527528 3.61427941137511e-07
-5.25525525525526 4.01606761563285e-07
-5.23523523523524 4.46073313973501e-07
-5.21521521521522 4.95264732746229e-07
-5.1951951951952 5.4966046193278e-07
-5.17517517517518 6.09786110324583e-07
-5.15515515515516 6.76217633231267e-07
-5.13513513513514 7.49585866251233e-07
-5.11511511511512 8.30581438046261e-07
-5.0950950950951 9.19960090959837e-07
-5.07507507507508 1.01854844024876e-06
-5.05505505505506 1.12725020473309e-06
-5.03503503503504 1.24705294381396e-06
-5.01501501501502 1.37903533806611e-06
-4.99499499499499 1.5243750529858e-06
-4.97497497497497 1.68435722796805e-06
-4.95495495495495 1.86038363520377e-06
-4.93493493493493 2.05398255592983e-06
-4.91491491491491 2.26681942433715e-06
-4.89489489489489 2.50070829244518e-06
-4.87487487487487 2.75762417238901e-06
-4.85485485485485 3.03971631583941e-06
-4.83483483483483 3.34932249368796e-06
-4.81481481481481 3.68898434268124e-06
-4.79479479479479 4.06146384937932e-06
-4.77477477477477 4.4697610456467e-06
-4.75475475475475 4.91713299385613e-06
-4.73473473473473 5.40711414409909e-06
-4.71471471471471 5.94353814994721e-06
-4.69469469469469 6.53056123369604e-06
-4.67467467467467 7.17268719654363e-06
-4.65465465465465 7.87479417380527e-06
-4.63463463463463 8.64216324004121e-06
-4.61461461461461 9.48050897386715e-06
-4.59459459459459 1.03960120972233e-05
-4.57457457457457 1.13953543089884e-05
-4.55455455455455 1.24857554380297e-05
-4.53453453453453 1.3675013046071e-05
-4.51451451451451 1.49715446161227e-05
-4.49449449449449 1.63844324676437e-05
-4.47447447447447 1.79234715450684e-05
-4.45445445445445 1.95992202318354e-05
-4.43443443443443 2.14230543475548e-05
-4.41441441441441 2.34072244914564e-05
-4.39439439439439 2.55649169007242e-05
-4.37437437437437 2.79103179977393e-05
-4.35435435435435 3.04586828055866e-05
-4.33433433433433 3.32264074164067e-05
-4.31431431431431 3.62311057022691e-05
-4.29429429429429 3.94916904631592e-05
-4.27427427427427 4.3028459211397e-05
-4.25425425425425 4.68631847962824e-05
-4.23423423423423 5.10192110769697e-05
-4.21421421421421 5.55215538554582e-05
-4.19419419419419 6.03970072851107e-05
-4.17417417417417 6.56742559732345e-05
-4.15415415415415 7.13839929989176e-05
-4.13413413413413 7.75590440694795e-05
-4.11411411411411 8.42344980404937e-05
-4.09409409409409 9.14478440253317e-05
-4.07407407407407 9.92391153205018e-05
-4.05405405405405 0.000107651040372646
-4.03403403403403 0.000116729201011866
-4.01401401401401 0.000126522198173995
-3.99399399399399 0.000137081825331481
-3.97397397397397 0.000148463249848567
-3.95395395395395 0.000160725202471485
-3.93393393393393 0.000173930175158222
-3.91391391391391 0.00018814462744512
-3.89389389389389 0.000203439201538965
-3.87387387387387 0.000219888946313312
-3.85385385385385 0.000237573550376443
-3.83383383383383 0.000256577584365551
-3.81381381381381 0.000276990752607344
-3.79379379379379 0.000298908154269281
-3.77377377377377 0.000322430554107926
-3.75375375375375 0.000347664662901427
-3.73373373373373 0.000374723427631836
-3.71371371371371 0.000403726331459719
-3.69369369369369 0.000434799703508357
-3.67367367367367 0.000468077038447599
-3.65365365365365 0.00050369932583812
-3.63363363363363 0.000541815389165405
-3.61361361361361 0.000582582234459159
-3.59359359359359 0.000626165408357979
-3.57357357357357 0.000672739365441021
-3.55355355355355 0.000722487844607978
-3.53353353353353 0.00077560425424601
-3.51351351351351 0.000832292065877155
-3.49349349349349 0.000892765215932443
-3.47347347347347 0.000957248515249216
-3.45345345345345 0.0010259780658361
-3.43343343343343 0.00109920168439588
-3.41341341341341 0.00117717933203981
-3.39339339339339 0.00126018354956833
-3.37337337337337 0.00134849989763212
-3.35335335335335 0.00144242740102448
-3.33333333333333 0.00154227899629111
-3.31331331331331 0.00164838198177652
-3.29329329329329 0.00176107846915772
-3.27327327327327 0.00188072583544552
-3.25325325325325 0.00200769717436226
-3.23323323323323 0.00214238174593163
-3.21321321321321 0.00228518542304204
-3.19319319319319 0.00243653113367012
-3.17317317317317 0.00259685929737497
-3.15315315315315 0.00276662825459747
-3.13313313313313 0.00294631468722261
-3.11311311311311 0.00313641402878609
-3.09309309309309 0.00333744086263052
-3.07307307307307 0.00354992930624086
-3.05305305305305 0.00377443337991422
-3.03303303303303 0.00401152735784579
-3.01301301301301 0.00426180609964128
-2.99299299299299 0.00452588536019618
-2.97297297297297 0.0048044020758154
-2.95295295295295 0.00509801462438215
-2.93293293293293 0.00540740305732385
-2.91291291291291 0.00573326930106519
-2.89289289289289 0.00607633732560526
-2.87287287287287 0.00643735327780636
-2.85285285285285 0.00681708557693873
-2.83283283283283 0.00721632496998623
-2.81281281281281 0.00763588454418632
-2.79279279279279 0.00807659969425075
-2.77277277277277 0.00853932804169477
-2.75275275275275 0.00902494930369032
-2.73273273273273 0.00953436510885489
-2.71271271271271 0.0100684987573917
-2.69269269269269 0.0106282949230102
-2.67267267267267 0.0112147192940778
-2.65265265265265 0.0118287581514852
-2.63263263263263 0.0124714178807513
-2.61261261261261 0.0131437244159435
-2.59259259259259 0.0138467226130541
-2.57257257257257 0.0145814755505492
-2.55255255255255 0.0153490637548887
-2.53253253253253 0.0161505843489182
-2.51251251251251 0.0169871501211409
-2.49249249249249 0.0178598885140022
-2.47247247247247 0.018769940529451
-2.45245245245245 0.0197184595501959
-2.43243243243243 0.0207066100752274
-2.41241241241241 0.0217355663683581
-2.39239239239239 0.0228065110187135
-2.37237237237237 0.0239206334123108
-2.35235235235235 0.0250791281140699
-2.33233233233233 0.0262831931598317
-2.31231231231231 0.0275340282581906
-2.29229229229229 0.0288328329022027
-2.27227227227227 0.0301808043912899
-2.25225225225225 0.0315791357639331
-2.23223223223223 0.033029013642035
-2.21221221221221 0.0345316159881232
-2.19219219219219 0.0360881097768736
-2.17217217217217 0.0376996485827434
-2.15215215215215 0.0393673700858293
-2.13213213213213 0.0410923934983949
-2.11211211211211 0.0428758169148479
-2.09209209209209 0.0447187145882915
-2.07207207207207 0.0466221341371235
-2.05205205205205 0.0485870936855041
-2.03203203203203 0.0506145789418747
-2.01201201201201 0.0527055402200587
-1.99199199199199 0.0548608894078376
-1.97197197197197 0.057081496888248
-1.95195195195195 0.059368188419199
-1.93193193193193 0.0617217419773594
-1.91191191191191 0.0641428845726061
-1.89189189189189 0.0666322890396674
-1.87187187187187 0.0691905708139176
-1.85185185185185 0.0718182846986055
-1.83183183183183 0.0745159216311036
-1.81181181181181 0.077283905456062
-1.79179179179179 0.0801225897136326
-1.77177177177177 0.083032254451193
-1.75175175175175 0.0860131030672496
-1.73173173173173 0.0890652591964251
-1.71171171171171 0.0921887636446459
-1.69169169169169 0.0953835713838294
-1.67167167167167 0.0986495486155338
-1.65165165165165 0.101986469913169
-1.63163163163163 0.10539401545248
-1.61161161161161 0.108871768340093
-1.59159159159159 0.112419212049971
-1.57157157157157 0.116035727977651
-1.55155155155155 0.119720593122119
-1.53153153153153 0.123472977905145
-1.51151151151151 0.127291944137829
-1.49149149149149 0.13117644314399
-1.47147147147147 0.135125314049902
-1.45145145145145 0.139137282249685
-1.43143143143143 0.143210958055468
-1.41141141141141 0.147344835541168
-1.39139139139139 0.151537291588457
-1.37137137137137 0.155786585143159
-1.35135135135135 0.160090856689972
-1.33133133133133 0.164448127952996
-1.31131131131131 0.168856301829129
-1.29129129129129 0.173313162560933
-1.27127127127127 0.17781637615506
-1.25125125125125 0.182363491051798
-1.23123123123123 0.186951939050736
-1.21121121121121 0.191579036496956
-1.19119119119119 0.1962419857315
-1.17117117117117 0.200937876809264
-1.15115115115115 0.205663689486728
-1.13113113113113 0.210416295481265
-1.11111111111111 0.215192461003031
-1.09109109109109 0.219988849559688
-1.07107107107107 0.224802025033432
-1.05105105105105 0.229628455029052
-1.03103103103103 0.234464514490888
-1.01101101101101 0.239306489585817
-0.990990990990991 0.24415058184851
-0.970970970970971 0.24899291258444
-0.950950950950951 0.253829527525259
-0.930930930930931 0.258656401730343
-0.910910910910911 0.2634694447275
-0.890890890890891 0.268264505884996
-0.870870870870871 0.273037380006279
-0.850850850850851 0.277783813137949
-0.830830830830831 0.282499508580786
-0.810810810810811 0.287180133092853
-0.790790790790791 0.291821323272996
-0.77077077077077 0.296418692112302
-0.75075075075075 0.300967835700437
-0.73073073073073 0.305464340073112
-0.71071071071071 0.309903788186304
-0.69069069069069 0.314281767002296
-0.67067067067067 0.318593874672039
-0.65065065065065 0.322835727797843
-0.63063063063063 0.327002968759958
-0.61061061061061 0.331091273090187
-0.59059059059059 0.33509635687531
-0.57057057057057 0.339013984172804
-0.55055055055055 0.34283997442106
-0.53053053053053 0.346570209826128
-0.51051051051051 0.350200642706842
-0.49049049049049 0.353727302780113
-0.47047047047047 0.357146304368113
-0.45045045045045 0.360453853509139
-0.43043043043043 0.363646254953996
-0.41041041041041 0.366719919029892
-0.39039039039039 0.369671368354051
-0.37037037037037 0.372497244379499
-0.35035035035035 0.375194313755802
-0.33033033033033 0.377759474487924
-0.31031031031031 0.38018976187679
-0.29029029029029 0.382482354225654
-0.27027027027027 0.384634578296894
-0.25025025025025 0.386643914504485
-0.23023023023023 0.388508001828027
-0.21021021021021 0.390224642434919
-0.19019019019019 0.391791805998011
-0.17017017017017 0.393207633696876
-0.15015015015015 0.394470441891644
-0.13013013013013 0.395578725459258
-0.11011011011011 0.396531160782876
-0.0900900900900901 0.397326608386124
-0.07007007007007 0.397964115204853
-0.05005005005005 0.398442916490068
-0.03003003003003 0.398762437336696
-0.01001001001001 0.398922293833933
0.01001001001001 0.398922293833933
0.03003003003003 0.398762437336696
0.05005005005005 0.398442916490068
0.07007007007007 0.397964115204853
0.0900900900900901 0.397326608386124
0.11011011011011 0.396531160782876
0.13013013013013 0.395578725459258
0.15015015015015 0.394470441891644
0.17017017017017 0.393207633696876
0.19019019019019 0.391791805998011
0.21021021021021 0.390224642434919
0.23023023023023 0.388508001828027
0.25025025025025 0.386643914504485
0.27027027027027 0.384634578296894
0.29029029029029 0.382482354225654
0.31031031031031 0.38018976187679
0.33033033033033 0.377759474487924
0.35035035035035 0.375194313755802
0.37037037037037 0.372497244379499
0.39039039039039 0.369671368354051
0.41041041041041 0.366719919029892
0.43043043043043 0.363646254953996
0.45045045045045 0.360453853509139
0.47047047047047 0.357146304368113
0.49049049049049 0.353727302780113
0.51051051051051 0.350200642706842
0.53053053053053 0.346570209826128
0.55055055055055 0.34283997442106
0.57057057057057 0.339013984172804
0.59059059059059 0.33509635687531
0.61061061061061 0.331091273090187
0.63063063063063 0.327002968759958
0.65065065065065 0.322835727797843
0.67067067067067 0.318593874672039
0.69069069069069 0.314281767002296
0.71071071071071 0.309903788186304
0.73073073073073 0.305464340073112
0.75075075075075 0.300967835700437
0.77077077077077 0.296418692112302
0.790790790790791 0.291821323272996
0.810810810810811 0.287180133092853
0.830830830830831 0.282499508580786
0.850850850850851 0.277783813137949
0.870870870870871 0.273037380006279
0.890890890890891 0.268264505884996
0.910910910910911 0.2634694447275
0.930930930930931 0.258656401730343
0.950950950950951 0.253829527525259
0.970970970970971 0.24899291258444
0.990990990990991 0.24415058184851
1.01101101101101 0.239306489585817
1.03103103103103 0.234464514490888
1.05105105105105 0.229628455029052
1.07107107107107 0.224802025033432
1.09109109109109 0.219988849559688
1.11111111111111 0.215192461003031
1.13113113113113 0.210416295481265
1.15115115115115 0.205663689486728
1.17117117117117 0.200937876809264
1.19119119119119 0.1962419857315
1.21121121121121 0.191579036496956
1.23123123123123 0.186951939050736
1.25125125125125 0.182363491051798
1.27127127127127 0.17781637615506
1.29129129129129 0.173313162560933
1.31131131131131 0.168856301829129
1.33133133133133 0.164448127952996
1.35135135135135 0.160090856689972
1.37137137137137 0.155786585143159
1.39139139139139 0.151537291588457
1.41141141141141 0.147344835541168
1.43143143143143 0.143210958055468
1.45145145145145 0.139137282249685
1.47147147147147 0.135125314049902
1.49149149149149 0.13117644314399
1.51151151151151 0.127291944137829
1.53153153153153 0.123472977905145
1.55155155155155 0.119720593122119
1.57157157157157 0.116035727977651
1.59159159159159 0.112419212049971
1.61161161161161 0.108871768340093
1.63163163163163 0.10539401545248
1.65165165165165 0.101986469913169
1.67167167167167 0.0986495486155338
1.69169169169169 0.0953835713838294
1.71171171171171 0.0921887636446459
1.73173173173173 0.0890652591964251
1.75175175175175 0.0860131030672496
1.77177177177177 0.083032254451193
1.79179179179179 0.0801225897136326
1.81181181181181 0.077283905456062
1.83183183183183 0.0745159216311036
1.85185185185185 0.0718182846986055
1.87187187187187 0.0691905708139176
1.89189189189189 0.0666322890396674
1.91191191191191 0.0641428845726061
1.93193193193193 0.0617217419773594
1.95195195195195 0.059368188419199
1.97197197197197 0.057081496888248
1.99199199199199 0.0548608894078376
2.01201201201201 0.0527055402200588
2.03203203203203 0.0506145789418748
2.05205205205205 0.0485870936855042
2.07207207207207 0.0466221341371236
2.09209209209209 0.0447187145882916
2.11211211211211 0.0428758169148479
2.13213213213213 0.041092393498395
2.15215215215215 0.0393673700858294
2.17217217217217 0.0376996485827434
2.19219219219219 0.0360881097768737
2.21221221221221 0.0345316159881233
2.23223223223223 0.033029013642035
2.25225225225225 0.0315791357639332
2.27227227227227 0.03018080439129
2.29229229229229 0.0288328329022028
2.31231231231231 0.0275340282581906
2.33233233233233 0.0262831931598317
2.35235235235235 0.02507912811407
2.37237237237237 0.0239206334123108
2.39239239239239 0.0228065110187136
2.41241241241241 0.0217355663683581
2.43243243243243 0.0207066100752275
2.45245245245245 0.0197184595501959
2.47247247247247 0.0187699405294511
2.49249249249249 0.0178598885140022
2.51251251251251 0.016987150121141
2.53253253253253 0.0161505843489182
2.55255255255255 0.0153490637548888
2.57257257257257 0.0145814755505493
2.59259259259259 0.0138467226130541
2.61261261261261 0.0131437244159435
2.63263263263263 0.0124714178807514
2.65265265265265 0.0118287581514852
2.67267267267267 0.0112147192940778
2.69269269269269 0.0106282949230103
2.71271271271271 0.0100684987573917
2.73273273273273 0.00953436510885491
2.75275275275275 0.00902494930369034
2.77277277277277 0.00853932804169479
2.79279279279279 0.00807659969425077
2.81281281281281 0.00763588454418634
2.83283283283283 0.00721632496998621
2.85285285285285 0.00681708557693871
2.87287287287287 0.00643735327780635
2.89289289289289 0.00607633732560524
2.91291291291291 0.00573326930106518
2.93293293293293 0.00540740305732384
2.95295295295295 0.00509801462438214
2.97297297297297 0.00480440207581539
2.99299299299299 0.00452588536019617
3.01301301301301 0.00426180609964127
3.03303303303303 0.00401152735784578
3.05305305305305 0.00377443337991421
3.07307307307307 0.00354992930624085
3.09309309309309 0.00333744086263051
3.11311311311311 0.00313641402878608
3.13313313313313 0.0029463146872226
3.15315315315315 0.00276662825459746
3.17317317317317 0.00259685929737496
3.19319319319319 0.00243653113367011
3.21321321321321 0.00228518542304203
3.23323323323323 0.00214238174593163
3.25325325325325 0.00200769717436225
3.27327327327327 0.00188072583544551
3.29329329329329 0.00176107846915771
3.31331331331331 0.00164838198177652
3.33333333333333 0.0015422789962911
3.35335335335335 0.00144242740102448
3.37337337337337 0.00134849989763212
3.39339339339339 0.00126018354956833
3.41341341341341 0.00117717933203981
3.43343343343343 0.00109920168439588
3.45345345345345 0.0010259780658361
3.47347347347347 0.000957248515249212
3.49349349349349 0.000892765215932441
3.51351351351351 0.000832292065877152
3.53353353353353 0.000775604254246008
3.55355355355355 0.000722487844607976
3.57357357357357 0.000672739365441019
3.59359359359359 0.000626165408357979
3.61361361361361 0.000582582234459159
3.63363363363363 0.000541815389165405
3.65365365365365 0.00050369932583812
3.67367367367367 0.000468077038447599
3.69369369369369 0.000434799703508357
3.71371371371371 0.000403726331459719
3.73373373373373 0.000374723427631836
3.75375375375375 0.000347664662901427
3.77377377377377 0.000322430554107926
3.79379379379379 0.000298908154269281
3.81381381381381 0.000276990752607344
3.83383383383383 0.000256577584365551
3.85385385385385 0.000237573550376443
3.87387387387387 0.000219888946313312
3.89389389389389 0.000203439201538965
3.91391391391391 0.00018814462744512
3.93393393393393 0.000173930175158222
3.95395395395395 0.000160725202471485
3.97397397397397 0.000148463249848567
3.99399399399399 0.000137081825331481
4.01401401401401 0.000126522198173995
4.03403403403403 0.000116729201011866
4.05405405405405 0.000107651040372646
4.07407407407407 9.92391153205018e-05
4.09409409409409 9.14478440253317e-05
4.11411411411411 8.42344980404937e-05
4.13413413413413 7.75590440694795e-05
4.15415415415415 7.13839929989176e-05
4.17417417417417 6.56742559732345e-05
4.19419419419419 6.03970072851107e-05
4.21421421421421 5.55215538554582e-05
4.23423423423423 5.10192110769697e-05
4.25425425425425 4.68631847962824e-05
4.27427427427427 4.3028459211397e-05
4.29429429429429 3.94916904631592e-05
4.31431431431431 3.62311057022691e-05
4.33433433433433 3.32264074164067e-05
4.35435435435435 3.04586828055866e-05
4.37437437437437 2.79103179977393e-05
4.39439439439439 2.55649169007242e-05
4.41441441441441 2.34072244914564e-05
4.43443443443443 2.14230543475548e-05
4.45445445445445 1.95992202318354e-05
4.47447447447447 1.79234715450684e-05
4.49449449449449 1.63844324676437e-05
4.51451451451451 1.49715446161227e-05
4.53453453453453 1.3675013046071e-05
4.55455455455455 1.24857554380297e-05
4.57457457457457 1.13953543089884e-05
4.59459459459459 1.03960120972233e-05
4.61461461461461 9.48050897386715e-06
4.63463463463463 8.64216324004121e-06
4.65465465465465 7.87479417380527e-06
4.67467467467467 7.17268719654363e-06
4.69469469469469 6.53056123369604e-06
4.71471471471471 5.94353814994721e-06
4.73473473473473 5.40711414409909e-06
4.75475475475475 4.91713299385613e-06
4.77477477477477 4.4697610456467e-06
4.79479479479479 4.06146384937932e-06
4.81481481481481 3.68898434268124e-06
4.83483483483483 3.34932249368796e-06
4.85485485485485 3.03971631583941e-06
4.87487487487487 2.75762417238901e-06
4.89489489489489 2.50070829244518e-06
4.91491491491491 2.26681942433715e-06
4.93493493493493 2.05398255592983e-06
4.95495495495495 1.86038363520377e-06
4.97497497497497 1.68435722796805e-06
4.99499499499499 1.5243750529858e-06
5.01501501501502 1.37903533806611e-06
5.03503503503504 1.24705294381396e-06
5.05505505505506 1.12725020473309e-06
5.07507507507508 1.01854844024876e-06
5.0950950950951 9.19960090959837e-07
5.11511511511512 8.30581438046261e-07
5.13513513513514 7.49585866251233e-07
5.15515515515516 6.76217633231267e-07
5.17517517517518 6.09786110324583e-07
5.1951951951952 5.4966046193278e-07
5.21521521521522 4.95264732746229e-07
5.23523523523524 4.46073313973501e-07
5.25525525525526 4.01606761563285e-07
5.27527527527528 3.61427941137511e-07
5.2952952952953 3.25138475990267e-07
5.31531531531532 2.92375476052561e-07
5.33533533533534 2.62808527181656e-07
5.35535535535536 2.36136921509202e-07
5.37537537537538 2.1208711087848e-07
5.3953953953954 1.90410366621162e-07
5.41541541541542 1.70880630071684e-07
5.43543543543544 1.5329253929596e-07
5.45545545545546 1.3745961852414e-07
5.47547547547548 1.23212617727566e-07
5.4954954954955 1.10397990671284e-07
5.51551551551552 9.8876500608364e-08
5.53553553553554 8.85219435638491e-08
5.55555555555556 7.92199798873018e-08
5.57557557557558 7.08670654362614e-08
5.5955955955956 6.33694743912418e-08
5.61561561561562 5.66424062986154e-08
5.63563563563564 5.06091704933412e-08
5.65565565565566 4.5200441571292e-08
5.67567567567568 4.03535800631662e-08
5.6956956956957 3.60120129107462e-08
5.71571571571572 3.2124668763613e-08
5.73573573573574 2.86454635022852e-08
5.75575575575576 2.55328317539416e-08
5.77577577577578 2.27493005011625e-08
5.7957957957958 2.02611011941336e-08
5.81581581581582 1.80378170640688e-08
5.83583583583584 1.60520626017053e-08
5.85585585585586 1.42791924110059e-08
5.87587587587588 1.2697036876002e-08
5.8958958958959 1.12856622892642e-08
5.91591591591592 1.00271532849868e-08
5.93593593593594 8.90541559921139e-09
5.95595595595596 7.90599734535251e-09
5.97597597597598 7.01592714588943e-09
5.995995995996 6.22356760178439e-09
6.01601601601602 5.5184827107339e-09
6.03603603603604 4.89131796456914e-09
6.05605605605606 4.33369196574694e-09
6.07607607607608 3.83809850362692e-09
6.0960960960961 3.3978181237675e-09
6.11611611611612 3.00683830841826e-09
6.13613613613614 2.65978146430924e-09
6.15615615615616 2.35183998527871e-09
6.17617617617618 2.07871772273778e-09
6.1961961961962 1.83657725691022e-09
6.21621621621622 1.6219924166386e-09
6.23623623623624 1.43190554571785e-09
6.25625625625626 1.26358905957511e-09
6.27627627627628 1.11461087800764e-09
6.2962962962963 9.82803357938119e-10
6.31631631631632 8.66235385045992e-10
6.33633633633634 7.63187314959515e-10
6.35635635635636 6.72128483698836e-10
6.37637637637638 5.91697033481532e-10
6.3963963963964 5.20681824054164e-10
6.41641641641642 4.58006221596989e-10
6.43643643643644 4.02713577147895e-10
6.45645645645646 3.53954224575805e-10
6.47647647647648 3.10973844559378e-10
6.4964964964965 2.73103055937435e-10
6.51651651651652 2.39748109325539e-10
6.53653653653654 2.10382570159619e-10
6.55655655655656 1.84539889444162e-10
6.57657657657658 1.61806770551276e-10
6.5965965965966 1.41817249531742e-10
6.61661661661662 1.24247414645767e-10
6.63663663663664 1.08810698278133e-10
6.65665665665666 9.52536811418137e-11
6.67667667667668 8.33523547614393e-11
6.6966966966967 7.29087937236022e-11
6.71671671671672 6.37481941394484e-11
6.73673673673674 5.57162392365998e-11
6.75675675675676 4.86767570277076e-11
6.77677677677678 4.25096386334907e-11
6.7967967967968 3.71089891068556e-11
6.81681681681682 3.23814855461048e-11
6.83683683683684 2.82449199306813e-11
6.85685685685686 2.46269064908966e-11
6.87687687687688 2.14637355595386e-11
6.8968968968969 1.86993577716892e-11
6.91691691691692 1.62844842008236e-11
6.93693693693694 1.41757895636779e-11
6.95695695695696 1.23352070109961e-11
6.97697697697698 1.07293042619725e-11
6.996996996997 9.32873195138548e-12
7.01701701701702 8.10773605306642e-12
7.03703703703704 7.04372713322415e-12
7.05705705705706 6.11689998287643e-12
7.07707707707708 5.30989788981512e-12
7.0970970970971 4.6075164458095e-12
7.11711711711712 3.99644235193917e-12
7.13713713713714 3.46502319108337e-12
7.15715715715716 3.00306458801651e-12
7.17717717717718 2.60165157997869e-12
7.1971971971972 2.25299137914217e-12
7.21721721721722 1.95027502769791e-12
7.23723723723724 1.68755573049414e-12
7.25725725725726 1.45964190299846e-12
7.27727727727728 1.262003197176e-12
7.2972972972973 1.0906879676822e-12
7.31731731731732 9.42250818252902e-13
7.33733733733734 8.13689025751829e-13
7.35735735735736 7.02386779168733e-13
7.37737737737738 6.06066294885245e-13
7.3973973973974 5.22744979473708e-13
7.41741741741742 4.50697908714381e-13
7.43743743743744 3.88424977793729e-13
7.45745745745746 3.34622154016963e-13
7.47747747747748 2.88156330935933e-13
7.4974974974975 2.48043342543511e-13
7.51751751751752 2.13428748996348e-13
7.53753753753754 1.83571051982055e-13
7.55755755755756 1.57827039041883e-13
7.57757757757758 1.35638992516663e-13
7.5975975975976 1.16523530854698e-13
7.61761761761762 1.000618782966e-13
7.63763763763764 8.58913838706873e-14
7.65765765765766 7.36981325813112e-14
7.67767767767768 6.32105109959248e-14
7.6976976976977 5.41936064405376e-14
7.71771771771772 4.64443339685915e-14
7.73773773773774 3.97871984155707e-14
7.75775775775776 3.40706104038536e-14
7.77777777777778 2.91636853079907e-14
7.7977977977978 2.4953463096687e-14
7.81781781781782 2.13424947819474e-14
7.83783783783784 1.82467480586654e-14
7.85785785785786 1.55937907247682e-14
7.87787787787788 1.33212157347804e-14
7.8978978978979 1.13752763482777e-14
7.91791791791792 9.70970386854701e-15
7.93793793793794 8.28468399583068e-15
7.95795795795796 7.06597090549298e-15
7.97797797797798 6.02412085866189e-15
7.997997997998 5.13382950919603e-15
8.01801801801802 4.3733591283238e-15
8.03803803803804 3.72404376402735e-15
8.05805805805806 3.16986191875719e-15
8.07807807807808 2.69706769497134e-15
8.0980980980981 2.29387254841619e-15
8.11811811811812 1.95017082606148e-15
8.13813813813814 1.65730316851105e-15
8.15815815815816 1.40785264250169e-15
8.17817817817818 1.19546915264237e-15
8.1981981981982 1.01471827585831e-15
8.21821821821822 8.60951178494166e-16
8.23823823823824 7.30192724674871e-16
8.25825825825826 6.19045274051723e-16
8.27827827827828 5.24606005103494e-16
8.2982982982983 4.44395893386326e-16
8.31831831831832 3.76298728354137e-16
8.33833833833834 3.18508772685576e-16
8.35835835835836 2.69485858889318e-16
8.37837837837838 2.27916883183252e-16
8.3983983983984 1.92682799625235e-16
8.41841841841842 1.62830341150177e-16
8.43843843843844 1.37547801096493e-16
8.45845845845846 1.16144301209627e-16
8.47847847847848 9.8032051926925e-17
8.4984984984985 8.27111796592674e-17
8.51851851851852 6.97567552529606e-17
8.53853853853854 5.88077091106193e-17
8.55855855855856 4.95573626747691e-17
8.57857857857858 4.17453440895294e-17
8.5985985985986 3.51506886838449e-17
8.61861861861862 2.95859531836935e-17
8.63863863863864 2.48921968838275e-17
8.65865865865866 2.09347039316773e-17
8.67867867867868 1.75993388643364e-17
8.6986986986987 1.47894429982973e-17
8.71871871871872 1.24231925503958e-17
8.73873873873874 1.04313507694241e-17
8.75875875875876 8.75535614211525e-18
8.77877877877878 7.34569713009337e-18
8.7987987987988 6.16053109048305e-18
8.81881881881882 5.16451119999019e-18
8.83883883883884 4.32779048512422e-18
8.85885885885886 3.62517658454128e-18
8.87887887887888 3.03541474067764e-18
8.8988988988989 2.54057982941549e-18
8.91891891891892 2.12556106807901e-18
8.93893893893894 1.77762546207239e-18
8.95895895895896 1.48604811781133e-18
8.97897897897898 1.24179931485728e-18
8.998998998999 1.03727973679138e-18
9.01901901901902 8.66096545670873e-19
9.03903903903904 7.22874080905507e-19
9.05905905905906 6.03093897535385e-19
9.07907907907908 5.0295965472299e-19
9.0990990990991 4.19283042962206e-19
9.11911911911912 3.49387515332417e-19
9.13913913913914 2.91027078875695e-19
9.15915915915916 2.42317819504618e-19
9.17917917917918 2.01680188581445e-19
9.1991991991992 1.67790380698338e-19
9.21921921921922 1.39539388139188e-19
9.23923923923924 1.1599853476858e-19
9.25925925925926 9.63904764362469e-20
9.27927927927928 8.00648113242436e-20
9.2992992992993 6.6477576193283e-20
9.31931931931932 5.51740167796855e-20
9.33933933933934 4.5774115701642e-20
9.35935935935936 3.79604417474457e-20
9.37937937937938 3.14679525475421e-20
9.3993993993994 2.60754402558043e-20
9.41941941941942 2.15983585814365e-20
9.43943943943944 1.78828106797973e-20
9.45945945945946 1.48005121824234e-20
9.47947947947948 1.22445730038445e-20
9.4994994994995 1.01259663374544e-20
9.51951951951952 8.37057415073758e-21
9.53953953953954 6.91671611023779e-21
9.55955955955956 5.71308371631411e-21
9.57957957957958 4.71701393695898e-21
9.5995995995996 3.89304716291232e-21
9.61961961961962 3.21172317125736e-21
9.63963963963964 2.64857624243904e-21
9.65965965965966 2.18329684678274e-21
9.67967967967968 1.79903258756111e-21
9.6996996996997 1.48180551599987e-21
9.71971971971972 1.22002665237565e-21
9.73973973973974 1.00409166885045e-21
9.75975975975976 8.26044308669654e-22
9.77977977977978 6.79296312742742e-22
9.7997997997998 5.58394465795474e-22
9.81981981981982 4.58826916995414e-22
9.83983983983984 3.7686222201397e-22
9.85985985985986 3.09415635142992e-22
9.87987987987988 2.53938085193762e-22
9.8998998998999 2.08324025950642e-22
9.91991991991992 1.70834984871876e-22
9.93993993993994 1.40036162642795e-22
9.95995995995996 1.14743877987917e-22
9.97997997997998 9.39820210218911e-23
10 7.69459862670642e-23
};
\addlegendentry{N(0,1)}; 
\legend{}; 
\end{axis}

\end{tikzpicture}

%% file: FinalFigs/Null_Dists_d_10_100_n_100_m_100_kernel__Dirichlet_Poly_5_2022_10_15_22_14_39cross.tex
\begin{tikzpicture}

\definecolor{darkorange25512714}{RGB}{255,127,14}
\definecolor{darkslategray38}{RGB}{38,38,38}
\definecolor{lightgray204}{RGB}{204,204,204}
\definecolor{steelblue31119180}{RGB}{31,119,180}

\begin{axis}[
axis line style={darkslategray38},
height=\figheight,
legend cell align={left},
legend style={fill opacity=0.8, draw opacity=1, text opacity=1, draw=none},
tick align=outside,
tick pos=left,
title={$\cmmd$~$(n/m=1)$},
width=\figwidth,
x grid style={lightgray204},
xmin=-6, xmax=6,
xtick style={color=darkslategray38},
y grid style={lightgray204},
ylabel=\textcolor{darkslategray38}{},
ymin=0, ymax=0.418868408525629,
ytick style={color=darkslategray38}, 
xticklabels=empty,
yticklabels=empty
]
\draw[draw=none,fill=steelblue31119180,fill opacity=0.8] (axis cs:-2.954021692276,0) rectangle (axis cs:-2.65674042701721,0.0215284267621705);
\addlegendimage{ybar,ybar legend,draw=none,fill=steelblue31119180,fill opacity=0.8}
\addlegendentry{d=10}

\draw[draw=none,fill=steelblue31119180,fill opacity=0.8] (axis cs:-2.21081829071045,0) rectangle (axis cs:-1.91353702545166,0.0753495057536032);
\draw[draw=none,fill=steelblue31119180,fill opacity=0.8] (axis cs:-1.46761500835419,0) rectangle (axis cs:-1.1703337430954,0.217975355930067);
\draw[draw=none,fill=steelblue31119180,fill opacity=0.8] (axis cs:-0.72441166639328,0) rectangle (axis cs:-0.427130401134491,0.382129605675184);
\draw[draw=none,fill=steelblue31119180,fill opacity=0.8] (axis cs:0.018791675567627,0) rectangle (axis cs:0.316072940826416,0.355219070064255);
\draw[draw=none,fill=steelblue31119180,fill opacity=0.8] (axis cs:0.761994957923889,0) rectangle (axis cs:1.05927622318268,0.207211140822409);
\draw[draw=none,fill=steelblue31119180,fill opacity=0.8] (axis cs:1.50519835948944,0) rectangle (axis cs:1.80247962474823,0.059203173595969);
\draw[draw=none,fill=steelblue31119180,fill opacity=0.8] (axis cs:2.2484016418457,0) rectangle (axis cs:2.54568290710449,0.0242194801074419);
\draw[draw=none,fill=steelblue31119180,fill opacity=0.8] (axis cs:2.99160504341125,0) rectangle (axis cs:3.28888630867004,0);
\draw[draw=none,fill=steelblue31119180,fill opacity=0.8] (axis cs:3.73480820655823,0) rectangle (axis cs:4.03208923339844,0.00269105334527132);
\draw[draw=none,fill=darkorange25512714,fill opacity=0.8] (axis cs:-2.65674018859863,0) rectangle (axis cs:-2.35945892333984,0.0349836934885271);
\addlegendimage{ybar,ybar legend,draw=none,fill=darkorange25512714,fill opacity=0.8}
\addlegendentry{d=100}

\draw[draw=none,fill=darkorange25512714,fill opacity=0.8] (axis cs:-1.91353690624237,0) rectangle (axis cs:-1.61625564098358,0.0645852906459456);
\draw[draw=none,fill=darkorange25512714,fill opacity=0.8] (axis cs:-1.17033362388611,0) rectangle (axis cs:-0.873052358627319,0.231430624814639);
\draw[draw=none,fill=darkorange25512714,fill opacity=0.8] (axis cs:-0.427130341529846,0) rectangle (axis cs:-0.129849076271057,0.320235373770048);
\draw[draw=none,fill=darkorange25512714,fill opacity=0.8] (axis cs:0.316073000431061,0) rectangle (axis cs:0.61335426568985,0.355219070064255);
\draw[draw=none,fill=darkorange25512714,fill opacity=0.8] (axis cs:1.05927634239197,0) rectangle (axis cs:1.35655760765076,0.212593248376238);
\draw[draw=none,fill=darkorange25512714,fill opacity=0.8] (axis cs:1.80247974395752,0) rectangle (axis cs:2.09976100921631,0.0968779204297674);
\draw[draw=none,fill=darkorange25512714,fill opacity=0.8] (axis cs:2.54568314552307,0) rectangle (axis cs:2.84296441078186,0.0242194801074419);
\draw[draw=none,fill=darkorange25512714,fill opacity=0.8] (axis cs:3.28888630867004,0) rectangle (axis cs:3.58616757392883,0.00538210841711525);
\draw[draw=none,fill=darkorange25512714,fill opacity=0.8] (axis cs:4.0320897102356,0) rectangle (axis cs:4.32937097549438,0);
\addplot [semithick, black]
table {%
-10 7.69459862670642e-23
-9.97997997997998 9.39820210218911e-23
-9.95995995995996 1.14743877987917e-22
-9.93993993993994 1.40036162642795e-22
-9.91991991991992 1.70834984871876e-22
-9.8998998998999 2.08324025950642e-22
-9.87987987987988 2.53938085193762e-22
-9.85985985985986 3.09415635142992e-22
-9.83983983983984 3.7686222201397e-22
-9.81981981981982 4.58826916995414e-22
-9.7997997997998 5.58394465795474e-22
-9.77977977977978 6.79296312742742e-22
-9.75975975975976 8.26044308669654e-22
-9.73973973973974 1.00409166885045e-21
-9.71971971971972 1.22002665237565e-21
-9.6996996996997 1.48180551599987e-21
-9.67967967967968 1.79903258756111e-21
-9.65965965965966 2.18329684678274e-21
-9.63963963963964 2.64857624243904e-21
-9.61961961961962 3.21172317125736e-21
-9.5995995995996 3.89304716291232e-21
-9.57957957957958 4.71701393695898e-21
-9.55955955955956 5.71308371631411e-21
-9.53953953953954 6.91671611023779e-21
-9.51951951951952 8.37057415073758e-21
-9.4994994994995 1.01259663374544e-20
-9.47947947947948 1.22445730038445e-20
-9.45945945945946 1.48005121824234e-20
-9.43943943943944 1.78828106797973e-20
-9.41941941941942 2.15983585814365e-20
-9.3993993993994 2.60754402558043e-20
-9.37937937937938 3.14679525475421e-20
-9.35935935935936 3.79604417474457e-20
-9.33933933933934 4.5774115701642e-20
-9.31931931931932 5.51740167796855e-20
-9.2992992992993 6.6477576193283e-20
-9.27927927927928 8.00648113242436e-20
-9.25925925925926 9.63904764362469e-20
-9.23923923923924 1.1599853476858e-19
-9.21921921921922 1.39539388139188e-19
-9.1991991991992 1.67790380698338e-19
-9.17917917917918 2.01680188581445e-19
-9.15915915915916 2.42317819504618e-19
-9.13913913913914 2.91027078875695e-19
-9.11911911911912 3.49387515332417e-19
-9.0990990990991 4.19283042962206e-19
-9.07907907907908 5.0295965472299e-19
-9.05905905905906 6.03093897535385e-19
-9.03903903903904 7.22874080905507e-19
-9.01901901901902 8.66096545670873e-19
-8.998998998999 1.03727973679138e-18
-8.97897897897898 1.24179931485728e-18
-8.95895895895896 1.48604811781133e-18
-8.93893893893894 1.77762546207239e-18
-8.91891891891892 2.12556106807901e-18
-8.8988988988989 2.54057982941549e-18
-8.87887887887888 3.03541474067764e-18
-8.85885885885886 3.62517658454128e-18
-8.83883883883884 4.32779048512422e-18
-8.81881881881882 5.16451119999019e-18
-8.7987987987988 6.16053109048305e-18
-8.77877877877878 7.34569713009337e-18
-8.75875875875876 8.75535614211525e-18
-8.73873873873874 1.04313507694241e-17
-8.71871871871872 1.24231925503958e-17
-8.6986986986987 1.47894429982973e-17
-8.67867867867868 1.75993388643364e-17
-8.65865865865866 2.09347039316773e-17
-8.63863863863864 2.48921968838275e-17
-8.61861861861862 2.95859531836935e-17
-8.5985985985986 3.51506886838449e-17
-8.57857857857858 4.17453440895294e-17
-8.55855855855856 4.95573626747691e-17
-8.53853853853854 5.88077091106193e-17
-8.51851851851852 6.97567552529606e-17
-8.4984984984985 8.27111796592674e-17
-8.47847847847848 9.8032051926925e-17
-8.45845845845846 1.16144301209627e-16
-8.43843843843844 1.37547801096493e-16
-8.41841841841842 1.62830341150177e-16
-8.3983983983984 1.92682799625235e-16
-8.37837837837838 2.27916883183252e-16
-8.35835835835836 2.69485858889318e-16
-8.33833833833834 3.18508772685576e-16
-8.31831831831832 3.76298728354137e-16
-8.2982982982983 4.44395893386326e-16
-8.27827827827828 5.24606005103494e-16
-8.25825825825826 6.19045274051723e-16
-8.23823823823824 7.30192724674871e-16
-8.21821821821822 8.60951178494166e-16
-8.1981981981982 1.01471827585831e-15
-8.17817817817818 1.19546915264237e-15
-8.15815815815816 1.40785264250169e-15
-8.13813813813814 1.65730316851105e-15
-8.11811811811812 1.95017082606148e-15
-8.0980980980981 2.29387254841619e-15
-8.07807807807808 2.69706769497134e-15
-8.05805805805806 3.16986191875719e-15
-8.03803803803804 3.72404376402735e-15
-8.01801801801802 4.3733591283238e-15
-7.997997997998 5.13382950919607e-15
-7.97797797797798 6.02412085866193e-15
-7.95795795795796 7.06597090549303e-15
-7.93793793793794 8.28468399583074e-15
-7.91791791791792 9.70970386854708e-15
-7.8978978978979 1.13752763482777e-14
-7.87787787787788 1.33212157347805e-14
-7.85785785785786 1.55937907247683e-14
-7.83783783783784 1.82467480586655e-14
-7.81781781781782 2.13424947819475e-14
-7.7977977977978 2.49534630966872e-14
-7.77777777777778 2.91636853079909e-14
-7.75775775775776 3.40706104038538e-14
-7.73773773773774 3.9787198415571e-14
-7.71771771771772 4.64443339685918e-14
-7.6976976976977 5.4193606440538e-14
-7.67767767767768 6.32105109959252e-14
-7.65765765765766 7.36981325813117e-14
-7.63763763763764 8.58913838706879e-14
-7.61761761761762 1.00061878296601e-13
-7.5975975975976 1.16523530854699e-13
-7.57757757757758 1.35638992516664e-13
-7.55755755755756 1.57827039041884e-13
-7.53753753753754 1.83571051982057e-13
-7.51751751751752 2.13428748996349e-13
-7.4974974974975 2.48043342543513e-13
-7.47747747747748 2.88156330935935e-13
-7.45745745745746 3.34622154016965e-13
-7.43743743743744 3.88424977793732e-13
-7.41741741741742 4.50697908714384e-13
-7.3973973973974 5.22744979473711e-13
-7.37737737737738 6.06066294885249e-13
-7.35735735735736 7.02386779168738e-13
-7.33733733733734 8.13689025751835e-13
-7.31731731731732 9.42250818252909e-13
-7.2972972972973 1.09068796768221e-12
-7.27727727727728 1.262003197176e-12
-7.25725725725726 1.45964190299847e-12
-7.23723723723724 1.68755573049416e-12
-7.21721721721722 1.95027502769792e-12
-7.1971971971972 2.25299137914218e-12
-7.17717717717718 2.60165157997871e-12
-7.15715715715716 3.00306458801653e-12
-7.13713713713714 3.4650231910834e-12
-7.11711711711712 3.99644235193919e-12
-7.0970970970971 4.60751644580953e-12
-7.07707707707708 5.30989788981514e-12
-7.05705705705706 6.11689998287646e-12
-7.03703703703704 7.0437271332242e-12
-7.01701701701702 8.10773605306645e-12
-6.996996996997 9.32873195138555e-12
-6.97697697697698 1.07293042619726e-11
-6.95695695695696 1.23352070109962e-11
-6.93693693693694 1.41757895636779e-11
-6.91691691691692 1.62844842008237e-11
-6.8968968968969 1.86993577716893e-11
-6.87687687687688 2.14637355595386e-11
-6.85685685685686 2.46269064908967e-11
-6.83683683683684 2.82449199306815e-11
-6.81681681681682 3.2381485546105e-11
-6.7967967967968 3.71089891068559e-11
-6.77677677677678 4.25096386334913e-11
-6.75675675675676 4.86767570277083e-11
-6.73673673673674 5.57162392366004e-11
-6.71671671671672 6.37481941394491e-11
-6.6966966966967 7.29087937236032e-11
-6.67667667667668 8.33523547614402e-11
-6.65665665665666 9.52536811418151e-11
-6.63663663663664 1.08810698278135e-10
-6.61661661661662 1.24247414645768e-10
-6.5965965965966 1.41817249531744e-10
-6.57657657657658 1.61806770551278e-10
-6.55655655655656 1.84539889444164e-10
-6.53653653653654 2.10382570159622e-10
-6.51651651651652 2.39748109325542e-10
-6.4964964964965 2.73103055937438e-10
-6.47647647647648 3.10973844559381e-10
-6.45645645645646 3.53954224575809e-10
-6.43643643643644 4.027135771479e-10
-6.41641641641642 4.58006221596996e-10
-6.3963963963964 5.20681824054169e-10
-6.37637637637638 5.91697033481538e-10
-6.35635635635636 6.72128483698846e-10
-6.33633633633634 7.63187314959523e-10
-6.31631631631632 8.66235385046001e-10
-6.2962962962963 9.8280335793813e-10
-6.27627627627628 1.11461087800766e-09
-6.25625625625626 1.26358905957513e-09
-6.23623623623624 1.43190554571787e-09
-6.21621621621622 1.62199241663862e-09
-6.1961961961962 1.83657725691024e-09
-6.17617617617618 2.0787177227378e-09
-6.15615615615616 2.35183998527873e-09
-6.13613613613614 2.65978146430928e-09
-6.11611611611612 3.00683830841829e-09
-6.0960960960961 3.39781812376754e-09
-6.07607607607608 3.83809850362696e-09
-6.05605605605606 4.33369196574699e-09
-6.03603603603604 4.89131796456919e-09
-6.01601601601602 5.51848271073395e-09
-5.995995995996 6.22356760178439e-09
-5.97597597597598 7.01592714588943e-09
-5.95595595595596 7.90599734535251e-09
-5.93593593593594 8.90541559921139e-09
-5.91591591591592 1.00271532849868e-08
-5.8958958958959 1.12856622892642e-08
-5.87587587587588 1.2697036876002e-08
-5.85585585585586 1.42791924110059e-08
-5.83583583583584 1.60520626017053e-08
-5.81581581581582 1.80378170640688e-08
-5.7957957957958 2.02611011941336e-08
-5.77577577577578 2.27493005011625e-08
-5.75575575575576 2.55328317539416e-08
-5.73573573573574 2.86454635022852e-08
-5.71571571571572 3.2124668763613e-08
-5.6956956956957 3.60120129107462e-08
-5.67567567567568 4.03535800631662e-08
-5.65565565565566 4.5200441571292e-08
-5.63563563563564 5.06091704933412e-08
-5.61561561561562 5.66424062986154e-08
-5.5955955955956 6.33694743912418e-08
-5.57557557557558 7.08670654362614e-08
-5.55555555555556 7.92199798873018e-08
-5.53553553553554 8.85219435638491e-08
-5.51551551551552 9.8876500608364e-08
-5.4954954954955 1.10397990671284e-07
-5.47547547547548 1.23212617727566e-07
-5.45545545545546 1.3745961852414e-07
-5.43543543543544 1.5329253929596e-07
-5.41541541541542 1.70880630071684e-07
-5.3953953953954 1.90410366621162e-07
-5.37537537537538 2.1208711087848e-07
-5.35535535535536 2.36136921509202e-07
-5.33533533533534 2.62808527181656e-07
-5.31531531531532 2.92375476052561e-07
-5.2952952952953 3.25138475990267e-07
-5.27527527527528 3.61427941137511e-07
-5.25525525525526 4.01606761563285e-07
-5.23523523523524 4.46073313973501e-07
-5.21521521521522 4.95264732746229e-07
-5.1951951951952 5.4966046193278e-07
-5.17517517517518 6.09786110324583e-07
-5.15515515515516 6.76217633231267e-07
-5.13513513513514 7.49585866251233e-07
-5.11511511511512 8.30581438046261e-07
-5.0950950950951 9.19960090959837e-07
-5.07507507507508 1.01854844024876e-06
-5.05505505505506 1.12725020473309e-06
-5.03503503503504 1.24705294381396e-06
-5.01501501501502 1.37903533806611e-06
-4.99499499499499 1.5243750529858e-06
-4.97497497497497 1.68435722796805e-06
-4.95495495495495 1.86038363520377e-06
-4.93493493493493 2.05398255592983e-06
-4.91491491491491 2.26681942433715e-06
-4.89489489489489 2.50070829244518e-06
-4.87487487487487 2.75762417238901e-06
-4.85485485485485 3.03971631583941e-06
-4.83483483483483 3.34932249368796e-06
-4.81481481481481 3.68898434268124e-06
-4.79479479479479 4.06146384937932e-06
-4.77477477477477 4.4697610456467e-06
-4.75475475475475 4.91713299385613e-06
-4.73473473473473 5.40711414409909e-06
-4.71471471471471 5.94353814994721e-06
-4.69469469469469 6.53056123369604e-06
-4.67467467467467 7.17268719654363e-06
-4.65465465465465 7.87479417380527e-06
-4.63463463463463 8.64216324004121e-06
-4.61461461461461 9.48050897386715e-06
-4.59459459459459 1.03960120972233e-05
-4.57457457457457 1.13953543089884e-05
-4.55455455455455 1.24857554380297e-05
-4.53453453453453 1.3675013046071e-05
-4.51451451451451 1.49715446161227e-05
-4.49449449449449 1.63844324676437e-05
-4.47447447447447 1.79234715450684e-05
-4.45445445445445 1.95992202318354e-05
-4.43443443443443 2.14230543475548e-05
-4.41441441441441 2.34072244914564e-05
-4.39439439439439 2.55649169007242e-05
-4.37437437437437 2.79103179977393e-05
-4.35435435435435 3.04586828055866e-05
-4.33433433433433 3.32264074164067e-05
-4.31431431431431 3.62311057022691e-05
-4.29429429429429 3.94916904631592e-05
-4.27427427427427 4.3028459211397e-05
-4.25425425425425 4.68631847962824e-05
-4.23423423423423 5.10192110769697e-05
-4.21421421421421 5.55215538554582e-05
-4.19419419419419 6.03970072851107e-05
-4.17417417417417 6.56742559732345e-05
-4.15415415415415 7.13839929989176e-05
-4.13413413413413 7.75590440694795e-05
-4.11411411411411 8.42344980404937e-05
-4.09409409409409 9.14478440253317e-05
-4.07407407407407 9.92391153205018e-05
-4.05405405405405 0.000107651040372646
-4.03403403403403 0.000116729201011866
-4.01401401401401 0.000126522198173995
-3.99399399399399 0.000137081825331481
-3.97397397397397 0.000148463249848567
-3.95395395395395 0.000160725202471485
-3.93393393393393 0.000173930175158222
-3.91391391391391 0.00018814462744512
-3.89389389389389 0.000203439201538965
-3.87387387387387 0.000219888946313312
-3.85385385385385 0.000237573550376443
-3.83383383383383 0.000256577584365551
-3.81381381381381 0.000276990752607344
-3.79379379379379 0.000298908154269281
-3.77377377377377 0.000322430554107926
-3.75375375375375 0.000347664662901427
-3.73373373373373 0.000374723427631836
-3.71371371371371 0.000403726331459719
-3.69369369369369 0.000434799703508357
-3.67367367367367 0.000468077038447599
-3.65365365365365 0.00050369932583812
-3.63363363363363 0.000541815389165405
-3.61361361361361 0.000582582234459159
-3.59359359359359 0.000626165408357979
-3.57357357357357 0.000672739365441021
-3.55355355355355 0.000722487844607978
-3.53353353353353 0.00077560425424601
-3.51351351351351 0.000832292065877155
-3.49349349349349 0.000892765215932443
-3.47347347347347 0.000957248515249216
-3.45345345345345 0.0010259780658361
-3.43343343343343 0.00109920168439588
-3.41341341341341 0.00117717933203981
-3.39339339339339 0.00126018354956833
-3.37337337337337 0.00134849989763212
-3.35335335335335 0.00144242740102448
-3.33333333333333 0.00154227899629111
-3.31331331331331 0.00164838198177652
-3.29329329329329 0.00176107846915772
-3.27327327327327 0.00188072583544552
-3.25325325325325 0.00200769717436226
-3.23323323323323 0.00214238174593163
-3.21321321321321 0.00228518542304204
-3.19319319319319 0.00243653113367012
-3.17317317317317 0.00259685929737497
-3.15315315315315 0.00276662825459747
-3.13313313313313 0.00294631468722261
-3.11311311311311 0.00313641402878609
-3.09309309309309 0.00333744086263052
-3.07307307307307 0.00354992930624086
-3.05305305305305 0.00377443337991422
-3.03303303303303 0.00401152735784579
-3.01301301301301 0.00426180609964128
-2.99299299299299 0.00452588536019618
-2.97297297297297 0.0048044020758154
-2.95295295295295 0.00509801462438215
-2.93293293293293 0.00540740305732385
-2.91291291291291 0.00573326930106519
-2.89289289289289 0.00607633732560526
-2.87287287287287 0.00643735327780636
-2.85285285285285 0.00681708557693873
-2.83283283283283 0.00721632496998623
-2.81281281281281 0.00763588454418632
-2.79279279279279 0.00807659969425075
-2.77277277277277 0.00853932804169477
-2.75275275275275 0.00902494930369032
-2.73273273273273 0.00953436510885489
-2.71271271271271 0.0100684987573917
-2.69269269269269 0.0106282949230102
-2.67267267267267 0.0112147192940778
-2.65265265265265 0.0118287581514852
-2.63263263263263 0.0124714178807513
-2.61261261261261 0.0131437244159435
-2.59259259259259 0.0138467226130541
-2.57257257257257 0.0145814755505492
-2.55255255255255 0.0153490637548887
-2.53253253253253 0.0161505843489182
-2.51251251251251 0.0169871501211409
-2.49249249249249 0.0178598885140022
-2.47247247247247 0.018769940529451
-2.45245245245245 0.0197184595501959
-2.43243243243243 0.0207066100752274
-2.41241241241241 0.0217355663683581
-2.39239239239239 0.0228065110187135
-2.37237237237237 0.0239206334123108
-2.35235235235235 0.0250791281140699
-2.33233233233233 0.0262831931598317
-2.31231231231231 0.0275340282581906
-2.29229229229229 0.0288328329022027
-2.27227227227227 0.0301808043912899
-2.25225225225225 0.0315791357639331
-2.23223223223223 0.033029013642035
-2.21221221221221 0.0345316159881232
-2.19219219219219 0.0360881097768736
-2.17217217217217 0.0376996485827434
-2.15215215215215 0.0393673700858293
-2.13213213213213 0.0410923934983949
-2.11211211211211 0.0428758169148479
-2.09209209209209 0.0447187145882915
-2.07207207207207 0.0466221341371235
-2.05205205205205 0.0485870936855041
-2.03203203203203 0.0506145789418747
-2.01201201201201 0.0527055402200587
-1.99199199199199 0.0548608894078376
-1.97197197197197 0.057081496888248
-1.95195195195195 0.059368188419199
-1.93193193193193 0.0617217419773594
-1.91191191191191 0.0641428845726061
-1.89189189189189 0.0666322890396674
-1.87187187187187 0.0691905708139176
-1.85185185185185 0.0718182846986055
-1.83183183183183 0.0745159216311036
-1.81181181181181 0.077283905456062
-1.79179179179179 0.0801225897136326
-1.77177177177177 0.083032254451193
-1.75175175175175 0.0860131030672496
-1.73173173173173 0.0890652591964251
-1.71171171171171 0.0921887636446459
-1.69169169169169 0.0953835713838294
-1.67167167167167 0.0986495486155338
-1.65165165165165 0.101986469913169
-1.63163163163163 0.10539401545248
-1.61161161161161 0.108871768340093
-1.59159159159159 0.112419212049971
-1.57157157157157 0.116035727977651
-1.55155155155155 0.119720593122119
-1.53153153153153 0.123472977905145
-1.51151151151151 0.127291944137829
-1.49149149149149 0.13117644314399
-1.47147147147147 0.135125314049902
-1.45145145145145 0.139137282249685
-1.43143143143143 0.143210958055468
-1.41141141141141 0.147344835541168
-1.39139139139139 0.151537291588457
-1.37137137137137 0.155786585143159
-1.35135135135135 0.160090856689972
-1.33133133133133 0.164448127952996
-1.31131131131131 0.168856301829129
-1.29129129129129 0.173313162560933
-1.27127127127127 0.17781637615506
-1.25125125125125 0.182363491051798
-1.23123123123123 0.186951939050736
-1.21121121121121 0.191579036496956
-1.19119119119119 0.1962419857315
-1.17117117117117 0.200937876809264
-1.15115115115115 0.205663689486728
-1.13113113113113 0.210416295481265
-1.11111111111111 0.215192461003031
-1.09109109109109 0.219988849559688
-1.07107107107107 0.224802025033432
-1.05105105105105 0.229628455029052
-1.03103103103103 0.234464514490888
-1.01101101101101 0.239306489585817
-0.990990990990991 0.24415058184851
-0.970970970970971 0.24899291258444
-0.950950950950951 0.253829527525259
-0.930930930930931 0.258656401730343
-0.910910910910911 0.2634694447275
-0.890890890890891 0.268264505884996
-0.870870870870871 0.273037380006279
-0.850850850850851 0.277783813137949
-0.830830830830831 0.282499508580786
-0.810810810810811 0.287180133092853
-0.790790790790791 0.291821323272996
-0.77077077077077 0.296418692112302
-0.75075075075075 0.300967835700437
-0.73073073073073 0.305464340073112
-0.71071071071071 0.309903788186304
-0.69069069069069 0.314281767002296
-0.67067067067067 0.318593874672039
-0.65065065065065 0.322835727797843
-0.63063063063063 0.327002968759958
-0.61061061061061 0.331091273090187
-0.59059059059059 0.33509635687531
-0.57057057057057 0.339013984172804
-0.55055055055055 0.34283997442106
-0.53053053053053 0.346570209826128
-0.51051051051051 0.350200642706842
-0.49049049049049 0.353727302780113
-0.47047047047047 0.357146304368113
-0.45045045045045 0.360453853509139
-0.43043043043043 0.363646254953996
-0.41041041041041 0.366719919029892
-0.39039039039039 0.369671368354051
-0.37037037037037 0.372497244379499
-0.35035035035035 0.375194313755802
-0.33033033033033 0.377759474487924
-0.31031031031031 0.38018976187679
-0.29029029029029 0.382482354225654
-0.27027027027027 0.384634578296894
-0.25025025025025 0.386643914504485
-0.23023023023023 0.388508001828027
-0.21021021021021 0.390224642434919
-0.19019019019019 0.391791805998011
-0.17017017017017 0.393207633696876
-0.15015015015015 0.394470441891644
-0.13013013013013 0.395578725459258
-0.11011011011011 0.396531160782876
-0.0900900900900901 0.397326608386124
-0.07007007007007 0.397964115204853
-0.05005005005005 0.398442916490068
-0.03003003003003 0.398762437336696
-0.01001001001001 0.398922293833933
0.01001001001001 0.398922293833933
0.03003003003003 0.398762437336696
0.05005005005005 0.398442916490068
0.07007007007007 0.397964115204853
0.0900900900900901 0.397326608386124
0.11011011011011 0.396531160782876
0.13013013013013 0.395578725459258
0.15015015015015 0.394470441891644
0.17017017017017 0.393207633696876
0.19019019019019 0.391791805998011
0.21021021021021 0.390224642434919
0.23023023023023 0.388508001828027
0.25025025025025 0.386643914504485
0.27027027027027 0.384634578296894
0.29029029029029 0.382482354225654
0.31031031031031 0.38018976187679
0.33033033033033 0.377759474487924
0.35035035035035 0.375194313755802
0.37037037037037 0.372497244379499
0.39039039039039 0.369671368354051
0.41041041041041 0.366719919029892
0.43043043043043 0.363646254953996
0.45045045045045 0.360453853509139
0.47047047047047 0.357146304368113
0.49049049049049 0.353727302780113
0.51051051051051 0.350200642706842
0.53053053053053 0.346570209826128
0.55055055055055 0.34283997442106
0.57057057057057 0.339013984172804
0.59059059059059 0.33509635687531
0.61061061061061 0.331091273090187
0.63063063063063 0.327002968759958
0.65065065065065 0.322835727797843
0.67067067067067 0.318593874672039
0.69069069069069 0.314281767002296
0.71071071071071 0.309903788186304
0.73073073073073 0.305464340073112
0.75075075075075 0.300967835700437
0.77077077077077 0.296418692112302
0.790790790790791 0.291821323272996
0.810810810810811 0.287180133092853
0.830830830830831 0.282499508580786
0.850850850850851 0.277783813137949
0.870870870870871 0.273037380006279
0.890890890890891 0.268264505884996
0.910910910910911 0.2634694447275
0.930930930930931 0.258656401730343
0.950950950950951 0.253829527525259
0.970970970970971 0.24899291258444
0.990990990990991 0.24415058184851
1.01101101101101 0.239306489585817
1.03103103103103 0.234464514490888
1.05105105105105 0.229628455029052
1.07107107107107 0.224802025033432
1.09109109109109 0.219988849559688
1.11111111111111 0.215192461003031
1.13113113113113 0.210416295481265
1.15115115115115 0.205663689486728
1.17117117117117 0.200937876809264
1.19119119119119 0.1962419857315
1.21121121121121 0.191579036496956
1.23123123123123 0.186951939050736
1.25125125125125 0.182363491051798
1.27127127127127 0.17781637615506
1.29129129129129 0.173313162560933
1.31131131131131 0.168856301829129
1.33133133133133 0.164448127952996
1.35135135135135 0.160090856689972
1.37137137137137 0.155786585143159
1.39139139139139 0.151537291588457
1.41141141141141 0.147344835541168
1.43143143143143 0.143210958055468
1.45145145145145 0.139137282249685
1.47147147147147 0.135125314049902
1.49149149149149 0.13117644314399
1.51151151151151 0.127291944137829
1.53153153153153 0.123472977905145
1.55155155155155 0.119720593122119
1.57157157157157 0.116035727977651
1.59159159159159 0.112419212049971
1.61161161161161 0.108871768340093
1.63163163163163 0.10539401545248
1.65165165165165 0.101986469913169
1.67167167167167 0.0986495486155338
1.69169169169169 0.0953835713838294
1.71171171171171 0.0921887636446459
1.73173173173173 0.0890652591964251
1.75175175175175 0.0860131030672496
1.77177177177177 0.083032254451193
1.79179179179179 0.0801225897136326
1.81181181181181 0.077283905456062
1.83183183183183 0.0745159216311036
1.85185185185185 0.0718182846986055
1.87187187187187 0.0691905708139176
1.89189189189189 0.0666322890396674
1.91191191191191 0.0641428845726061
1.93193193193193 0.0617217419773594
1.95195195195195 0.059368188419199
1.97197197197197 0.057081496888248
1.99199199199199 0.0548608894078376
2.01201201201201 0.0527055402200588
2.03203203203203 0.0506145789418748
2.05205205205205 0.0485870936855042
2.07207207207207 0.0466221341371236
2.09209209209209 0.0447187145882916
2.11211211211211 0.0428758169148479
2.13213213213213 0.041092393498395
2.15215215215215 0.0393673700858294
2.17217217217217 0.0376996485827434
2.19219219219219 0.0360881097768737
2.21221221221221 0.0345316159881233
2.23223223223223 0.033029013642035
2.25225225225225 0.0315791357639332
2.27227227227227 0.03018080439129
2.29229229229229 0.0288328329022028
2.31231231231231 0.0275340282581906
2.33233233233233 0.0262831931598317
2.35235235235235 0.02507912811407
2.37237237237237 0.0239206334123108
2.39239239239239 0.0228065110187136
2.41241241241241 0.0217355663683581
2.43243243243243 0.0207066100752275
2.45245245245245 0.0197184595501959
2.47247247247247 0.0187699405294511
2.49249249249249 0.0178598885140022
2.51251251251251 0.016987150121141
2.53253253253253 0.0161505843489182
2.55255255255255 0.0153490637548888
2.57257257257257 0.0145814755505493
2.59259259259259 0.0138467226130541
2.61261261261261 0.0131437244159435
2.63263263263263 0.0124714178807514
2.65265265265265 0.0118287581514852
2.67267267267267 0.0112147192940778
2.69269269269269 0.0106282949230103
2.71271271271271 0.0100684987573917
2.73273273273273 0.00953436510885491
2.75275275275275 0.00902494930369034
2.77277277277277 0.00853932804169479
2.79279279279279 0.00807659969425077
2.81281281281281 0.00763588454418634
2.83283283283283 0.00721632496998621
2.85285285285285 0.00681708557693871
2.87287287287287 0.00643735327780635
2.89289289289289 0.00607633732560524
2.91291291291291 0.00573326930106518
2.93293293293293 0.00540740305732384
2.95295295295295 0.00509801462438214
2.97297297297297 0.00480440207581539
2.99299299299299 0.00452588536019617
3.01301301301301 0.00426180609964127
3.03303303303303 0.00401152735784578
3.05305305305305 0.00377443337991421
3.07307307307307 0.00354992930624085
3.09309309309309 0.00333744086263051
3.11311311311311 0.00313641402878608
3.13313313313313 0.0029463146872226
3.15315315315315 0.00276662825459746
3.17317317317317 0.00259685929737496
3.19319319319319 0.00243653113367011
3.21321321321321 0.00228518542304203
3.23323323323323 0.00214238174593163
3.25325325325325 0.00200769717436225
3.27327327327327 0.00188072583544551
3.29329329329329 0.00176107846915771
3.31331331331331 0.00164838198177652
3.33333333333333 0.0015422789962911
3.35335335335335 0.00144242740102448
3.37337337337337 0.00134849989763212
3.39339339339339 0.00126018354956833
3.41341341341341 0.00117717933203981
3.43343343343343 0.00109920168439588
3.45345345345345 0.0010259780658361
3.47347347347347 0.000957248515249212
3.49349349349349 0.000892765215932441
3.51351351351351 0.000832292065877152
3.53353353353353 0.000775604254246008
3.55355355355355 0.000722487844607976
3.57357357357357 0.000672739365441019
3.59359359359359 0.000626165408357979
3.61361361361361 0.000582582234459159
3.63363363363363 0.000541815389165405
3.65365365365365 0.00050369932583812
3.67367367367367 0.000468077038447599
3.69369369369369 0.000434799703508357
3.71371371371371 0.000403726331459719
3.73373373373373 0.000374723427631836
3.75375375375375 0.000347664662901427
3.77377377377377 0.000322430554107926
3.79379379379379 0.000298908154269281
3.81381381381381 0.000276990752607344
3.83383383383383 0.000256577584365551
3.85385385385385 0.000237573550376443
3.87387387387387 0.000219888946313312
3.89389389389389 0.000203439201538965
3.91391391391391 0.00018814462744512
3.93393393393393 0.000173930175158222
3.95395395395395 0.000160725202471485
3.97397397397397 0.000148463249848567
3.99399399399399 0.000137081825331481
4.01401401401401 0.000126522198173995
4.03403403403403 0.000116729201011866
4.05405405405405 0.000107651040372646
4.07407407407407 9.92391153205018e-05
4.09409409409409 9.14478440253317e-05
4.11411411411411 8.42344980404937e-05
4.13413413413413 7.75590440694795e-05
4.15415415415415 7.13839929989176e-05
4.17417417417417 6.56742559732345e-05
4.19419419419419 6.03970072851107e-05
4.21421421421421 5.55215538554582e-05
4.23423423423423 5.10192110769697e-05
4.25425425425425 4.68631847962824e-05
4.27427427427427 4.3028459211397e-05
4.29429429429429 3.94916904631592e-05
4.31431431431431 3.62311057022691e-05
4.33433433433433 3.32264074164067e-05
4.35435435435435 3.04586828055866e-05
4.37437437437437 2.79103179977393e-05
4.39439439439439 2.55649169007242e-05
4.41441441441441 2.34072244914564e-05
4.43443443443443 2.14230543475548e-05
4.45445445445445 1.95992202318354e-05
4.47447447447447 1.79234715450684e-05
4.49449449449449 1.63844324676437e-05
4.51451451451451 1.49715446161227e-05
4.53453453453453 1.3675013046071e-05
4.55455455455455 1.24857554380297e-05
4.57457457457457 1.13953543089884e-05
4.59459459459459 1.03960120972233e-05
4.61461461461461 9.48050897386715e-06
4.63463463463463 8.64216324004121e-06
4.65465465465465 7.87479417380527e-06
4.67467467467467 7.17268719654363e-06
4.69469469469469 6.53056123369604e-06
4.71471471471471 5.94353814994721e-06
4.73473473473473 5.40711414409909e-06
4.75475475475475 4.91713299385613e-06
4.77477477477477 4.4697610456467e-06
4.79479479479479 4.06146384937932e-06
4.81481481481481 3.68898434268124e-06
4.83483483483483 3.34932249368796e-06
4.85485485485485 3.03971631583941e-06
4.87487487487487 2.75762417238901e-06
4.89489489489489 2.50070829244518e-06
4.91491491491491 2.26681942433715e-06
4.93493493493493 2.05398255592983e-06
4.95495495495495 1.86038363520377e-06
4.97497497497497 1.68435722796805e-06
4.99499499499499 1.5243750529858e-06
5.01501501501502 1.37903533806611e-06
5.03503503503504 1.24705294381396e-06
5.05505505505506 1.12725020473309e-06
5.07507507507508 1.01854844024876e-06
5.0950950950951 9.19960090959837e-07
5.11511511511512 8.30581438046261e-07
5.13513513513514 7.49585866251233e-07
5.15515515515516 6.76217633231267e-07
5.17517517517518 6.09786110324583e-07
5.1951951951952 5.4966046193278e-07
5.21521521521522 4.95264732746229e-07
5.23523523523524 4.46073313973501e-07
5.25525525525526 4.01606761563285e-07
5.27527527527528 3.61427941137511e-07
5.2952952952953 3.25138475990267e-07
5.31531531531532 2.92375476052561e-07
5.33533533533534 2.62808527181656e-07
5.35535535535536 2.36136921509202e-07
5.37537537537538 2.1208711087848e-07
5.3953953953954 1.90410366621162e-07
5.41541541541542 1.70880630071684e-07
5.43543543543544 1.5329253929596e-07
5.45545545545546 1.3745961852414e-07
5.47547547547548 1.23212617727566e-07
5.4954954954955 1.10397990671284e-07
5.51551551551552 9.8876500608364e-08
5.53553553553554 8.85219435638491e-08
5.55555555555556 7.92199798873018e-08
5.57557557557558 7.08670654362614e-08
5.5955955955956 6.33694743912418e-08
5.61561561561562 5.66424062986154e-08
5.63563563563564 5.06091704933412e-08
5.65565565565566 4.5200441571292e-08
5.67567567567568 4.03535800631662e-08
5.6956956956957 3.60120129107462e-08
5.71571571571572 3.2124668763613e-08
5.73573573573574 2.86454635022852e-08
5.75575575575576 2.55328317539416e-08
5.77577577577578 2.27493005011625e-08
5.7957957957958 2.02611011941336e-08
5.81581581581582 1.80378170640688e-08
5.83583583583584 1.60520626017053e-08
5.85585585585586 1.42791924110059e-08
5.87587587587588 1.2697036876002e-08
5.8958958958959 1.12856622892642e-08
5.91591591591592 1.00271532849868e-08
5.93593593593594 8.90541559921139e-09
5.95595595595596 7.90599734535251e-09
5.97597597597598 7.01592714588943e-09
5.995995995996 6.22356760178439e-09
6.01601601601602 5.5184827107339e-09
6.03603603603604 4.89131796456914e-09
6.05605605605606 4.33369196574694e-09
6.07607607607608 3.83809850362692e-09
6.0960960960961 3.3978181237675e-09
6.11611611611612 3.00683830841826e-09
6.13613613613614 2.65978146430924e-09
6.15615615615616 2.35183998527871e-09
6.17617617617618 2.07871772273778e-09
6.1961961961962 1.83657725691022e-09
6.21621621621622 1.6219924166386e-09
6.23623623623624 1.43190554571785e-09
6.25625625625626 1.26358905957511e-09
6.27627627627628 1.11461087800764e-09
6.2962962962963 9.82803357938119e-10
6.31631631631632 8.66235385045992e-10
6.33633633633634 7.63187314959515e-10
6.35635635635636 6.72128483698836e-10
6.37637637637638 5.91697033481532e-10
6.3963963963964 5.20681824054164e-10
6.41641641641642 4.58006221596989e-10
6.43643643643644 4.02713577147895e-10
6.45645645645646 3.53954224575805e-10
6.47647647647648 3.10973844559378e-10
6.4964964964965 2.73103055937435e-10
6.51651651651652 2.39748109325539e-10
6.53653653653654 2.10382570159619e-10
6.55655655655656 1.84539889444162e-10
6.57657657657658 1.61806770551276e-10
6.5965965965966 1.41817249531742e-10
6.61661661661662 1.24247414645767e-10
6.63663663663664 1.08810698278133e-10
6.65665665665666 9.52536811418137e-11
6.67667667667668 8.33523547614393e-11
6.6966966966967 7.29087937236022e-11
6.71671671671672 6.37481941394484e-11
6.73673673673674 5.57162392365998e-11
6.75675675675676 4.86767570277076e-11
6.77677677677678 4.25096386334907e-11
6.7967967967968 3.71089891068556e-11
6.81681681681682 3.23814855461048e-11
6.83683683683684 2.82449199306813e-11
6.85685685685686 2.46269064908966e-11
6.87687687687688 2.14637355595386e-11
6.8968968968969 1.86993577716892e-11
6.91691691691692 1.62844842008236e-11
6.93693693693694 1.41757895636779e-11
6.95695695695696 1.23352070109961e-11
6.97697697697698 1.07293042619725e-11
6.996996996997 9.32873195138548e-12
7.01701701701702 8.10773605306642e-12
7.03703703703704 7.04372713322415e-12
7.05705705705706 6.11689998287643e-12
7.07707707707708 5.30989788981512e-12
7.0970970970971 4.6075164458095e-12
7.11711711711712 3.99644235193917e-12
7.13713713713714 3.46502319108337e-12
7.15715715715716 3.00306458801651e-12
7.17717717717718 2.60165157997869e-12
7.1971971971972 2.25299137914217e-12
7.21721721721722 1.95027502769791e-12
7.23723723723724 1.68755573049414e-12
7.25725725725726 1.45964190299846e-12
7.27727727727728 1.262003197176e-12
7.2972972972973 1.0906879676822e-12
7.31731731731732 9.42250818252902e-13
7.33733733733734 8.13689025751829e-13
7.35735735735736 7.02386779168733e-13
7.37737737737738 6.06066294885245e-13
7.3973973973974 5.22744979473708e-13
7.41741741741742 4.50697908714381e-13
7.43743743743744 3.88424977793729e-13
7.45745745745746 3.34622154016963e-13
7.47747747747748 2.88156330935933e-13
7.4974974974975 2.48043342543511e-13
7.51751751751752 2.13428748996348e-13
7.53753753753754 1.83571051982055e-13
7.55755755755756 1.57827039041883e-13
7.57757757757758 1.35638992516663e-13
7.5975975975976 1.16523530854698e-13
7.61761761761762 1.000618782966e-13
7.63763763763764 8.58913838706873e-14
7.65765765765766 7.36981325813112e-14
7.67767767767768 6.32105109959248e-14
7.6976976976977 5.41936064405376e-14
7.71771771771772 4.64443339685915e-14
7.73773773773774 3.97871984155707e-14
7.75775775775776 3.40706104038536e-14
7.77777777777778 2.91636853079907e-14
7.7977977977978 2.4953463096687e-14
7.81781781781782 2.13424947819474e-14
7.83783783783784 1.82467480586654e-14
7.85785785785786 1.55937907247682e-14
7.87787787787788 1.33212157347804e-14
7.8978978978979 1.13752763482777e-14
7.91791791791792 9.70970386854701e-15
7.93793793793794 8.28468399583068e-15
7.95795795795796 7.06597090549298e-15
7.97797797797798 6.02412085866189e-15
7.997997997998 5.13382950919603e-15
8.01801801801802 4.3733591283238e-15
8.03803803803804 3.72404376402735e-15
8.05805805805806 3.16986191875719e-15
8.07807807807808 2.69706769497134e-15
8.0980980980981 2.29387254841619e-15
8.11811811811812 1.95017082606148e-15
8.13813813813814 1.65730316851105e-15
8.15815815815816 1.40785264250169e-15
8.17817817817818 1.19546915264237e-15
8.1981981981982 1.01471827585831e-15
8.21821821821822 8.60951178494166e-16
8.23823823823824 7.30192724674871e-16
8.25825825825826 6.19045274051723e-16
8.27827827827828 5.24606005103494e-16
8.2982982982983 4.44395893386326e-16
8.31831831831832 3.76298728354137e-16
8.33833833833834 3.18508772685576e-16
8.35835835835836 2.69485858889318e-16
8.37837837837838 2.27916883183252e-16
8.3983983983984 1.92682799625235e-16
8.41841841841842 1.62830341150177e-16
8.43843843843844 1.37547801096493e-16
8.45845845845846 1.16144301209627e-16
8.47847847847848 9.8032051926925e-17
8.4984984984985 8.27111796592674e-17
8.51851851851852 6.97567552529606e-17
8.53853853853854 5.88077091106193e-17
8.55855855855856 4.95573626747691e-17
8.57857857857858 4.17453440895294e-17
8.5985985985986 3.51506886838449e-17
8.61861861861862 2.95859531836935e-17
8.63863863863864 2.48921968838275e-17
8.65865865865866 2.09347039316773e-17
8.67867867867868 1.75993388643364e-17
8.6986986986987 1.47894429982973e-17
8.71871871871872 1.24231925503958e-17
8.73873873873874 1.04313507694241e-17
8.75875875875876 8.75535614211525e-18
8.77877877877878 7.34569713009337e-18
8.7987987987988 6.16053109048305e-18
8.81881881881882 5.16451119999019e-18
8.83883883883884 4.32779048512422e-18
8.85885885885886 3.62517658454128e-18
8.87887887887888 3.03541474067764e-18
8.8988988988989 2.54057982941549e-18
8.91891891891892 2.12556106807901e-18
8.93893893893894 1.77762546207239e-18
8.95895895895896 1.48604811781133e-18
8.97897897897898 1.24179931485728e-18
8.998998998999 1.03727973679138e-18
9.01901901901902 8.66096545670873e-19
9.03903903903904 7.22874080905507e-19
9.05905905905906 6.03093897535385e-19
9.07907907907908 5.0295965472299e-19
9.0990990990991 4.19283042962206e-19
9.11911911911912 3.49387515332417e-19
9.13913913913914 2.91027078875695e-19
9.15915915915916 2.42317819504618e-19
9.17917917917918 2.01680188581445e-19
9.1991991991992 1.67790380698338e-19
9.21921921921922 1.39539388139188e-19
9.23923923923924 1.1599853476858e-19
9.25925925925926 9.63904764362469e-20
9.27927927927928 8.00648113242436e-20
9.2992992992993 6.6477576193283e-20
9.31931931931932 5.51740167796855e-20
9.33933933933934 4.5774115701642e-20
9.35935935935936 3.79604417474457e-20
9.37937937937938 3.14679525475421e-20
9.3993993993994 2.60754402558043e-20
9.41941941941942 2.15983585814365e-20
9.43943943943944 1.78828106797973e-20
9.45945945945946 1.48005121824234e-20
9.47947947947948 1.22445730038445e-20
9.4994994994995 1.01259663374544e-20
9.51951951951952 8.37057415073758e-21
9.53953953953954 6.91671611023779e-21
9.55955955955956 5.71308371631411e-21
9.57957957957958 4.71701393695898e-21
9.5995995995996 3.89304716291232e-21
9.61961961961962 3.21172317125736e-21
9.63963963963964 2.64857624243904e-21
9.65965965965966 2.18329684678274e-21
9.67967967967968 1.79903258756111e-21
9.6996996996997 1.48180551599987e-21
9.71971971971972 1.22002665237565e-21
9.73973973973974 1.00409166885045e-21
9.75975975975976 8.26044308669654e-22
9.77977977977978 6.79296312742742e-22
9.7997997997998 5.58394465795474e-22
9.81981981981982 4.58826916995414e-22
9.83983983983984 3.7686222201397e-22
9.85985985985986 3.09415635142992e-22
9.87987987987988 2.53938085193762e-22
9.8998998998999 2.08324025950642e-22
9.91991991991992 1.70834984871876e-22
9.93993993993994 1.40036162642795e-22
9.95995995995996 1.14743877987917e-22
9.97997997997998 9.39820210218911e-23
10 7.69459862670642e-23
};
\addlegendentry{N(0,1)}; 
\legend{}; 
\end{axis}

\end{tikzpicture}

%% file: FinalFigs/Null_Dists_d_10_100_n_100_m_20_kernel__Dirichlet_RBF_2022_10_15_22_20_23mmd.tex
\begin{tikzpicture}

\definecolor{darkorange25512714}{RGB}{255,127,14}
\definecolor{darkslategray38}{RGB}{38,38,38}
\definecolor{lightgray204}{RGB}{204,204,204}
\definecolor{steelblue31119180}{RGB}{31,119,180}

\begin{axis}[
axis line style={darkslategray38},
height=\figheight,
legend cell align={left},
legend style={fill opacity=0.8, draw opacity=1, text opacity=1, draw=none},
tick align=outside,
tick pos=left,
title={$\dmmd$~$(n/m=5)$},
width=\figwidth,
x grid style={lightgray204},
xmin=-6, xmax=6,
xtick style={color=darkslategray38},
y grid style={lightgray204},
ylabel=\textcolor{darkslategray38}{Probability density},
ymin=0, ymax=1.22692691761745,
ytick style={color=darkslategray38}, 
xticklabels=empty,
yticklabels=empty
]
\draw[draw=none,fill=steelblue31119180,fill opacity=0.8] (axis cs:-1.17582952976227,0) rectangle (axis cs:-1.05122554302216,0.205450850903063);
\addlegendimage{ybar,ybar legend,draw=none,fill=steelblue31119180,fill opacity=0.8}
\addlegendentry{mmd (d=10)}

\draw[draw=none,fill=steelblue31119180,fill opacity=0.8] (axis cs:-0.864319562911987,0) rectangle (axis cs:-0.739715576171875,0.404481440109294);
\draw[draw=none,fill=steelblue31119180,fill opacity=0.8] (axis cs:-0.552809596061707,0) rectangle (axis cs:-0.428205609321594,0.603511874527746);
\draw[draw=none,fill=steelblue31119180,fill opacity=0.8] (axis cs:-0.241299569606781,0) rectangle (axis cs:-0.116695582866669,0.712657707250323);
\draw[draw=none,fill=steelblue31119180,fill opacity=0.8] (axis cs:0.0702104270458221,0) rectangle (axis cs:0.194814413785934,0.500786496986714);
\draw[draw=none,fill=steelblue31119180,fill opacity=0.8] (axis cs:0.381720423698425,0) rectangle (axis cs:0.506324410438538,0.385220345443242);
\draw[draw=none,fill=steelblue31119180,fill opacity=0.8] (axis cs:0.693230450153351,0) rectangle (axis cs:0.817834436893463,0.23755259181022);
\draw[draw=none,fill=steelblue31119180,fill opacity=0.8] (axis cs:1.00474035739899,0) rectangle (axis cs:1.1293443441391,0.0770440838303417);
\draw[draw=none,fill=steelblue31119180,fill opacity=0.8] (axis cs:1.31625044345856,0) rectangle (axis cs:1.44085454940796,0.0706237164847141);
\draw[draw=none,fill=steelblue31119180,fill opacity=0.8] (axis cs:1.62776052951813,0) rectangle (axis cs:1.75236451625824,0.0128406806383903);
\draw[draw=none,fill=darkorange25512714,fill opacity=0.8] (axis cs:-1.05122554302216,0) rectangle (axis cs:-0.926621556282043,0.0256813563628828);
\addlegendimage{ybar,ybar legend,draw=none,fill=darkorange25512714,fill opacity=0.8}
\addlegendentry{mmd (d=100)}

\draw[draw=none,fill=darkorange25512714,fill opacity=0.8] (axis cs:-0.739715576171875,0) rectangle (axis cs:-0.615111589431763,0.134827146703098);
\draw[draw=none,fill=darkorange25512714,fill opacity=0.8] (axis cs:-0.428205579519272,0) rectangle (axis cs:-0.30360159277916,0.571410179074143);
\draw[draw=none,fill=darkorange25512714,fill opacity=0.8] (axis cs:-0.116695553064346,0) rectangle (axis cs:0.00790843367576599,1.16850182630233);
\draw[draw=none,fill=darkorange25512714,fill opacity=0.8] (axis cs:0.194814443588257,0) rectangle (axis cs:0.319418430328369,0.866745860169312);
\draw[draw=none,fill=darkorange25512714,fill opacity=0.8] (axis cs:0.506324410438538,0) rectangle (axis cs:0.63092839717865,0.333857632717477);
\draw[draw=none,fill=darkorange25512714,fill opacity=0.8] (axis cs:0.817834436893463,0) rectangle (axis cs:0.942438423633575,0.0963051047879271);
\draw[draw=none,fill=darkorange25512714,fill opacity=0.8] (axis cs:1.1293443441391,0) rectangle (axis cs:1.25394833087921,0.0128406806383903);
\draw[draw=none,fill=darkorange25512714,fill opacity=0.8] (axis cs:1.44085443019867,0) rectangle (axis cs:1.56545853614807,0);
\draw[draw=none,fill=darkorange25512714,fill opacity=0.8] (axis cs:1.75236451625824,0) rectangle (axis cs:1.87696850299835,0);
\addplot [semithick, black]
table {%
-10 7.69459862670642e-23
-9.97997997997998 9.39820210218911e-23
-9.95995995995996 1.14743877987917e-22
-9.93993993993994 1.40036162642795e-22
-9.91991991991992 1.70834984871876e-22
-9.8998998998999 2.08324025950642e-22
-9.87987987987988 2.53938085193762e-22
-9.85985985985986 3.09415635142992e-22
-9.83983983983984 3.7686222201397e-22
-9.81981981981982 4.58826916995414e-22
-9.7997997997998 5.58394465795474e-22
-9.77977977977978 6.79296312742742e-22
-9.75975975975976 8.26044308669654e-22
-9.73973973973974 1.00409166885045e-21
-9.71971971971972 1.22002665237565e-21
-9.6996996996997 1.48180551599987e-21
-9.67967967967968 1.79903258756111e-21
-9.65965965965966 2.18329684678274e-21
-9.63963963963964 2.64857624243904e-21
-9.61961961961962 3.21172317125736e-21
-9.5995995995996 3.89304716291232e-21
-9.57957957957958 4.71701393695898e-21
-9.55955955955956 5.71308371631411e-21
-9.53953953953954 6.91671611023779e-21
-9.51951951951952 8.37057415073758e-21
-9.4994994994995 1.01259663374544e-20
-9.47947947947948 1.22445730038445e-20
-9.45945945945946 1.48005121824234e-20
-9.43943943943944 1.78828106797973e-20
-9.41941941941942 2.15983585814365e-20
-9.3993993993994 2.60754402558043e-20
-9.37937937937938 3.14679525475421e-20
-9.35935935935936 3.79604417474457e-20
-9.33933933933934 4.5774115701642e-20
-9.31931931931932 5.51740167796855e-20
-9.2992992992993 6.6477576193283e-20
-9.27927927927928 8.00648113242436e-20
-9.25925925925926 9.63904764362469e-20
-9.23923923923924 1.1599853476858e-19
-9.21921921921922 1.39539388139188e-19
-9.1991991991992 1.67790380698338e-19
-9.17917917917918 2.01680188581445e-19
-9.15915915915916 2.42317819504618e-19
-9.13913913913914 2.91027078875695e-19
-9.11911911911912 3.49387515332417e-19
-9.0990990990991 4.19283042962206e-19
-9.07907907907908 5.0295965472299e-19
-9.05905905905906 6.03093897535385e-19
-9.03903903903904 7.22874080905507e-19
-9.01901901901902 8.66096545670873e-19
-8.998998998999 1.03727973679138e-18
-8.97897897897898 1.24179931485728e-18
-8.95895895895896 1.48604811781133e-18
-8.93893893893894 1.77762546207239e-18
-8.91891891891892 2.12556106807901e-18
-8.8988988988989 2.54057982941549e-18
-8.87887887887888 3.03541474067764e-18
-8.85885885885886 3.62517658454128e-18
-8.83883883883884 4.32779048512422e-18
-8.81881881881882 5.16451119999019e-18
-8.7987987987988 6.16053109048305e-18
-8.77877877877878 7.34569713009337e-18
-8.75875875875876 8.75535614211525e-18
-8.73873873873874 1.04313507694241e-17
-8.71871871871872 1.24231925503958e-17
-8.6986986986987 1.47894429982973e-17
-8.67867867867868 1.75993388643364e-17
-8.65865865865866 2.09347039316773e-17
-8.63863863863864 2.48921968838275e-17
-8.61861861861862 2.95859531836935e-17
-8.5985985985986 3.51506886838449e-17
-8.57857857857858 4.17453440895294e-17
-8.55855855855856 4.95573626747691e-17
-8.53853853853854 5.88077091106193e-17
-8.51851851851852 6.97567552529606e-17
-8.4984984984985 8.27111796592674e-17
-8.47847847847848 9.8032051926925e-17
-8.45845845845846 1.16144301209627e-16
-8.43843843843844 1.37547801096493e-16
-8.41841841841842 1.62830341150177e-16
-8.3983983983984 1.92682799625235e-16
-8.37837837837838 2.27916883183252e-16
-8.35835835835836 2.69485858889318e-16
-8.33833833833834 3.18508772685576e-16
-8.31831831831832 3.76298728354137e-16
-8.2982982982983 4.44395893386326e-16
-8.27827827827828 5.24606005103494e-16
-8.25825825825826 6.19045274051723e-16
-8.23823823823824 7.30192724674871e-16
-8.21821821821822 8.60951178494166e-16
-8.1981981981982 1.01471827585831e-15
-8.17817817817818 1.19546915264237e-15
-8.15815815815816 1.40785264250169e-15
-8.13813813813814 1.65730316851105e-15
-8.11811811811812 1.95017082606148e-15
-8.0980980980981 2.29387254841619e-15
-8.07807807807808 2.69706769497134e-15
-8.05805805805806 3.16986191875719e-15
-8.03803803803804 3.72404376402735e-15
-8.01801801801802 4.3733591283238e-15
-7.997997997998 5.13382950919607e-15
-7.97797797797798 6.02412085866193e-15
-7.95795795795796 7.06597090549303e-15
-7.93793793793794 8.28468399583074e-15
-7.91791791791792 9.70970386854708e-15
-7.8978978978979 1.13752763482777e-14
-7.87787787787788 1.33212157347805e-14
-7.85785785785786 1.55937907247683e-14
-7.83783783783784 1.82467480586655e-14
-7.81781781781782 2.13424947819475e-14
-7.7977977977978 2.49534630966872e-14
-7.77777777777778 2.91636853079909e-14
-7.75775775775776 3.40706104038538e-14
-7.73773773773774 3.9787198415571e-14
-7.71771771771772 4.64443339685918e-14
-7.6976976976977 5.4193606440538e-14
-7.67767767767768 6.32105109959252e-14
-7.65765765765766 7.36981325813117e-14
-7.63763763763764 8.58913838706879e-14
-7.61761761761762 1.00061878296601e-13
-7.5975975975976 1.16523530854699e-13
-7.57757757757758 1.35638992516664e-13
-7.55755755755756 1.57827039041884e-13
-7.53753753753754 1.83571051982057e-13
-7.51751751751752 2.13428748996349e-13
-7.4974974974975 2.48043342543513e-13
-7.47747747747748 2.88156330935935e-13
-7.45745745745746 3.34622154016965e-13
-7.43743743743744 3.88424977793732e-13
-7.41741741741742 4.50697908714384e-13
-7.3973973973974 5.22744979473711e-13
-7.37737737737738 6.06066294885249e-13
-7.35735735735736 7.02386779168738e-13
-7.33733733733734 8.13689025751835e-13
-7.31731731731732 9.42250818252909e-13
-7.2972972972973 1.09068796768221e-12
-7.27727727727728 1.262003197176e-12
-7.25725725725726 1.45964190299847e-12
-7.23723723723724 1.68755573049416e-12
-7.21721721721722 1.95027502769792e-12
-7.1971971971972 2.25299137914218e-12
-7.17717717717718 2.60165157997871e-12
-7.15715715715716 3.00306458801653e-12
-7.13713713713714 3.4650231910834e-12
-7.11711711711712 3.99644235193919e-12
-7.0970970970971 4.60751644580953e-12
-7.07707707707708 5.30989788981514e-12
-7.05705705705706 6.11689998287646e-12
-7.03703703703704 7.0437271332242e-12
-7.01701701701702 8.10773605306645e-12
-6.996996996997 9.32873195138555e-12
-6.97697697697698 1.07293042619726e-11
-6.95695695695696 1.23352070109962e-11
-6.93693693693694 1.41757895636779e-11
-6.91691691691692 1.62844842008237e-11
-6.8968968968969 1.86993577716893e-11
-6.87687687687688 2.14637355595386e-11
-6.85685685685686 2.46269064908967e-11
-6.83683683683684 2.82449199306815e-11
-6.81681681681682 3.2381485546105e-11
-6.7967967967968 3.71089891068559e-11
-6.77677677677678 4.25096386334913e-11
-6.75675675675676 4.86767570277083e-11
-6.73673673673674 5.57162392366004e-11
-6.71671671671672 6.37481941394491e-11
-6.6966966966967 7.29087937236032e-11
-6.67667667667668 8.33523547614402e-11
-6.65665665665666 9.52536811418151e-11
-6.63663663663664 1.08810698278135e-10
-6.61661661661662 1.24247414645768e-10
-6.5965965965966 1.41817249531744e-10
-6.57657657657658 1.61806770551278e-10
-6.55655655655656 1.84539889444164e-10
-6.53653653653654 2.10382570159622e-10
-6.51651651651652 2.39748109325542e-10
-6.4964964964965 2.73103055937438e-10
-6.47647647647648 3.10973844559381e-10
-6.45645645645646 3.53954224575809e-10
-6.43643643643644 4.027135771479e-10
-6.41641641641642 4.58006221596996e-10
-6.3963963963964 5.20681824054169e-10
-6.37637637637638 5.91697033481538e-10
-6.35635635635636 6.72128483698846e-10
-6.33633633633634 7.63187314959523e-10
-6.31631631631632 8.66235385046001e-10
-6.2962962962963 9.8280335793813e-10
-6.27627627627628 1.11461087800766e-09
-6.25625625625626 1.26358905957513e-09
-6.23623623623624 1.43190554571787e-09
-6.21621621621622 1.62199241663862e-09
-6.1961961961962 1.83657725691024e-09
-6.17617617617618 2.0787177227378e-09
-6.15615615615616 2.35183998527873e-09
-6.13613613613614 2.65978146430928e-09
-6.11611611611612 3.00683830841829e-09
-6.0960960960961 3.39781812376754e-09
-6.07607607607608 3.83809850362696e-09
-6.05605605605606 4.33369196574699e-09
-6.03603603603604 4.89131796456919e-09
-6.01601601601602 5.51848271073395e-09
-5.995995995996 6.22356760178439e-09
-5.97597597597598 7.01592714588943e-09
-5.95595595595596 7.90599734535251e-09
-5.93593593593594 8.90541559921139e-09
-5.91591591591592 1.00271532849868e-08
-5.8958958958959 1.12856622892642e-08
-5.87587587587588 1.2697036876002e-08
-5.85585585585586 1.42791924110059e-08
-5.83583583583584 1.60520626017053e-08
-5.81581581581582 1.80378170640688e-08
-5.7957957957958 2.02611011941336e-08
-5.77577577577578 2.27493005011625e-08
-5.75575575575576 2.55328317539416e-08
-5.73573573573574 2.86454635022852e-08
-5.71571571571572 3.2124668763613e-08
-5.6956956956957 3.60120129107462e-08
-5.67567567567568 4.03535800631662e-08
-5.65565565565566 4.5200441571292e-08
-5.63563563563564 5.06091704933412e-08
-5.61561561561562 5.66424062986154e-08
-5.5955955955956 6.33694743912418e-08
-5.57557557557558 7.08670654362614e-08
-5.55555555555556 7.92199798873018e-08
-5.53553553553554 8.85219435638491e-08
-5.51551551551552 9.8876500608364e-08
-5.4954954954955 1.10397990671284e-07
-5.47547547547548 1.23212617727566e-07
-5.45545545545546 1.3745961852414e-07
-5.43543543543544 1.5329253929596e-07
-5.41541541541542 1.70880630071684e-07
-5.3953953953954 1.90410366621162e-07
-5.37537537537538 2.1208711087848e-07
-5.35535535535536 2.36136921509202e-07
-5.33533533533534 2.62808527181656e-07
-5.31531531531532 2.92375476052561e-07
-5.2952952952953 3.25138475990267e-07
-5.27527527527528 3.61427941137511e-07
-5.25525525525526 4.01606761563285e-07
-5.23523523523524 4.46073313973501e-07
-5.21521521521522 4.95264732746229e-07
-5.1951951951952 5.4966046193278e-07
-5.17517517517518 6.09786110324583e-07
-5.15515515515516 6.76217633231267e-07
-5.13513513513514 7.49585866251233e-07
-5.11511511511512 8.30581438046261e-07
-5.0950950950951 9.19960090959837e-07
-5.07507507507508 1.01854844024876e-06
-5.05505505505506 1.12725020473309e-06
-5.03503503503504 1.24705294381396e-06
-5.01501501501502 1.37903533806611e-06
-4.99499499499499 1.5243750529858e-06
-4.97497497497497 1.68435722796805e-06
-4.95495495495495 1.86038363520377e-06
-4.93493493493493 2.05398255592983e-06
-4.91491491491491 2.26681942433715e-06
-4.89489489489489 2.50070829244518e-06
-4.87487487487487 2.75762417238901e-06
-4.85485485485485 3.03971631583941e-06
-4.83483483483483 3.34932249368796e-06
-4.81481481481481 3.68898434268124e-06
-4.79479479479479 4.06146384937932e-06
-4.77477477477477 4.4697610456467e-06
-4.75475475475475 4.91713299385613e-06
-4.73473473473473 5.40711414409909e-06
-4.71471471471471 5.94353814994721e-06
-4.69469469469469 6.53056123369604e-06
-4.67467467467467 7.17268719654363e-06
-4.65465465465465 7.87479417380527e-06
-4.63463463463463 8.64216324004121e-06
-4.61461461461461 9.48050897386715e-06
-4.59459459459459 1.03960120972233e-05
-4.57457457457457 1.13953543089884e-05
-4.55455455455455 1.24857554380297e-05
-4.53453453453453 1.3675013046071e-05
-4.51451451451451 1.49715446161227e-05
-4.49449449449449 1.63844324676437e-05
-4.47447447447447 1.79234715450684e-05
-4.45445445445445 1.95992202318354e-05
-4.43443443443443 2.14230543475548e-05
-4.41441441441441 2.34072244914564e-05
-4.39439439439439 2.55649169007242e-05
-4.37437437437437 2.79103179977393e-05
-4.35435435435435 3.04586828055866e-05
-4.33433433433433 3.32264074164067e-05
-4.31431431431431 3.62311057022691e-05
-4.29429429429429 3.94916904631592e-05
-4.27427427427427 4.3028459211397e-05
-4.25425425425425 4.68631847962824e-05
-4.23423423423423 5.10192110769697e-05
-4.21421421421421 5.55215538554582e-05
-4.19419419419419 6.03970072851107e-05
-4.17417417417417 6.56742559732345e-05
-4.15415415415415 7.13839929989176e-05
-4.13413413413413 7.75590440694795e-05
-4.11411411411411 8.42344980404937e-05
-4.09409409409409 9.14478440253317e-05
-4.07407407407407 9.92391153205018e-05
-4.05405405405405 0.000107651040372646
-4.03403403403403 0.000116729201011866
-4.01401401401401 0.000126522198173995
-3.99399399399399 0.000137081825331481
-3.97397397397397 0.000148463249848567
-3.95395395395395 0.000160725202471485
-3.93393393393393 0.000173930175158222
-3.91391391391391 0.00018814462744512
-3.89389389389389 0.000203439201538965
-3.87387387387387 0.000219888946313312
-3.85385385385385 0.000237573550376443
-3.83383383383383 0.000256577584365551
-3.81381381381381 0.000276990752607344
-3.79379379379379 0.000298908154269281
-3.77377377377377 0.000322430554107926
-3.75375375375375 0.000347664662901427
-3.73373373373373 0.000374723427631836
-3.71371371371371 0.000403726331459719
-3.69369369369369 0.000434799703508357
-3.67367367367367 0.000468077038447599
-3.65365365365365 0.00050369932583812
-3.63363363363363 0.000541815389165405
-3.61361361361361 0.000582582234459159
-3.59359359359359 0.000626165408357979
-3.57357357357357 0.000672739365441021
-3.55355355355355 0.000722487844607978
-3.53353353353353 0.00077560425424601
-3.51351351351351 0.000832292065877155
-3.49349349349349 0.000892765215932443
-3.47347347347347 0.000957248515249216
-3.45345345345345 0.0010259780658361
-3.43343343343343 0.00109920168439588
-3.41341341341341 0.00117717933203981
-3.39339339339339 0.00126018354956833
-3.37337337337337 0.00134849989763212
-3.35335335335335 0.00144242740102448
-3.33333333333333 0.00154227899629111
-3.31331331331331 0.00164838198177652
-3.29329329329329 0.00176107846915772
-3.27327327327327 0.00188072583544552
-3.25325325325325 0.00200769717436226
-3.23323323323323 0.00214238174593163
-3.21321321321321 0.00228518542304204
-3.19319319319319 0.00243653113367012
-3.17317317317317 0.00259685929737497
-3.15315315315315 0.00276662825459747
-3.13313313313313 0.00294631468722261
-3.11311311311311 0.00313641402878609
-3.09309309309309 0.00333744086263052
-3.07307307307307 0.00354992930624086
-3.05305305305305 0.00377443337991422
-3.03303303303303 0.00401152735784579
-3.01301301301301 0.00426180609964128
-2.99299299299299 0.00452588536019618
-2.97297297297297 0.0048044020758154
-2.95295295295295 0.00509801462438215
-2.93293293293293 0.00540740305732385
-2.91291291291291 0.00573326930106519
-2.89289289289289 0.00607633732560526
-2.87287287287287 0.00643735327780636
-2.85285285285285 0.00681708557693873
-2.83283283283283 0.00721632496998623
-2.81281281281281 0.00763588454418632
-2.79279279279279 0.00807659969425075
-2.77277277277277 0.00853932804169477
-2.75275275275275 0.00902494930369032
-2.73273273273273 0.00953436510885489
-2.71271271271271 0.0100684987573917
-2.69269269269269 0.0106282949230102
-2.67267267267267 0.0112147192940778
-2.65265265265265 0.0118287581514852
-2.63263263263263 0.0124714178807513
-2.61261261261261 0.0131437244159435
-2.59259259259259 0.0138467226130541
-2.57257257257257 0.0145814755505492
-2.55255255255255 0.0153490637548887
-2.53253253253253 0.0161505843489182
-2.51251251251251 0.0169871501211409
-2.49249249249249 0.0178598885140022
-2.47247247247247 0.018769940529451
-2.45245245245245 0.0197184595501959
-2.43243243243243 0.0207066100752274
-2.41241241241241 0.0217355663683581
-2.39239239239239 0.0228065110187135
-2.37237237237237 0.0239206334123108
-2.35235235235235 0.0250791281140699
-2.33233233233233 0.0262831931598317
-2.31231231231231 0.0275340282581906
-2.29229229229229 0.0288328329022027
-2.27227227227227 0.0301808043912899
-2.25225225225225 0.0315791357639331
-2.23223223223223 0.033029013642035
-2.21221221221221 0.0345316159881232
-2.19219219219219 0.0360881097768736
-2.17217217217217 0.0376996485827434
-2.15215215215215 0.0393673700858293
-2.13213213213213 0.0410923934983949
-2.11211211211211 0.0428758169148479
-2.09209209209209 0.0447187145882915
-2.07207207207207 0.0466221341371235
-2.05205205205205 0.0485870936855041
-2.03203203203203 0.0506145789418747
-2.01201201201201 0.0527055402200587
-1.99199199199199 0.0548608894078376
-1.97197197197197 0.057081496888248
-1.95195195195195 0.059368188419199
-1.93193193193193 0.0617217419773594
-1.91191191191191 0.0641428845726061
-1.89189189189189 0.0666322890396674
-1.87187187187187 0.0691905708139176
-1.85185185185185 0.0718182846986055
-1.83183183183183 0.0745159216311036
-1.81181181181181 0.077283905456062
-1.79179179179179 0.0801225897136326
-1.77177177177177 0.083032254451193
-1.75175175175175 0.0860131030672496
-1.73173173173173 0.0890652591964251
-1.71171171171171 0.0921887636446459
-1.69169169169169 0.0953835713838294
-1.67167167167167 0.0986495486155338
-1.65165165165165 0.101986469913169
-1.63163163163163 0.10539401545248
-1.61161161161161 0.108871768340093
-1.59159159159159 0.112419212049971
-1.57157157157157 0.116035727977651
-1.55155155155155 0.119720593122119
-1.53153153153153 0.123472977905145
-1.51151151151151 0.127291944137829
-1.49149149149149 0.13117644314399
-1.47147147147147 0.135125314049902
-1.45145145145145 0.139137282249685
-1.43143143143143 0.143210958055468
-1.41141141141141 0.147344835541168
-1.39139139139139 0.151537291588457
-1.37137137137137 0.155786585143159
-1.35135135135135 0.160090856689972
-1.33133133133133 0.164448127952996
-1.31131131131131 0.168856301829129
-1.29129129129129 0.173313162560933
-1.27127127127127 0.17781637615506
-1.25125125125125 0.182363491051798
-1.23123123123123 0.186951939050736
-1.21121121121121 0.191579036496956
-1.19119119119119 0.1962419857315
-1.17117117117117 0.200937876809264
-1.15115115115115 0.205663689486728
-1.13113113113113 0.210416295481265
-1.11111111111111 0.215192461003031
-1.09109109109109 0.219988849559688
-1.07107107107107 0.224802025033432
-1.05105105105105 0.229628455029052
-1.03103103103103 0.234464514490888
-1.01101101101101 0.239306489585817
-0.990990990990991 0.24415058184851
-0.970970970970971 0.24899291258444
-0.950950950950951 0.253829527525259
-0.930930930930931 0.258656401730343
-0.910910910910911 0.2634694447275
-0.890890890890891 0.268264505884996
-0.870870870870871 0.273037380006279
-0.850850850850851 0.277783813137949
-0.830830830830831 0.282499508580786
-0.810810810810811 0.287180133092853
-0.790790790790791 0.291821323272996
-0.77077077077077 0.296418692112302
-0.75075075075075 0.300967835700437
-0.73073073073073 0.305464340073112
-0.71071071071071 0.309903788186304
-0.69069069069069 0.314281767002296
-0.67067067067067 0.318593874672039
-0.65065065065065 0.322835727797843
-0.63063063063063 0.327002968759958
-0.61061061061061 0.331091273090187
-0.59059059059059 0.33509635687531
-0.57057057057057 0.339013984172804
-0.55055055055055 0.34283997442106
-0.53053053053053 0.346570209826128
-0.51051051051051 0.350200642706842
-0.49049049049049 0.353727302780113
-0.47047047047047 0.357146304368113
-0.45045045045045 0.360453853509139
-0.43043043043043 0.363646254953996
-0.41041041041041 0.366719919029892
-0.39039039039039 0.369671368354051
-0.37037037037037 0.372497244379499
-0.35035035035035 0.375194313755802
-0.33033033033033 0.377759474487924
-0.31031031031031 0.38018976187679
-0.29029029029029 0.382482354225654
-0.27027027027027 0.384634578296894
-0.25025025025025 0.386643914504485
-0.23023023023023 0.388508001828027
-0.21021021021021 0.390224642434919
-0.19019019019019 0.391791805998011
-0.17017017017017 0.393207633696876
-0.15015015015015 0.394470441891644
-0.13013013013013 0.395578725459258
-0.11011011011011 0.396531160782876
-0.0900900900900901 0.397326608386124
-0.07007007007007 0.397964115204853
-0.05005005005005 0.398442916490068
-0.03003003003003 0.398762437336696
-0.01001001001001 0.398922293833933
0.01001001001001 0.398922293833933
0.03003003003003 0.398762437336696
0.05005005005005 0.398442916490068
0.07007007007007 0.397964115204853
0.0900900900900901 0.397326608386124
0.11011011011011 0.396531160782876
0.13013013013013 0.395578725459258
0.15015015015015 0.394470441891644
0.17017017017017 0.393207633696876
0.19019019019019 0.391791805998011
0.21021021021021 0.390224642434919
0.23023023023023 0.388508001828027
0.25025025025025 0.386643914504485
0.27027027027027 0.384634578296894
0.29029029029029 0.382482354225654
0.31031031031031 0.38018976187679
0.33033033033033 0.377759474487924
0.35035035035035 0.375194313755802
0.37037037037037 0.372497244379499
0.39039039039039 0.369671368354051
0.41041041041041 0.366719919029892
0.43043043043043 0.363646254953996
0.45045045045045 0.360453853509139
0.47047047047047 0.357146304368113
0.49049049049049 0.353727302780113
0.51051051051051 0.350200642706842
0.53053053053053 0.346570209826128
0.55055055055055 0.34283997442106
0.57057057057057 0.339013984172804
0.59059059059059 0.33509635687531
0.61061061061061 0.331091273090187
0.63063063063063 0.327002968759958
0.65065065065065 0.322835727797843
0.67067067067067 0.318593874672039
0.69069069069069 0.314281767002296
0.71071071071071 0.309903788186304
0.73073073073073 0.305464340073112
0.75075075075075 0.300967835700437
0.77077077077077 0.296418692112302
0.790790790790791 0.291821323272996
0.810810810810811 0.287180133092853
0.830830830830831 0.282499508580786
0.850850850850851 0.277783813137949
0.870870870870871 0.273037380006279
0.890890890890891 0.268264505884996
0.910910910910911 0.2634694447275
0.930930930930931 0.258656401730343
0.950950950950951 0.253829527525259
0.970970970970971 0.24899291258444
0.990990990990991 0.24415058184851
1.01101101101101 0.239306489585817
1.03103103103103 0.234464514490888
1.05105105105105 0.229628455029052
1.07107107107107 0.224802025033432
1.09109109109109 0.219988849559688
1.11111111111111 0.215192461003031
1.13113113113113 0.210416295481265
1.15115115115115 0.205663689486728
1.17117117117117 0.200937876809264
1.19119119119119 0.1962419857315
1.21121121121121 0.191579036496956
1.23123123123123 0.186951939050736
1.25125125125125 0.182363491051798
1.27127127127127 0.17781637615506
1.29129129129129 0.173313162560933
1.31131131131131 0.168856301829129
1.33133133133133 0.164448127952996
1.35135135135135 0.160090856689972
1.37137137137137 0.155786585143159
1.39139139139139 0.151537291588457
1.41141141141141 0.147344835541168
1.43143143143143 0.143210958055468
1.45145145145145 0.139137282249685
1.47147147147147 0.135125314049902
1.49149149149149 0.13117644314399
1.51151151151151 0.127291944137829
1.53153153153153 0.123472977905145
1.55155155155155 0.119720593122119
1.57157157157157 0.116035727977651
1.59159159159159 0.112419212049971
1.61161161161161 0.108871768340093
1.63163163163163 0.10539401545248
1.65165165165165 0.101986469913169
1.67167167167167 0.0986495486155338
1.69169169169169 0.0953835713838294
1.71171171171171 0.0921887636446459
1.73173173173173 0.0890652591964251
1.75175175175175 0.0860131030672496
1.77177177177177 0.083032254451193
1.79179179179179 0.0801225897136326
1.81181181181181 0.077283905456062
1.83183183183183 0.0745159216311036
1.85185185185185 0.0718182846986055
1.87187187187187 0.0691905708139176
1.89189189189189 0.0666322890396674
1.91191191191191 0.0641428845726061
1.93193193193193 0.0617217419773594
1.95195195195195 0.059368188419199
1.97197197197197 0.057081496888248
1.99199199199199 0.0548608894078376
2.01201201201201 0.0527055402200588
2.03203203203203 0.0506145789418748
2.05205205205205 0.0485870936855042
2.07207207207207 0.0466221341371236
2.09209209209209 0.0447187145882916
2.11211211211211 0.0428758169148479
2.13213213213213 0.041092393498395
2.15215215215215 0.0393673700858294
2.17217217217217 0.0376996485827434
2.19219219219219 0.0360881097768737
2.21221221221221 0.0345316159881233
2.23223223223223 0.033029013642035
2.25225225225225 0.0315791357639332
2.27227227227227 0.03018080439129
2.29229229229229 0.0288328329022028
2.31231231231231 0.0275340282581906
2.33233233233233 0.0262831931598317
2.35235235235235 0.02507912811407
2.37237237237237 0.0239206334123108
2.39239239239239 0.0228065110187136
2.41241241241241 0.0217355663683581
2.43243243243243 0.0207066100752275
2.45245245245245 0.0197184595501959
2.47247247247247 0.0187699405294511
2.49249249249249 0.0178598885140022
2.51251251251251 0.016987150121141
2.53253253253253 0.0161505843489182
2.55255255255255 0.0153490637548888
2.57257257257257 0.0145814755505493
2.59259259259259 0.0138467226130541
2.61261261261261 0.0131437244159435
2.63263263263263 0.0124714178807514
2.65265265265265 0.0118287581514852
2.67267267267267 0.0112147192940778
2.69269269269269 0.0106282949230103
2.71271271271271 0.0100684987573917
2.73273273273273 0.00953436510885491
2.75275275275275 0.00902494930369034
2.77277277277277 0.00853932804169479
2.79279279279279 0.00807659969425077
2.81281281281281 0.00763588454418634
2.83283283283283 0.00721632496998621
2.85285285285285 0.00681708557693871
2.87287287287287 0.00643735327780635
2.89289289289289 0.00607633732560524
2.91291291291291 0.00573326930106518
2.93293293293293 0.00540740305732384
2.95295295295295 0.00509801462438214
2.97297297297297 0.00480440207581539
2.99299299299299 0.00452588536019617
3.01301301301301 0.00426180609964127
3.03303303303303 0.00401152735784578
3.05305305305305 0.00377443337991421
3.07307307307307 0.00354992930624085
3.09309309309309 0.00333744086263051
3.11311311311311 0.00313641402878608
3.13313313313313 0.0029463146872226
3.15315315315315 0.00276662825459746
3.17317317317317 0.00259685929737496
3.19319319319319 0.00243653113367011
3.21321321321321 0.00228518542304203
3.23323323323323 0.00214238174593163
3.25325325325325 0.00200769717436225
3.27327327327327 0.00188072583544551
3.29329329329329 0.00176107846915771
3.31331331331331 0.00164838198177652
3.33333333333333 0.0015422789962911
3.35335335335335 0.00144242740102448
3.37337337337337 0.00134849989763212
3.39339339339339 0.00126018354956833
3.41341341341341 0.00117717933203981
3.43343343343343 0.00109920168439588
3.45345345345345 0.0010259780658361
3.47347347347347 0.000957248515249212
3.49349349349349 0.000892765215932441
3.51351351351351 0.000832292065877152
3.53353353353353 0.000775604254246008
3.55355355355355 0.000722487844607976
3.57357357357357 0.000672739365441019
3.59359359359359 0.000626165408357979
3.61361361361361 0.000582582234459159
3.63363363363363 0.000541815389165405
3.65365365365365 0.00050369932583812
3.67367367367367 0.000468077038447599
3.69369369369369 0.000434799703508357
3.71371371371371 0.000403726331459719
3.73373373373373 0.000374723427631836
3.75375375375375 0.000347664662901427
3.77377377377377 0.000322430554107926
3.79379379379379 0.000298908154269281
3.81381381381381 0.000276990752607344
3.83383383383383 0.000256577584365551
3.85385385385385 0.000237573550376443
3.87387387387387 0.000219888946313312
3.89389389389389 0.000203439201538965
3.91391391391391 0.00018814462744512
3.93393393393393 0.000173930175158222
3.95395395395395 0.000160725202471485
3.97397397397397 0.000148463249848567
3.99399399399399 0.000137081825331481
4.01401401401401 0.000126522198173995
4.03403403403403 0.000116729201011866
4.05405405405405 0.000107651040372646
4.07407407407407 9.92391153205018e-05
4.09409409409409 9.14478440253317e-05
4.11411411411411 8.42344980404937e-05
4.13413413413413 7.75590440694795e-05
4.15415415415415 7.13839929989176e-05
4.17417417417417 6.56742559732345e-05
4.19419419419419 6.03970072851107e-05
4.21421421421421 5.55215538554582e-05
4.23423423423423 5.10192110769697e-05
4.25425425425425 4.68631847962824e-05
4.27427427427427 4.3028459211397e-05
4.29429429429429 3.94916904631592e-05
4.31431431431431 3.62311057022691e-05
4.33433433433433 3.32264074164067e-05
4.35435435435435 3.04586828055866e-05
4.37437437437437 2.79103179977393e-05
4.39439439439439 2.55649169007242e-05
4.41441441441441 2.34072244914564e-05
4.43443443443443 2.14230543475548e-05
4.45445445445445 1.95992202318354e-05
4.47447447447447 1.79234715450684e-05
4.49449449449449 1.63844324676437e-05
4.51451451451451 1.49715446161227e-05
4.53453453453453 1.3675013046071e-05
4.55455455455455 1.24857554380297e-05
4.57457457457457 1.13953543089884e-05
4.59459459459459 1.03960120972233e-05
4.61461461461461 9.48050897386715e-06
4.63463463463463 8.64216324004121e-06
4.65465465465465 7.87479417380527e-06
4.67467467467467 7.17268719654363e-06
4.69469469469469 6.53056123369604e-06
4.71471471471471 5.94353814994721e-06
4.73473473473473 5.40711414409909e-06
4.75475475475475 4.91713299385613e-06
4.77477477477477 4.4697610456467e-06
4.79479479479479 4.06146384937932e-06
4.81481481481481 3.68898434268124e-06
4.83483483483483 3.34932249368796e-06
4.85485485485485 3.03971631583941e-06
4.87487487487487 2.75762417238901e-06
4.89489489489489 2.50070829244518e-06
4.91491491491491 2.26681942433715e-06
4.93493493493493 2.05398255592983e-06
4.95495495495495 1.86038363520377e-06
4.97497497497497 1.68435722796805e-06
4.99499499499499 1.5243750529858e-06
5.01501501501502 1.37903533806611e-06
5.03503503503504 1.24705294381396e-06
5.05505505505506 1.12725020473309e-06
5.07507507507508 1.01854844024876e-06
5.0950950950951 9.19960090959837e-07
5.11511511511512 8.30581438046261e-07
5.13513513513514 7.49585866251233e-07
5.15515515515516 6.76217633231267e-07
5.17517517517518 6.09786110324583e-07
5.1951951951952 5.4966046193278e-07
5.21521521521522 4.95264732746229e-07
5.23523523523524 4.46073313973501e-07
5.25525525525526 4.01606761563285e-07
5.27527527527528 3.61427941137511e-07
5.2952952952953 3.25138475990267e-07
5.31531531531532 2.92375476052561e-07
5.33533533533534 2.62808527181656e-07
5.35535535535536 2.36136921509202e-07
5.37537537537538 2.1208711087848e-07
5.3953953953954 1.90410366621162e-07
5.41541541541542 1.70880630071684e-07
5.43543543543544 1.5329253929596e-07
5.45545545545546 1.3745961852414e-07
5.47547547547548 1.23212617727566e-07
5.4954954954955 1.10397990671284e-07
5.51551551551552 9.8876500608364e-08
5.53553553553554 8.85219435638491e-08
5.55555555555556 7.92199798873018e-08
5.57557557557558 7.08670654362614e-08
5.5955955955956 6.33694743912418e-08
5.61561561561562 5.66424062986154e-08
5.63563563563564 5.06091704933412e-08
5.65565565565566 4.5200441571292e-08
5.67567567567568 4.03535800631662e-08
5.6956956956957 3.60120129107462e-08
5.71571571571572 3.2124668763613e-08
5.73573573573574 2.86454635022852e-08
5.75575575575576 2.55328317539416e-08
5.77577577577578 2.27493005011625e-08
5.7957957957958 2.02611011941336e-08
5.81581581581582 1.80378170640688e-08
5.83583583583584 1.60520626017053e-08
5.85585585585586 1.42791924110059e-08
5.87587587587588 1.2697036876002e-08
5.8958958958959 1.12856622892642e-08
5.91591591591592 1.00271532849868e-08
5.93593593593594 8.90541559921139e-09
5.95595595595596 7.90599734535251e-09
5.97597597597598 7.01592714588943e-09
5.995995995996 6.22356760178439e-09
6.01601601601602 5.5184827107339e-09
6.03603603603604 4.89131796456914e-09
6.05605605605606 4.33369196574694e-09
6.07607607607608 3.83809850362692e-09
6.0960960960961 3.3978181237675e-09
6.11611611611612 3.00683830841826e-09
6.13613613613614 2.65978146430924e-09
6.15615615615616 2.35183998527871e-09
6.17617617617618 2.07871772273778e-09
6.1961961961962 1.83657725691022e-09
6.21621621621622 1.6219924166386e-09
6.23623623623624 1.43190554571785e-09
6.25625625625626 1.26358905957511e-09
6.27627627627628 1.11461087800764e-09
6.2962962962963 9.82803357938119e-10
6.31631631631632 8.66235385045992e-10
6.33633633633634 7.63187314959515e-10
6.35635635635636 6.72128483698836e-10
6.37637637637638 5.91697033481532e-10
6.3963963963964 5.20681824054164e-10
6.41641641641642 4.58006221596989e-10
6.43643643643644 4.02713577147895e-10
6.45645645645646 3.53954224575805e-10
6.47647647647648 3.10973844559378e-10
6.4964964964965 2.73103055937435e-10
6.51651651651652 2.39748109325539e-10
6.53653653653654 2.10382570159619e-10
6.55655655655656 1.84539889444162e-10
6.57657657657658 1.61806770551276e-10
6.5965965965966 1.41817249531742e-10
6.61661661661662 1.24247414645767e-10
6.63663663663664 1.08810698278133e-10
6.65665665665666 9.52536811418137e-11
6.67667667667668 8.33523547614393e-11
6.6966966966967 7.29087937236022e-11
6.71671671671672 6.37481941394484e-11
6.73673673673674 5.57162392365998e-11
6.75675675675676 4.86767570277076e-11
6.77677677677678 4.25096386334907e-11
6.7967967967968 3.71089891068556e-11
6.81681681681682 3.23814855461048e-11
6.83683683683684 2.82449199306813e-11
6.85685685685686 2.46269064908966e-11
6.87687687687688 2.14637355595386e-11
6.8968968968969 1.86993577716892e-11
6.91691691691692 1.62844842008236e-11
6.93693693693694 1.41757895636779e-11
6.95695695695696 1.23352070109961e-11
6.97697697697698 1.07293042619725e-11
6.996996996997 9.32873195138548e-12
7.01701701701702 8.10773605306642e-12
7.03703703703704 7.04372713322415e-12
7.05705705705706 6.11689998287643e-12
7.07707707707708 5.30989788981512e-12
7.0970970970971 4.6075164458095e-12
7.11711711711712 3.99644235193917e-12
7.13713713713714 3.46502319108337e-12
7.15715715715716 3.00306458801651e-12
7.17717717717718 2.60165157997869e-12
7.1971971971972 2.25299137914217e-12
7.21721721721722 1.95027502769791e-12
7.23723723723724 1.68755573049414e-12
7.25725725725726 1.45964190299846e-12
7.27727727727728 1.262003197176e-12
7.2972972972973 1.0906879676822e-12
7.31731731731732 9.42250818252902e-13
7.33733733733734 8.13689025751829e-13
7.35735735735736 7.02386779168733e-13
7.37737737737738 6.06066294885245e-13
7.3973973973974 5.22744979473708e-13
7.41741741741742 4.50697908714381e-13
7.43743743743744 3.88424977793729e-13
7.45745745745746 3.34622154016963e-13
7.47747747747748 2.88156330935933e-13
7.4974974974975 2.48043342543511e-13
7.51751751751752 2.13428748996348e-13
7.53753753753754 1.83571051982055e-13
7.55755755755756 1.57827039041883e-13
7.57757757757758 1.35638992516663e-13
7.5975975975976 1.16523530854698e-13
7.61761761761762 1.000618782966e-13
7.63763763763764 8.58913838706873e-14
7.65765765765766 7.36981325813112e-14
7.67767767767768 6.32105109959248e-14
7.6976976976977 5.41936064405376e-14
7.71771771771772 4.64443339685915e-14
7.73773773773774 3.97871984155707e-14
7.75775775775776 3.40706104038536e-14
7.77777777777778 2.91636853079907e-14
7.7977977977978 2.4953463096687e-14
7.81781781781782 2.13424947819474e-14
7.83783783783784 1.82467480586654e-14
7.85785785785786 1.55937907247682e-14
7.87787787787788 1.33212157347804e-14
7.8978978978979 1.13752763482777e-14
7.91791791791792 9.70970386854701e-15
7.93793793793794 8.28468399583068e-15
7.95795795795796 7.06597090549298e-15
7.97797797797798 6.02412085866189e-15
7.997997997998 5.13382950919603e-15
8.01801801801802 4.3733591283238e-15
8.03803803803804 3.72404376402735e-15
8.05805805805806 3.16986191875719e-15
8.07807807807808 2.69706769497134e-15
8.0980980980981 2.29387254841619e-15
8.11811811811812 1.95017082606148e-15
8.13813813813814 1.65730316851105e-15
8.15815815815816 1.40785264250169e-15
8.17817817817818 1.19546915264237e-15
8.1981981981982 1.01471827585831e-15
8.21821821821822 8.60951178494166e-16
8.23823823823824 7.30192724674871e-16
8.25825825825826 6.19045274051723e-16
8.27827827827828 5.24606005103494e-16
8.2982982982983 4.44395893386326e-16
8.31831831831832 3.76298728354137e-16
8.33833833833834 3.18508772685576e-16
8.35835835835836 2.69485858889318e-16
8.37837837837838 2.27916883183252e-16
8.3983983983984 1.92682799625235e-16
8.41841841841842 1.62830341150177e-16
8.43843843843844 1.37547801096493e-16
8.45845845845846 1.16144301209627e-16
8.47847847847848 9.8032051926925e-17
8.4984984984985 8.27111796592674e-17
8.51851851851852 6.97567552529606e-17
8.53853853853854 5.88077091106193e-17
8.55855855855856 4.95573626747691e-17
8.57857857857858 4.17453440895294e-17
8.5985985985986 3.51506886838449e-17
8.61861861861862 2.95859531836935e-17
8.63863863863864 2.48921968838275e-17
8.65865865865866 2.09347039316773e-17
8.67867867867868 1.75993388643364e-17
8.6986986986987 1.47894429982973e-17
8.71871871871872 1.24231925503958e-17
8.73873873873874 1.04313507694241e-17
8.75875875875876 8.75535614211525e-18
8.77877877877878 7.34569713009337e-18
8.7987987987988 6.16053109048305e-18
8.81881881881882 5.16451119999019e-18
8.83883883883884 4.32779048512422e-18
8.85885885885886 3.62517658454128e-18
8.87887887887888 3.03541474067764e-18
8.8988988988989 2.54057982941549e-18
8.91891891891892 2.12556106807901e-18
8.93893893893894 1.77762546207239e-18
8.95895895895896 1.48604811781133e-18
8.97897897897898 1.24179931485728e-18
8.998998998999 1.03727973679138e-18
9.01901901901902 8.66096545670873e-19
9.03903903903904 7.22874080905507e-19
9.05905905905906 6.03093897535385e-19
9.07907907907908 5.0295965472299e-19
9.0990990990991 4.19283042962206e-19
9.11911911911912 3.49387515332417e-19
9.13913913913914 2.91027078875695e-19
9.15915915915916 2.42317819504618e-19
9.17917917917918 2.01680188581445e-19
9.1991991991992 1.67790380698338e-19
9.21921921921922 1.39539388139188e-19
9.23923923923924 1.1599853476858e-19
9.25925925925926 9.63904764362469e-20
9.27927927927928 8.00648113242436e-20
9.2992992992993 6.6477576193283e-20
9.31931931931932 5.51740167796855e-20
9.33933933933934 4.5774115701642e-20
9.35935935935936 3.79604417474457e-20
9.37937937937938 3.14679525475421e-20
9.3993993993994 2.60754402558043e-20
9.41941941941942 2.15983585814365e-20
9.43943943943944 1.78828106797973e-20
9.45945945945946 1.48005121824234e-20
9.47947947947948 1.22445730038445e-20
9.4994994994995 1.01259663374544e-20
9.51951951951952 8.37057415073758e-21
9.53953953953954 6.91671611023779e-21
9.55955955955956 5.71308371631411e-21
9.57957957957958 4.71701393695898e-21
9.5995995995996 3.89304716291232e-21
9.61961961961962 3.21172317125736e-21
9.63963963963964 2.64857624243904e-21
9.65965965965966 2.18329684678274e-21
9.67967967967968 1.79903258756111e-21
9.6996996996997 1.48180551599987e-21
9.71971971971972 1.22002665237565e-21
9.73973973973974 1.00409166885045e-21
9.75975975975976 8.26044308669654e-22
9.77977977977978 6.79296312742742e-22
9.7997997997998 5.58394465795474e-22
9.81981981981982 4.58826916995414e-22
9.83983983983984 3.7686222201397e-22
9.85985985985986 3.09415635142992e-22
9.87987987987988 2.53938085193762e-22
9.8998998998999 2.08324025950642e-22
9.91991991991992 1.70834984871876e-22
9.93993993993994 1.40036162642795e-22
9.95995995995996 1.14743877987917e-22
9.97997997997998 9.39820210218911e-23
10 7.69459862670642e-23
};
\addlegendentry{N(0,1)}; 
\legend{}; 
\end{axis}

\end{tikzpicture}

%% file: FinalFigs/Null_Dists_d_10_100_n_100_m_100_kernel__Dirichlet_RBF_2022_10_15_22_22_56mmd.tex
\begin{tikzpicture}

\definecolor{darkorange25512714}{RGB}{255,127,14}
\definecolor{darkslategray38}{RGB}{38,38,38}
\definecolor{lightgray204}{RGB}{204,204,204}
\definecolor{steelblue31119180}{RGB}{31,119,180}

\begin{axis}[
axis line style={darkslategray38},
height=\figheight,
legend cell align={left},
legend style={fill opacity=0.8, draw opacity=1, text opacity=1, draw=none},
tick align=outside,
tick pos=left,
title={$\dmmd$~$(n/m=1)$},
width=\figwidth,
x grid style={lightgray204},
xmin=-6, xmax=6,
xtick style={color=darkslategray38},
y grid style={lightgray204},
ylabel=\textcolor{darkslategray38}{},
ymin=0, ymax=0.72587073292044,
ytick style={color=darkslategray38}, 
xticklabels=empty,
yticklabels=empty
]
\draw[draw=none,fill=steelblue31119180,fill opacity=0.8] (axis cs:-1.7765257358551,0) rectangle (axis cs:-1.62955737113953,0.0272167503907176);
\addlegendimage{ybar,ybar legend,draw=none,fill=steelblue31119180,fill opacity=0.8}
\addlegendentry{mmd (d=10)}

\draw[draw=none,fill=steelblue31119180,fill opacity=0.8] (axis cs:-1.40910494327545,0) rectangle (axis cs:-1.26213657855988,0.0816502511721529);
\draw[draw=none,fill=steelblue31119180,fill opacity=0.8] (axis cs:-1.0416841506958,0) rectangle (axis cs:-0.894715785980225,0.255837453672746);
\draw[draw=none,fill=steelblue31119180,fill opacity=0.8] (axis cs:-0.674263417720795,0) rectangle (axis cs:-0.527295053005219,0.484458156954774);
\draw[draw=none,fill=steelblue31119180,fill opacity=0.8] (axis cs:-0.306842595338821,0) rectangle (axis cs:-0.159874230623245,0.642315309220936);
\draw[draw=none,fill=steelblue31119180,fill opacity=0.8] (axis cs:0.0605781972408295,0) rectangle (axis cs:0.207546561956406,0.691305459924228);
\draw[draw=none,fill=steelblue31119180,fill opacity=0.8] (axis cs:0.427998960018158,0) rectangle (axis cs:0.574967324733734,0.337487704844899);
\draw[draw=none,fill=steelblue31119180,fill opacity=0.8] (axis cs:0.795419752597809,0) rectangle (axis cs:0.942388117313385,0.168743852422449);
\draw[draw=none,fill=steelblue31119180,fill opacity=0.8] (axis cs:1.1628406047821,0) rectangle (axis cs:1.30980896949768,0.0272167503907176);
\draw[draw=none,fill=steelblue31119180,fill opacity=0.8] (axis cs:1.53026139736176,0) rectangle (axis cs:1.67722976207733,0.00544335007814353);
\draw[draw=none,fill=darkorange25512714,fill opacity=0.8] (axis cs:-1.62955749034882,0) rectangle (axis cs:-1.48258912563324,0.00544335007814353);
\addlegendimage{ybar,ybar legend,draw=none,fill=darkorange25512714,fill opacity=0.8}
\addlegendentry{mmd (d=100)}

\draw[draw=none,fill=darkorange25512714,fill opacity=0.8] (axis cs:-1.26213669776917,0) rectangle (axis cs:-1.11516833305359,0.0435468006251482);
\draw[draw=none,fill=darkorange25512714,fill opacity=0.8] (axis cs:-0.894715905189514,0) rectangle (axis cs:-0.747747540473938,0.250394103594602);
\draw[draw=none,fill=darkorange25512714,fill opacity=0.8] (axis cs:-0.527295112609863,0) rectangle (axis cs:-0.380326747894287,0.587881808439501);
\draw[draw=none,fill=darkorange25512714,fill opacity=0.8] (axis cs:-0.15987429022789,0) rectangle (axis cs:-0.0129059255123138,0.674975409689798);
\draw[draw=none,fill=darkorange25512714,fill opacity=0.8] (axis cs:0.207546502351761,0) rectangle (axis cs:0.354514867067337,0.593325158517645);
\draw[draw=none,fill=darkorange25512714,fill opacity=0.8] (axis cs:0.574967265129089,0) rectangle (axis cs:0.721935629844666,0.37014780531376);
\draw[draw=none,fill=darkorange25512714,fill opacity=0.8] (axis cs:0.94238805770874,0) rectangle (axis cs:1.08935642242432,0.152413802188019);
\draw[draw=none,fill=darkorange25512714,fill opacity=0.8] (axis cs:1.30980885028839,0) rectangle (axis cs:1.45677721500397,0.0272167503907176);
\draw[draw=none,fill=darkorange25512714,fill opacity=0.8] (axis cs:1.67722964286804,0) rectangle (axis cs:1.82419800758362,0.0163300502344306);
\addplot [semithick, black]
table {%
-10 7.69459862670642e-23
-9.97997997997998 9.39820210218911e-23
-9.95995995995996 1.14743877987917e-22
-9.93993993993994 1.40036162642795e-22
-9.91991991991992 1.70834984871876e-22
-9.8998998998999 2.08324025950642e-22
-9.87987987987988 2.53938085193762e-22
-9.85985985985986 3.09415635142992e-22
-9.83983983983984 3.7686222201397e-22
-9.81981981981982 4.58826916995414e-22
-9.7997997997998 5.58394465795474e-22
-9.77977977977978 6.79296312742742e-22
-9.75975975975976 8.26044308669654e-22
-9.73973973973974 1.00409166885045e-21
-9.71971971971972 1.22002665237565e-21
-9.6996996996997 1.48180551599987e-21
-9.67967967967968 1.79903258756111e-21
-9.65965965965966 2.18329684678274e-21
-9.63963963963964 2.64857624243904e-21
-9.61961961961962 3.21172317125736e-21
-9.5995995995996 3.89304716291232e-21
-9.57957957957958 4.71701393695898e-21
-9.55955955955956 5.71308371631411e-21
-9.53953953953954 6.91671611023779e-21
-9.51951951951952 8.37057415073758e-21
-9.4994994994995 1.01259663374544e-20
-9.47947947947948 1.22445730038445e-20
-9.45945945945946 1.48005121824234e-20
-9.43943943943944 1.78828106797973e-20
-9.41941941941942 2.15983585814365e-20
-9.3993993993994 2.60754402558043e-20
-9.37937937937938 3.14679525475421e-20
-9.35935935935936 3.79604417474457e-20
-9.33933933933934 4.5774115701642e-20
-9.31931931931932 5.51740167796855e-20
-9.2992992992993 6.6477576193283e-20
-9.27927927927928 8.00648113242436e-20
-9.25925925925926 9.63904764362469e-20
-9.23923923923924 1.1599853476858e-19
-9.21921921921922 1.39539388139188e-19
-9.1991991991992 1.67790380698338e-19
-9.17917917917918 2.01680188581445e-19
-9.15915915915916 2.42317819504618e-19
-9.13913913913914 2.91027078875695e-19
-9.11911911911912 3.49387515332417e-19
-9.0990990990991 4.19283042962206e-19
-9.07907907907908 5.0295965472299e-19
-9.05905905905906 6.03093897535385e-19
-9.03903903903904 7.22874080905507e-19
-9.01901901901902 8.66096545670873e-19
-8.998998998999 1.03727973679138e-18
-8.97897897897898 1.24179931485728e-18
-8.95895895895896 1.48604811781133e-18
-8.93893893893894 1.77762546207239e-18
-8.91891891891892 2.12556106807901e-18
-8.8988988988989 2.54057982941549e-18
-8.87887887887888 3.03541474067764e-18
-8.85885885885886 3.62517658454128e-18
-8.83883883883884 4.32779048512422e-18
-8.81881881881882 5.16451119999019e-18
-8.7987987987988 6.16053109048305e-18
-8.77877877877878 7.34569713009337e-18
-8.75875875875876 8.75535614211525e-18
-8.73873873873874 1.04313507694241e-17
-8.71871871871872 1.24231925503958e-17
-8.6986986986987 1.47894429982973e-17
-8.67867867867868 1.75993388643364e-17
-8.65865865865866 2.09347039316773e-17
-8.63863863863864 2.48921968838275e-17
-8.61861861861862 2.95859531836935e-17
-8.5985985985986 3.51506886838449e-17
-8.57857857857858 4.17453440895294e-17
-8.55855855855856 4.95573626747691e-17
-8.53853853853854 5.88077091106193e-17
-8.51851851851852 6.97567552529606e-17
-8.4984984984985 8.27111796592674e-17
-8.47847847847848 9.8032051926925e-17
-8.45845845845846 1.16144301209627e-16
-8.43843843843844 1.37547801096493e-16
-8.41841841841842 1.62830341150177e-16
-8.3983983983984 1.92682799625235e-16
-8.37837837837838 2.27916883183252e-16
-8.35835835835836 2.69485858889318e-16
-8.33833833833834 3.18508772685576e-16
-8.31831831831832 3.76298728354137e-16
-8.2982982982983 4.44395893386326e-16
-8.27827827827828 5.24606005103494e-16
-8.25825825825826 6.19045274051723e-16
-8.23823823823824 7.30192724674871e-16
-8.21821821821822 8.60951178494166e-16
-8.1981981981982 1.01471827585831e-15
-8.17817817817818 1.19546915264237e-15
-8.15815815815816 1.40785264250169e-15
-8.13813813813814 1.65730316851105e-15
-8.11811811811812 1.95017082606148e-15
-8.0980980980981 2.29387254841619e-15
-8.07807807807808 2.69706769497134e-15
-8.05805805805806 3.16986191875719e-15
-8.03803803803804 3.72404376402735e-15
-8.01801801801802 4.3733591283238e-15
-7.997997997998 5.13382950919607e-15
-7.97797797797798 6.02412085866193e-15
-7.95795795795796 7.06597090549303e-15
-7.93793793793794 8.28468399583074e-15
-7.91791791791792 9.70970386854708e-15
-7.8978978978979 1.13752763482777e-14
-7.87787787787788 1.33212157347805e-14
-7.85785785785786 1.55937907247683e-14
-7.83783783783784 1.82467480586655e-14
-7.81781781781782 2.13424947819475e-14
-7.7977977977978 2.49534630966872e-14
-7.77777777777778 2.91636853079909e-14
-7.75775775775776 3.40706104038538e-14
-7.73773773773774 3.9787198415571e-14
-7.71771771771772 4.64443339685918e-14
-7.6976976976977 5.4193606440538e-14
-7.67767767767768 6.32105109959252e-14
-7.65765765765766 7.36981325813117e-14
-7.63763763763764 8.58913838706879e-14
-7.61761761761762 1.00061878296601e-13
-7.5975975975976 1.16523530854699e-13
-7.57757757757758 1.35638992516664e-13
-7.55755755755756 1.57827039041884e-13
-7.53753753753754 1.83571051982057e-13
-7.51751751751752 2.13428748996349e-13
-7.4974974974975 2.48043342543513e-13
-7.47747747747748 2.88156330935935e-13
-7.45745745745746 3.34622154016965e-13
-7.43743743743744 3.88424977793732e-13
-7.41741741741742 4.50697908714384e-13
-7.3973973973974 5.22744979473711e-13
-7.37737737737738 6.06066294885249e-13
-7.35735735735736 7.02386779168738e-13
-7.33733733733734 8.13689025751835e-13
-7.31731731731732 9.42250818252909e-13
-7.2972972972973 1.09068796768221e-12
-7.27727727727728 1.262003197176e-12
-7.25725725725726 1.45964190299847e-12
-7.23723723723724 1.68755573049416e-12
-7.21721721721722 1.95027502769792e-12
-7.1971971971972 2.25299137914218e-12
-7.17717717717718 2.60165157997871e-12
-7.15715715715716 3.00306458801653e-12
-7.13713713713714 3.4650231910834e-12
-7.11711711711712 3.99644235193919e-12
-7.0970970970971 4.60751644580953e-12
-7.07707707707708 5.30989788981514e-12
-7.05705705705706 6.11689998287646e-12
-7.03703703703704 7.0437271332242e-12
-7.01701701701702 8.10773605306645e-12
-6.996996996997 9.32873195138555e-12
-6.97697697697698 1.07293042619726e-11
-6.95695695695696 1.23352070109962e-11
-6.93693693693694 1.41757895636779e-11
-6.91691691691692 1.62844842008237e-11
-6.8968968968969 1.86993577716893e-11
-6.87687687687688 2.14637355595386e-11
-6.85685685685686 2.46269064908967e-11
-6.83683683683684 2.82449199306815e-11
-6.81681681681682 3.2381485546105e-11
-6.7967967967968 3.71089891068559e-11
-6.77677677677678 4.25096386334913e-11
-6.75675675675676 4.86767570277083e-11
-6.73673673673674 5.57162392366004e-11
-6.71671671671672 6.37481941394491e-11
-6.6966966966967 7.29087937236032e-11
-6.67667667667668 8.33523547614402e-11
-6.65665665665666 9.52536811418151e-11
-6.63663663663664 1.08810698278135e-10
-6.61661661661662 1.24247414645768e-10
-6.5965965965966 1.41817249531744e-10
-6.57657657657658 1.61806770551278e-10
-6.55655655655656 1.84539889444164e-10
-6.53653653653654 2.10382570159622e-10
-6.51651651651652 2.39748109325542e-10
-6.4964964964965 2.73103055937438e-10
-6.47647647647648 3.10973844559381e-10
-6.45645645645646 3.53954224575809e-10
-6.43643643643644 4.027135771479e-10
-6.41641641641642 4.58006221596996e-10
-6.3963963963964 5.20681824054169e-10
-6.37637637637638 5.91697033481538e-10
-6.35635635635636 6.72128483698846e-10
-6.33633633633634 7.63187314959523e-10
-6.31631631631632 8.66235385046001e-10
-6.2962962962963 9.8280335793813e-10
-6.27627627627628 1.11461087800766e-09
-6.25625625625626 1.26358905957513e-09
-6.23623623623624 1.43190554571787e-09
-6.21621621621622 1.62199241663862e-09
-6.1961961961962 1.83657725691024e-09
-6.17617617617618 2.0787177227378e-09
-6.15615615615616 2.35183998527873e-09
-6.13613613613614 2.65978146430928e-09
-6.11611611611612 3.00683830841829e-09
-6.0960960960961 3.39781812376754e-09
-6.07607607607608 3.83809850362696e-09
-6.05605605605606 4.33369196574699e-09
-6.03603603603604 4.89131796456919e-09
-6.01601601601602 5.51848271073395e-09
-5.995995995996 6.22356760178439e-09
-5.97597597597598 7.01592714588943e-09
-5.95595595595596 7.90599734535251e-09
-5.93593593593594 8.90541559921139e-09
-5.91591591591592 1.00271532849868e-08
-5.8958958958959 1.12856622892642e-08
-5.87587587587588 1.2697036876002e-08
-5.85585585585586 1.42791924110059e-08
-5.83583583583584 1.60520626017053e-08
-5.81581581581582 1.80378170640688e-08
-5.7957957957958 2.02611011941336e-08
-5.77577577577578 2.27493005011625e-08
-5.75575575575576 2.55328317539416e-08
-5.73573573573574 2.86454635022852e-08
-5.71571571571572 3.2124668763613e-08
-5.6956956956957 3.60120129107462e-08
-5.67567567567568 4.03535800631662e-08
-5.65565565565566 4.5200441571292e-08
-5.63563563563564 5.06091704933412e-08
-5.61561561561562 5.66424062986154e-08
-5.5955955955956 6.33694743912418e-08
-5.57557557557558 7.08670654362614e-08
-5.55555555555556 7.92199798873018e-08
-5.53553553553554 8.85219435638491e-08
-5.51551551551552 9.8876500608364e-08
-5.4954954954955 1.10397990671284e-07
-5.47547547547548 1.23212617727566e-07
-5.45545545545546 1.3745961852414e-07
-5.43543543543544 1.5329253929596e-07
-5.41541541541542 1.70880630071684e-07
-5.3953953953954 1.90410366621162e-07
-5.37537537537538 2.1208711087848e-07
-5.35535535535536 2.36136921509202e-07
-5.33533533533534 2.62808527181656e-07
-5.31531531531532 2.92375476052561e-07
-5.2952952952953 3.25138475990267e-07
-5.27527527527528 3.61427941137511e-07
-5.25525525525526 4.01606761563285e-07
-5.23523523523524 4.46073313973501e-07
-5.21521521521522 4.95264732746229e-07
-5.1951951951952 5.4966046193278e-07
-5.17517517517518 6.09786110324583e-07
-5.15515515515516 6.76217633231267e-07
-5.13513513513514 7.49585866251233e-07
-5.11511511511512 8.30581438046261e-07
-5.0950950950951 9.19960090959837e-07
-5.07507507507508 1.01854844024876e-06
-5.05505505505506 1.12725020473309e-06
-5.03503503503504 1.24705294381396e-06
-5.01501501501502 1.37903533806611e-06
-4.99499499499499 1.5243750529858e-06
-4.97497497497497 1.68435722796805e-06
-4.95495495495495 1.86038363520377e-06
-4.93493493493493 2.05398255592983e-06
-4.91491491491491 2.26681942433715e-06
-4.89489489489489 2.50070829244518e-06
-4.87487487487487 2.75762417238901e-06
-4.85485485485485 3.03971631583941e-06
-4.83483483483483 3.34932249368796e-06
-4.81481481481481 3.68898434268124e-06
-4.79479479479479 4.06146384937932e-06
-4.77477477477477 4.4697610456467e-06
-4.75475475475475 4.91713299385613e-06
-4.73473473473473 5.40711414409909e-06
-4.71471471471471 5.94353814994721e-06
-4.69469469469469 6.53056123369604e-06
-4.67467467467467 7.17268719654363e-06
-4.65465465465465 7.87479417380527e-06
-4.63463463463463 8.64216324004121e-06
-4.61461461461461 9.48050897386715e-06
-4.59459459459459 1.03960120972233e-05
-4.57457457457457 1.13953543089884e-05
-4.55455455455455 1.24857554380297e-05
-4.53453453453453 1.3675013046071e-05
-4.51451451451451 1.49715446161227e-05
-4.49449449449449 1.63844324676437e-05
-4.47447447447447 1.79234715450684e-05
-4.45445445445445 1.95992202318354e-05
-4.43443443443443 2.14230543475548e-05
-4.41441441441441 2.34072244914564e-05
-4.39439439439439 2.55649169007242e-05
-4.37437437437437 2.79103179977393e-05
-4.35435435435435 3.04586828055866e-05
-4.33433433433433 3.32264074164067e-05
-4.31431431431431 3.62311057022691e-05
-4.29429429429429 3.94916904631592e-05
-4.27427427427427 4.3028459211397e-05
-4.25425425425425 4.68631847962824e-05
-4.23423423423423 5.10192110769697e-05
-4.21421421421421 5.55215538554582e-05
-4.19419419419419 6.03970072851107e-05
-4.17417417417417 6.56742559732345e-05
-4.15415415415415 7.13839929989176e-05
-4.13413413413413 7.75590440694795e-05
-4.11411411411411 8.42344980404937e-05
-4.09409409409409 9.14478440253317e-05
-4.07407407407407 9.92391153205018e-05
-4.05405405405405 0.000107651040372646
-4.03403403403403 0.000116729201011866
-4.01401401401401 0.000126522198173995
-3.99399399399399 0.000137081825331481
-3.97397397397397 0.000148463249848567
-3.95395395395395 0.000160725202471485
-3.93393393393393 0.000173930175158222
-3.91391391391391 0.00018814462744512
-3.89389389389389 0.000203439201538965
-3.87387387387387 0.000219888946313312
-3.85385385385385 0.000237573550376443
-3.83383383383383 0.000256577584365551
-3.81381381381381 0.000276990752607344
-3.79379379379379 0.000298908154269281
-3.77377377377377 0.000322430554107926
-3.75375375375375 0.000347664662901427
-3.73373373373373 0.000374723427631836
-3.71371371371371 0.000403726331459719
-3.69369369369369 0.000434799703508357
-3.67367367367367 0.000468077038447599
-3.65365365365365 0.00050369932583812
-3.63363363363363 0.000541815389165405
-3.61361361361361 0.000582582234459159
-3.59359359359359 0.000626165408357979
-3.57357357357357 0.000672739365441021
-3.55355355355355 0.000722487844607978
-3.53353353353353 0.00077560425424601
-3.51351351351351 0.000832292065877155
-3.49349349349349 0.000892765215932443
-3.47347347347347 0.000957248515249216
-3.45345345345345 0.0010259780658361
-3.43343343343343 0.00109920168439588
-3.41341341341341 0.00117717933203981
-3.39339339339339 0.00126018354956833
-3.37337337337337 0.00134849989763212
-3.35335335335335 0.00144242740102448
-3.33333333333333 0.00154227899629111
-3.31331331331331 0.00164838198177652
-3.29329329329329 0.00176107846915772
-3.27327327327327 0.00188072583544552
-3.25325325325325 0.00200769717436226
-3.23323323323323 0.00214238174593163
-3.21321321321321 0.00228518542304204
-3.19319319319319 0.00243653113367012
-3.17317317317317 0.00259685929737497
-3.15315315315315 0.00276662825459747
-3.13313313313313 0.00294631468722261
-3.11311311311311 0.00313641402878609
-3.09309309309309 0.00333744086263052
-3.07307307307307 0.00354992930624086
-3.05305305305305 0.00377443337991422
-3.03303303303303 0.00401152735784579
-3.01301301301301 0.00426180609964128
-2.99299299299299 0.00452588536019618
-2.97297297297297 0.0048044020758154
-2.95295295295295 0.00509801462438215
-2.93293293293293 0.00540740305732385
-2.91291291291291 0.00573326930106519
-2.89289289289289 0.00607633732560526
-2.87287287287287 0.00643735327780636
-2.85285285285285 0.00681708557693873
-2.83283283283283 0.00721632496998623
-2.81281281281281 0.00763588454418632
-2.79279279279279 0.00807659969425075
-2.77277277277277 0.00853932804169477
-2.75275275275275 0.00902494930369032
-2.73273273273273 0.00953436510885489
-2.71271271271271 0.0100684987573917
-2.69269269269269 0.0106282949230102
-2.67267267267267 0.0112147192940778
-2.65265265265265 0.0118287581514852
-2.63263263263263 0.0124714178807513
-2.61261261261261 0.0131437244159435
-2.59259259259259 0.0138467226130541
-2.57257257257257 0.0145814755505492
-2.55255255255255 0.0153490637548887
-2.53253253253253 0.0161505843489182
-2.51251251251251 0.0169871501211409
-2.49249249249249 0.0178598885140022
-2.47247247247247 0.018769940529451
-2.45245245245245 0.0197184595501959
-2.43243243243243 0.0207066100752274
-2.41241241241241 0.0217355663683581
-2.39239239239239 0.0228065110187135
-2.37237237237237 0.0239206334123108
-2.35235235235235 0.0250791281140699
-2.33233233233233 0.0262831931598317
-2.31231231231231 0.0275340282581906
-2.29229229229229 0.0288328329022027
-2.27227227227227 0.0301808043912899
-2.25225225225225 0.0315791357639331
-2.23223223223223 0.033029013642035
-2.21221221221221 0.0345316159881232
-2.19219219219219 0.0360881097768736
-2.17217217217217 0.0376996485827434
-2.15215215215215 0.0393673700858293
-2.13213213213213 0.0410923934983949
-2.11211211211211 0.0428758169148479
-2.09209209209209 0.0447187145882915
-2.07207207207207 0.0466221341371235
-2.05205205205205 0.0485870936855041
-2.03203203203203 0.0506145789418747
-2.01201201201201 0.0527055402200587
-1.99199199199199 0.0548608894078376
-1.97197197197197 0.057081496888248
-1.95195195195195 0.059368188419199
-1.93193193193193 0.0617217419773594
-1.91191191191191 0.0641428845726061
-1.89189189189189 0.0666322890396674
-1.87187187187187 0.0691905708139176
-1.85185185185185 0.0718182846986055
-1.83183183183183 0.0745159216311036
-1.81181181181181 0.077283905456062
-1.79179179179179 0.0801225897136326
-1.77177177177177 0.083032254451193
-1.75175175175175 0.0860131030672496
-1.73173173173173 0.0890652591964251
-1.71171171171171 0.0921887636446459
-1.69169169169169 0.0953835713838294
-1.67167167167167 0.0986495486155338
-1.65165165165165 0.101986469913169
-1.63163163163163 0.10539401545248
-1.61161161161161 0.108871768340093
-1.59159159159159 0.112419212049971
-1.57157157157157 0.116035727977651
-1.55155155155155 0.119720593122119
-1.53153153153153 0.123472977905145
-1.51151151151151 0.127291944137829
-1.49149149149149 0.13117644314399
-1.47147147147147 0.135125314049902
-1.45145145145145 0.139137282249685
-1.43143143143143 0.143210958055468
-1.41141141141141 0.147344835541168
-1.39139139139139 0.151537291588457
-1.37137137137137 0.155786585143159
-1.35135135135135 0.160090856689972
-1.33133133133133 0.164448127952996
-1.31131131131131 0.168856301829129
-1.29129129129129 0.173313162560933
-1.27127127127127 0.17781637615506
-1.25125125125125 0.182363491051798
-1.23123123123123 0.186951939050736
-1.21121121121121 0.191579036496956
-1.19119119119119 0.1962419857315
-1.17117117117117 0.200937876809264
-1.15115115115115 0.205663689486728
-1.13113113113113 0.210416295481265
-1.11111111111111 0.215192461003031
-1.09109109109109 0.219988849559688
-1.07107107107107 0.224802025033432
-1.05105105105105 0.229628455029052
-1.03103103103103 0.234464514490888
-1.01101101101101 0.239306489585817
-0.990990990990991 0.24415058184851
-0.970970970970971 0.24899291258444
-0.950950950950951 0.253829527525259
-0.930930930930931 0.258656401730343
-0.910910910910911 0.2634694447275
-0.890890890890891 0.268264505884996
-0.870870870870871 0.273037380006279
-0.850850850850851 0.277783813137949
-0.830830830830831 0.282499508580786
-0.810810810810811 0.287180133092853
-0.790790790790791 0.291821323272996
-0.77077077077077 0.296418692112302
-0.75075075075075 0.300967835700437
-0.73073073073073 0.305464340073112
-0.71071071071071 0.309903788186304
-0.69069069069069 0.314281767002296
-0.67067067067067 0.318593874672039
-0.65065065065065 0.322835727797843
-0.63063063063063 0.327002968759958
-0.61061061061061 0.331091273090187
-0.59059059059059 0.33509635687531
-0.57057057057057 0.339013984172804
-0.55055055055055 0.34283997442106
-0.53053053053053 0.346570209826128
-0.51051051051051 0.350200642706842
-0.49049049049049 0.353727302780113
-0.47047047047047 0.357146304368113
-0.45045045045045 0.360453853509139
-0.43043043043043 0.363646254953996
-0.41041041041041 0.366719919029892
-0.39039039039039 0.369671368354051
-0.37037037037037 0.372497244379499
-0.35035035035035 0.375194313755802
-0.33033033033033 0.377759474487924
-0.31031031031031 0.38018976187679
-0.29029029029029 0.382482354225654
-0.27027027027027 0.384634578296894
-0.25025025025025 0.386643914504485
-0.23023023023023 0.388508001828027
-0.21021021021021 0.390224642434919
-0.19019019019019 0.391791805998011
-0.17017017017017 0.393207633696876
-0.15015015015015 0.394470441891644
-0.13013013013013 0.395578725459258
-0.11011011011011 0.396531160782876
-0.0900900900900901 0.397326608386124
-0.07007007007007 0.397964115204853
-0.05005005005005 0.398442916490068
-0.03003003003003 0.398762437336696
-0.01001001001001 0.398922293833933
0.01001001001001 0.398922293833933
0.03003003003003 0.398762437336696
0.05005005005005 0.398442916490068
0.07007007007007 0.397964115204853
0.0900900900900901 0.397326608386124
0.11011011011011 0.396531160782876
0.13013013013013 0.395578725459258
0.15015015015015 0.394470441891644
0.17017017017017 0.393207633696876
0.19019019019019 0.391791805998011
0.21021021021021 0.390224642434919
0.23023023023023 0.388508001828027
0.25025025025025 0.386643914504485
0.27027027027027 0.384634578296894
0.29029029029029 0.382482354225654
0.31031031031031 0.38018976187679
0.33033033033033 0.377759474487924
0.35035035035035 0.375194313755802
0.37037037037037 0.372497244379499
0.39039039039039 0.369671368354051
0.41041041041041 0.366719919029892
0.43043043043043 0.363646254953996
0.45045045045045 0.360453853509139
0.47047047047047 0.357146304368113
0.49049049049049 0.353727302780113
0.51051051051051 0.350200642706842
0.53053053053053 0.346570209826128
0.55055055055055 0.34283997442106
0.57057057057057 0.339013984172804
0.59059059059059 0.33509635687531
0.61061061061061 0.331091273090187
0.63063063063063 0.327002968759958
0.65065065065065 0.322835727797843
0.67067067067067 0.318593874672039
0.69069069069069 0.314281767002296
0.71071071071071 0.309903788186304
0.73073073073073 0.305464340073112
0.75075075075075 0.300967835700437
0.77077077077077 0.296418692112302
0.790790790790791 0.291821323272996
0.810810810810811 0.287180133092853
0.830830830830831 0.282499508580786
0.850850850850851 0.277783813137949
0.870870870870871 0.273037380006279
0.890890890890891 0.268264505884996
0.910910910910911 0.2634694447275
0.930930930930931 0.258656401730343
0.950950950950951 0.253829527525259
0.970970970970971 0.24899291258444
0.990990990990991 0.24415058184851
1.01101101101101 0.239306489585817
1.03103103103103 0.234464514490888
1.05105105105105 0.229628455029052
1.07107107107107 0.224802025033432
1.09109109109109 0.219988849559688
1.11111111111111 0.215192461003031
1.13113113113113 0.210416295481265
1.15115115115115 0.205663689486728
1.17117117117117 0.200937876809264
1.19119119119119 0.1962419857315
1.21121121121121 0.191579036496956
1.23123123123123 0.186951939050736
1.25125125125125 0.182363491051798
1.27127127127127 0.17781637615506
1.29129129129129 0.173313162560933
1.31131131131131 0.168856301829129
1.33133133133133 0.164448127952996
1.35135135135135 0.160090856689972
1.37137137137137 0.155786585143159
1.39139139139139 0.151537291588457
1.41141141141141 0.147344835541168
1.43143143143143 0.143210958055468
1.45145145145145 0.139137282249685
1.47147147147147 0.135125314049902
1.49149149149149 0.13117644314399
1.51151151151151 0.127291944137829
1.53153153153153 0.123472977905145
1.55155155155155 0.119720593122119
1.57157157157157 0.116035727977651
1.59159159159159 0.112419212049971
1.61161161161161 0.108871768340093
1.63163163163163 0.10539401545248
1.65165165165165 0.101986469913169
1.67167167167167 0.0986495486155338
1.69169169169169 0.0953835713838294
1.71171171171171 0.0921887636446459
1.73173173173173 0.0890652591964251
1.75175175175175 0.0860131030672496
1.77177177177177 0.083032254451193
1.79179179179179 0.0801225897136326
1.81181181181181 0.077283905456062
1.83183183183183 0.0745159216311036
1.85185185185185 0.0718182846986055
1.87187187187187 0.0691905708139176
1.89189189189189 0.0666322890396674
1.91191191191191 0.0641428845726061
1.93193193193193 0.0617217419773594
1.95195195195195 0.059368188419199
1.97197197197197 0.057081496888248
1.99199199199199 0.0548608894078376
2.01201201201201 0.0527055402200588
2.03203203203203 0.0506145789418748
2.05205205205205 0.0485870936855042
2.07207207207207 0.0466221341371236
2.09209209209209 0.0447187145882916
2.11211211211211 0.0428758169148479
2.13213213213213 0.041092393498395
2.15215215215215 0.0393673700858294
2.17217217217217 0.0376996485827434
2.19219219219219 0.0360881097768737
2.21221221221221 0.0345316159881233
2.23223223223223 0.033029013642035
2.25225225225225 0.0315791357639332
2.27227227227227 0.03018080439129
2.29229229229229 0.0288328329022028
2.31231231231231 0.0275340282581906
2.33233233233233 0.0262831931598317
2.35235235235235 0.02507912811407
2.37237237237237 0.0239206334123108
2.39239239239239 0.0228065110187136
2.41241241241241 0.0217355663683581
2.43243243243243 0.0207066100752275
2.45245245245245 0.0197184595501959
2.47247247247247 0.0187699405294511
2.49249249249249 0.0178598885140022
2.51251251251251 0.016987150121141
2.53253253253253 0.0161505843489182
2.55255255255255 0.0153490637548888
2.57257257257257 0.0145814755505493
2.59259259259259 0.0138467226130541
2.61261261261261 0.0131437244159435
2.63263263263263 0.0124714178807514
2.65265265265265 0.0118287581514852
2.67267267267267 0.0112147192940778
2.69269269269269 0.0106282949230103
2.71271271271271 0.0100684987573917
2.73273273273273 0.00953436510885491
2.75275275275275 0.00902494930369034
2.77277277277277 0.00853932804169479
2.79279279279279 0.00807659969425077
2.81281281281281 0.00763588454418634
2.83283283283283 0.00721632496998621
2.85285285285285 0.00681708557693871
2.87287287287287 0.00643735327780635
2.89289289289289 0.00607633732560524
2.91291291291291 0.00573326930106518
2.93293293293293 0.00540740305732384
2.95295295295295 0.00509801462438214
2.97297297297297 0.00480440207581539
2.99299299299299 0.00452588536019617
3.01301301301301 0.00426180609964127
3.03303303303303 0.00401152735784578
3.05305305305305 0.00377443337991421
3.07307307307307 0.00354992930624085
3.09309309309309 0.00333744086263051
3.11311311311311 0.00313641402878608
3.13313313313313 0.0029463146872226
3.15315315315315 0.00276662825459746
3.17317317317317 0.00259685929737496
3.19319319319319 0.00243653113367011
3.21321321321321 0.00228518542304203
3.23323323323323 0.00214238174593163
3.25325325325325 0.00200769717436225
3.27327327327327 0.00188072583544551
3.29329329329329 0.00176107846915771
3.31331331331331 0.00164838198177652
3.33333333333333 0.0015422789962911
3.35335335335335 0.00144242740102448
3.37337337337337 0.00134849989763212
3.39339339339339 0.00126018354956833
3.41341341341341 0.00117717933203981
3.43343343343343 0.00109920168439588
3.45345345345345 0.0010259780658361
3.47347347347347 0.000957248515249212
3.49349349349349 0.000892765215932441
3.51351351351351 0.000832292065877152
3.53353353353353 0.000775604254246008
3.55355355355355 0.000722487844607976
3.57357357357357 0.000672739365441019
3.59359359359359 0.000626165408357979
3.61361361361361 0.000582582234459159
3.63363363363363 0.000541815389165405
3.65365365365365 0.00050369932583812
3.67367367367367 0.000468077038447599
3.69369369369369 0.000434799703508357
3.71371371371371 0.000403726331459719
3.73373373373373 0.000374723427631836
3.75375375375375 0.000347664662901427
3.77377377377377 0.000322430554107926
3.79379379379379 0.000298908154269281
3.81381381381381 0.000276990752607344
3.83383383383383 0.000256577584365551
3.85385385385385 0.000237573550376443
3.87387387387387 0.000219888946313312
3.89389389389389 0.000203439201538965
3.91391391391391 0.00018814462744512
3.93393393393393 0.000173930175158222
3.95395395395395 0.000160725202471485
3.97397397397397 0.000148463249848567
3.99399399399399 0.000137081825331481
4.01401401401401 0.000126522198173995
4.03403403403403 0.000116729201011866
4.05405405405405 0.000107651040372646
4.07407407407407 9.92391153205018e-05
4.09409409409409 9.14478440253317e-05
4.11411411411411 8.42344980404937e-05
4.13413413413413 7.75590440694795e-05
4.15415415415415 7.13839929989176e-05
4.17417417417417 6.56742559732345e-05
4.19419419419419 6.03970072851107e-05
4.21421421421421 5.55215538554582e-05
4.23423423423423 5.10192110769697e-05
4.25425425425425 4.68631847962824e-05
4.27427427427427 4.3028459211397e-05
4.29429429429429 3.94916904631592e-05
4.31431431431431 3.62311057022691e-05
4.33433433433433 3.32264074164067e-05
4.35435435435435 3.04586828055866e-05
4.37437437437437 2.79103179977393e-05
4.39439439439439 2.55649169007242e-05
4.41441441441441 2.34072244914564e-05
4.43443443443443 2.14230543475548e-05
4.45445445445445 1.95992202318354e-05
4.47447447447447 1.79234715450684e-05
4.49449449449449 1.63844324676437e-05
4.51451451451451 1.49715446161227e-05
4.53453453453453 1.3675013046071e-05
4.55455455455455 1.24857554380297e-05
4.57457457457457 1.13953543089884e-05
4.59459459459459 1.03960120972233e-05
4.61461461461461 9.48050897386715e-06
4.63463463463463 8.64216324004121e-06
4.65465465465465 7.87479417380527e-06
4.67467467467467 7.17268719654363e-06
4.69469469469469 6.53056123369604e-06
4.71471471471471 5.94353814994721e-06
4.73473473473473 5.40711414409909e-06
4.75475475475475 4.91713299385613e-06
4.77477477477477 4.4697610456467e-06
4.79479479479479 4.06146384937932e-06
4.81481481481481 3.68898434268124e-06
4.83483483483483 3.34932249368796e-06
4.85485485485485 3.03971631583941e-06
4.87487487487487 2.75762417238901e-06
4.89489489489489 2.50070829244518e-06
4.91491491491491 2.26681942433715e-06
4.93493493493493 2.05398255592983e-06
4.95495495495495 1.86038363520377e-06
4.97497497497497 1.68435722796805e-06
4.99499499499499 1.5243750529858e-06
5.01501501501502 1.37903533806611e-06
5.03503503503504 1.24705294381396e-06
5.05505505505506 1.12725020473309e-06
5.07507507507508 1.01854844024876e-06
5.0950950950951 9.19960090959837e-07
5.11511511511512 8.30581438046261e-07
5.13513513513514 7.49585866251233e-07
5.15515515515516 6.76217633231267e-07
5.17517517517518 6.09786110324583e-07
5.1951951951952 5.4966046193278e-07
5.21521521521522 4.95264732746229e-07
5.23523523523524 4.46073313973501e-07
5.25525525525526 4.01606761563285e-07
5.27527527527528 3.61427941137511e-07
5.2952952952953 3.25138475990267e-07
5.31531531531532 2.92375476052561e-07
5.33533533533534 2.62808527181656e-07
5.35535535535536 2.36136921509202e-07
5.37537537537538 2.1208711087848e-07
5.3953953953954 1.90410366621162e-07
5.41541541541542 1.70880630071684e-07
5.43543543543544 1.5329253929596e-07
5.45545545545546 1.3745961852414e-07
5.47547547547548 1.23212617727566e-07
5.4954954954955 1.10397990671284e-07
5.51551551551552 9.8876500608364e-08
5.53553553553554 8.85219435638491e-08
5.55555555555556 7.92199798873018e-08
5.57557557557558 7.08670654362614e-08
5.5955955955956 6.33694743912418e-08
5.61561561561562 5.66424062986154e-08
5.63563563563564 5.06091704933412e-08
5.65565565565566 4.5200441571292e-08
5.67567567567568 4.03535800631662e-08
5.6956956956957 3.60120129107462e-08
5.71571571571572 3.2124668763613e-08
5.73573573573574 2.86454635022852e-08
5.75575575575576 2.55328317539416e-08
5.77577577577578 2.27493005011625e-08
5.7957957957958 2.02611011941336e-08
5.81581581581582 1.80378170640688e-08
5.83583583583584 1.60520626017053e-08
5.85585585585586 1.42791924110059e-08
5.87587587587588 1.2697036876002e-08
5.8958958958959 1.12856622892642e-08
5.91591591591592 1.00271532849868e-08
5.93593593593594 8.90541559921139e-09
5.95595595595596 7.90599734535251e-09
5.97597597597598 7.01592714588943e-09
5.995995995996 6.22356760178439e-09
6.01601601601602 5.5184827107339e-09
6.03603603603604 4.89131796456914e-09
6.05605605605606 4.33369196574694e-09
6.07607607607608 3.83809850362692e-09
6.0960960960961 3.3978181237675e-09
6.11611611611612 3.00683830841826e-09
6.13613613613614 2.65978146430924e-09
6.15615615615616 2.35183998527871e-09
6.17617617617618 2.07871772273778e-09
6.1961961961962 1.83657725691022e-09
6.21621621621622 1.6219924166386e-09
6.23623623623624 1.43190554571785e-09
6.25625625625626 1.26358905957511e-09
6.27627627627628 1.11461087800764e-09
6.2962962962963 9.82803357938119e-10
6.31631631631632 8.66235385045992e-10
6.33633633633634 7.63187314959515e-10
6.35635635635636 6.72128483698836e-10
6.37637637637638 5.91697033481532e-10
6.3963963963964 5.20681824054164e-10
6.41641641641642 4.58006221596989e-10
6.43643643643644 4.02713577147895e-10
6.45645645645646 3.53954224575805e-10
6.47647647647648 3.10973844559378e-10
6.4964964964965 2.73103055937435e-10
6.51651651651652 2.39748109325539e-10
6.53653653653654 2.10382570159619e-10
6.55655655655656 1.84539889444162e-10
6.57657657657658 1.61806770551276e-10
6.5965965965966 1.41817249531742e-10
6.61661661661662 1.24247414645767e-10
6.63663663663664 1.08810698278133e-10
6.65665665665666 9.52536811418137e-11
6.67667667667668 8.33523547614393e-11
6.6966966966967 7.29087937236022e-11
6.71671671671672 6.37481941394484e-11
6.73673673673674 5.57162392365998e-11
6.75675675675676 4.86767570277076e-11
6.77677677677678 4.25096386334907e-11
6.7967967967968 3.71089891068556e-11
6.81681681681682 3.23814855461048e-11
6.83683683683684 2.82449199306813e-11
6.85685685685686 2.46269064908966e-11
6.87687687687688 2.14637355595386e-11
6.8968968968969 1.86993577716892e-11
6.91691691691692 1.62844842008236e-11
6.93693693693694 1.41757895636779e-11
6.95695695695696 1.23352070109961e-11
6.97697697697698 1.07293042619725e-11
6.996996996997 9.32873195138548e-12
7.01701701701702 8.10773605306642e-12
7.03703703703704 7.04372713322415e-12
7.05705705705706 6.11689998287643e-12
7.07707707707708 5.30989788981512e-12
7.0970970970971 4.6075164458095e-12
7.11711711711712 3.99644235193917e-12
7.13713713713714 3.46502319108337e-12
7.15715715715716 3.00306458801651e-12
7.17717717717718 2.60165157997869e-12
7.1971971971972 2.25299137914217e-12
7.21721721721722 1.95027502769791e-12
7.23723723723724 1.68755573049414e-12
7.25725725725726 1.45964190299846e-12
7.27727727727728 1.262003197176e-12
7.2972972972973 1.0906879676822e-12
7.31731731731732 9.42250818252902e-13
7.33733733733734 8.13689025751829e-13
7.35735735735736 7.02386779168733e-13
7.37737737737738 6.06066294885245e-13
7.3973973973974 5.22744979473708e-13
7.41741741741742 4.50697908714381e-13
7.43743743743744 3.88424977793729e-13
7.45745745745746 3.34622154016963e-13
7.47747747747748 2.88156330935933e-13
7.4974974974975 2.48043342543511e-13
7.51751751751752 2.13428748996348e-13
7.53753753753754 1.83571051982055e-13
7.55755755755756 1.57827039041883e-13
7.57757757757758 1.35638992516663e-13
7.5975975975976 1.16523530854698e-13
7.61761761761762 1.000618782966e-13
7.63763763763764 8.58913838706873e-14
7.65765765765766 7.36981325813112e-14
7.67767767767768 6.32105109959248e-14
7.6976976976977 5.41936064405376e-14
7.71771771771772 4.64443339685915e-14
7.73773773773774 3.97871984155707e-14
7.75775775775776 3.40706104038536e-14
7.77777777777778 2.91636853079907e-14
7.7977977977978 2.4953463096687e-14
7.81781781781782 2.13424947819474e-14
7.83783783783784 1.82467480586654e-14
7.85785785785786 1.55937907247682e-14
7.87787787787788 1.33212157347804e-14
7.8978978978979 1.13752763482777e-14
7.91791791791792 9.70970386854701e-15
7.93793793793794 8.28468399583068e-15
7.95795795795796 7.06597090549298e-15
7.97797797797798 6.02412085866189e-15
7.997997997998 5.13382950919603e-15
8.01801801801802 4.3733591283238e-15
8.03803803803804 3.72404376402735e-15
8.05805805805806 3.16986191875719e-15
8.07807807807808 2.69706769497134e-15
8.0980980980981 2.29387254841619e-15
8.11811811811812 1.95017082606148e-15
8.13813813813814 1.65730316851105e-15
8.15815815815816 1.40785264250169e-15
8.17817817817818 1.19546915264237e-15
8.1981981981982 1.01471827585831e-15
8.21821821821822 8.60951178494166e-16
8.23823823823824 7.30192724674871e-16
8.25825825825826 6.19045274051723e-16
8.27827827827828 5.24606005103494e-16
8.2982982982983 4.44395893386326e-16
8.31831831831832 3.76298728354137e-16
8.33833833833834 3.18508772685576e-16
8.35835835835836 2.69485858889318e-16
8.37837837837838 2.27916883183252e-16
8.3983983983984 1.92682799625235e-16
8.41841841841842 1.62830341150177e-16
8.43843843843844 1.37547801096493e-16
8.45845845845846 1.16144301209627e-16
8.47847847847848 9.8032051926925e-17
8.4984984984985 8.27111796592674e-17
8.51851851851852 6.97567552529606e-17
8.53853853853854 5.88077091106193e-17
8.55855855855856 4.95573626747691e-17
8.57857857857858 4.17453440895294e-17
8.5985985985986 3.51506886838449e-17
8.61861861861862 2.95859531836935e-17
8.63863863863864 2.48921968838275e-17
8.65865865865866 2.09347039316773e-17
8.67867867867868 1.75993388643364e-17
8.6986986986987 1.47894429982973e-17
8.71871871871872 1.24231925503958e-17
8.73873873873874 1.04313507694241e-17
8.75875875875876 8.75535614211525e-18
8.77877877877878 7.34569713009337e-18
8.7987987987988 6.16053109048305e-18
8.81881881881882 5.16451119999019e-18
8.83883883883884 4.32779048512422e-18
8.85885885885886 3.62517658454128e-18
8.87887887887888 3.03541474067764e-18
8.8988988988989 2.54057982941549e-18
8.91891891891892 2.12556106807901e-18
8.93893893893894 1.77762546207239e-18
8.95895895895896 1.48604811781133e-18
8.97897897897898 1.24179931485728e-18
8.998998998999 1.03727973679138e-18
9.01901901901902 8.66096545670873e-19
9.03903903903904 7.22874080905507e-19
9.05905905905906 6.03093897535385e-19
9.07907907907908 5.0295965472299e-19
9.0990990990991 4.19283042962206e-19
9.11911911911912 3.49387515332417e-19
9.13913913913914 2.91027078875695e-19
9.15915915915916 2.42317819504618e-19
9.17917917917918 2.01680188581445e-19
9.1991991991992 1.67790380698338e-19
9.21921921921922 1.39539388139188e-19
9.23923923923924 1.1599853476858e-19
9.25925925925926 9.63904764362469e-20
9.27927927927928 8.00648113242436e-20
9.2992992992993 6.6477576193283e-20
9.31931931931932 5.51740167796855e-20
9.33933933933934 4.5774115701642e-20
9.35935935935936 3.79604417474457e-20
9.37937937937938 3.14679525475421e-20
9.3993993993994 2.60754402558043e-20
9.41941941941942 2.15983585814365e-20
9.43943943943944 1.78828106797973e-20
9.45945945945946 1.48005121824234e-20
9.47947947947948 1.22445730038445e-20
9.4994994994995 1.01259663374544e-20
9.51951951951952 8.37057415073758e-21
9.53953953953954 6.91671611023779e-21
9.55955955955956 5.71308371631411e-21
9.57957957957958 4.71701393695898e-21
9.5995995995996 3.89304716291232e-21
9.61961961961962 3.21172317125736e-21
9.63963963963964 2.64857624243904e-21
9.65965965965966 2.18329684678274e-21
9.67967967967968 1.79903258756111e-21
9.6996996996997 1.48180551599987e-21
9.71971971971972 1.22002665237565e-21
9.73973973973974 1.00409166885045e-21
9.75975975975976 8.26044308669654e-22
9.77977977977978 6.79296312742742e-22
9.7997997997998 5.58394465795474e-22
9.81981981981982 4.58826916995414e-22
9.83983983983984 3.7686222201397e-22
9.85985985985986 3.09415635142992e-22
9.87987987987988 2.53938085193762e-22
9.8998998998999 2.08324025950642e-22
9.91991991991992 1.70834984871876e-22
9.93993993993994 1.40036162642795e-22
9.95995995995996 1.14743877987917e-22
9.97997997997998 9.39820210218911e-23
10 7.69459862670642e-23
};
\addlegendentry{N(0,1)}; 
\legend{}; 
\end{axis}

\end{tikzpicture}

%% file: FinalFigs/Null_Dists_d_10_100_n_20_m_100_kernel__Dirichlet_Poly_5_2022_10_15_22_11_41mmd.tex
\begin{tikzpicture}

\definecolor{darkorange25512714}{RGB}{255,127,14}
\definecolor{darkslategray38}{RGB}{38,38,38}
\definecolor{lightgray204}{RGB}{204,204,204}
\definecolor{steelblue31119180}{RGB}{31,119,180}

\begin{axis}[
axis line style={darkslategray38},
height=\figheight,
legend cell align={left},
legend style={fill opacity=0.8, draw opacity=1, text opacity=1, draw=none},
tick align=outside,
tick pos=left,
title={$\dmmd$~$(n/m=0.2)$},
width=\figwidth,
x grid style={lightgray204},
xmin=-6, xmax=6,
xtick style={color=darkslategray38},
y grid style={lightgray204},
ylabel=\textcolor{darkslategray38}{},
ymin=0, ymax=1.03851378592425,
ytick style={color=darkslategray38}, 
xticklabels=empty,
yticklabels=empty
]
\draw[draw=none,fill=steelblue31119180,fill opacity=0.8] (axis cs:-1.37188339233398,0) rectangle (axis cs:-1.24004113674164,0.109221452100811);
\addlegendimage{ybar,ybar legend,draw=none,fill=steelblue31119180,fill opacity=0.8}
\addlegendentry{mmd (d=10)}

\draw[draw=none,fill=steelblue31119180,fill opacity=0.8] (axis cs:-1.04227781295776,0) rectangle (axis cs:-0.910435557365417,0.242714250218721);
\draw[draw=none,fill=steelblue31119180,fill opacity=0.8] (axis cs:-0.712672114372253,0) rectangle (axis cs:-0.580829858779907,0.436885768900847);
\draw[draw=none,fill=steelblue31119180,fill opacity=0.8] (axis cs:-0.383066475391388,0) rectangle (axis cs:-0.251224219799042,0.637125022039417);
\draw[draw=none,fill=steelblue31119180,fill opacity=0.8] (axis cs:-0.0534608513116837,0) rectangle (axis cs:0.0783814042806625,0.685667880861468);
\draw[draw=none,fill=steelblue31119180,fill opacity=0.8] (axis cs:0.276144772768021,0) rectangle (axis cs:0.407987028360367,0.461157158809483);
\draw[draw=none,fill=steelblue31119180,fill opacity=0.8] (axis cs:0.605750381946564,0) rectangle (axis cs:0.73759263753891,0.285189295579549);
\draw[draw=none,fill=steelblue31119180,fill opacity=0.8] (axis cs:0.935356080532074,0) rectangle (axis cs:1.06719827651978,0.127425027450946);
\draw[draw=none,fill=steelblue31119180,fill opacity=0.8] (axis cs:1.26496171951294,0) rectangle (axis cs:1.39680397510529,0.0424749937882762);
\draw[draw=none,fill=steelblue31119180,fill opacity=0.8] (axis cs:1.59456729888916,0) rectangle (axis cs:1.72640955448151,0.00606785845004506);
\draw[draw=none,fill=darkorange25512714,fill opacity=0.8] (axis cs:-1.24004101753235,0) rectangle (axis cs:-1.10819876194,0);
\addlegendimage{ybar,ybar legend,draw=none,fill=darkorange25512714,fill opacity=0.8}
\addlegendentry{mmd (d=100)}

\draw[draw=none,fill=darkorange25512714,fill opacity=0.8] (axis cs:-0.910435497760773,0) rectangle (axis cs:-0.778593242168427,0.0485428500437443);
\draw[draw=none,fill=darkorange25512714,fill opacity=0.8] (axis cs:-0.580829858779907,0) rectangle (axis cs:-0.448987603187561,0.412614337295245);
\draw[draw=none,fill=darkorange25512714,fill opacity=0.8] (axis cs:-0.251224219799042,0) rectangle (axis cs:-0.119381964206696,0.989060748499285);
\draw[draw=none,fill=darkorange25512714,fill opacity=0.8] (axis cs:0.0783814042806625,0) rectangle (axis cs:0.210223659873009,0.782753598505569);
\draw[draw=none,fill=darkorange25512714,fill opacity=0.8] (axis cs:0.407987028360367,0) rectangle (axis cs:0.539829254150391,0.570378591159097);
\draw[draw=none,fill=darkorange25512714,fill opacity=0.8] (axis cs:0.73759263753891,0) rectangle (axis cs:0.869434893131256,0.175967863229934);
\draw[draw=none,fill=darkorange25512714,fill opacity=0.8] (axis cs:1.06719827651978,0) rectangle (axis cs:1.19904053211212,0.0485428676003605);
\draw[draw=none,fill=darkorange25512714,fill opacity=0.8] (axis cs:1.396803855896,0) rectangle (axis cs:1.52864611148834,0.00606785625546803);
\draw[draw=none,fill=darkorange25512714,fill opacity=0.8] (axis cs:1.7264096736908,0) rectangle (axis cs:1.85825192928314,0);
\addplot [semithick, black]
table {%
-10 7.69459862670642e-23
-9.97997997997998 9.39820210218911e-23
-9.95995995995996 1.14743877987917e-22
-9.93993993993994 1.40036162642795e-22
-9.91991991991992 1.70834984871876e-22
-9.8998998998999 2.08324025950642e-22
-9.87987987987988 2.53938085193762e-22
-9.85985985985986 3.09415635142992e-22
-9.83983983983984 3.7686222201397e-22
-9.81981981981982 4.58826916995414e-22
-9.7997997997998 5.58394465795474e-22
-9.77977977977978 6.79296312742742e-22
-9.75975975975976 8.26044308669654e-22
-9.73973973973974 1.00409166885045e-21
-9.71971971971972 1.22002665237565e-21
-9.6996996996997 1.48180551599987e-21
-9.67967967967968 1.79903258756111e-21
-9.65965965965966 2.18329684678274e-21
-9.63963963963964 2.64857624243904e-21
-9.61961961961962 3.21172317125736e-21
-9.5995995995996 3.89304716291232e-21
-9.57957957957958 4.71701393695898e-21
-9.55955955955956 5.71308371631411e-21
-9.53953953953954 6.91671611023779e-21
-9.51951951951952 8.37057415073758e-21
-9.4994994994995 1.01259663374544e-20
-9.47947947947948 1.22445730038445e-20
-9.45945945945946 1.48005121824234e-20
-9.43943943943944 1.78828106797973e-20
-9.41941941941942 2.15983585814365e-20
-9.3993993993994 2.60754402558043e-20
-9.37937937937938 3.14679525475421e-20
-9.35935935935936 3.79604417474457e-20
-9.33933933933934 4.5774115701642e-20
-9.31931931931932 5.51740167796855e-20
-9.2992992992993 6.6477576193283e-20
-9.27927927927928 8.00648113242436e-20
-9.25925925925926 9.63904764362469e-20
-9.23923923923924 1.1599853476858e-19
-9.21921921921922 1.39539388139188e-19
-9.1991991991992 1.67790380698338e-19
-9.17917917917918 2.01680188581445e-19
-9.15915915915916 2.42317819504618e-19
-9.13913913913914 2.91027078875695e-19
-9.11911911911912 3.49387515332417e-19
-9.0990990990991 4.19283042962206e-19
-9.07907907907908 5.0295965472299e-19
-9.05905905905906 6.03093897535385e-19
-9.03903903903904 7.22874080905507e-19
-9.01901901901902 8.66096545670873e-19
-8.998998998999 1.03727973679138e-18
-8.97897897897898 1.24179931485728e-18
-8.95895895895896 1.48604811781133e-18
-8.93893893893894 1.77762546207239e-18
-8.91891891891892 2.12556106807901e-18
-8.8988988988989 2.54057982941549e-18
-8.87887887887888 3.03541474067764e-18
-8.85885885885886 3.62517658454128e-18
-8.83883883883884 4.32779048512422e-18
-8.81881881881882 5.16451119999019e-18
-8.7987987987988 6.16053109048305e-18
-8.77877877877878 7.34569713009337e-18
-8.75875875875876 8.75535614211525e-18
-8.73873873873874 1.04313507694241e-17
-8.71871871871872 1.24231925503958e-17
-8.6986986986987 1.47894429982973e-17
-8.67867867867868 1.75993388643364e-17
-8.65865865865866 2.09347039316773e-17
-8.63863863863864 2.48921968838275e-17
-8.61861861861862 2.95859531836935e-17
-8.5985985985986 3.51506886838449e-17
-8.57857857857858 4.17453440895294e-17
-8.55855855855856 4.95573626747691e-17
-8.53853853853854 5.88077091106193e-17
-8.51851851851852 6.97567552529606e-17
-8.4984984984985 8.27111796592674e-17
-8.47847847847848 9.8032051926925e-17
-8.45845845845846 1.16144301209627e-16
-8.43843843843844 1.37547801096493e-16
-8.41841841841842 1.62830341150177e-16
-8.3983983983984 1.92682799625235e-16
-8.37837837837838 2.27916883183252e-16
-8.35835835835836 2.69485858889318e-16
-8.33833833833834 3.18508772685576e-16
-8.31831831831832 3.76298728354137e-16
-8.2982982982983 4.44395893386326e-16
-8.27827827827828 5.24606005103494e-16
-8.25825825825826 6.19045274051723e-16
-8.23823823823824 7.30192724674871e-16
-8.21821821821822 8.60951178494166e-16
-8.1981981981982 1.01471827585831e-15
-8.17817817817818 1.19546915264237e-15
-8.15815815815816 1.40785264250169e-15
-8.13813813813814 1.65730316851105e-15
-8.11811811811812 1.95017082606148e-15
-8.0980980980981 2.29387254841619e-15
-8.07807807807808 2.69706769497134e-15
-8.05805805805806 3.16986191875719e-15
-8.03803803803804 3.72404376402735e-15
-8.01801801801802 4.3733591283238e-15
-7.997997997998 5.13382950919607e-15
-7.97797797797798 6.02412085866193e-15
-7.95795795795796 7.06597090549303e-15
-7.93793793793794 8.28468399583074e-15
-7.91791791791792 9.70970386854708e-15
-7.8978978978979 1.13752763482777e-14
-7.87787787787788 1.33212157347805e-14
-7.85785785785786 1.55937907247683e-14
-7.83783783783784 1.82467480586655e-14
-7.81781781781782 2.13424947819475e-14
-7.7977977977978 2.49534630966872e-14
-7.77777777777778 2.91636853079909e-14
-7.75775775775776 3.40706104038538e-14
-7.73773773773774 3.9787198415571e-14
-7.71771771771772 4.64443339685918e-14
-7.6976976976977 5.4193606440538e-14
-7.67767767767768 6.32105109959252e-14
-7.65765765765766 7.36981325813117e-14
-7.63763763763764 8.58913838706879e-14
-7.61761761761762 1.00061878296601e-13
-7.5975975975976 1.16523530854699e-13
-7.57757757757758 1.35638992516664e-13
-7.55755755755756 1.57827039041884e-13
-7.53753753753754 1.83571051982057e-13
-7.51751751751752 2.13428748996349e-13
-7.4974974974975 2.48043342543513e-13
-7.47747747747748 2.88156330935935e-13
-7.45745745745746 3.34622154016965e-13
-7.43743743743744 3.88424977793732e-13
-7.41741741741742 4.50697908714384e-13
-7.3973973973974 5.22744979473711e-13
-7.37737737737738 6.06066294885249e-13
-7.35735735735736 7.02386779168738e-13
-7.33733733733734 8.13689025751835e-13
-7.31731731731732 9.42250818252909e-13
-7.2972972972973 1.09068796768221e-12
-7.27727727727728 1.262003197176e-12
-7.25725725725726 1.45964190299847e-12
-7.23723723723724 1.68755573049416e-12
-7.21721721721722 1.95027502769792e-12
-7.1971971971972 2.25299137914218e-12
-7.17717717717718 2.60165157997871e-12
-7.15715715715716 3.00306458801653e-12
-7.13713713713714 3.4650231910834e-12
-7.11711711711712 3.99644235193919e-12
-7.0970970970971 4.60751644580953e-12
-7.07707707707708 5.30989788981514e-12
-7.05705705705706 6.11689998287646e-12
-7.03703703703704 7.0437271332242e-12
-7.01701701701702 8.10773605306645e-12
-6.996996996997 9.32873195138555e-12
-6.97697697697698 1.07293042619726e-11
-6.95695695695696 1.23352070109962e-11
-6.93693693693694 1.41757895636779e-11
-6.91691691691692 1.62844842008237e-11
-6.8968968968969 1.86993577716893e-11
-6.87687687687688 2.14637355595386e-11
-6.85685685685686 2.46269064908967e-11
-6.83683683683684 2.82449199306815e-11
-6.81681681681682 3.2381485546105e-11
-6.7967967967968 3.71089891068559e-11
-6.77677677677678 4.25096386334913e-11
-6.75675675675676 4.86767570277083e-11
-6.73673673673674 5.57162392366004e-11
-6.71671671671672 6.37481941394491e-11
-6.6966966966967 7.29087937236032e-11
-6.67667667667668 8.33523547614402e-11
-6.65665665665666 9.52536811418151e-11
-6.63663663663664 1.08810698278135e-10
-6.61661661661662 1.24247414645768e-10
-6.5965965965966 1.41817249531744e-10
-6.57657657657658 1.61806770551278e-10
-6.55655655655656 1.84539889444164e-10
-6.53653653653654 2.10382570159622e-10
-6.51651651651652 2.39748109325542e-10
-6.4964964964965 2.73103055937438e-10
-6.47647647647648 3.10973844559381e-10
-6.45645645645646 3.53954224575809e-10
-6.43643643643644 4.027135771479e-10
-6.41641641641642 4.58006221596996e-10
-6.3963963963964 5.20681824054169e-10
-6.37637637637638 5.91697033481538e-10
-6.35635635635636 6.72128483698846e-10
-6.33633633633634 7.63187314959523e-10
-6.31631631631632 8.66235385046001e-10
-6.2962962962963 9.8280335793813e-10
-6.27627627627628 1.11461087800766e-09
-6.25625625625626 1.26358905957513e-09
-6.23623623623624 1.43190554571787e-09
-6.21621621621622 1.62199241663862e-09
-6.1961961961962 1.83657725691024e-09
-6.17617617617618 2.0787177227378e-09
-6.15615615615616 2.35183998527873e-09
-6.13613613613614 2.65978146430928e-09
-6.11611611611612 3.00683830841829e-09
-6.0960960960961 3.39781812376754e-09
-6.07607607607608 3.83809850362696e-09
-6.05605605605606 4.33369196574699e-09
-6.03603603603604 4.89131796456919e-09
-6.01601601601602 5.51848271073395e-09
-5.995995995996 6.22356760178439e-09
-5.97597597597598 7.01592714588943e-09
-5.95595595595596 7.90599734535251e-09
-5.93593593593594 8.90541559921139e-09
-5.91591591591592 1.00271532849868e-08
-5.8958958958959 1.12856622892642e-08
-5.87587587587588 1.2697036876002e-08
-5.85585585585586 1.42791924110059e-08
-5.83583583583584 1.60520626017053e-08
-5.81581581581582 1.80378170640688e-08
-5.7957957957958 2.02611011941336e-08
-5.77577577577578 2.27493005011625e-08
-5.75575575575576 2.55328317539416e-08
-5.73573573573574 2.86454635022852e-08
-5.71571571571572 3.2124668763613e-08
-5.6956956956957 3.60120129107462e-08
-5.67567567567568 4.03535800631662e-08
-5.65565565565566 4.5200441571292e-08
-5.63563563563564 5.06091704933412e-08
-5.61561561561562 5.66424062986154e-08
-5.5955955955956 6.33694743912418e-08
-5.57557557557558 7.08670654362614e-08
-5.55555555555556 7.92199798873018e-08
-5.53553553553554 8.85219435638491e-08
-5.51551551551552 9.8876500608364e-08
-5.4954954954955 1.10397990671284e-07
-5.47547547547548 1.23212617727566e-07
-5.45545545545546 1.3745961852414e-07
-5.43543543543544 1.5329253929596e-07
-5.41541541541542 1.70880630071684e-07
-5.3953953953954 1.90410366621162e-07
-5.37537537537538 2.1208711087848e-07
-5.35535535535536 2.36136921509202e-07
-5.33533533533534 2.62808527181656e-07
-5.31531531531532 2.92375476052561e-07
-5.2952952952953 3.25138475990267e-07
-5.27527527527528 3.61427941137511e-07
-5.25525525525526 4.01606761563285e-07
-5.23523523523524 4.46073313973501e-07
-5.21521521521522 4.95264732746229e-07
-5.1951951951952 5.4966046193278e-07
-5.17517517517518 6.09786110324583e-07
-5.15515515515516 6.76217633231267e-07
-5.13513513513514 7.49585866251233e-07
-5.11511511511512 8.30581438046261e-07
-5.0950950950951 9.19960090959837e-07
-5.07507507507508 1.01854844024876e-06
-5.05505505505506 1.12725020473309e-06
-5.03503503503504 1.24705294381396e-06
-5.01501501501502 1.37903533806611e-06
-4.99499499499499 1.5243750529858e-06
-4.97497497497497 1.68435722796805e-06
-4.95495495495495 1.86038363520377e-06
-4.93493493493493 2.05398255592983e-06
-4.91491491491491 2.26681942433715e-06
-4.89489489489489 2.50070829244518e-06
-4.87487487487487 2.75762417238901e-06
-4.85485485485485 3.03971631583941e-06
-4.83483483483483 3.34932249368796e-06
-4.81481481481481 3.68898434268124e-06
-4.79479479479479 4.06146384937932e-06
-4.77477477477477 4.4697610456467e-06
-4.75475475475475 4.91713299385613e-06
-4.73473473473473 5.40711414409909e-06
-4.71471471471471 5.94353814994721e-06
-4.69469469469469 6.53056123369604e-06
-4.67467467467467 7.17268719654363e-06
-4.65465465465465 7.87479417380527e-06
-4.63463463463463 8.64216324004121e-06
-4.61461461461461 9.48050897386715e-06
-4.59459459459459 1.03960120972233e-05
-4.57457457457457 1.13953543089884e-05
-4.55455455455455 1.24857554380297e-05
-4.53453453453453 1.3675013046071e-05
-4.51451451451451 1.49715446161227e-05
-4.49449449449449 1.63844324676437e-05
-4.47447447447447 1.79234715450684e-05
-4.45445445445445 1.95992202318354e-05
-4.43443443443443 2.14230543475548e-05
-4.41441441441441 2.34072244914564e-05
-4.39439439439439 2.55649169007242e-05
-4.37437437437437 2.79103179977393e-05
-4.35435435435435 3.04586828055866e-05
-4.33433433433433 3.32264074164067e-05
-4.31431431431431 3.62311057022691e-05
-4.29429429429429 3.94916904631592e-05
-4.27427427427427 4.3028459211397e-05
-4.25425425425425 4.68631847962824e-05
-4.23423423423423 5.10192110769697e-05
-4.21421421421421 5.55215538554582e-05
-4.19419419419419 6.03970072851107e-05
-4.17417417417417 6.56742559732345e-05
-4.15415415415415 7.13839929989176e-05
-4.13413413413413 7.75590440694795e-05
-4.11411411411411 8.42344980404937e-05
-4.09409409409409 9.14478440253317e-05
-4.07407407407407 9.92391153205018e-05
-4.05405405405405 0.000107651040372646
-4.03403403403403 0.000116729201011866
-4.01401401401401 0.000126522198173995
-3.99399399399399 0.000137081825331481
-3.97397397397397 0.000148463249848567
-3.95395395395395 0.000160725202471485
-3.93393393393393 0.000173930175158222
-3.91391391391391 0.00018814462744512
-3.89389389389389 0.000203439201538965
-3.87387387387387 0.000219888946313312
-3.85385385385385 0.000237573550376443
-3.83383383383383 0.000256577584365551
-3.81381381381381 0.000276990752607344
-3.79379379379379 0.000298908154269281
-3.77377377377377 0.000322430554107926
-3.75375375375375 0.000347664662901427
-3.73373373373373 0.000374723427631836
-3.71371371371371 0.000403726331459719
-3.69369369369369 0.000434799703508357
-3.67367367367367 0.000468077038447599
-3.65365365365365 0.00050369932583812
-3.63363363363363 0.000541815389165405
-3.61361361361361 0.000582582234459159
-3.59359359359359 0.000626165408357979
-3.57357357357357 0.000672739365441021
-3.55355355355355 0.000722487844607978
-3.53353353353353 0.00077560425424601
-3.51351351351351 0.000832292065877155
-3.49349349349349 0.000892765215932443
-3.47347347347347 0.000957248515249216
-3.45345345345345 0.0010259780658361
-3.43343343343343 0.00109920168439588
-3.41341341341341 0.00117717933203981
-3.39339339339339 0.00126018354956833
-3.37337337337337 0.00134849989763212
-3.35335335335335 0.00144242740102448
-3.33333333333333 0.00154227899629111
-3.31331331331331 0.00164838198177652
-3.29329329329329 0.00176107846915772
-3.27327327327327 0.00188072583544552
-3.25325325325325 0.00200769717436226
-3.23323323323323 0.00214238174593163
-3.21321321321321 0.00228518542304204
-3.19319319319319 0.00243653113367012
-3.17317317317317 0.00259685929737497
-3.15315315315315 0.00276662825459747
-3.13313313313313 0.00294631468722261
-3.11311311311311 0.00313641402878609
-3.09309309309309 0.00333744086263052
-3.07307307307307 0.00354992930624086
-3.05305305305305 0.00377443337991422
-3.03303303303303 0.00401152735784579
-3.01301301301301 0.00426180609964128
-2.99299299299299 0.00452588536019618
-2.97297297297297 0.0048044020758154
-2.95295295295295 0.00509801462438215
-2.93293293293293 0.00540740305732385
-2.91291291291291 0.00573326930106519
-2.89289289289289 0.00607633732560526
-2.87287287287287 0.00643735327780636
-2.85285285285285 0.00681708557693873
-2.83283283283283 0.00721632496998623
-2.81281281281281 0.00763588454418632
-2.79279279279279 0.00807659969425075
-2.77277277277277 0.00853932804169477
-2.75275275275275 0.00902494930369032
-2.73273273273273 0.00953436510885489
-2.71271271271271 0.0100684987573917
-2.69269269269269 0.0106282949230102
-2.67267267267267 0.0112147192940778
-2.65265265265265 0.0118287581514852
-2.63263263263263 0.0124714178807513
-2.61261261261261 0.0131437244159435
-2.59259259259259 0.0138467226130541
-2.57257257257257 0.0145814755505492
-2.55255255255255 0.0153490637548887
-2.53253253253253 0.0161505843489182
-2.51251251251251 0.0169871501211409
-2.49249249249249 0.0178598885140022
-2.47247247247247 0.018769940529451
-2.45245245245245 0.0197184595501959
-2.43243243243243 0.0207066100752274
-2.41241241241241 0.0217355663683581
-2.39239239239239 0.0228065110187135
-2.37237237237237 0.0239206334123108
-2.35235235235235 0.0250791281140699
-2.33233233233233 0.0262831931598317
-2.31231231231231 0.0275340282581906
-2.29229229229229 0.0288328329022027
-2.27227227227227 0.0301808043912899
-2.25225225225225 0.0315791357639331
-2.23223223223223 0.033029013642035
-2.21221221221221 0.0345316159881232
-2.19219219219219 0.0360881097768736
-2.17217217217217 0.0376996485827434
-2.15215215215215 0.0393673700858293
-2.13213213213213 0.0410923934983949
-2.11211211211211 0.0428758169148479
-2.09209209209209 0.0447187145882915
-2.07207207207207 0.0466221341371235
-2.05205205205205 0.0485870936855041
-2.03203203203203 0.0506145789418747
-2.01201201201201 0.0527055402200587
-1.99199199199199 0.0548608894078376
-1.97197197197197 0.057081496888248
-1.95195195195195 0.059368188419199
-1.93193193193193 0.0617217419773594
-1.91191191191191 0.0641428845726061
-1.89189189189189 0.0666322890396674
-1.87187187187187 0.0691905708139176
-1.85185185185185 0.0718182846986055
-1.83183183183183 0.0745159216311036
-1.81181181181181 0.077283905456062
-1.79179179179179 0.0801225897136326
-1.77177177177177 0.083032254451193
-1.75175175175175 0.0860131030672496
-1.73173173173173 0.0890652591964251
-1.71171171171171 0.0921887636446459
-1.69169169169169 0.0953835713838294
-1.67167167167167 0.0986495486155338
-1.65165165165165 0.101986469913169
-1.63163163163163 0.10539401545248
-1.61161161161161 0.108871768340093
-1.59159159159159 0.112419212049971
-1.57157157157157 0.116035727977651
-1.55155155155155 0.119720593122119
-1.53153153153153 0.123472977905145
-1.51151151151151 0.127291944137829
-1.49149149149149 0.13117644314399
-1.47147147147147 0.135125314049902
-1.45145145145145 0.139137282249685
-1.43143143143143 0.143210958055468
-1.41141141141141 0.147344835541168
-1.39139139139139 0.151537291588457
-1.37137137137137 0.155786585143159
-1.35135135135135 0.160090856689972
-1.33133133133133 0.164448127952996
-1.31131131131131 0.168856301829129
-1.29129129129129 0.173313162560933
-1.27127127127127 0.17781637615506
-1.25125125125125 0.182363491051798
-1.23123123123123 0.186951939050736
-1.21121121121121 0.191579036496956
-1.19119119119119 0.1962419857315
-1.17117117117117 0.200937876809264
-1.15115115115115 0.205663689486728
-1.13113113113113 0.210416295481265
-1.11111111111111 0.215192461003031
-1.09109109109109 0.219988849559688
-1.07107107107107 0.224802025033432
-1.05105105105105 0.229628455029052
-1.03103103103103 0.234464514490888
-1.01101101101101 0.239306489585817
-0.990990990990991 0.24415058184851
-0.970970970970971 0.24899291258444
-0.950950950950951 0.253829527525259
-0.930930930930931 0.258656401730343
-0.910910910910911 0.2634694447275
-0.890890890890891 0.268264505884996
-0.870870870870871 0.273037380006279
-0.850850850850851 0.277783813137949
-0.830830830830831 0.282499508580786
-0.810810810810811 0.287180133092853
-0.790790790790791 0.291821323272996
-0.77077077077077 0.296418692112302
-0.75075075075075 0.300967835700437
-0.73073073073073 0.305464340073112
-0.71071071071071 0.309903788186304
-0.69069069069069 0.314281767002296
-0.67067067067067 0.318593874672039
-0.65065065065065 0.322835727797843
-0.63063063063063 0.327002968759958
-0.61061061061061 0.331091273090187
-0.59059059059059 0.33509635687531
-0.57057057057057 0.339013984172804
-0.55055055055055 0.34283997442106
-0.53053053053053 0.346570209826128
-0.51051051051051 0.350200642706842
-0.49049049049049 0.353727302780113
-0.47047047047047 0.357146304368113
-0.45045045045045 0.360453853509139
-0.43043043043043 0.363646254953996
-0.41041041041041 0.366719919029892
-0.39039039039039 0.369671368354051
-0.37037037037037 0.372497244379499
-0.35035035035035 0.375194313755802
-0.33033033033033 0.377759474487924
-0.31031031031031 0.38018976187679
-0.29029029029029 0.382482354225654
-0.27027027027027 0.384634578296894
-0.25025025025025 0.386643914504485
-0.23023023023023 0.388508001828027
-0.21021021021021 0.390224642434919
-0.19019019019019 0.391791805998011
-0.17017017017017 0.393207633696876
-0.15015015015015 0.394470441891644
-0.13013013013013 0.395578725459258
-0.11011011011011 0.396531160782876
-0.0900900900900901 0.397326608386124
-0.07007007007007 0.397964115204853
-0.05005005005005 0.398442916490068
-0.03003003003003 0.398762437336696
-0.01001001001001 0.398922293833933
0.01001001001001 0.398922293833933
0.03003003003003 0.398762437336696
0.05005005005005 0.398442916490068
0.07007007007007 0.397964115204853
0.0900900900900901 0.397326608386124
0.11011011011011 0.396531160782876
0.13013013013013 0.395578725459258
0.15015015015015 0.394470441891644
0.17017017017017 0.393207633696876
0.19019019019019 0.391791805998011
0.21021021021021 0.390224642434919
0.23023023023023 0.388508001828027
0.25025025025025 0.386643914504485
0.27027027027027 0.384634578296894
0.29029029029029 0.382482354225654
0.31031031031031 0.38018976187679
0.33033033033033 0.377759474487924
0.35035035035035 0.375194313755802
0.37037037037037 0.372497244379499
0.39039039039039 0.369671368354051
0.41041041041041 0.366719919029892
0.43043043043043 0.363646254953996
0.45045045045045 0.360453853509139
0.47047047047047 0.357146304368113
0.49049049049049 0.353727302780113
0.51051051051051 0.350200642706842
0.53053053053053 0.346570209826128
0.55055055055055 0.34283997442106
0.57057057057057 0.339013984172804
0.59059059059059 0.33509635687531
0.61061061061061 0.331091273090187
0.63063063063063 0.327002968759958
0.65065065065065 0.322835727797843
0.67067067067067 0.318593874672039
0.69069069069069 0.314281767002296
0.71071071071071 0.309903788186304
0.73073073073073 0.305464340073112
0.75075075075075 0.300967835700437
0.77077077077077 0.296418692112302
0.790790790790791 0.291821323272996
0.810810810810811 0.287180133092853
0.830830830830831 0.282499508580786
0.850850850850851 0.277783813137949
0.870870870870871 0.273037380006279
0.890890890890891 0.268264505884996
0.910910910910911 0.2634694447275
0.930930930930931 0.258656401730343
0.950950950950951 0.253829527525259
0.970970970970971 0.24899291258444
0.990990990990991 0.24415058184851
1.01101101101101 0.239306489585817
1.03103103103103 0.234464514490888
1.05105105105105 0.229628455029052
1.07107107107107 0.224802025033432
1.09109109109109 0.219988849559688
1.11111111111111 0.215192461003031
1.13113113113113 0.210416295481265
1.15115115115115 0.205663689486728
1.17117117117117 0.200937876809264
1.19119119119119 0.1962419857315
1.21121121121121 0.191579036496956
1.23123123123123 0.186951939050736
1.25125125125125 0.182363491051798
1.27127127127127 0.17781637615506
1.29129129129129 0.173313162560933
1.31131131131131 0.168856301829129
1.33133133133133 0.164448127952996
1.35135135135135 0.160090856689972
1.37137137137137 0.155786585143159
1.39139139139139 0.151537291588457
1.41141141141141 0.147344835541168
1.43143143143143 0.143210958055468
1.45145145145145 0.139137282249685
1.47147147147147 0.135125314049902
1.49149149149149 0.13117644314399
1.51151151151151 0.127291944137829
1.53153153153153 0.123472977905145
1.55155155155155 0.119720593122119
1.57157157157157 0.116035727977651
1.59159159159159 0.112419212049971
1.61161161161161 0.108871768340093
1.63163163163163 0.10539401545248
1.65165165165165 0.101986469913169
1.67167167167167 0.0986495486155338
1.69169169169169 0.0953835713838294
1.71171171171171 0.0921887636446459
1.73173173173173 0.0890652591964251
1.75175175175175 0.0860131030672496
1.77177177177177 0.083032254451193
1.79179179179179 0.0801225897136326
1.81181181181181 0.077283905456062
1.83183183183183 0.0745159216311036
1.85185185185185 0.0718182846986055
1.87187187187187 0.0691905708139176
1.89189189189189 0.0666322890396674
1.91191191191191 0.0641428845726061
1.93193193193193 0.0617217419773594
1.95195195195195 0.059368188419199
1.97197197197197 0.057081496888248
1.99199199199199 0.0548608894078376
2.01201201201201 0.0527055402200588
2.03203203203203 0.0506145789418748
2.05205205205205 0.0485870936855042
2.07207207207207 0.0466221341371236
2.09209209209209 0.0447187145882916
2.11211211211211 0.0428758169148479
2.13213213213213 0.041092393498395
2.15215215215215 0.0393673700858294
2.17217217217217 0.0376996485827434
2.19219219219219 0.0360881097768737
2.21221221221221 0.0345316159881233
2.23223223223223 0.033029013642035
2.25225225225225 0.0315791357639332
2.27227227227227 0.03018080439129
2.29229229229229 0.0288328329022028
2.31231231231231 0.0275340282581906
2.33233233233233 0.0262831931598317
2.35235235235235 0.02507912811407
2.37237237237237 0.0239206334123108
2.39239239239239 0.0228065110187136
2.41241241241241 0.0217355663683581
2.43243243243243 0.0207066100752275
2.45245245245245 0.0197184595501959
2.47247247247247 0.0187699405294511
2.49249249249249 0.0178598885140022
2.51251251251251 0.016987150121141
2.53253253253253 0.0161505843489182
2.55255255255255 0.0153490637548888
2.57257257257257 0.0145814755505493
2.59259259259259 0.0138467226130541
2.61261261261261 0.0131437244159435
2.63263263263263 0.0124714178807514
2.65265265265265 0.0118287581514852
2.67267267267267 0.0112147192940778
2.69269269269269 0.0106282949230103
2.71271271271271 0.0100684987573917
2.73273273273273 0.00953436510885491
2.75275275275275 0.00902494930369034
2.77277277277277 0.00853932804169479
2.79279279279279 0.00807659969425077
2.81281281281281 0.00763588454418634
2.83283283283283 0.00721632496998621
2.85285285285285 0.00681708557693871
2.87287287287287 0.00643735327780635
2.89289289289289 0.00607633732560524
2.91291291291291 0.00573326930106518
2.93293293293293 0.00540740305732384
2.95295295295295 0.00509801462438214
2.97297297297297 0.00480440207581539
2.99299299299299 0.00452588536019617
3.01301301301301 0.00426180609964127
3.03303303303303 0.00401152735784578
3.05305305305305 0.00377443337991421
3.07307307307307 0.00354992930624085
3.09309309309309 0.00333744086263051
3.11311311311311 0.00313641402878608
3.13313313313313 0.0029463146872226
3.15315315315315 0.00276662825459746
3.17317317317317 0.00259685929737496
3.19319319319319 0.00243653113367011
3.21321321321321 0.00228518542304203
3.23323323323323 0.00214238174593163
3.25325325325325 0.00200769717436225
3.27327327327327 0.00188072583544551
3.29329329329329 0.00176107846915771
3.31331331331331 0.00164838198177652
3.33333333333333 0.0015422789962911
3.35335335335335 0.00144242740102448
3.37337337337337 0.00134849989763212
3.39339339339339 0.00126018354956833
3.41341341341341 0.00117717933203981
3.43343343343343 0.00109920168439588
3.45345345345345 0.0010259780658361
3.47347347347347 0.000957248515249212
3.49349349349349 0.000892765215932441
3.51351351351351 0.000832292065877152
3.53353353353353 0.000775604254246008
3.55355355355355 0.000722487844607976
3.57357357357357 0.000672739365441019
3.59359359359359 0.000626165408357979
3.61361361361361 0.000582582234459159
3.63363363363363 0.000541815389165405
3.65365365365365 0.00050369932583812
3.67367367367367 0.000468077038447599
3.69369369369369 0.000434799703508357
3.71371371371371 0.000403726331459719
3.73373373373373 0.000374723427631836
3.75375375375375 0.000347664662901427
3.77377377377377 0.000322430554107926
3.79379379379379 0.000298908154269281
3.81381381381381 0.000276990752607344
3.83383383383383 0.000256577584365551
3.85385385385385 0.000237573550376443
3.87387387387387 0.000219888946313312
3.89389389389389 0.000203439201538965
3.91391391391391 0.00018814462744512
3.93393393393393 0.000173930175158222
3.95395395395395 0.000160725202471485
3.97397397397397 0.000148463249848567
3.99399399399399 0.000137081825331481
4.01401401401401 0.000126522198173995
4.03403403403403 0.000116729201011866
4.05405405405405 0.000107651040372646
4.07407407407407 9.92391153205018e-05
4.09409409409409 9.14478440253317e-05
4.11411411411411 8.42344980404937e-05
4.13413413413413 7.75590440694795e-05
4.15415415415415 7.13839929989176e-05
4.17417417417417 6.56742559732345e-05
4.19419419419419 6.03970072851107e-05
4.21421421421421 5.55215538554582e-05
4.23423423423423 5.10192110769697e-05
4.25425425425425 4.68631847962824e-05
4.27427427427427 4.3028459211397e-05
4.29429429429429 3.94916904631592e-05
4.31431431431431 3.62311057022691e-05
4.33433433433433 3.32264074164067e-05
4.35435435435435 3.04586828055866e-05
4.37437437437437 2.79103179977393e-05
4.39439439439439 2.55649169007242e-05
4.41441441441441 2.34072244914564e-05
4.43443443443443 2.14230543475548e-05
4.45445445445445 1.95992202318354e-05
4.47447447447447 1.79234715450684e-05
4.49449449449449 1.63844324676437e-05
4.51451451451451 1.49715446161227e-05
4.53453453453453 1.3675013046071e-05
4.55455455455455 1.24857554380297e-05
4.57457457457457 1.13953543089884e-05
4.59459459459459 1.03960120972233e-05
4.61461461461461 9.48050897386715e-06
4.63463463463463 8.64216324004121e-06
4.65465465465465 7.87479417380527e-06
4.67467467467467 7.17268719654363e-06
4.69469469469469 6.53056123369604e-06
4.71471471471471 5.94353814994721e-06
4.73473473473473 5.40711414409909e-06
4.75475475475475 4.91713299385613e-06
4.77477477477477 4.4697610456467e-06
4.79479479479479 4.06146384937932e-06
4.81481481481481 3.68898434268124e-06
4.83483483483483 3.34932249368796e-06
4.85485485485485 3.03971631583941e-06
4.87487487487487 2.75762417238901e-06
4.89489489489489 2.50070829244518e-06
4.91491491491491 2.26681942433715e-06
4.93493493493493 2.05398255592983e-06
4.95495495495495 1.86038363520377e-06
4.97497497497497 1.68435722796805e-06
4.99499499499499 1.5243750529858e-06
5.01501501501502 1.37903533806611e-06
5.03503503503504 1.24705294381396e-06
5.05505505505506 1.12725020473309e-06
5.07507507507508 1.01854844024876e-06
5.0950950950951 9.19960090959837e-07
5.11511511511512 8.30581438046261e-07
5.13513513513514 7.49585866251233e-07
5.15515515515516 6.76217633231267e-07
5.17517517517518 6.09786110324583e-07
5.1951951951952 5.4966046193278e-07
5.21521521521522 4.95264732746229e-07
5.23523523523524 4.46073313973501e-07
5.25525525525526 4.01606761563285e-07
5.27527527527528 3.61427941137511e-07
5.2952952952953 3.25138475990267e-07
5.31531531531532 2.92375476052561e-07
5.33533533533534 2.62808527181656e-07
5.35535535535536 2.36136921509202e-07
5.37537537537538 2.1208711087848e-07
5.3953953953954 1.90410366621162e-07
5.41541541541542 1.70880630071684e-07
5.43543543543544 1.5329253929596e-07
5.45545545545546 1.3745961852414e-07
5.47547547547548 1.23212617727566e-07
5.4954954954955 1.10397990671284e-07
5.51551551551552 9.8876500608364e-08
5.53553553553554 8.85219435638491e-08
5.55555555555556 7.92199798873018e-08
5.57557557557558 7.08670654362614e-08
5.5955955955956 6.33694743912418e-08
5.61561561561562 5.66424062986154e-08
5.63563563563564 5.06091704933412e-08
5.65565565565566 4.5200441571292e-08
5.67567567567568 4.03535800631662e-08
5.6956956956957 3.60120129107462e-08
5.71571571571572 3.2124668763613e-08
5.73573573573574 2.86454635022852e-08
5.75575575575576 2.55328317539416e-08
5.77577577577578 2.27493005011625e-08
5.7957957957958 2.02611011941336e-08
5.81581581581582 1.80378170640688e-08
5.83583583583584 1.60520626017053e-08
5.85585585585586 1.42791924110059e-08
5.87587587587588 1.2697036876002e-08
5.8958958958959 1.12856622892642e-08
5.91591591591592 1.00271532849868e-08
5.93593593593594 8.90541559921139e-09
5.95595595595596 7.90599734535251e-09
5.97597597597598 7.01592714588943e-09
5.995995995996 6.22356760178439e-09
6.01601601601602 5.5184827107339e-09
6.03603603603604 4.89131796456914e-09
6.05605605605606 4.33369196574694e-09
6.07607607607608 3.83809850362692e-09
6.0960960960961 3.3978181237675e-09
6.11611611611612 3.00683830841826e-09
6.13613613613614 2.65978146430924e-09
6.15615615615616 2.35183998527871e-09
6.17617617617618 2.07871772273778e-09
6.1961961961962 1.83657725691022e-09
6.21621621621622 1.6219924166386e-09
6.23623623623624 1.43190554571785e-09
6.25625625625626 1.26358905957511e-09
6.27627627627628 1.11461087800764e-09
6.2962962962963 9.82803357938119e-10
6.31631631631632 8.66235385045992e-10
6.33633633633634 7.63187314959515e-10
6.35635635635636 6.72128483698836e-10
6.37637637637638 5.91697033481532e-10
6.3963963963964 5.20681824054164e-10
6.41641641641642 4.58006221596989e-10
6.43643643643644 4.02713577147895e-10
6.45645645645646 3.53954224575805e-10
6.47647647647648 3.10973844559378e-10
6.4964964964965 2.73103055937435e-10
6.51651651651652 2.39748109325539e-10
6.53653653653654 2.10382570159619e-10
6.55655655655656 1.84539889444162e-10
6.57657657657658 1.61806770551276e-10
6.5965965965966 1.41817249531742e-10
6.61661661661662 1.24247414645767e-10
6.63663663663664 1.08810698278133e-10
6.65665665665666 9.52536811418137e-11
6.67667667667668 8.33523547614393e-11
6.6966966966967 7.29087937236022e-11
6.71671671671672 6.37481941394484e-11
6.73673673673674 5.57162392365998e-11
6.75675675675676 4.86767570277076e-11
6.77677677677678 4.25096386334907e-11
6.7967967967968 3.71089891068556e-11
6.81681681681682 3.23814855461048e-11
6.83683683683684 2.82449199306813e-11
6.85685685685686 2.46269064908966e-11
6.87687687687688 2.14637355595386e-11
6.8968968968969 1.86993577716892e-11
6.91691691691692 1.62844842008236e-11
6.93693693693694 1.41757895636779e-11
6.95695695695696 1.23352070109961e-11
6.97697697697698 1.07293042619725e-11
6.996996996997 9.32873195138548e-12
7.01701701701702 8.10773605306642e-12
7.03703703703704 7.04372713322415e-12
7.05705705705706 6.11689998287643e-12
7.07707707707708 5.30989788981512e-12
7.0970970970971 4.6075164458095e-12
7.11711711711712 3.99644235193917e-12
7.13713713713714 3.46502319108337e-12
7.15715715715716 3.00306458801651e-12
7.17717717717718 2.60165157997869e-12
7.1971971971972 2.25299137914217e-12
7.21721721721722 1.95027502769791e-12
7.23723723723724 1.68755573049414e-12
7.25725725725726 1.45964190299846e-12
7.27727727727728 1.262003197176e-12
7.2972972972973 1.0906879676822e-12
7.31731731731732 9.42250818252902e-13
7.33733733733734 8.13689025751829e-13
7.35735735735736 7.02386779168733e-13
7.37737737737738 6.06066294885245e-13
7.3973973973974 5.22744979473708e-13
7.41741741741742 4.50697908714381e-13
7.43743743743744 3.88424977793729e-13
7.45745745745746 3.34622154016963e-13
7.47747747747748 2.88156330935933e-13
7.4974974974975 2.48043342543511e-13
7.51751751751752 2.13428748996348e-13
7.53753753753754 1.83571051982055e-13
7.55755755755756 1.57827039041883e-13
7.57757757757758 1.35638992516663e-13
7.5975975975976 1.16523530854698e-13
7.61761761761762 1.000618782966e-13
7.63763763763764 8.58913838706873e-14
7.65765765765766 7.36981325813112e-14
7.67767767767768 6.32105109959248e-14
7.6976976976977 5.41936064405376e-14
7.71771771771772 4.64443339685915e-14
7.73773773773774 3.97871984155707e-14
7.75775775775776 3.40706104038536e-14
7.77777777777778 2.91636853079907e-14
7.7977977977978 2.4953463096687e-14
7.81781781781782 2.13424947819474e-14
7.83783783783784 1.82467480586654e-14
7.85785785785786 1.55937907247682e-14
7.87787787787788 1.33212157347804e-14
7.8978978978979 1.13752763482777e-14
7.91791791791792 9.70970386854701e-15
7.93793793793794 8.28468399583068e-15
7.95795795795796 7.06597090549298e-15
7.97797797797798 6.02412085866189e-15
7.997997997998 5.13382950919603e-15
8.01801801801802 4.3733591283238e-15
8.03803803803804 3.72404376402735e-15
8.05805805805806 3.16986191875719e-15
8.07807807807808 2.69706769497134e-15
8.0980980980981 2.29387254841619e-15
8.11811811811812 1.95017082606148e-15
8.13813813813814 1.65730316851105e-15
8.15815815815816 1.40785264250169e-15
8.17817817817818 1.19546915264237e-15
8.1981981981982 1.01471827585831e-15
8.21821821821822 8.60951178494166e-16
8.23823823823824 7.30192724674871e-16
8.25825825825826 6.19045274051723e-16
8.27827827827828 5.24606005103494e-16
8.2982982982983 4.44395893386326e-16
8.31831831831832 3.76298728354137e-16
8.33833833833834 3.18508772685576e-16
8.35835835835836 2.69485858889318e-16
8.37837837837838 2.27916883183252e-16
8.3983983983984 1.92682799625235e-16
8.41841841841842 1.62830341150177e-16
8.43843843843844 1.37547801096493e-16
8.45845845845846 1.16144301209627e-16
8.47847847847848 9.8032051926925e-17
8.4984984984985 8.27111796592674e-17
8.51851851851852 6.97567552529606e-17
8.53853853853854 5.88077091106193e-17
8.55855855855856 4.95573626747691e-17
8.57857857857858 4.17453440895294e-17
8.5985985985986 3.51506886838449e-17
8.61861861861862 2.95859531836935e-17
8.63863863863864 2.48921968838275e-17
8.65865865865866 2.09347039316773e-17
8.67867867867868 1.75993388643364e-17
8.6986986986987 1.47894429982973e-17
8.71871871871872 1.24231925503958e-17
8.73873873873874 1.04313507694241e-17
8.75875875875876 8.75535614211525e-18
8.77877877877878 7.34569713009337e-18
8.7987987987988 6.16053109048305e-18
8.81881881881882 5.16451119999019e-18
8.83883883883884 4.32779048512422e-18
8.85885885885886 3.62517658454128e-18
8.87887887887888 3.03541474067764e-18
8.8988988988989 2.54057982941549e-18
8.91891891891892 2.12556106807901e-18
8.93893893893894 1.77762546207239e-18
8.95895895895896 1.48604811781133e-18
8.97897897897898 1.24179931485728e-18
8.998998998999 1.03727973679138e-18
9.01901901901902 8.66096545670873e-19
9.03903903903904 7.22874080905507e-19
9.05905905905906 6.03093897535385e-19
9.07907907907908 5.0295965472299e-19
9.0990990990991 4.19283042962206e-19
9.11911911911912 3.49387515332417e-19
9.13913913913914 2.91027078875695e-19
9.15915915915916 2.42317819504618e-19
9.17917917917918 2.01680188581445e-19
9.1991991991992 1.67790380698338e-19
9.21921921921922 1.39539388139188e-19
9.23923923923924 1.1599853476858e-19
9.25925925925926 9.63904764362469e-20
9.27927927927928 8.00648113242436e-20
9.2992992992993 6.6477576193283e-20
9.31931931931932 5.51740167796855e-20
9.33933933933934 4.5774115701642e-20
9.35935935935936 3.79604417474457e-20
9.37937937937938 3.14679525475421e-20
9.3993993993994 2.60754402558043e-20
9.41941941941942 2.15983585814365e-20
9.43943943943944 1.78828106797973e-20
9.45945945945946 1.48005121824234e-20
9.47947947947948 1.22445730038445e-20
9.4994994994995 1.01259663374544e-20
9.51951951951952 8.37057415073758e-21
9.53953953953954 6.91671611023779e-21
9.55955955955956 5.71308371631411e-21
9.57957957957958 4.71701393695898e-21
9.5995995995996 3.89304716291232e-21
9.61961961961962 3.21172317125736e-21
9.63963963963964 2.64857624243904e-21
9.65965965965966 2.18329684678274e-21
9.67967967967968 1.79903258756111e-21
9.6996996996997 1.48180551599987e-21
9.71971971971972 1.22002665237565e-21
9.73973973973974 1.00409166885045e-21
9.75975975975976 8.26044308669654e-22
9.77977977977978 6.79296312742742e-22
9.7997997997998 5.58394465795474e-22
9.81981981981982 4.58826916995414e-22
9.83983983983984 3.7686222201397e-22
9.85985985985986 3.09415635142992e-22
9.87987987987988 2.53938085193762e-22
9.8998998998999 2.08324025950642e-22
9.91991991991992 1.70834984871876e-22
9.93993993993994 1.40036162642795e-22
9.95995995995996 1.14743877987917e-22
9.97997997997998 9.39820210218911e-23
10 7.69459862670642e-23
};
\addlegendentry{N(0,1)}; 
\legend{}; 
\end{axis}

\end{tikzpicture}

%% file: FinalFigs/Null_Dists_d_10_100_n_100_m_100_kernel__Dirichlet_Poly_5_2022_10_15_22_14_39mmd.tex
\begin{tikzpicture}

\definecolor{darkorange25512714}{RGB}{255,127,14}
\definecolor{darkslategray38}{RGB}{38,38,38}
\definecolor{lightgray204}{RGB}{204,204,204}
\definecolor{steelblue31119180}{RGB}{31,119,180}

\begin{axis}[
axis line style={darkslategray38},
height=\figheight,
legend cell align={left},
legend style={fill opacity=0.8, draw opacity=1, text opacity=1, draw=none},
tick align=outside,
tick pos=left,
title={$\dmmd$~$(n/m=1)$},
width=\figwidth,
x grid style={lightgray204},
xmin=-6, xmax=6,
xtick style={color=darkslategray38},
y grid style={lightgray204},
ylabel=\textcolor{darkslategray38}{},
ymin=0, ymax=0.806078259905132,
ytick style={color=darkslategray38}, 
xticklabels=empty,
yticklabels=empty
]
\draw[draw=none,fill=steelblue31119180,fill opacity=0.8] (axis cs:-1.61746501922607,0) rectangle (axis cs:-1.4840784072876,0.0539784500802375);
\addlegendimage{ybar,ybar legend,draw=none,fill=steelblue31119180,fill opacity=0.8}
\addlegendentry{mmd (d=10)}

\draw[draw=none,fill=steelblue31119180,fill opacity=0.8] (axis cs:-1.28399848937988,0) rectangle (axis cs:-1.1506119966507,0.149940165912462);
\draw[draw=none,fill=steelblue31119180,fill opacity=0.8] (axis cs:-0.950532257556915,0) rectangle (axis cs:-0.817145764827728,0.329868365007417);
\draw[draw=none,fill=steelblue31119180,fill opacity=0.8] (axis cs:-0.617065966129303,0) rectangle (axis cs:-0.483679354190826,0.593763003947977);
\draw[draw=none,fill=steelblue31119180,fill opacity=0.8] (axis cs:-0.283599555492401,0) rectangle (axis cs:-0.150212943553925,0.767693580862031);
\draw[draw=none,fill=steelblue31119180,fill opacity=0.8] (axis cs:0.0498668104410172,0) rectangle (axis cs:0.183253422379494,0.6237510344504);
\draw[draw=none,fill=steelblue31119180,fill opacity=0.8] (axis cs:0.383333206176758,0) rectangle (axis cs:0.516719818115234,0.263894668421323);
\draw[draw=none,fill=steelblue31119180,fill opacity=0.8] (axis cs:0.716799557209015,0) rectangle (axis cs:0.850186169147491,0.1439425335473);
\draw[draw=none,fill=steelblue31119180,fill opacity=0.8] (axis cs:1.05026602745056,0) rectangle (axis cs:1.18365252017975,0.0539784693767387);
\draw[draw=none,fill=steelblue31119180,fill opacity=0.8] (axis cs:1.38373231887817,0) rectangle (axis cs:1.51711893081665,0.0179928166934125);
\draw[draw=none,fill=darkorange25512714,fill opacity=0.8] (axis cs:-1.48407852649689,0) rectangle (axis cs:-1.35069191455841,0.0179928166934125);
\addlegendimage{ybar,ybar legend,draw=none,fill=darkorange25512714,fill opacity=0.8}
\addlegendentry{mmd (d=100)}

\draw[draw=none,fill=darkorange25512714,fill opacity=0.8] (axis cs:-1.15061211585999,0) rectangle (axis cs:-1.0172256231308,0.113954526093471);
\draw[draw=none,fill=darkorange25512714,fill opacity=0.8] (axis cs:-0.817145705223083,0) rectangle (axis cs:-0.683759212493896,0.263894692005934);
\draw[draw=none,fill=darkorange25512714,fill opacity=0.8] (axis cs:-0.483679354190826,0) rectangle (axis cs:-0.350292861461639,0.515794124641677);
\draw[draw=none,fill=darkorange25512714,fill opacity=0.8] (axis cs:-0.150212958455086,0) rectangle (axis cs:-0.0168264657258987,0.641743852751854);
\draw[draw=none,fill=darkorange25512714,fill opacity=0.8] (axis cs:0.183253422379494,0) rectangle (axis cs:0.31663990020752,0.671731883254277);
\draw[draw=none,fill=darkorange25512714,fill opacity=0.8] (axis cs:0.516719818115234,0) rectangle (axis cs:0.650106310844421,0.449820457536346);
\draw[draw=none,fill=darkorange25512714,fill opacity=0.8] (axis cs:0.850186109542847,0) rectangle (axis cs:0.983572721481323,0.251899433707775);
\draw[draw=none,fill=darkorange25512714,fill opacity=0.8] (axis cs:1.18365263938904,0) rectangle (axis cs:1.31703913211823,0.0539784693767387);
\draw[draw=none,fill=darkorange25512714,fill opacity=0.8] (axis cs:1.51711881160736,0) rectangle (axis cs:1.65050542354584,0.0179928166934125);
\addplot [semithick, black]
table {%
-10 7.69459862670642e-23
-9.97997997997998 9.39820210218911e-23
-9.95995995995996 1.14743877987917e-22
-9.93993993993994 1.40036162642795e-22
-9.91991991991992 1.70834984871876e-22
-9.8998998998999 2.08324025950642e-22
-9.87987987987988 2.53938085193762e-22
-9.85985985985986 3.09415635142992e-22
-9.83983983983984 3.7686222201397e-22
-9.81981981981982 4.58826916995414e-22
-9.7997997997998 5.58394465795474e-22
-9.77977977977978 6.79296312742742e-22
-9.75975975975976 8.26044308669654e-22
-9.73973973973974 1.00409166885045e-21
-9.71971971971972 1.22002665237565e-21
-9.6996996996997 1.48180551599987e-21
-9.67967967967968 1.79903258756111e-21
-9.65965965965966 2.18329684678274e-21
-9.63963963963964 2.64857624243904e-21
-9.61961961961962 3.21172317125736e-21
-9.5995995995996 3.89304716291232e-21
-9.57957957957958 4.71701393695898e-21
-9.55955955955956 5.71308371631411e-21
-9.53953953953954 6.91671611023779e-21
-9.51951951951952 8.37057415073758e-21
-9.4994994994995 1.01259663374544e-20
-9.47947947947948 1.22445730038445e-20
-9.45945945945946 1.48005121824234e-20
-9.43943943943944 1.78828106797973e-20
-9.41941941941942 2.15983585814365e-20
-9.3993993993994 2.60754402558043e-20
-9.37937937937938 3.14679525475421e-20
-9.35935935935936 3.79604417474457e-20
-9.33933933933934 4.5774115701642e-20
-9.31931931931932 5.51740167796855e-20
-9.2992992992993 6.6477576193283e-20
-9.27927927927928 8.00648113242436e-20
-9.25925925925926 9.63904764362469e-20
-9.23923923923924 1.1599853476858e-19
-9.21921921921922 1.39539388139188e-19
-9.1991991991992 1.67790380698338e-19
-9.17917917917918 2.01680188581445e-19
-9.15915915915916 2.42317819504618e-19
-9.13913913913914 2.91027078875695e-19
-9.11911911911912 3.49387515332417e-19
-9.0990990990991 4.19283042962206e-19
-9.07907907907908 5.0295965472299e-19
-9.05905905905906 6.03093897535385e-19
-9.03903903903904 7.22874080905507e-19
-9.01901901901902 8.66096545670873e-19
-8.998998998999 1.03727973679138e-18
-8.97897897897898 1.24179931485728e-18
-8.95895895895896 1.48604811781133e-18
-8.93893893893894 1.77762546207239e-18
-8.91891891891892 2.12556106807901e-18
-8.8988988988989 2.54057982941549e-18
-8.87887887887888 3.03541474067764e-18
-8.85885885885886 3.62517658454128e-18
-8.83883883883884 4.32779048512422e-18
-8.81881881881882 5.16451119999019e-18
-8.7987987987988 6.16053109048305e-18
-8.77877877877878 7.34569713009337e-18
-8.75875875875876 8.75535614211525e-18
-8.73873873873874 1.04313507694241e-17
-8.71871871871872 1.24231925503958e-17
-8.6986986986987 1.47894429982973e-17
-8.67867867867868 1.75993388643364e-17
-8.65865865865866 2.09347039316773e-17
-8.63863863863864 2.48921968838275e-17
-8.61861861861862 2.95859531836935e-17
-8.5985985985986 3.51506886838449e-17
-8.57857857857858 4.17453440895294e-17
-8.55855855855856 4.95573626747691e-17
-8.53853853853854 5.88077091106193e-17
-8.51851851851852 6.97567552529606e-17
-8.4984984984985 8.27111796592674e-17
-8.47847847847848 9.8032051926925e-17
-8.45845845845846 1.16144301209627e-16
-8.43843843843844 1.37547801096493e-16
-8.41841841841842 1.62830341150177e-16
-8.3983983983984 1.92682799625235e-16
-8.37837837837838 2.27916883183252e-16
-8.35835835835836 2.69485858889318e-16
-8.33833833833834 3.18508772685576e-16
-8.31831831831832 3.76298728354137e-16
-8.2982982982983 4.44395893386326e-16
-8.27827827827828 5.24606005103494e-16
-8.25825825825826 6.19045274051723e-16
-8.23823823823824 7.30192724674871e-16
-8.21821821821822 8.60951178494166e-16
-8.1981981981982 1.01471827585831e-15
-8.17817817817818 1.19546915264237e-15
-8.15815815815816 1.40785264250169e-15
-8.13813813813814 1.65730316851105e-15
-8.11811811811812 1.95017082606148e-15
-8.0980980980981 2.29387254841619e-15
-8.07807807807808 2.69706769497134e-15
-8.05805805805806 3.16986191875719e-15
-8.03803803803804 3.72404376402735e-15
-8.01801801801802 4.3733591283238e-15
-7.997997997998 5.13382950919607e-15
-7.97797797797798 6.02412085866193e-15
-7.95795795795796 7.06597090549303e-15
-7.93793793793794 8.28468399583074e-15
-7.91791791791792 9.70970386854708e-15
-7.8978978978979 1.13752763482777e-14
-7.87787787787788 1.33212157347805e-14
-7.85785785785786 1.55937907247683e-14
-7.83783783783784 1.82467480586655e-14
-7.81781781781782 2.13424947819475e-14
-7.7977977977978 2.49534630966872e-14
-7.77777777777778 2.91636853079909e-14
-7.75775775775776 3.40706104038538e-14
-7.73773773773774 3.9787198415571e-14
-7.71771771771772 4.64443339685918e-14
-7.6976976976977 5.4193606440538e-14
-7.67767767767768 6.32105109959252e-14
-7.65765765765766 7.36981325813117e-14
-7.63763763763764 8.58913838706879e-14
-7.61761761761762 1.00061878296601e-13
-7.5975975975976 1.16523530854699e-13
-7.57757757757758 1.35638992516664e-13
-7.55755755755756 1.57827039041884e-13
-7.53753753753754 1.83571051982057e-13
-7.51751751751752 2.13428748996349e-13
-7.4974974974975 2.48043342543513e-13
-7.47747747747748 2.88156330935935e-13
-7.45745745745746 3.34622154016965e-13
-7.43743743743744 3.88424977793732e-13
-7.41741741741742 4.50697908714384e-13
-7.3973973973974 5.22744979473711e-13
-7.37737737737738 6.06066294885249e-13
-7.35735735735736 7.02386779168738e-13
-7.33733733733734 8.13689025751835e-13
-7.31731731731732 9.42250818252909e-13
-7.2972972972973 1.09068796768221e-12
-7.27727727727728 1.262003197176e-12
-7.25725725725726 1.45964190299847e-12
-7.23723723723724 1.68755573049416e-12
-7.21721721721722 1.95027502769792e-12
-7.1971971971972 2.25299137914218e-12
-7.17717717717718 2.60165157997871e-12
-7.15715715715716 3.00306458801653e-12
-7.13713713713714 3.4650231910834e-12
-7.11711711711712 3.99644235193919e-12
-7.0970970970971 4.60751644580953e-12
-7.07707707707708 5.30989788981514e-12
-7.05705705705706 6.11689998287646e-12
-7.03703703703704 7.0437271332242e-12
-7.01701701701702 8.10773605306645e-12
-6.996996996997 9.32873195138555e-12
-6.97697697697698 1.07293042619726e-11
-6.95695695695696 1.23352070109962e-11
-6.93693693693694 1.41757895636779e-11
-6.91691691691692 1.62844842008237e-11
-6.8968968968969 1.86993577716893e-11
-6.87687687687688 2.14637355595386e-11
-6.85685685685686 2.46269064908967e-11
-6.83683683683684 2.82449199306815e-11
-6.81681681681682 3.2381485546105e-11
-6.7967967967968 3.71089891068559e-11
-6.77677677677678 4.25096386334913e-11
-6.75675675675676 4.86767570277083e-11
-6.73673673673674 5.57162392366004e-11
-6.71671671671672 6.37481941394491e-11
-6.6966966966967 7.29087937236032e-11
-6.67667667667668 8.33523547614402e-11
-6.65665665665666 9.52536811418151e-11
-6.63663663663664 1.08810698278135e-10
-6.61661661661662 1.24247414645768e-10
-6.5965965965966 1.41817249531744e-10
-6.57657657657658 1.61806770551278e-10
-6.55655655655656 1.84539889444164e-10
-6.53653653653654 2.10382570159622e-10
-6.51651651651652 2.39748109325542e-10
-6.4964964964965 2.73103055937438e-10
-6.47647647647648 3.10973844559381e-10
-6.45645645645646 3.53954224575809e-10
-6.43643643643644 4.027135771479e-10
-6.41641641641642 4.58006221596996e-10
-6.3963963963964 5.20681824054169e-10
-6.37637637637638 5.91697033481538e-10
-6.35635635635636 6.72128483698846e-10
-6.33633633633634 7.63187314959523e-10
-6.31631631631632 8.66235385046001e-10
-6.2962962962963 9.8280335793813e-10
-6.27627627627628 1.11461087800766e-09
-6.25625625625626 1.26358905957513e-09
-6.23623623623624 1.43190554571787e-09
-6.21621621621622 1.62199241663862e-09
-6.1961961961962 1.83657725691024e-09
-6.17617617617618 2.0787177227378e-09
-6.15615615615616 2.35183998527873e-09
-6.13613613613614 2.65978146430928e-09
-6.11611611611612 3.00683830841829e-09
-6.0960960960961 3.39781812376754e-09
-6.07607607607608 3.83809850362696e-09
-6.05605605605606 4.33369196574699e-09
-6.03603603603604 4.89131796456919e-09
-6.01601601601602 5.51848271073395e-09
-5.995995995996 6.22356760178439e-09
-5.97597597597598 7.01592714588943e-09
-5.95595595595596 7.90599734535251e-09
-5.93593593593594 8.90541559921139e-09
-5.91591591591592 1.00271532849868e-08
-5.8958958958959 1.12856622892642e-08
-5.87587587587588 1.2697036876002e-08
-5.85585585585586 1.42791924110059e-08
-5.83583583583584 1.60520626017053e-08
-5.81581581581582 1.80378170640688e-08
-5.7957957957958 2.02611011941336e-08
-5.77577577577578 2.27493005011625e-08
-5.75575575575576 2.55328317539416e-08
-5.73573573573574 2.86454635022852e-08
-5.71571571571572 3.2124668763613e-08
-5.6956956956957 3.60120129107462e-08
-5.67567567567568 4.03535800631662e-08
-5.65565565565566 4.5200441571292e-08
-5.63563563563564 5.06091704933412e-08
-5.61561561561562 5.66424062986154e-08
-5.5955955955956 6.33694743912418e-08
-5.57557557557558 7.08670654362614e-08
-5.55555555555556 7.92199798873018e-08
-5.53553553553554 8.85219435638491e-08
-5.51551551551552 9.8876500608364e-08
-5.4954954954955 1.10397990671284e-07
-5.47547547547548 1.23212617727566e-07
-5.45545545545546 1.3745961852414e-07
-5.43543543543544 1.5329253929596e-07
-5.41541541541542 1.70880630071684e-07
-5.3953953953954 1.90410366621162e-07
-5.37537537537538 2.1208711087848e-07
-5.35535535535536 2.36136921509202e-07
-5.33533533533534 2.62808527181656e-07
-5.31531531531532 2.92375476052561e-07
-5.2952952952953 3.25138475990267e-07
-5.27527527527528 3.61427941137511e-07
-5.25525525525526 4.01606761563285e-07
-5.23523523523524 4.46073313973501e-07
-5.21521521521522 4.95264732746229e-07
-5.1951951951952 5.4966046193278e-07
-5.17517517517518 6.09786110324583e-07
-5.15515515515516 6.76217633231267e-07
-5.13513513513514 7.49585866251233e-07
-5.11511511511512 8.30581438046261e-07
-5.0950950950951 9.19960090959837e-07
-5.07507507507508 1.01854844024876e-06
-5.05505505505506 1.12725020473309e-06
-5.03503503503504 1.24705294381396e-06
-5.01501501501502 1.37903533806611e-06
-4.99499499499499 1.5243750529858e-06
-4.97497497497497 1.68435722796805e-06
-4.95495495495495 1.86038363520377e-06
-4.93493493493493 2.05398255592983e-06
-4.91491491491491 2.26681942433715e-06
-4.89489489489489 2.50070829244518e-06
-4.87487487487487 2.75762417238901e-06
-4.85485485485485 3.03971631583941e-06
-4.83483483483483 3.34932249368796e-06
-4.81481481481481 3.68898434268124e-06
-4.79479479479479 4.06146384937932e-06
-4.77477477477477 4.4697610456467e-06
-4.75475475475475 4.91713299385613e-06
-4.73473473473473 5.40711414409909e-06
-4.71471471471471 5.94353814994721e-06
-4.69469469469469 6.53056123369604e-06
-4.67467467467467 7.17268719654363e-06
-4.65465465465465 7.87479417380527e-06
-4.63463463463463 8.64216324004121e-06
-4.61461461461461 9.48050897386715e-06
-4.59459459459459 1.03960120972233e-05
-4.57457457457457 1.13953543089884e-05
-4.55455455455455 1.24857554380297e-05
-4.53453453453453 1.3675013046071e-05
-4.51451451451451 1.49715446161227e-05
-4.49449449449449 1.63844324676437e-05
-4.47447447447447 1.79234715450684e-05
-4.45445445445445 1.95992202318354e-05
-4.43443443443443 2.14230543475548e-05
-4.41441441441441 2.34072244914564e-05
-4.39439439439439 2.55649169007242e-05
-4.37437437437437 2.79103179977393e-05
-4.35435435435435 3.04586828055866e-05
-4.33433433433433 3.32264074164067e-05
-4.31431431431431 3.62311057022691e-05
-4.29429429429429 3.94916904631592e-05
-4.27427427427427 4.3028459211397e-05
-4.25425425425425 4.68631847962824e-05
-4.23423423423423 5.10192110769697e-05
-4.21421421421421 5.55215538554582e-05
-4.19419419419419 6.03970072851107e-05
-4.17417417417417 6.56742559732345e-05
-4.15415415415415 7.13839929989176e-05
-4.13413413413413 7.75590440694795e-05
-4.11411411411411 8.42344980404937e-05
-4.09409409409409 9.14478440253317e-05
-4.07407407407407 9.92391153205018e-05
-4.05405405405405 0.000107651040372646
-4.03403403403403 0.000116729201011866
-4.01401401401401 0.000126522198173995
-3.99399399399399 0.000137081825331481
-3.97397397397397 0.000148463249848567
-3.95395395395395 0.000160725202471485
-3.93393393393393 0.000173930175158222
-3.91391391391391 0.00018814462744512
-3.89389389389389 0.000203439201538965
-3.87387387387387 0.000219888946313312
-3.85385385385385 0.000237573550376443
-3.83383383383383 0.000256577584365551
-3.81381381381381 0.000276990752607344
-3.79379379379379 0.000298908154269281
-3.77377377377377 0.000322430554107926
-3.75375375375375 0.000347664662901427
-3.73373373373373 0.000374723427631836
-3.71371371371371 0.000403726331459719
-3.69369369369369 0.000434799703508357
-3.67367367367367 0.000468077038447599
-3.65365365365365 0.00050369932583812
-3.63363363363363 0.000541815389165405
-3.61361361361361 0.000582582234459159
-3.59359359359359 0.000626165408357979
-3.57357357357357 0.000672739365441021
-3.55355355355355 0.000722487844607978
-3.53353353353353 0.00077560425424601
-3.51351351351351 0.000832292065877155
-3.49349349349349 0.000892765215932443
-3.47347347347347 0.000957248515249216
-3.45345345345345 0.0010259780658361
-3.43343343343343 0.00109920168439588
-3.41341341341341 0.00117717933203981
-3.39339339339339 0.00126018354956833
-3.37337337337337 0.00134849989763212
-3.35335335335335 0.00144242740102448
-3.33333333333333 0.00154227899629111
-3.31331331331331 0.00164838198177652
-3.29329329329329 0.00176107846915772
-3.27327327327327 0.00188072583544552
-3.25325325325325 0.00200769717436226
-3.23323323323323 0.00214238174593163
-3.21321321321321 0.00228518542304204
-3.19319319319319 0.00243653113367012
-3.17317317317317 0.00259685929737497
-3.15315315315315 0.00276662825459747
-3.13313313313313 0.00294631468722261
-3.11311311311311 0.00313641402878609
-3.09309309309309 0.00333744086263052
-3.07307307307307 0.00354992930624086
-3.05305305305305 0.00377443337991422
-3.03303303303303 0.00401152735784579
-3.01301301301301 0.00426180609964128
-2.99299299299299 0.00452588536019618
-2.97297297297297 0.0048044020758154
-2.95295295295295 0.00509801462438215
-2.93293293293293 0.00540740305732385
-2.91291291291291 0.00573326930106519
-2.89289289289289 0.00607633732560526
-2.87287287287287 0.00643735327780636
-2.85285285285285 0.00681708557693873
-2.83283283283283 0.00721632496998623
-2.81281281281281 0.00763588454418632
-2.79279279279279 0.00807659969425075
-2.77277277277277 0.00853932804169477
-2.75275275275275 0.00902494930369032
-2.73273273273273 0.00953436510885489
-2.71271271271271 0.0100684987573917
-2.69269269269269 0.0106282949230102
-2.67267267267267 0.0112147192940778
-2.65265265265265 0.0118287581514852
-2.63263263263263 0.0124714178807513
-2.61261261261261 0.0131437244159435
-2.59259259259259 0.0138467226130541
-2.57257257257257 0.0145814755505492
-2.55255255255255 0.0153490637548887
-2.53253253253253 0.0161505843489182
-2.51251251251251 0.0169871501211409
-2.49249249249249 0.0178598885140022
-2.47247247247247 0.018769940529451
-2.45245245245245 0.0197184595501959
-2.43243243243243 0.0207066100752274
-2.41241241241241 0.0217355663683581
-2.39239239239239 0.0228065110187135
-2.37237237237237 0.0239206334123108
-2.35235235235235 0.0250791281140699
-2.33233233233233 0.0262831931598317
-2.31231231231231 0.0275340282581906
-2.29229229229229 0.0288328329022027
-2.27227227227227 0.0301808043912899
-2.25225225225225 0.0315791357639331
-2.23223223223223 0.033029013642035
-2.21221221221221 0.0345316159881232
-2.19219219219219 0.0360881097768736
-2.17217217217217 0.0376996485827434
-2.15215215215215 0.0393673700858293
-2.13213213213213 0.0410923934983949
-2.11211211211211 0.0428758169148479
-2.09209209209209 0.0447187145882915
-2.07207207207207 0.0466221341371235
-2.05205205205205 0.0485870936855041
-2.03203203203203 0.0506145789418747
-2.01201201201201 0.0527055402200587
-1.99199199199199 0.0548608894078376
-1.97197197197197 0.057081496888248
-1.95195195195195 0.059368188419199
-1.93193193193193 0.0617217419773594
-1.91191191191191 0.0641428845726061
-1.89189189189189 0.0666322890396674
-1.87187187187187 0.0691905708139176
-1.85185185185185 0.0718182846986055
-1.83183183183183 0.0745159216311036
-1.81181181181181 0.077283905456062
-1.79179179179179 0.0801225897136326
-1.77177177177177 0.083032254451193
-1.75175175175175 0.0860131030672496
-1.73173173173173 0.0890652591964251
-1.71171171171171 0.0921887636446459
-1.69169169169169 0.0953835713838294
-1.67167167167167 0.0986495486155338
-1.65165165165165 0.101986469913169
-1.63163163163163 0.10539401545248
-1.61161161161161 0.108871768340093
-1.59159159159159 0.112419212049971
-1.57157157157157 0.116035727977651
-1.55155155155155 0.119720593122119
-1.53153153153153 0.123472977905145
-1.51151151151151 0.127291944137829
-1.49149149149149 0.13117644314399
-1.47147147147147 0.135125314049902
-1.45145145145145 0.139137282249685
-1.43143143143143 0.143210958055468
-1.41141141141141 0.147344835541168
-1.39139139139139 0.151537291588457
-1.37137137137137 0.155786585143159
-1.35135135135135 0.160090856689972
-1.33133133133133 0.164448127952996
-1.31131131131131 0.168856301829129
-1.29129129129129 0.173313162560933
-1.27127127127127 0.17781637615506
-1.25125125125125 0.182363491051798
-1.23123123123123 0.186951939050736
-1.21121121121121 0.191579036496956
-1.19119119119119 0.1962419857315
-1.17117117117117 0.200937876809264
-1.15115115115115 0.205663689486728
-1.13113113113113 0.210416295481265
-1.11111111111111 0.215192461003031
-1.09109109109109 0.219988849559688
-1.07107107107107 0.224802025033432
-1.05105105105105 0.229628455029052
-1.03103103103103 0.234464514490888
-1.01101101101101 0.239306489585817
-0.990990990990991 0.24415058184851
-0.970970970970971 0.24899291258444
-0.950950950950951 0.253829527525259
-0.930930930930931 0.258656401730343
-0.910910910910911 0.2634694447275
-0.890890890890891 0.268264505884996
-0.870870870870871 0.273037380006279
-0.850850850850851 0.277783813137949
-0.830830830830831 0.282499508580786
-0.810810810810811 0.287180133092853
-0.790790790790791 0.291821323272996
-0.77077077077077 0.296418692112302
-0.75075075075075 0.300967835700437
-0.73073073073073 0.305464340073112
-0.71071071071071 0.309903788186304
-0.69069069069069 0.314281767002296
-0.67067067067067 0.318593874672039
-0.65065065065065 0.322835727797843
-0.63063063063063 0.327002968759958
-0.61061061061061 0.331091273090187
-0.59059059059059 0.33509635687531
-0.57057057057057 0.339013984172804
-0.55055055055055 0.34283997442106
-0.53053053053053 0.346570209826128
-0.51051051051051 0.350200642706842
-0.49049049049049 0.353727302780113
-0.47047047047047 0.357146304368113
-0.45045045045045 0.360453853509139
-0.43043043043043 0.363646254953996
-0.41041041041041 0.366719919029892
-0.39039039039039 0.369671368354051
-0.37037037037037 0.372497244379499
-0.35035035035035 0.375194313755802
-0.33033033033033 0.377759474487924
-0.31031031031031 0.38018976187679
-0.29029029029029 0.382482354225654
-0.27027027027027 0.384634578296894
-0.25025025025025 0.386643914504485
-0.23023023023023 0.388508001828027
-0.21021021021021 0.390224642434919
-0.19019019019019 0.391791805998011
-0.17017017017017 0.393207633696876
-0.15015015015015 0.394470441891644
-0.13013013013013 0.395578725459258
-0.11011011011011 0.396531160782876
-0.0900900900900901 0.397326608386124
-0.07007007007007 0.397964115204853
-0.05005005005005 0.398442916490068
-0.03003003003003 0.398762437336696
-0.01001001001001 0.398922293833933
0.01001001001001 0.398922293833933
0.03003003003003 0.398762437336696
0.05005005005005 0.398442916490068
0.07007007007007 0.397964115204853
0.0900900900900901 0.397326608386124
0.11011011011011 0.396531160782876
0.13013013013013 0.395578725459258
0.15015015015015 0.394470441891644
0.17017017017017 0.393207633696876
0.19019019019019 0.391791805998011
0.21021021021021 0.390224642434919
0.23023023023023 0.388508001828027
0.25025025025025 0.386643914504485
0.27027027027027 0.384634578296894
0.29029029029029 0.382482354225654
0.31031031031031 0.38018976187679
0.33033033033033 0.377759474487924
0.35035035035035 0.375194313755802
0.37037037037037 0.372497244379499
0.39039039039039 0.369671368354051
0.41041041041041 0.366719919029892
0.43043043043043 0.363646254953996
0.45045045045045 0.360453853509139
0.47047047047047 0.357146304368113
0.49049049049049 0.353727302780113
0.51051051051051 0.350200642706842
0.53053053053053 0.346570209826128
0.55055055055055 0.34283997442106
0.57057057057057 0.339013984172804
0.59059059059059 0.33509635687531
0.61061061061061 0.331091273090187
0.63063063063063 0.327002968759958
0.65065065065065 0.322835727797843
0.67067067067067 0.318593874672039
0.69069069069069 0.314281767002296
0.71071071071071 0.309903788186304
0.73073073073073 0.305464340073112
0.75075075075075 0.300967835700437
0.77077077077077 0.296418692112302
0.790790790790791 0.291821323272996
0.810810810810811 0.287180133092853
0.830830830830831 0.282499508580786
0.850850850850851 0.277783813137949
0.870870870870871 0.273037380006279
0.890890890890891 0.268264505884996
0.910910910910911 0.2634694447275
0.930930930930931 0.258656401730343
0.950950950950951 0.253829527525259
0.970970970970971 0.24899291258444
0.990990990990991 0.24415058184851
1.01101101101101 0.239306489585817
1.03103103103103 0.234464514490888
1.05105105105105 0.229628455029052
1.07107107107107 0.224802025033432
1.09109109109109 0.219988849559688
1.11111111111111 0.215192461003031
1.13113113113113 0.210416295481265
1.15115115115115 0.205663689486728
1.17117117117117 0.200937876809264
1.19119119119119 0.1962419857315
1.21121121121121 0.191579036496956
1.23123123123123 0.186951939050736
1.25125125125125 0.182363491051798
1.27127127127127 0.17781637615506
1.29129129129129 0.173313162560933
1.31131131131131 0.168856301829129
1.33133133133133 0.164448127952996
1.35135135135135 0.160090856689972
1.37137137137137 0.155786585143159
1.39139139139139 0.151537291588457
1.41141141141141 0.147344835541168
1.43143143143143 0.143210958055468
1.45145145145145 0.139137282249685
1.47147147147147 0.135125314049902
1.49149149149149 0.13117644314399
1.51151151151151 0.127291944137829
1.53153153153153 0.123472977905145
1.55155155155155 0.119720593122119
1.57157157157157 0.116035727977651
1.59159159159159 0.112419212049971
1.61161161161161 0.108871768340093
1.63163163163163 0.10539401545248
1.65165165165165 0.101986469913169
1.67167167167167 0.0986495486155338
1.69169169169169 0.0953835713838294
1.71171171171171 0.0921887636446459
1.73173173173173 0.0890652591964251
1.75175175175175 0.0860131030672496
1.77177177177177 0.083032254451193
1.79179179179179 0.0801225897136326
1.81181181181181 0.077283905456062
1.83183183183183 0.0745159216311036
1.85185185185185 0.0718182846986055
1.87187187187187 0.0691905708139176
1.89189189189189 0.0666322890396674
1.91191191191191 0.0641428845726061
1.93193193193193 0.0617217419773594
1.95195195195195 0.059368188419199
1.97197197197197 0.057081496888248
1.99199199199199 0.0548608894078376
2.01201201201201 0.0527055402200588
2.03203203203203 0.0506145789418748
2.05205205205205 0.0485870936855042
2.07207207207207 0.0466221341371236
2.09209209209209 0.0447187145882916
2.11211211211211 0.0428758169148479
2.13213213213213 0.041092393498395
2.15215215215215 0.0393673700858294
2.17217217217217 0.0376996485827434
2.19219219219219 0.0360881097768737
2.21221221221221 0.0345316159881233
2.23223223223223 0.033029013642035
2.25225225225225 0.0315791357639332
2.27227227227227 0.03018080439129
2.29229229229229 0.0288328329022028
2.31231231231231 0.0275340282581906
2.33233233233233 0.0262831931598317
2.35235235235235 0.02507912811407
2.37237237237237 0.0239206334123108
2.39239239239239 0.0228065110187136
2.41241241241241 0.0217355663683581
2.43243243243243 0.0207066100752275
2.45245245245245 0.0197184595501959
2.47247247247247 0.0187699405294511
2.49249249249249 0.0178598885140022
2.51251251251251 0.016987150121141
2.53253253253253 0.0161505843489182
2.55255255255255 0.0153490637548888
2.57257257257257 0.0145814755505493
2.59259259259259 0.0138467226130541
2.61261261261261 0.0131437244159435
2.63263263263263 0.0124714178807514
2.65265265265265 0.0118287581514852
2.67267267267267 0.0112147192940778
2.69269269269269 0.0106282949230103
2.71271271271271 0.0100684987573917
2.73273273273273 0.00953436510885491
2.75275275275275 0.00902494930369034
2.77277277277277 0.00853932804169479
2.79279279279279 0.00807659969425077
2.81281281281281 0.00763588454418634
2.83283283283283 0.00721632496998621
2.85285285285285 0.00681708557693871
2.87287287287287 0.00643735327780635
2.89289289289289 0.00607633732560524
2.91291291291291 0.00573326930106518
2.93293293293293 0.00540740305732384
2.95295295295295 0.00509801462438214
2.97297297297297 0.00480440207581539
2.99299299299299 0.00452588536019617
3.01301301301301 0.00426180609964127
3.03303303303303 0.00401152735784578
3.05305305305305 0.00377443337991421
3.07307307307307 0.00354992930624085
3.09309309309309 0.00333744086263051
3.11311311311311 0.00313641402878608
3.13313313313313 0.0029463146872226
3.15315315315315 0.00276662825459746
3.17317317317317 0.00259685929737496
3.19319319319319 0.00243653113367011
3.21321321321321 0.00228518542304203
3.23323323323323 0.00214238174593163
3.25325325325325 0.00200769717436225
3.27327327327327 0.00188072583544551
3.29329329329329 0.00176107846915771
3.31331331331331 0.00164838198177652
3.33333333333333 0.0015422789962911
3.35335335335335 0.00144242740102448
3.37337337337337 0.00134849989763212
3.39339339339339 0.00126018354956833
3.41341341341341 0.00117717933203981
3.43343343343343 0.00109920168439588
3.45345345345345 0.0010259780658361
3.47347347347347 0.000957248515249212
3.49349349349349 0.000892765215932441
3.51351351351351 0.000832292065877152
3.53353353353353 0.000775604254246008
3.55355355355355 0.000722487844607976
3.57357357357357 0.000672739365441019
3.59359359359359 0.000626165408357979
3.61361361361361 0.000582582234459159
3.63363363363363 0.000541815389165405
3.65365365365365 0.00050369932583812
3.67367367367367 0.000468077038447599
3.69369369369369 0.000434799703508357
3.71371371371371 0.000403726331459719
3.73373373373373 0.000374723427631836
3.75375375375375 0.000347664662901427
3.77377377377377 0.000322430554107926
3.79379379379379 0.000298908154269281
3.81381381381381 0.000276990752607344
3.83383383383383 0.000256577584365551
3.85385385385385 0.000237573550376443
3.87387387387387 0.000219888946313312
3.89389389389389 0.000203439201538965
3.91391391391391 0.00018814462744512
3.93393393393393 0.000173930175158222
3.95395395395395 0.000160725202471485
3.97397397397397 0.000148463249848567
3.99399399399399 0.000137081825331481
4.01401401401401 0.000126522198173995
4.03403403403403 0.000116729201011866
4.05405405405405 0.000107651040372646
4.07407407407407 9.92391153205018e-05
4.09409409409409 9.14478440253317e-05
4.11411411411411 8.42344980404937e-05
4.13413413413413 7.75590440694795e-05
4.15415415415415 7.13839929989176e-05
4.17417417417417 6.56742559732345e-05
4.19419419419419 6.03970072851107e-05
4.21421421421421 5.55215538554582e-05
4.23423423423423 5.10192110769697e-05
4.25425425425425 4.68631847962824e-05
4.27427427427427 4.3028459211397e-05
4.29429429429429 3.94916904631592e-05
4.31431431431431 3.62311057022691e-05
4.33433433433433 3.32264074164067e-05
4.35435435435435 3.04586828055866e-05
4.37437437437437 2.79103179977393e-05
4.39439439439439 2.55649169007242e-05
4.41441441441441 2.34072244914564e-05
4.43443443443443 2.14230543475548e-05
4.45445445445445 1.95992202318354e-05
4.47447447447447 1.79234715450684e-05
4.49449449449449 1.63844324676437e-05
4.51451451451451 1.49715446161227e-05
4.53453453453453 1.3675013046071e-05
4.55455455455455 1.24857554380297e-05
4.57457457457457 1.13953543089884e-05
4.59459459459459 1.03960120972233e-05
4.61461461461461 9.48050897386715e-06
4.63463463463463 8.64216324004121e-06
4.65465465465465 7.87479417380527e-06
4.67467467467467 7.17268719654363e-06
4.69469469469469 6.53056123369604e-06
4.71471471471471 5.94353814994721e-06
4.73473473473473 5.40711414409909e-06
4.75475475475475 4.91713299385613e-06
4.77477477477477 4.4697610456467e-06
4.79479479479479 4.06146384937932e-06
4.81481481481481 3.68898434268124e-06
4.83483483483483 3.34932249368796e-06
4.85485485485485 3.03971631583941e-06
4.87487487487487 2.75762417238901e-06
4.89489489489489 2.50070829244518e-06
4.91491491491491 2.26681942433715e-06
4.93493493493493 2.05398255592983e-06
4.95495495495495 1.86038363520377e-06
4.97497497497497 1.68435722796805e-06
4.99499499499499 1.5243750529858e-06
5.01501501501502 1.37903533806611e-06
5.03503503503504 1.24705294381396e-06
5.05505505505506 1.12725020473309e-06
5.07507507507508 1.01854844024876e-06
5.0950950950951 9.19960090959837e-07
5.11511511511512 8.30581438046261e-07
5.13513513513514 7.49585866251233e-07
5.15515515515516 6.76217633231267e-07
5.17517517517518 6.09786110324583e-07
5.1951951951952 5.4966046193278e-07
5.21521521521522 4.95264732746229e-07
5.23523523523524 4.46073313973501e-07
5.25525525525526 4.01606761563285e-07
5.27527527527528 3.61427941137511e-07
5.2952952952953 3.25138475990267e-07
5.31531531531532 2.92375476052561e-07
5.33533533533534 2.62808527181656e-07
5.35535535535536 2.36136921509202e-07
5.37537537537538 2.1208711087848e-07
5.3953953953954 1.90410366621162e-07
5.41541541541542 1.70880630071684e-07
5.43543543543544 1.5329253929596e-07
5.45545545545546 1.3745961852414e-07
5.47547547547548 1.23212617727566e-07
5.4954954954955 1.10397990671284e-07
5.51551551551552 9.8876500608364e-08
5.53553553553554 8.85219435638491e-08
5.55555555555556 7.92199798873018e-08
5.57557557557558 7.08670654362614e-08
5.5955955955956 6.33694743912418e-08
5.61561561561562 5.66424062986154e-08
5.63563563563564 5.06091704933412e-08
5.65565565565566 4.5200441571292e-08
5.67567567567568 4.03535800631662e-08
5.6956956956957 3.60120129107462e-08
5.71571571571572 3.2124668763613e-08
5.73573573573574 2.86454635022852e-08
5.75575575575576 2.55328317539416e-08
5.77577577577578 2.27493005011625e-08
5.7957957957958 2.02611011941336e-08
5.81581581581582 1.80378170640688e-08
5.83583583583584 1.60520626017053e-08
5.85585585585586 1.42791924110059e-08
5.87587587587588 1.2697036876002e-08
5.8958958958959 1.12856622892642e-08
5.91591591591592 1.00271532849868e-08
5.93593593593594 8.90541559921139e-09
5.95595595595596 7.90599734535251e-09
5.97597597597598 7.01592714588943e-09
5.995995995996 6.22356760178439e-09
6.01601601601602 5.5184827107339e-09
6.03603603603604 4.89131796456914e-09
6.05605605605606 4.33369196574694e-09
6.07607607607608 3.83809850362692e-09
6.0960960960961 3.3978181237675e-09
6.11611611611612 3.00683830841826e-09
6.13613613613614 2.65978146430924e-09
6.15615615615616 2.35183998527871e-09
6.17617617617618 2.07871772273778e-09
6.1961961961962 1.83657725691022e-09
6.21621621621622 1.6219924166386e-09
6.23623623623624 1.43190554571785e-09
6.25625625625626 1.26358905957511e-09
6.27627627627628 1.11461087800764e-09
6.2962962962963 9.82803357938119e-10
6.31631631631632 8.66235385045992e-10
6.33633633633634 7.63187314959515e-10
6.35635635635636 6.72128483698836e-10
6.37637637637638 5.91697033481532e-10
6.3963963963964 5.20681824054164e-10
6.41641641641642 4.58006221596989e-10
6.43643643643644 4.02713577147895e-10
6.45645645645646 3.53954224575805e-10
6.47647647647648 3.10973844559378e-10
6.4964964964965 2.73103055937435e-10
6.51651651651652 2.39748109325539e-10
6.53653653653654 2.10382570159619e-10
6.55655655655656 1.84539889444162e-10
6.57657657657658 1.61806770551276e-10
6.5965965965966 1.41817249531742e-10
6.61661661661662 1.24247414645767e-10
6.63663663663664 1.08810698278133e-10
6.65665665665666 9.52536811418137e-11
6.67667667667668 8.33523547614393e-11
6.6966966966967 7.29087937236022e-11
6.71671671671672 6.37481941394484e-11
6.73673673673674 5.57162392365998e-11
6.75675675675676 4.86767570277076e-11
6.77677677677678 4.25096386334907e-11
6.7967967967968 3.71089891068556e-11
6.81681681681682 3.23814855461048e-11
6.83683683683684 2.82449199306813e-11
6.85685685685686 2.46269064908966e-11
6.87687687687688 2.14637355595386e-11
6.8968968968969 1.86993577716892e-11
6.91691691691692 1.62844842008236e-11
6.93693693693694 1.41757895636779e-11
6.95695695695696 1.23352070109961e-11
6.97697697697698 1.07293042619725e-11
6.996996996997 9.32873195138548e-12
7.01701701701702 8.10773605306642e-12
7.03703703703704 7.04372713322415e-12
7.05705705705706 6.11689998287643e-12
7.07707707707708 5.30989788981512e-12
7.0970970970971 4.6075164458095e-12
7.11711711711712 3.99644235193917e-12
7.13713713713714 3.46502319108337e-12
7.15715715715716 3.00306458801651e-12
7.17717717717718 2.60165157997869e-12
7.1971971971972 2.25299137914217e-12
7.21721721721722 1.95027502769791e-12
7.23723723723724 1.68755573049414e-12
7.25725725725726 1.45964190299846e-12
7.27727727727728 1.262003197176e-12
7.2972972972973 1.0906879676822e-12
7.31731731731732 9.42250818252902e-13
7.33733733733734 8.13689025751829e-13
7.35735735735736 7.02386779168733e-13
7.37737737737738 6.06066294885245e-13
7.3973973973974 5.22744979473708e-13
7.41741741741742 4.50697908714381e-13
7.43743743743744 3.88424977793729e-13
7.45745745745746 3.34622154016963e-13
7.47747747747748 2.88156330935933e-13
7.4974974974975 2.48043342543511e-13
7.51751751751752 2.13428748996348e-13
7.53753753753754 1.83571051982055e-13
7.55755755755756 1.57827039041883e-13
7.57757757757758 1.35638992516663e-13
7.5975975975976 1.16523530854698e-13
7.61761761761762 1.000618782966e-13
7.63763763763764 8.58913838706873e-14
7.65765765765766 7.36981325813112e-14
7.67767767767768 6.32105109959248e-14
7.6976976976977 5.41936064405376e-14
7.71771771771772 4.64443339685915e-14
7.73773773773774 3.97871984155707e-14
7.75775775775776 3.40706104038536e-14
7.77777777777778 2.91636853079907e-14
7.7977977977978 2.4953463096687e-14
7.81781781781782 2.13424947819474e-14
7.83783783783784 1.82467480586654e-14
7.85785785785786 1.55937907247682e-14
7.87787787787788 1.33212157347804e-14
7.8978978978979 1.13752763482777e-14
7.91791791791792 9.70970386854701e-15
7.93793793793794 8.28468399583068e-15
7.95795795795796 7.06597090549298e-15
7.97797797797798 6.02412085866189e-15
7.997997997998 5.13382950919603e-15
8.01801801801802 4.3733591283238e-15
8.03803803803804 3.72404376402735e-15
8.05805805805806 3.16986191875719e-15
8.07807807807808 2.69706769497134e-15
8.0980980980981 2.29387254841619e-15
8.11811811811812 1.95017082606148e-15
8.13813813813814 1.65730316851105e-15
8.15815815815816 1.40785264250169e-15
8.17817817817818 1.19546915264237e-15
8.1981981981982 1.01471827585831e-15
8.21821821821822 8.60951178494166e-16
8.23823823823824 7.30192724674871e-16
8.25825825825826 6.19045274051723e-16
8.27827827827828 5.24606005103494e-16
8.2982982982983 4.44395893386326e-16
8.31831831831832 3.76298728354137e-16
8.33833833833834 3.18508772685576e-16
8.35835835835836 2.69485858889318e-16
8.37837837837838 2.27916883183252e-16
8.3983983983984 1.92682799625235e-16
8.41841841841842 1.62830341150177e-16
8.43843843843844 1.37547801096493e-16
8.45845845845846 1.16144301209627e-16
8.47847847847848 9.8032051926925e-17
8.4984984984985 8.27111796592674e-17
8.51851851851852 6.97567552529606e-17
8.53853853853854 5.88077091106193e-17
8.55855855855856 4.95573626747691e-17
8.57857857857858 4.17453440895294e-17
8.5985985985986 3.51506886838449e-17
8.61861861861862 2.95859531836935e-17
8.63863863863864 2.48921968838275e-17
8.65865865865866 2.09347039316773e-17
8.67867867867868 1.75993388643364e-17
8.6986986986987 1.47894429982973e-17
8.71871871871872 1.24231925503958e-17
8.73873873873874 1.04313507694241e-17
8.75875875875876 8.75535614211525e-18
8.77877877877878 7.34569713009337e-18
8.7987987987988 6.16053109048305e-18
8.81881881881882 5.16451119999019e-18
8.83883883883884 4.32779048512422e-18
8.85885885885886 3.62517658454128e-18
8.87887887887888 3.03541474067764e-18
8.8988988988989 2.54057982941549e-18
8.91891891891892 2.12556106807901e-18
8.93893893893894 1.77762546207239e-18
8.95895895895896 1.48604811781133e-18
8.97897897897898 1.24179931485728e-18
8.998998998999 1.03727973679138e-18
9.01901901901902 8.66096545670873e-19
9.03903903903904 7.22874080905507e-19
9.05905905905906 6.03093897535385e-19
9.07907907907908 5.0295965472299e-19
9.0990990990991 4.19283042962206e-19
9.11911911911912 3.49387515332417e-19
9.13913913913914 2.91027078875695e-19
9.15915915915916 2.42317819504618e-19
9.17917917917918 2.01680188581445e-19
9.1991991991992 1.67790380698338e-19
9.21921921921922 1.39539388139188e-19
9.23923923923924 1.1599853476858e-19
9.25925925925926 9.63904764362469e-20
9.27927927927928 8.00648113242436e-20
9.2992992992993 6.6477576193283e-20
9.31931931931932 5.51740167796855e-20
9.33933933933934 4.5774115701642e-20
9.35935935935936 3.79604417474457e-20
9.37937937937938 3.14679525475421e-20
9.3993993993994 2.60754402558043e-20
9.41941941941942 2.15983585814365e-20
9.43943943943944 1.78828106797973e-20
9.45945945945946 1.48005121824234e-20
9.47947947947948 1.22445730038445e-20
9.4994994994995 1.01259663374544e-20
9.51951951951952 8.37057415073758e-21
9.53953953953954 6.91671611023779e-21
9.55955955955956 5.71308371631411e-21
9.57957957957958 4.71701393695898e-21
9.5995995995996 3.89304716291232e-21
9.61961961961962 3.21172317125736e-21
9.63963963963964 2.64857624243904e-21
9.65965965965966 2.18329684678274e-21
9.67967967967968 1.79903258756111e-21
9.6996996996997 1.48180551599987e-21
9.71971971971972 1.22002665237565e-21
9.73973973973974 1.00409166885045e-21
9.75975975975976 8.26044308669654e-22
9.77977977977978 6.79296312742742e-22
9.7997997997998 5.58394465795474e-22
9.81981981981982 4.58826916995414e-22
9.83983983983984 3.7686222201397e-22
9.85985985985986 3.09415635142992e-22
9.87987987987988 2.53938085193762e-22
9.8998998998999 2.08324025950642e-22
9.91991991991992 1.70834984871876e-22
9.93993993993994 1.40036162642795e-22
9.95995995995996 1.14743877987917e-22
9.97997997997998 9.39820210218911e-23
10 7.69459862670642e-23
};
\addlegendentry{N(0,1)}; 
\legend{}; 
\end{axis}

\end{tikzpicture}

%% file: FinalFigs/PowerCurveAllMethods_RBFd_10_eps_0_2.tex
\begin{tikzpicture}

\definecolor{crimson2143940}{RGB}{214,39,40}
\definecolor{darkorange25512714}{RGB}{255,127,14}
\definecolor{darkslategray38}{RGB}{38,38,38}
\definecolor{forestgreen4416044}{RGB}{44,160,44}
\definecolor{lightgray204}{RGB}{204,204,204}
\definecolor{mediumpurple148103189}{RGB}{148,103,189}
\definecolor{sienna1408675}{RGB}{140,86,75}
\definecolor{steelblue31119180}{RGB}{31,119,180}

\begin{axis}[
axis line style={darkslategray38},
height=\figheight,
legend cell align={left},
legend style={
  fill opacity=0.4,
  draw opacity=1,
  text opacity=1,
  at={(0.01,0.95)},
  anchor=north west,
  draw=lightgray204, 
  nodes={scale=0.7, transform shape}
},
tick align=outside,
tick pos=left,
title={($d=10, j=5,  \epsilon=0.2)$},
width=\figwidth,
x grid style={lightgray204},
xlabel=\textcolor{darkslategray38}{Sample-Size (n+m)},
xmin=-8, xmax=900,
xtick style={color=darkslategray38},
y grid style={lightgray204},
ylabel=\textcolor{darkslategray38}{Power},
ymin=-0.00951058496303117, ymax=1.04935222789267,
ytick style={color=darkslategray38}
]
\path [draw=steelblue31119180, fill=steelblue31119180, opacity=0.3]
(axis cs:40,0.174838679514016)
--(axis cs:40,0.125161320485984)
--(axis cs:90,0.239496065827504)
--(axis cs:140,0.41026471534592)
--(axis cs:190,0.465064067566472)
--(axis cs:242,0.558843534464774)
--(axis cs:292,0.660384252138658)
--(axis cs:342,0.803717032606648)
--(axis cs:392,0.839910161419411)
--(axis cs:444,0.884146478115196)
--(axis cs:494,0.871690184470914)
--(axis cs:544,0.964177598464296)
--(axis cs:594,0.963926410925088)
--(axis cs:646,0.964460817156914)
--(axis cs:696,0.969506222081633)
--(axis cs:746,0.969270630260821)
--(axis cs:796,1)
--(axis cs:848,0.989951237775454)
--(axis cs:898,1)
--(axis cs:948,1)
--(axis cs:1000,1)
--(axis cs:1000,1)
--(axis cs:1000,1)
--(axis cs:948,1)
--(axis cs:898,1)
--(axis cs:848,1.00004876222455)
--(axis cs:796,1)
--(axis cs:746,0.990729369739179)
--(axis cs:696,0.990493777918367)
--(axis cs:646,0.985539182843086)
--(axis cs:594,0.986073589074911)
--(axis cs:544,0.985822401535704)
--(axis cs:494,0.918309815529086)
--(axis cs:444,0.925853521884804)
--(axis cs:392,0.890089838580589)
--(axis cs:342,0.856282967393352)
--(axis cs:292,0.729615747861342)
--(axis cs:242,0.631156465535226)
--(axis cs:190,0.534935932433528)
--(axis cs:140,0.47973528465408)
--(axis cs:90,0.300503934172497)
--(axis cs:40,0.174838679514016)
--cycle;

\path [draw=darkorange25512714, fill=darkorange25512714, opacity=0.3]
(axis cs:40,0.184369537849537)
--(axis cs:40,0.135630462150463)
--(axis cs:90,0.159065900825361)
--(axis cs:140,0.274080654585586)
--(axis cs:190,0.287760670369873)
--(axis cs:242,0.365018085029547)
--(axis cs:292,0.526951702010542)
--(axis cs:342,0.667536019960578)
--(axis cs:392,0.61576756837442)
--(axis cs:444,0.720919981086664)
--(axis cs:494,0.749195465025422)
--(axis cs:544,0.786491404804866)
--(axis cs:594,0.889168293396843)
--(axis cs:646,0.894263558646672)
--(axis cs:696,0.911006843337665)
--(axis cs:746,0.930553979786806)
--(axis cs:796,0.96327633269834)
--(axis cs:848,0.976430431749499)
--(axis cs:898,0.97647430354751)
--(axis cs:948,0.989521006023)
--(axis cs:1000,0.976290845333788)
--(axis cs:1000,0.993709154666212)
--(axis cs:1000,0.993709154666212)
--(axis cs:948,1.000478993977)
--(axis cs:898,0.99352569645249)
--(axis cs:848,0.993569568250501)
--(axis cs:796,0.98672366730166)
--(axis cs:746,0.959446020213194)
--(axis cs:696,0.948993156662335)
--(axis cs:646,0.935736441353328)
--(axis cs:594,0.930831706603157)
--(axis cs:544,0.843508595195134)
--(axis cs:494,0.810804534974578)
--(axis cs:444,0.779080018913336)
--(axis cs:392,0.68423243162558)
--(axis cs:342,0.732463980039422)
--(axis cs:292,0.593048297989458)
--(axis cs:242,0.434981914970453)
--(axis cs:190,0.352239329630127)
--(axis cs:140,0.345919345414414)
--(axis cs:90,0.210934099174639)
--(axis cs:40,0.184369537849537)
--cycle;

\path [draw=forestgreen4416044, fill=forestgreen4416044, opacity=0.3]
(axis cs:40,0.264155318494544)
--(axis cs:40,0.204262881976661)
--(axis cs:90,0.247068687667636)
--(axis cs:140,0.456703675496643)
--(axis cs:190,0.472585426628284)
--(axis cs:242,0.592463124850524)
--(axis cs:292,0.787077210556462)
--(axis cs:342,0.903729350323494)
--(axis cs:392,0.867817642273713)
--(axis cs:444,0.933438921687398)
--(axis cs:494,0.948382071974656)
--(axis cs:544,0.963183285393053)
--(axis cs:594,0.99004718923164)
--(axis cs:646,0.990962875699993)
--(axis cs:696,0.993435996694691)
--(axis cs:746,0.995511706352707)
--(axis cs:796,0.998546107066721)
--(axis cs:848,0.999247206052825)
--(axis cs:898,0.999247206052825)
--(axis cs:948,0.999796664172607)
--(axis cs:1000,0.999247206052825)
--(axis cs:1000,1.00057616512638)
--(axis cs:1000,1.00057616512638)
--(axis cs:948,1.00018801978831)
--(axis cs:898,1.00057616512638)
--(axis cs:848,1.00057616512638)
--(axis cs:796,1.00089976761037)
--(axis cs:746,1.00122210003559)
--(axis cs:696,1.00093085180097)
--(axis cs:646,1.0002932703593)
--(axis cs:594,0.999999140193895)
--(axis cs:544,0.985536271941862)
--(axis cs:494,0.975449980849763)
--(axis cs:444,0.964552531983674)
--(axis cs:392,0.912076283494903)
--(axis cs:342,0.941515650046391)
--(axis cs:292,0.842041285974959)
--(axis cs:242,0.660867208866711)
--(axis cs:190,0.543287196755591)
--(axis cs:140,0.527405425097203)
--(axis cs:90,0.310481549427104)
--(axis cs:40,0.264155318494544)
--cycle;

\path [draw=crimson2143940, fill=crimson2143940, opacity=0.3]
(axis cs:40,0.0878450973659434)
--(axis cs:40,0.0521549026340566)
--(axis cs:90,0.179710573214404)
--(axis cs:140,0.356340863053251)
--(axis cs:190,0.436125267528732)
--(axis cs:242,0.575106599836072)
--(axis cs:292,0.638983562825495)
--(axis cs:342,0.783081094004399)
--(axis cs:392,0.828845698250575)
--(axis cs:444,0.873073075911109)
--(axis cs:494,0.872350841627116)
--(axis cs:544,0.952471258044001)
--(axis cs:594,0.969918333471097)
--(axis cs:646,0.96355778102814)
--(axis cs:696,0.970486325630967)
--(axis cs:746,0.971045985537202)
--(axis cs:796,1)
--(axis cs:848,1)
--(axis cs:898,1)
--(axis cs:948,1)
--(axis cs:1000,1)
--(axis cs:1000,1)
--(axis cs:1000,1)
--(axis cs:948,1)
--(axis cs:898,1)
--(axis cs:848,1)
--(axis cs:796,1)
--(axis cs:746,0.988954014462798)
--(axis cs:696,0.989513674369033)
--(axis cs:646,0.98644221897186)
--(axis cs:594,0.990081666528903)
--(axis cs:544,0.977528741955999)
--(axis cs:494,0.917649158372884)
--(axis cs:444,0.916926924088891)
--(axis cs:392,0.881154301749425)
--(axis cs:342,0.836918905995601)
--(axis cs:292,0.701016437174505)
--(axis cs:242,0.644893400163928)
--(axis cs:190,0.503874732471268)
--(axis cs:140,0.423659136946749)
--(axis cs:90,0.240289426785596)
--(axis cs:40,0.0878450973659434)
--cycle;

\path [draw=mediumpurple148103189, fill=mediumpurple148103189, opacity=0.3]
(axis cs:40,0.192730297059354)
--(axis cs:40,0.137269702940646)
--(axis cs:90,0.135083188406219)
--(axis cs:140,0.132848584291054)
--(axis cs:190,0.204695555193998)
--(axis cs:242,0.250506123731885)
--(axis cs:292,0.258569958638271)
--(axis cs:342,0.307742597748734)
--(axis cs:392,0.326850801291736)
--(axis cs:444,0.29715567020017)
--(axis cs:494,0.375395087054003)
--(axis cs:544,0.377402310281248)
--(axis cs:594,0.471802946290371)
--(axis cs:646,0.495036483300446)
--(axis cs:696,0.45887068364887)
--(axis cs:746,0.561931170265641)
--(axis cs:796,0.532047041104022)
--(axis cs:848,0.636360820833439)
--(axis cs:898,0.604108357518776)
--(axis cs:948,0.637937960529623)
--(axis cs:1000,0.639556382803105)
--(axis cs:1000,0.710443617196895)
--(axis cs:1000,0.710443617196895)
--(axis cs:948,0.702062039470377)
--(axis cs:898,0.675891642481224)
--(axis cs:848,0.703639179166561)
--(axis cs:796,0.597952958895978)
--(axis cs:746,0.628068829734359)
--(axis cs:696,0.53112931635113)
--(axis cs:646,0.564963516699554)
--(axis cs:594,0.538197053709629)
--(axis cs:544,0.442597689718752)
--(axis cs:494,0.444604912945997)
--(axis cs:444,0.36284432979983)
--(axis cs:392,0.393149198708264)
--(axis cs:342,0.372257402251266)
--(axis cs:292,0.321430041361729)
--(axis cs:242,0.319493876268115)
--(axis cs:190,0.265304444806002)
--(axis cs:140,0.187151415708946)
--(axis cs:90,0.184916811593781)
--(axis cs:40,0.192730297059354)
--cycle;

\path [draw=sienna1408675, fill=sienna1408675, opacity=0.3]
(axis cs:40,0.10124705861996)
--(axis cs:40,0.0587529413800404)
--(axis cs:90,0.038619542894046)
--(axis cs:140,0.101280229765025)
--(axis cs:190,0.0670507834432809)
--(axis cs:242,0.074410090456731)
--(axis cs:292,0.058124055581983)
--(axis cs:342,0.0680171089253483)
--(axis cs:392,0.0699731586364699)
--(axis cs:444,0.0651541093674282)
--(axis cs:494,0.101503404501928)
--(axis cs:544,0.110414447840246)
--(axis cs:594,0.155026213743207)
--(axis cs:646,0.0866498929338643)
--(axis cs:696,0.12434518807319)
--(axis cs:746,0.140094980425625)
--(axis cs:796,0.127217344865007)
--(axis cs:848,0.127648839515587)
--(axis cs:898,0.10140128446292)
--(axis cs:948,0.139417791827131)
--(axis cs:1000,0.0662611533172396)
--(axis cs:1000,0.10373884668276)
--(axis cs:1000,0.10373884668276)
--(axis cs:948,0.190582208172869)
--(axis cs:898,0.14859871553708)
--(axis cs:848,0.172351160484413)
--(axis cs:796,0.172782655134992)
--(axis cs:746,0.189905019574375)
--(axis cs:696,0.17565481192681)
--(axis cs:646,0.133350107066136)
--(axis cs:594,0.204973786256793)
--(axis cs:544,0.159585552159754)
--(axis cs:494,0.148496595498072)
--(axis cs:444,0.104845890632572)
--(axis cs:392,0.11002684136353)
--(axis cs:342,0.111982891074652)
--(axis cs:292,0.091875944418017)
--(axis cs:242,0.115589909543269)
--(axis cs:190,0.102949216556719)
--(axis cs:140,0.148719770234975)
--(axis cs:90,0.071380457105954)
--(axis cs:40,0.10124705861996)
--cycle;

\addplot [semithick, steelblue31119180]
table {%
40 0.15
90 0.27
140 0.445
190 0.5
242 0.595
292 0.695
342 0.83
392 0.865
444 0.905
494 0.895
544 0.975
594 0.975
646 0.975
696 0.98
746 0.98
796 1
848 0.995
898 1
948 1
1000 1
};
\addlegendentry{$\dmmd$-perm}
\addplot [semithick, darkorange25512714]
table {%
40 0.16
90 0.185
140 0.31
190 0.32
242 0.4
292 0.56
342 0.7
392 0.65
444 0.75
494 0.78
544 0.815
594 0.91
646 0.915
696 0.93
746 0.945
796 0.975
848 0.985
898 0.985
948 0.995
1000 0.985
};
\addlegendentry{$\cmmd$}
\addplot [semithick, forestgreen4416044, dashed]
table {%
40 0.234209100235603
90 0.27877511854737
140 0.492054550296923
190 0.507936311691937
242 0.626665166858618
292 0.814559248265711
342 0.922622500184942
392 0.889946962884308
444 0.948995726835536
494 0.961916026412209
544 0.974359778667457
594 0.995023164712768
646 0.995628073029646
696 0.997183424247829
746 0.998366903194148
796 0.999722937338545
848 0.999911685589605
898 0.999911685589605
948 0.999992341980458
1000 0.999911685589605
};
\addlegendentry{predicted}
\addplot [semithick, crimson2143940]
table {%
40 0.07
90 0.21
140 0.39
190 0.47
242 0.61
292 0.67
342 0.81
392 0.855
444 0.895
494 0.895
544 0.965
594 0.98
646 0.975
696 0.98
746 0.98
796 1
848 1
898 1
948 1
1000 1
};
\addlegendentry{$\dmmd$-spectral}
\addplot [semithick, mediumpurple148103189]
table {%
40 0.165
90 0.16
140 0.16
190 0.235
242 0.285
292 0.29
342 0.34
392 0.36
444 0.33
494 0.41
544 0.41
594 0.505
646 0.53
696 0.495
746 0.595
796 0.565
848 0.67
898 0.64
948 0.67
1000 0.675
};
\addlegendentry{B-$\dmmd$}
\addplot [semithick, sienna1408675]
table {%
40 0.08
90 0.055
140 0.125
190 0.085
242 0.095
292 0.075
342 0.09
392 0.09
444 0.085
494 0.125
544 0.135
594 0.18
646 0.11
696 0.15
746 0.165
796 0.15
848 0.15
898 0.125
948 0.165
1000 0.085
};
\addlegendentry{L-$\dmmd$}
\end{axis}

\end{tikzpicture}

%% file: FinalFigs/PowerCurveAllMethods_RBFd_50_eps_0_125.tex
\begin{tikzpicture}

\definecolor{crimson2143940}{RGB}{214,39,40}
\definecolor{darkorange25512714}{RGB}{255,127,14}
\definecolor{darkslategray38}{RGB}{38,38,38}
\definecolor{forestgreen4416044}{RGB}{44,160,44}
\definecolor{lightgray204}{RGB}{204,204,204}
\definecolor{mediumpurple148103189}{RGB}{148,103,189}
\definecolor{sienna1408675}{RGB}{140,86,75}
\definecolor{steelblue31119180}{RGB}{31,119,180}

\begin{axis}[
axis line style={darkslategray38},
height=\figheight,
legend cell align={left},
legend style={
  fill opacity=0.8,
  draw opacity=1,
  text opacity=1,
  at={(0.03,0.97)},
  anchor=north west,
  draw=lightgray204
},
tick align=outside,
tick pos=left,
title={($d=50, j=15, \epsilon=0.125)$},
width=\figwidth,
x grid style={lightgray204},
xlabel=\textcolor{darkslategray38}{Sample-Size (n+m)},
xmin=-3, xmax=943,
xtick style={color=darkslategray38},
y grid style={lightgray204},
ylabel=\textcolor{darkslategray38}{Power},
ymin=-0.0471488784622712, ymax=1.05022961378031,
ytick style={color=darkslategray38}
]
\path [draw=steelblue31119180, fill=steelblue31119180, opacity=0.3]
(axis cs:40,0.171405065801849)
--(axis cs:40,0.118594934198151)
--(axis cs:84,0.126317319091792)
--(axis cs:130,0.190413696158526)
--(axis cs:174,0.254738638497252)
--(axis cs:220,0.334280502933804)
--(axis cs:266,0.400097645136753)
--(axis cs:310,0.46320193619224)
--(axis cs:356,0.549003472389687)
--(axis cs:402,0.595244029289919)
--(axis cs:446,0.67241957067502)
--(axis cs:492,0.749881909506079)
--(axis cs:536,0.756368898030289)
--(axis cs:582,0.824843502330412)
--(axis cs:628,0.867019151995629)
--(axis cs:672,0.90377927015864)
--(axis cs:718,0.922461185900979)
--(axis cs:764,0.951968211941564)
--(axis cs:808,0.946028958521284)
--(axis cs:854,0.953007189653797)
--(axis cs:900,0.976044868789348)
--(axis cs:900,0.993955131210652)
--(axis cs:900,0.993955131210652)
--(axis cs:854,0.976992810346203)
--(axis cs:808,0.973971041478716)
--(axis cs:764,0.978031788058436)
--(axis cs:718,0.95753881409902)
--(axis cs:672,0.94622072984136)
--(axis cs:628,0.912980848004371)
--(axis cs:582,0.875156497669588)
--(axis cs:536,0.813631101969711)
--(axis cs:492,0.810118090493921)
--(axis cs:446,0.73758042932498)
--(axis cs:402,0.664755970710081)
--(axis cs:356,0.620996527610313)
--(axis cs:310,0.53679806380776)
--(axis cs:266,0.469902354863247)
--(axis cs:220,0.395719497066196)
--(axis cs:174,0.315261361502748)
--(axis cs:130,0.249586303841474)
--(axis cs:84,0.173682680908208)
--(axis cs:40,0.171405065801849)
--cycle;

\path [draw=darkorange25512714, fill=darkorange25512714, opacity=0.3]
(axis cs:40,0.132439905859874)
--(axis cs:40,0.0875600941401262)
--(axis cs:84,0.0868047553364919)
--(axis cs:130,0.134043895226749)
--(axis cs:174,0.139039501160417)
--(axis cs:220,0.191617082602382)
--(axis cs:266,0.294576570253574)
--(axis cs:310,0.296234485047611)
--(axis cs:356,0.368582705001607)
--(axis cs:402,0.372841175394614)
--(axis cs:446,0.434380798717546)
--(axis cs:492,0.495966927849517)
--(axis cs:536,0.524512053102497)
--(axis cs:582,0.655308081907698)
--(axis cs:628,0.661972445216153)
--(axis cs:672,0.750331930042552)
--(axis cs:718,0.784767732470091)
--(axis cs:764,0.793484073182331)
--(axis cs:808,0.888938245562157)
--(axis cs:854,0.836156552262728)
--(axis cs:900,0.873629751990208)
--(axis cs:900,0.916370248009792)
--(axis cs:900,0.916370248009792)
--(axis cs:854,0.883843447737272)
--(axis cs:808,0.931061754437843)
--(axis cs:764,0.846515926817669)
--(axis cs:718,0.845232267529909)
--(axis cs:672,0.809668069957448)
--(axis cs:628,0.728027554783847)
--(axis cs:582,0.714691918092302)
--(axis cs:536,0.595487946897503)
--(axis cs:492,0.564033072150483)
--(axis cs:446,0.505619201282454)
--(axis cs:402,0.437158824605386)
--(axis cs:356,0.441417294998393)
--(axis cs:310,0.363765514952389)
--(axis cs:266,0.365423429746426)
--(axis cs:220,0.248382917397618)
--(axis cs:174,0.190960498839583)
--(axis cs:130,0.185956104773251)
--(axis cs:84,0.133195244663508)
--(axis cs:40,0.132439905859874)
--cycle;

\path [draw=forestgreen4416044, fill=forestgreen4416044, opacity=0.3]
(axis cs:40,0.171089121235566)
--(axis cs:40,0.121136380583771)
--(axis cs:84,0.121136380583771)
--(axis cs:130,0.204262881976661)
--(axis cs:174,0.212792604536382)
--(axis cs:220,0.30712548992801)
--(axis cs:266,0.488283483666098)
--(axis cs:310,0.488283483666098)
--(axis cs:356,0.599493164317601)
--(axis cs:402,0.599493164317601)
--(axis cs:446,0.685469948423024)
--(axis cs:492,0.755513971899651)
--(axis cs:536,0.787077210556462)
--(axis cs:582,0.893615853794394)
--(axis cs:628,0.900420522556035)
--(axis cs:672,0.948382071974656)
--(axis cs:718,0.963183285393053)
--(axis cs:764,0.965072522085565)
--(axis cs:808,0.99004718923164)
--(axis cs:854,0.978232045934359)
--(axis cs:900,0.987017792600458)
--(axis cs:900,0.998859612858613)
--(axis cs:900,0.998859612858613)
--(axis cs:854,0.99460186340913)
--(axis cs:808,0.999999140193895)
--(axis cs:764,0.98675484858091)
--(axis cs:718,0.985536271941862)
--(axis cs:672,0.975449980849763)
--(axis cs:628,0.938865248729917)
--(axis cs:582,0.933370909642669)
--(axis cs:536,0.842041285974959)
--(axis cs:492,0.813653805129203)
--(axis cs:446,0.749152874008371)
--(axis cs:402,0.667634301561307)
--(axis cs:356,0.667634301561307)
--(axis cs:310,0.558915355705637)
--(axis cs:266,0.558915355705637)
--(axis cs:220,0.374148383640913)
--(axis cs:174,0.27345815135575)
--(axis cs:130,0.264155318494544)
--(axis cs:84,0.171089121235566)
--(axis cs:40,0.171089121235566)
--cycle;

\path [draw=crimson2143940, fill=crimson2143940, opacity=0.3]
(axis cs:40,0.01726803790579)
--(axis cs:40,0.00273196209421002)
--(axis cs:84,0.051204322305381)
--(axis cs:130,0.0693854420372398)
--(axis cs:174,0.148651802718212)
--(axis cs:220,0.254232982351225)
--(axis cs:266,0.331774445076116)
--(axis cs:310,0.41163348422445)
--(axis cs:356,0.490294092721843)
--(axis cs:402,0.549241819117858)
--(axis cs:446,0.628365455985584)
--(axis cs:492,0.68340237350686)
--(axis cs:536,0.714352814158556)
--(axis cs:582,0.8119591369605)
--(axis cs:628,0.814664858891255)
--(axis cs:672,0.918447073521579)
--(axis cs:718,0.928597427793178)
--(axis cs:764,0.970866134443731)
--(axis cs:808,0.958196425329588)
--(axis cs:854,0.946170683314061)
--(axis cs:900,0.989651226776166)
--(axis cs:900,1.00034877322383)
--(axis cs:900,1.00034877322383)
--(axis cs:854,0.973829316685939)
--(axis cs:808,0.981803574670412)
--(axis cs:764,0.989133865556269)
--(axis cs:718,0.961402572206822)
--(axis cs:672,0.951552926478421)
--(axis cs:628,0.865335141108745)
--(axis cs:582,0.8680408630395)
--(axis cs:536,0.775647185841444)
--(axis cs:492,0.74659762649314)
--(axis cs:446,0.691634544014416)
--(axis cs:402,0.620758180882142)
--(axis cs:356,0.559705907278157)
--(axis cs:310,0.47836651577555)
--(axis cs:266,0.398225554923884)
--(axis cs:220,0.315767017648774)
--(axis cs:174,0.201348197281788)
--(axis cs:130,0.11061455796276)
--(axis cs:84,0.0887956776946191)
--(axis cs:40,0.01726803790579)
--cycle;

\path [draw=mediumpurple148103189, fill=mediumpurple148103189, opacity=0.3]
(axis cs:40,0.0933966708944852)
--(axis cs:40,0.0566033291055148)
--(axis cs:84,0.079487808503234)
--(axis cs:130,0.0832863090424497)
--(axis cs:174,0.115212099725874)
--(axis cs:220,0.0735337619737412)
--(axis cs:266,0.0866242112432543)
--(axis cs:310,0.100985017697279)
--(axis cs:356,0.194480835938709)
--(axis cs:402,0.128947756814431)
--(axis cs:446,0.138964267342746)
--(axis cs:492,0.199754173841669)
--(axis cs:536,0.207960961312798)
--(axis cs:582,0.214865727568099)
--(axis cs:628,0.209190098993992)
--(axis cs:672,0.226623964776594)
--(axis cs:718,0.247848260155942)
--(axis cs:764,0.190446668969471)
--(axis cs:808,0.272607572258319)
--(axis cs:854,0.289091101604878)
--(axis cs:900,0.280781227739733)
--(axis cs:900,0.349218772260267)
--(axis cs:900,0.349218772260267)
--(axis cs:854,0.350908898395122)
--(axis cs:808,0.337392427741681)
--(axis cs:764,0.249553331030529)
--(axis cs:718,0.312151739844058)
--(axis cs:672,0.283376035223406)
--(axis cs:628,0.270809901006008)
--(axis cs:582,0.275134272431901)
--(axis cs:536,0.272039038687202)
--(axis cs:492,0.260245826158331)
--(axis cs:446,0.191035732657254)
--(axis cs:402,0.181052243185569)
--(axis cs:356,0.255519164061291)
--(axis cs:310,0.149014982302721)
--(axis cs:266,0.133375788756746)
--(axis cs:220,0.116466238026259)
--(axis cs:174,0.164787900274126)
--(axis cs:130,0.12671369095755)
--(axis cs:84,0.120512191496766)
--(axis cs:40,0.0933966708944852)
--cycle;

\path [draw=sienna1408675, fill=sienna1408675, opacity=0.3]
(axis cs:40,0.0712754416222725)
--(axis cs:40,0.0387245583777275)
--(axis cs:84,0.0392655036623348)
--(axis cs:130,0.051639189151892)
--(axis cs:174,0.0440023634558101)
--(axis cs:220,0.0506016914397157)
--(axis cs:266,0.0523924022081375)
--(axis cs:310,0.0614852086158121)
--(axis cs:356,0.0573512748335751)
--(axis cs:402,0.0841280091989288)
--(axis cs:446,0.0393382631869898)
--(axis cs:492,0.0566326546556123)
--(axis cs:536,0.0565297705753285)
--(axis cs:582,0.0527518116893397)
--(axis cs:628,0.0743899448084194)
--(axis cs:672,0.0744379597315831)
--(axis cs:718,0.0709671993653062)
--(axis cs:764,0.0836299859616331)
--(axis cs:808,0.051826943019954)
--(axis cs:854,0.0647365014126385)
--(axis cs:900,0.055409201777365)
--(axis cs:900,0.094590798222635)
--(axis cs:900,0.094590798222635)
--(axis cs:854,0.105263498587361)
--(axis cs:808,0.088173056980046)
--(axis cs:764,0.126370014038367)
--(axis cs:718,0.109032800634694)
--(axis cs:672,0.115562040268417)
--(axis cs:628,0.115610055191581)
--(axis cs:582,0.0872481883106603)
--(axis cs:536,0.0934702294246715)
--(axis cs:492,0.0933673453443877)
--(axis cs:446,0.0706617368130102)
--(axis cs:402,0.125871990801071)
--(axis cs:356,0.0926487251664249)
--(axis cs:310,0.0985147913841879)
--(axis cs:266,0.0876075977918625)
--(axis cs:220,0.0893983085602843)
--(axis cs:174,0.0759976365441899)
--(axis cs:130,0.088360810848108)
--(axis cs:84,0.0707344963376652)
--(axis cs:40,0.0712754416222725)
--cycle;

\addplot [semithick, steelblue31119180]
table {%
40 0.145
84 0.15
130 0.22
174 0.285
220 0.365
266 0.435
310 0.5
356 0.585
402 0.63
446 0.705
492 0.78
536 0.785
582 0.85
628 0.89
672 0.925
718 0.94
764 0.965
808 0.96
854 0.965
900 0.985
};
\addlegendentry{$\dmmd$-perm}
\addplot [semithick, darkorange25512714]
table {%
40 0.11
84 0.11
130 0.16
174 0.165
220 0.22
266 0.33
310 0.33
356 0.405
402 0.405
446 0.47
492 0.53
536 0.56
582 0.685
628 0.695
672 0.78
718 0.815
764 0.82
808 0.91
854 0.86
900 0.895
};
\addlegendentry{$\cmmd$}
\addplot [semithick, forestgreen4416044, dashed]
table {%
40 0.146112750909668
84 0.146112750909668
130 0.234209100235603
174 0.243125377946066
220 0.340636936784461
266 0.523599419685867
310 0.523599419685867
356 0.633563732939454
402 0.633563732939454
446 0.717311411215698
492 0.784583888514427
536 0.814559248265711
582 0.913493381718532
628 0.919642885642976
672 0.961916026412209
718 0.974359778667457
764 0.975913685333238
808 0.995023164712768
854 0.986416954671745
900 0.992938702729535
};
\addlegendentry{predicted}
\addplot [semithick, crimson2143940]
table {%
40 0.01
84 0.07
130 0.09
174 0.175
220 0.285
266 0.365
310 0.445
356 0.525
402 0.585
446 0.66
492 0.715
536 0.745
582 0.84
628 0.84
672 0.935
718 0.945
764 0.98
808 0.97
854 0.96
900 0.995
};
\addlegendentry{$\dmmd$-spectral}
\addplot [semithick, mediumpurple148103189]
table {%
40 0.075
84 0.1
130 0.105
174 0.14
220 0.095
266 0.11
310 0.125
356 0.225
402 0.155
446 0.165
492 0.23
536 0.24
582 0.245
628 0.24
672 0.255
718 0.28
764 0.22
808 0.305
854 0.32
900 0.315
};
\addlegendentry{B-$\dmmd$}
\addplot [semithick, sienna1408675]
table {%
40 0.055
84 0.055
130 0.07
174 0.06
220 0.07
266 0.07
310 0.08
356 0.075
402 0.105
446 0.055
492 0.075
536 0.075
582 0.07
628 0.095
672 0.095
718 0.09
764 0.105
808 0.07
854 0.085
900 0.075
};
\addlegendentry{L-$\dmmd$}; 
\legend{}
\end{axis}

\end{tikzpicture}

%% file: FinalFigs/PowerCurveAllMethods_RBFd_100_eps_0_15.tex
\begin{tikzpicture}

\definecolor{crimson2143940}{RGB}{214,39,40}
\definecolor{darkorange25512714}{RGB}{255,127,14}
\definecolor{darkslategray38}{RGB}{38,38,38}
\definecolor{forestgreen4416044}{RGB}{44,160,44}
\definecolor{lightgray204}{RGB}{204,204,204}
\definecolor{mediumpurple148103189}{RGB}{148,103,189}
\definecolor{sienna1408675}{RGB}{140,86,75}
\definecolor{steelblue31119180}{RGB}{31,119,180}

\begin{axis}[
axis line style={darkslategray38},
height=\figheight,
legend cell align={left},
legend style={
  fill opacity=0.8,
  draw opacity=1,
  text opacity=1,
  at={(0.01,0.95)},
  anchor=north west,
  draw=lightgray204
},
tick align=outside,
tick pos=left,
title={$(d=100, j=20,  \epsilon=0.15)$},
width=\figwidth,
x grid style={lightgray204},
xlabel=\textcolor{darkslategray38}{Sample-Size (n+m)},
xmin=-8, xmax=801,
xtick style={color=darkslategray38},
y grid style={lightgray204},
ylabel=\textcolor{darkslategray38}{Power},
ymin=-0.0463459862329632, ymax=1.05108140268629,
ytick style={color=darkslategray38}
]
\path [draw=steelblue31119180, fill=steelblue31119180, opacity=0.3]
(axis cs:40,0.132380292670115)
--(axis cs:40,0.0876197073298851)
--(axis cs:90,0.138194788286604)
--(axis cs:140,0.286376976050738)
--(axis cs:190,0.380131532869941)
--(axis cs:242,0.520307097267597)
--(axis cs:292,0.611884897705126)
--(axis cs:342,0.715907056267885)
--(axis cs:392,0.835047244641122)
--(axis cs:444,0.866762436121658)
--(axis cs:494,0.932561267964671)
--(axis cs:544,0.945333733433488)
--(axis cs:594,0.970096591495854)
--(axis cs:646,0.990191218969427)
--(axis cs:696,0.99033738270925)
--(axis cs:746,1)
--(axis cs:796,1)
--(axis cs:848,1)
--(axis cs:898,1)
--(axis cs:948,1)
--(axis cs:1000,1)
--(axis cs:1000,1)
--(axis cs:1000,1)
--(axis cs:948,1)
--(axis cs:898,1)
--(axis cs:848,1)
--(axis cs:796,1)
--(axis cs:746,1)
--(axis cs:696,0.99966261729075)
--(axis cs:646,0.999808781030573)
--(axis cs:594,0.989903408504146)
--(axis cs:544,0.974666266566512)
--(axis cs:494,0.967438732035329)
--(axis cs:444,0.913237563878342)
--(axis cs:392,0.884952755358878)
--(axis cs:342,0.774092943732115)
--(axis cs:292,0.678115102294874)
--(axis cs:242,0.589692902732403)
--(axis cs:190,0.449868467130059)
--(axis cs:140,0.343623023949262)
--(axis cs:90,0.191805211713396)
--(axis cs:40,0.132380292670115)
--cycle;

\path [draw=darkorange25512714, fill=darkorange25512714, opacity=0.3]
(axis cs:40,0.126813069476807)
--(axis cs:40,0.0831869305231932)
--(axis cs:90,0.0736546404340428)
--(axis cs:140,0.17497735229864)
--(axis cs:190,0.240671046728531)
--(axis cs:242,0.350368944572826)
--(axis cs:292,0.404415602084621)
--(axis cs:342,0.497360393200779)
--(axis cs:392,0.605605996162122)
--(axis cs:444,0.715313313169031)
--(axis cs:494,0.732935855305341)
--(axis cs:544,0.80654657314136)
--(axis cs:594,0.858102511559542)
--(axis cs:646,0.894493430930553)
--(axis cs:696,0.893848699803558)
--(axis cs:746,0.94611196558184)
--(axis cs:796,0.945451890157137)
--(axis cs:848,0.98305022482378)
--(axis cs:898,0.982536086817225)
--(axis cs:948,1)
--(axis cs:1000,0.990467064968478)
--(axis cs:1000,0.999532935031522)
--(axis cs:1000,0.999532935031522)
--(axis cs:948,1)
--(axis cs:898,0.997463913182775)
--(axis cs:848,0.99694977517622)
--(axis cs:796,0.974548109842863)
--(axis cs:746,0.97388803441816)
--(axis cs:696,0.936151300196442)
--(axis cs:646,0.935506569069447)
--(axis cs:594,0.901897488440458)
--(axis cs:544,0.86345342685864)
--(axis cs:494,0.797064144694659)
--(axis cs:444,0.774686686830969)
--(axis cs:392,0.674394003837879)
--(axis cs:342,0.572639606799222)
--(axis cs:292,0.475584397915379)
--(axis cs:242,0.419631055427174)
--(axis cs:190,0.299328953271469)
--(axis cs:140,0.23502264770136)
--(axis cs:90,0.116345359565957)
--(axis cs:40,0.126813069476807)
--cycle;

\path [draw=forestgreen4416044, fill=forestgreen4416044, opacity=0.3]
(axis cs:40,0.161882087215446)
--(axis cs:40,0.113175874036059)
--(axis cs:90,0.0975491307710998)
--(axis cs:140,0.281424360170552)
--(axis cs:190,0.391514873128113)
--(axis cs:242,0.571030817920761)
--(axis cs:292,0.647058932949559)
--(axis cs:342,0.760935756393305)
--(axis cs:392,0.85987546354705)
--(axis cs:444,0.930740946722414)
--(axis cs:494,0.941175511739009)
--(axis cs:544,0.970410475060232)
--(axis cs:594,0.983553111467559)
--(axis cs:646,0.990962875699993)
--(axis cs:696,0.990962875699993)
--(axis cs:746,0.997208546895715)
--(axis cs:796,0.997208546895715)
--(axis cs:848,0.999541505553267)
--(axis cs:898,0.999541505553267)
--(axis cs:948,1)
--(axis cs:1000,0.999796664172607)
--(axis cs:1000,1.00018801978831)
--(axis cs:1000,1.00018801978831)
--(axis cs:948,1)
--(axis cs:898,1.00038700496571)
--(axis cs:848,1.00038700496571)
--(axis cs:796,1.0011983395536)
--(axis cs:746,1.0011983395536)
--(axis cs:696,1.0002932703593)
--(axis cs:646,1.0002932703593)
--(axis cs:594,0.997317656038901)
--(axis cs:544,0.990088171757436)
--(axis cs:494,0.970267713983327)
--(axis cs:444,0.962527119906727)
--(axis cs:392,0.905392744331919)
--(axis cs:342,0.818563055770458)
--(axis cs:292,0.713026297118769)
--(axis cs:242,0.640146807764285)
--(axis cs:190,0.461457090584579)
--(axis cs:140,0.347074450275596)
--(axis cs:90,0.143600523401369)
--(axis cs:40,0.161882087215446)
--cycle;

\path [draw=crimson2143940, fill=crimson2143940, opacity=0.3]
(axis cs:40,0.0164629231002697)
--(axis cs:40,0.00353707689973028)
--(axis cs:90,0.0301397341881109)
--(axis cs:140,0.107184448834183)
--(axis cs:190,0.162113444099351)
--(axis cs:242,0.406546384724517)
--(axis cs:292,0.492036234380662)
--(axis cs:342,0.597418908858051)
--(axis cs:392,0.749975853467584)
--(axis cs:444,0.798640182094711)
--(axis cs:494,0.86378421872756)
--(axis cs:544,0.922970687036759)
--(axis cs:594,0.934896109772645)
--(axis cs:646,0.989629304048822)
--(axis cs:696,0.989965680284289)
--(axis cs:746,0.983367928302559)
--(axis cs:796,1)
--(axis cs:848,1)
--(axis cs:898,1)
--(axis cs:948,1)
--(axis cs:1000,1)
--(axis cs:1000,1)
--(axis cs:1000,1)
--(axis cs:948,1)
--(axis cs:898,1)
--(axis cs:848,1)
--(axis cs:796,1)
--(axis cs:746,0.996632071697441)
--(axis cs:696,1.00003431971571)
--(axis cs:646,1.00037069595118)
--(axis cs:594,0.965103890227355)
--(axis cs:544,0.957029312963241)
--(axis cs:494,0.90621578127244)
--(axis cs:444,0.851359817905289)
--(axis cs:392,0.810024146532416)
--(axis cs:342,0.662581091141949)
--(axis cs:292,0.567963765619338)
--(axis cs:242,0.473453615275483)
--(axis cs:190,0.217886555900649)
--(axis cs:140,0.152815551165817)
--(axis cs:90,0.0598602658118891)
--(axis cs:40,0.0164629231002697)
--cycle;

\path [draw=mediumpurple148103189, fill=mediumpurple148103189, opacity=0.3]
(axis cs:40,0.0729116582146936)
--(axis cs:40,0.0370883417853064)
--(axis cs:90,0.0357193312481523)
--(axis cs:140,0.0744169973036002)
--(axis cs:190,0.118991802195333)
--(axis cs:242,0.145435302261172)
--(axis cs:292,0.105075727192153)
--(axis cs:342,0.115168681569437)
--(axis cs:392,0.200429575586407)
--(axis cs:444,0.224519360653031)
--(axis cs:494,0.254084146461726)
--(axis cs:544,0.218576372345001)
--(axis cs:594,0.32490478130856)
--(axis cs:646,0.322664116294742)
--(axis cs:696,0.305523640345883)
--(axis cs:746,0.355128234917056)
--(axis cs:796,0.363631778982194)
--(axis cs:848,0.390479073375704)
--(axis cs:898,0.448644335035651)
--(axis cs:948,0.443274880871534)
--(axis cs:1000,0.46166279293342)
--(axis cs:1000,0.528337207066579)
--(axis cs:1000,0.528337207066579)
--(axis cs:948,0.516725119128466)
--(axis cs:898,0.521355664964349)
--(axis cs:848,0.459520926624296)
--(axis cs:796,0.436368221017806)
--(axis cs:746,0.424871765082944)
--(axis cs:696,0.374476359654117)
--(axis cs:646,0.387335883705258)
--(axis cs:594,0.39509521869144)
--(axis cs:544,0.281423627654999)
--(axis cs:494,0.315915853538274)
--(axis cs:444,0.285480639346969)
--(axis cs:392,0.259570424413593)
--(axis cs:342,0.164831318430563)
--(axis cs:292,0.154924272807847)
--(axis cs:242,0.194564697738828)
--(axis cs:190,0.161008197804667)
--(axis cs:140,0.1155830026964)
--(axis cs:90,0.0642806687518477)
--(axis cs:40,0.0729116582146936)
--cycle;

\path [draw=sienna1408675, fill=sienna1408675, opacity=0.3]
(axis cs:40,0.0299786960571008)
--(axis cs:40,0.0100213039428992)
--(axis cs:90,0.0140855886553603)
--(axis cs:140,0.0306255652285038)
--(axis cs:190,0.0349071415894801)
--(axis cs:242,0.0482168425199547)
--(axis cs:292,0.0430855830724201)
--(axis cs:342,0.0695356407381027)
--(axis cs:392,0.0525389003782694)
--(axis cs:444,0.0351849949375641)
--(axis cs:494,0.078193808218765)
--(axis cs:544,0.05096687098767)
--(axis cs:594,0.0792961960017006)
--(axis cs:646,0.0663758389450692)
--(axis cs:696,0.088850546697373)
--(axis cs:746,0.0513677698597296)
--(axis cs:796,0.0419230983019766)
--(axis cs:848,0.0786962591313169)
--(axis cs:898,0.0783125036023058)
--(axis cs:948,0.0853229194238576)
--(axis cs:1000,0.078676831848902)
--(axis cs:1000,0.121323168151098)
--(axis cs:1000,0.121323168151098)
--(axis cs:948,0.124677080576142)
--(axis cs:898,0.121687496397694)
--(axis cs:848,0.121303740868683)
--(axis cs:796,0.0780769016980234)
--(axis cs:746,0.0886322301402704)
--(axis cs:696,0.131149453302627)
--(axis cs:646,0.103624161054931)
--(axis cs:594,0.120703803998299)
--(axis cs:544,0.0890331290123301)
--(axis cs:494,0.121806191781235)
--(axis cs:444,0.0648150050624359)
--(axis cs:392,0.0874610996217306)
--(axis cs:342,0.110464359261897)
--(axis cs:292,0.0769144169275799)
--(axis cs:242,0.0817831574800453)
--(axis cs:190,0.0650928584105199)
--(axis cs:140,0.0593744347714962)
--(axis cs:90,0.0359144113446397)
--(axis cs:40,0.0299786960571008)
--cycle;

\addplot [semithick, steelblue31119180]
table {%
40 0.11
90 0.165
140 0.315
190 0.415
242 0.555
292 0.645
342 0.745
392 0.86
444 0.89
494 0.95
544 0.96
594 0.98
646 0.995
696 0.995
746 1
796 1
848 1
898 1
948 1
1000 1
};
\addlegendentry{$\dmmd$-perm}
\addplot [semithick, darkorange25512714]
table {%
40 0.105
90 0.095
140 0.205
190 0.27
242 0.385
292 0.44
342 0.535
392 0.64
444 0.745
494 0.765
544 0.835
594 0.88
646 0.915
696 0.915
746 0.96
796 0.96
848 0.99
898 0.99
948 1
1000 0.995
};
\addlegendentry{$\cmmd$}
\addplot [semithick, forestgreen4416044, dashed]
table {%
40 0.137528980625752
90 0.120574827086234
140 0.314249405223074
190 0.426485981856346
242 0.605588812842523
292 0.680042615034164
342 0.789749406081881
392 0.882634103939485
444 0.946634033314571
494 0.955721612861168
544 0.980249323408834
594 0.99043538375323
646 0.995628073029646
696 0.995628073029646
746 0.999203443224658
796 0.999203443224658
848 0.999964255259491
898 0.999964255259491
948 1
1000 0.999992341980458
};
\addlegendentry{predicted}
\addplot [semithick, crimson2143940]
table {%
40 0.01
90 0.045
140 0.13
190 0.19
242 0.44
292 0.53
342 0.63
392 0.78
444 0.825
494 0.885
544 0.94
594 0.95
646 0.995
696 0.995
746 0.99
796 1
848 1
898 1
948 1
1000 1
};
\addlegendentry{$\dmmd$-spectral}
\addplot [semithick, mediumpurple148103189]
table {%
40 0.055
90 0.05
140 0.095
190 0.14
242 0.17
292 0.13
342 0.14
392 0.23
444 0.255
494 0.285
544 0.25
594 0.36
646 0.355
696 0.34
746 0.39
796 0.4
848 0.425
898 0.485
948 0.48
1000 0.495
};
\addlegendentry{B-$\dmmd$}
\addplot [semithick, sienna1408675]
table {%
40 0.02
90 0.025
140 0.045
190 0.05
242 0.065
292 0.06
342 0.09
392 0.07
444 0.05
494 0.1
544 0.07
594 0.1
646 0.085
696 0.11
746 0.07
796 0.06
848 0.1
898 0.1
948 0.105
1000 0.1
};
\addlegendentry{L-$\dmmd$}; 
\legend{}; 
\end{axis}

\end{tikzpicture}

%% file: FinalFigs/PowerCurveAllMethods_Polynomiald_10_eps_0_3.tex
\begin{tikzpicture}

\definecolor{crimson2143940}{RGB}{214,39,40}
\definecolor{darkorange25512714}{RGB}{255,127,14}
\definecolor{darkslategray38}{RGB}{38,38,38}
\definecolor{forestgreen4416044}{RGB}{44,160,44}
\definecolor{lightgray204}{RGB}{204,204,204}
\definecolor{mediumpurple148103189}{RGB}{148,103,189}
\definecolor{sienna1408675}{RGB}{140,86,75}
\definecolor{steelblue31119180}{RGB}{31,119,180}

\begin{axis}[
axis line style={darkslategray38},
height=\figheight,
legend cell align={left},
legend style={
  fill opacity=0.4,
  draw opacity=1,
  text opacity=1,
  at={(0.01,0.95)},
  anchor=north west,
  draw=lightgray204, 
  nodes={scale=0.7, transform shape}
},
tick align=outside,
tick pos=left,
title={$(d=10, j=5,  \epsilon=0.3)$},
width=\figwidth,
x grid style={lightgray204},
xlabel=\textcolor{darkslategray38}{Sample-Size (n+m)},
xmin=27, xmax=313,
xtick style={color=darkslategray38},
y grid style={lightgray204},
ylabel=\textcolor{darkslategray38}{Power},
ymin=-0.0309943101001777, ymax=1.04170846674983,
ytick style={color=darkslategray38}
]
\path [draw=steelblue31119180, fill=steelblue31119180, opacity=0.3]
(axis cs:40,0.296086321348786)
--(axis cs:40,0.233913678651214)
--(axis cs:52,0.247049516316145)
--(axis cs:66,0.33017221834512)
--(axis cs:80,0.472659197598081)
--(axis cs:94,0.453615671299308)
--(axis cs:108,0.5974)
--(axis cs:122,0.635648881022593)
--(axis cs:134,0.665403612905391)
--(axis cs:148,0.734661740326775)
--(axis cs:162,0.677735623359501)
--(axis cs:176,0.818537006971999)
--(axis cs:190,0.856132710669202)
--(axis cs:204,0.851908348370028)
--(axis cs:216,0.840262629484927)
--(axis cs:230,0.912595420861164)
--(axis cs:244,0.900533682936929)
--(axis cs:258,0.939106310214428)
--(axis cs:272,0.929458141198685)
--(axis cs:286,0.913489113439915)
--(axis cs:300,0.934599107168739)
--(axis cs:300,0.965400892831261)
--(axis cs:300,0.965400892831261)
--(axis cs:286,0.946510886560085)
--(axis cs:272,0.960541858801314)
--(axis cs:258,0.970893689785572)
--(axis cs:244,0.939466317063071)
--(axis cs:230,0.947404579138836)
--(axis cs:216,0.889737370515073)
--(axis cs:204,0.898091651629972)
--(axis cs:190,0.903867289330798)
--(axis cs:176,0.871462993028)
--(axis cs:162,0.742264376640499)
--(axis cs:148,0.795338259673225)
--(axis cs:134,0.734596387094609)
--(axis cs:122,0.704351118977407)
--(axis cs:108,0.6626)
--(axis cs:94,0.526384328700692)
--(axis cs:80,0.53734080240192)
--(axis cs:66,0.39982778165488)
--(axis cs:52,0.312950483683855)
--(axis cs:40,0.296086321348786)
--cycle;

\path [draw=darkorange25512714, fill=darkorange25512714, opacity=0.3]
(axis cs:40,0.235414624360659)
--(axis cs:40,0.174585375639341)
--(axis cs:52,0.211913626524594)
--(axis cs:66,0.235914877084668)
--(axis cs:80,0.35524393038619)
--(axis cs:94,0.295364072121567)
--(axis cs:108,0.43557195910308)
--(axis cs:122,0.422573037793591)
--(axis cs:134,0.49983410743348)
--(axis cs:148,0.564639674326516)
--(axis cs:162,0.52013317057336)
--(axis cs:176,0.605221495503688)
--(axis cs:190,0.612896884886354)
--(axis cs:204,0.659799289779892)
--(axis cs:216,0.668315776796645)
--(axis cs:230,0.733610999139826)
--(axis cs:244,0.76601483137879)
--(axis cs:258,0.76416108796795)
--(axis cs:272,0.819658532007794)
--(axis cs:286,0.8081860111136)
--(axis cs:300,0.825966286699721)
--(axis cs:300,0.874033713300279)
--(axis cs:300,0.874033713300279)
--(axis cs:286,0.8618139888864)
--(axis cs:272,0.870341467992206)
--(axis cs:258,0.81583891203205)
--(axis cs:244,0.82398516862121)
--(axis cs:230,0.796389000860174)
--(axis cs:216,0.731684223203355)
--(axis cs:204,0.730200710220108)
--(axis cs:190,0.677103115113646)
--(axis cs:176,0.674778504496312)
--(axis cs:162,0.57986682942664)
--(axis cs:148,0.625360325673484)
--(axis cs:134,0.57016589256652)
--(axis cs:122,0.497426962206409)
--(axis cs:108,0.50442804089692)
--(axis cs:94,0.364635927878433)
--(axis cs:80,0.42475606961381)
--(axis cs:66,0.294085122915333)
--(axis cs:52,0.268086373475406)
--(axis cs:40,0.235414624360659)
--cycle;

\path [draw=forestgreen4416044, fill=forestgreen4416044, opacity=0.3]
(axis cs:40,0.347074450275596)
--(axis cs:40,0.281424360170552)
--(axis cs:52,0.341175949291156)
--(axis cs:66,0.383200032165871)
--(axis cs:80,0.578231432897049)
--(axis cs:94,0.488283483666098)
--(axis cs:108,0.685469948423024)
--(axis cs:122,0.67291306241403)
--(axis cs:134,0.760935756393305)
--(axis cs:148,0.820959735845155)
--(axis cs:162,0.776814357360355)
--(axis cs:176,0.85987546354705)
--(axis cs:190,0.863878467639374)
--(axis cs:204,0.900420522556035)
--(axis cs:216,0.903729350323494)
--(axis cs:230,0.941175511739009)
--(axis cs:244,0.95506678489173)
--(axis cs:258,0.952895999831797)
--(axis cs:272,0.973698270297606)
--(axis cs:286,0.970410475060232)
--(axis cs:300,0.975262152520286)
--(axis cs:300,0.992949249620289)
--(axis cs:300,0.992949249620289)
--(axis cs:286,0.990088171757436)
--(axis cs:272,0.992047096705784)
--(axis cs:258,0.978614494940319)
--(axis cs:244,0.980110997355926)
--(axis cs:230,0.970267713983327)
--(axis cs:216,0.941515650046391)
--(axis cs:204,0.938865248729917)
--(axis cs:190,0.908768175011978)
--(axis cs:176,0.905392744331919)
--(axis cs:162,0.832863146012774)
--(axis cs:148,0.871944246771236)
--(axis cs:134,0.818563055770458)
--(axis cs:122,0.73739737616524)
--(axis cs:108,0.749152874008371)
--(axis cs:94,0.558915355705637)
--(axis cs:80,0.647123206231149)
--(axis cs:66,0.452955134872112)
--(axis cs:52,0.409656439264845)
--(axis cs:40,0.347074450275596)
--cycle;

\path [draw=crimson2143940, fill=crimson2143940, opacity=0.3]
(axis cs:40,0.0422350929706316)
--(axis cs:40,0.0177649070293684)
--(axis cs:52,0.0418101848552548)
--(axis cs:66,0.114276712885014)
--(axis cs:80,0.19362322280869)
--(axis cs:94,0.223456627082698)
--(axis cs:108,0.408611265479547)
--(axis cs:122,0.351229783832495)
--(axis cs:134,0.446388283292875)
--(axis cs:148,0.568308175297486)
--(axis cs:162,0.519643812422717)
--(axis cs:176,0.711306316025996)
--(axis cs:190,0.715091901180449)
--(axis cs:204,0.704400255311523)
--(axis cs:216,0.804463947055193)
--(axis cs:230,0.831078722860182)
--(axis cs:244,0.81249573896648)
--(axis cs:258,0.862593541667635)
--(axis cs:272,0.893992917860874)
--(axis cs:286,0.873321958114258)
--(axis cs:300,0.9180203062454)
--(axis cs:300,0.9519796937546)
--(axis cs:300,0.9519796937546)
--(axis cs:286,0.916678041885742)
--(axis cs:272,0.936007082139126)
--(axis cs:258,0.907406458332365)
--(axis cs:244,0.86750426103352)
--(axis cs:230,0.878921277139818)
--(axis cs:216,0.855536052944807)
--(axis cs:204,0.765599744688477)
--(axis cs:190,0.774908098819551)
--(axis cs:176,0.768693683974004)
--(axis cs:162,0.590356187577283)
--(axis cs:148,0.641691824702514)
--(axis cs:134,0.513611716707125)
--(axis cs:122,0.418770216167505)
--(axis cs:108,0.481388734520453)
--(axis cs:94,0.286543372917302)
--(axis cs:80,0.24637677719131)
--(axis cs:66,0.165723287114986)
--(axis cs:52,0.0781898151447452)
--(axis cs:40,0.0422350929706316)
--cycle;

\path [draw=mediumpurple148103189, fill=mediumpurple148103189, opacity=0.3]
(axis cs:40,0.258766028140847)
--(axis cs:40,0.201233971859153)
--(axis cs:52,0.214041317857506)
--(axis cs:66,0.241846858523426)
--(axis cs:80,0.193855641143834)
--(axis cs:94,0.237943643376079)
--(axis cs:108,0.271639179041876)
--(axis cs:122,0.253648165922868)
--(axis cs:134,0.304921596743865)
--(axis cs:148,0.303554835711716)
--(axis cs:162,0.401758669283091)
--(axis cs:176,0.337730083669151)
--(axis cs:190,0.401123837658318)
--(axis cs:204,0.455448670430792)
--(axis cs:216,0.437189940566954)
--(axis cs:230,0.505855317031784)
--(axis cs:244,0.52872095129417)
--(axis cs:258,0.460748366827259)
--(axis cs:272,0.585600227108308)
--(axis cs:286,0.534211594056175)
--(axis cs:300,0.564158412772311)
--(axis cs:300,0.635841587227688)
--(axis cs:300,0.635841587227688)
--(axis cs:286,0.605788405943825)
--(axis cs:272,0.654399772891692)
--(axis cs:258,0.529251633172741)
--(axis cs:244,0.60127904870583)
--(axis cs:230,0.574144682968216)
--(axis cs:216,0.502810059433046)
--(axis cs:204,0.524551329569208)
--(axis cs:190,0.468876162341682)
--(axis cs:176,0.402269916330849)
--(axis cs:162,0.478241330716909)
--(axis cs:148,0.376445164288284)
--(axis cs:134,0.375078403256135)
--(axis cs:122,0.316351834077132)
--(axis cs:108,0.338360820958124)
--(axis cs:94,0.302056356623921)
--(axis cs:80,0.246144358856166)
--(axis cs:66,0.298153141476574)
--(axis cs:52,0.275958682142494)
--(axis cs:40,0.258766028140847)
--cycle;

\path [draw=sienna1408675, fill=sienna1408675, opacity=0.3]
(axis cs:40,0.217426891457108)
--(axis cs:40,0.162573108542892)
--(axis cs:52,0.0925813135978042)
--(axis cs:66,0.176336532397492)
--(axis cs:80,0.150208600234296)
--(axis cs:94,0.144213424034975)
--(axis cs:108,0.119003077105165)
--(axis cs:122,0.152107393452744)
--(axis cs:134,0.133730067091825)
--(axis cs:148,0.162915687197198)
--(axis cs:162,0.182159213822164)
--(axis cs:176,0.135998663823886)
--(axis cs:190,0.160033534991914)
--(axis cs:204,0.159480705828726)
--(axis cs:216,0.181826441917997)
--(axis cs:230,0.155274684610588)
--(axis cs:244,0.175232946064483)
--(axis cs:258,0.173533145729044)
--(axis cs:272,0.183917539109652)
--(axis cs:286,0.148520810057708)
--(axis cs:300,0.165895704784345)
--(axis cs:300,0.224104295215655)
--(axis cs:300,0.224104295215655)
--(axis cs:286,0.201479189942292)
--(axis cs:272,0.246082460890348)
--(axis cs:258,0.226466854270956)
--(axis cs:244,0.234767053935517)
--(axis cs:230,0.214725315389412)
--(axis cs:216,0.238173558082003)
--(axis cs:204,0.210519294171274)
--(axis cs:190,0.209966465008086)
--(axis cs:176,0.194001336176114)
--(axis cs:162,0.237840786177836)
--(axis cs:148,0.217084312802802)
--(axis cs:134,0.186269932908175)
--(axis cs:122,0.207892606547255)
--(axis cs:108,0.170996922894835)
--(axis cs:94,0.195786575965025)
--(axis cs:80,0.209791399765704)
--(axis cs:66,0.233663467602508)
--(axis cs:52,0.137418686402196)
--(axis cs:40,0.217426891457108)
--cycle;

\addplot [semithick, steelblue31119180]
table {%
40 0.265
52 0.28
66 0.365
80 0.505
94 0.49
108 0.63
122 0.67
134 0.7
148 0.765
162 0.71
176 0.845
190 0.88
204 0.875
216 0.865
230 0.93
244 0.92
258 0.955
272 0.945
286 0.93
300 0.95
};
\addlegendentry{$\dmmd$-perm}
\addplot [semithick, darkorange25512714]
table {%
40 0.205
52 0.24
66 0.265
80 0.39
94 0.33
108 0.47
122 0.46
134 0.535
148 0.595
162 0.55
176 0.64
190 0.645
204 0.695
216 0.7
230 0.765
244 0.795
258 0.79
272 0.845
286 0.835
300 0.85
};
\addlegendentry{$\cmmd$}
\addplot [semithick, forestgreen4416044, dashed]
table {%
40 0.314249405223074
52 0.375416194278
66 0.418077583518991
80 0.612677319564099
94 0.523599419685867
108 0.717311411215698
122 0.705155219289635
134 0.789749406081881
148 0.846451991308196
162 0.804838751686564
176 0.882634103939485
190 0.886323321325676
204 0.919642885642976
216 0.922622500184942
230 0.955721612861168
244 0.967588891123828
258 0.965755247386058
272 0.982872683501695
286 0.980249323408834
300 0.984105701070288
};
\addlegendentry{predicted}
\addplot [semithick, crimson2143940]
table {%
40 0.03
52 0.06
66 0.14
80 0.22
94 0.255
108 0.445
122 0.385
134 0.48
148 0.605
162 0.555
176 0.74
190 0.745
204 0.735
216 0.83
230 0.855
244 0.84
258 0.885
272 0.915
286 0.895
300 0.935
};
\addlegendentry{$\dmmd$-spectral}
\addplot [semithick, mediumpurple148103189]
table {%
40 0.23
52 0.245
66 0.27
80 0.22
94 0.27
108 0.305
122 0.285
134 0.34
148 0.34
162 0.44
176 0.37
190 0.435
204 0.49
216 0.47
230 0.54
244 0.565
258 0.495
272 0.62
286 0.57
300 0.6
};
\addlegendentry{B-$\dmmd$}
\addplot [semithick, sienna1408675]
table {%
40 0.19
52 0.115
66 0.205
80 0.18
94 0.17
108 0.145
122 0.18
134 0.16
148 0.19
162 0.21
176 0.165
190 0.185
204 0.185
216 0.21
230 0.185
244 0.205
258 0.2
272 0.215
286 0.175
300 0.195
};
\addlegendentry{L-$\dmmd$}
\end{axis}

\end{tikzpicture}

%% file: FinalFigs/PowerCurveAllMethods_Polynomiald_50_eps_0_4.tex
\begin{tikzpicture}

\definecolor{crimson2143940}{RGB}{214,39,40}
\definecolor{darkorange25512714}{RGB}{255,127,14}
\definecolor{darkslategray38}{RGB}{38,38,38}
\definecolor{forestgreen4416044}{RGB}{44,160,44}
\definecolor{lightgray204}{RGB}{204,204,204}
\definecolor{mediumpurple148103189}{RGB}{148,103,189}
\definecolor{sienna1408675}{RGB}{140,86,75}
\definecolor{steelblue31119180}{RGB}{31,119,180}

\begin{axis}[
axis line style={darkslategray38},
height=\figheight,
legend cell align={left},
legend style={
  fill opacity=0.8,
  draw opacity=1,
  text opacity=1,
  at={(0.03,0.97)},
  anchor=north west,
  draw=lightgray204
},
tick align=outside,
tick pos=left,
title={$(d=50, j=5,  \epsilon=0.4)$},
width=\figwidth,
x grid style={lightgray204},
xlabel=\textcolor{darkslategray38}{Sample-Size (n+m)},
xmin=27, xmax=313,
xtick style={color=darkslategray38},
y grid style={lightgray204},
ylabel=\textcolor{darkslategray38}{Power},
ymin=-0.0499641568717913, ymax=1.04924729430762,
ytick style={color=darkslategray38}
]
\path [draw=steelblue31119180, fill=steelblue31119180, opacity=0.3]
(axis cs:40,0.319186839573146)
--(axis cs:40,0.250813160426854)
--(axis cs:52,0.263879518400256)
--(axis cs:66,0.330679461324157)
--(axis cs:80,0.414142861589357)
--(axis cs:94,0.450477579748807)
--(axis cs:108,0.523278071946043)
--(axis cs:122,0.593804083448544)
--(axis cs:134,0.608669513648844)
--(axis cs:148,0.70078742779213)
--(axis cs:162,0.72339328591897)
--(axis cs:176,0.799011011947365)
--(axis cs:190,0.866602363901454)
--(axis cs:204,0.860097703419162)
--(axis cs:216,0.884506159461926)
--(axis cs:230,0.895307504602006)
--(axis cs:244,0.917538900378269)
--(axis cs:258,0.94632666829189)
--(axis cs:272,0.940715742931465)
--(axis cs:286,0.963719513529993)
--(axis cs:300,0.982807816743158)
--(axis cs:300,0.997192183256842)
--(axis cs:300,0.997192183256842)
--(axis cs:286,0.986280486470007)
--(axis cs:272,0.969284257068535)
--(axis cs:258,0.97367333170811)
--(axis cs:244,0.952461099621731)
--(axis cs:230,0.934692495397994)
--(axis cs:216,0.925493840538074)
--(axis cs:204,0.909902296580838)
--(axis cs:190,0.913397636098546)
--(axis cs:176,0.850988988052635)
--(axis cs:162,0.78660671408103)
--(axis cs:148,0.75921257220787)
--(axis cs:134,0.671330486351156)
--(axis cs:122,0.666195916551456)
--(axis cs:108,0.596721928053957)
--(axis cs:94,0.519522420251193)
--(axis cs:80,0.485857138410643)
--(axis cs:66,0.399320538675842)
--(axis cs:52,0.326120481599744)
--(axis cs:40,0.319186839573146)
--cycle;

\path [draw=darkorange25512714, fill=darkorange25512714, opacity=0.3]
(axis cs:40,0.237986603938313)
--(axis cs:40,0.182013396061687)
--(axis cs:52,0.173421107999015)
--(axis cs:66,0.218873202863128)
--(axis cs:80,0.282270044301894)
--(axis cs:94,0.290527300768338)
--(axis cs:108,0.33551421488497)
--(axis cs:122,0.414231438385085)
--(axis cs:134,0.458460372538845)
--(axis cs:148,0.519822103886105)
--(axis cs:162,0.540412285418085)
--(axis cs:176,0.595369531117815)
--(axis cs:190,0.65399017292857)
--(axis cs:204,0.678798878225294)
--(axis cs:216,0.662456577085377)
--(axis cs:230,0.770380338033664)
--(axis cs:244,0.740337829226437)
--(axis cs:258,0.8198914855079)
--(axis cs:272,0.805908559611347)
--(axis cs:286,0.858375997132816)
--(axis cs:300,0.879029783024489)
--(axis cs:300,0.920970216975511)
--(axis cs:300,0.920970216975511)
--(axis cs:286,0.901624002867184)
--(axis cs:272,0.854091440388652)
--(axis cs:258,0.8701085144921)
--(axis cs:244,0.799662170773563)
--(axis cs:230,0.829619661966336)
--(axis cs:216,0.727543422914623)
--(axis cs:204,0.741201121774705)
--(axis cs:190,0.71600982707143)
--(axis cs:176,0.664630468882185)
--(axis cs:162,0.609587714581915)
--(axis cs:148,0.590177896113895)
--(axis cs:134,0.531539627461155)
--(axis cs:122,0.485768561614915)
--(axis cs:108,0.40448578511503)
--(axis cs:94,0.349472699231662)
--(axis cs:80,0.347729955698106)
--(axis cs:66,0.281126797136872)
--(axis cs:52,0.226578892000985)
--(axis cs:40,0.237986603938313)
--cycle;

\path [draw=forestgreen4416044, fill=forestgreen4416044, opacity=0.3]
(axis cs:40,0.35613777375121)
--(axis cs:40,0.29000219047407)
--(axis cs:52,0.272839235903184)
--(axis cs:66,0.358068797124078)
--(axis cs:80,0.464667016249198)
--(axis cs:94,0.472585426628284)
--(axis cs:108,0.549096627581411)
--(axis cs:122,0.660108625235377)
--(axis cs:134,0.715766275892613)
--(axis cs:148,0.781978102598149)
--(axis cs:162,0.801986450833689)
--(axis cs:176,0.851677515653231)
--(axis cs:190,0.893615853794394)
--(axis cs:204,0.910160968027803)
--(axis cs:216,0.900420522556035)
--(axis cs:230,0.957180559954521)
--(axis cs:244,0.943636199174637)
--(axis cs:258,0.973698270297606)
--(axis cs:272,0.968685728078371)
--(axis cs:286,0.983553111467559)
--(axis cs:300,0.988075223447559)
--(axis cs:300,0.999283137435825)
--(axis cs:300,0.999283137435825)
--(axis cs:286,0.997317656038901)
--(axis cs:272,0.989030268987846)
--(axis cs:258,0.992047096705784)
--(axis cs:244,0.972053786216349)
--(axis cs:230,0.981550947421621)
--(axis cs:216,0.938865248729917)
--(axis cs:204,0.946624550861967)
--(axis cs:190,0.933370909642669)
--(axis cs:176,0.898438897513574)
--(axis cs:162,0.855278247408959)
--(axis cs:148,0.837487658992389)
--(axis cs:134,0.777284599749359)
--(axis cs:122,0.7253549966369)
--(axis cs:108,0.618803497685276)
--(axis cs:94,0.543287196755591)
--(axis cs:80,0.535377694297176)
--(axis cs:66,0.427128915033684)
--(axis cs:52,0.337975035137333)
--(axis cs:40,0.35613777375121)
--cycle;

\path [draw=crimson2143940, fill=crimson2143940, opacity=0.3]
(axis cs:40,0)
--(axis cs:40,0)
--(axis cs:52,0)
--(axis cs:66,0)
--(axis cs:80,0.00309149075414963)
--(axis cs:94,0.0100707502801068)
--(axis cs:108,0.0171151493993915)
--(axis cs:122,0.0261392866345199)
--(axis cs:134,0.0606558956009848)
--(axis cs:148,0.117746376387717)
--(axis cs:162,0.139763568001796)
--(axis cs:176,0.199942762934694)
--(axis cs:190,0.300715610330648)
--(axis cs:204,0.335483192499885)
--(axis cs:216,0.329336862518842)
--(axis cs:230,0.434502676157209)
--(axis cs:244,0.552592284869186)
--(axis cs:258,0.610816524459909)
--(axis cs:272,0.71125647899091)
--(axis cs:286,0.726195747308955)
--(axis cs:300,0.746769874247535)
--(axis cs:300,0.803230125752465)
--(axis cs:300,0.803230125752465)
--(axis cs:286,0.793804252691045)
--(axis cs:272,0.76874352100909)
--(axis cs:258,0.679183475540091)
--(axis cs:244,0.617407715130814)
--(axis cs:230,0.505497323842791)
--(axis cs:216,0.400663137481158)
--(axis cs:204,0.404516807500115)
--(axis cs:190,0.369284389669352)
--(axis cs:176,0.260057237065306)
--(axis cs:162,0.190236431998205)
--(axis cs:148,0.172253623612283)
--(axis cs:134,0.0993441043990152)
--(axis cs:122,0.0538607133654802)
--(axis cs:108,0.0428848506006085)
--(axis cs:94,0.0299292497198932)
--(axis cs:80,0.0169085092458504)
--(axis cs:66,0)
--(axis cs:52,0)
--(axis cs:40,0)
--cycle;

\path [draw=mediumpurple148103189, fill=mediumpurple148103189, opacity=0.3]
(axis cs:40,0.0826586381128331)
--(axis cs:40,0.0473413618871669)
--(axis cs:52,0.0513758389450692)
--(axis cs:66,0.0303846185475712)
--(axis cs:80,0.0344771136704542)
--(axis cs:94,0.0423385306387039)
--(axis cs:108,0.0950786703506016)
--(axis cs:122,0.0703996332942468)
--(axis cs:134,0.0708639737667404)
--(axis cs:148,0.101070742280631)
--(axis cs:162,0.128474175319135)
--(axis cs:176,0.10997006791859)
--(axis cs:190,0.133869031782194)
--(axis cs:204,0.133798866913815)
--(axis cs:216,0.130018056520759)
--(axis cs:230,0.138867692409586)
--(axis cs:244,0.164698556140012)
--(axis cs:258,0.160286959414107)
--(axis cs:272,0.214933448817664)
--(axis cs:286,0.150382448193681)
--(axis cs:300,0.236845899574864)
--(axis cs:300,0.303154100425136)
--(axis cs:300,0.303154100425136)
--(axis cs:286,0.209617551806319)
--(axis cs:272,0.275066551182335)
--(axis cs:258,0.209713040585893)
--(axis cs:244,0.225301443859988)
--(axis cs:230,0.191132307590414)
--(axis cs:216,0.179981943479241)
--(axis cs:204,0.186201133086185)
--(axis cs:190,0.186130968217806)
--(axis cs:176,0.16002993208141)
--(axis cs:162,0.181525824680865)
--(axis cs:148,0.148929257719369)
--(axis cs:134,0.10913602623326)
--(axis cs:122,0.109600366705753)
--(axis cs:108,0.134921329649398)
--(axis cs:94,0.0776614693612961)
--(axis cs:80,0.0655228863295458)
--(axis cs:66,0.0596153814524288)
--(axis cs:52,0.0886241610549308)
--(axis cs:40,0.0826586381128331)
--cycle;

\path [draw=sienna1408675, fill=sienna1408675, opacity=0.3]
(axis cs:40,0.0650192376637431)
--(axis cs:40,0.0349807623362569)
--(axis cs:52,0.035322125494473)
--(axis cs:66,0.0525667164595995)
--(axis cs:80,0.080105025760258)
--(axis cs:94,0.0566571539830919)
--(axis cs:108,0.0605481524013784)
--(axis cs:122,0.0554427922494033)
--(axis cs:134,0.0693741818344096)
--(axis cs:148,0.0516489782300821)
--(axis cs:162,0.0903868029123246)
--(axis cs:176,0.0429087741808845)
--(axis cs:190,0.0698537999861016)
--(axis cs:204,0.0754523818586509)
--(axis cs:216,0.0666694244498434)
--(axis cs:230,0.0510686932569355)
--(axis cs:244,0.104327085957708)
--(axis cs:258,0.074856824977179)
--(axis cs:272,0.0921320229359919)
--(axis cs:286,0.0841074175842238)
--(axis cs:300,0.0927627002538528)
--(axis cs:300,0.137237299746147)
--(axis cs:300,0.137237299746147)
--(axis cs:286,0.125892582415776)
--(axis cs:272,0.137867977064008)
--(axis cs:258,0.115143175022821)
--(axis cs:244,0.145672914042292)
--(axis cs:230,0.0889313067430645)
--(axis cs:216,0.103330575550157)
--(axis cs:204,0.114547618141349)
--(axis cs:190,0.110146200013898)
--(axis cs:176,0.0770912258191155)
--(axis cs:162,0.129613197087675)
--(axis cs:148,0.0883510217699179)
--(axis cs:134,0.11062581816559)
--(axis cs:122,0.0945572077505967)
--(axis cs:108,0.0994518475986216)
--(axis cs:94,0.0933428460169081)
--(axis cs:80,0.119894974239742)
--(axis cs:66,0.0874332835404005)
--(axis cs:52,0.064677874505527)
--(axis cs:40,0.0650192376637431)
--cycle;

\addplot [semithick, steelblue31119180]
table {%
40 0.285
52 0.295
66 0.365
80 0.45
94 0.485
108 0.56
122 0.63
134 0.64
148 0.73
162 0.755
176 0.825
190 0.89
204 0.885
216 0.905
230 0.915
244 0.935
258 0.96
272 0.955
286 0.975
300 0.99
};
\addlegendentry{$\dmmd$-perm}
\addplot [semithick, darkorange25512714]
table {%
40 0.21
52 0.2
66 0.25
80 0.315
94 0.32
108 0.37
122 0.45
134 0.495
148 0.555
162 0.575
176 0.63
190 0.685
204 0.71
216 0.695
230 0.8
244 0.77
258 0.845
272 0.83
286 0.88
300 0.9
};
\addlegendentry{$\cmmd$}
\addplot [semithick, forestgreen4416044, dashed]
table {%
40 0.32306998211264
52 0.305407135520259
66 0.392598856078881
80 0.500022355273187
94 0.507936311691937
108 0.583950062633343
122 0.692731810936138
134 0.746525437820986
148 0.809732880795269
162 0.828632349121324
176 0.875058206583402
190 0.913493381718532
204 0.928392759444885
216 0.919642885642976
230 0.969365753688071
244 0.957844992695493
258 0.982872683501695
272 0.978857998533108
286 0.99043538375323
300 0.993679180441692
};
\addlegendentry{predicted}
\addplot [semithick, crimson2143940]
table {%
40 0
52 0
66 0
80 0.01
94 0.02
108 0.03
122 0.04
134 0.08
148 0.145
162 0.165
176 0.23
190 0.335
204 0.37
216 0.365
230 0.47
244 0.585
258 0.645
272 0.74
286 0.76
300 0.775
};
\addlegendentry{$\dmmd$-spectral}
\addplot [semithick, mediumpurple148103189]
table {%
40 0.065
52 0.07
66 0.045
80 0.05
94 0.06
108 0.115
122 0.09
134 0.09
148 0.125
162 0.155
176 0.135
190 0.16
204 0.16
216 0.155
230 0.165
244 0.195
258 0.185
272 0.245
286 0.18
300 0.27
};
\addlegendentry{B-$\dmmd$}
\addplot [semithick, sienna1408675]
table {%
40 0.05
52 0.05
66 0.07
80 0.1
94 0.075
108 0.08
122 0.075
134 0.09
148 0.07
162 0.11
176 0.06
190 0.09
204 0.095
216 0.085
230 0.07
244 0.125
258 0.095
272 0.115
286 0.105
300 0.115
};
\addlegendentry{L-$\dmmd$}; 
\legend{}; 
\end{axis}

\end{tikzpicture}

%% file: FinalFigs/PowerCurveAllMethods_Polynomiald_100_eps_0_5.tex
\begin{tikzpicture}

\definecolor{crimson2143940}{RGB}{214,39,40}
\definecolor{darkorange25512714}{RGB}{255,127,14}
\definecolor{darkslategray38}{RGB}{38,38,38}
\definecolor{forestgreen4416044}{RGB}{44,160,44}
\definecolor{lightgray204}{RGB}{204,204,204}
\definecolor{mediumpurple148103189}{RGB}{148,103,189}
\definecolor{sienna1408675}{RGB}{140,86,75}
\definecolor{steelblue31119180}{RGB}{31,119,180}

\begin{axis}[
axis line style={darkslategray38},
height=\figheight,
legend cell align={left},
legend style={
  fill opacity=0.8,
  draw opacity=1,
  text opacity=1,
  at={(0.91,0.5)},
  anchor=east,
  draw=lightgray204
},
tick align=outside,
tick pos=left,
title={$(d=100, j=10,  \epsilon=0.5)$},
width=\figwidth,
x grid style={lightgray204},
xlabel=\textcolor{darkslategray38}{Sample-Size (n+m)},
xmin=27, xmax=313,
xtick style={color=darkslategray38},
y grid style={lightgray204},
ylabel=\textcolor{darkslategray38}{Power},
ymin=-0.0500532498693353, ymax=1.05111824725604,
ytick style={color=darkslategray38}
]
\path [draw=steelblue31119180, fill=steelblue31119180, opacity=0.3]
(axis cs:40,0.582020188748844)
--(axis cs:40,0.507979811251157)
--(axis cs:52,0.708637641989162)
--(axis cs:66,0.845963569316556)
--(axis cs:80,0.891276368541333)
--(axis cs:94,0.945390948867226)
--(axis cs:108,0.96976391188002)
--(axis cs:122,0.990412040213777)
--(axis cs:134,1)
--(axis cs:148,1)
--(axis cs:162,1)
--(axis cs:176,1)
--(axis cs:190,1)
--(axis cs:204,1)
--(axis cs:216,1)
--(axis cs:230,1)
--(axis cs:244,1)
--(axis cs:258,1)
--(axis cs:272,1)
--(axis cs:286,1)
--(axis cs:300,1)
--(axis cs:300,1)
--(axis cs:300,1)
--(axis cs:286,1)
--(axis cs:272,1)
--(axis cs:258,1)
--(axis cs:244,1)
--(axis cs:230,1)
--(axis cs:216,1)
--(axis cs:204,1)
--(axis cs:190,1)
--(axis cs:176,1)
--(axis cs:162,1)
--(axis cs:148,1)
--(axis cs:134,1)
--(axis cs:122,0.999587959786223)
--(axis cs:108,0.99023608811998)
--(axis cs:94,0.974609051132774)
--(axis cs:80,0.928723631458667)
--(axis cs:66,0.894036430683444)
--(axis cs:52,0.771362358010838)
--(axis cs:40,0.582020188748844)
--cycle;

\path [draw=darkorange25512714, fill=darkorange25512714, opacity=0.3]
(axis cs:40,0.420039040726024)
--(axis cs:40,0.349960959273976)
--(axis cs:52,0.516130037274895)
--(axis cs:66,0.6791)
--(axis cs:80,0.720236442164956)
--(axis cs:94,0.819431464648909)
--(axis cs:108,0.908233683320419)
--(axis cs:122,0.911622449156648)
--(axis cs:134,0.91761841851269)
--(axis cs:148,0.964328659877972)
--(axis cs:162,0.989815949942371)
--(axis cs:176,0.957790270273262)
--(axis cs:190,1)
--(axis cs:204,1)
--(axis cs:216,1)
--(axis cs:230,1)
--(axis cs:244,1)
--(axis cs:258,1)
--(axis cs:272,1)
--(axis cs:286,1)
--(axis cs:300,1)
--(axis cs:300,1)
--(axis cs:300,1)
--(axis cs:286,1)
--(axis cs:272,1)
--(axis cs:258,1)
--(axis cs:244,1)
--(axis cs:230,1)
--(axis cs:216,1)
--(axis cs:204,1)
--(axis cs:190,1)
--(axis cs:176,0.982209729726738)
--(axis cs:162,1.00018405005763)
--(axis cs:148,0.985671340122028)
--(axis cs:134,0.95238158148731)
--(axis cs:122,0.948377550843352)
--(axis cs:108,0.941766316679581)
--(axis cs:94,0.870568535351091)
--(axis cs:80,0.779763557835044)
--(axis cs:66,0.7409)
--(axis cs:52,0.583869962725105)
--(axis cs:40,0.420039040726024)
--cycle;

\path [draw=forestgreen4416044, fill=forestgreen4416044, opacity=0.3]
(axis cs:40,0.640146807764285)
--(axis cs:40,0.571030817920761)
--(axis cs:52,0.776814357360355)
--(axis cs:66,0.910160968027803)
--(axis cs:80,0.933438921687398)
--(axis cs:94,0.973698270297606)
--(axis cs:108,0.992656654456372)
--(axis cs:122,0.993435996694691)
--(axis cs:134,0.994171170107054)
--(axis cs:148,0.998546107066721)
--(axis cs:162,0.999796664172607)
--(axis cs:176,0.998138969045341)
--(axis cs:190,1)
--(axis cs:204,1)
--(axis cs:216,1)
--(axis cs:230,1)
--(axis cs:244,1)
--(axis cs:258,1)
--(axis cs:272,1)
--(axis cs:286,1)
--(axis cs:300,1)
--(axis cs:300,1)
--(axis cs:300,1)
--(axis cs:286,1)
--(axis cs:272,1)
--(axis cs:258,1)
--(axis cs:244,1)
--(axis cs:230,1)
--(axis cs:216,1)
--(axis cs:204,1)
--(axis cs:190,1)
--(axis cs:176,1.00102675320798)
--(axis cs:162,1.00018801978831)
--(axis cs:148,1.00089976761037)
--(axis cs:134,1.00106499738671)
--(axis cs:122,1.00093085180097)
--(axis cs:108,1.00075818448694)
--(axis cs:94,0.992047096705784)
--(axis cs:80,0.964552531983674)
--(axis cs:66,0.946624550861967)
--(axis cs:52,0.832863146012774)
--(axis cs:40,0.640146807764285)
--cycle;

\path [draw=crimson2143940, fill=crimson2143940, opacity=0.3]
(axis cs:40,0)
--(axis cs:40,0)
--(axis cs:52,0)
--(axis cs:66,0)
--(axis cs:80,0.00309275742426835)
--(axis cs:94,0.0561127688635946)
--(axis cs:108,0.107378232275085)
--(axis cs:122,0.222821629935623)
--(axis cs:134,0.338533996520054)
--(axis cs:148,0.556264716171344)
--(axis cs:162,0.690266012712722)
--(axis cs:176,0.826016724577323)
--(axis cs:190,0.923576617279013)
--(axis cs:204,0.928469743649901)
--(axis cs:216,0.990222513736283)
--(axis cs:230,1)
--(axis cs:244,0.990102041241497)
--(axis cs:258,1)
--(axis cs:272,1)
--(axis cs:286,1)
--(axis cs:300,1)
--(axis cs:300,1)
--(axis cs:300,1)
--(axis cs:286,1)
--(axis cs:272,1)
--(axis cs:258,1)
--(axis cs:244,0.999897958758503)
--(axis cs:230,1)
--(axis cs:216,0.999777486263717)
--(axis cs:204,0.961530256350099)
--(axis cs:190,0.956423382720986)
--(axis cs:176,0.873983275422677)
--(axis cs:162,0.749733987287278)
--(axis cs:148,0.623735283828656)
--(axis cs:134,0.401466003479946)
--(axis cs:122,0.287178370064377)
--(axis cs:108,0.152621767724915)
--(axis cs:94,0.0938872311364054)
--(axis cs:80,0.0169072425757316)
--(axis cs:66,0)
--(axis cs:52,0)
--(axis cs:40,0)
--cycle;

\path [draw=mediumpurple148103189, fill=mediumpurple148103189, opacity=0.3]
(axis cs:40,0.0667593071157492)
--(axis cs:40,0.0332406928842509)
--(axis cs:52,0.0298182881400021)
--(axis cs:66,0.0215632035067878)
--(axis cs:80,0.0933248990775129)
--(axis cs:94,0.0639253707031417)
--(axis cs:108,0.101716368410405)
--(axis cs:122,0.0865882641608957)
--(axis cs:134,0.169077880085609)
--(axis cs:148,0.244396905793695)
--(axis cs:162,0.203805288909817)
--(axis cs:176,0.260777383209345)
--(axis cs:190,0.277317062937979)
--(axis cs:204,0.416750789483057)
--(axis cs:216,0.422110269459906)
--(axis cs:230,0.419204085428641)
--(axis cs:244,0.567408292849254)
--(axis cs:258,0.511496502048294)
--(axis cs:272,0.646412139469743)
--(axis cs:286,0.612751056218847)
--(axis cs:300,0.725390299309179)
--(axis cs:300,0.784609700690821)
--(axis cs:300,0.784609700690821)
--(axis cs:286,0.677248943781154)
--(axis cs:272,0.713587860530257)
--(axis cs:258,0.578503497951707)
--(axis cs:244,0.632591707150746)
--(axis cs:230,0.490795914571359)
--(axis cs:216,0.487889730540094)
--(axis cs:204,0.483249210516943)
--(axis cs:190,0.342682937062021)
--(axis cs:176,0.329222616790655)
--(axis cs:162,0.266194711090183)
--(axis cs:148,0.305603094206305)
--(axis cs:134,0.230922119914391)
--(axis cs:122,0.133411735839104)
--(axis cs:108,0.148283631589595)
--(axis cs:94,0.106074629296858)
--(axis cs:80,0.136675100922487)
--(axis cs:66,0.0484367964932122)
--(axis cs:52,0.0601817118599979)
--(axis cs:40,0.0667593071157492)
--cycle;

\path [draw=sienna1408675, fill=sienna1408675, opacity=0.3]
(axis cs:40,0.0593759999652198)
--(axis cs:40,0.0306240000347802)
--(axis cs:52,0.105577175122439)
--(axis cs:66,0.0766779504421585)
--(axis cs:80,0.0570506963923388)
--(axis cs:94,0.0689435431518026)
--(axis cs:108,0.0929889942301584)
--(axis cs:122,0.0794413035432691)
--(axis cs:134,0.124207026150519)
--(axis cs:148,0.141616069687233)
--(axis cs:162,0.116799191824421)
--(axis cs:176,0.162103270801042)
--(axis cs:190,0.134251517034979)
--(axis cs:204,0.112071428849577)
--(axis cs:216,0.161818844949151)
--(axis cs:230,0.168557840802515)
--(axis cs:244,0.168122604757901)
--(axis cs:258,0.122065310117641)
--(axis cs:272,0.149722786941595)
--(axis cs:286,0.235516625447551)
--(axis cs:300,0.23899218848419)
--(axis cs:300,0.30100781151581)
--(axis cs:300,0.30100781151581)
--(axis cs:286,0.294483374552449)
--(axis cs:272,0.200277213058405)
--(axis cs:258,0.167934689882359)
--(axis cs:244,0.221877395242099)
--(axis cs:230,0.231442159197485)
--(axis cs:216,0.218181155050849)
--(axis cs:204,0.157928571150423)
--(axis cs:190,0.185748482965021)
--(axis cs:176,0.217896729198958)
--(axis cs:162,0.163200808175579)
--(axis cs:148,0.198383930312767)
--(axis cs:134,0.175792973849481)
--(axis cs:122,0.120558696456731)
--(axis cs:108,0.137011005769842)
--(axis cs:94,0.111056456848197)
--(axis cs:80,0.0929493036076612)
--(axis cs:66,0.113322049557842)
--(axis cs:52,0.154422824877561)
--(axis cs:40,0.0593759999652198)
--cycle;

\addplot [semithick, steelblue31119180]
table {%
40 0.545
52 0.74
66 0.87
80 0.91
94 0.96
108 0.98
122 0.995
134 1
148 1
162 1
176 1
190 1
204 1
216 1
230 1
244 1
258 1
272 1
286 1
300 1
};
\addlegendentry{$\dmmd$-perm}
\addplot [semithick, darkorange25512714]
table {%
40 0.385
52 0.55
66 0.71
80 0.75
94 0.845
108 0.925
122 0.93
134 0.935
148 0.975
162 0.995
176 0.97
190 1
204 1
216 1
230 1
244 1
258 1
272 1
286 1
300 1
};
\addlegendentry{$\cmmd$}
\addplot [semithick, forestgreen4416044, dashed]
table {%
40 0.605588812842523
52 0.804838751686564
66 0.928392759444885
80 0.948995726835536
94 0.982872683501695
108 0.996707419471656
122 0.997183424247829
134 0.99761808374688
148 0.999722937338545
162 0.999992341980458
176 0.999582861126661
190 1
204 1
216 1
230 1
244 1
258 1
272 1
286 1
300 1
};
\addlegendentry{predicted}
\addplot [semithick, crimson2143940]
table {%
40 0
52 0
66 0
80 0.01
94 0.075
108 0.13
122 0.255
134 0.37
148 0.59
162 0.72
176 0.85
190 0.94
204 0.945
216 0.995
230 1
244 0.995
258 1
272 1
286 1
300 1
};
\addlegendentry{$\dmmd$-spectral}
\addplot [semithick, mediumpurple148103189]
table {%
40 0.05
52 0.045
66 0.035
80 0.115
94 0.085
108 0.125
122 0.11
134 0.2
148 0.275
162 0.235
176 0.295
190 0.31
204 0.45
216 0.455
230 0.455
244 0.6
258 0.545
272 0.68
286 0.645
300 0.755
};
\addlegendentry{B-$\dmmd$}
\addplot [semithick, sienna1408675]
table {%
40 0.045
52 0.13
66 0.095
80 0.075
94 0.09
108 0.115
122 0.1
134 0.15
148 0.17
162 0.14
176 0.19
190 0.16
204 0.135
216 0.19
230 0.2
244 0.195
258 0.145
272 0.175
286 0.265
300 0.27
};
\addlegendentry{L-$\dmmd$}; 
\legend{}; 
\end{axis}

\end{tikzpicture}

%% file: FinalFigs/Type_I_comparison_d_10_seed_5541_.tex
\begin{tikzpicture}

\definecolor{darkorange25512714}{RGB}{255,127,14}
\definecolor{darkslategray38}{RGB}{38,38,38}
\definecolor{forestgreen4416044}{RGB}{44,160,44}
\definecolor{lightgray204}{RGB}{204,204,204}
\definecolor{steelblue31119180}{RGB}{31,119,180}

\begin{axis}[
axis line style={darkslategray38},
height=\figheight,
legend cell align={left},
legend style={fill opacity=0.8, draw opacity=1, text opacity=1, draw=none},
tick align=outside,
tick pos=left,
title={GMD Source: (d=10, 200 trials)},
width=\figwidth,
x grid style={lightgray204},
xlabel=\textcolor{darkslategray38}{Sample-Size (n+m)},
xmin=32, xmax=208,
xtick style={color=darkslategray38},
y grid style={lightgray204},
ylabel=\textcolor{darkslategray38}{Type-I error},
ymin=6.93889390390723e-18, ymax=0.77,
ytick style={color=darkslategray38}
]
\addplot [semithick, steelblue31119180]
table {%
40 0.065
56 0.09
74 0.08
92 0.04
110 0.045
128 0.07
146 0.035
164 0.06
182 0.07
200 0.055
};
\addlegendentry{x-MMD}
\addplot [semithick, darkorange25512714]
table {%
40 0.43
56 0.29
74 0.21
92 0.15
110 0.1
128 0.12
146 0.095
164 0.115
182 0.08
200 0.085
};
\addlegendentry{ME}
\addplot [semithick, forestgreen4416044]
table {%
40 0.515
56 0.735
74 0.53
92 0.325
110 0.24
128 0.15
146 0.13
164 0.115
182 0.1
200 0.11
};
\addlegendentry{SCF}
\addplot [semithick, black, opacity=0.5, dashed, forget plot]
table {%
40 0.05
56 0.05
74 0.05
92 0.05
110 0.05
128 0.05
146 0.05
164 0.05
182 0.05
200 0.05
};
\end{axis}

\end{tikzpicture}

%% file: FinalFigs/Power_comparison_my_1_0d_10_seed_9876_.tex
\begin{tikzpicture}

\definecolor{darkorange25512714}{RGB}{255,127,14}
\definecolor{darkslategray38}{RGB}{38,38,38}
\definecolor{forestgreen4416044}{RGB}{44,160,44}
\definecolor{lightgray204}{RGB}{204,204,204}
\definecolor{steelblue31119180}{RGB}{31,119,180}

\begin{axis}[
axis line style={darkslategray38},
height=\figheight,
legend cell align={left},
legend style={
  fill opacity=0.8,
  draw opacity=1,
  text opacity=1,
  at={(0.97,0.03)},
  anchor=south east,
  draw=none
},
tick align=outside,
tick pos=left,
title={GMD Source: (\(\displaystyle \epsilon\)= 1.0, d=10)},
width=\figwidth,
x grid style={lightgray204},
xlabel=\textcolor{darkslategray38}{Sample-Size (n+m)},
xmin=22, xmax=418,
xtick style={color=darkslategray38},
y grid style={lightgray204},
ylabel=\textcolor{darkslategray38}{Power},
ymin=0.38575, ymax=1.02925,
ytick style={color=darkslategray38}
]
\addplot [semithick, steelblue31119180]
table {%
40 0.415
80 0.76
120 0.905
160 0.94
200 0.995
240 1
280 1
320 1
360 1
400 1
};
\addlegendentry{x-MMD}
\addplot [semithick, darkorange25512714]
table {%
40 0.555
80 0.53
120 0.56
160 0.64
200 0.705
240 0.79
280 0.845
320 0.875
360 0.935
400 0.945
};
\addlegendentry{ME}
\addplot [semithick, forestgreen4416044]
table {%
40 0.52
80 0.515
120 0.645
160 0.65
200 0.755
240 0.775
280 0.845
320 0.785
360 0.855
400 0.9
};
\addlegendentry{SCF}
\end{axis}

\end{tikzpicture}

%% file: FinalFigs/Type_I_comparison_d_100_seed_1417_.tex
\begin{tikzpicture}

\definecolor{darkorange25512714}{RGB}{255,127,14}
\definecolor{darkslategray38}{RGB}{38,38,38}
\definecolor{forestgreen4416044}{RGB}{44,160,44}
\definecolor{lightgray204}{RGB}{204,204,204}
\definecolor{steelblue31119180}{RGB}{31,119,180}

\begin{axis}[
axis line style={darkslategray38},
height=\figheight,
legend cell align={left},
legend style={fill opacity=0.8, draw opacity=1, text opacity=1, draw=none},
tick align=outside,
tick pos=left,
title={GMD Source: (d=100, 200 trials)},
width=\figwidth,
x grid style={lightgray204},
xlabel=\textcolor{darkslategray38}{Sample-Size (n+m)},
xmin=35.5, xmax=105,
xtick style={color=darkslategray38},
y grid style={lightgray204},
ylabel=\textcolor{darkslategray38}{Type-I error},
ymin=0.0085, ymax=0.9215,
ytick style={color=darkslategray38}
]
\addplot [semithick, steelblue31119180]
table {%
40 0.08
46 0.095
52 0.075
60 0.05
66 0.075
72 0.05
80 0.055
86 0.05
92 0.06
100 0.055
};
\addlegendentry{x-MMD}
\addplot [semithick, darkorange25512714]
table {%
40 0.425
46 0.29
52 0.305
60 0.235
66 0.225
72 0.24
80 0.175
86 0.15
92 0.185
100 0.15
};
\addlegendentry{ME}
\addplot [semithick, forestgreen4416044]
table {%
40 0.605
46 0.82
52 0.88
60 0.79
66 0.655
72 0.565
80 0.5
86 0.43
92 0.435
100 0.42
};
\addlegendentry{SCF}
\addplot [semithick, black, opacity=0.5, dashed, forget plot]
table {%
40 0.05
46 0.05
52 0.05
60 0.05
66 0.05
72 0.05
80 0.05
86 0.05
92 0.05
100 0.05
};
\end{axis}

\end{tikzpicture}

%% file: FinalFigs/Power_FR_d_10_seed_1484.tex
\begin{tikzpicture}

\definecolor{darkorange25512714}{RGB}{255,127,14}
\definecolor{darkslategray38}{RGB}{38,38,38}
\definecolor{lightgray204}{RGB}{204,204,204}
\definecolor{steelblue31119180}{RGB}{31,119,180}

\begin{axis}[
axis line style={darkslategray38},
height=\figheight,
legend cell align={left},
legend style={
  fill opacity=0.8,
  draw opacity=1,
  text opacity=1,
  at={(0.03,0.97)},
  anchor=north west,
  draw=none
},
tick align=outside,
tick pos=left,
title={GMD Source $(\epsilon=1.2, d=10)$},
width=\figwidth,
x grid style={lightgray204},
xlabel=\textcolor{darkslategray38}{Sample-Size (n+m)},
xmin=10.5, xmax=210.5,
xtick style={color=darkslategray38},
y grid style={lightgray204},
ylabel=\textcolor{darkslategray38}{Power},
ymin=0.01825, ymax=1.04675,
ytick style={color=darkslategray38}
]
\addplot [semithick, steelblue31119180]
table {%
20 0.335
40 0.56
60 0.78
80 0.91
100 0.93
120 0.98
140 1
160 1
180 0.995
200 1
};
\addlegendentry{$\cmmd$}
\addplot [semithick, darkorange25512714]
table {%
20 0.065
40 0.11
60 0.185
80 0.215
100 0.24
120 0.275
140 0.345
160 0.38
180 0.38
200 0.425
};
\addlegendentry{$\mathrm{FR}$}
\end{axis}

\end{tikzpicture}

%% file: FinalFigs/Power_FR_d_100_seed_2404.tex
\begin{tikzpicture}

\definecolor{darkorange25512714}{RGB}{255,127,14}
\definecolor{darkslategray38}{RGB}{38,38,38}
\definecolor{lightgray204}{RGB}{204,204,204}
\definecolor{steelblue31119180}{RGB}{31,119,180}

\begin{axis}[
axis line style={darkslategray38},
height=\figheight,
legend cell align={left},
legend style={
  fill opacity=0.8,
  draw opacity=1,
  text opacity=1,
  at={(0.03,0.97)},
  anchor=north west,
  draw=none
},
tick align=outside,
tick pos=left,
title={GMD Source $(\epsilon=2.0, d=100)$},
width=\figwidth,
x grid style={lightgray204},
xlabel=\textcolor{darkslategray38}{Sample-Size $(n+m)$},
xmin=10.5, xmax=210.5,
xtick style={color=darkslategray38},
y grid style={lightgray204},
ylabel=\textcolor{darkslategray38}{Power},
ymin=-0.00275, ymax=1.04775,
ytick style={color=darkslategray38}
]
\addplot [semithick, steelblue31119180]
table {%
20 0.335
40 0.58
60 0.81
80 0.955
100 1
120 0.995
140 1
160 1
180 1
200 1
};
\addlegendentry{$\cmmd$}
\addplot [semithick, darkorange25512714]
table {%
20 0.045
40 0.09
60 0.125
80 0.17
100 0.26
120 0.32
140 0.375
160 0.385
180 0.455
200 0.515
};
\addlegendentry{$\mathrm{FR}$}
\end{axis}

\end{tikzpicture}